\documentclass[11pt]{article}
\usepackage[T1]{fontenc}
\usepackage[utf8]{inputenc}
\usepackage{lmodern}
\usepackage[margin=.9in]{geometry}
\usepackage{amsmath,amssymb,amsthm,mathtools,mathrsfs}
\usepackage{booktabs,array,longtable,tabularx,multirow}
\usepackage{microtype,graphicx,adjustbox,tikz,xcolor,float,needspace,placeins}
\usepackage{enumitem,fancyhdr,xurl,aliascnt,multicol}
\usepackage[numbers,sort&compress]{natbib}
\usepackage[colorlinks=true,allcolors=blue]{hyperref}
\usepackage{bookmark}
\usepackage[nameinlink,noabbrev,capitalise]{cleveref}
\usetikzlibrary{arrows.meta,positioning,fit,backgrounds,calc,decorations.pathreplacing,matrix}
\newtheoremstyle{manuscript}{7pt}{7pt}{\itshape}{}{\bfseries}{.}{.5em}{}

\theoremstyle{manuscript}
\newtheorem{theorem}{Theorem}[section]
\newaliascnt{lemma}{theorem}\newtheorem{lemma}[lemma]{Lemma}\aliascntresetthe{lemma}
\newaliascnt{proposition}{theorem}\newtheorem{proposition}[proposition]{Proposition}\aliascntresetthe{proposition}
\newaliascnt{corollary}{theorem}\newtheorem{corollary}[corollary]{Corollary}\aliascntresetthe{corollary}
\theoremstyle{remark}
\crefname{lemma}{Lemma}{Lemmas}\crefname{proposition}{Proposition}{Propositions}
\crefname{corollary}{Corollary}{Corollaries}\crefname{definition}{Definition}{Definitions}

\newtheoremstyle{manuscriptdef}{7pt}{7pt}{\normalfont}{}{\bfseries}{.}{.5em}{}
\theoremstyle{manuscriptdef}
\newaliascnt{definition}{theorem}\newtheorem{definition}[definition]{Definition}\aliascntresetthe{definition}

\numberwithin{equation}{section}
\newcommand{\Be}[1]{\textnormal{\textsc{Be}}\!\left[#1\right]}
\newcommand{\Id}{\mathcal{I}}
\newcommand{\oracle}{\widehat{\boldsymbol{O}}}
\newcommand{\OTuple}{\boldsymbol{O}}
\newcommand{\catalyst}{\vec}
\newcommand{\vect}[1]{\boldsymbol{#1}}
\newcommand{\Bz}{\vect{z}}
\newcommand{\Bone}{\vect{1}}
\newcommand{\ket}[1]{\lvert#1\rangle}
\newcommand{\bra}[1]{\langle#1\rvert}
\newcommand{\ketbra}[1]{\lvert#1\rangle\langle#1\rvert}
\newcommand{\norm}[1]{\left\lVert#1\right\rVert}
\newcommand{\nuc}[1]{\left\lVert#1\right\rVert_*}
\newcommand{\mathval}[1]{\mathcal{#1}}
\newcommand{\ii}{\mathrm i}
\newcommand{\module}[1]{\IIRTag{#1}}
\newcommand{\modref}[2]{\hyperref[#1]{\texttt{#2}}}
\newcommand{\kgNorm}[1]{\norm{#1}}
\newcommand{\appRe}{\operatorname{Re}}
\newcommand{\kgRe}{\operatorname{Re}}
\newcommand{\appT}{\mathsf T}
\newcommand{\kgSens}{\mathcal C_{\mathrm{sens}}}
\newcommand{\kgFb}{\mathcal C_{\mathrm{fb}}}
\NewDocumentCommand{\SOSPhaseDiagram}{O{\linewidth}m}{\IIRDiagram[#1]{#2}}
\DeclareMathOperator{\artanh}{artanh}
\DeclareMathOperator{\sech}{sech}
\DeclareMathOperator{\sgn}{sgn}
\DeclareMathOperator{\diag}{diag}
\DeclareMathOperator{\erf}{erf}
\DeclareMathOperator{\ran}{ran}
\DeclareMathOperator{\polylog}{polylog}
\newcommand{\cK}{\mathcal K}
\newcommand{\cL}{\mathcal L}
\newcommand{\cP}{\mathcal P}
\newcommand{\cM}{\mathcal M}

\newsavebox{\TDfigurebox}
\providecommand{\TDfit}[1]{\sbox{\TDfigurebox}{#1}\ifdim\wd\TDfigurebox>\linewidth \resizebox{\linewidth}{!}{\usebox{\TDfigurebox}}\else\usebox{\TDfigurebox}\fi}

\makeatletter
\def\input@path{{libraries/IIR/}}
\makeatother
\usepackage{iir-tikz}
\renewcommand{\IIRRoot}{libraries/IIR/}
\NewDocumentEnvironment{mqspdiagram}{O{}}{\begin{tikzpicture}[iir,#1]}{\end{tikzpicture}}
\tikzset{mqsp/system/.style={iir/system}}
\NewDocumentCommand{\MQSPSystem}{O{} m m m}{\node[iir/system,#1] (#2) at #3 {#4}; \coordinate (#2-pi) at ($(#2.west)+(0,.20)$); \coordinate (#2-po) at ($(#2.east)+(0,.20)$); \coordinate (#2-li) at ($(#2.west)+(0,-.20)$); \coordinate (#2-lo) at ($(#2.east)+(0,-.20)$);}
\NewDocumentCommand{\MQSPGate}{O{} m m m}{\node[iir/gate,#1] (#2) at #3 {#4};}
\NewDocumentCommand{\MQSPPublicWire}{O{} m}{\draw[iir/public,#1] #2;}
\NewDocumentCommand{\MQSPPrivateWire}{O{} m}{\draw[iir/private,#1] #2;}
\NewDocumentCommand{\MQSPTaggedSystem}{O{} m m m m}{\MQSPSystem[#1]{#2}{#3}{\IIRTag{#4}\IfBlankTF{#5}{}{\\#5}}}
\NewDocumentCommand{\MQSPTaggedGate}{O{} m m m m}{\MQSPGate[#1]{#2}{#3}{\IIRTag{#4}\IfBlankTF{#5}{}{\\#5}}}
\NewDocumentCommand{\MQSPTaggedFeedback}{O{} m m m m}{\MQSPTaggedGate[#1]{#2-q}{(#2.center |- 0,#3)}{#4}{#5}\path let \p1=(#2.east),\p2=(#2-q.east) in coordinate (#2-return-r) at ({max(\x1,\x2)+7pt},\y2); \path let \p1=(#2.west),\p2=(#2-q.west) in coordinate (#2-return-l) at ({min(\x1,\x2)-7pt},\y2); \MQSPPrivateWire{(#2-lo) -- (#2-return-r |- #2-lo) -- (#2-return-r) -- (#2-q.east)}\MQSPPrivateWire{(#2-q.west) -- (#2-return-l) -- (#2-return-l |- #2-li) -- (#2-li)}}

\newcommand{\ModularityModuleName}[2]{\expandafter\def\csname#1\endcsname{\IIRTag{#2}}}

\ModularityModuleName{IntroIIR}{IntroIIR}

\ModularityModuleName{EWire}{Wire}
\ModularityModuleName{EQuery}{Query}
\ModularityModuleName{ESeries}{Series}
\ModularityModuleName{EDirectSum}{DirectSum}
\ModularityModuleName{ESpectator}{Spectator}
\ModularityModuleName{EClose}{Close}
\ModularityModuleName{ESubstitute}{Substitute}
\ModularityModuleName{EInverse}{Inverse}
\ModularityModuleName{ESelect}{Project}
\ModularityModuleName{EDelay}{Delay}
\ModularityModuleName{EBuffered}{BufferedSeries}
\ModularityModuleName{FPAATapIIR}{FPAA}
\ModularityModuleName{GroverIIR}{Grover}
\ModularityModuleName{FPAAIIR}{FPAA}
\ModularityModuleName{OAAIIR}{OAA}
\ModularityModuleName{SignIIR}{Sign}
\ModularityModuleName{ThresholdIIR}{Threshold}
\ModularityModuleName{CayleyIIR}{Cayley}
\ModularityModuleName{ExpIIR}{Exp}
\ModularityModuleName{HamSimIIR}{HamSim}
\ModularityModuleName{ReflectTwoIIR}{Chebyshev2}
\ModularityModuleName{ReflectionWalkIIR}{ReflectionWalk}
\ModularityModuleName{StatePrepIIR}{StatePrep}
\ModularityModuleName{PrepReturnIIR}{PreparationQuery}
\ModularityModuleName{AverageCostIIR}{AverageCost}
\ModularityModuleName{WeightedCayleyIIR}{WeightedCayley}
\ModularityModuleName{EvenIIR}{Even}
\ModularityModuleName{OddIIR}{Odd}
\ModularityModuleName{PairsIIR}{Pairs}
\ModularityModuleName{LiftIIR}{Lift}
\IIRName{Wire}{Wire}\IIRName{Query}{Query}
\IIRName{Series}{Series}\IIRName{DirectSum}{DirectSum}
\IIRName{Spectator}{Spectator}\IIRName{Close}{Close}
\IIRName{Substitute}{Substitute}\IIRName{Inverse}{Inverse}
\IIRName{Project}{Project}\IIRName{Delay}{Delay}
\IIRName{BufferedSeries}{BufferedSeries}
\IIRName{TapFPAA}{FPAA}\IIRName{Lift}{Quditization}
\IIRName{ReflectionWalk}{ReflectionWalk}
\IIRName{Chebyshev2}{Chebyshev2}
\IIRName{StatePrep}{StatePrep}\IIRName{PrepReturn}{PreparationQuery}
\newcommand{\RuleRow}[2]{\par\medskip\noindent \begin{minipage}[c]{.50\linewidth}#1\end{minipage}\hfill \begin{minipage}[c]{.47\linewidth}\centering\IIRDiagram{#2}\end{minipage} \par\medskip}

\newcolumntype{Y}{>{\centering\arraybackslash}X}

\IIRTarget{Query}{eq:mod-query-system}
\IIRTarget{AP1}{eq:ach-blaschke}
\IIRTarget{Cayley}{eq:sp-cayley-warmup-target}
\IIRTarget{Exp}{eq:sp-ham-response}
\IIRTarget{FPAA}{eq:sp-fpaa-response}
\IIRTarget{TapFPAA}{eq:sp-fpaa-response}
\IIRTarget{Grover}{eq:sp-grover-iterate}
\IIRTarget{HamSim}{eq:sp-ham-response}
\IIRTarget{WeightedCayley}{eq:sp-weighted-cayley}
\IIRTarget{Wire}{iir:Wire}
\IIRTarget{Series}{iir:Series}
\IIRTarget{DirectSum}{iir:DirectSum}
\IIRTarget{Spectator}{iir:Spectator}
\IIRTarget{Close}{iir:Close}
\IIRTarget{Substitute}{iir:Substitute}
\IIRTarget{BufferedSeries}{iir:BufferedSeries}
\IIRTarget{Delay}{iir:Delay}
\IIRTarget{Inverse}{iir:Inverse}
\IIRTarget{Project}{iir:Project}
\IIRTarget{Chebyshev2}{iir:Chebyshev2}
\IIRTarget{ReflectionWalk}{iir:ReflectionWalk}
\IIRTarget{PrepReturn}{eq:mod-preparation-query}
\IIRTarget{PreparationQuery}{eq:mod-preparation-query}
\IIRTarget{OAA}{iir:OAA}
\IIRTarget{Sign}{eq:mod-sign-response}
\IIRTarget{Even}{eq:mod-sign-matchings}
\IIRTarget{Odd}{eq:mod-sign-matchings}
\IIRTarget{Pairs}{eq:mod-sign-matchings}
\IIRTarget{Threshold}{iir:Threshold}
\IIRTarget{AverageCost}{subsec:mod-average}
\IIRTarget{StatePrep}{subsec:mod-stateprep}
\IIRTarget{QLSP}{wql:eq:scalar-response}
\IIRTarget{Reciprocal}{eq:mod-reciprocal-completion}
\IIRTarget{Lift}{thm:iir-lift}
\IIRTarget{Sparse2}{eq:s6sp-junction}
\IIRTarget{Normalize}{mqspgsp:eq:gsp-normalization}
\IIRTarget{Choice}{mqspgsp:eq:gsp-choice}
\IIRTarget{RealPart}{mqspgsp:eq:gsp-real-system}
\IIRTarget{Ground}{mqspgsp:eq:gsp-ground-response}
\IIRTarget{Filter}{mqspgsp:eq:gsp-simple-filter-response}
\IIRTarget{Round}{mqspgsp:eq:gsp-simple-round-system}
\IIRTarget{Select}{mqspgsp:eq:gsp-selection-response}
\IIRTarget{Median}{mqspgsp:eq:gsp-median-system}

\IIRTarget{IntroIIR}{eq:intro-feedback}

\IIRTarget{FeedbackPath}{eq:intro-system}

\newcommand{\SPInlineFPAA}{\vcenter{\hbox{\begin{mqspdiagram}[scale=.88,transform shape] \MQSPTaggedSystem[minimum width=1.65cm]{fpt}{(0,0)}{TapFPAA}{$S_{\rm FPAA}$} \MQSPPublicWire{(fpt-pi) ++(-.45,0) -- (fpt-pi)} \MQSPPublicWire{(fpt-po) -- ++(.45,0)} \MQSPTaggedFeedback{fpt}{-1.18}{Grover}{$zO_\theta$} \end{mqspdiagram} }}}
\newcommand{\SPSimulationDiagram}[5]{\vcenter{\hbox{\begin{mqspdiagram}[scale=.88,transform shape] \MQSPTaggedSystem[minimum width=1.7cm]{#1exp}{(0,0)}{Exp}{#5} \MQSPTaggedSystem[minimum width=2.6cm]{#1c}{(0,-1.32)}{#2}{#3} \coordinate (#1c-pi) at ($(#1c.east)+(0,.20)$); \coordinate (#1c-po) at ($(#1c.west)+(0,.20)$); \coordinate (#1c-li) at ($(#1c.east)+(0,-.20)$); \coordinate (#1c-lo) at ($(#1c.west)+(0,-.20)$); \MQSPPublicWire{(#1exp-pi) ++(-.65,0) -- (#1exp-pi)} \MQSPPublicWire{(#1exp-po) -- ++(.65,0)} \MQSPPrivateWire{(#1exp-lo) -- (1.68,-.2) -- (1.68,-1.12) -- (#1c-pi)} \MQSPPrivateWire{(#1c-po) -- (-1.68,-1.12) -- (-1.68,-.2) -- (#1exp-li)} \MQSPTaggedGate[minimum width=2.6cm]{#1o}{(0,-2.54)}{FeedbackPath}{#4} \path let \p1=(#1o.west) in coordinate (#1left) at (\x1-9pt,\y1); \path let \p1=(#1o.east) in coordinate (#1right) at (\x1+9pt,\y1); \MQSPPrivateWire{(#1c-lo) -- (#1left |- #1c-lo) -- (#1left) -- (#1o.west)} \MQSPPrivateWire{(#1o.east) -- (#1right) -- (#1right |- #1c-li) -- (#1c-li)} \end{mqspdiagram} }}}

\tikzset{
td/public/.style={iir/public,draw=black},
td/private/.style={iir/private,draw=black},
td/classical/.style={iir/classical,draw=black},
td/record/.style={iir/box,draw=black!65,fill=white,minimum width=1.8cm,minimum height=.55cm,align=center},
td/current/.style={draw=black,fill=white,line width=.9pt},
td/oracle/.style={iir/gate,draw=black,fill=white,minimum width=.75cm,minimum height=.48cm},
td/system/.style={iir/system,draw=black,fill=white,minimum width=.92cm,minimum height=1.75cm},
td/clock/.style={iir/gate,draw=black,fill=white,minimum width=.83cm,minimum height=.55cm},
td/future/.style={draw=black!45,text=black!55,dashed,fill=black!2}}
\newcommand{\TDChronologyDiagram}{\begin{mqspdiagram}[font=\small] \fill[black!3] (11.40,-1.62) rectangle (16.12,2.68); \node[anchor=west,font=\small] at (-.45,3.08) {Online execution of the unitary lift}; \foreach \x/\id/\lab in {3.00/z/0,6.50/o/1,10.00/j/j}{ \node[mqsp/system,td/system] (s\id) at (\x,0) {$\widehat S_{\lab}$}; \coordinate (s\id-pi) at ($(s\id.west)+(0,.60)$); \coordinate (s\id-po) at ($(s\id.east)+(0,.60)$); \coordinate (s\id-li) at ($(s\id.west)+(0,-.60)$); \coordinate (s\id-lo) at ($(s\id.east)+(0,-.60)$); \MQSPGate[td/oracle]{q\id}{(\x-1.15,-.60)}{$O_{\lab}$} \draw[td/private] (q\id.east)--(s\id-li); } \node[td/record] (r0) at (3.00,2.23) {$\chi_0$}; \node[td/record] (r1) at (6.50,2.23) {$\chi_{\le1}$}; \node[td/record,td/current] (rj) at (10.00,2.23) {new $\chi_j$}; \draw[td/classical] (r0.south)--(sz.north); \draw[td/classical] (r1.south)--(so.north); \draw[td/classical,line width=1pt] (rj.south)--(sj.north); \node[font=\scriptsize,text=black!55] at (13.55,2.23) {not yet executed}; \node[anchor=west,font=\scriptsize,text=black] at (-.43,1.44) {public clock + system}; \MQSPGate[td/clock]{pg}{(.65,.60)}{$P_{\rm in}$} \draw[td/public] (-.48,.60) node[left] {$|0\rangle_c|\psi\rangle$}--(pg.west); \draw[td/public] (pg.east)--node[above,font=\scriptsize] {$\widehat{\boldsymbol{u}}_0$}(sz-pi); \draw[td/public] (sz-po)--node[above,font=\small] {$\widehat{\boldsymbol{v}}_0=\widehat{\boldsymbol{u}}_1$}(so-pi); \draw[td/public] (so-po)--node[above,font=\small] {$\cdots\quad\widehat{\boldsymbol{u}}_j$}(sj-pi); \draw[td/public] (sj-po)--node[above,font=\scriptsize] {$\widehat{\boldsymbol{v}}_j$}(11.38,.60); \node[anchor=east,font=\small,text=black] at (1.04,-.60) {$0_{\mathcal M}$}; \draw[td/private] (1.10,-.60)--(qz.west); \draw[td/private] (sz-lo)--node[below,font=\scriptsize] {$\boldsymbol{m}_1$}(qo.west); \draw[td/private] (so-lo)--node[below,font=\scriptsize] {$\cdots\quad \boldsymbol{m}_j$}(qj.west); \draw[td/private] (sj-lo)--node[below,font=\scriptsize] {$\boldsymbol{m}_{j+1}$}(11.38,-.60); \node[anchor=west,font=\scriptsize,text=black] at (-.43,-1.22) {private memory: zero amplitude initially; unmeasured thereafter}; \node[mqsp/system,td/system,td/future,minimum width=.65cm] (future) at (12.10,0) {$\cdots$}; \coordinate (fpi) at ($(future.west)+(0,.60)$); \coordinate (fpo) at ($(future.east)+(0,.60)$); \coordinate (fli) at ($(future.west)+(0,-.60)$); \coordinate (flo) at ($(future.east)+(0,-.60)$); \draw[td/public,draw=black!45] (11.42,.60)--(fpi); \draw[td/private,dashed,draw=black!45] (11.42,-.60)--(fli); \MQSPGate[td/clock,td/future]{ph}{(13.58,.60)}{$P_{\rm out}^\dagger$} \draw[td/public,draw=black!45] (fpo)--(ph.west); \draw[td/public,draw=black!45] (ph.east)--(14.42,.60); \node[anchor=west,align=left,font=\scriptsize,text=black!60] at (14.43,.60) {$B_{h,g}|\psi\rangle$\\$\oplus\;\text{other public}$}; \node[font=\scriptsize,text=black!55] at (13.58,1.35) {final clock selection}; \draw[td/private,dashed,draw=black!45] (flo)--(14.42,-.60); \node[anchor=west,font=\scriptsize,text=black!60] at (14.43,-.60) {$\boldsymbol{m}_N$ unmeasured}; \draw[black!55,densely dashed] (11.40,-1.62)--(11.40,2.69); \draw[-{Latex[length=2mm]},black,line width=.7pt] (-.45,-1.80)--(15.93,-1.80); \foreach \x/\lab in {3.00/0,6.50/1,10.00/j,12.10/{N-1}}{ \draw (\x,-1.73)--(\x,-1.90); \node[below,font=\scriptsize] at (\x,-1.90) {$\tau_{\lab}$};} \node[anchor=east,font=\scriptsize] at (15.95,-2.30) {physical execution time}; \node[anchor=west,font=\small] at (-.45,-2.86) {$|\Psi_j\rangle=\widehat{\boldsymbol{u}}_j\oplus \boldsymbol{m}_j,\quad \widehat{\boldsymbol{v}}_j=\widehat{\boldsymbol{u}}_{j+1},\quad \widehat{\boldsymbol{u}}_0=|c_{\rm in}\rangle\otimes|\psi\rangle,\quad \boldsymbol{m}_0=0.$}; \node[anchor=west,font=\scriptsize] at (-.45,-3.37) {The rails are components of one direct-sum state, evolved coherently until the final selection.}; \end{mqspdiagram}}

\tikzset{qlsp box/.style={iir/box,minimum height=6.5mm},qlsp wire/.style={iir/public},qlsp load/.style={Latex-Latex,line width=.45pt},qlsp private/.style={iir/private},
box/.style={iir/box},signal/.style={iir/public},oracle/.style={iir/private},port/.style={circle,draw,fill=white,inner sep=1.4pt}}
\allowdisplaybreaks[2]
\setlist{nosep,leftmargin=*}

\makeatletter
\newcommand{\settocdepth}[1]{\if@filesw \immediate\write\@auxout{\string\@writefile{toc}{\string\setcounter{tocdepth}{#1}}}\fi \bookmarksetup{depth=#1}}

\g@addto@macro\appendix{
\settocdepth{1}
}
\makeatother

\makeatletter
\renewcommand*\l@section[2]{\ifnum \c@tocdepth >\z@ \addpenalty\@secpenalty \addvspace{2pt}\setlength\@tempdima{1.5em}\begingroup \parindent \z@ \rightskip \@pnumwidth \parfillskip -\@pnumwidth \leavevmode \bfseries \advance\leftskip\@tempdima \hskip -\leftskip #1\nobreak\hfil \nobreak\hb@xt@\@pnumwidth{\hss #2}\par \endgroup \fi}
\makeatother

\hypersetup{pdftitle={Multivariate Quantum Signal Processing},pdfauthor={Guang Hao Low}}

\tikzset{qsys/.style={iir/box,minimum height=7mm,align=center},
qwire/.style={iir/public},qfeedback/.style={iir/private},
qcoupling/.style={Latex-Latex,line width=.6pt},
qlabel/.style={font=\scriptsize,align=center}}

\makeatletter
\def\@maketitle{\newpage \null \vspace{-1.5em}\begin{center}\let\footnote\thanks {\LARGE\bfseries \@title \par}\vskip 0.5em{\large \lineskip .3em\begin{tabular}[t]{c}\@author \end{tabular}\par}\ifx\@date\@empty\else \vskip 0.3em{\normalsize \@date\par}\fi \end{center}\par \vspace{-0.6em}}
\makeatother

\renewenvironment{abstract}{\par\vspace{0.2em}{\centering\bfseries\abstractname\par\vspace{0.25em}}\small \setlength{\parskip}{2pt}}{\par\vspace{0.4em}}

\usepackage{pgfplots}
\pgfplotsset{compat=1.18}
\usepgfplotslibrary{groupplots}

\title{Multivariate Quantum Signal Processing}
\author{Guang Hao Low\thanks{Guang Hao Low conceptualized the project and is responsible for its synthesis, interpretation, verification, presentation, manuscript writing, and editing.
Large language models performed technical derivations, computational checks, drafted and edited much of the manuscript, and generated formal Lean 4 proofs of the MQSP formalism in Sections 2, 3}\\[2pt]\small Google Quantum AI, Venice, CA 90291, United States}
\date{Draft v8.5 --- 30 September 2026}
\begin{document}
\maketitle
\begin{abstract}
We develop Multivariate Quantum Signal Processing for optimizing the overall weighted query cost of matrix-valued transformations of multiple noncommuting nonnormal matrices, each block-encoded with nonuniform query cost.
Our method equips the quantum-transducer framework with the analytic structure of multiport Infinite-Impulse-Response (IIR) filters, leading to a methodology for designing quantum algorithms by first specifying the IIR transfer functions realizing the these transformations.
Uniform analytic continuation and growth bounds of the transfer function give feasible per-oracle schedules and clock weights for unrolling the IIR into a finite response, typically with logarithmic precision dependence and near-unity clock normalization.
Quditization then block-encodes this finite response by projecting a unitary Toeplitz lift of IIR with normalization-optimized input and output clock states.
This connection between functional design and construction of a finite-query circuit directly parallels the approximation-theoretic approach of QSP/QSVT.
\\ \\ Every finite causally ordered quantum query circuit has a finite analytic MQSP realization, and we provide a complete characterization of achievable quantum transfer functions.
We construct multi-oracle quantum algorithms from a library of analytic IIR modules that guarantee achievability and gate efficiency.
In this manner, we significantly improve or resolve the multi-oracle query complexity of several long-standing problems: \textbf{(1)} We achieve Hamiltonian time evolution under $H=\sum_jH_j$ with weighted query cost $\le (1+\delta)t\langle\boldsymbol\lambda,\boldsymbol C\rangle_{1/2}+\mathcal{O}\big(\frac{\|\boldsymbol C\|_1}{\delta^2}\log(\frac{1}{\epsilon})\big)$ for any clock normalization $1+\delta$, where each $H_j$ is block-encoded with normalization $\lambda_j$ and cost $C_j$.
We present similar bounds for spectrum-amplified low-energy simulation, black-box state preparation, and other structured block-encodings.
\textbf{(2)} In the $d$-sparse access model, we achieve Hamiltonian simulation with optimal time--norm dependence $\mathcal{O}(t\sqrt d\,\norm H_{1\to2}+\sqrt{d\norm H_{\rm max}t\log 1/\epsilon}+\log1/\epsilon)$.
\textbf{(3)} We solve systems of linear equations with worst-case optimal complexity $\Theta(\kappa\log(\frac{1}{\epsilon}))$ to the matrix block encoding and $\Theta(1/\sqrt p)$ to state preparation simultaneously.
\textbf{(4)} We prepare quantum ground states and estimate their energies with optimal per-oracle complexities under a variety of gap and energy promises
\textbf{(5)} We prepare solution states of inhomogeneous linear differential equations for arbitrary time-dependent nonnormal matrices with complexity parameters to all oracles that encompass many prior approaches.
\textbf{(6)} We obtain per-oracle complexities for estimating the eigenvalues of the three-oracle generalized eigenvalue problem.
\\ \\ We then apply these building blocks to create novel quantum applications.
\textbf{(Dynamic correlation)} We show that time evolution by the core Hamiltonian exactly downfolded from a hierarchy of block-encoded Hamiltonians at different energy scales can be simulated with leading query cost $\approx t\langle\boldsymbol\lambda_\text{eff},\boldsymbol C_\text{eff}\rangle_{1/2}$ and additive logarithmic error dependence, where the normalizations are further weighted by the correlation energy contribution of coupled blocks that may also be coarse-grained to lower cost $C_\text{eff}$, without any requirement for an initial wavefunction on the virtual space.
\textbf{(Multiscale QML)} The Kron reduction is the analogue of downfolding for graph problems.
We apply this to multiscale graph classification problems where features at different scales are block-encoded by a hierarchy of quantum oracles with non-uniform costs and normalizations to obtain similarly correlation-weighted costs.
\textbf{(Online MQSP)} The same analytic variables extend to IIRs and oracles that vary with time.
As an example, we give an anisotropic Kalman filter construction in which a low-dimensional covariance and gain are updated classically while high-dimensional unitary oracles and arriving quantum observations are each queried once online, in event order.
\end{abstract}

\clearpage
\begingroup
\fontsize{9}{10.2}\selectfont \setlength{\parskip}{0pt} \setlength{\columnsep}{18pt}
\makeatletter
\def\@pnumwidth{2.3em}
\def\@tocrmarg{3em}
\@starttoc{toc}
\makeatother
\endgroup
\clearpage
\settocdepth{3}
\section{Introduction}\label{sec:introduction}

Many quantum algorithms obtain their advantage by manipulating exponentially large matrices without ever writing them down explicitly.
The input is supplied through quantum oracles~\cite{Bennett1997QuantumOracles} or block encodings~\cite{Low2016Qubitization} that can be queried coherently.
More generally, suppose that we are given several such oracles $\OTuple=(O_1,\ldots,O_m)$ with different costs and wish to approximate a function of them or decide a promised property with bounded error.
After finitely many queries interspersed by arbitrary unitary operations, quantum amplitudes in the resulting computation are polynomials of the oracles.
This polynomial structure underlies the quantum polynomial method \cite{Beals2001polynomial}.
Quantum query complexity also admits exact semidefinite characterizations: first for finite-query function evaluation \cite{Barnum2004QuerySDP}, and more generally for state-conversion problems \cite{Lee2011StateConversion}.
Related work showed that the general adversary SDP is tight and can be converted directly into quantum query algorithms \cite{Reichardt2011Reflections}.
An equivalent characterization of quantum query algorithms is completely bounded forms of these polynomials \cite{Arunachalam2019-fc}.
These results characterize, in principle, the full set of achievable query algorithms, but the corresponding quantum gate implementations, which are required to work uniformly across a promised set of oracles, can be exponentially large in the number of oracle calls.

The success of QSP and QSVT can be viewed as a sequence of deliberate structural restrictions.
General quantum query algorithms admit much broader completely bounded characterizations \cite{Arunachalam2019-fc}, but QSP does not attempt to parameterize this full space, even for a single oracle.
Instead it restricts attention to spectral transformations of a structured signal, trading universality for a compact and highly optimizable design language.
A block encoding first embeds the operator into a unitary matrix, and qubitization encodes the operator spectral information in invariant-subspace rotations \cite{Low2016Qubitization,Gilyen2018singular}.
QSP then provides an exact synthesis language for bounded polynomial transformations of this operator spectral signal through the spectral mapping theorem \cite{Low2016Methodology,Low2016HamSim}.
In applications, however, one usually designs a function rather than a polynomial---for example an exponential, inverse, sign function, or spectral filter.
Approximation theory turns smoothness, analyticity, or spectral gaps of the desired function into bounds on the required polynomial degree, after which QSP/QSVT realizes the polynomial exactly.
Thus the practical strength of QSP comes from moving from a universal but unwieldy query space to a structured hierarchy: spectral encoding, polynomial synthesis, and ultimately functional design.

For algorithm design, we distinguish three increasingly demanding goals:
\begin{enumerate}
\item[\textbf{(i)}] \textbf{Achievability:}
Identify a feasible complexity--error--normalization region permitted by the oracle model, using, for example, a semidefinite characterization that holds for every promised oracle and input state.

\item[\textbf{(ii)}] \textbf{Efficient realization:}
Represent useful points in this region by succinct circuits with controlled gate and space overheads, as achieved by structured constructions that encode matrices into oracles such as Linear-Combination-of-Unitaries (LCU)~\cite{Childs2012LCU}, Block-Encodings~\cite{Low2016Qubitization}, and Qubitization~\cite{Low2016Qubitization}, whose resources can be analyzed at the level of these algorithmic primitives.

\item[\textbf{(iii)}] \textbf{Design and optimization:}
Provide an expressive structured language that generates families of algorithms that are efficiently realizable by construction, while exposing parameters through which competing achievable resources can be efficiently optimized.
This connection from functional design to circuit synthesis is exemplified by Quantum Signal Processing (QSP)~\cite{Low2016HamSim,Motlagh2024gqsp} and Quantum Singular Value Transformations (QSVT)~\cite{Gilyen2018singular,Martyn2021GrandUnification}, and later extensions~\cite{Low2024Eigenvalue}.
\end{enumerate}
Extending QSP beyond a single signal has proved substantially more difficult than the univariate theory.
Rossi and Chuang introduced multivariable QSP as an alternating sequence of different queries for coherently transforming several signals \cite{Rossi2022MultivariableQSP}.
Subsequent work developed sharper realizability conditions and additional multivariate architectures \cite{Mori2024MultivariableComment,Nemeth2023MultivariateVariants,Laneve2025MultivariateQSP}, while more recent results gave constructive decision procedures, adversary-based characterizations, and modular constructions \cite{Ito2026ConstructiveDecision,Laneve2026QSPAdversary,Rossi2025ModularQSP}.
These works substantially clarify which multivariable transformations can be realized within prescribed QSP-like architectures, but they do not by themselves provide a general design space in which oracle costs, schedules, normalization, approximation error, and realization choices can be explored and optimized jointly.

In this work we develop \emph{Multivariate Quantum Signal Processing} (MQSP) based on the structure of analytic multiport Infinite-Impulse-Response (IIR) functions.
Our key result is a methodology that satisfies points \textbf{(i)}, \textbf{(ii)}, and \textbf{(iii)} for general multi-oracle problems.
Underlying the implementation of our IIR transfer functions is the quantum transducer framework~\cite{Belovs2024TamingQuantum} whereby explicit realizations are parameterized by a set of unitary oracles $\OTuple$ and their promises.
We introduce additional analytic variables $(\Bz;\OTuple)$, which elevates transducers to \emph{generating functions} of quantum query algorithms.
We provide a complete characterization of achievable analytic transfer functions in the quantum query setting, then design algorithms by first specifying a valid transfer function in $(\Bz;\OTuple)$ coordinates, which reduces to the unitary transducer model when $\Bz=\boldsymbol{1}$.
Sophisticated quantum algorithms are constructed from networks of primitive IIR modules that satisfy the same desirable quantum transducer properties of \emph{thriftiness} and \emph{lossless composition}, and are implemented in the standard quantum gate model by unrolling the IIR to a quantum circuit of finite query depth, or horizon.

Many of the primitive analytic IIR modules we use have an exact correspondence to classical lossless IIR filters~\cite{Regalia1988AllpassFilters,Smith2007DigitalFilters,Smith2010PhysicalAudioSignal}.
This unlocks up to seventy years of templates and IIR network design principles that are transferrable to the quantum realm.
Whenever the primitive modules are, by construction, achievable and gate-efficient, we prove that these properties are also inherited by the overall IIR network.
Moreover, analytic properties in $\Bz$ provide constructive feasible query-complexity, realizability, and correctness guarantees when unrolling the IIR.
These in turn facilitate the rapid optimization of complex IIR networks by globally allocating per-oracle complexities, errors, and normalizations, without the need to separately approximate individual components or numerically solve any kind of semidefinite program characterization of completely bounded complexity over some uniform oracle promise.
This completes the analogy to QSP/QSVT from function design to gate implementation, as summarized in~\cref{tab:qsp-mqsp-analogues}.
\begin{table}[!b]
\centering
\footnotesize
\renewcommand{\tabularxcolumn}[1]{m{#1}}
\renewcommand{\arraystretch}{1.05}
\setlength{\tabcolsep}{5pt}
\begin{tabularx}{\linewidth}{
  @{}>{\raggedright\arraybackslash}m{0.15\linewidth}|
  >{\raggedright\arraybackslash}X|
  >{\raggedright\arraybackslash}m{0.49\linewidth}@{}
}
\hline\hline
 & \centering QSP/QSVT & \centering\arraybackslash MQSP
\\
\hline
Target function & Scalar spectral function $f(x)$ & Matrix-valued response $F(\OTuple)=F(\Bone;\OTuple)$ with analytic extension $F(\Bz;\OTuple)$
\\
\hline
Achievable functions & Bounded polynomials with the required parity and completion conditions & Schur--Agler realization conditions uniform over $\mathfrak O$; includes every finite causally ordered query circuit
\\
Modular design & Compose separately compiled $f_j(A_j/\lambda_j)$ & Compose realizable IIR modules before compilation
\\
Matrix substitution & Spectral mapping $f(x)\mapsto f(A/\lambda)$ & Substitute $x_j\mapsto A_j/\lambda_j$ through realizable oracle modules, preserving multiplication order
\\
Compilation & Degree-$N$ approximation $f_N$; $O(N)$ phase angles & Horizon-$N$ FIR approximation $\tilde F_N$ selected by clock weights
\\
$\epsilon$-approximation & $N$ from polynomial approximation on the spectral promise & $(N,\alpha_{\rm clk})$ from analytic continuation, growth bounds, and the query schedule
\\
Unitary lift & Qubitization: two-dimensional invariant subspaces & Quditization: unitary Toeplitz lift and input/output clock states
\\
Block-encoded output & $\Be{f_N(A/\lambda)}$ & $\Be{\tilde F_N(\OTuple)/\alpha_{\rm clk}}$
\\
Circuit cost & Oracle queries interleaved with $O(N)$ phase rotations & $N$ system-matrix applications, scheduled oracle queries, clock preparation, and routing
\\
Exact $\epsilon$--$\alpha$ tradeoffs & Real polynomial minimax: semi-infinite LP with $O(N)$ variables & Semi-infinite SDP with causal Gram constraints; operator dimension can be exponential in the qubit count
\\
\hline\hline
\end{tabularx}
\caption{Structural analogies between QSP/QSVT, including generalized variants~\cite{Motlagh2024gqsp}, and MQSP for a uniform oracle promise $\OTuple\in\mathfrak O$.
The notation $f_N(A/\lambda)$ refers to the Hermitian case; general QSVT transforms singular values.}
\label{tab:qsp-mqsp-analogues}
\end{table}

We now provide a self-contained explanation of IIRs~\cite{Helton1974,Ball1998UnitaryColligations} with a mix of quantum transducer nomenclature.
Let $S$ be a unitary system matrix acting on a direct sum of an input quantum state $\ket{\psi}$ and a fictitious \emph{catalyst} state $\catalyst{f}$.
These states live in \emph{public} and \emph{private} Hilbert spaces $\ket{\psi}\in\mathcal{P}$ and $\catalyst{f}\in\mathcal{L}$ respectively.
Following the transducer formalism~\cite{Belovs2024TamingQuantum}, we define a layer $\oracle$ comprised solely of unitary oracle queries.
\begin{align}
S=\left(\begin{array}{c|c}A&B
\\
\hline
C&D\end{array}\right)
\quad\text{on }(\ket{\psi}\oplus\catalyst{f}) \in \mathcal P\oplus\mathcal L,
\quad \mathcal L=\bigoplus_{j=1}^m\mathcal L_j,
\quad \oracle(\Bz)=\bigoplus_{j=1}^m z_jO_j.
\label{eq:intro-system}
\end{align}
This demonstrates three concepts: \textbf{(a)} The lines in $S$ demark public and private spaces; \textbf{(b)} the private space is a direct sum of oracle subspaces $O_j:\mathcal{L}_j\to\mathcal{L}_j$; \textbf{(c)} our key insight in the quantum-algorithmic setting, which is to append analytic variables $z_j$ to their respective oracle queries and leave explicit any functional dependence on the multidimensional $\Bz=(z_1,z_2,\ldots,z_m)$.
These define the IIR, where classical wires carrying scalar voltages are replaced by vectors carrying quantum amplitudes, and reduce to a quantum transducer at the \emph{evaluation point} where $\Bz=\boldsymbol{1}$.

We now describe the next key concept of \emph{feedback}, which is a resource unavailable to the standard gate model of quantum computation.
Feedback here is identical to that in classical analog electrical circuits with antecedents dating back a century~\cite{Campbell1922WaveFilter,Butterworth1930FilterAmplifiers,Black1934FeedbackAmplifiers}.
Through feedback, the combined input $\ket{\psi}\oplus\catalyst{f}$ is transformed by the product $(\Id\oplus \oracle(\Bz))\cdot S$ as follows.
\begin{align}
\left(\begin{array}{c|c}\Id & 0
\\
\hline
0 & \oracle(\Bz)\end{array}\right)\cdot S\begin{pmatrix}\ket{\psi}
\\
\hline
\catalyst{f}\end{pmatrix} = \begin{pmatrix}F(\Bz;\OTuple)\ket{\psi}
\\
\hline
\catalyst{f}\end{pmatrix} = \begin{pmatrix}\vec{\psi}_{\rm out}
\\
\hline
\oracle(\Bz)\catalyst{g}\end{pmatrix},
\quad \IIRDiagram{s1_intro_simplest}
\label{eq:intro-feedback}
\end{align}
Above, dashed lines connecting inputs and outputs in a loop represent feedback, which is defined to equate input and output catalyst states $\catalyst{f}=\oracle(\Bz)\catalyst{g}$.
This defines the following self-consistent system of linear equations that we can solve to obtain the IIR \emph{transfer function} $F(\Bz)$.
\begin{subequations}
\label{eq:intro-response}
\begin{align}
\catalyst{g}&=C\ket{\psi}+D\oracle(\Bz)\catalyst{g}
\quad\Rightarrow
\quad \vec{g} =\Gamma(\Bz;\OTuple)\ket{\psi},
\quad \Gamma(\Bz;\OTuple)\doteq(\Id-D\oracle(\Bz))^{-1}C,
\\
\vec{\psi}_{\rm out}&=A\ket{\psi}+B\oracle(\Bz)\catalyst{g}
\quad\Rightarrow
\quad \vec{\psi}_{\rm out}=F(\Bz)\ket{\psi},
\quad F(\Bz;\OTuple)\doteq A+B\oracle(\Bz)\Gamma(\Bz;\OTuple).
\end{align}
\end{subequations}
The transfer function maps the public input $\ket{\psi}$ to the output $\catalyst{\psi}_{\rm {out}}$ and the catalyst state $\vec{g}$ is defined from $\ket{\psi}$ through the \emph{catalyst operator} $\Gamma(\Bz;\OTuple)$.
At the evaluation point, $\Bz=\Bone$, we drop the argument $\Bone$.
Whenever the context is clear, we also drop the tuple of oracles $\OTuple$.
There, $F$ is unitary so the normalized output quantum state is $\ket{\psi_{\rm {out}}}=F\ket{\psi}$.
The combined overall unitary system matrix has a classical correspondence in lossless digital filters and rational inner functions~\cite{Vaidyanathan1985Lossless,Regalia1988AllpassFilters}.
Related quantum formulations describe feedback through additive traces, such as in quantum control~\cite{Gough2008QuantumFeedback}, quantum Turing automata~\cite{Bartha2014QuantumTuring}, and with coherent-loop semantics~\cite{AndresMartinez2022Loops}.

We may interpret the transfer function $F(\Bz)$ as a generating function of quantum query algorithms through its Taylor series.
For $\Bz$ in the polydisk $\mathbb{D}^{m}$, the inverse in \cref{eq:intro-response} expands as $(I-D\oracle(\Bz))^{-1}=\sum_{k\ge0}[D\oracle(\Bz)]^k$ as $D$, being a sub-block of a unitary, must be a contraction with $\|D\|\le1$.
By collecting powers of $\Bz$, the transfer function has the convergent expansion\footnote{The transfer function exists at the evaluation point $F(\Bone)$ from the radial Abel limit $\lim_{r\uparrow1}F(r\boldsymbol1)$ even when the Taylor series fails to converge.}
\begin{align}
F(\Bz) =A+\sum_{k\ge0}B\oracle(\Bz)[D\oracle(\Bz)]^kC =\sum_{\boldsymbol{n}\in\mathbb N_0^m}F_{\boldsymbol{n}}\prod_{j=1}^m z_j^{n_j},
\quad \max_j|z_j|<1.
\label{eq:intro-feedback-series}
\end{align}
Each power describes another round trip through the feedback loop and $F_{\boldsymbol{n}}$ sums ordered paths containing exactly $n_j$ queries to the corresponding oracle $O_j$.
The analyticity of this series in the open polydisk follows from the simple fact that query counts $n_j\ge 0$ are nonnegative.
Hence, we may obtain different algorithms by truncating the sum to a finite query depth, or horizon, $\|\boldsymbol{n}\|_1=n_1+\ldots+n_m\le N$.

Implementing the abstract IIR model of quantum computation as a finite quantum circuit requires unrolling the feedback loop to finite query depth, or horizon, $N$ and leads to the Finite-Impulse-Response (FIR) transfer function $\tilde{F}_N$ that approximates $F$.
This introduces three key challenges.
First \textbf{($\boldsymbol \alpha$)}, unrolling truncates~\cref{eq:intro-feedback-series} to depth $N$, which introduces a tail approximation error that has been studied classically~\cite{Knese2021Bidisk}.
However, the remaining challenges have no established classical analogue.
Second \textbf{($\boldsymbol \beta$)}, the catalyst state $\Gamma\ket{\psi}$ is in general not available as an input.
As quantum states are normalized, an input $\ket{\psi}$ constrains the initial private state input to the trivial all-zeroes vector $\catalyst{0}$.
Omitting this state introduces an error in the actual implemented response function $\tilde{F}_N$.
Third \textbf{($\boldsymbol \gamma$)}, unrolling the IIR introduces a dimension $N$ clock register that controls the different system matrix applications.
As this clock state can become entangled with the input state $\ket{\psi}$, projecting out the desired response function introduces a clock normalization factor $\alpha_{\rm clk}>1$ that reduces the success probability of the algorithm.
The catalyst state involved in unrolling quantum transfer functions~\cite{ClementPerdrix2020PBS,Carette2021DelayedTrace} was given a quantum-algorithmic interpretation~\cite{Belovs2023LasVegas} and culminated in the seminal quantum transducer result: any unitary IIR may be unrolled to form an $\epsilon$-approximation to $F\ket{\psi}$ with constant success probability and query horizon $N=\mathcal{O}(W/\epsilon^2)$~\cite{Belovs2024TamingQuantum}, where the Las Vegas complexity, or \emph{catalyst} weight of the catalyst state, is $W=\|\Gamma\ket{\psi}\|^2$.

Quantum transducers describe a unitary transformation on an input state using a fictitious catalyst state that participates in the computation and returns unchanged.
Its squared norm determines a state-dependent Las Vegas query weight $W$ \cite{Belovs2023LasVegas} with particularly clean composition rules~\cite{Jeffery2022SubroutineComposition} paralleling the classical case~\cite{Mason1953SignalFlow,Redheffer1962ScatteringTransfer}: a computational branch contributes in proportion to the probability of entering it.
This permits resource bounds that follow the amplitude through an algorithm, including branches with very different oracle costs.
More recently, bespoke methods for unrolling transducers provide query-optimal and gate-efficient quantum algorithms for time-dependent Hamiltonian simulation \cite{Chen2026OptimaTimeDependent,Chen2026GateEfficientSimulation} and guided ground-state energy estimation \cite{Jeffery2026GuidedEnergy} with a logarithmic, rather than polynomial, dependence on error.
These results represent important progress towards satisfying points \textbf{(i)} and \textbf{(ii)} above in important applications with up to two oracles, while leaving open the broader goal of a general multi-oracle language for systematically designing transfer functions in a manner that also makes the exact characterization, tradeoffs, and optimization of quantum resources transparent.

The starting point of multivariate quantum signal processing is an observation relating the transfer function $F(\Bz)$ to the catalyst weight $W$.
In~\cref{thm:iir-energy}, we prove that the weight contribution $w_j$ to $W=w_1+w_2+\ldots+w_m$ from oracle $j$ is the first-order boundary derivative
\begin{align}
w_j=\bra{\psi}\Gamma_j^\dagger\Gamma_j\ket{\psi},
\quad \Gamma_j^\dagger\Gamma_j=F^\dagger\left.\partial_{z_j}F(\Bz)\right|_{\Bz=\boldsymbol1}.
\label{eq:s1_catalyst_weight_boundary}
\end{align}
This provides a tantalizing hint: Perhaps the functional dependence of $F$ on the multidimensional variable $\Bz$, such as through higher-order derivatives, provides the information necessary to address the seemingly disparate challenges on \textbf{($\boldsymbol \alpha$)} truncation error; \textbf{($\boldsymbol \beta$)} catalyst omission error; and \textbf{($\boldsymbol \gamma$)} clock normalization.
There is relevant classical work~\cite{Agler1990Polydics}, even in the non-commuting case~\cite{Ball1998UnitaryColligations} for \textbf{($\boldsymbol \alpha$)}, but these do not address the quantum concerns of \textbf{($\boldsymbol \beta$)} and \textbf{($\boldsymbol \gamma$)}.
Nevertheless, the fact that different analytic extensions of $F$ to $F(\Bz)$ can have different catalyst weights suggests that other analytic structure like singularities and pole placements are relevant quantities to optimize and analyze before setting $\Bz=\Bone$, which is information erased in the transducer picture.
MQSP formalizes this intuition and provides a quantum-algorithmic operational interpretation of the full dependence of $F$ on the multi-dimensional (in general non-commuting) analytic variable $\Bz$, which generalizes the catalyst weight interpretation by Belovs, Jeffery, and Yolcu, to the full analytic domain.
Combined with the classical inspiration of building sophisticated IIR design from modular IIR components, this provides a method for designing gate-efficient multi-oracle quantum algorithms through their transfer functions.

We now provide a high-level overview of how MQSP accomplishes the three goals \textbf{(i)}-\textbf{(ii)}-\textbf{(iii)} of algorithm design by connecting the design of a transfer function to a finite quantum circuit in four broad steps.
\begin{enumerate}[label=\textbf{(\arabic*)},leftmargin=*,itemsep=1ex]

\item \textbf{Quditization:} \cref{sec:quditization} compiles the transfer function $F$ into a quantum circuit that queries different oracles at different rates.
We assume that the specification of $F$ comes with three additional pieces of information: integer port delays $r_j>0$, clock weights $c_n$, and horizon $N$.
MQSP uses integer port delays $r_j>0$ to define different analytic continuations
\begin{align}
G(\Bz)=F_{\boldsymbol r}(\Bz) \doteq F(z_1^{r_1},\ldots,z_m^{r_m}),
\quad G(z\Bone)=\sum_{n\ge0}G_nz^n,
\quad G_{n<0}=0,
\quad G(\Bone)=F(\Bone),
\label{eq:intro-delays}
\end{align}
through a delay substitution $z_j\rightarrow z_j^{r_j}$.
We write $G(z)$ for the diagonal response $G(z\Bone)$.
Although the quantum transducer formalism already describes how each oracle may be queried every $r_j$ steps~\cite{Belovs2024TamingQuantum}, \cref{eq:intro-delays} lets the dependence of query counts, clock normalization, and approximation error on the query schedule be analyzed within the same functional description.
For instance, the catalyst weight $w_{\boldsymbol{r},j}$ of $G$ to oracle $j$ under a port delay follows from a simple chain rule:
\begin{align}
w_{\boldsymbol{r},j}=\bra{\psi}G^\dagger\left.\partial_{z_j}G(\Bz)\right|_{\Bz=\boldsymbol1}\ket{\psi} = \bra{\psi}F^\dagger\left.r_j[z_j^{r_j-1}\partial_{z_j}F(\Bz)]\right|_{\Bz=\boldsymbol1}\ket{\psi} = r_j\bra{\psi}\Gamma^\dagger_j\Gamma_j\ket{\psi}=r_jw_j.
\label{eq:intro_delay_weight}
\end{align}

Quditization then converts the specification $(F,\boldsymbol r,\boldsymbol{c})$ to a block-encoding $\Be{\tilde{G}_N/\alpha_{\rm {clk}}}$ of the desired FIR $\tilde{G}_N=\sum_{n=0}^{N-1}c_nG_n$ with the prescribed per-oracle query counts and the smallest possible clock normalization for those clock weights $\alpha_{\rm {clk}}$.
This is performed in two distinct steps that can be independently optimized, and provides a conceptually clearer view of prior quantum transducer implementations~\cite{Belovs2024TamingQuantum,Chen2026OptimaTimeDependent}, which left these steps implicit and may be viewed as particular unrollings and clock choices.
\begin{itemize}
\item \textbf{Unitary Toeplitz lift} (\cref{subsec:iir-lift}):
The unitary Toeplitz lift $\mathcal{W}_N[G]$ is a quantum circuit that unrolls $G$ to finite horizon $N$ and makes $Q_j=\big\lfloor\frac{N-1}{r_j}\big\rfloor$ queries to oracle $O_j$.
It acts on the space $\mathcal H_{N,\boldsymbol{r}}$ and block-encodes the Toeplitz matrix $\mathcal T_N[G]$ of transfer function coefficients.
\begin{align}
\mathcal{W}_N[G] = { \left[ \begin{array}{cccc|c} G_0 & 0 & \cdots & 0 & \ast
\\
G_1 & G_0 & \cdots & 0 & \ast
\\
G_2 & G_1 & \cdots & 0 & \ast
\\
\vdots & \vdots & \ddots & \vdots & \ast
\\
G_{N-1} & G_{N-2} & \cdots & G_0 & \ast
\\
\hline
\ast & \ast & \cdots & \ast & \ast \end{array} \right] },
\quad \mathcal H_{N,\boldsymbol{r}} =(\mathbb C^N\otimes\mathcal P) \oplus\bigoplus_{j=1}^m(\mathbb C^{r_j}\otimes\mathcal L_j).
\label{intro:eq:expanded-lift}
\end{align}
Structurally, $\mathcal T_N$ resembles the Toeplitz structure of Hamiltonians that arise in the Dyson-series simulation construction~\cite{Low2018IntPicSim,Low2018SpectralNorm} that suffered from a poor block-normalization of $N$.
The crucial difference is that $\mathcal T_N[G]$ is already a contraction with normalization one as the unitary system matrix $S$ enforces correlations among the different impulse response coefficients $G_n$ to be precisely those required to fit simultaneously inside a unitary.
The unitary Toeplitz lift $\mathcal{W}_N[G]$ acts on $b=\lceil\log_2\dim\mathcal H_{N,\boldsymbol{r}}\rceil$ qubits and we provide in \cref{thm:iir-lift} a space- and gate-efficient circuit implementation that uses only $\mathcal{O}(1)$ additional ancilla, $N$ system matrix applications, a query cost of $\sum_{j=1}^{m}Q_jC_j$, and almost-linear scaling $\mathcal{O}(Nmb \log b)$ additional primitive quantum gates.

\item \textbf{Clock state optimization} (\cref{subsec:iir-clock-body}):
The desired clock weights of $\tilde{G}_N$ are implemented using an input and output clock state.
These clock states define a clock matrix $X$ as follows.
\begin{align}
\ket{c_{\rm in}}=\sum_{i\in[N],\ell\in[R]}g_{i\ell}|i,\ell\rangle, \;\; \ket{c_{\rm out}}=\sum_{o\in[N],\ell\in[R]}h_{o\ell}|o,\ell\rangle, \;\; X=\alpha_{\rm clk}\sum_{o,i\in[N]}\ket{o}\bra{i}\sum_{\ell\in[R]}\overline{h_{o\ell}}g_{i\ell},
\label{eq:intro-framework-clock-states}
\end{align}
where the label $\ell$ indexes optional ancillas on which $\mathcal{W}_N[G]$ acts trivially, and $R$ is a constant in all our constructions.
As these states have dimension $RN$, they may always be prepared by clock-state preparation unitaries $P_{\rm in/\rm out}\ket{0}=\ket{c_{{\rm in}/{\rm out}}}$ using $\mathcal{O}(RN)$ primitive quantum gates~\cite{shende2006synthesis}.
Their overlap then determines the clock weights in the block-encoding
\begin{align}
\Be{\frac{\tilde{G}_N}{\alpha_{\rm {clk}}}}=(P_{\rm out}^{\dagger}\otimes \Id)(\Id\otimes\mathcal W_N[G])(P_{\rm in}\otimes \Id),
\quad c_n(X)=\sum_{i=0}^{N-1-n}X_{i+n,i}.
\label{eq:intro-clock-selection}
\end{align}
\cref{thm:iir-clock-body} proves that $X$ is feasible precisely when its nuclear norm is at most $\alpha_{\rm{clk}}$.
Hence, even when feasible clock states realizing the prescribed clock weights are provided, clock normalization may be further improved to $\|\boldsymbol c\|_{{\rm clk},N}=\min_{X:\ c_n(X)=c_n}\|X\|_*$ by convex optimization over a scalar matrix of size $N\times N$.
\end{itemize}

\item \textbf{Clock shaping:} \cref{sec:iir} tells us how to choose feasible quditization parameters $(F,\boldsymbol r,\boldsymbol{c})$ for clock weights $\boldsymbol{c}$, port delays $\boldsymbol{r}$, and horizon $N$ with both a small error and a small normalization.
For a fixed unitary Toeplitz lift, \cref{cor:iir-clock-frontier} gives the exact error--normalization frontier:
\begin{align}
\epsilon_{{\rm clk},N}[\alpha_{\rm clk}] =\min_{\|X\|_*\le\alpha_{\rm clk}}\; \sup_{\boldsymbol O\in\mathfrak O} \left\|\sum_{n=0}^{N-1}c_n(X)G_n-G\right\|.
\label{eq:intro-framework-frontier}
\end{align}
Although the clock matrix $X$ in~\cref{eq:intro-framework-clock-states} is small, these error constraints involve exponentially large operators and must hold throughout the oracle promise.
Instead of explicitly solving this, we use analytic bounds to construct feasible clocks.

\begin{itemize}
\item \textbf{Catalyst omission error and normalization}~\cref{subsec:clock-design-kernel}: \cref{eq:intro-framework-frontier} has a direct interpretation in terms of a transient error from replacing the catalyst state with a $\vect{0}$ input.
Due to the $\vect{0}$ input, the public output at step $n$ is the sum $\sum_{k=0}^nG_k$, instead of the steady-state value $G$.
The difference (multiplied by the unitary $G^\dagger$ for convenience) is a transient error $K_n=G^\dagger \big(G-\sum_{k=0}^{n}G_k\big)$ that is equivalently generated by the zero-catalyst transient generating function $K(z)$.
Then for any choice of clock weights, the error of the FIR $\tilde{G}_N=\sum_{n}c_nG_n$ with $c_N=0$ is
\begin{align}
G^\dagger(G-\widetilde G_N) =(1-c_0)I+\sum_{n=0}^{N-1}(c_n-c_{n+1})K_n,
\quad K(z)=\sum_{n\ge0}K_nz^n=\frac{I-G^\dagger G(z)}{1-z}.
\label{eq:intro-framework-error-identity}
\end{align}
Where adjacent clock weights agree, the corresponding transient cancels.
This suggests a flat plateau with smooth edges.
The optimal flat clock in \cref{lem:iir-optimal-flat} achieves normalization $1+\delta$ by extending the clock support by a factor $O(\delta^{-1/2})$ relative to the plateau width.
Then \cref{lem:iir-flat-smoothing} shows that positive averages of translated clocks smooth the edges without increasing their nuclear norm.

When $W$ is a scalar, it refers to the state-dependent catalyst weight $W=\bra{\psi}G^\dagger G'\ket{\psi}$.
We overload notation and also allow it to be a state-independent operator, which we call the group-delay operator $W=G^\dagger G'$.
In the absence of any analytic information, we improve in~\cref{prop:iir-uniform-catalyst-amplification} the state-independent error scaling to $N=\mathcal{O}(\|W\|/\epsilon)$, over the $N=O(\bra{\psi}W\ket{\psi}/\epsilon^2)$~\cite{Belovs2024TamingQuantum} uniform-clock baseline of \cref{thm:iir-uniform}.
Importantly, this improvement is not generic as catalyst weight and group-delay operator norm are inequivalent.
\item \textbf{Analytic clock shaping}~\cref{subsec:clock-design-flat,subsec:clock-design-smoothing}: Analytic bounds determine how wide those edges must be to suppress the remaining transient.
At the evaluation point, the transient generating function equals the group-delay operator, which describes the displacement that the clock must accommodate:
\begin{align}
W=K=G^\dagger G'=\sum_jr_j\Gamma_j^\dagger\Gamma_j,
\quad G^\dagger G(e^{i\theta})=I+i\theta W+O(\theta^2).
\label{eq:intro-framework-delay-expansion}
\end{align}
For instance, a pure delay has transfer function $z^dG$, so each eigenvalue of $W$ describes a clock displacement to first order in frequency.
A spread of eigenvalues requires a wider clock to cover different arrival times and frequency variation of the group delay produces dispersion.
\cref{thm:intro-master-two} converts bounds on these delays and local analytic growth into a horizon $N$ that is feasible for desired error $\epsilon$ and normalization $\alpha_{\rm clk}$.
\begin{theorem}[Analytic MQSP compilation]
\label{thm:intro-master-two}
Let the transfer function be analytic and contractive in the unit disk, with unitary value at $z=1$.
Assume it continues analytically to an open neighborhood of the closed disk $|z-1|\le\zeta$, where $0<\zeta\le1/16$.
Assume the following bounds hold uniformly over the oracle promise:
\begin{align}
aI\preceq W\preceq wI,
\quad 0\le a\le w,
\quad \|G(z)\|\le M_{\rm loc} \begin{cases} |z|^{a-h_-},&|z|\le1,
\\
|z|^{w+h_+},&|z|\ge1, \end{cases}
\quad|z-1|\le\zeta.
\label{eq:intro-master-cap}
\end{align}
The nonnegative radial excesses $h_\pm$ allow for departures from the first-order delay bounds over this neighborhood.
Assume $M_{\rm loc}\ge1$ and $h_-\le a$.
For $0<\delta\le1$ and $0<\epsilon\le\min\{\delta,1/4\}$, there is an explicit clock of dimension $N$ that bounds the error in~\cref{eq:intro-framework-frontier}
\begin{align}
\epsilon_{{\rm clk},N}[1+\delta]\le\epsilon,
\quad N=w+O\!\left( \frac{(w-a)+h_++h_-+ \zeta^{-1}\log(M_{\rm loc}/\epsilon)}{\sqrt\delta} \right).
\label{eq:intro-master-two}
\end{align}
Quditization implements $\Be{\widetilde G_N/(1+\delta)}$ with error $\|\widetilde G_N-G\|\le\epsilon$.
\end{theorem}
Setting $\delta=1$ gives the normalization-two bound used in the warm-up examples.
A common delay $w$ contributes once to the horizon; the additional clock width pays for delay spread and the precision needed at the edges.
Equalizing the group delay and controlling dispersion can therefore reduce the cost of approaching unit normalization.
Even tighter bounds, further multi-oracle clock-shaping bounds, and optimizations exploiting other analytic structure are presented in~\cref{subsec:analytic-comparison}.
\end{itemize}

\item\textbf{Achievable modular design:} MQSP starts from the transformation we want to implement and designs a transfer function $F(\OTuple)$ and analytic continuation $F(\Bz;\OTuple)$ that realizes it.
Without loss of generality, we assume that these oracles are block-encodings~\cite{Low2016Qubitization}.
A matrix $A_j$ is block-encoded by oracle $O_j=\Be{A_j/\lambda_j}$ with block-encoding normalization $\lambda_j$ if
\begin{align}
(\langle0|_{{\rm out},j}\otimes \Id_{\rm out,j})\Be{\frac{A_j}{\lambda_j}}(\ket{0}_{{\rm {in}}, j}\otimes \Id_{\rm{in},j})=\frac{A_j}{\lambda_j},
\quad \Pi_{\rm {in/out},j}=\ketbra{0}_{\rm in/out,j}\otimes \Id_{\rm in/out,j},
\label{eq:intro-block-encoding}
\end{align}
and we assign $O_j$ a cost $C_j$.
Whenever $A_j$ is square, we drop the labels $(\rm{in},\rm{out})$.
We equivalently represent the transformation $F(\OTuple)\equiv F(\boldsymbol{A})=F(A_1,A_2,\ldots,A_m)$, which implicitly absorbs any factors of $\Pi_{\rm in/out,j}$ and $\lambda_j$ into the system matrix.
The challenge is to prove that a single FIR is a uniform approximation across an entire promised oracle family $\mathfrak O$ without knowing the $A_j$, and that its realization as a block-encoding circuit has small normalization.
\begin{align}
\sup_{\OTuple\in\mathfrak O} \left\|\widetilde F_N(\OTuple)-F{(\boldsymbol O)}\right\| \le\epsilon,
\quad \text{and}
\quad \Be{\frac{\tilde{F}_N(\OTuple)}{\alpha_{\rm {clk}}}}
\quad \text{such that}
\quad \alpha_{\rm {clk}}=\Theta(1).
\label{eq:intro-uniform-approximation}
\end{align}
\begin{itemize}
\item \textbf{Achievability}:
\cref{sec:achievable} shows that MQSP is complete for finite causally ordered quantum query algorithms.
We prove in \cref{thm:ach-analytic-completeness} that every such circuit admits an analytic transfer-function realization that preserves its oracle order and per-oracle query counts.
Moreover, \cref{thm:ach-clock-completeness} recovers the full achievable exact error-normalization Pareto frontier in~\cref{eq:intro-uniform-approximation} for a fixed query schedule.
Analytic transfer functions therefore provide a representation of the entire finite quantum query model.
In the case of one Hermitian operator, such as in QSP, verifying~\cref{eq:intro-uniform-approximation} is easy as spectral mapping reduces this to a scalar approximation problem.
However, this is in general exponentially harder for even only two noncommuting operators.
Without a common eigenbasis, a horizon-$N$ expansion over $m$ oracle types has $m^N$ ordered paths, and each coefficient $F_n$ comprised of a sum of paths in~\cref{eq:intro-feedback-series} may itself act on the exponentially many basis states of $\boldsymbol{A}$.
To turn this characterization into an algorithm-design method, we need structured ways to choose the transfer function and its analytic continuation.
\item \textbf{Modularity:} \cref{sec:modularity} imposes structural IIR restrictions using achievability theory to elucidate a complete set of achievable primitive IIR modules that query quantum oracles encoding different types of structured $A_j$.
We prove that any transfer function built from this set using standard rules of IIR connection and composition is guaranteed to be achievable.
The overall approximation error in~\cref{eq:intro-uniform-approximation} is then bounded from analytic properties of the combined transfer function.
Sophisticated transfer functions may be built from simple transfer functions, and simple transfer functions with scalar arguments $F(\Bone;\boldsymbol{x})$ may be exact classical IIR modules, such as all-pass filters \cite{Regalia1988AllpassFilters}, wave-digital junctions \cite{Fettweis1986WaveDigitalFilters}, lossless synthesis \cite{Darlington1939NetworkSynthesis}, or similarly inspired by specifying a rational unitary $F(\boldsymbol{x})$ followed by an analytic extension whose poles, derivatives, and oracle dependence are later optimized to give favorable query costs.
As the rules of IIR composition preserve the order of $\boldsymbol{x}$ needed for noncommuting input, oracles may be substituted into the arguments $\boldsymbol{x}\rightarrow\OTuple\rightarrow \boldsymbol{A}$, which lifts the dimension of the transfer function by taking tensor products of public and private spaces with appropriate factors of $\Pi_{\rm in/out,j}$ and identity.
Analyticity in $\Bz$ provides uniform approximation guarantees as a function of $N$ without expanding the computation into paths, and spectral promises on the $A_j$ enter through operator inequalities in the multiplication order specified by the network.
This is how a compact functional description of a small-dimension $F(\Bz,\boldsymbol{x})$ can support approximation guarantees uniformly across $\mathfrak{O}$ in a quantum computation of exponentially larger dimension.
\end{itemize}
\item\textbf{Cost optimization}: The combined elements of \textbf{(1)}, \textbf{(2)}, and \textbf{(3)} provide the tools that allow us to optimize the overall resource cost of implementing an IIR.
In addition to an explicit error dependence, the analytic clock horizon bound in~\cref{eq:intro-master-two} depend on analytic growth exponents, and these in turn depend on the query schedule $\boldsymbol{r}$, for instance through~\cref{eq:intro_delay_weight,eq:intro-framework-delay-expansion}.
The analytic description gives concrete ways to improve an algorithm: move its poles, adjust its couplings, equalize its group delay, or change the port delays to reduce the cost of expensive queries.
For each candidate, uniform analytic bounds supply feasible clocks and a weighted query bound.
We can compare these bounds while keeping the network in its functional form.

For example, a longer delay reduces how often its oracle is queried, but also changes the analytic bounds and the required horizon.
The inequality version (\cref{cor:iir-mixed-scale}) of the analytic clock horizon bound shows these effects explicitly, and for many of our applications, a continuous upper bound on the weighted query cost $\langle\boldsymbol{C},\boldsymbol{Q}\rangle=\sum_{j}C_jQ_j$ is convex in the logarithms of the delays and analytic radius (\cref{cor:iir-scale-allocation}).
Here, $Q_j=\lfloor\frac{N-1}{r_j}\rfloor$ queries are made to $O_j$, each with cost $C_j$.
Hence, this upper bound can be optimized using standard convex methods without constructing any of the IIR coefficient matrices $G_n$.
After rounding the delays, the analytic-domain conditions must still hold and the horizon and actual query cost are then finalized.
For other applications using different analytic clock shaping theorems, we have for example~\cref{cor:iir-mixed-scale} that balances dispersion against error by optimizing the radius; or~\cref{thm:iir-mixed} that uses a fixed finite polynomial degree in one port to keep its query count independent of precision under a uniform analytic bound on the other ports.

Once an application has proved analytic estimates as functions of the delays, this tradeoff becomes an optimization over scalar parameters.
For example, suppose the leading horizon is a weighted sum of delays, with positive block-encoding normalization coefficients $\lambda_j$.
Using the cost bound $\langle\boldsymbol{C},\boldsymbol{Q}\rangle\le N\sum_j\frac{C_j}{r_j}$, \cref{lem:common-delay-allocation} gives the weighted query cost
\begin{align}
\min_{r_j>0} \left(\sum_j\lambda_jr_j\right) \left(\sum_j\frac{C_j}{r_j}\right) =\left(\sum_j\sqrt{C_j\lambda_j}\right)^2=\langle\boldsymbol{C},\boldsymbol{\lambda}\rangle_{1/2},
\quad r_j\propto\sqrt{C_j/\lambda_j},
\label{eq:intro-framework-allocation}
\end{align}
where expensive, weakly weighted oracles are queried less often.
The error, gate, and qubit cost of a final implementation (\cref{app:resources}) follows from a triangle inequality with separate contributions from clock approximation, finite realization, and gate synthesis, such as described in~ \cref{eq:qudit-error-budget} :
\begin{align}
\epsilon_{\rm final}\le \underbrace{\|\widetilde G_N-G\|}_{\text{clock approximation error}} +\underbrace{\alpha_{\rm clk}\|T_N[\widehat G]-T_N[G]\|}_{\text{finite-realization error}} +\underbrace{\alpha_{\rm clk}\epsilon_{\rm circuit}}_{\text{synthesis error}}.
\label{eq:sp-error-breakdown}
\end{align}

\end{enumerate}

Next, the warm-up examples in \cref{sec:intro-simple-proofs} only use what we have described thus far in this introduction.
These examples preview how MQSP is used in the later more detailed algorithm constructions and analysis.
They use the Quditization and analytic clock shaping bounds to derive the standard optimal amplitude and time scalings for amplitude amplification and Hamiltonian simulation respectively, with logarithmic precision dependence.
We even provide the first highly non-trivial result enabled by MQSP: Hamiltonian simulation with leading $t\langle\boldsymbol{C},\boldsymbol{\lambda}\rangle_{1/2}$ scaling and additive dependence on $\log1/\epsilon$.
\cref{sec:intro-algorithmic-results} summarizes the multioracle algorithms developed in \cref{sec:algorithms} and makes a comparison with prior art.
\cref{intro:subsec:applications} introduces downfolding and Kron reduction from \cref{sec:downfolding} and online Kalman filtering from \cref{sec:time-dependent-mqsp}.
We conclude and give our perspective on MQSP in~\cref{sec:intro-discussion}.

\FloatBarrier

In the remainder of the paper, \cref{sec:quditization} implements quditization and clock weights, including gate and qubit costs.
\cref{sec:iir} selects the weights, horizons, and schedules.
\cref{sec:achievable} characterizes achievable transfer functions and uses it to obtain a principled set of IIR modules in \cref{sec:modularity} that are later combined to form the multioracle algorithms of \cref{sec:algorithms}.
\cref{sec:downfolding} applies the framework to downfolding and Kron reduction, and \cref{sec:time-dependent-mqsp} develops causally ordered MQSP processing and online quantum Kalman filtering.

\subsection{Warm-up examples}
\label{sec:intro-simple-proofs}

Fixed-point amplitude amplification (\cref{sec:s1_1_FPAA}), Hamiltonian simulation (\cref{sec:s1_1_HamSim}) and multi-oracle Hamiltonian simulation (\cref{sec:s1_1_WeightedHamSim}) illustrate how classical all-pass filters lead to new quantum algorithms for familiar problems with short, self-contained proofs.
In each case, we first design a scalar transfer function with the desired value at $z=1$, verify its analytic and group-delay bounds to obtain a finite clock from \cref{thm:intro-master-two}, and then realize the transfer function by connecting a unitary system matrix $S$ to the oracle feedback ports as in \cref{eq:intro-system,eq:intro-response}.

As a first example, consider a self-inverse unitary oracle that block-encodes an unknown real scalar $x\in[-1,1]$.
We assume controlled access to
\begin{align}
O(x)=\Be{x} =\begin{pmatrix} x & \sqrt{1-x^2}
\\
\sqrt{1-x^2} & -x \end{pmatrix},
\quad \bra{0}_a O(x)\ket{0}_a=x,
\quad O(x)^2=I.
\label{eq:sp-cayley-warmup-oracle}
\end{align}
A single query sends amplitude $x$ into the system subspace of $O$ and amplitude $\sqrt{1-x^2}$ into the orthogonal subspace.
Suppose we wish to use this signal to set a quantum phase.
We choose the Cayley transformation
\begin{align}
\IIRTag{Cayley}(x) =\frac{1-ix}{1+ix} =e^{-2i\arctan x},
\quad \bigl|\IIRTag{Cayley}(x)\bigr|=1.
\label{eq:sp-cayley-warmup-target}
\end{align}
For a Hermitian matrix, this transformation will apply a different phase to each eigenspace while preserving its amplitude.

A first-order all-pass filter \cite[Section~4.1.2]{Smith2010PhysicalAudioSignal}, or Blaschke factor~\cite{Regalia1988AllpassFilters}, gives an analytic continuation with the desired value at $z=1$.
The factor $(z-ix)/(1+ixz)$ has unit modulus on the unit circle, but its value at $z=0$ is $-ix$.
Because $z$ counts oracle calls, the zero-query coefficient of an oracle-independent system matrix cannot depend on $x$.
Multiplying by one delay preserves the desired value at $z=1$ and gives the transfer function
\begin{align}
\IIRTag{Cayley}(z;x) =z\,\frac{z-ix}{1+ixz},
\quad \IIRTag{Cayley}(0;x)=0.
\label{eq:sp-cayley-warmup-transfer}
\end{align}
As elsewhere, we omit the analytic variable when it is evaluated at $z=1$.
The unitarity identity in~\cref{eq:iir-energy} gives
\begin{align}
1-\bigl|\IIRTag{Cayley}(z;x)\bigr|^2 =(1-|z|^2) \left(1+\frac{|z|^2(1-x^2)}{|1+ixz|^2}\right).
\label{eq:sp-cayley-warmup-modulus}
\end{align}
Thus the transfer function is contractive inside the unit disk and has unit modulus on the unit circle wherever the denominator is nonzero.
At $x=\pm1$, the removable continuation is $\IIRTag{Cayley}(z;\pm1)=\mp iz$.

We realize this transfer function by connecting the oracle to the private ports of a three-dimensional unitary system matrix.
In the order system $\ket{\psi}$, private $\ket{0}_a$, private $\ket{1}_a$, let
\begin{align}
S_{\IIRTag{Cayley}} =\begin{array}{ccc} & \begin{array}{ccc} \scriptstyle\ket{\psi} & \scriptstyle\ket{0}_a & \scriptstyle\ket{1}_a \end{array}
\\[-2pt]
\begin{array}{r} \scriptstyle\ket{\psi}
\\
\hline
\scriptstyle\ket{0}_a
\\
\scriptstyle\ket{1}_a \end{array} & \left(\begin{array}{c|cc} 0 & -i & 0
\\
\hline
1 & 0 & 0
\\
0 & 0 & i \end{array}\right) \end{array}
\quad \IIRDiagram[0.49\linewidth]{WarmupCayley}
\label{eq:sp-cayley-warmup-system}
\end{align}
The public space $\mathcal P$ is one-dimensional in this scalar example, and the private space $\mathcal L$ consists of the two oracle coordinates.
The diagram in \cref{eq:sp-cayley-warmup-system} uses the same public and private labels as \cref{eq:intro-system}, with feedback $\catalyst f=\oracle(z)\catalyst g$ and $\oracle(z)=zO(x)$.
The system matrix injects the public input into the first oracle state, extracts that state with phase $-i$, and feeds the second state back with phase $i$.
Writing its blocks as $S_{\IIRTag{Cayley}}= \left(\begin{smallmatrix}A&B\\ C&D\end{smallmatrix}\right)$, the feedback equations of \cref{eq:intro-response} give
\begin{align}
\Gamma_{\IIRTag{Cayley}}(z;x) =(I-D\oracle(z))^{-1}C =\begin{pmatrix} 1
\\[2pt]
\dfrac{iz\sqrt{1-x^2}}{1+ixz} \end{pmatrix}.
\label{eq:sp-cayley-warmup-private}
\end{align}
Substituting these private amplitudes into the public output yields the transfer function
\begin{align}
F=A+B\oracle(z)\Gamma_{\IIRTag{Cayley}}(z;x) =-izx+\frac{z^2(1-x^2)}{1+ixz} =z\,\frac{z-ix}{1+ixz}.
\label{eq:sp-cayley-warmup-elimination}
\end{align}

The squared catalyst norm equals the group delay:
\begin{align}
W(x) =\overline{\IIRTag{Cayley}(x)} \left.\partial_z\IIRTag{Cayley}(z;x)\right|_{z=1} =\bigl\|\Gamma_{\IIRTag{Cayley}}(x)\bigr\|^2 =1+\frac{1-x^2}{1+x^2} =\frac{2}{1+x^2}.
\label{eq:sp-cayley-warmup-weight}
\end{align}
The first-order clock displacement in~\cref{eq:intro-framework-delay-expansion} therefore lies between one and two steps throughout the promised signal interval.
For $x\ne0$, the possible pole at $z=i/x$ stays away from the evaluation point.
More explicitly, throughout $|z-1|\le1/16$,
\begin{align}
|1+ixz| \ge \sqrt{1+x^2}-|x|\,|z-1| \ge \frac{15}{16}.
\label{eq:sp-cayley-warmup-denominator}
\end{align}
On the exterior part of this neighborhood, \cref{eq:sp-cayley-warmup-modulus} therefore gives
\begin{align}
\bigl|\IIRTag{Cayley}(z;x)\bigr|^2 \le |z|^2\bigl[1+2(|z|^2-1)\bigr] \le |z|^6,
\quad |z|\ge1.
\label{eq:sp-cayley-warmup-growth}
\end{align}
With $G(z)=\IIRTag{Cayley}(z;x)$, \cref{thm:intro-master-two} consequently gives clock states that implement
\begin{align}
\Be{\widetilde G_N/2},
\quad \sup_{x\in[-1,1]} \bigl|\widetilde G_N(x)-\IIRTag{Cayley}(x)\bigr| \le\epsilon,
\quad N=O\left(\log\frac1\epsilon\right).
\label{eq:sp-cayley-warmup-error}
\end{align}
Then $Q_O=N-1$ counts controlled queries to the original oracle.
Oblivious amplitude amplification in \cref{lem:iir-cubic} gives an approximate unitary and triples the query cost.

For a Hermitian matrix, the invariant signal subspaces of a self-inverse block encoding have the scalar action in \cref{eq:sp-cayley-warmup-oracle}.
Substituting $x\mapsto H/\lambda$ therefore gives the unitary transformation
\begin{align}
O=\Be{H/\lambda},
\quad \IIRTag{Cayley}(H/\lambda) =(I-iH/\lambda)(I+iH/\lambda)^{-1} =e^{-2i\arctan(H/\lambda)},
\label{eq:sp-cayley-warmup-matrix}
\end{align}
where $H=H^\dagger$, $\|H\|\le\lambda$, and $O^2=I$.
The same clock gives a uniform operator-norm approximation.
Chen et al.~\cite{Chen2026OptimaTimeDependent} achieve Hamiltonian simulation in the transducer framework by cascading evaluated Cayley transformations over a very large number of short time steps.
We instead solve this by a direct functional in~\cref{sec:s1_1_HamSim} by defining an exact analytic exponential transform that linearizes the $\arctan$ phase.

\subsubsection{Fixed-point amplitude amplification}\label{sec:s1_1_FPAA}
We seek the state $|\psi\rangle$ through the preparation encoding $A=\Be{\lambda|\psi\rangle\langle0|}$.
Inverse access and a positive lower amplitude bound $\lambda\ge\lambda_0>0$ are given.
Write $a$ for the signal ancillas and $s$ for the system register.
The good component of the prepared state is \begin{align}
(|0\rangle\langle0|_a\otimes I)A|0\rangle_{as} =\lambda|0\rangle_a|\psi\rangle_s.
\end{align} Without loss of generality, we assume $\lambda=\sin\theta\in[\lambda_0,1/2]$, with $\lambda_0=\sin\theta_0$; a single-qubit rotation can reduce the good amplitude by a fixed factor if necessary.
The goal is failure amplitude at most $\epsilon$.
We repeatedly transmit part of the good amplitude, while the Grover iterate mixes the retained good and bad amplitudes.

Recall that the Grover iterate is
\begin{align}
O_G=A(2|0\rangle\langle0|_{as}-I)A^\dagger (2|0\rangle\langle0|_a-I) =\Be{2\lambda^2|\psi\rangle\langle\psi|-I}.
\label{eq:sp-grover-iterate}
\end{align}
Its encoded matrix acts on $|\psi\rangle$ as the scalar $T_2[\lambda]=2\lambda^2-1=-\cos2\theta$.
We first design a scalar all-pass filter: its modulus is one on the unit circle, as required for a unitary quantum transformation.
A first-order all-pass \cite[Section 4.1.2]{Smith2010PhysicalAudioSignal}, or Blaschke factor~\cite{Regalia1988AllpassFilters} satisfies
\begin{align}
b_a(z)=\frac{z-a}{1-\overline a z},
\quad 1-|b_a(z)|^2= \frac{(1-|a|^2)(1-|z|^2)}{|1-\overline a z|^2},
\quad |a|<1.
\label{eq:sp-blaschke-modulus}
\end{align}
The identity gives unit modulus on the unit circle and contractivity inside it.
The two-dimensional Grover rotation suggests a second-order filter, called a BiQuad in classical IIR design \cite[Section B.1.6]{Smith2007DigitalFilters}.
Its target value is one throughout the promised signal interval.
We seek poles outside the unit disk, with enough separation to permit a short finite clock.
This filter has a transfer function of the form
\begin{subequations}
\label{eq:sp-fpaa-response}
\begin{align}
\module{FPAA}(z;\lambda)= \frac{(z-\sqrt c\,e^{i\varphi})(z-\sqrt c\,e^{-i\varphi})} {(1-\sqrt c\,e^{i\varphi}z)(1-\sqrt c\,e^{-i\varphi}z)},
\quad \module{FPAA}(\lambda)=1,
\\
c=\frac{1-\sin2\theta_0}{1+\sin2\theta_0} =1-4\lambda_0+O(\lambda_0^2),
\quad \cos\varphi=\frac{\cos2\theta}{\cos2\theta_0}.
\end{align}
\end{subequations}
In the realization below, the denominator has the form
\begin{align}
1-(1+c)\cos(2\theta)z+cz^2 =1-2\sqrt c\cos\varphi\,z+cz^2.
\label{eq:sp-fpaa-denominator}
\end{align}
A classical critically damped filter relaxes to its steady state fastest.
This intuition from classical IIR design tells us to choose filter parameters that achieve critical damping across the promised interval containing $\lambda$.
The limiting choice is
\begin{align}
\frac{2\sqrt c}{1+c}=\cos(2\theta_0),
\quad c=\frac{1-\sin(2\theta_0)}{1+\sin(2\theta_0)}.
\label{eq:sp-fpaa-critical}
\end{align}
At the threshold, the two poles coalesce on the positive real axis.
This is critical damping, and it fixes the value of $c$ in this family.
The circuit realizes the filter by mixing the public state with the good feedback state through one rotation.
It leaves the bad state unchanged.
With $O_\theta=-O_G$, a known phase change, we obtain the following system matrix in public/good/bad order and feedback in good/bad order:
\begin{align}
S_{\module{FPAA}}=\left(\begin{array}{c|cc} c&-\sqrt{1-c^2}&0
\\
\hline
\sqrt{1-c^2}&c&0
\\
0&0&1 \end{array}\right),
\quad\SPInlineFPAA
\quad \oracle(z)=zO_\theta =z\left(\begin{smallmatrix}\cos2\theta&\sin2\theta
\\
-\sin2\theta&\cos2\theta\end{smallmatrix}\right).
\label{eq:sp-fpaa-system}
\end{align}
Eliminating the two private amplitudes gives \cref{eq:sp-fpaa-response}.
The diagram shows the one-dimensional public port above the two-dimensional feedback connection.
Each Grover iterate uses one call each to $A$ and $A^\dagger$.
Denote the respective counts by $Q_A$ and $Q_{A^\dagger}$.

Write the lower-left and lower-right blocks of $S_{\module{FPAA}}$ as $C,D$.
The defining catalyst equation $\Gamma(z)=(I-D\oracle(z))^{-1}C$ gives
\begin{align}
\Gamma=\sqrt{\frac{1-c}{1+c}} \begin{pmatrix}1
\\
-\cot\theta\end{pmatrix},
\quad \|\Gamma\|^2=\frac{\sin2\theta_0}{\lambda^2} \le\frac2{\lambda_0}.
\label{eq:sp-fpaa-weight}
\end{align}
To obtain a finite clock, we use the analytic properties of \cref{eq:sp-fpaa-response}.
The two factors in \cref{eq:sp-fpaa-response} are Blaschke factors with parameters $a=\sqrt c\,e^{\pm i\varphi}$.
The modulus identity \cref{eq:sp-blaschke-modulus} therefore makes their product contractive in the unit disk.
Its poles have radius $c^{-1/2}=1+2\lambda_0+O(\lambda_0^2)$.
The transfer function is therefore analytic on $|z-1|\le\zeta=(1-\sqrt c)/16$.
On the exterior part of this neighborhood, the all-pass modulus identity and the two denominator bounds give
\begin{align}
|\module{FPAA}(z;\lambda)|\le|z|^{16/(1-\sqrt c)},
\quad |z|\ge1.
\end{align}
These are the hypotheses of \cref{thm:intro-master-two}, which gives clock states with $\alpha_{\rm clk}=2$ and $N=O(\lambda_0^{-1}\log(1/\epsilon))$.

For state preparation, feed $A|0\rangle_{as}$ into the private space and store each transmission at a fresh time address.
The quditization lift gives the $N$ orthogonal public time slots and retains the final two-dimensional private update.
Applying each tap before the Grover iterate gives
\begin{align}
\boldsymbol{r}_N=K^N\binom{\lambda}{\sqrt{1-\lambda^2}},
\quad K=O_\theta\begin{pmatrix}c&0
\\
0&1\end{pmatrix},
\quad \operatorname{spec}K=\{\sqrt c\,e^{i\varphi},\sqrt c\,e^{-i\varphi}\}.
\label{eq:sp-fpaa-residual}
\end{align}
These eigenvalues are the reciprocals of the two poles of \cref{eq:sp-fpaa-response}.
In a unitary Schur basis, $K$ has these diagonal entries and an off-diagonal entry of magnitude $1-c$.
The off-diagonal entry of $K^N$ therefore has magnitude $(1-c)c^{(N-1)/2}|\sin(N\varphi)/\sin\varphi|$.
The ratio is interpreted by continuity at $\varphi=0$ and is at most $N$.
After $N$ rounds,
\begin{subequations}
\label{eq:sp-fpaa-cost}
\begin{align}
A|0\rangle_{as}\ \mapsto\ \left(\sum_{k<N}\beta_k|k\rangle\right)\otimes|\psi\rangle \oplus\boldsymbol{r}_N,
\quad \|\boldsymbol{r}_N\|\le \left[1+\frac{N(1-c)}{\sqrt c}\right]c^{N/2},
\\
Q_A+Q_{A^\dagger}=2N+1 =O\left(\frac{\log(1/\epsilon)}{\lambda_0}\right).
\end{align}
\end{subequations}
Here $\beta_k$ is the amplitude transmitted at time $k$.
Choosing $N=O(\lambda_0^{-1}\log(1/\epsilon))$ makes the private residual at most $\epsilon$, and every successful address contains the target state.

Ancilla-controlled attenuation already appears in measurement-based fixed-point search~\cite{Tulsi2006FixedPointSearch}.
In particular, Mizel's critically damped search uses the same coefficient $c$ when his Grover rotation angle is $2\theta_0$~\cite{Mizel2009DampedSearch}.
The contribution here is a transfer-function analysis that supports coherent composition and general clock selection.
A modification and more detailed analysis in \cref{sec:modularity} recovers the constant factor of the optimal QSVT small-amplitude query count $\log(2/\epsilon)/\lambda_0$ \cite{Yoder2014}.

\subsubsection{Hamiltonian simulation}\label{sec:s1_1_HamSim}
Hamiltonian simulation targets a transfer function $e^{-it\lambda x}$ for $x\in[-1,1]$, and $t,\lambda>0$.
Substituting a Hermitian contraction $H/\lambda$ will then give $e^{-itH}$.
In other words we seek a transfer function $F_t(z;x)$ that satisfies $F_t(1;x)=e^{-it\lambda x}$, is contractive in the disk, and has a regular analytic neighborhood with small delay spread and controlled dispersion.
The first-order all-pass filter in~\cref{eq:sp-cayley-warmup-target} followed by a unit delay in~\cref{eq:sp-cayley-warmup-transfer} is the first step.

To make the resulting phase linear in $x$, we compose the classical pure-delay transfer function \cite{Smith2007DigitalFilters,Smith2010PhysicalAudioSignal} with the bilinear transformation $(1-w)/(1+w)$~\cite[Section 7.3.2]{Smith2010PhysicalAudioSignal}.
We call this the exponential transform for the Cayley transform, or $\module{Exp}_\tau$.
Substituting the Cayley transfer function for its feedback connection cancels the nonlinear phase exactly:
\begin{align}
\begin{gathered} \module{Exp}_{\tau}(z;x)=e^{-\tau(1-zx)/(1+zx)},\quad \module{Exp}_{\tau}(x)=e^{-\tau(1-x)/(1+x)},
\\
F_t(z;x)=\module{Exp}_{t\lambda}(\module{Cayley}(z;x)) =\exp\left[-t\lambda\frac{1-z^2+2izx}{1+z^2}\right],
\\
F_t(x)=e^{-it\lambda x},
\\
\|\Gamma_t\|^2=\overline{F_t(x)}F_t'(x)=t\lambda. \end{gathered}
\quad\SPSimulationDiagram{single}{Cayley}{$\module{Cayley}(z;x)$}{$zO(x)$}{$\module{Exp}_{t\lambda}$}
\label{eq:sp-ham-response}
\end{align}
The diagram feeds the exponential transform through the Cayley transformation, whose feedback connection applies $zO(x)$.
The last line is computed using~\cref{eq:s1_catalyst_weight_boundary} and the product $\overline{F_t(x)}F_t'(x)$ is the scalar group delay and squared catalyst norm, giving query weight $t\lambda$.

The query analysis below depends on this transfer function; different lossless realizations can have different gate and qubit costs.
To obtain the spectral realization, take the positive-real transform of the exponential:
\begin{align}
\frac{1+\module{Exp}_\tau(w)}{1-\module{Exp}_\tau(w)} =\coth\left(\frac{\tau}{2}\frac{1-w}{1+w}\right).
\label{eq:sp-exp-positive-real}
\end{align}
In the variable $(1-w)/(1+w)$, its poles occur at integer multiples of $2\pi i/\tau$.
Their positive spectral weights give the Cauchy amplitudes in \cref{eq:sp-exp-system}; \cref{app:mod-exp} derives the representation.
For positive $\tau$, the Exp system couples the public state to the normalized private state $|v_\tau\rangle$.
It then applies the diagonal unitary $U_\tau$.
The private modes are indexed by $k\in\mathbb Z$:
\begin{subequations}
\label{eq:sp-exp-system}
\begin{align}
S_{\module{Exp}_\tau}=\left(\begin{array}{c|c} e^{-\tau}&\sqrt{1-e^{-2\tau}}\langle v_\tau|
\\
\hline
\sqrt{1-e^{-2\tau}}U_\tau|v_\tau\rangle& U_\tau[I-(1+e^{-\tau})|v_\tau\rangle\langle v_\tau|] \end{array}\right),
\\
\langle k|v_\tau\rangle= \sqrt{\frac{2\tau\tanh(\tau/2)}{\tau^2+4\pi^2k^2}},
\quad U_\tau|k\rangle=\frac{\tau-2\pi i k}{\tau+2\pi i k}|k\rangle,
\quad k\in\mathbb Z.
\end{align}
\end{subequations}
These coefficients give $\module{Exp}_\tau(w)$ when the feedback operation is $wI$.
The implementation theorem, \cref{thm:exp-implementation}, proves this identity and gives a finite realization that approximates the horizon-$N$ Toeplitz block, together with its gate and qubit costs.

The query proof only requires contractivity for $z$ in the unit disc and a growth bound near $z=1$.
The modulus of $F_t$ is the exponential of the real part of its exponent, which is nonpositive because
\begin{align}
\operatorname{Re}\frac{1-z^2+2izx}{1+z^2} =\frac{1-|z|^2}{2} \left(\frac{1+x}{|1-iz|^2}+\frac{1-x}{|1+iz|^2}\right)\ge0
\quad (|z|<1).
\label{eq:sp-scalar-certificate}
\end{align}
The exponent's only possible poles are at $\pm i$.
Thus $F_t$ is analytic through $|z-1|\le1/16$.
On the exterior part of this neighborhood, $|1\pm iz|\ge\sqrt2-1/16$ and $(|z|^2-1)/(\sqrt2-1/16)^2\le2\log|z|$.
The same identity therefore gives $|F_t(z;x)|\le|z|^{2t\lambda}$.
Thus the analytic radius is $\zeta=1/16$ and the exterior radial exponent is $2t\lambda$.
Choose the growth parameters
\begin{align}
a=h_-=0,
\quad w=h_+=t\lambda,
\quad M=1.
\end{align}
Substitution into \cref{thm:intro-master-two} gives
\begin{align}
\alpha_{\rm clk}=2,
\quad N=O\left(t\lambda+\log\frac1\epsilon\right),
\quad Q_O=N-1.
\label{eq:sp-ham-cost}
\end{align}
Here $Q_O$ counts controlled calls to the self-inverse oracle $O(x)$.
Use a controlled self-inverse block encoding to substitute the normalized Hamiltonian for the scalar signal: \begin{align}
x\mapsto H/\lambda,
\quad O=\Be{H/\lambda}.
\end{align} Its invariant subspaces have the scalar action $O(x)$, so the same uniform calculation approximates $e^{-itH}$ in operator norm.
Oblivious amplitude amplification in \cref{lem:iir-cubic} gives Hamiltonian simulation and triples the query cost.
A refined construction has clock normalization and leading horizon coefficient arbitrarily close to one; see \cref{prop:mod-ham}.
The joint time--precision refinement is treated by Chen et al. \cite{Chen2026OptimaTimeDependent}.

\subsubsection{Hamiltonian simulation with weighted query cost}\label{sec:s1_1_WeightedHamSim}
The target is the evolution $e^{-itH}$ generated by a sum of separately queried terms $H=\sum_{j=1}^mH_j$.
Each component has a controlled self-inverse encoding $O_j=\Be{H_j/\lambda_j}$, with positive normalization and query cost $C_j>0$.
Write $Q_j$ for the number of calls to port $j$.
Collect the normalizations and costs in vectors, with \begin{align}
\boldsymbol{\lambda}=(\lambda_j)_j,
\quad \boldsymbol{C}=(C_j)_j,
\quad \Lambda= \|\boldsymbol{\lambda}\|_1=\sum_j\lambda_j,
\quad \|\boldsymbol{C}\|_1=\sum_jC_j.
\end{align} We minimize $\sum_jC_jQ_j$ subject to a uniform operator-norm error guarantee.
With LCU, qubitization, and QSP, one usually first block-encodes the sum.
Implementing its SELECT operation from the separate oracles already queries every $O_j$ once.
Here, positive integer delays $r_j$ allow different query periods.
We use the same \IIRTag{Cayley}--\IIRTag{Exp} construction, with the individual signals combined inside the \IIRTag{Cayley} transformation.

The classical \emph{parallel wave-digital adaptor} connects several components losslessly~\cite{Fettweis1986WaveDigitalFilters,Smith2010PhysicalAudioSignal}.
In normalized amplitudes, with a fixed phase on each private input, it gives our \emph{multivariate Cayley transformation}.
The public input is distributed among the signal states with amplitudes proportional to $\sqrt{\lambda_j}$; the orthogonal states are retained with phase $i$.
For scalar signals $\boldsymbol{x}\in[-1,1]^m$, introduce one delay variable per oracle.
The realization has one public state and $2m$ private states:
\begin{subequations}
\label{eq:sp-weighted-cayley}
\begin{align}
|\omega\rangle=\sum_{j=1}^m \sqrt{\frac{\lambda_j}{\|\boldsymbol{\lambda}\|_1}}\, |j\rangle|0\rangle_a,
\quad Z(\boldsymbol{z})=\bigoplus_{j=1}^m z_jO(x_j),
\\
S_C^{(m)}=\left(\begin{array}{c|c} 0&-i\langle\omega|
\\
\hline
|\omega\rangle&i(I-|\omega\rangle\langle\omega|) \end{array}\right),
\quad C(\boldsymbol{z};\boldsymbol{x})= \frac{1-\sum_j\dfrac{\lambda_j}{\|\boldsymbol{\lambda}\|_1} \dfrac{1-z_j^2+2iz_jx_j}{1+z_j^2}} {1+\sum_j\dfrac{\lambda_j}{\|\boldsymbol{\lambda}\|_1} \dfrac{1-z_j^2+2iz_jx_j}{1+z_j^2}}.
\end{align}
\end{subequations}
Eliminating the private states adds the terms displayed in the Cayley fraction.
Substitution into the exponential transform of \cref{eq:sp-exp-system}, with $\tau=t\|\boldsymbol{\lambda}\|_1$, therefore gives the exponential of their sum.
The diagram uses $z_j=z^{r_j}$ to implement the chosen delays.
At $\boldsymbol{z}=\boldsymbol{1}$ $\boldsymbol{1}=(1,\ldots,1)$, the transfer function is the desired evolution:
\begin{align}
\begin{gathered} F_t(\boldsymbol{z};\boldsymbol{x}) =\module{Exp}_{t\|\boldsymbol{\lambda}\|_1}(C(\boldsymbol{z};\boldsymbol{x})) =\exp\left[-t\sum_j\lambda_j \frac{1-z_j^2+2iz_jx_j}{1+z_j^2}\right],
\\
F_t(\boldsymbol{x})=e^{-it\sum_j\lambda_jx_j}. \end{gathered}
\quad\SPSimulationDiagram{average}{WeightedCayley}{$C(\boldsymbol{z};\boldsymbol{x})$} {$\bigoplus_jz_jO(x_j)$; $z_j=z^{r_j}$}{$\module{Exp}_{t\|\boldsymbol{\lambda}\|_1}$}
\label{eq:sp-average-response}
\end{align}

For the chosen delays, set $G(z)=F_t(z^{r_1},\ldots,z^{r_m};\boldsymbol{x})$.
The derivative identity at $z=1$ gives its catalyst weight $\overline{G}G'=t\sum_j\lambda_jr_j$.
The delayed states remain in the local analytic neighborhood whenever \begin{align}
|z-1|\le\frac1{32\max_jr_j}
\quad\Longrightarrow
\quad |z^{r_j}-1|\le\frac1{16}
\quad\text{for every }j.
\end{align} Thus \cref{eq:sp-scalar-certificate} proves contractivity for $z$ in the unit disc, analyticity on this neighborhood, and exterior growth at most $|z|^{2t\sum_j\lambda_jr_j}$.
Substituting radius $\zeta=1/(32\max_jr_j)$ and exterior radial exponent $2t\sum_j\lambda_jr_j$ into \cref{thm:intro-master-two} gives
\begin{align}
N=O\left(t\sum_j\lambda_jr_j +\max_jr_j\log\frac1\epsilon\right),
\quad Q_j=\left\lfloor\frac{N-1}{r_j}\right\rfloor,
\quad\alpha_{\rm clk}=2.
\label{eq:sp-average-horizon}
\end{align}
To minimize the leading cost, choose relative delays $r_j\propto\sqrt{C_j/\lambda_j}$.
Including the precision term gives the regularized choice
\begin{align}
r_j\ \propto\ \sqrt{\frac{C_j} {t\lambda_j+\dfrac{C_j}{\|\boldsymbol{C}\|_1}\log(2/\epsilon)}}
\quad\text{before rescaling and rounding.}
\label{eq:sp-allocation}
\end{align}
Rescale so the smallest delay is one and round upward; this changes the cost by at most a constant factor.
Define the paired quantity $\langle\boldsymbol{C},\boldsymbol{\lambda}\rangle_{1/2} =(\sum_j\sqrt{C_j\lambda_j})^2$.
Substitution into \cref{eq:sp-average-horizon} gives
\begin{align}
\sum_jC_jq_j= O\left(t\langle\boldsymbol{C},\boldsymbol{\lambda}\rangle_{1/2} +\|\boldsymbol{C}\|_1\log\frac1\epsilon\right).
\label{eq:sp-average-cost}
\end{align}
Cauchy--Schwarz also gives $\langle\boldsymbol{C},\boldsymbol{\lambda}\rangle_{1/2}\le m\langle\boldsymbol{C},\boldsymbol{\lambda}\rangle$, where $\langle\boldsymbol{C},\boldsymbol{\lambda}\rangle=\sum_jC_j\lambda_j$.
The leading cost is at most the LCU--QSP cost $t\|\boldsymbol{\lambda}\|_1\|\boldsymbol{C}\|_1$ in the same separate-oracle model, and can be much smaller when expensive terms have small weights.

For noncommuting components, make the operator substitutions with their signal isometries: \begin{align}
x_j\mapsto H_j/\lambda_j,
\quad |0\rangle_{a_j}\otimes I.
\end{align} Eliminating the private variables gives one matrix exponential of the sum in \cref{eq:sp-average-response}.
In \cref{eq:sp-scalar-certificate}, $1\pm x_j$ becomes the positive operator $I\pm H_j/\lambda_j$.
The matrix exponential is controlled by the Hermitian part of its exponent: $\|e^B\|\le e^{\lambda_{\max}((B+B^\dagger)/2)}$.
Applying this inequality to the sum gives the same analytic bounds and the target value $e^{-itH}$ at the evaluation point.
This proves \cref{eq:sp-average-cost} for matrix oracles with constant success probability; Hamiltonian simulation again costs only a constant factor.
We may also apply tighter analytic clock shaping~\cref{thm:iir-analytic-clock-selection} to obtain a smaller clock normalization $1+\delta$ for any $0<\delta\le1/4$.
Its leading query-cost coefficient is also $1+\delta$, with additive remainder $O(\delta^{-2}\|\boldsymbol{C}\|_1\log(1/\epsilon))$, as proved in \cref{thm:multi-hs}.

\subsection{Overview of new algorithms}
\label{sec:intro-algorithmic-results}
Multivariate quantum signal processing extends the above examples to other multi-oracle algorithms in~\cref{sec:algorithms}, including other variants of Hamiltonian simulation (\cref{sec:multi-hs}), such as simulating sparse matrices and its extension to black-box quantum state preparation in~\cref{subsec:wsp}.
We also investigate quantum algorithms for solving systems of linear equations (\cref{wql:sec:qlsp}), simulating linear differential equations (\cref{sec:alg-volterra}), quantum ground state preparation and estimation (\cref{mqspgsp:sec:ground-state}), and eigenvalue estimation of generalized eigenvalue problems (\cref{subsec:app-generalized-eigenvalues}).
In the following, we show that MQSP resolves or significantly improves the query complexity of problems where there can be multiple oracles $O_j$.
Unless otherwise stated, we assume the cost model (\cref{def:resource-accounting}) where controlled-$O_j$ and also the inverse $O_j^\dagger$ all have equal cost $C_j$.
If an oracle $O_j$ is queried $q_j$ times, we evaluate the total query cost of the algorithm with the sum $\langle\boldsymbol{C},\boldsymbol{q}\rangle=\sum_{j}C_jq_j$.
Typically, any additional one- and two-qubit quantum gate overhead is almost linear in $\langle\boldsymbol{C},\boldsymbol{q}\rangle$ and any qubit overhead is logarithmic in $\max_jq_j$.
These gate and qubit overheads are summarized in~\cref{app:resources}.
We also express our costs using the following products.
\begin{align}
\langle\boldsymbol{u},\boldsymbol{v}\rangle=\sum_ju_jv_j,
\quad \langle\boldsymbol{u},\boldsymbol{v}\rangle_{1/2} =\left(\sum_j\sqrt{u_jv_j}\right)^2,
\quad v_{\Sigma}=\|\boldsymbol{v}\|_1=\sum_jv_j.
\label{eq:intro-weighted-norm}
\end{align}
By Cauchy--Schwarz, the half-quasinorm on $m$ terms satisfies
\begin{align}
\langle\boldsymbol{C},\boldsymbol{\lambda}\rangle_{1/2} \le\min\{C_\Sigma\|\boldsymbol{\lambda}\|_1, m\langle\boldsymbol{C},\boldsymbol{\lambda}\rangle\}.
\end{align}

\subsubsection{Hamiltonian simulation and spectral amplification}
\label{sec:intro-alg-hamiltonian}
The Hamiltonian simulation problem is the original motivation~\cite{Feynman1982} for quantum computation and, today, the simulation of quantum systems remains one of its most compelling applications.
The problem is to approximate the real time-evolution operator $e^{-iHt}$ for some target evolution time $t$ and some target error $\epsilon$.
The challenge is that the Hamiltonian $H=\sum_jH_j$ is a sum of non-commuting parts.
Hence, while time-evolution by a single term $H_j$ could be easy, their combination into $e^{-iHt}$ is non-trivial.

Hamiltonian simulation has received tremendous attention over the years.
Lloyd~\cite{Lloyd1996universal} provided the first algorithm with polynomial quantum gate scaling in time $t$, error $\epsilon$, and number of $k$-local Hamiltonian terms, using first-order Trotterization~\cite{Trotter1959Product}.
Subsequent developments in this framework focused on higher-order Trotter-Suzuki methods~\cite{Wiebe2010} and multiproduct formulas~\cite{Childs2012LCU,Low2019Multiproduct} to achieve nearly linear scaling in $t$, polylogarithmic scaling in error, and nearly linear scaling in system size for geometrically-local interactions by exploiting commutators~\cite{childs2019theory} and information lightcones~\cite{Haah2019Lattice}.
In parallel, an approach was developed based on querying Boolean oracles~\cite{Aharonov2003Adiabatic} specifying the entries of a sparse Hamiltonian with at most $d$ non-zero entries in every row.
Following a long series of work based on quantum walk~\cite{Childs2010,Berry2014Exponential,Berry2015Hamiltonian}, this culminated in an algorithm using Quantum Signal Processing~\cite{Low2016HamSim} with optimal query complexity with respect to all parameters of $td\|H\|_{\rm{max}}$ and $\epsilon$ in terms of the matrix max-norm $\|H\|_{\rm{max}}$.
The quantum walk picture was generalized by Qubitization~\cite{Low2016Qubitization} to any block-encoded matrix $\Be{H/\lambda}$, which is today the standard oracular model for matrix access on quantum computers.

The block-encoding framework assumes that each term $H_j$ is block-encoded in the unitary $\Be{H_j/\lambda_j}$ with block-encoding normalization $\lambda_j$, and each such unitary has a cost $C_j$.
Prior approaches based on the combination of LCU~\cite{Childs2012LCU}, Qubitization~\cite{Low2016Qubitization}, and QSP~\cite{Low2016HamSim} have a leading order cost $C_\Sigma\big(t\Lambda+\mathcal{O}(\log\frac{1}{\epsilon})\big)$, which is dominated by the sum of all the costs multiplied by the sum of all the norms $\Lambda=\|\boldsymbol{\lambda}\|_1$.
Standard block encodings by LCUs are effective for long-range interactions~\cite{vonBurg2020carbon,Lee2020hypercontraction,Low2025-tn}, while product formulas can exploit short-range structure more directly.
This comparison depends on the supplied access and on the implementation of each step.
Recursive block encodings, average-cost simulation, and low-rank structure can also reduce the gate cost of Trotter steps for structured Hamiltonians~\cite{Low2023TrotterSteps}.
Improving this system-size dependence $\Lambda$ in the block-encoding picture has proven to be exceptionally challenging.
Underlying this challenge is its nature as a quantum oracle.
Hence a quantum query algorithm is required to approximate $e^{-iHt}$ uniformly across all possible ways to block-encode all possible Hamiltonians, which makes it independent of any finer Hamiltonian structure by definition.

However, there are notable successful examples of exploiting Hamiltonian structure even within the block-encoding framework.
Sum-of-Squares-Spectral Amplification (SOSSA) instead assumes that $H=\sum_{j}S_j^\dagger S_j-\beta I$ has a positive semidefinite component, which can be efficiently computed for electronic structure problems~\cite{King2025SOSSA,Low2025-tn}.
Concrete lower-bound certificates arise from near-frustration-free representations~\cite{Rubin2026LowerBoundCertificates}, and a first-quantized plane-wave construction gives a further application of spectral amplification~\cite{Dutkiewicz2026FirstQuantizedAmplification}.
This enables roughly a $\sqrt{\Lambda}$ speed up using spectral amplification, where one instead may choose to simulate a unitary $e^{it\lambda f(H/\lambda)}$, where the function $f(x)\approx \sqrt{x+\beta/\lambda}$ has a quadratically amplified gradient for low-lying eigenvalues $(E+\beta)$ of the positive semidefinite term $\sum_{j}S_j^\dagger S_j$.
Another approach, based on moving to the interaction picture~\cite{Low2018IntPicSim}, is able to achieve a weighted cost scaling like $\tilde{O}\left(t(C_0\lambda_0+C_1\lambda_1)\right)$.
By recursing the interaction picture simulation algorithm $m$ times, one is in fact able to obtain a weighted cost $\tilde{O}\left(\big(\sum_{j}C_j\lambda_j\big)t\log^m(t\Lambda/\epsilon)\right)$~\cite{Low2018SpectralNorm} across multiple block-encodings, though with a completely impractical algorithm using an extremely large number of qubits and with exponential scaling in the number of terms $m$.
Notably this algorithm combined with uniform spectral amplification~\cite{Low2017USA} improves the sparse simulation result with respect to other matrix norms, achieving query complexity $\mathcal{O}((\sqrt{d}\|H\|_{1\rightarrow 2}t)^{1+o(1)}/\epsilon^{o(1)})$ in the Euclidean row norm $\|H\|_{1\rightarrow 2}$~\cite{Low2018SpectralNorm}.

We use MQSP in~\cref{sec:multi-hs} to obtain a Hamiltonian simulation algorithm for block-encoded terms with a significantly improved dependence on system size.
Specifically we obtain a weighted cost $\approx t\langle\boldsymbol{C},\boldsymbol{\lambda}\rangle_{1/2}+\mathcal{O}(C_\Sigma\log\frac{1}{\epsilon})$ that can be significantly smaller than the prior standard approach of qubitization and quantum signal processing.
Moreover, our algorithm is practical, as reflected by the near-unity constant factor, unlike the recursive interaction picture simulation algorithm.
MQSP is also able to exploit other finer structure in how the block-encoding is formed.
For instance, if it is known that each block-encoding has an implementation like $\Be{H_j/\lambda_j}=P_j\cdot \textsc{S} \cdot P_j^\dagger$, where the middle unitary is shared across all terms, a leading weighted cost $<2t(C_S\Lambda+\langle\boldsymbol{c},\boldsymbol{\lambda}\rangle_{1/2})$ is possible.
This factorization is motivated by electronic structure simulation problems~\cite{vonBurg2020carbon,Lee2020hypercontraction,Low2026DenserSurfaceCode} where $\textsc{S}$ with cost $C_S$ can sometimes be much more expensive~\cite{Low2025-tn} than the costs $c_j$ for invoking $\{P_j,P_j^\dagger\}$.
We also obtain similar weighted-cost improvements for SOSSA, and obtain a sparse Hamiltonian simulation result with optimal time-norm complexity $\Theta(t\sqrt{d}\|H\|_{1\rightarrow 2})$ and additive dependence on error.
We compare our results with prior art in~\cref{tab:intro-alg-structured-hs}.
Overall, the block-encoding access models we consider are:
\begin{itemize}[itemsep=0.5em]
\item \textbf{Standard LCU}: Each $\Be{H_j/\lambda_j}$ costs $C_j$.
\item \textbf{Shared-}\textsc{S}: Each $\Be{H_j/\lambda_j}=P_j\cdot \textsc{S} \cdot P_j^\dagger$.
$\textsc{S}$ costs $C_S$ and $\{P_j,P_j^\dagger\}$ together cost $c_j$.
\item \textbf{Sum-of-Squares (SOS)}: Each $H_j=A_j^\dagger A_j$, and $\{\Be{A_j/\sqrt{\lambda_j}},\Be{A_j/\sqrt{\lambda_j}}^\dagger\}$ together cost $c_j=2C_j$.
\item \textbf{Sparse access}: The $l^{\text{th}}$ non-zero column index of a given row of a $d$-sparse Hamiltonian is computed by a Boolean oracle $O_f$, and another Boolean oracle $O_H$ computes the value of a matrix entry.
\end{itemize}
\begin{table}[!htbp]
\centering\small
\begin{tabularx}{\linewidth}{lXl}
\hline
\hline
\multicolumn{3}{c}{\textbf{Hamiltonian simulation}}
\\
\hline
Year & Construction & Weighted query cost
\\
\hline
2015~\cite{Berry2015Truncated} & Truncated Taylor series & $C_\Sigma\times O\left(t\Lambda \frac{\log(t\Lambda/\epsilon)} {\log\log(t\Lambda/\epsilon)}\right)$
\\
2016~\cite{Low2016HamSim} & Sparse quantum walk + QSP & $C_\Sigma\times \mathcal{O}\left(td\|H\|_{\rm{max}}+\frac{\log(1/\epsilon)}{\log(e+\log(1/\epsilon)/d\|H\|_{\rm max} t)}\right)$
\\
2016~\cite{Low2016Qubitization} & Qubitization + QSP & $C_\Sigma\times \left(2t\Lambda+\mathcal{O}\left(\frac{\log(1/\epsilon)}{\log(e+\log(1/\epsilon)/\Lambda t)}\right)\right)$
\\
2018~\cite{Low2018IntPicSim} & Sparse access + Interaction picture & $C_\Sigma\times \mathcal{O}\left((\sqrt{d}\|H\|_{1\rightarrow 2}t)^{1+o(1)}/\epsilon^{o(1)}\right)$
\\
2024~\cite{Motlagh2024gqsp} & Qubitization + Generalized QSP & $C_\Sigma\times \left(t\Lambda+\mathcal{O}\left(\frac{\log(1/\epsilon)}{\log(e+\log(1/\epsilon)/\Lambda t)}\right)\right)$
\\
2026~\cite{Chen2026OptimaTimeDependent} & Quantum Transducers&$C_\Sigma\times \left(48t\Lambda+\mathcal{O}\left(\frac{\log(1/\epsilon)}{\log(e+\log(1/\epsilon)/\Lambda t)}\right)\right)$
\\
2025~\cite{King2025SOSSA} & SOS-Spectral-Amplification & $C_\Sigma\times t\sqrt{2E\Lambda}$
\\
\hline
\cref{thm:multi-hs} & Standard LCU & $(1+\delta)t\langle\boldsymbol{C},\boldsymbol{\lambda}\rangle_{1/2}+O(\frac{C_\Sigma}{\delta^2}\log(1/\epsilon))$
\\
\cref{thm:multi-hs-select}& Shared-\textsc{S} & $(2-\gamma)t(C_S\Lambda+\langle\boldsymbol{c},\boldsymbol{\lambda}\rangle_{1/2})+O_{\rm inst,\delta}(\log(1/\epsilon))$
\\
\cref{thm:multi-sos-phase}& SOS-Spectral-Amplification & $\frac{(1+\delta)k}{\sqrt\Lambda} \big(\sum_jc_j^{2/3}\lambda_j^{1/3}\big)^{3/2} +\mathcal{O}(\frac{c_\Sigma}{\delta^3}\log(1/\epsilon))$
\\
\cref{thm:multi-hs-sparse} & Sparse second-moment cancellation & $C_\Sigma\times \mathcal{O}\left(\sqrt{d}\|H\|_{1\rightarrow 2}t+\sqrt{d\|H\|_{\rm max}t\log\frac{1}{\epsilon}}+\log\frac{1}{\epsilon}\right)$
\\
\hline
\hline
\end{tabularx}
\caption{Query complexity to approximate the time-evolution operator $e^{-iHt}$ for $t\ge0$ with constant success probability.
MQSP block-encodes $\Be{\widetilde G/(1+\delta)}$ with error $\|\widetilde G-e^{-itH}\|\le\epsilon$ for any $0<\epsilon\le\delta\le1/4$.
The SOSSA rows give phase-amplification costs: the earlier row gives the leading low-energy scaling, and $k$ is the gain in \cref{thm:multi-sos-phase}.
In the shared-oracle row, $\gamma>0$ and the additive coefficient depend on the instance.}
\label{tab:intro-alg-structured-hs}
\end{table}

\subsubsection{Quantum linear systems problem}
\label{sec:intro-alg-linear-systems}
Quantum algorithms for systems of linear equations output the normalized solution state $\ket{x}\propto H^{-1}|b\rangle$.
The matrix $H$ to be inverted is block-encoded in $O_H=\Be{H/\Lambda}$ and, at constant overhead, $H$ may be assumed to be Hermitian and $O_H=O_H^\dagger$ is self-inverse.
The block-encoding normalization $\Lambda$ is usually set to one by convention.
The initial state $O_b|0\rangle=|b\rangle$ is prepared by a state preparation unitary $O_b$.
We are provided with an inverse-norm upper bound $\kappa$ and a lower bound on the solution state norm $\sqrt{p}$.
\begin{align}
\kappa\ge\left\|H^{-1}\right\|,
\quad \frac{1}{\kappa}\le\sqrt{p}\le\frac{\|H^{-1}|b\rangle\|}{\kappa}\le1.
\end{align}
Overall, the goal is to prepare $\ket{x}$ to error $\epsilon$ with constant success probability using as few queries as possible to both the block-encoding and state preparation oracles.

The seminal algorithm by Harrow, Hassidim, and Lloyd~\cite{Harrow2009}, or HHL algorithm, proved that this could be accomplished using quantum phase estimation and amplitude amplification using $\mathcal{O}(\kappa/(\epsilon^2\sqrt{p}))$ queries to $O_H$ and $\Theta(1/\sqrt{p})$ queries to $O_b$, which is optimal with respect to the state preparation oracle.
A sequence of improvements such as Variable-Time-Amplitude-Amplification (VTAA)~\cite{Ambainis2012VTAA}, linear-combination-of-quantum walks~\cite{Childs2015LinearSystems}, and QSVT to approximate the inverse operator $H^{-1}$~\cite{Gilyen2018singular} improved the $O_H$ complexity to nearly-linear scaling in condition number and polylogarithmic dependence on error.
This culminated in an algorithm based on discrete adiabatic evolution~\cite{Costa2021linearsystems} which achieved the optimal $O_H$ complexity of $\Theta(\kappa\log\frac{1}{\epsilon})$, which has since been replicated by other methods such as kernel reflections~\cite{Dalzell2024LinearSystems} and block-preconditioning~\cite{Low2024QLSP}.
However, these methods all make the same number of $\mathcal{O}(\kappa\log\frac{1}{\epsilon})$ queries to initial state preparation, departing from the optimal $\Theta(1/\sqrt{p})$ scaling of the original algorithm.
The algorithm with best overall performance~\cite{Low2024QLSP} is based on VTAA with a deterministic schedule of gapped-phase-estimation followed by a single QSVT correction.
This algorithm has optimal scaling to $O_b$, and $\mathcal{O}(\kappa\log\frac{2}{p}\log\frac{\log(4/p)}{\epsilon})$ queries to $O_H$, which is optimal for any constant $p$, but not for general $p$.

We use MQSP in~\cref{wql:sec:qlsp} to attain both optimal query bounds with $\Theta(\kappa\log\frac{1}{\epsilon})$ to $O_H$ and $\Theta(1/\sqrt{p})$ to $O_b$ simultaneously and tabulate a comparison in~\cref{tab:intro-alg-qlsp}.
In the multi-oracle setting,
\begin{align}
H=\sum_jH_j,
\quad\boldsymbol{b}=\sum_ja_j\ket{b_j},
\quad \kappa\ge\|H^{-1}\|,
\quad0<\sqrt{p}\le\frac{\|H^{-1}\boldsymbol b\|}{\kappa}.
\end{align}
The $H_j$ do not commute and each is block-encoded in $\Be{H_j/\lambda_j}$ with cost $C_j$ and each normalized state $\ket{b_j}$ is in general nonorthogonal to the others and is prepared by $O_{b,j}\ket{0}=\ket{b_j}$ with cost $c_j$.
The sum of normalizations is $\Lambda=\sum_{j}{\lambda_j}$, which is set to one above.
Note that more care is required to lower bound $p$ as it can be smaller due to sign cancellations in the amplitudes of the initial states $\ket{b_j}$.
Similar to the setting of multi-oracle Hamiltonian simulation, we obtain a leading weighted cost $\kappa\langle\boldsymbol{C},\boldsymbol{\lambda}\rangle_{1/2}$ for queries to the block-encoding, and additionally obtain a weighted cost for state preparation queries.
We also obtain similar improvements for the less-studied case where $H=A^2$ with Hermitian $A$ is positive definite~\cite{Orsucci2021PositiveSystems} and we are provided with the block-encoding $\Be{A/\sqrt{\Lambda}}$, which makes it amenable to spectral amplification.
\begin{table}[!htbp]
\centering\small
\begin{tabularx}{\linewidth}{lXll}
\hline
\hline
\multicolumn{4}{c}{\textbf{Quantum linear systems problem}}
\\
\hline
Year & Construction & $Q_H=\mathcal{O}(\cdots)$ & $Q_b=\mathcal{O}(\cdots)$
\\
\hline
2009 \cite{Harrow2009} & Phase estimation & $\frac{\kappa}{\epsilon^2\sqrt p}$ & $\frac{1}{\sqrt{p}}$
\\
2017 \cite{Childs2015LinearSystems} & LCU of quantum walks + VTAA & $\kappa\polylog\frac{\kappa}{\epsilon}$ & $\kappa\polylog\frac{\kappa}{\epsilon}$
\\
2019 \cite{Gilyen2018singular} & QSVT reciprocal & $\kappa \frac{1}{\sqrt{p}}\log\frac{1}{\epsilon\sqrt p}$ & $\frac{1}{\sqrt{p}}$
\\
2022 \cite{Costa2021linearsystems} & Discrete adiabatic evolution & $\kappa\log\frac{1}{\epsilon}$ & $\kappa\log\frac{1}{\epsilon}$
\\
2024 \cite{Dalzell2024LinearSystems} & Kernel reflection & $\kappa\log\frac{1}{\epsilon}$ & $\kappa\log\frac{1}{\epsilon}$
\\
2024 \cite{Low2024QLSP} & Deterministic VTAA + GPE & $\kappa\log\frac2p\log\frac{\log(4/p)}\epsilon$ & $\frac{1}{\sqrt{p}}$
\\
\hline
\multirow{2}{*}{This work} & Two-oracle~\cref{wql:cor:single} & $\kappa\log\frac{1}{\epsilon}$ & $\frac{1}{\sqrt{p}}$
\\
& Multi-oracle~\cref{wql:thm:main} & $(\kappa\langle\boldsymbol C,{\boldsymbol\lambda}\rangle_{1/2}+C_\Sigma)\log\frac{1}{\epsilon}$ & $\frac{1}{\sqrt{p}} \langle\boldsymbol c,|\boldsymbol a|\rangle_{1/2} +c_\Sigma$
\\
\hline
\hline
2021 \cite{Orsucci2021PositiveSystems} & Square-root-access+QSVT & $\frac{\sqrt\kappa}{\sqrt{p}}\log\frac{\kappa}{\epsilon}$ & $\frac{1}{\sqrt{p}}$
\\
2021 \cite{Orsucci2021PositiveSystems} & Square-root-access+QSP+VTAA & $ \frac{\sqrt\kappa}{p^{1/4}}\polylog\frac{\kappa}{\epsilon}$ & $\sqrt{\log\kappa}+\frac{1}{\sqrt{p}}$
\\
\hline
This work & Square-root-access~\cref{wql:cor:single} & $\sqrt\kappa(\frac{1}{p^{1/4}}+\log\frac{1}{\epsilon})$ & $\frac{1}{\sqrt{p}}$
\\
\hline
\hline
\end{tabularx}
\caption{Query complexity $Q_H$ to the block-encoding and $Q_b$ to state preparation to obtain the normalized quantum state $\propto H^{-1}\ket{b}$ with constant success probability and error $\epsilon$.
Note that the single-oracle MQSP complexity is a specialization of the general multi-oracle result that accesses independent block-encodings $\Be{H_j/\lambda_j}$ and $O_{b,j}$ with costs $C_j$ and $c_j$ respectively.
To simplify the presentation, we normalize $\Lambda=C_\Sigma=c_\Sigma=1$ so $\lambda_j,C_j$, and $c_j$ are fractions summing to one.
The multi-oracle row gives weighted costs; the square-root-access MQSP row counts queries to a Hermitian factor.
}
\label{tab:intro-alg-qlsp}
\end{table}

The query complexity can also depend on the weights of the solution along the right singular vectors, rather than the smallest singular value of the whole matrix.
For one block encoding and one state preparation, let $\vect x=H^{-1}|b\rangle$, $\|H\|\le1$, and suppose $1\le r_0\le\|\vect x\|$ is given.
Writing $|H|=(H^\dagger H)^{1/2}$, \cref{cert:thm:tail} gives
\begin{align}
\left\|\mathbf1_{[0,1/K)}(|H|)\vect x\right\| \le c\epsilon r_0
\quad\Longrightarrow
\quad Q_H=O\left(K\log\frac1\epsilon\right),
\quad Q_b=O\left(\frac K{r_0}\right),
\label{eq:intro-qlsp-tail}
\end{align}
for $K\ge2$ and a sufficiently small universal constant $c>0$.
Neither the singular projector nor a modified preparation is implemented.
Known inverse-moment bounds give sufficient cutoffs through
\begin{align}
\left\|\mathbf1_{[0,1/K)}(|H|)\vect x\right\| \le \frac{\bigl\||H|^{-q}\vect x\bigr\|}{K^q},
\quad q>0,
\label{eq:intro-qlsp-moments}
\end{align}
while spectral-band information can give tighter bounds.
These results build on state-dependent truncation~\cite{Li2025BeyondConditionNumber,Dalzell2026BeyondConditionNumber} where the cutoff itself may depend on $\epsilon$.
Classical algebraic information can also avoid singular truncation: we give a nonnormal example with a constant-degree, constant-normalization inverse approximation on the specified input, although every admissible inverse cutoff parameter $K$ grows exponentially with the matrix dimension.

Our general QLSP results can be further specialized to exploit structure in the equations and the requested output.
Dalzell, Li, and Su~\cite{Dalzell2026BeyondConditionNumber} develop the block-encoding interpretation of Li's row-weighted method~\cite{Li2025BeyondConditionNumber} through row normalization and affine dilation.
Our solver applies to these transformed systems once their encoding normalization, inverse-state bounds, and output-selection probability have been established.
Separately encoded components also allow weighted scheduling to reduce calls to expensive terms with small normalization coefficients.
Combining this saving with a restricted inverse bound requires the common invariance condition in~\cref{cert:eq:weighted-invariance}, or a corresponding stability analysis.
Other applications may need only a readout $C\vect x$, or may leave a residual of the linear equation after the coarse finite-clock computation on which the inverse is easier to approximate.
Bounds for that output or residual, and rational approximations adapted to the available classical information, can then guide a different implementation.
Each application could be significantly improved with a bespoke analysis of its normalization, error, and success analysis.

\subsubsection{Linear differential equations}
\label{sec:intro-alg-autonomous}
Quantum algorithms for linear differential equations prepare with constant success probability the normalized solution state $\ket{x(T)}$ to error $\epsilon$ at final time $T$ of the inhomogeneous differential equation
\begin{align}
\boldsymbol{x}'(t)=-H(t)\boldsymbol{x}(t)+\boldsymbol b(t),
\quad \boldsymbol x(0)=\boldsymbol x_0=\ket{\boldsymbol{x}_0},
\quad \|\boldsymbol x(0)\|=1.
\label{eq:intro_linear_DE}
\end{align}
In general, the matrix $H(t)$ is a time-dependent arbitrary square matrix, and the inhomogeneous forcing term $\boldsymbol b(t)$ is also time-dependent.
Note that $H(t)$ can be a non-normal matrix, compared to the Hermitian case of Hamiltonian simulation.
Following convention, $H(t)$ is accessed via the $\textsc{Ham-T}$ block-encoding, and time-controlled preparation of $\boldsymbol b(t)/\beta$ as a successful output is assumed.
\begin{align}
\textsc{Ham-T}[H/\lambda]=\Be{\sum_{t}\ketbra{t}\otimes \frac{H(t)}{\lambda}},
\quad O_x\ket{0}=\ket{\boldsymbol{x}_0},
\quad (\bra{0}_f\otimes I)O_b\ket{t}\ket{0}=\ket{t}\boldsymbol b(t)/\beta.
\end{align}
In practice, a finite discretization of the time variable $t$ is required, whose fineness does not affect the stated query bounds and contributes only a logarithmic factor to the quantum gate and ancilla qubit complexity.
Note that the block-encoding normalization $\lambda$ and upper bound on state normalization $\|\boldsymbol{b}(t)\|\le\beta$ are assumed to be known.
In the time-independent case, we simply drop the control by $\ket{t}$.

Algorithms for solving~\cref{eq:intro_linear_DE} tend to be sensitive to the manner in which $H$ is not a normal matrix.
For instance, algorithms based on a Linear-Combination-of-Hamiltonian-Simulation (LCHS)~\cite{an2023lchs} typically require that time-evolution by $H(t)$ is contractive, with $H^\dagger(t)+H(t)\succeq0$~\cite{Krovi2022LinearDE}.
For time-independent problems, Taylor~\cite{Berry2017Differential} and Faber~\cite{Low2024Eigenvalue} expansions admit bounds in terms of an eigendecomposition $H=S\Sigma S^{-1}$ with $\operatorname{Re}\Sigma\succeq0$ and condition number $\kappa=\|S\|\|S^{-1}\|$.
Eigenvalue processing also admits analyses using numerical-range or resolvent bounds, which can apply to defective matrices without assuming such an eigendecomposition~\cite{Low2024Eigenvalue}.
There are also algorithms treating the more general class of arbitrary $H(t)$, but require additional bounds on time-evolved matrix and state norms.

The comparison~\cref{tab:intro-alg-de-prior} of these different quantum algorithms for linear differential equations is simplified by focusing on the homogeneous time-independent case, so $H(t)=H$ and $\boldsymbol{b}(t)=0$.
This comparison remains representative as algorithms for the inhomogeneous case tend to use the algorithm for the homogeneous case as a subroutine.
Moreover, the time-dependent cases are typically obtained by elevating an internal Taylor series or Hamiltonian simulation subroutine to their time-dependent analogues.
Within this framework, the complexities are expressed in terms of the following matrix and state bounds.
\begin{align}
\max_{t\in[0,T]}\|e^{-Ht}\|\le C,
\quad \max_{t\in[0,T]}\|e^{-Ht}\ket{\boldsymbol{x}_0}\|\le g,
\quad \|e^{-HT}\ket{\boldsymbol{x}_0}\|\ge\sqrt{p}.
\end{align}
The seminal algorithm by Berry~\cite{Berry2014Differential} proved that $\ket{x(T)}$ could be prepared using a polynomial number of queries to both Hamiltonian and state preparation oracles in time, error, and condition number $C=\kappa$.
Subsequent improvements reduced this to nearly linear scaling in time and logarithmic scaling in error~\cite{Berry2017Differential}, and improved QLSP solvers combined with block-preconditioning~\cite{Low2024QLSP} provided a separation in query complexity between block-encoding and state preparation oracles~\cite{Low2024Eigenvalue}.
A significant improvement to the overall two-oracle complexity was discovered by specializing~\cite{Krovi2022LinearDE} to semi-contractive time-evolution $\max_{t\ge0}\|e^{-Ht}\|\le C=1$.
In this case, it was found that LCHS, and the related Schr\"{o}dingerization~\cite{Jin2024Schrodingerization}, approach provided optimal $\Theta(1/\sqrt{p})$~\cite{an2023lchs} scaling to the state preparation oracle.
This was improved using different kernel functions~\cite{an2023lchsoptimal,Jin2025Schrodingerization} to block-encode the time-evolution $\Be{e^{-HT}}$ using $\mathcal{O}(\lambda T\log\frac{1}{\epsilon})$ queries~\cite{Low2025NonunitaryDynamics} to the block-encoding of $H$, which was conjectured to be optimal.
Another important approach is time-marching~\cite{Fang2023Marching} and further improvements to the Taylor series method~\cite{Krovi2022LinearDE}, which are applicable to arbitrary matrices with bounded $C$, but tend to have worse complexity.

We use MQSP in~\cref{sec:alg-volterra} to obtain general bounds for solving linear differential equations that separate the costs of matrix and state-preparation queries.
Our method works for arbitrary non-normal matrices and makes $\mathcal{O}(C/\sqrt{p})$ queries to state preparation, which is optimal in the contractive case, and $\mathcal{O}(\frac{\lambda TCg}{\sqrt{p}}\log\frac{1}{\epsilon})$ queries to the matrix block-encoding.
By substituting appropriate values of $C,g$, and $p$, this accommodates the matrix assumptions of the disparate algorithms for different forms of matrix non-normality.
In the contractive case with a single block-encoding of $H$, this removes the additional $\log\frac{1}{p}$ dependence incurred by applying amplitude amplification to an approximation of the final time-evolved state when $\|\boldsymbol{x}(T)\|\ge\sqrt p$ is small, and additional structure can favor more specialized solvers.
The general time-dependent complexity is expressed in terms of additional generalized parameters as in~\cref{thm:vde-constant}.
\begin{subequations}
\label{eq:intro-de-parameters}
\begin{align}
\max_{0\le s\le t\le T}\left\|\mathcal{T}\left[e^{-\int_{s}^tH(s')\mathrm{d}s'}\right]\right\|&\le C,
\quad \max_{t\in[0,T]}\|\boldsymbol{x}(t)\|\le g,
\quad \|\boldsymbol{x}(T)\|\ge\sqrt{p},
\\
\quad \max_{t\in[0,T]}\|\boldsymbol b(t)\|&\le\beta,
\quad B=\sqrt{1+T^2\beta^2},
\quad g\ge\beta/(2\lambda).
\end{align}
\end{subequations}
Tighter bounds are possible when additional time-dependent upper bounds are given.
A two-time envelope retains the dependence of propagation on both endpoints:
\begin{align}
\left\|\mathcal{T}\left[e^{-\int_{t_i}^{t_f}H(s)\mathrm{d}s}\right]\right\| \le C(t_i,t_f),
\quad 0\le t_i\le t_f\le T.
\label{eq:intro-de-time-dependent-envelope}
\end{align}
Time-dependent oracle normalizations can also reduce costs through norm rescaling~\cite{Berry2020TimeDependentL1}.
More powerful state preparation methods~\cite{an2023lchs} can use a known bound $\|\boldsymbol{b}(t)\|\le\beta(t)$, with coherent access to its time-dependent magnitude.
We give both refinements below.
As with our Hamiltonian simulation and quantum linear systems results, we also provide the multi-oracle complexity where $H=\sum_jH_j$ and $\boldsymbol{x}_0=\sum_{j}a_j\ket{\boldsymbol{x}_j}$ are themselves a linear combination of terms encoded by independent oracles with non-uniform costs.
In this most general case, the forcing columns have norm at most one and $\beta=\sum_k\beta_k$;
\begin{subequations}
\label{eq:intro-vde-multi-input}
\begin{align}
H(t)&=\sum_jH_j(t)
\quad\text{in}
\quad\textsc{Ham-T}[H_j/\lambda_j],
\quad \boldsymbol x_0=\sum_{k=0}^{s-1}\xi_k|v_k\rangle,
\quad\boldsymbol b(t)=\sum_{k=0}^{q-1}\beta_k\boldsymbol v_k(t),
\\
\boldsymbol\omega&= (\xi_0/\sqrt2,\ldots,\xi_{s-1}/\sqrt2,T\beta_0,\ldots,T\beta_{q-1}),
\quad \boldsymbol c= (c_{0,0},\ldots,c_{0,s-1},c_{b,0},\ldots,c_{b,q-1}).
\end{align}
\end{subequations}
Like the Taylor and Dyson approaches, our algorithms are based on embedding into a system of linear equations, but this time solved with our weighted-cost multi-oracle variant (\cref{sec:intro-alg-linear-systems}).
\begin{table}[!htbp]
\centering\small
\begin{tabularx}{\linewidth}{llXcc}
\hline
\hline
\multicolumn{5}{c}{\textbf{Linear differential equations}}
\\
\hline
Year&Construction & Constraint & $O_H$ & $O_x, O_b$
\\
\hline
2017~\cite{Berry2017Differential} & Taylor series & \multirow{2}{*}{$\operatorname{Re}\Sigma\succeq0$} & \multicolumn{2}{c}{$\frac{\kappa Tg}{\sqrt{p}}\polylog(\kappa g T/(\sqrt p\epsilon))$}
\\
2024~\cite{Low2024Eigenvalue} & Faber series&&\multicolumn{2}{c}{$\widetilde O(\frac{\kappa^2}{\sqrt p}(T+\log\frac{\kappa}{\sqrt p\epsilon}))$}
\\
\hline
2023~\cite{an2023lchsoptimal}&LCHS, any $\nu>1$ &\multirow{3}{*}{$H+H^\dagger\succeq0$} & $\widetilde O( \frac{T}{\sqrt{p}}\log^{\nu}\frac{1}{\epsilon})$ & $\frac{1}{\sqrt{p}}$
\\
2024~\cite{Berry2024Dyson}&Dyson series & & $\widetilde O( \frac{T}{\sqrt{p}}\log^2\frac{1}{\epsilon})$ & $ \frac{T}{\sqrt{p}}\log\frac{1}{p\epsilon}$
\\
2025~\cite{Low2025NonunitaryDynamics}&LCHS & & $ \frac{T}{\sqrt{p}}\log\frac{1}{p\epsilon}$ & $\frac{1}{\sqrt{p}}$
\\
\hline
2023~\cite{Fang2023Marching}&Time marching & \multirow{2}{*}{None} & $\widetilde O(T^2\mathcal A_T\log\frac{1}{\epsilon})$ & $\mathcal A_T$
\\
2023~\cite{Krovi2022LinearDE} &Taylor series&& \multicolumn{2}{c}{$\widetilde O(\frac{CTg}{\sqrt{p}}\polylog(1/\epsilon))$}
\\
\hline
\multirow{2}{*}{MQSP}& Homogeneous & \multirow{4}{*}{None} & $\frac{CTg}{\sqrt{p}}\log\frac{1}{\epsilon}$ & $\frac{C}{\sqrt{p}}$
\\
& Inhomogeneous & & \multirow{2}{*}{$\frac{CTg}{\sqrt{p}}\log\frac{1}{\epsilon}$} & \multirow{2}{*}{$\frac{CB}{\sqrt{p}}$}
\\
\cref{thm:vde-constant}& Time-dependent & & &
\\
\cref{thm:vde-multi}& General multi-oracle & & $\big(\frac{CTg}{\sqrt{p}}\langle\boldsymbol C,\boldsymbol\lambda\rangle_{1/2} +C_\Sigma\big)\log\frac{1}{\epsilon}$ & $\frac{C}{\sqrt{p}}\langle\boldsymbol c,|\boldsymbol\omega|\rangle_{1/2} +c_\Sigma$
\\
\hline
\hline
\end{tabularx}
\caption{Query complexity $Q_H$ to the block-encoding and $Q_x$ to state preparation of $\ket{\boldsymbol{x}_0}$ to obtain the normalized solution state of~\cref{eq:intro_linear_DE} with constant success probability and error $\epsilon$.
Note that the single-oracle MQSP complexity for the homogeneous, inhomogeneous, and the combination with time-dependence are all specializations of the general time-dependent multi-oracle result that accesses independent block-encodings $\textsc{Ham-T}[H_j/\lambda_j]$ and similarly for the initial state and forcing terms.
To simplify the presentation, we use $\Lambda=C_\Sigma=c_\Sigma=1$ and $T\ge1$; the multi-oracle row gives weighted costs.
The Taylor rows quote sparse-oracle bounds at fixed sparsity.
For time marching, $\mathcal A_T=\frac{1}{\sqrt p}\prod_l\big\|\mathcal{T}[e^{-\int_{t_l}^{t_{l+1}}H(s)\mathrm{d}s}]\big\|$ depends on the chosen mesh.
The notation $\tilde{\mathcal{O}}$ hides polylogarithmic factors in its arguments.}
\label{tab:intro-alg-de-prior}
\end{table}

Classical information about the trajectory and the encoded linear system can sharpen these bounds.
Quantum history-state algorithms already use root-mean-square solution norms and classical stability information like Lyapunov bounds~\cite{Jennings2024CostLinearDE}, building on the Taylor-history approach \cite{Berry2017Differential,Krovi2022LinearDE}.
We use these kinds of bounds to balance the final-state output while retaining separate costs for matrix and state-preparation queries.
With $m=\Theta(\lambda T)$, a bound on the integrated solution norm supplies a mean-square history bound
\begin{align}
W_2=O\left(\max\left\{ \sqrt p,\, \left(\frac1T\int_0^T\|x(t)\|^2\,dt\right)^{1/2},\, \frac{B}{m} \right\}\right).
\end{align}
After balancing the final-state output using $W_2$, a certified inverse cutoff parameter $K_\epsilon$ gives
\begin{align}
Q_H=O\left(K_\epsilon\log\frac1\epsilon\right),
\;\; Q_x+Q_b=O\left(\frac{BK_\epsilon}{mW_2}\right),
\;\; K_\epsilon=O\left(\frac{CmW_2}{\sqrt p}\right)
\;\;\text{always suffices}.
\end{align}
A smaller inverse cutoff parameter is valid when the component of the unnormalized inverse state below singular value $1/K_\epsilon$ has norm at most a sufficiently small constant times $\epsilon mW_2/B$.
Thus a smaller history norm reduces matrix queries, while an additional bound on the small-singular-value component of the inverse state can reduce both oracle counts.
Two-time bounds $\|U(t,s)\|\le f(t,s)$ retain more information than a single worst-case constant $C$.
Their values on the coarse mesh give a scalar lower-triangular Green matrix.
Bounds $k\ge\|G\|$ and $\gamma\ge\|PG\|$ on the history inverse and its final row give
\begin{subequations}
\label{eq:intro-de-two-time}
\begin{align}
Q_H&=O\left(\left[k+(\gamma+1) \frac{\sqrt m W_2}{\sqrt p}\right] \log\frac1\epsilon\right),
\\
Q_x+Q_b&=O\left(\frac{B(k+1)}{mW_2} +\frac{B(\gamma+1)}{\sqrt{mp}}\right).
\end{align}
\end{subequations}
The construction and its discretization allowance are given in \cref{vdc:thm:green,vdc:prop:envelope}.
Decay with the separation $t-s$ can reduce both costs through these two norms.
If every generator term preserves a common reducing subspace containing all initial states and forcing vectors, the analysis can also be restricted to that subspace, without implementing its projector.

The constant $\beta$ above is the normalization supplied by the black-box preparation oracle.
Stronger access can replace $T\beta$ by $I_\beta=\int_0^T\beta(t)\,dt$.
Specifically, suppose the successful preparation column is $\boldsymbol b(t)/\beta(t)$, and the magnitude envelope and the cumulative time change can be evaluated coherently.
\Cref{vde:cor:source-integral} chooses a clock density proportional to $\max\{1/T,\beta(t)/I_\beta\}$ and obtains
\begin{align}
Q_H=O\left(\frac{C\lambda TW_\rho}{\sqrt p} \log\frac1\epsilon\right),
\quad Q_x+Q_b=O\left(\frac{C\sqrt{1+I_\beta^2}}{\sqrt p}\right).
\label{eq:intro-de-source-integral}
\end{align}
Here $W_\rho$ is a history bound in the new clock, including the lower bound accounting for the orthogonal completion amplitudes of the preparation.
A peak trajectory bound remains valid, while an integrated bound is reweighted by the clock density.
This $L^1$ dependence on the forcing normalization is also present in LCHS \cite{an2023lchs,Low2025NonunitaryDynamics} and rescaling time realizes it in the  solver, following the norm-rescaling principle of \cite{Berry2020TimeDependentL1}.

Although our general bounds significant improve upon many generic quantum linear differential equation solver, they do not supersede every bespoke case.
For example, LCHS with two separate block-encoding oracles for the Hamiltonian and dissipative terms can have better complexity than our general single-matrix bound, since the precision-dependent factor need not multiply the Hamiltonian block-encoding normalization~\cite{Low2025NonunitaryDynamics}.
Eigenvalue processing can also give better bounds for particular defective matrices when useful numerical-range or resolvent information is available~\cite{Low2024Eigenvalue}.
For driven systems with strict dissipation, exponential decay allows contributions from the distant past to be discarded~\cite{Yang2025Dissipative}.
Access to the square root of a positive semidefinite generator permits square-root dependence on time~\cite{Low2025NonunitaryDynamics}.
An efficiently accessible eigensystem allows the exponential to be implemented directly in the eigenbasis~\cite{An2022Differential}.
We expect similarly bespoke MQSP constructions to improve on these specialized bounds.

\subsubsection{Ground-state preparation and energy estimation}
\label{mqspgsp:sec:intro}
The quantum ground state preparation and ground state energy estimation problems are related.
In these problems, we are provided with a block-encoding of the Hamiltonian $\Be{H/\lambda}$ and a unitary $B$ that prepares a trial state, or guiding state, with overlap at least $\sqrt{p_0}$ with the ground state $\ket{\psi_{\rm g}}$.
The ground state is the lowest energy eigenstate of $H$.
These problems may also provide additional assumptions, such as a lower bound on the gap $E_1-E_0\ge\gamma_0$ between the ground and first excited states, or also an energy threshold $E_{\rm T}$ that is guaranteed to lie between $E_0$ and $E_1$.
In the preparation problem, the goal is to prepare $\ket{\psi_{\rm g}}$ with error $\epsilon$ and constant success probability, given either a gap promise $\gamma_0>0$ or threshold promise $E_{\rm T}$.
In the estimation problem, the goal is to estimate $E_0$ to error $\epsilon_E$ and confidence $1-q$.
Typically, this proceeds by first preparing the ground state, followed by a high-confidence variant of quantum phase estimation, such as with Kaiser windows~\cite{Berry2024Topological}, or a fuzzy binary search by quantum signal processing~\cite{King2025SOSSA,Low2024QLSP} which uses only a few ancilla qubits.
Energy estimation does not require a spectral gap: states within the requested energy resolution of the ground energy are also acceptable.
Our gap-free result uses this weaker target directly.

Quantum algorithms for this problem are typically some combination of thresholded ternary search, amplitude amplification, and eigenstate filtering~\cite{Ge2019GroundStatePreparation,Lin2020Ground}.
For instance, one may use quantum phase estimation or a shifted sign function in QSP to flag, or filter, all eigenstates below some known upper bound on the ground state energy.
The target accuracy of the filter is finite, say $\epsilon_j$ at iteration $j$, which costs $Q_{H,j}=\mathcal{O}(\frac{1}{\epsilon_j}\log\frac{1}{q})$ queries to $\Be{H/\lambda}$, which splits the spectrum into three parts: a lower and upper portion that flags all states as $\ket{0}$ and $\ket{1}$ respectively with high confidence $1-q$, and a middle portion of width $\epsilon_j$ where this flag is in an indeterminate superposition.
Then it suffices to use at most $\mathcal{O}(\frac{1}{\sqrt{p_0}}\log\frac{1}{q})$ rounds of amplitude amplification and estimation to decide whether the ground state is not contained in either the lower or middle portions.
This uses the same number of queries $B,B^\dagger$, and a factor $Q_{H,j}$ more to the block-encoding.
The interval is then narrowed by, for instance a factor of three, and the fuzzy search proceeds to the next iteration $j+1$.
The search can stop at the requested energy resolution for energy estimation.
Preparing the individual ground state additionally requires a separating threshold or gap and a filter accurate enough for the desired state error.
Other alternatives exist, such as by using a generic quantum minimum-finding algorithm as a subroutine~\cite{Mande2026QPEBounds}, or by exploiting additional structure in Hamiltonians, such as if it has a sum-of-squares representation with a good sum-of-squares gap $E_{\rm g}>E_0$~\cite{King2025SOSSA}.

The overall cost can be optimized, such as by a non-uniform sequence of confidence parameters $q_j$~\cite{Berry2024Rapid,King2025SOSSA}, but the iterative filter-amplification approach generally suffers from the logarithmic overheads of both composing two quantum algorithms and a classical outer loop.
This makes ground state problems especially suited to the quantum transducer approach due to its properties of exactness and lossless composition.
However, the challenge remains finding an appropriate transducer that solves the problem.
A compelling construction was recently demonstrated by Jeffery and Witteveen~\cite{Jeffery2026GuidedEnergy} using a quantum transducer for the phase-threshold decision problem.
This was combined with a classical interval search to obtain the current best quantum algorithms for estimating the ground state energy.
We use MQSP in~\cref{mqspgsp:sec:ground-state} to further improve the per-oracle query complexity for ground state preparation and estimation under various promises of finite gap $\gamma_0$, energy threshold $E_{\rm T}$, or gapless $\gamma_0=0$, and tabulate our results in comparison to prior art in~\cref{mqspgsp:tab:comparison}.
Overall, the algorithm works similarly to the previous transducer construction: A coarse candidate state with constant overlap with the ground state is prepared, followed by filtering to the desired accuracy.
With a gap promise, coherent selection of coarse energy labels supplies a candidate and a threshold; a median of $O(\log(1/\epsilon))$ samples gives the required error.
Once preparation succeeds, a high-confidence phase measurement has additive cost $O((\lambda/\epsilon_E)\log(1/q))$ to the block-encoding.
If only the energy is required, the coarse-label sampler itself is sufficient.
For gap-free energy estimation, the block-encoding-query bound matches the joint lower bound of Somma and de Wolf~\cite{Somma2026GroundLowerBounds}.

\begin{table}[!b]
\centering
\small
\begin{tabularx}{\textwidth}{ccXcc}
\hline
\hline
Year & Input & Method & Queries to $\Be{H/\lambda}$ & Queries to $B,B^\dagger$
\\
\hline
\multicolumn{5}{c}{\textbf{Ground-state preparation}}
\\
2019~\cite{Ge2019GroundStatePreparation} & $\gamma_0>0$ & Filtering and min-finding & $\widetilde O\big(\frac{\lambda^{3/2}}{\gamma_0^{3/2}\sqrt{p_0}}\big)$ & $\widetilde O\big(\frac{\sqrt{\lambda/\gamma_0}}{\sqrt{p_0}}\big)$
\\
2020~\cite{Lin2020Ground} &$E_{\rm T}$& Polynomial filtering & $\frac{\lambda}{\gamma_0\sqrt{p_0}} \log\frac1{\epsilon\sqrt{p_0}}$ & $\frac{1}{\sqrt{p_0}}$
\\
2020~\cite{Lin2020Ground} &$\gamma_0>0$& Search \& filtering & $ \tilde{\mathcal{O}}\big(\frac{\lambda\log\frac{1}{\epsilon}}{\gamma_0\sqrt{p_0}} \big)+\frac{\lambda \log\frac1\epsilon}{\gamma_0\sqrt{p_0}} $ &$ \frac{\log\frac\lambda{\gamma_0}}{\sqrt{p_0}} \log\frac{\log(\lambda/\gamma_0)}{\epsilon} $
\\
\hline
\multirow{2}{*}{\cref{mqspgsp:thm:gsp-main}}&$E_{\rm T}$& Coarse $\ket{\psi_{\rm g}}$ \& filtering. & $\frac\lambda{\gamma_0}\left(\frac1{\sqrt{p_0}}+\log\frac1\epsilon\right)$ & $\frac{1}{\sqrt{p_0}}$
\\
&$\gamma_0>0$& Binning \& median & $\frac{\lambda}{\gamma_0\sqrt{p_0}}\log\frac1\epsilon$ & $\frac1{\sqrt{p_0}}\log\frac1\epsilon$
\\
\hline
\hline
\multicolumn{5}{c}{\textbf{Ground-state energy estimation}}
\\
2020~\cite{Lin2020Ground} &$\gamma_0=0$& Thresholded search & $\frac{\lambda\log\frac\lambda{\epsilon_E}}{\epsilon_E\sqrt{p_0}} \log\frac1{p_0} \log\frac{\log(\lambda/\epsilon_E)}q$ & $\frac{\log\frac\lambda{\epsilon_E}}{\sqrt{p_0}} \log\frac{\log(\lambda/\epsilon_E)}q$
\\
2024~\cite{Berry2024Rapid} & $\gamma_0=0$ & Thresholded search & $\frac{\lambda}{\epsilon_E \sqrt{p_0}}\log\frac1{p_0}\log\frac1q$ & $\frac1{\sqrt{p_0}}\log\frac{\lambda}{\epsilon_E}\log\frac{\log\frac{\lambda}{\epsilon_E}}q$
\\
2025~\cite{King2025SOSSA} & $\gamma_0=0$ & Spectrum-amplified search & $\frac{\sqrt{\max\{\epsilon_E,\lambda+E_{\rm g}\}\lambda}}{\epsilon_E \sqrt{p_0}}\log\frac1{p_0}\log\frac1q$ & $\frac1{\sqrt{p_0}}\log\frac{\lambda}{\epsilon_E}\log\frac{\log\frac{\lambda}{\epsilon_E}}q$
\\
2026~\cite{Mande2026QPEBounds} &$\gamma_0=0$& QPE \& min-finding & $\frac\lambda{\epsilon_E\sqrt{p_0}}\log\frac1{p_0}\log\frac1q$ & $\frac1{\sqrt{p_0}}\log\frac1q$
\\
2026~\cite{Jeffery2026GuidedEnergy} &$E_{\rm T}$& Decision \& search$^{\dagger}$ & $\frac\lambda{\epsilon_E\sqrt{p_0}}\log\frac1q$ & $\frac{\log\frac\lambda{\epsilon_E}}{\sqrt{p_0}} \log\log\frac\lambda{\epsilon_E}\log\frac1q$
\\
\hline
\cref{mqspgsp:prop:gsp-gap-free-energy}&$\gamma_0=0$& Binning \& median & $\frac\lambda{\epsilon_E\sqrt{p_0}}\log\frac1q$ & $\frac1{\sqrt{p_0}}\log\frac1q$
\\
\multirow{2}{*}{\cref{mqspgsp:prop:gsp-energy-after-preparation}} &$E_{\rm T}$&\multirow{2}{*}{Coarse $\ket{\psi_{\rm g}}$ \& QPE} & $\frac\lambda{\gamma_0} \left(\frac1{\sqrt{p_0}}+\log\frac1q\right) +\frac\lambda{\epsilon_E}\log\frac1q$ & $\frac{1}{\sqrt{p_0}}$
\\
&$\gamma_0>0$& & $\left(\frac\lambda{\gamma_0\sqrt{p_0}} +\frac\lambda{\epsilon_E}\right)\log\frac1q$ & $\frac1{\sqrt{p_0}}\log\frac1q$
\\
\hline
\hline
\end{tabularx}
\caption{Separate oracle costs for ground-state preparation and energy estimation.
Let $H$ have ground and first excited energies $E_0,E_1$, let $\lambda\ge\|H\|$, and let $B|0\rangle$ have ground-state probability at least $p_0>0$.
Preparation is with constant success probability and error $\epsilon$; energy precision $\epsilon_E$ is an additive error, with confidence $1-q$.
All entries are big-$\mathcal{O}$ bounds.
We assume either a supplied energy threshold $E_{\rm T}$ with $E_0\le E_{\rm T}-\gamma_0/2<E_{\rm T}+\gamma_0/2\le E_1$; or a gap $E_1-E_0\ge\gamma_0$.
$^{\dagger}$ counts queries to $e^{iH/\lambda}$.}
\label{mqspgsp:tab:comparison}
\end{table}

\subsubsection{Generalized eigenvalue estimation}
\label{sec:intro-alg-eigenvalues}
The quantum phase estimation algorithm~\cite{Kitaev1997Diamond} underlies Shor's factoring algorithm~\cite{sho_polynomialtime_1997}.
In the case where the quantum eigenphase of the unitary is generated by Hamiltonian time-evolution of block-encoded Hermitian matrices, the problem reduces to estimating the eigenvalue $E$ of a target eigenstate $H\ket{\psi}=E\ket{\psi}$.
This problem underlies many other applications of quantum computation, such as estimating ground state energies of highly-correlated quantum states.
Typically, one is not provided with the target eigenstate.
Instead, it is assumed that there is a state preparation unitary $O_\psi\ket{0}=\ket{\psi_{\rm{trial}}}$ that prepares a trial state with finite overlap $|\langle{\psi}|\psi_{\rm{trial}}\rangle|\ge\sqrt{p}>0$, and one is provided with a lower bound $\sqrt{p}$.
The reduction to phase estimation uses Qubitization~\cite{Low2016Qubitization}, which converts a block-encoding $O_H=\Be{H/\Lambda}$ to a unitary quantum walk whose eigenphases, by the spectral mapping theorem, are $\pm\arccos(E/\Lambda)$.
Both branches recover the same signed energy through $E=\Lambda\cos\varphi$.
Using modern phase estimation techniques, such as Kaiser-window phase estimation~\cite{Berry2024Topological}, an exact eigenstate gives an estimate of $E$ to error $\epsilon$ and confidence $1-q$ using $\mathcal{O}(\frac{\Lambda}{\epsilon}\log\frac{1}{q})$ queries to $O_H$ and one eigenstate preparation.
With a trial state, selecting a designated eigenvalue requires an additional promise, as in ground-state estimation by fuzzy binary search~\cite{Lin2020Ground,Berry2024Rapid,King2025SOSSA}; otherwise one obtains a distribution of estimates.

The generalized eigenvalue problem is significantly more challenging.
Given Hermitian $A$ and positive-definite $B$ block-encoded in $\Be{A/\lambda_A}$ and $\Be{B/\lambda_B}$, the goal is to estimate the eigenvalue $\lambda$ of a target eigenstate $\boldsymbol{\psi}$ under the following promises:
\begin{align}
A\boldsymbol{\psi}=\lambda B\boldsymbol{\psi},
\quad\beta I\preceq B\preceq b_{\max}I,
\quad \kappa_B=b_{\max}/\beta,
\quad \operatorname{spec}(B^{-1/2}AB^{-1/2})\subseteq[-R,R].
\end{align}
Here the transformation must account for access to both $A$ and $B$.
A variety of techniques can be found in prior art for solving this problem.
One approach rescales the state $\boldsymbol{\psi}= B^{-1/2}\ket{\psi'}$ and $A\rightarrow A'=B^{-1/2}AB^{-1/2}$.
This then becomes a single-matrix eigenvalue problem $A'\ket{\psi'}=\lambda\ket{\psi'}$.
However, this incurs additional cost, such as by using QSVT to block-encode factors of $B^{-1/2}$.
The transformed matrix $A'$ is Hermitian, so Hamiltonian simulation applies once its block-encoding has been constructed.
Other prior approaches rely on general algorithms for linear differential equations~\cite{Shao2020GeneralizedEigenvalue}, such as described in~\cref{sec:intro-alg-autonomous}, or complicated eigenvalue processing algorithms~\cite{Low2024Eigenvalue}.
Note that even using our improved weighted-cost linear differential equation solver at this point would not avoid the multiplicative cost of block-encoding $B^{-1/2}$ from $B$.

We use MQSP in~\cref{subsec:app-generalized-eigenvalues} to solve this problem with weighted cost to each oracle.
For the purposes of comparison with prior art in~\cref{tab:gep-comparison}, we assume that we start with the eigenstate $\ket{\psi}$.
However, we also provide an analysis of the finite-overlap situation, comparable to the Hermitian case.
Overall, we estimate $\lambda$ using the following query complexities to each oracle:
\begin{align}
Q_A=\mathcal{O}\left(\frac{\lambda_A}{\beta\epsilon}\log\left(\frac{1}{q}\right)\right),
\quad Q_B=\mathcal{O}\left(\frac{R\lambda_B}{\beta\epsilon}\log\left(\frac{1}{q}\right)\right),
\quad Q_\psi=\mathcal{O}\left(\sqrt{\kappa_B}\log\left(\frac{1}{q}\right)\right).
\end{align}
Observe that this directly generalizes the complexity of Hermitian phase estimation, where $\beta=b_{\rm max}=\kappa_B=\lambda_B=1$ as $B$ is trivially the identity matrix.
For arbitrary trial states, let $p$ lower-bound the squared overlap of $B^{1/2}|\psi_{\rm trial}\rangle/\sqrt{\langle\psi_{\rm trial}|B|\psi_{\rm trial}\rangle}$ with the target eigenspace of $B^{-1/2}AB^{-1/2}$.
Without a test recognizing that eigenspace, let the number of repetitions be $m=\lceil(4\kappa_B/p)\log(2/q)\rceil$.
Then the resulting list contains at least one estimate correct to error $\epsilon$ with probability $1-q$.
\begin{align}
Q_A=\mathcal{O}\left(m\left(\frac{\lambda_A}{\beta\epsilon}\right) \log\left(\frac{\kappa_B}{p}\right)\right),
\quad Q_B=\mathcal{O}\left(m\left(\frac{R\lambda_B}{\beta\epsilon}\right) \log\left(\frac{\kappa_B}{p}\right)\right),
\quad Q_\psi=\mathcal{O}\left(m\kappa_B\right).
\end{align}

If the desired eigenvalue is known to lie in an interval $I$ whose endpoints are separated from the spectrum by $g>0$, its estimates can instead be recognized coherently and amplified.
With $h=\min\{\epsilon,g/2\}$ and squared $B$-metric weight at least $p$ in $I$, \cref{thm:gepamp-main} gives
\begin{subequations}
\label{eq:intro-gep-amplified}
\begin{align}
Q_A&=O\left[\left(\frac{\lambda_A}{\beta h}\right) \frac{\log^2(1/(pq))\log(1/q)}{\sqrt p}\right],
\quad
Q_B=O\left[\left(\frac{R\lambda_B}{\beta h}\right) \frac{\log^2(1/(pq))\log(1/q)}{\sqrt p}\right],
\\
Q_\psi&=O\left(\sqrt{\frac{\kappa_B}{p}}\log\frac 1 q\right).
\end{align}
\end{subequations}

\begin{table}[H]
\centering
\small
\begin{tabularx}{\linewidth}{Xccc}
\hline
\hline
\multicolumn{4}{c}{\textbf{Generalized eigenvalue estimation}}
\\
Method & $Q_A$ & $Q_B$ & $Q_\psi$
\\
\hline
QEP & $O(\frac{\lambda_A\sqrt{\kappa_B}}{\beta\epsilon})$ & $O(\frac{\lambda_A\lambda_B\sqrt{\kappa_B}}{\beta^2\epsilon}\log(\frac{\lambda_A\kappa_B}{\beta\epsilon}))$ & $O(\frac{\lambda_A\sqrt{\kappa_B}}{\beta\epsilon})$
\\
Improved QEP & $\widetilde{O}(\frac{\lambda_A\sqrt{\kappa_B}}{\beta\epsilon}\log(\kappa_B))$ & $\widetilde{O}(\frac{\lambda_A\lambda_B\sqrt{\kappa_B}}{\beta^2\epsilon}\log(\kappa_B)\log(\frac{\lambda_A\kappa_B}{\beta\epsilon}))$ & $O(\sqrt{\kappa_B})$
\\
Hermitian + QPE & $O(\frac{\lambda_A}{\beta\epsilon})$ & $O(\frac{\lambda_A\lambda_B}{\beta^2\epsilon}\log(\frac{\lambda_A}{\beta\epsilon})+\frac{\lambda_B\sqrt{\kappa_B}}{\beta}\log(\kappa_B))$ & $O(\sqrt{\kappa_B})$
\\
Cayley + LCU & $\widetilde{O}(\frac{\lambda_A+R\lambda_B}{\beta\epsilon}\log(\kappa_B))$ & $\widetilde{O}(\frac{\lambda_A+R\lambda_B}{\beta\epsilon}\log(\kappa_B))$ & $O(\sqrt{\kappa_B})$
\\
\hline
\cref{thm:gep-eigenstate} & $O(\frac{\lambda_A}{\beta\epsilon})$ & $O(\frac{R\lambda_B}{\beta\epsilon})$ & $O(\sqrt{\kappa_B})$
\\
\hline
\hline
\end{tabularx}
\caption{The first two query columns count $\Be{A/\lambda_A}$ and $\Be{B/\lambda_B}$; the third counts state preparation.
All rows assume the same exact generalized eigenstate, additive error $\epsilon$, constant success, and a classical estimate of $\lambda$.
We assume $0<\epsilon<\min\{R,\lambda_A/\beta\}$.
Confidence $1-\delta$ multiplies the bounds by $\log(2/\delta)$.}
\label{tab:gep-comparison}
\end{table}

\subsection{Overview of new applications}
\label{intro:subsec:applications}

Our Hamiltonian simulation with weighted query cost reduces the cost of accessing weak, expensive interactions in correlated molecules and materials~\cite{Bauer2020QuantumChemistry,Reiher2016Reaction}.
We develop this in \cref{sec:downfolding} and extend coherent quantum signal processing to the online case in \cref{sec:online-qkf}.

\subsubsection{Correlation-weighted simulation}
In a correlated molecule, material, or multiscale graph, an active space interacts with internal degrees of freedom that may be costly to access.
We want to simulate the effective Hamiltonian acting on that active space while accounting for the internal response.
MQSP closes the internal sectors as feedback connections, then compiles the coupled transfer function with one unitary Toeplitz lift and one pair of clock states.
Eliminating the private amplitude in \cref{eq:intro-response} produces the Schur complement that describes this response.

Electronic downfolding and Kron reduction incorporate the influence of eliminated variables through a Schur complement~\cite{Dusson2021SchurPerturbation,Dorfler2013KronReduction}.
Partition the Hermitian model $\mathcal H=\sum_jH_j$ into an active block, an invertible internal block, and their couplings.
The corresponding block decomposition is
\begin{align}
\mathcal H=\left(\begin{array}{c|c}A&B
\\
\hline
B^\dagger&K\end{array}\right),
\quad B=(B_1,B_2,\ldots).
\end{align}
The block $K_{ik}$ couples the corresponding pair of internal sectors.
The effective Hamiltonian $H_{\rm eff}$ is
\begin{align}
\mathcal H= \left[ \begin{array}{c|ccc} A&B_1&B_2&\cdots
\\
\hline
B_1^\dagger&K_{11}&K_{12}&\cdots
\\
B_2^\dagger&K_{12}^\dagger&K_{22}&\cdots
\\
\vdots&\vdots&\vdots&\ddots \end{array}\right] =\left(\begin{array}{c|c}A&B
\\
\hline
B^\dagger&K\end{array}\right),
\quad H_{\rm eff}=A-BK^{-1}B^\dagger.
\label{intro:eq:downfolded-model}
\end{align}
Applying $B$ to the eliminated amplitude $-K^{-1}B^\dagger|\psi\rangle$ produces the correction to the active block.
Let $Q_i$ project onto the corresponding internal sector.
The active-sector projector is $Q_0=I-\sum_{i\ge1}Q_i$.
Given positive known upper bounds $\varrho_i\ge\|Q_iK^{-1}B^\dagger\|$ on the internal response amplitudes, set $\varrho_0=1$.
The sensitivity weight depends on whether a component is diagonal or couples two distinct sectors:
\begin{align}
\begin{array}{ll} H_j=Q_iH_jQ_i:&\chi_j=\varrho_i^2,
\\
H_j=Q_iH_jQ_k+Q_kH_jQ_i:&\chi_j=2\varrho_i\varrho_k\quad(i\ne k). \end{array}
\end{align}
Let $\kappa_j\ge1$ be the dimensionless coarse-graining parameters of \cref{sec:downfolding}.
They control the stability of the coupled inverse under component errors: smaller $\kappa_j$ permit coarser representations and can reduce the block-encoding cost $C_j$.
Write $\boldsymbol{\chi}=(\chi_j)_j$.
The product $\lambda_j\chi_j$ weights each component by its influence on the active system.
Under the response and inverse-stability bounds in \cref{sec:downfolding}, simulation of the effective Hamiltonian for time $t$ and error $\epsilon$ has the following total weighted query cost:
\begin{align}
\sum_j C_jq_j =O\!\left( |t|\left(\sum_j\sqrt{C_j\lambda_j\chi_j}\right)^2 +\log(1/\epsilon)\sum_j C_j\kappa_j \right).
\label{intro:eq:correlation-weighted-cost}
\end{align}
The correlation-weighted normalizations $\lambda_j\chi_j$ enter the same weighted query allocation as the original component strengths.

This reduction addresses two costs: accurately solving the retained interacting problem and representing its coupling to a larger model~\cite{Bauer2020QuantumChemistry,Bauman2019Downfolding}.
Quantum downfolding and inverse-based matrix arithmetic provide important precedents~\cite{Bauman2019Downfolding,Gilyen2018singular}.
Related quantum matrix-equation algorithms construct a block encoding of the solution of a Sylvester equation \cite{Somma2025LinearMatrix}, illustrating the use of matrix-valued responses beyond vector-state preparation.
Directly nesting an inverse encoding inside a simulation repeats its query and normalization costs at each outer call.
MQSP closes the feedback before compilation and resolves the coupled response with one finite lift and clock extraction.
For a positive internal block with a given lower bound $K\succeq\Delta I$ for $\Delta>0$, the relaxation operator $\Sigma=BK^{-1}B^\dagger=X^\dagger KX$ controls the internal response: $\|X\|^2\le\|\Sigma\|/\Delta$ for $X=-K^{-1}B^\dagger$.
A uniformly small correlation or relaxation response can therefore reduce the cost of resolving a large external space.

To allocate data precision using these sensitivities, let each Hermitian approximation $\widehat H_j$ change only the same diagonal block, or the same two mutually adjoint coupling blocks, as $H_j$, with every other block zero.
Given normalized component-error bounds, invertibility of the internal block is preserved by the sufficient condition $\eta_j\ge\|\widehat H_j-H_j\|/\lambda_j$ for all $j$ and $\max_j\kappa_j\eta_j<1$.
Thus $1/\kappa_j$ sets a sufficient normalized data-error scale, subject also to the response-error bound.
Let $\widehat H_{\rm eff}$ be the effective Hamiltonian of the approximated model, and write $\boldsymbol{\eta}=(\eta_j)_j$.
If invertibility and the response bounds hold throughout the model interpolation, then
\begin{align}
\|\widehat H_{\rm eff}-H_{\rm eff}\| \le\sum_j\lambda_j\chi_j\eta_j,
\quad \|e^{it\widehat H_{\rm eff}}-e^{itH_{\rm eff}}\| \le |t|\sum_j\lambda_j\chi_j\eta_j.
\label{intro:eq:coarse-graining}
\end{align}
Weakly influential components can be accessed less frequently and at lower resolution.
The cost $C_j$ in \cref{intro:eq:correlation-weighted-cost} is the cost of the chosen block encoding, with its corresponding response and stability bounds; \cref{intro:eq:coarse-graining} controls the model error.
The classical principle of output-sensitive approximation~\cite{Becker2001ErrorEstimation} thus controls quantum query allocation as well as data resolution.
Our residual bounds also control the error from removing internal sectors, given known upper bounds on their contributions.

In \cref{sec:downfolding}, we apply this construction to external correlation around an electronic active space, a central obstacle to accurate quantum-chemical predictions~\cite{Bauman2019Downfolding,Reiher2016Reaction}.
The target is the downfolded Hamiltonian at a given energy parameter.
In \cref{subsec:kron-graph-learning}, Kron reduction preserves selected graph responses while allowing task-dependent data resolution.
This connects to graph signal processing and semi-supervised learning, where predictions depend on propagation through a much larger network~\cite{Shuman2013GraphSignals,Zhou2003GraphLearning}.
The savings include response, loading, and readout costs under the stated bounds, with accuracy allocated according to the desired output.

\subsubsection{Online multivariate quantum signal processing}
The analytic variables also describe time-varying quantum systems.
In \cref{sec:time-dependent-mqsp}, the available oracles and system matrix change from one event to the next, producing a response indexed separately by input and output time.
The variables $z_j$ count oracle uses, while $t$ records the arrival time.
An online unitary lift and two-time clock factorization combine these events as they arrive, without replay.

This setting models online quantum sensor fusion, where coherent observations are combined with an evolving model and a classical prior.
Temporal smoothness can reduce the cost of approximating a drifting model, and analytic continuation in physical time gives geometric convergence under the hypotheses of \cref{sec:time-dependent-mqsp}.
This is an optional sampling assumption, distinct from the structural analyticity in the analytic variables established in \cref{sec:achievable}.

We demonstrate the construction through online quantum Kalman filtering \cite{Kalman1960LinearFiltering,Shi2024QuantumKalman}.
To keep the covariance update classical while the transported mode space may be large, we use the structured anisotropic model
\begin{align}
A_t=a_t\otimes U_t,
\quad C_t=c_t\otimes I_N,
\quad P_t=p_t\otimes I_N ,
\label{intro:eq:online-kalman}
\end{align}
with analogous tensor forms for the process and observation covariances.
The small matrices $p_t$ and the Kalman gain are updated classically, while the high-dimensional $U_t$ is queried once at its event.
The posterior mean obeys
\begin{align}
\vect\mu_t=(f_t\otimes U_t)\vect\mu_{t-1} +(k_t\otimes I_N)\vect y_t ,
\label{intro:eq:online-kalman-mean}
\end{align}
where $f_t$ and $k_t$ are small matrices determined by the classical Kalman recursion.

A $d\times d$ weighted controllability Gramian $g_t$ gives a unitary realization of this update and determines its preparation scale.
One online pass uses each scheduled transport and observation-preparation oracle once and selects
\begin{align}
(\langle0|_{\rm ok}\otimes I)U_{\rm alg}|0\rangle =\frac{\vect\mu_T}{Z_T},
\quad Z_T=\sqrt{\lambda_{\max}(g_T)},
\quad p_{\rm succ}=\frac{\|\vect\mu_T\|^2}{Z_T^2}.
\label{intro:eq:online-output}
\end{align}
The clock over input times can be optimized using only the small Gramian when the model is known in advance; known envelopes give a forward-only choice when future gains are revealed online.
The circuit retains the estimate coherently with logarithmic time-address overhead and no intermediate postselection.
Under a covariance-metric contraction, old observations are exponentially suppressed, so only a logarithmic quantum history is needed for fixed stability margin and target precision.
Increasing the success probability without replay uses independent live copies and their corresponding oracle calls.

\subsection{Discussion and conclusion}\label{sec:intro-discussion}

MQSP provides a signal-processing language in which the signal itself is a wavefunction interacting coherently with an environment.
A wire carrying a scalar voltage in a classical IIR filter becomes a quantum signal space carrying complex amplitudes, with the $2^n$ basis states of an $n$-qubit data register serving as virtual channels \cite{Childs2009Universal}.
Structured quantum oracles and block encodings organize coherent access to information encoded in large matrices \cite{Low2016Qubitization,Gilyen2018singular}.
When these operations and the known filter connections have efficient circuit realizations, a gate-efficient MQSP network can organize propagation, transformation, and interference across these exponentially many channels.
The large matrices are accessed through their quantum implementations, and the filter network specifies how their actions contribute to the computation in a structured manner.

Reversibility gives this picture its physical structure.
Just as a Toffoli gate computes an AND into a target while retaining its inputs \cite{Toffoli1980Reversible}, an MQSP realization retains the internal amplitudes accompanying its selected transformation.
Private feedback and finite clock registers preserve coherent propagation, and the clock preparations specify how contributions from different propagation times interfere at the output.
The clock-norm characterization and quantitative analytic continuation and growth bounds developed here make this description operational: they relate the chosen filter to approximation error, normalization, and individual oracle costs.
Efficient preparation and output extraction complete the requirements for an efficient algorithm.

Together, these ingredients connect the lossless composition rules of quantum transducers \cite{Belovs2024TamingQuantum} with up to a century of classical filter and network-synthesis theory \cite{Darlington1939NetworkSynthesis,Regalia1988AllpassFilters}, and provide a common design perspective on quantum computation based on coherent wave propagation.
The exponential oracle separation obtained by welded-tree quantum walks \cite{Childs2003Exponential}, the simulation of exponentially many classical oscillators \cite{Babbush2023ClassicalOscillators}, and the compressed evolution of selected observables in shadow Hamiltonian simulation \cite{Somma2025ShadowSimulation} illustrate different ways in which structured wave dynamics make large computations accessible.
MQSP develops this perspective into a modular filter-design framework, bringing established templates for signal flow, feedback, and amplitude and phase control into quantum algorithm design under explicit realization and compilation conditions.
We expect this connection to suggest applications well beyond those developed here.
The broader opportunity is to design quantum algorithms by designing how a wavefunction propagates and interferes to make the desired information observable.

\settocdepth{2}
\section{Quditization}
\label{sec:quditization}

Quditization implements a finite part of an IIR computation and optimizes the clock normalization for clock weights.
Its output is the block encoding
\begin{align}
P_{\rm out}^{\dagger}(I_{\rm label}\otimes\mathcal W_N[G])P_{\rm in} =\Be{\widetilde G_N/\alpha_{\rm clk}},
\quad \widetilde G_N=\sum_{n=0}^{N-1}c_nG_n\approx G.
\label{eq:qudit-destination}
\end{align}
Steady-state feedback first identifies the target transformation and its catalyst weights.
The unitary Toeplitz lift then unrolls the causal recurrence while keeping the final private registers.
Finally, input and output clock preparations combine its public outputs.
Their least normalization for given clock weights is the nuclear-norm problem in \cref{thm:iir-clock-body}; \cref{sec:iir} chooses those weights, the horizon, and the oracle delays.
The resource tables in \cref{tab:resources-modules,tab:resources-algorithms,tab:resources-applications} collect the module, algorithm, and application costs, and domain and precision conventions are collected in \cref{app:conditions}.

\subsection{Steady-state feedback and catalyst weights}
\label{subsec:qudit-system}

We use the public/private decomposition, feedback equations, and transfer function of \cref{eq:intro-response}.
Each private subspace contains the oracle's signal and work registers, as specified in \cref{def:resource-accounting}.
A realization may connect several copies of the same oracle space.
On the direct sum of copies, the query preserves their label: $I_{\rm copy}\otimes O_j: |\ell\rangle\otimes|\phi\rangle\mapsto|\ell\rangle\otimes O_j|\phi\rangle$.
The copy label records which feedback connection is used and is retained by the query.
We include this identity action in the notation $O_j$.
For a fixed oracle tuple, define the analytic feedback operator and its evaluation at $\Bz=\Bone$ by
\begin{align}
\boldsymbol{O}=(O_1,\ldots,O_m),
\quad \oracle(\Bz)=\bigoplus_jz_jO_j,
\quad Q=\oracle(\boldsymbol{1})=\bigoplus_jO_j.
\end{align}
Thus $\oracle(\Bz)$ includes the delay variables, and $Q$ is the unitary feedback operation with every delay variable set to one.
Independent controlled oracles are counted separately; an identity feedback connection has zero query cost but still occupies storage.

For $\max_j|z_j|<1$, the transfer function has the convergent expansion
\begin{align}
F(\Bz)=A+\sum_{k\ge0}B\oracle(\Bz)[D\oracle(\Bz)]^kC =\sum_{\boldsymbol{n}\in\mathbb N_0^m}F_{\boldsymbol{n}} \prod_{j=1}^m z_j^{n_j}.
\label{eq:qudit-response-series}
\end{align}
Each $F_{\boldsymbol n}$ sums ordered paths with exactly $n_j$ calls to port $j$.
The analytic variables commute, while the oracle operators retain their order.
Because these query counts are nonnegative and $\|D\|\le1$, the series converges throughout the open polydisk $\max_j|z_j|<1$ \cite{Helton1974,Ball1998UnitaryColligations}.
\Cref{prop:ach-query-markers,thm:ach-analytic-completeness} connects this expansion to the full finite-query model.

For example, swapping equal public and private spaces gives
\begin{align}
S=\left(\begin{array}{c|c}0&I
\\
\hline
I&0\end{array}\right),
\quad F(z)=zO,
\quad\Gamma(z)=I.
\end{align}
A single feedback connection applies $O$ and transmits the input publicly.

Fix an oracle promise set $\mathfrak O$.
All known circuit choices are common to this set, and uniform bounds hold for every unitary satisfying the oracle promises.
For the steady-state catalyst identities, assume that $\Gamma(\Bz)$ has a finite analytic continuation to a neighborhood of $\Bone$; its value there refers to this continuation.
The delayed function $G$ is defined in \cref{eq:intro-delays}.
We retain the Abel-limit convention at $z=1$:
\begin{align}
G(1)=G=F=\lim_{x\uparrow1}G(x) =\lim_{x\uparrow1}\sum_{n\ge0}x^nG_n.
\label{eq:iir-ideal-response}
\end{align}
This convention also applies when the coefficient series fails to converge at the evaluation point \cite{Hardy1949DivergentSeries}.
To illustrate the distinction for an analytic function, consider
\begin{align}
f(z)=\frac{1-z}{1+z}=1+2\sum_{n\ge1}(-z)^n
\quad(|z|<1),
\quad \lim_{x\uparrow1}f(x)=0.
\label{eq:qudit-abel-example}
\end{align}
The Taylor series oscillates at $z=1$ even though the function is analytic there.
Its obstruction is the pole at the opposite point of the circle.
This example explains the Abel limit for general analytic functions; the MQSP transfer functions additionally satisfy disk contractivity.
The steady-state feedback equation is then
\begin{align}
S\bigl(|\psi\rangle\oplus Q\Gamma|\psi\rangle\bigr) =G|\psi\rangle\oplus\Gamma|\psi\rangle.
\label{eq:iir-transducer}
\end{align}
The catalyst is $\Gamma|\psi\rangle$: querying it and applying $S$ returns the same private vector.
Its weight in port $j$ is
\begin{align}
W_j(\psi)=\|\Gamma_j|\psi\rangle\|^2 =\langle\psi|\Gamma_j^\dagger\Gamma_j|\psi\rangle,
\label{eq:qudit-lv-weight}
\end{align}
where $\Gamma_j$ is the catalyst operator projected onto the private space $\mathcal L_j$.
For a fixed input we abbreviate this weight by $W_j$.

These are Las Vegas query weights: a controlled call is weighted by the squared norm of the state on which it acts~\cite{Belovs2023LasVegas,Belovs2024TamingQuantum}.
More generally, let $V_s$ be the circuit prefix before the $s$th query.
Write $\Pi_s$ for the projector onto its active branch.
The weight operator for an oracle type is
\begin{align}
L_j(\mathcal A)=\sum_{s:j_s=j}V_s^\dagger\Pi_sV_s,
\quad 0\preceq L_j(\mathcal A)\preceq q_jI.
\label{eq:qudit-circuit-lv}
\end{align}
Its expectation in the input state is the circuit's Las Vegas weight, whereas $q_j$ counts the scheduled calls.
A disabled branch contributes zero weight to a call that still occupies a query slot.
For the finite circuit constructed below, these weights are computed from its own prefixes $V_s$.

The system matrix maps the column $|\psi\rangle\oplus\oracle(\Bz)\Gamma(\Bz)|\psi\rangle$ to $F(\Bz)|\psi\rangle\oplus\Gamma(\Bz)|\psi\rangle$.
Equating their squared norms gives the conservation identity below; equating cross inner products at two points gives \cref{thm:iir-energy}.
\begin{align}
1+\|\oracle(\Bz)\Gamma(\Bz)|\psi\rangle\|^2 =\|F(\Bz)|\psi\rangle\|^2+\|\Gamma(\Bz)|\psi\rangle\|^2.
\label{eq:qudit-norm-conservation}
\end{align}
At $\Bz=\Bone$ the private norms agree, so the public output has unit norm.
Taking inner products at two different values of the variables also gives the following expression for the catalyst weights in terms of derivatives of $F$.

\begin{theorem}[Unitary kernel and catalyst identities]
\label{thm:iir-energy}
For a fixed oracle tuple, wherever the private transfer functions are defined,
\begin{align}
I-F(\boldsymbol{w})^\dagger F(\Bz) =\sum_{j=1}^m(1-\overline w_jz_j) \Gamma_j(\boldsymbol{w})^\dagger\Gamma_j(\Bz).
\label{eq:iir-kernel}
\end{align}
The transfer function $F$ is contractive on the open polydisk.
Under the analytic continuation assumption above, $G=F$ is unitary and
\begin{align}
\Gamma_j^\dagger\Gamma_j =G^\dagger\partial_{z_j}F\succeq0,
\quad W_j(\psi)=\langle\psi|G^\dagger\partial_{z_j}F|\psi\rangle.
\label{eq:iir-energy}
\end{align}
For the delays in \cref{eq:intro-delays}, let $R_{\boldsymbol{r}}=\bigoplus_jr_jI_{\mathcal L_j}$.
Then
\begin{subequations}
\label{eq:qudit-delayed-energy}
\begin{align}
G^\dagger G'&=\Gamma^\dagger R_{\boldsymbol{r}}\Gamma,
\\
W(\psi)&:=\langle\psi|G^\dagger G'|\psi\rangle =\sum_jr_jW_j(\psi).
\end{align}
\end{subequations}
To compare two oracle substitutions, use the same system matrix with another tuple $\widehat{\boldsymbol{O}}$.
Write $\widehat G,\widehat\Gamma$ for its transfer function and catalyst operator evaluated at $\Bz=\boldsymbol{1}$.
Then
\begin{align}
I-\widehat G^\dagger G =\sum_j\widehat\Gamma_j^\dagger (I-\widehat O_j^\dagger O_j)\Gamma_j.
\label{eq:qudit-cross-oracle}
\end{align}
\end{theorem}
\begin{proof}
The feedback equations give
\begin{align}
S\begin{pmatrix}I
\\
\oracle(\Bz)\Gamma(\Bz)\end{pmatrix} =\begin{pmatrix}F(\Bz)
\\
\Gamma(\Bz)\end{pmatrix}.
\end{align}
Unitarity of the system matrix preserves inner products of these columns evaluated at $\boldsymbol{w}$ and $\Bz$.
Using $\oracle(\boldsymbol{w})^\dagger \oracle(\Bz) =\bigoplus_j\overline w_jz_jI$ and rearranging proves \cref{eq:iir-kernel}.
At equal interior points its right side is positive semidefinite.
At $\boldsymbol{w}=\Bz=\Bone$ it vanishes, so the square finite-dimensional transfer function $G$ is unitary.

Fix the left argument at $\boldsymbol{w}=\boldsymbol{1}$ and differentiate in $z_j$.
Then evaluate the right argument at $\Bz=\Bone$.
Every derivative of $\Gamma_k(\Bz)$ is multiplied by $1-z_k$, which vanishes at the evaluation point; the remaining term proves \cref{eq:iir-energy}.
Applying the chain rule to $G(z)=F(z^{r_1},\ldots,z^{r_m})$ proves \cref{eq:qudit-delayed-energy}.
Finally, take the same inner product for the two oracle tuples.
Their private contribution contains $\widehat Q^\dagger Q=\bigoplus_j\widehat O_j^\dagger O_j$, which gives \cref{eq:qudit-cross-oracle}.
\end{proof}

At equal arguments, the unitarity identity gives the probability balance
\begin{align}
I-F(\Bz)^\dagger F(\Bz) =\sum_j(1-|z_j|^2)\Gamma_j(\Bz)^\dagger\Gamma_j(\Bz) \succeq0,
\quad \max_j|z_j|<1.
\label{eq:qudit-damped-probability}
\end{align}
For a normalized public input, the missing output probability is exactly that damped in the private ports.
The two-point identity is the standard unitary-realization identity~\cite{Ball1998UnitaryColligations}; its form with cross-products of the oracles describes quantum state conversion~\cite{Belovs2023LasVegas,Belovs2024TamingQuantum}.
The positive kernels are $\Gamma_j(\boldsymbol{w})^\dagger\Gamma_j(\Bz)$.
Between successive queries, port $j$ stores its catalyst in $r_j$ private positions.
Its contribution to the stored squared norm is multiplied by that delay.

Each intermediate branch contributes its catalyst weight multiplied by its actual squared amplitude before any worst-case norm is taken.
This is the transducer framework's thrifty composition \cite{Belovs2023LasVegas,Belovs2024TamingQuantum}.
If two networks with disjoint private coordinates are connected through a known unitary $V$, their transfer function is $F_2VF_1$.
Evaluating at $\Bz=\boldsymbol{1}$ gives
\begin{align}
W_j^{(21)}(\psi)=W_j^{(1)}(\psi)+W_j^{(2)}(VF_1|\psi\rangle).
\label{eq:qudit-composed-energy}
\end{align}
Here the argument of the second weight is the normalized intermediate state.
Concatenating $\Gamma_{1j}$ and $\Gamma_{2j}VF_1$, or differentiating the product transfer function, proves the identity before any worst-case norm is taken.
A uniform catalyst bound can therefore be specified as
\begin{align}
w\ge \sup_{\boldsymbol{O}\in\mathfrak O} \|R_{\boldsymbol{r}}^{1/2}\Gamma\|^2.
\label{eq:qudit-uniform-catalyst}
\end{align}
The operator norm maximizes over normalized public inputs, and the supremum ranges over all unitaries satisfying the oracle promises.

\subsection{Unitary Toeplitz lift}
\label{subsec:iir-lift}

Causal Toeplitz input--output maps and unitary state-space realizations are classical constructions \cite{Helton1974,Ball1998UnitaryColligations}.
Quantum transducer unrolling retains the catalyst in an enlarged clock space; Belovs, Jeffery, and Yolcu's finite automaton \cite[Lemma 4.6 and Theorem 5.5]{Belovs2024TamingQuantum} reuses a shared internal state in a direct sum, and permits different query rates.
Chen et al.'s reuse circuit \cite[Figure 1 and Section 5]{Chen2026OptimaTimeDependent} also starts with zero private amplitude and retains the finite outputs.
Our construction makes the finite public block and integer-delay schedule explicit, with a realization using a binary address register whose gate and qubit costs are given below.
Belovs, Jeffery, and Yolcu's implementation \cite[Lemma 4.6]{Belovs2024TamingQuantum} uses $O(N)$ routing gates and a logarithmic auxiliary register in its setting.
\cref{thm:iir-gates} implements the binary-address routing with constant additional routing workspace.
For address width $b$, its overhead is
\begin{align}
O(Rb\log(2+b)),
\quad R=(m+1)N+\sum_jq_j,
\end{align}
where the latter quantity bounds the number of routed operations.
The alternative in \cref{lem:iir-routing} uses more scratch to reduce the arithmetic overhead.
These are gate--space tradeoffs at a fixed horizon; analytic clock shaping separately reduces the required horizon and determines nonuniform endpoint preparation.
Our periodic convention queries port $j$ at $r_j,2r_j,\ldots<N$.
Shifting these query events while keeping one per write--read interval can change the finite count by one and the unmeasured private output.

We now implement a finite part of the transfer function by interpreting each factor $z_j^{r_j}$ as a delay of $r_j$ time steps.
With one port and unit delay, public inputs $\boldsymbol{u}_k$ and initial private amplitude $\boldsymbol{g}_{-1}=0$ evolve by
\begin{align}
\boldsymbol{y}_k=A\boldsymbol{u}_k+BO\boldsymbol{g}_{k-1},
\quad \boldsymbol{g}_k=C\boldsymbol{u}_k+DO\boldsymbol{g}_{k-1}.
\label{eq:qudit-one-port}
\end{align}
Consequently,
\begin{align}
G_{0}=A,
\quad G_{n}=BO(DO)^{n-1}C\ (n\ge1),
\quad \boldsymbol{y}_k=\sum_{i=0}^kG_{k-i}\boldsymbol{u}_i.
\label{eq:qudit-causal-convolution}
\end{align}
Each output combines the present and earlier inputs, with a coefficient that depends only on their time difference.
This is a causal convolution.
Including the final private amplitude gives the unitary Toeplitz lift constructed in \cref{thm:iir-lift}.

\begin{figure}[!htb]
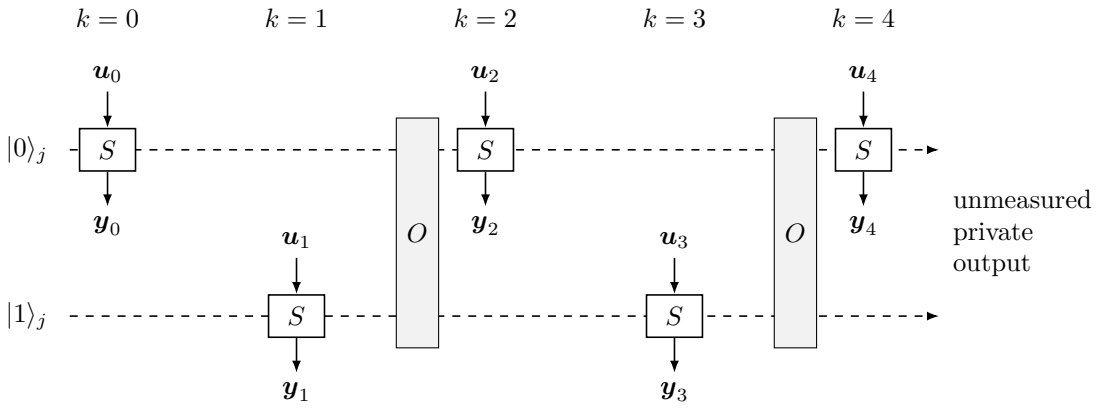

\centering
\IIRDiagram{Quditization1}
\caption{One oracle port with delay $r=2$ and horizon $N=5$.
At each time, the system matrix couples the corresponding public clock subspace to one private position.
These subspaces are $\ket k_{\rm clk}\otimes\mathcal P$ and $\ket{k\bmod2}_j\otimes\mathcal L_j$.
Each oracle rectangle denotes a controlled call on both private positions.
The calls occur at times $2$ and $4$, before the corresponding system operation.
An amplitude written at time $k$ is read at $k+2$ after one oracle call.
The lines denote direct-sum subspaces represented by the binary address register.
The input consists of the public amplitudes $(\boldsymbol{u}_0,\ldots,\boldsymbol{u}_4)$ and two initial private amplitudes.
The output consists of $(\boldsymbol{y}_0,\ldots,\boldsymbol{y}_4)$ and both unmeasured private amplitudes.
With zero private input, the circuit block-encodes $\mathcal T_5[G]$.}
\label{fig:qudit-delay}
\end{figure}

\begin{figure}[!htb]
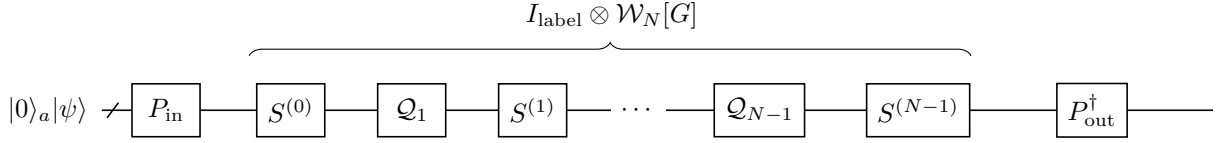

\centering
\IIRDiagram{Quditization2}
\caption{Circuit for quditization on the binary address register and clock label.
The system operation $S^{(k)}$ acts through the embedding $J_k$ in \cref{eq:qudit-step-embedding}, coupling the indicated public and private basis subspaces.
Routing updates the address during these couplings.
The operation $\mathcal Q_k$ applies the scheduled oracles to all their private positions.
A port is scheduled precisely when $r_j\mid k$.
Projecting the encoding register onto $|0\rangle_a$ after input-clock preparation and output-clock unpreparation gives $\widetilde G_N/\alpha_{\rm clk}$, as defined in \cref{eq:iir-clock-matrix,eq:iir-complete-be}.}
\label{fig:qudit-circuit}
\end{figure}

\begin{theorem}[Unitary Toeplitz lift]
\label{thm:iir-lift}
Use the finite-dimensional system matrix and controlled oracle ports from \cref{eq:intro-response}.
For positive integer port delays $\boldsymbol{r}$ and an integer horizon $N\ge1$, an oracle-independent circuit constructs a unitary lift.
This lift, denoted $\mathcal W_N[G]$, acts on
\begin{align}
\mathcal H_{N,\boldsymbol{r}} =(\mathbb C^N\otimes\mathcal P) \oplus\bigoplus_{j=1}^m(\mathbb C^{r_j}\otimes\mathcal L_j)
\label{eq:iir-buffer-space}
\end{align}
whose public block is the Toeplitz matrix
\begin{align}
T_N[G]=[G_{o-i}]_{o,i=0}^{N-1} =\begin{pmatrix} G_{0}&0&\cdots&0
\\
G_{1}&G_{0}&\ddots&\vdots
\\
\vdots&\ddots&\ddots&0
\\
G_{N-1}&\cdots&G_{1}&G_{0} \end{pmatrix},
\quad G_{n}=0
\quad(n<0).
\label{eq:iir-toeplitz}
\end{align}
The circuit uses $N$ applications of $S$ and exactly
\begin{align}
q_j=\left\lfloor\frac{N-1}{r_j}\right\rfloor
\label{eq:iir-exact-count}
\end{align}
controlled calls to $O_j$.
Its known gates and query order are independent of the oracle tuple.
It is unitary on arbitrary public and private inputs and requires no intermediate measurements.
\end{theorem}
\begin{proof}
Let $\{\ket{x}_{\rm clk}:0\le x<N\}$ be an orthonormal clock basis for the public summand in \cref{eq:iir-buffer-space}.
For port $j$, let $\{\ket{x}_j:0\le x<r_j\}$ be the basis of its private delay register.
At step $k$, the isometry
\begin{align}
J_k:\ket\psi\oplus\bigoplus_j\ket{\phi_j} \mapsto\ket{k}_{\rm clk}\ket\psi \oplus\bigoplus_j\ket{k\bmod r_j}_j\ket{\phi_j}
\label{eq:qudit-step-embedding}
\end{align}
embeds $\mathcal P\oplus\bigoplus_j\mathcal L_j$ in the fixed Hilbert space.
Apply the unitary $S^{(k)}=J_kSJ_k^\dagger+(I-J_kJ_k^\dagger)$ after the scheduled oracle operations.
At a positive multiple of the port delay, apply $I_{r_j}\otimes O_j$ on that port's private summand and the identity elsewhere.
Disjoint port operations commute.
\cref{fig:qudit-delay,fig:qudit-circuit} show these basis subspaces and their implementation on a common system register.

An amplitude deposited in $\ket{k\bmod r_j}_j\otimes\mathcal L_j$ is next coupled to the public register at step $k+r_j$.
The interval $(k,k+r_j]$ contains exactly one positive multiple of the delay.
The private state therefore passes through $O_j$ once between these couplings.
Starting with zero private amplitude, the private input at step $k$ is consequently $O_j\boldsymbol{g}_{k-r_j,j}$, with negative-time terms zero.
The generating functions obey $\boldsymbol{f}_j(z)=z^{r_j}O_j\boldsymbol{g}_j(z)$ and $\boldsymbol{y}(z)=G(z)\boldsymbol{u}(z)$, proving \cref{eq:iir-toeplitz} on the public basis states.

Every factor acts unitarily on \cref{eq:iir-buffer-space}.
Their product $\mathcal W_N[G]$ retains every private basis state, including amplitude queried after its last coupling to a public state.
There are $N$ system-matrix factors.
Counting positive multiples of each port delay below the horizon gives $q_j=\lfloor(N-1)/r_j\rfloor$, which proves \cref{eq:iir-exact-count}.
\end{proof}

As discussed in \cref{intro:eq:expanded-lift} of the introduction, $T_N[G]$ is a contraction.
If $R_N$ maps a public input to the retained private output, unitarity gives
\begin{align}
T_N[G]^\dagger T_N[G]+R_N^\dagger R_N=I.
\label{eq:qudit-energy-balance}
\end{align}
For the pure delay $G(z)=zI$, the public block shifts the input forward one position, and the last input exits privately.
For example,
\begin{align}
\mathcal T_3[zI]=\begin{pmatrix}0&0&0
\\
I&0&0
\\
0&I&0\end{pmatrix},
\quad \mathcal W_3[zI]= \left(\begin{array}{ccc|c} 0&0&0&I
\\
I&0&0&0
\\
0&I&0&0
\\
\hline
0&0&I&0 \end{array}\right).
\label{eq:qudit-pure-delay}
\end{align}
The final row and column are the private space.
They complete the truncated shift to a unitary, while the public matrix alone loses the last input.
The construction of the lift uses only finite-time coefficients and does not require a steady-state catalyst.

For one private space and $O=I$, the three-step lift acts on $(\ket0_{\rm clk}\otimes\mathcal P)\oplus (\ket1_{\rm clk}\otimes\mathcal P)\oplus (\ket2_{\rm clk}\otimes\mathcal P)\oplus\mathcal L$ as
\begin{align}
\mathcal W_3[G]=S^{(2)}S^{(1)}S^{(0)} =\left(\begin{array}{ccc|c} A&0&0&B
\\
BC&A&0&BD
\\
BDC&BC&A&BD^2
\\
\hline
D^2C&DC&C&D^3 \end{array}\right).
\label{eq:qudit-three-step}
\end{align}
The upper-left block is $\mathcal T_3[G]$.
Its last column carries initial private amplitude to all public outputs and the final private output.

For constant-label clocks prepared without additional scratch, \cref{thm:iir-gates} uses
\begin{align}
n_{\rm total}=\left\lceil\log_2\dim\mathcal H_{N,\boldsymbol r}\right\rceil+O(1),
\quad G_{\rm routing}=O\!\left(\left[(m+1)N+\sum_jq_j\right]b\log(2+b)\right),
\label{eq:qudit-resource-summary}
\end{align}
where $b$ is the binary address width.
General clock labels and restored scratch contribute the additional terms in \cref{eq:qudit-main-cost}.
The resource convention is \cref{def:resource-accounting}.
For unitary targets at normalization two, oblivious amplitude amplification in \cref{lem:iir-cubic} gives an approximate unitary and triples the query costs; \cref{eq:qudit-error-budget} combines clock, realization, and synthesis errors.

\subsection{Clock state optimization}
\label{subsec:iir-clock-body}

Prepare the clocks from \cref{eq:intro-clock-selection}.
The label $\ell$ allows a sum of outer products in the clock matrix.
For $r_{\rm clk}$ label values, its ancilla uses $\lceil\log_2r_{\rm clk}\rceil$ qubits, included in the resource count.
The lift acts as the identity on this label.
Explicit analytic clocks need only a constant number of labels (one pair before normalization padding; the signed sine construction uses two).
General numerical factorizations may need more.
Write
\begin{align}
|c_{\rm in}\rangle=\sum_{i,\ell}g_{i\ell}|i,\ell\rangle,
\quad |c_{\rm out}\rangle=\sum_{o,\ell}h_{o\ell}|o,\ell\rangle,
\quad P_{\rm in}|0\rangle=|c_{\rm in}\rangle,
\quad P_{\rm out}|0\rangle=|c_{\rm out}\rangle.
\label{eq:iir-clock-amplitudes}
\end{align}
Let $P_{\rm in}$ and $P_{\rm out}$ be known unitaries preparing these states on the public positions, with initially zero private amplitude.
Applying these operations gives $P_{\rm out}^\dagger(I_{\rm label}\otimes\mathcal W_N[G])P_{\rm in}=\Be{B_N}$, where
\begin{align}
B_N=\sum_{n=0}^{N-1} \left(\sum_{o-i=n,\ell}\overline{h_{o\ell}}g_{i\ell}\right)G_{n}.
\label{eq:qudit-extraction}
\end{align}
For a prescribed normalization $\alpha_{\rm clk}>0$, define
\begin{align}
X_{oi}=\alpha_{\rm clk}\sum_\ell\overline{h_{o\ell}}g_{i\ell},
\quad c_n(X)=\sum_{o-i=n}X_{oi},
\quad \widetilde G_N=\sum_{n=0}^{N-1}c_n(X)G_{n}. \
\label{eq:iir-clock-matrix}
\end{align}
Thus quditization implements $\Be{\widetilde G_N/\alpha_{\rm clk}}$.
The approximation error is $\|\widetilde G_N-G\|$, and the error in the encoded block is this quantity divided by $\alpha_{\rm clk}$.

The clock definition also applies to a finite kernel $E=[E_{oi}]$ whose entries depend separately on the input and output times.
In particular, the label $\ell$ can store a second time index, so long as the lift acts identically on each label.
We call this a clock with two time registers; its preparation and label qubits are included in the costs above.
The resulting circuit block-encodes
\begin{align}
B_{h,g}[E]=\sum_{o,i,\ell}\overline{h_{o\ell}}g_{i\ell}E_{oi},
\quad \mathcal L_X[E]=\sum_{o,i}X_{oi}E_{oi}.
\label{eq:qudit-general-clock}
\end{align}
For any scalar coefficient matrix $X$, the factorization in \cref{thm:iir-clock-body} gives a block encoding $\Be{\mathcal L_X[E]/\alpha_{\rm clk}}$ whenever $\|X\|_*\le\alpha_{\rm clk}$.
This formulation will also describe the online kernels in \cref{sec:time-dependent-mqsp}.

For input index $k$, the subsequent output indices contribute the corresponding causal coefficients: $o=k,\ldots,N-1: G_0,\ldots,G_{N-1-k}$.
Each has clock overlap $1/N$.
At a fixed step offset $n=o-i$, counting the input--output pairs gives $k=0,\ldots,N-1-n$ where the number of pairs is $N-n$ and $c_n=1-n/N$.
Other clock states change these weights and their normalization.

For target clock weights $\boldsymbol{c}=(c_0,\ldots,c_{N-1})$, define
\begin{align}
\|\boldsymbol{c}\|_{{\rm clk},N} =\min\left\{\|X\|_*: \sum_{o-i=n}X_{oi}=c_n,\quad0\le n<N\right\}.
\label{eq:qudit-clock-norm}
\end{align}
The nuclear norm $\|X\|_*$ is the sum of the singular values.
Only $\sum_{i=0}^{N-1-n}X_{i+n,i}=c_n$ for $0\le n<N$ is constrained; entries strictly above the main diagonal are unconstrained.
Placing $\boldsymbol{c}$ in the first column gives a feasible matrix.
Restricting to matrices whose nuclear norm is no larger than that feasible value gives a closed bounded set in finite-dimensional space, hence a compact set; the continuous nuclear norm attains its minimum there.
The following theorem shows that this clock norm is the least normalization needed to implement the specified coefficients.

\begin{theorem}[Exact clock characterization]
\label{thm:iir-clock-body}
For fixed finite scalar clock supports, an overlap matrix $X/\alpha_{\rm clk}$ is attainable by normalized clocks with spectator labels if and only if $\|X\|_*\le\alpha_{\rm clk}$.
Consequently, the coefficient vector is attainable at clock normalization $\alpha_{\rm clk}$ exactly when $\|\boldsymbol{c}\|_{{\rm clk},N}\le\alpha_{\rm clk}$.
\end{theorem}
\begin{proof}
Each label contributes $\overline h_\ell g_\ell^T$, so normalized clocks satisfy
\begin{align}
\left\|\sum_\ell\overline h_\ell g_\ell^T\right\|_* \le\sum_\ell\|h_\ell\|\|g_\ell\| \le\left(\sum_\ell\|h_\ell\|^2\right)^{1/2} \left(\sum_\ell\|g_\ell\|^2\right)^{1/2}=1.
\end{align}
Conversely, let $X=\sum_{\ell=1}^rs_\ell a_\ell b_\ell^\dagger$ be a singular-value decomposition, with $s=\sum_\ell s_\ell\le\alpha_{\rm clk}$.
Set $g_\ell=\sqrt{s_\ell/\alpha_{\rm clk}}\,\overline b_\ell$ and $h_\ell=\sqrt{s_\ell/\alpha_{\rm clk}}\,\overline a_\ell$.
Their overlap matrix is $X/\alpha_{\rm clk}$, and each clock has squared norm $s/\alpha_{\rm clk}$.
Complete their norms on two distinct unused labels, which contribute zero to the overlap.
At most $r+2$ labels suffice.
Minimizing over matrices with the desired clock weights proves the coefficient statement.
\end{proof}

\cref{lem:clock-extraction} extends the clock factorization to any finite public kernel; we will use this form for online evolution.

\begin{lemma}[Clock extraction of a finite kernel]
\label{lem:clock-extraction}
For a block matrix $E=[E_{oi}]$ and a scalar matrix $X$ with matching index sets, define
\begin{align}
\mathcal L_X(E)=\sum_{o,i}X_{oi}E_{oi}.
\end{align}
Then
\begin{align}
\|\mathcal L_X(E)\|\le\|X\|_*\|E\|.
\label{eq:master-nuclear-perturb}
\end{align}
The clock factorization in \cref{thm:iir-clock-body} implements this extraction for arbitrary finite kernels, including non-Toeplitz ones.
\end{lemma}
\begin{proof}
Take a singular-value decomposition of the coefficient matrix.
Each rank-one term extracts a compression of the block operator:
\begin{align}
X=\sum_\ell s_\ell a_\ell b_\ell^\dagger,
\quad \mathcal L_X(E)=\sum_\ell s_\ell (\langle\overline{a_\ell}|\otimes I) E(|\overline{b_\ell}\rangle\otimes I).
\end{align}
Both singular vectors have unit norm, so each compression has norm at most $\|E\|$.
The triangle inequality yields
\begin{align}
\|\mathcal L_X(E)\|\le\sum_\ell s_\ell\|E\| =\|X\|_*\|E\|.
\end{align}
The clock construction in \cref{thm:iir-clock-body} depends only on this rank-one factorization and therefore applies to every finite block kernel.
\end{proof}

Classically computing the matrix factorization and preparing its clock states have separate costs.
A construction at normalization $\widehat\alpha\le\alpha_{\rm clk}$ can be attenuated by $\widehat\alpha/\alpha_{\rm clk}$ to obtain the prescribed normalization without changing the lift or query counts.
Later constructions use $\alpha_{\rm clk}=2$, or $1+\delta$ when the normalization overhead is kept explicit.

Standard state preparation \cite{shende2006synthesis,Mottonen2005StatePreparation} implements a constant-label clock pair with $O(N)$ gates.
\cref{eq:qudit-clock-preparation} gives the rotation and CNOT counts, and \cref{thm:iir-gates} includes their workspace.

\section{Clock shaping}
\label{sec:iir}

We now choose the clock weights, horizon, and oracle delays to approximate the target transformation uniformly over its oracle promise.
At a fixed lift, the normalization constrains which finite approximations can be implemented.

Causal query counting makes finite histories polynomial in the analytic variables and finite-dimensional IIR responses analytic in the open polydisk.
This includes every finite fixed-order query algorithm by \cref{thm:ach-analytic-completeness}.
Analytic clock shaping uses a quantitative continuation through the point $\Bz=\Bone$: growth bounds in that neighborhood control the horizon and give logarithmic precision dependence.

The promise set $\mathfrak O$ consists of all allowed tuples $(O_1,\ldots,O_m)$, with all matrix blocks unrestricted subject to the stated promises.
A uniform approximation uses a single choice of clocks and known gates for every tuple in this set.
A clock chosen in advance must satisfy this black-box promise \cite{Belovs2023LasVegas,Gilyen2018singular}.
\cref{cor:iir-clock-frontier} gives the least uniform error at a fixed normalization.
\begin{corollary}[Error--normalization frontier for a fixed lift]
\label{cor:iir-clock-frontier}
For fixed $S$, public input and output spaces, delays, and horizon, the least uniform operator error at normalization $\alpha_{\rm clk}$ is
\begin{align}
\epsilon_{{\rm clk},N}(\alpha_{\rm clk}) =\min_{\|X\|_*\le\alpha_{\rm clk}} \sup_{\boldsymbol{O}\in\mathfrak O} \left\|\sum_{n=0}^{N-1}c_n(X)G_{n}-G\right\|.
\label{eq:iir-joint-frontier}
\end{align}
The same clock is used for every allowed oracle tuple.
The minimum is attained.
It is nonincreasing in the horizon and normalization, and convex in the normalization.
For a finite explicit promise set it has a semidefinite representation.
\end{corollary}
\begin{proof}
\cref{thm:iir-clock-body} identifies the feasible clocks.
The nuclear-norm ball is compact, and the error is a continuous convex function of $X$: each finite Toeplitz block is contractive, giving a uniform Lipschitz bound for the supremum.
Increasing the normalization enlarges the feasible set.
Increasing the horizon permits zero-padding an old clock without changing its coefficients.
Convex combinations of feasible matrices prove convexity in the normalization.

The semidefinite representation and its block-matrix proof are given in \cref{app:clock-frontier-sdp}.
\end{proof}
This optimization is exact only after the realization has been fixed.
\Cref{thm:ach-clock-completeness} later shows that optimizing this error--normalization tradeoff for the chosen lift over all schedule-preserving unitary realizations recovers the optimal error--normalization tradeoff over circuits with the same online query schedule.
The fixed-lift SDP is therefore the exact inner problem of a more general query-synthesis problem.

For oracle costs $C_j$, the outer optimization minimizes
\begin{align}
\sum_j C_j\left\lfloor\frac{N-1}{r_j}\right\rfloor
\quad\text{subject to}
\quad \epsilon_N(\alpha_{\rm clk};\boldsymbol{r})\le\epsilon_{\rm clk}.
\label{eq:clock-design-outer}
\end{align}
For a finite list of $J$ oracle instances, the constraints in \cref{eq:iir-joint-frontier} have sizes
\begin{align}
\begin{array}{ll} \text{clock-norm PSD block:}&2N,
\\
\text{error PSD blocks:}&J\text{ blocks of order }2d,
\\
\text{scalar clock variables:}&O(N^2). \end{array}
\end{align}
The public-space dimension $d$ is the order of each coefficient matrix $G_n$.
It is generally exponential in the system-qubit count, as is the dimension of an oracle acting on that register.
The first $N$ coefficient matrices contain $O(JNd^2)$ complex entries in a dense representation.
The private dimension affects computing these coefficients, while the clock-norm block still has order $2N$.
A continuous promise generally imposes infinitely many operator constraints.
These are representation obstacles; structured promises may admit smaller exact descriptions.
A succinct block encoding alone gives no efficient classical separation procedure for those constraints.
The evaluated operator $F$ also leaves its realization and analytic query coordinates unspecified.

The analytic theorem avoids constructing the coefficient matrices by using known bounds uniform over the full promise.
For example, the noncommuting Hamiltonian construction in \cref{sec:multi-hs-response} has
\begin{align}
G(z)=\exp[-tM(z^{r_1},\ldots,z^{r_m})],
\quad W=\Omega I,
\quad\Omega=t\sum_j\lambda_jr_j.
\end{align}
\cref{eq:multi-hs-radial} gives the additional two-sided estimates from the Hermitian part of the generator.
The application proves the analytic bounds; clock shaping computes scalar weights from them; quditization implements those weights.
Deriving these bounds is generally difficult; a succinct realization alone does not give them or an efficient procedure for finding them.

The startup-transient identity in \cref{subsec:clock-design-kernel} expresses the approximation error through adjacent clock-weight differences.
With only a catalyst-norm bound, \cref{subsec:clock-design-flat} gives inverse-square precision dependence for the full uniform-clock output, and inverse-linear dependence after a sharper public-block estimate and amplitude amplification.
Local analytic continuation then gives the logarithmic precision dependence in \cref{subsec:clock-design-smoothing}.
The final refinements allocate queries among ports and allow approximate realizations or additional coefficient structure.

\subsection{Catalyst omission error and normalization}
\label{subsec:clock-design-kernel}

To compile the transfer function, we must control its difference from the desired transformation at $\Bz=\boldsymbol{1}$.
Multiplication by the unitary $G^\dagger$ preserves this error norm and makes the target the identity: $\|F(\Bz)-G\|=\|I-G^\dagger F(\Bz)\|$.
The difference vanishes at $\Bz=\boldsymbol{1}$.
Fixing the first argument of the unitarity identity \cref{eq:iir-kernel} there factors this difference into the displacement $1-z_j$ of each oracle coordinate and a matrix coefficient:
\begin{align}
K_j(\Bz)=\Gamma_j^\dagger\Gamma_j(\Bz),
\quad I-G^\dagger F(\Bz)=\sum_j(1-z_j)K_j(\Bz).
\label{eq:qudit-polarized-kernel}
\end{align}
We call $K_j(\Bz)$ the port-$j$ transient generating function.
It measures the overlap between the steady-state private amplitude and its analytically continued value at port $j$.
The left factor is fixed, so the product is analytic in $\Bz$ and admits Cauchy coefficient estimates.
Conjugating a varying left factor would also introduce $\overline{\Bz}$ into the product.
At $\Bz=\boldsymbol{1}$, $K_j=\Gamma_j^\dagger\Gamma_j\succeq0$: its expectation is precisely that port's Las Vegas query weight.
Away from $\Bz=\boldsymbol{1}$, the analytic matrix $K_j(\Bz)$ gives the higher-order information needed to bound a finite approximation.

The delayed slice combines these port contributions into the \emph{zero-catalyst transient generating function} $K(z)$.
Each factor $1-z_j$ becomes $1-z^{r_j}=(1-z)\sum_{s=0}^{r_j-1}z^s$, so define
\begin{align}
K(z)=\sum_j\sum_{s=0}^{r_j-1}z^s K_j(z^{r_1},\ldots,z^{r_m}) =\frac{I-G^\dagger G(z)}{1-z},
\quad K=\sum_jr_j\Gamma_j^\dagger\Gamma_j=G^\dagger G'=W.
\label{eq:qudit-polarized-delayed}
\end{align}
Thus $K_j(\Bz)$ describes one oracle port in the multivariate system, whereas $K(z)$ combines them along the chosen one-variable delay slice.
The geometric sum accounts for the $r_j$ private clock positions between successive uses of port $j$.
It therefore gives the same delay-weighted energy as the explicit delay line.
The quotient removes the known first-order zero of the error at $z=1$, leaving an analytic function whose value at $z=1$ is the group-delay matrix.

Expanding $K(z)$ about $z=1$ gives the higher derivatives.
Suppressing arguments at this evaluation point,
\begin{align}
K(z)=\sum_{k\ge0}\frac{G^\dagger G^{(k+1)}}{(k+1)!}(z-1)^k =W+\frac{z-1}{2}G^\dagger G''+O((z-1)^2).
\label{eq:clock-design-kernel-boundary-series}
\end{align}
In terms of the matrix group delay $W(\theta)$ from \cref{eq:intro-framework-delay-expansion}, differentiation gives
\begin{align}
G^\dagger G''=W^2-W-i\,\left.\frac{dW(\theta)}{d\theta}\right|_{\theta=0}.
\label{eq:qudit-kernel-dispersion}
\end{align}
Indeed, differentiating $W(\theta)=-iG(e^{i\theta})^\dagger\partial_\theta G(e^{i\theta})$ gives $dW/d\theta=i(W+G^\dagger G''-W^2)$ at zero.
Thus $K$ separates the delay at $z=1$ from its analytic variation, which includes both the curvature of a pure delay and frequency dispersion.
For example, a pure delay $d\ge1$ has $G(z)=z^dG$ and $K(z)=(1+z+\cdots+z^{d-1})I$.
Its delay is the constant $dI$ on the unit circle.

The zero-private-input evolution in \cref{eq:qudit-causal-convolution} has step response $Y_n\ket\psi$, where \begin{align}
Y_n=\sum_{k=0}^nG_k
\quad(n\ge0),
\quad Y_n=0
\quad(n<0).
\label{eq:qudit-step-response}
\end{align} For $|z|<1$, contractivity bounds the Taylor coefficients, so summing the absolutely convergent series gives
\begin{align}
\sum_{n\ge0}Y_nz^n =\sum_{k\ge0}G_k\sum_{n\ge k}z^n =\frac{G(z)}{1-z}.
\label{eq:clock-step-generating}
\end{align}
Since $G^\dagger G=I$, the zero-catalyst transient generating function is therefore
\begin{align}
K(z)=G^\dagger\sum_{n\ge0}(G-Y_n)z^n =\frac{I-G^\dagger G(z)}{1-z}.
\label{eq:qudit-polarized-delayed-series}
\end{align}
The fixed factor $G^\dagger$ expresses the output difference in the input frame and preserves its norm.
The varying argument remains analytic, as in \cref{eq:qudit-polarized-kernel}.
The startup transient at step $n$ is the difference $G-Y_n$ from the steady-state output.
Its coefficients satisfy
\begin{subequations}
\label{eq:clock-design-kernel-coefficients}
\begin{align}
K_n=G^\dagger(G-Y_n),
\quad \|K_n\boldsymbol{\psi}\|=\|(G-Y_n)\boldsymbol{\psi}\|,
\\
G^\dagger G_0=I-K_0,
\quad G^\dagger G_n=K_{n-1}-K_n
\quad(n\ge1).
\end{align}
\end{subequations}
For the system matrix in \cref{eq:intro-response}, the private amplitude just before step $n$ is $\Gamma\boldsymbol{\psi}$ in the steady-state solution and $[I-(DO)^n]\Gamma\boldsymbol{\psi}$ with zero initial private amplitude.
Their difference is $(DO)^n\Gamma\boldsymbol{\psi}$.
Similarly,
\begin{align}
Y_n-Y_{n-1}=G_n,
\quad (G-Y_{n-1})-(G-Y_n)=G_n
\quad(n\ge1).
\label{eq:clock-step-differences}
\end{align}
For the single unit-delay oracle, this gives $Y_n=G-BO(DO)^n\Gamma$.
The delayed system satisfies the same identity in its buffer realization.
This identifies the transient directly with the missing initial condition.
Because these identities use only finite sums, they remain valid when the coefficient series fails to converge at $z=1$ or the transient oscillates.

For any finite clock weights $c_0,\ldots,c_{N-1}$, set $c_N=0$.
Multiplying by the weights, summing, and subtracting from $I$ gives
\begin{subequations}
\label{eq:clock-design-residual}
\begin{align}
G^\dagger\widetilde G_N &=c_0(I-K_0)+\sum_{n=1}^{N-1}c_n(K_{n-1}-K_n),
\\
G^\dagger(G-\widetilde G_N) &=(1-c_0)I+\sum_{n=0}^{N-1}(c_n-c_{n+1})K_n.
\end{align}
\end{subequations}
Each change in a clock weight multiplies the corresponding startup-transient coefficient.
On a flat interval $c_n=c_{n+1}$, these terms vanish, so the transition positions and widths must be chosen subject to the clock-norm constraint.

In the three-step example of \cref{eq:qudit-three-step}, absorb the oracle into the one-step blocks.
A steady-state catalyst then obeys $C+D\Gamma=\Gamma$ and $A+B\Gamma=G$.
The normalized input $\ket{c_{\rm in}}\otimes\ket\psi$, where $\ket{c_{\rm in}}=(\ket0+\ket1+\ket2)/\sqrt3$, occupies the public space and has zero private amplitude.
Its evolution is
\begin{align}
\mathcal W_3[G]\frac1{\sqrt3} \begin{pmatrix}\ket\psi
\\
\ket\psi
\\
\ket\psi
\\
0\end{pmatrix} =\frac1{\sqrt3} \begin{pmatrix}Y_0\ket\psi
\\
Y_1\ket\psi
\\
Y_2\ket\psi
\\
(I-D^3)\Gamma\ket\psi\end{pmatrix},
\quad Y_n=G-BD^n\Gamma.
\label{eq:qudit-normalized-step}
\end{align}
The complete output has norm one.
If the comparison input also contained $\Gamma\ket\psi/\sqrt3$ in its private component, each public output would instead equal $G\ket\psi/\sqrt3$.
The missing column contributes $(B\Gamma,BD\Gamma,BD^2\Gamma,D^3\Gamma)^{\mathsf T}\ket\psi/\sqrt3$.
This comparison input is an unnormalized vector used to calculate the error; the implemented circuit starts with zero private amplitude.

Every input envelope is a sum of shifted constant inputs.
Setting $g_{-1}=0$ and $\Delta g_j=g_j-g_{j-1}$ in the one-port example gives
\begin{align}
\boldsymbol{y}_n=g_nG\ket\psi -\sum_{j=0}^n\Delta g_jBD^{n-j}\Gamma\ket\psi.
\label{eq:qudit-envelope-transient}
\end{align}
Indeed, $g_i=\sum_{j=0}^i\Delta g_j$ in \cref{eq:qudit-causal-convolution} yields $\boldsymbol{y}_n=\sum_{j=0}^n\Delta g_jY_{n-j}\ket\psi$.
Clock shaping controls the residual of the weighted public output; individual public outputs can have larger errors.

\subsection{Flat clock shaping with polynomial convergence}
\label{subsec:clock-design-flat}

We first analyze uniform clock-state amplitudes, which give triangular clock weights and polynomial convergence at normalization one.
A sharper public-block estimate and one amplification step improve the precision dependence from inverse square to inverse linear.
We then construct a flat clock-weight plateau that cancels early transients exactly and provides the starting point for analytic smoothing.
Uniform clock amplitudes $g_i=h_i=1/\sqrt N$ give a block-encoding of
\begin{align}
\Be{\widetilde G^{\rm unif}_N/\alpha_{\rm clk}},
\quad \widetilde G^{\rm unif}_N=\sum_{n=0}^{N-1}\left(1-\frac{n}{N}\right)G_n,
\quad \alpha_{\rm clk}=1,
\quad \|\widetilde G^{\rm unif}_N-G\|\le\frac{2w_{\rm all}}{N}.
\label{eq:iir-fejer}
\end{align}

Thus the encoded-operator residual for the uniform clock is
\begin{align}
G^\dagger(G-\widetilde G_N^{\rm unif}) =\frac1N\sum_{n=0}^{N-1}K_n.
\end{align}
\cref{thm:iir-uniform} first bounds the complete output isometry using only the steady-state catalyst norm; \cref{lem:iir-uniform-terminal-residual} then gives a sharper bound on the encoded operator.

For a fixed oracle tuple $\boldsymbol O$, let $W_{\boldsymbol O} =\sum_j r_j\Gamma_{j,\boldsymbol O}^{\dagger} \Gamma_{j,\boldsymbol O}$.
For prescribed normalized inputs $|\psi_{\boldsymbol O}\rangle$, the two bounds used below are
\begin{align}
w_{\rm in}\ge \sup_{\boldsymbol O\in\mathfrak O} \langle\psi_{\boldsymbol O}|W_{\boldsymbol O} |\psi_{\boldsymbol O}\rangle,
\quad w_{\rm all}\ge \sup_{\boldsymbol O\in\mathfrak O}\|W_{\boldsymbol O}\|.
\label{eq:catalyst-domain-bounds}
\end{align}
Both bounds are uniform over the oracle promise.
Only $w_{\rm all}$ controls every normalized public input.
A statement on a promised input subspace instead uses the operator norm of the corresponding compression of $W_{\boldsymbol O}$ and retains that input restriction.
Choosing the horizon requires only the given scalar bound $w_{\rm in}$ or $w_{\rm all}$; the steady-state catalyst vector is used in the analysis.

\begin{theorem}[Uniform-clock baseline]
\label{thm:iir-uniform}
Let $G=F$ be regular and suppose the following delay-weighted catalyst bound is finite:
\begin{align}
w =\sup_{\boldsymbol{O}\in\mathfrak O} \left\|\sum_jr_j\Gamma_j^\dagger \Gamma_j\right\|.
\label{eq:uniform-delayed-energy}
\end{align}
The delayed lift with uniform public input and output clocks has isometry error
\begin{align}
\epsilon_N\le2\sqrt{w/N}
\label{eq:uniform-clean-error}
\end{align}
and uses the exact counts
\begin{align}
q_j=\left\lfloor\frac{N-1}{r_j}\right\rfloor.
\label{eq:uniform-counts}
\end{align}
In particular, $N\ge4w/\epsilon^2$ suffices for isometry error at most $\epsilon$.
\end{theorem}

\begin{proof}
Fix a normalized public input $|\psi\rangle$.
For the analysis, add the steady-state private wave to the finite lift.
With the schedule of \cref{thm:iir-lift}, initialize every buffer position in port $j$ to $O_j\Gamma_j|\psi\rangle/\sqrt N$.
Its first read occurs before the first scheduled query.
After a read, the system matrix leaves $\Gamma_j|\psi\rangle/\sqrt N$ in that position, and the next scheduled query restores $O_j\Gamma_j|\psi\rangle/\sqrt N$ before the next read.
The total squared norm of the initial private wave is
\begin{align}
\frac1N \left\langle\psi\left| \sum_jr_j\Gamma_j^\dagger\Gamma_j \right|\psi\right\rangle \le\frac{w}N.
\label{eq:uniform-analysis-catalyst}
\end{align}
At every tick the input and output amplitudes satisfy the steady-state transducer relation.
Each public clock slice therefore transmits $G|\psi\rangle/\sqrt N$, and every scheduled oracle call restores the steady-state amplitude in the next private buffer.
The final private wave has the same norm as the added initial wave.

The actual circuit starts with zero private buffers.
By unitarity, removing the analysis-only initial private vector changes the final state by at most $\sqrt{w/N}$.
Removing the final steady-state private vector from the ideal comparison state costs the same amount.
The triangle inequality gives \cref{eq:uniform-clean-error}.
The query counts are those of the exact delayed lift in \cref{thm:iir-lift}.
No catalyst is prepared by the implemented circuit.
\end{proof}

A catalyst bound on the whole public space gives a sharper operator error after clock projection.
\cref{lem:iir-uniform-terminal-residual} derives this estimate, and \cref{prop:iir-uniform-catalyst-amplification} gives the error guarantee for the joint system and ancillary state after oblivious amplitude amplification.

For this calculation, absorb the oracle feedback into a finite one-step unitary.
For fixed delays, include their buffers in the private space.
Suppress the oracle tuple in the notation and write
\begin{align}
\widehat S= \left(\begin{array}{c|c} A&B
\\
\hline
C&D_{\rm buf} \end{array}\right),
\quad \widehat S\binom{I}{\Gamma_{\rm buf}} =\binom{G}{\Gamma_{\rm buf}},
\quad G^{\dagger}G=GG^{\dagger}=I.
\label{eq:uniform-buffered-steady-state}
\end{align}
Thus $\Gamma_{\rm buf}^{\dagger}\Gamma_{\rm buf}=W_{\boldsymbol O}$ for the delayed realization.
The buffered matrix is used to analyze the public block; the circuit remains the scheduled unitary Toeplitz lift.
Its query costs are those of the scheduled implementation in \cref{thm:iir-lift}.
A bounded steady-state map suffices; strict contractivity of $D_{\rm buf}$ is not assumed.

\begin{lemma}[Terminal formula for the uniform-clock residual]
\label{lem:iir-uniform-terminal-residual}
The uniform-clock circuit block-encodes $\widetilde G_N$ at normalization one.
Then
\begin{align}
\widetilde G_N =\frac1N\sum_{n=0}^{N-1}Y_n,
\quad Y_n=A+B\sum_{\ell=0}^{n-1}D_{\rm buf}^{\ell}C,
\quad G^{\dagger}(G-\widetilde G_N) =\frac1N\Gamma_{\rm buf}^{\dagger} (I-D_{\rm buf}^{N})\Gamma_{\rm buf}.
\label{eq:uniform-terminal-residual}
\end{align}
The empty sum at $n=0$ is zero.
Consequently,
\begin{align}
\|\widetilde G_N-G\| \le\frac{2\|\Gamma_{\rm buf}\|^2}{N},
\quad \|(\widetilde G_N-G)|\psi\rangle\| \le\frac{2\|\Gamma_{\rm buf}\| \|\Gamma_{\rm buf}|\psi\rangle\|}{N}, \quad(\|\psi\|=1).
\label{eq:uniform-terminal-state-bound}
\end{align}
\end{lemma}
\begin{proof}
The steady-state equations give $C=(I-D_{\rm buf})\Gamma_{\rm buf}$ and $G=A+B\Gamma_{\rm buf}$.
Conservation of the cross inner product between the steady-state column and the private-input column gives
\begin{align}
\Gamma_{\rm buf}^{\dagger} =G^{\dagger}B+\Gamma_{\rm buf}^{\dagger}D_{\rm buf},
\quad G^{\dagger}B=\Gamma_{\rm buf}^{\dagger}(I-D_{\rm buf}).
\label{eq:uniform-cross-column}
\end{align}
Finite summation, without an infinite-series limit, now yields
\begin{subequations}
\label{eq:uniform-telescoping-proof}
\begin{align}
G-Y_n&=BD_{\rm buf}^{n}\Gamma_{\rm buf},
\\
G^{\dagger}(G-\widetilde G_N) &=\frac1N\Gamma_{\rm buf}^{\dagger}(I-D_{\rm buf}) \sum_{n=0}^{N-1}D_{\rm buf}^{n}\Gamma_{\rm buf} =\frac1N\Gamma_{\rm buf}^{\dagger} (I-D_{\rm buf}^{N})\Gamma_{\rm buf}.
\end{align}
\end{subequations}
A block of a unitary is a contraction, so $\|I-D_{\rm buf}^{N}\|\le2$.
Applying this bound proves the two norm estimates.
\end{proof}
The left factor $\Gamma_{\rm buf}^{\dagger}$ in \cref{eq:uniform-terminal-residual} explains the distinction between the two bounds.
The state-specific quantity $\|\Gamma_{\rm buf}|\psi\rangle\|^2$ cannot replace $\|\Gamma_{\rm buf}\|\,\|\Gamma_{\rm buf}|\psi\rangle\|$ in \cref{eq:uniform-terminal-state-bound} without another argument.

For a unitary target, oblivious amplitude amplification in \cref{lem:iir-cubic} gives a complete isometry estimate from this operator estimate.
The hypothesis on the target matters because the same cubic transforms the singular values of a general nonunitary operator.

\begin{proposition}[Oblivious amplitude amplification with a uniform catalyst bound]
\label{prop:iir-uniform-catalyst-amplification}
Suppose the target is unitary on the full public space and a known bound $w_{\rm all}$ in \cref{eq:catalyst-domain-bounds} holds for the executed finite realization.
Assume access to the complete uniform-clock circuit and its inverse.
For $0<\epsilon\le1$, let $\eta_\epsilon>0$ be the unique solution of $\eta_\epsilon^2(3+\eta_\epsilon)=\epsilon^2$, and choose
\begin{align}
N=\max\left\{1,\left\lceil\frac{2w_{\rm all}}{\eta_\epsilon} \right\rceil\right\}.
\label{eq:uniform-catalyst-horizon}
\end{align}
Uniform-clock extraction, known attenuation to normalization two, and one application of \cref{lem:iir-cubic} have complete isometry error at most $\epsilon$, uniformly over the oracle promise and every public input.
No catalyst preparation or quantitative analytic-continuation bound is needed.

Oblivious amplitude amplification triples the query costs by \cref{lem:iir-cubic}.
Thus a black-box one-step transducer requires $3N\le6w_{\rm all}/\eta_\epsilon+3$ calls.
A scheduled lift has the total actual query count
\begin{align}
Q_j=3\left\lfloor\frac{N-1}{r_j}\right\rfloor
\label{eq:uniform-catalyst-oracle-counts}
\end{align}
for calls to $O_j$ and its inverse.
\end{proposition}
\begin{proof}
By \cref{lem:iir-uniform-terminal-residual}, the uniform-clock encoding satisfies $\|\widetilde G_N-G\|\le2w_{\rm all}/N$.
A one-qubit rotation gives $\Be{\widetilde G_N/2}$ without additional oracle calls.
Applying \cref{lem:iir-cubic} gives isometry error at most $\eta\sqrt{3+\eta}$, where $\eta=2w_{\rm all}/N\le\eta_\epsilon<1$.
The defining equation for $\eta_\epsilon$ gives the claimed error.
Reversing the entire circuit gives the stated counts.
\end{proof}
The simpler choice $N=\max\{1,\lceil4w_{\rm all}/\epsilon\rceil\}$ also suffices.
The sharper choice has leading horizon $2\sqrt3\,w_{\rm all}/\epsilon$ as $\epsilon\to0$.

If one lift contains $q_j^+$ forward and $q_j^-$ inverse calls, including calls inside state-preparation reflections, its total cost after oblivious amplitude amplification is
\begin{align}
Q_j=3(q_j^++q_j^-).
\label{eq:uniform-catalyst-expanded-counts}
\end{align}
For approximate circuits, allocating half the error to oblivious amplitude amplification uses $\eta_{\epsilon/2}$ in the horizon above (or the simpler numerator $8$).
Let the attenuated lift and each reflection have operator errors $\eta_A$ and $\eta_R$, respectively.
The three circuit calls and two reflections then add at most $3\eta_A+2\eta_R$.
Finite-realization and synthesis errors follow \cref{eq:qudit-error-budget}.

Oblivious amplitude amplification requires a catalyst bound on the entire reflected public space and inverse access.
With only a state-specific catalyst bound, \cref{thm:iir-uniform} still controls the forward output on that state.
Oblivious amplitude amplification improves the precision scaling; the analytic clocks below instead control precision together with normalization and leading query cost.

A transfer function supported on a known interval of delays would be reproduced exactly by assigning weight one to every coefficient in that interval.
For a pure delay, the clock error depends on only one weight: $G(z)=z^pU$ and $\widetilde G=c_pU$.
Choosing $c_p=1$ removes that error.
For uniform clocks, even the pure delay has error $p/N$ when $N>p$.
A flat clock-weight plateau eliminates these early transients exactly.
More generally, \cref{eq:clock-design-residual} removes every term whose adjacent weights agree; the group delay is $K(1)$, while the first Taylor coefficient is $K_0=I-G^\dagger G_0$.

Two normalized finite clock states cannot have unit correlation both at displacement zero and at a positive displacement: equality at zero identifies the states, while equality at a positive shift would require a nonzero finite sequence to be invariant under that shift.
We must allow a normalization larger than one.
\cref{lem:iir-optimal-flat} computes the least normalization for a plateau of width $D$ on $KD$ positions.
Its quadratic convergence to one gives the $\delta^{-1/2}$ dependence in the analytic clock theorem.

\begin{lemma}[Optimal flat clock]
\label{lem:iir-optimal-flat}
Let the plateau width and flat-clock block count be integers $D\ge1$ and $K\ge2$.
Denote the lower truncated shift on the block coordinates by $S_K$.
Define
\begin{align}
\alpha_K= \left[\min_{0\le s\le1}\|(1-s)I_K+sS_K\|\right]^{-1}.
\label{eq:analytic-alpha-K}
\end{align}
On $KD$ clock positions there is a rank-one clock matrix $X$ whose diagonal sums, extended to all integer displacements, satisfy
\begin{align}
c_n(X)=1
\quad(0\le n\le D),
\quad 0\le c_n(X)\le1
\quad(n\in\mathbb Z),
\label{eq:analytic-flat-coefficients}
\end{align}
with coefficients nondecreasing for $n\le0$ and nonincreasing for $n\ge D$, and
\begin{align}
\|X\|_*=\alpha_K.
\label{eq:analytic-flat-norm}
\end{align}
This is the minimum possible nuclear norm for the prescribed plateau on $KD$ positions.
Moreover,
\begin{align}
\alpha_K=1+\frac{\pi^2}{8K^2}+O(K^{-3}),
\quad \sec\!\frac{\pi}{2K+1}\le\alpha_K\le\sec\!\frac{\pi}{2K}.
\label{eq:analytic-alpha-asymptotic}
\end{align}
Consequently, \begin{align}
K\ge\frac{\pi}{2\arccos(1/(1+\delta))}
\label{eq:analytic-K-sufficient}
\end{align} is sufficient for $\alpha_K\le1+\delta$.
\end{lemma}

\begin{proof}
The lower bound follows from trace duality: an operator of small norm tests how large the nuclear norm of a feasible clock must be.
For any test matrix, the singular-value decomposition gives $\operatorname{Re}\operatorname{tr}(Y^\dagger X) \le\|X\|_*\|Y\|$.
Choose a convex combination of the identity and the shift through one plateau width: $Y(s)=(1-s)I_{KD}+sS_{KD}^{D} $ for $0\le s\le1$.
The flat clock-weight conditions make its trace pairing equal to one.
Thus
\begin{subequations}
\label{eq:analytic-flat-dual}
\begin{align}
1&=(1-s)\operatorname{Re}c_0(X)+s\operatorname{Re}c_D(X)
\\
&=\operatorname{Re}\operatorname{tr}(Y(s)^\dagger X) \le\|X\|_*\|Y(s)\|.
\end{align}
\end{subequations}
Dividing by the test norm gives a lower bound on every feasible clock.
Minimizing that denominator makes the lower bound strongest.
The shift $S_{KD}^{D}$ decomposes into $D$ independent chains.
Each chain has the flat-clock block count $K$ as its length.
Therefore every feasible clock obeys
\begin{align}
\|X\|_*\ge \left[\min_{0\le s\le1}\|(1-s)I_K+sS_K\|\right]^{-1} =\alpha_K.
\label{eq:analytic-flat-dual-bound}
\end{align}

A matching pair of vectors both evaluates the scalar minimum and constructs the optimal clock.
Put $s_*=(K+1)/(2K+1)$ and let $\theta$ be the root in $(\pi/(K+1),\pi/K)$ of
\begin{align}
K\sin((K+1)\theta)+(K+1)\sin(K\theta)=0.
\label{eq:analytic-flat-recurrence}
\end{align}
Normalize the positive sine sequences $u_j\propto\sin((K-j)\theta)$ and $v_j\propto\sin((j+1)\theta)$, for $0\le j<K$.
They are reversals of one another.
The tridiagonal recurrence for $B_*^\dagger B_*$, where $B_*=(1-s_*)I_K+s_*S_K$, and finite sine sums give
\begin{align}
B_*v=\sigma_*u,
\quad B_*^\dagger u=\sigma_*v,
\quad u^\dagger v=u^\dagger S_Kv=\sigma_*=\|B_*\|.
\label{eq:analytic-flat-two-correlations}
\end{align}
Positivity identifies the top singular vectors.
For every real $s$, $u^\dagger[(1-s)I_K+sS_K]v=\sigma_*$, so the operator norm is at least $\sigma_*$.
Equality at $s_*$ proves the minimum directly.

For completeness, put $t=\theta/2$ and $\nu=(2K+1)t-\pi$.
The same recurrence gives
\begin{align}
\tan\nu=\frac{\tan t}{2K+1},
\quad \alpha_K=\frac{\cos\nu}{\cos t},
\quad 0<\nu<t<\frac{\pi}{2K}.
\label{eq:analytic-flat-scalar-form}
\end{align}
Thus $\alpha_K\le\sec(\pi/(2K))$.
Evaluating the norm at $s=1/2$ gives $\|B(1/2)\|=\cos(\pi/(2K+1))$ and the lower secant bound.
These bounds also yield $|\alpha_K-1-\pi^2/(8K^2)|\le12/K^3$ for $K\ge2$, by the elementary Taylor bounds for the secant on $[0,\pi/4]$.

Replace each entry of the two singular vectors by a constant block of length $D$, divided by $\sqrt D$.
Denote the resulting normalized vectors by $\widetilde u,\widetilde v$.
Set
\begin{align}
X_*=\frac{\widetilde u\widetilde v^\dagger}{\|B_*\|}.
\label{eq:analytic-flat-primal}
\end{align}
At displacement $bD+r$, the fine-grid correlation interpolates linearly between the two adjacent coarse correlations.
The interpolation parameter is $r/D\in[0,1]$.
The two equalities in \cref{eq:analytic-flat-two-correlations} therefore imply $c_n(X_*)=1 $ for $0\le n\le D$ and $\|X_*\|_*=\|B_*\|^{-1}=\alpha_K$.
This matches \cref{eq:analytic-flat-dual-bound}.

The coarse singular vectors are positive sine sequences.
Their logarithms are concave because
\begin{align}
\frac{d^2}{dx^2}\log\sin x=-\csc^2x<0
\quad(0<x<\pi).
\end{align}
The sampled sequences are therefore log-concave and have no internal zeros.
Convolution preserves these properties \cite{Hoggar1974LogarithmicConcavity}; reversing one sequence identifies their cross-correlation as a convolution.
Its consecutive ratios are nonincreasing wherever they are defined.
The equal correlations at shifts zero and one are consequently global maxima.
Block interpolation proves the asserted monotonicity and the interval of values in \cref{eq:analytic-flat-coefficients}.

\end{proof}

The two secant bounds reduce exact block-count selection to at most two candidates.
For $x_\delta=\pi/[2\arccos(1/(1+\delta))]$, set
\begin{align}
K_-=\max\{2,\lceil x_\delta-\tfrac12\rceil\},
\quad K_+=\max\{2,\lceil x_\delta\rceil\}.
\label{eq:analytic-K-candidates}
\end{align}
The least feasible count $K_\delta$ lies between these consecutive integers; test $K_-$ using \cref{eq:analytic-alpha-K}, and use $K_+$ if needed.
For example, $\alpha_2=5/4$, so normalization $5/4$ needs only two blocks, where the upper secant bound alone would prescribe three.

\subsection{Analytic clock shaping with exponential convergence}
\label{subsec:clock-design-smoothing}

The flat-clock lemma preserves a finite band of coefficients at the least normalization cost, but a locally analytic transfer function need not have a finite or absolutely summable impulse tail.
Continuation near $z=1$ controls a local part of the circle; singularities elsewhere can obstruct a global Taylor-tail estimate.
Smooth clock edges suppress their contribution.

To control this residual, convolve the rising and falling edges of the flat window with repeated boxcar averages.
Each boxcar is a uniform average over a short interval of integer displacements.
The resulting coefficients can be expressed as an expectation over the sum of those displacements, but they are computed classically and define one deterministic clock.
In a contour representation, each averaging step contributes a factor smaller than one away from $z=1$.
Repeating it logarithmically many times suppresses the error.
Positive weights give a sufficient condition for preserving normalization: averaging shifted copies of one endpoint vector cannot increase its norm.
\cref{lem:iir-flat-smoothing} bounds the resulting approximation error at this normalization.

\begin{lemma}[Smoothed flat-window estimate]
\label{lem:iir-flat-smoothing}
Let $G(z)=\sum_{n\ge0}G_nz^n$ be an analytic, contractive operator-valued function in the unit disk, and write $G=G(1)$.
The input and output spaces may differ.
Assume analytic continuation to a neighborhood of $|z-1|\le\zeta$ for $0<\zeta\le1/16$.
Suppose known numbers $0\le A_-\le A_+$ and $M_{\rm loc}\ge1$ satisfy
\begin{align}
\|G(z)\|\le M_{\rm loc} \begin{cases} |z|^{A_-},& |z|\le1,
\\
|z|^{A_+},& |z|\ge1, \end{cases}
\quad |z-1|\le\zeta.
\label{eq:analytic-two-sided-cap}
\end{align}
For $0<\epsilon_{\rm clk}<1$, choose
\begin{align}
\rho&=\zeta,
\quad M_{\rm sm}=\left\lceil\frac{2e(1+\rho)}{\rho}\right\rceil,
\quad \beta=\frac{2(1+\rho)}{M_{\rm sm}\rho}\le e^{-1},
\label{eq:analytic-smoothing-parameters-a}
\\
q_{\rm sm}&=\max\left\{1, \left\lceil\frac{\log(3M_{\rm loc}/\epsilon_{\rm clk})} {\log(1/\beta)}\right\rceil\right\},
\quad d=q_{\rm sm}(M_{\rm sm}-1).
\label{eq:analytic-smoothing-parameters-b}
\end{align}
Let $c_n^0$ be a finitely supported real sequence on the integers, with values in $[0,1]$.
It equals one at indices $0,\ldots,D$, is nondecreasing before this plateau, and is nonincreasing afterwards.
Let $U_1,\ldots,U_{q_{\rm sm}}$ be independent and set
\begin{align}
V=U_1+\cdots+U_{q_{\rm sm}},
\quad U_i\sim\operatorname{Unif}\{0,\ldots,M_{\rm sm}-1\},
\label{eq:analytic-smoothing-random-variable}
\end{align}
and define the smoothed coefficients
\begin{align}
c_n=\mathbb E\! \left[c^0_{n-s_{\rm off}-V}\right],
\quad s_{\rm off}=\lfloor A_-\rfloor-d.
\label{eq:analytic-flat-smoothing}
\end{align}
If
\begin{align}
D\ge\lceil A_+\rceil-\lfloor A_-\rfloor+d,
\label{eq:analytic-smoothing-width}
\end{align}
then
\begin{align}
\left\|\sum_{n\ge0}c_nG_{n}-G\right\| \le3M_{\rm loc}\beta^{q_{\rm sm}} \le\epsilon_{\rm clk}.
\label{eq:analytic-flat-error}
\end{align}
Only analyticity and the norm bounds enter this estimate; $G(1)$ may be nonunitary.
If the base window is realized by a clock matrix, this positive smoothing can be realized without increasing its nuclear norm.
No absolute summability of the impulse coefficients and no continuation around the full unit circle are required.
\end{lemma}

\begin{proof}
Use the step-response operators $Y_n$ from \cref{eq:qudit-step-response}.
The expectation in \cref{eq:analytic-flat-smoothing} defines deterministic scalar weights; the circuit prepares the resulting clock vectors without sampling $V$.
For any integer cutoff $J$, finite summation gives
\begin{align}
\mathbb E[Y_{J+V-1}] =\sum_vp_v\sum_{0\le n<J+v}G_n =\sum_{n\ge0}\Pr(n<J+V)G_n.
\label{eq:analytic-averaged-step}
\end{align}
For a uniform average over three shifts and a positive threshold index, this reduces to $Y_{J-1}+\tfrac23G_J+\tfrac13G_{J+1}$ for $J\ge1$.
A finite window has both a rising and a falling edge.
The two transition sequences of the base window are
\begin{align}
\ell_j=c^0_j-c^0_{j-1}
\quad(j\le0),
\quad u_j=c^0_{j-1}-c^0_j
\quad(j\ge D+1).
\label{eq:analytic-transition-distributions}
\end{align}
They are nonnegative and each has total mass one.
Finite summation by parts gives
\begin{align}
\sum_{n\ge0}c_nG_{n} =\sum_{j,v}u_jp_vY_{j+s_{\rm off}+v-1} -\sum_{j,v}\ell_jp_vY_{j+s_{\rm off}+v-1},
\label{eq:analytic-two-transitions}
\end{align}
where $p_v=\Pr[V=v]$.
The width condition implies
\begin{align}
j+s_{\rm off}\ge A_++1
\quad\text{on the upper transition},
\quad j+s_{\rm off}+d\le A_-
\quad\text{on the lower transition}.
\label{eq:analytic-transition-placement}
\end{align}

The generating function in \cref{eq:clock-step-generating} converts these partial sums into contour integrals.
A positively oriented contour $\mathcal C_-$ surrounds zero while excluding the evaluation point.
Cauchy's coefficient formula extracts the appropriate step response.
Deforming the contour outward near $z=1$ gives a second contour $\mathcal C_+$ that also encloses the simple pole there.
Its residue is
\begin{align}
\operatorname*{Res}_{z=1}\left[-\frac{G(z)z^{-J}}{z-1}\right]=-G.
\end{align}
The added residue subtracts the steady-state transformation, giving
\begin{align}
Y_{J-1}-G &=-\frac1{2\pi i}\oint_{\mathcal C_+} \frac{G(z)z^{-J}}{z-1}\,dz,
\label{eq:analytic-upper-contour}
\\
Y_{J-1} &=-\frac1{2\pi i}\oint_{\mathcal C_-} \frac{G(z)z^{-J}}{z-1}\,dz.
\label{eq:analytic-lower-contour}
\end{align}
The second identity also holds for $J\le0$, since the integrand is then analytic at zero.
Define
\begin{align}
H_M(z)=\frac{1+z+\cdots+z^{M-1}}{M}.
\end{align}
The independence in \cref{eq:analytic-smoothing-random-variable} gives
\begin{align}
\mathbb E[z^{-V}] =\prod_{i=1}^{q_{\rm sm}}\frac1{M_{\rm sm}} \sum_{u=0}^{M_{\rm sm}-1}z^{-u} =H_{M_{\rm sm}}(1/z)^{q_{\rm sm}}.
\label{eq:analytic-smoothing-transform}
\end{align}
This normalized geometric-sum polynomial multiplies the upper-contour integrand when the displacement is averaged.
Its small modulus on the contour away from the evaluation point suppresses singularities elsewhere on the unit circle.
For the lower contour use
\begin{align}
H_{M_{\rm sm}}(1/z)^{q_{\rm sm}} =z^{-d}H_{M_{\rm sm}}(z)^{q_{\rm sm}}.
\label{eq:analytic-smoothing-reversal}
\end{align}

Near $z=1$, analytic continuation allows the outer contour to leave the unit disk, and radial growth determines the required cutoff.
Away from $1$, disk contractivity and the smoothing polynomial control the integral.
Local continuation alone gives no uniform exponential bound on every unsmoothed transient, since other points of the unit circle can support a long-lived response.
Take the outer and inner detours on $|z-1|=\rho$.
On the outer detour, \cref{eq:analytic-transition-placement,eq:analytic-two-sided-cap} imply that the norm of the transfer function together with the residual power of $z$ is at most $M_{\rm loc}$.
On the inner detour the same is true after the factor $z^{-d}$ in \cref{eq:analytic-smoothing-reversal}.
The geometric series identity gives
\begin{align}
|H_{M_{\rm sm}}(1/z)| \le\frac{2(1+\rho)}{M_{\rm sm}\rho}=\beta,
\quad |H_{M_{\rm sm}}(z)| \le\frac{2}{M_{\rm sm}\rho}\le\beta.
\label{eq:analytic-smoothing-detour-bound}
\end{align}
Each detour has length at most $2\pi\rho$, so each detour integral is at most $M_{\rm loc}\beta^{q_{\rm sm}}$.

It remains to bound the unit-circle arcs that lie outside the local neighborhood.
Put $t=|z-1|$.
Disk contractivity gives an integrand norm at most
\begin{align}
t^{-1}\left(\frac{2}{M_{\rm sm}t}\right)^{q_{\rm sm}}.
\label{eq:analytic-arc-integrand}
\end{align}
On $\rho\le t\le1$, each arc Jacobian is at most $2/\sqrt3$.
On the remaining part $t\ge1$, their total normalized length is at most one.
Hence each remaining contour contribution is at most
\begin{align}
\left[ \frac{2}{\pi\sqrt3\,q_{\rm sm}}+\rho^{q_{\rm sm}} \right] \left(\frac{2}{M_{\rm sm}\rho}\right)^{q_{\rm sm}} <0.431\,\beta^{q_{\rm sm}}.
\label{eq:analytic-flat-arc-bound}
\end{align}
The two contours together therefore contribute less than $3M_{\rm loc}\beta^{q_{\rm sm}}$.
The definition of $q_{\rm sm}$ gives \cref{eq:analytic-flat-error}.
If analytic continuation is available only near the evaluation point, first use the boundaries of $\{|z|<r\}\cup\{|z-1|<\rho\}$ and $\{|z|<r\}\setminus\{|z-1|\le\rho\}$.
Then let $r\uparrow1$; disk contractivity controls the remaining arcs.

On a clock space extended to contain every shift, let $T$ translate the address and replace an output factor by
\begin{align}
h'=\sum_vp_vT^{s_{\rm off}+v}h,
\quad \|h'\|\le\sum_vp_v\|T^{s_{\rm off}+v}h\|=\|h\|.
\label{eq:analytic-smoothed-clock-factor}
\end{align}
The input factor is unchanged.
Thus smoothing cannot increase the nuclear norm of a rank-one clock matrix.
For a general clock the same conclusion follows by applying the operation to each singular-vector term and using convexity.
\end{proof}

For the one-port example in \cref{eq:qudit-normalized-step}, $G-\mathbb E[Y_{J+V-1}]=BD^{J-1}H_{M_{\rm sm}}(D)^{q_{\rm sm}}\Gamma$ when $J\ge1$.
This identity expresses cancellation of the transient in the observable public output.
The contour proof bounds that output; local analyticity need not bound $H_{M_{\rm sm}}(D)$ on every private state.
Other probability weights replace $H_M$ by their generating polynomial and require corresponding contour bounds.
Signed or complex weights incur the norm factor $\sum_v|p_v|$ in \cref{eq:analytic-smoothed-clock-factor}.

\subsubsection{Choosing the analytic scale and oracle delays}
\label{subsec:clock-scale}

A smaller analytic neighborhood can give a narrower interval of radial exponents, at the cost of a longer smoothing filter.
We keep that choice free in \cref{thm:iir-scale}.
The clock geometry is unchanged: the input clock is fixed, and the output clock is a positive average of translations.
The following definition specifies the two contour estimates needed to use a different averaging distribution in the same construction.

Fix a local radius $0<\rho\le1/16$ and radial exponents as in \cref{eq:analytic-two-sided-cap}.
Write their integer endpoints as $\ell=\lfloor A_-\rfloor$ and $u=\lceil A_+\rceil$.
Approach the unit circle with $r<1$.
Let the positively oriented contours $\mathcal C_{+,r}$ and $\mathcal C_{-,r}$ be the boundaries of $\{|z|<r\}\cup\{|z-1|<\rho\}$ and $\{|z|<r\}\setminus\{|z-1|\le\rho\}$, respectively.
The first contour encloses the evaluation point; the second excludes it.
The complete integrals, including their arcs away from the local neighborhood, enter \cref{def:iir-positive-smoother}.

\begin{definition}[Positive smoothing distribution]
\label{def:iir-positive-smoother}
Let $p_v\ge0$ be a probability distribution supported on the integer interval $0,\ldots,d$.
The support bound is a positive integer.
Its generating and reversed polynomials are
\begin{align}
P(z)=\sum_{v=0}^dp_vz^v,
\quad P^\vee(z)=z^dP(1/z)=\sum_{v=0}^dp_vz^{d-v}.
\label{eq:scale-smoothing-polynomial}
\end{align}
The distribution is admissible at error $\epsilon_{\rm clk}$ if the following bounds hold uniformly over the promised oracle family:
\begin{align}
E_+&=\sup_{J\in\mathbb Z,\ J\ge u+1}\limsup_{r\uparrow1} \frac1{2\pi}\int_{\mathcal C_{+,r}} \frac{\|G(z)\|\,|z|^{-J}|P(1/z)|}{|z-1|}\,|dz|,
\label{eq:scale-upper-integral}
\\
E_-&=\sup_{J\in\mathbb Z,\ J+d\le\ell}\limsup_{r\uparrow1} \frac1{2\pi}\int_{\mathcal C_{-,r}} \frac{\|G(z)\|\,|z|^{-(J+d)}|P^\vee(z)|}{|z-1|}\,|dz|,
\label{eq:scale-lower-integral}
\\
E_++E_-&\le\epsilon_{\rm clk}.
\label{eq:scale-complete-smoothing}
\end{align}
\end{definition}

The upper integral bounds the late-time tail after subtracting the steady-state output.
The lower integral bounds the missing early-time contribution.
This follows from the two coefficient identities in \cref{eq:analytic-upper-contour,eq:analytic-lower-contour}; averaging a cutoff inserts the corresponding polynomial in \cref{eq:scale-smoothing-polynomial}.
A bound restricted to the unit circle does not control either local detour.
The repeated uniform average in \cref{lem:iir-flat-smoothing} satisfies \cref{eq:scale-complete-smoothing}, with its stated constants.
Positive weights do not increase the nuclear norm by \cref{eq:analytic-smoothed-clock-factor}.
Signed or complex weights instead multiply that bound by their coefficient one-norm.

For each admissible radius, let the radial endpoints and smoothing support be denoted by $A_\pm(\rho)$ and $d(\rho,\epsilon_{\rm clk})$.
\cref{thm:iir-scale} separates their contribution to the finite horizon from the choice of radius.

\begin{theorem}[Scale-optimized analytic clock shaping]
\label{thm:iir-scale}
Let $G$ be analytic and contractive in the unit disk.
At every radius $\rho$ in a nonempty set $\mathcal R\subseteq(0,1/16]$, assume analytic continuation and the radial bounds in \cref{eq:analytic-two-sided-cap}, with known $0\le A_-(\rho)\le A_+(\rho)$.
Set
\begin{align}
\ell(\rho)=\lfloor A_-(\rho)\rfloor,
\quad u(\rho)=\lceil A_+(\rho)\rceil.
\label{eq:scale-radial-endpoints}
\end{align}
Use an admissible positive smoother from \cref{def:iir-positive-smoother}, with $d\ge1$, $0<\epsilon_{\rm clk}<1$, and $0<\delta\le1$.
Choose a flat-clock block count $K\ge2$ satisfying $\alpha_K\le1+\delta$.
The least such block count is
\begin{align}
K_\delta=\min\{K\in\mathbb Z:K\ge2,\ \alpha_K\le1+\delta\}.
\label{eq:scale-block-count}
\end{align}
For every admissible radius, a rank-one clock matrix has nuclear norm at most $1+\delta$ and approximation error at most $\epsilon_{\rm clk}$.
Its horizon is
\begin{align}
N(\rho)=K\bigl(u(\rho)-\ell(\rho)+d(\rho,\epsilon_{\rm clk})\bigr) +\max\{\ell(\rho),d(\rho,\epsilon_{\rm clk})\}.
\label{eq:analytic-horizon-exact}
\end{align}
If a unitary circuit with public block $T_N[G]$ is available, these clocks implement $\Be{\widetilde G_N/(1+\delta)}$.
This includes rectangular compressions of a unitary Toeplitz lift.
For a delayed IIR with port delays $r_j$, the smallest feasible horizon has exact query counts
\begin{align}
N=\min_{\rho\in\mathcal R}N(\rho),
\quad q_j=\left\lfloor\frac{N-1}{r_j}\right\rfloor.
\label{eq:scale-optimized-counts}
\end{align}
\end{theorem}

\begin{proof}
At any admissible radius, apply \cref{lem:iir-optimal-flat} with plateau width $D=u-\ell+d$.
Average the output clock with the given distribution and translate it by $\ell-d$.
The two transition distributions in \cref{eq:analytic-two-transitions} have mass one, and their cutoffs satisfy the respective conditions in \cref{def:iir-positive-smoother}.
Their errors therefore sum to at most $\epsilon_{\rm clk}$.
The input support is $[0,KD-1]$ and the output support lies in $[\ell-d,KD-1+\ell]$.
A common translation to nonnegative addresses gives \cref{eq:analytic-horizon-exact}.
Positive averaging preserves rank one and does not increase its nuclear norm; known attenuation pads the normalization to $1+\delta$.
The feasible horizons form a nonempty subset of the positive integers, so their minimum is attained.
Finally, \cref{thm:iir-lift} gives the counts.
\end{proof}

The minimum in \cref{eq:scale-optimized-counts} is over this construction.
The full error--normalization tradeoff for the chosen lift remains \cref{eq:iir-joint-frontier}.
In an implementation, any finite list of radii with proved growth and smoothing bounds gives a computable choice; the theorem does not assert an efficient global search for arbitrary functions of the radius.

To interpret the exact horizon, suppress the radius arguments and write $A_-=a-h_-$ and $A_+=w+h_+$ for the specialization obtained from the group-delay bounds in \cref{eq:analytic-growth-certificate}.
Its equivalent form and a bound including rounding are
\begin{subequations}
\label{eq:analytic-horizon-catalyst}
\begin{align}
N&=u+(K-1)(u-\ell)+Kd+(d-\ell)_+,
\\
N&\le w+(K-1)(w-a)+Kh_++(K-1)h_-+Kd+(d-\ell)_++2K-1.
\end{align}
\end{subequations}
The common displacement enters once.
Each coarse block covers the delay spread and the additional smoothing width.
The last positive-part term keeps the earliest clock address nonnegative.

\subsubsection{A fixed analytic neighborhood}
\label{subsec:clock-design-proof}

\cref{thm:iir-analytic-clock-selection} chooses the radial interval from the group-delay bounds and controls the remaining transients by smoothing.
Its statement and proof use the scale construction above at one fixed radius.

\begin{theorem}[Analytic clock shaping]
\label{thm:iir-analytic-clock-selection}
Fix positive integer port delays.
Let $G,W$ be the delayed transfer function and group-delay operator from \cref{eq:intro-delays,eq:qudit-polarized-delayed}.
Suppose $G(z)$ is analytic and contractive in the unit disk, with unitary value $G=G(1)$.
Assume that it continues analytically to a neighborhood of the closed disk $|z-1|\le\zeta$ for $0<\zeta\le1/16$.
Suppose known bounds, uniform over $\mathfrak O$, satisfy
\begin{align}
aI\preceq W\preceq wI,
\quad 0\le a\le w,
\label{eq:analytic-catalyst-bracket}
\end{align}
and, for $|z-1|\le\zeta$,
\begin{align}
\|G(z)\|\le M_{\rm loc} \begin{cases} |z|^{a-h_-},&|z|\le1,
\\
|z|^{w+h_+},&|z|\ge1, \end{cases}
\quad M_{\rm loc}\ge1,
\quad h_+\ge0,
\quad 0\le h_-\le a.
\label{eq:analytic-growth-certificate}
\end{align}
For parameters $0<\delta\le1$ and $0<\epsilon_{\rm clk}<1$, an explicit rank-one clock exists.
Its matrix $X$ has nuclear norm at most $1+\delta$ and satisfies
\begin{align}
\|\widetilde G_N-G\|\le\epsilon_{\rm clk},
\quad N=w+O\!\left( \frac{(w-a)+h_++h_-+ \zeta^{-1}\log(3M_{\rm loc}/\epsilon_{\rm clk})} {\sqrt\delta}\right).
\label{eq:analytic-horizon-bigO}
\end{align}
Quditization implements $\Be{\widetilde G_N/(1+\delta)}$ with exactly
\begin{align}
q_j=\left\lfloor\frac{N-1}{r_j}\right\rfloor
\label{eq:analytic-native-counts}
\end{align}
calls to $O_j$.
An explicit finite choice of $N$ is given in \cref{eq:analytic-horizon-exact}; the constants in the displayed asymptotic bound are universal.
\end{theorem}

\begin{proof}[Proof of \cref{thm:iir-analytic-clock-selection}]
Use \cref{thm:iir-scale} at the single radius $\rho=\zeta$.
The repeated uniform average of \cref{lem:iir-flat-smoothing} is admissible and has support $d=O\!\left(\zeta^{-1}\log\frac{3M_{\rm loc}}{\epsilon_{\rm clk}}\right)$.
Choose $K=K_\delta$ by \cref{eq:analytic-K-candidates} and set $\ell=\lfloor a-h_-\rfloor$ and $u=\lceil w+h_+\rceil$.
The secant bounds give the requested normalization and $K=O(\delta^{-1/2})$.
The plateau width is $D=u-\ell+d$.
Substituting these choices into \cref{eq:analytic-horizon-catalyst} proves \cref{eq:analytic-horizon-bigO}; the exact counts follow from \cref{thm:iir-lift}.
Rank-one clock preparation costs $O(N)$ gates by \cref{eq:qudit-clock-preparation}.
\end{proof}

In the exact horizon bound, shifting the clock accounts for a common delay, whereas a broad range of possible delays must be covered by its plateau.
Accordingly, the penalty for normalization near one multiplies the spread $w-a$, while the common displacement enters the leading horizon.
The rounding and the need to keep all clock addresses nonnegative are retained in \cref{eq:analytic-horizon-exact}.

The normalization penalty in \cref{eq:analytic-horizon-catalyst} depends on which analytic parameters remain fixed.
At a fixed continuation radius and fixed radial excesses, the flat-clock block count is $K=\Theta(\delta^{-1/2})$.
It multiplies the delay spread and smoothing width.
A sharp leading time coefficient imposes an additional constraint.
For single-term Hamiltonian simulation, the group delay is scalar and the proof in \cref{app:multi-hs-delays} bounds the radial correction by $O(Ks\Omega)$.
The local log-radius is $s$, and $\Omega$ is the common delay.
Keeping that correction within the leading-cost allowance gives
\begin{align}
s=\Theta(\delta/K),
\quad d=O\!\left(s^{-1}\log(1/\epsilon)\right),
\quad Kd=O\!\left(\delta^{-2}\log(1/\epsilon)\right).
\label{eq:analytic-normalization-hierarchy}
\end{align}
With several ports, the radial term instead contains the delay-weighted second moment in \cref{eq:multi-hs-app-horizon}; its allocation proof gives the same precision order in the weighted cost.
SOS amplification has that order at fixed delays as well.
Optimizing its clipped cubic-moment allocation adds another inverse power of the normalization allowance, giving the bound in \cref{eq:multi-sos-weighted-normalized}.
These are sufficient costs for the constructions proved here.

For many applications a constant normalization is sufficient.
Normalization two is especially useful because one step of oblivious amplitude amplification then gives unitary output with the stated isometry error; \cref{lem:iir-cubic} proves the error bound and triples the query costs.
\cref{cor:iir-analytic-two} gives the corresponding horizon.

\begin{corollary}[Normalization-two analytic selection]
\label{cor:iir-analytic-two}
Under the hypotheses of \cref{thm:iir-analytic-clock-selection}, a clock with normalization two can be chosen with
\begin{align}
N=O\!\left( w+(w-a)+h_++h_-+ \zeta^{-1}\log\frac{3M_{\rm loc}}{\epsilon_{\rm clk}} \right).
\label{eq:analytic-two-horizon}
\end{align}
If no useful lower catalyst bound is given, take $a=h_-=0$ and obtain
\begin{align}
N=O\!\left( w+h_++ \zeta^{-1}\log\frac{3M_{\rm loc}}{\epsilon_{\rm clk}} \right).
\label{eq:analytic-two-simple}
\end{align}
The circuit implements $\Be{\widetilde G/2}$ with the same query, gate, and qubit formulas as \cref{thm:iir-analytic-clock-selection}.
\end{corollary}

\begin{proof}
Take $K=2$.
The exact flat-clock normalization is $\alpha_2=5/4<2$.
Apply \cref{thm:iir-analytic-clock-selection}, then compose with a known attenuation to obtain $\Be{\widetilde G/2}$.
Since $K$ is a constant, \cref{eq:analytic-horizon-catalyst} gives \cref{eq:analytic-two-horizon}.
Disk contractivity gives the trivial interior exponent zero when no lower exponent is known.
\end{proof}

\subsection{Multi-oracle and structural refinements}
\label{subsec:analytic-comparison}

\subsubsection{Energy estimates for radial growth}
\label{subsec:radial-energy}
A bound on radial growth often follows from an energy estimate after setting $z=e^s$.
The next lemma isolates this step so that each application need only bound the Hermitian part of its logarithmic derivative.
\begin{lemma}[Radial growth from an energy inequality]
\label{lem:iir-radial-energy}
Let $F(s)=G(e^s)$ be differentiable along the horizontal segments from $iy$ to $x+iy$ in a given logarithmic neighborhood, with $F(iy)$ unitary.
If $F$ is invertible there and
\begin{align}
A_-I\preceq\operatorname{Re}\bigl(F'(s)F(s)^{-1}\bigr) \preceq A_+I,
\label{eq:iir-logarithmic-energy}
\end{align}
then $\|F(x+iy)\|\le e^{A_+x}$ for $x\ge0$ and $\|F(x+iy)\|\le e^{A_-x}$ for $x\le0$.
The same conclusions hold directly from the corresponding quadratic-form inequalities for $\frac{d}{dx}\|F(x+iy)v\|^2$.
\end{lemma}
\begin{proof}
For $f(x)=F(x+iy)v$, $\frac{d}{dx}\|f(x)\|^2 =2\operatorname{Re}\langle f(x),F'(x+iy)F(x+iy)^{-1}f(x)\rangle$.
Integrate the upper inequality forward from zero and the lower inequality backward from zero.
Since $\|F(iy)v\|=\|v\|$, taking the supremum over unit vectors proves both bounds.
\end{proof}
The order $F'F^{-1}$ ensures that the generator acts on the evolving output vector.
The quadratic-form version also applies when an inverse is inconvenient; the needed differential inequality must then be proved directly.

\subsubsection{Delay allocation and analytic domains}
\cref{lem:common-delay-allocation} chooses unequal delays by balancing oracle costs against their contributions to the horizon.

\begin{lemma}[Weighted delay allocation]
\label{lem:common-delay-allocation}
For $a_j,C_j>0$ and $p>0$,
\begin{align}
\inf_{r_j>0}\left(\sum_j a_jr_j^p\right)^{1/p} \sum_j\frac{C_j}{r_j} =\left(\sum_j a_j^{1/(p+1)}C_j^{p/(p+1)}\right)^{(p+1)/p}.
\label{eq:common-delay-allocation}
\end{align}
Equality holds at $r_j\propto(C_j/a_j)^{1/(p+1)}$.
\end{lemma}
\begin{proof}
Apply H\"older's inequality to
$\sum_j(a_jr_j^p)^{1/(p+1)}(C_j/r_j)^{p/(p+1)}$
with conjugate exponents $p+1$ and $(p+1)/p$.
Its equality condition gives the stated rates.
\end{proof}
The cases $p=1$ and $p=2$ give the allocations used for Hamiltonian simulation and SOS phase amplification, respectively.
Integer delays can be obtained by scaling and rounding these rates; the precision term must be included when bounding that rounding.
The refined Hamiltonian and SOS estimates retain their second moments and clipping conditions, while \cref{lem:app-allocation} includes the neighborhood allowed by the feedback perturbation bound.

When the radial excess decreases linearly with the local radius, \cref{cor:iir-mixed-scale} balances that decrease against the smoothing width.

\begin{corollary}[Mixed dispersion and precision bound]
\label{cor:iir-mixed-scale}
Under \cref{thm:iir-scale}, suppose all radii $0<\rho\le\rho_0$ are admissible and the following known bounds hold:
\begin{align}
h_+(\rho)+h_-(\rho)\le D_{\rm rad}\rho,
\quad d(\rho,\epsilon_{\rm clk})\le \frac{c_{\rm sm}}\rho\log\frac{M_0}{\epsilon_{\rm clk}},
\label{eq:scale-linear-data}
\end{align}
Here the radial-dispersion constant is nonnegative and the smoothing constant is positive; the given growth factor satisfies $M_0\ge1$.
The support bound includes integer rounding and the complete contour error.
Then
\begin{align}
N\le{}
\quad w+(K-1)(w-a)+K D_{\rm rad}\rho +\frac{(K+1)c_{\rm sm}}\rho\log\frac{M_0}{\epsilon_{\rm clk}} +2K-1.
\label{eq:scale-rho-bound}
\end{align}
If $D_{\rm rad}>0$, define
\begin{align}
\rho_*= \sqrt{\frac{(K+1)c_{\rm sm}\log(M_0/\epsilon_{\rm clk})} {K D_{\rm rad}}}.
\label{eq:scale-rho-star}
\end{align}
When $\rho_*\le\rho_0$, the horizon satisfies
\begin{align}
N\le w+(K-1)(w-a) +2\sqrt{K(K+1)c_{\rm sm}D_{\rm rad} \log\frac{M_0}{\epsilon_{\rm clk}}} +2K-1.
\label{eq:scale-mixed-horizon}
\end{align}
Otherwise use $\rho=\rho_0$ in \cref{eq:scale-rho-bound}; this also applies when $D_{\rm rad}=0$.
\end{corollary}

\begin{proof}
Since $\ell\ge0$, the positive part in \cref{eq:analytic-horizon-catalyst} is at most $d$.
Insert \cref{eq:scale-linear-data}.
Minimizing the remaining function $A\rho+B/\rho$ gives $\rho=\sqrt{B/A}$ when that radius is allowed.
If it exceeds the domain, the function decreases up to the endpoint.
This proves all three cases.
\end{proof}

The square-root term in \cref{eq:scale-mixed-horizon} couples dispersion and precision.
It can be smaller than a bound that separates the leading propagation cost from a pure precision remainder.
Suppose the local growth prefactor is bounded by $M_*$.
The box-average construction permits $M_0=3M_*$ and $c_{\rm sm}=12$.
Indeed, its block length minus one is less than $2e(1+\rho)/\rho$, and the repetition count is at most twice $\log(3M_*/\epsilon_{\rm clk})$.
Their product gives the stated support bound for $\rho\le1/16$.
These constants come from both contour integrals in \cref{def:iir-positive-smoother}.

Applications often express the group delay and radial excess through first and second moments of the port delays.
In that case \cref{cor:iir-scale-allocation} gives a scalar optimization using the known strengths and query costs.

\begin{corollary}[Weighted delay allocation]
\label{cor:iir-scale-allocation}
Suppose an application proves the assumptions of \cref{cor:iir-mixed-scale} with
\begin{align}
w=a=\sum_j\mu_jr_j,
\quad D_{\rm rad}=\sum_j\nu_jr_j^2,
\label{eq:scale-delay-moments}
\end{align}
for given nonnegative coefficients.
For positive integer delays and any admissible radius, the actual query cost obeys
\begin{align}
\sum_j C_jq_j\le \left[\sum_j\mu_jr_j+K\rho\sum_j\nu_jr_j^2 +\frac{(K+1)c_{\rm sm}}\rho\log\frac{M_0}{\epsilon_{\rm clk}} +2K-1\right]\sum_j\frac{C_j}{r_j}.
\label{eq:scale-weighted-objective}
\end{align}
Its continuous relaxation is convex in logarithmic variables under log-convex analytic-domain restrictions, for example
\begin{align}
r_j=e^{y_j},
\quad\rho=e^\tau,
\quad \tau+y_j\le\log u_j,
\quad y_j\ge0,
\quad\tau\le\log\rho_0.
\label{eq:scale-log-variables}
\end{align}
\end{corollary}

\begin{proof}
Multiply \cref{eq:scale-rho-bound} by the sum of reciprocal-delay costs and use the exact integer query counts.
Expanding the objective gives positive multiples of exponentials with affine exponents $y_j-y_i,\tau+2y_j-y_i,-\tau-y_i,-y_i$.
Each term is convex, and the displayed domain restrictions are linear.
\end{proof}

The relaxed value is a lower bound on the minimum of this integer upper-bound objective.
To obtain a integer query schedule, round the delays, recheck the analytic-domain restrictions, and evaluate \cref{eq:analytic-horizon-exact,eq:scale-optimized-counts} again.
The rounding cost is included in this evaluation.
The scalar program depends only on the delay variables and analytic radius, using the uniform certificate bounds already proved for the oracle family.

For $\nu_j=\mu_j>0$, the continuous square-root allocation exposes the paired half-quasinorm.
Let its free common scale be $\eta>0$, so that
\begin{align}
r_j=\eta\sqrt{C_j/\mu_j},
\quad T=\langle\boldsymbol{C},\boldsymbol{\mu}\rangle_{1/2}.
\label{eq:scale-half-allocation}
\end{align}
If the optimizing radius and these delays satisfy the required domain, the relaxed objective in \cref{eq:scale-weighted-objective} becomes
\begin{align}
T+2\sqrt{K(K+1)c_{\rm sm}T\|\boldsymbol C\|_1 \log\frac{M_0}{\epsilon_{\rm clk}}} +(2K-1)\frac{\sqrt T}{\eta}.
\label{eq:scale-half-mixed}
\end{align}
The final term retains the integer-horizon allowance; analytic-radius restrictions and integer rounding still require the preceding check.
Other transfer functions must supply their own delay moments and domain bounds before using this specialization.

To prescribe a leading coefficient, one may separate a mixed resource bound by Young's inequality.
Let $T$ be the leading cost and $B$ the dispersion--precision coefficient.
Write the remaining cost as $R$ and choose any $\eta_{\rm lead}>0$.
Then
\begin{align}
Q\le T+2\sqrt{BT}+R \le(1+\eta_{\rm lead})T+\frac B{\eta_{\rm lead}}+R.
\label{eq:scale-young}
\end{align}
The extra inverse power of the leading-cost allowance enters in the second inequality.
For example, the product of flat-clock block counts in \cref{eq:scale-half-mixed} is of order $\delta^{-1}$.
Taking the leading-cost allowance equal to the normalization slack can turn this into a $\delta^{-2}$ separated precision cost, even though the unexpanded estimate contains a mixed square root.

The Hamiltonian simulation proof in \cref{app:multi-hs-delays} supplies the needed first and second moments and retains the mixed query-cost bound in \cref{eq:multi-hs-physical-certificate} before applying \cref{eq:scale-young}.
Factor-access linear systems, SOS amplification, shared SELECT, and downfolding require their own radial estimates; the present optimization changes no application bound without those estimates.
Online transfer kernels also require their separate two-time clock analysis in \cref{sec:time-dependent-mqsp}.
After a horizon is chosen, gate and qubit costs follow from \cref{thm:iir-gates}, including the preparation of the resulting clocks.

To obtain individual oracle counts, \cref{thm:iir-polydisc} pulls a joint analytic neighborhood back along the delayed slice $z\mapsto(z^{r_1},\ldots,z^{r_m})$ and applies the one-variable theorem.
The joint neighborhood excludes singularities involving several coordinates; bounds on this neighborhood can be propagated through modular constructions.

\begin{corollary}[Joint analytic domain]
\label{thm:iir-polydisc}
Suppose $F(\Bz)$ has unitary value at $\Bz=\boldsymbol{1}$ and continues jointly through the neighborhood $|z_j-1|\le\zeta_j$ for $0<\zeta_j\le1/2$.
Assume the uniform exterior estimate
\begin{align}
\|F(\Bz)\|\le\prod_j|z_j|^{a_j},
\quad a_j\ge0,
\label{eq:iir-product-envelope}
\end{align}
when every $|z_j|\ge1$ in that neighborhood.
For the delayed transfer function one may take
\begin{align}
A_+=\sum_j a_jr_j,
\quad \zeta=\min\left\{\frac1{16}, \frac14\min_j\frac{\zeta_j}{r_j}\right\}.
\label{eq:iir-product-pullback}
\end{align}
At normalization two,
\begin{align}
N=O\!\left( \sum_j a_jr_j+ \log(1/\epsilon)\max_j\frac{r_j}{\zeta_j} \right),
\quad q_j=\left\lfloor\frac{N-1}{r_j}\right\rfloor.
\label{eq:iir-anisotropic-count}
\end{align}
For positive query costs $C_j$, the exact weighted query cost of this chosen schedule is therefore
\begin{align}
\sum_j C_j\left\lfloor\frac{N-1}{r_j}\right\rfloor.
\label{eq:iir-polydisc-priced-cost}
\end{align}
The delays may be optimized against this expression.
The closed-form weighted allocations in the applications use the stronger moment or conditioning estimates proved there.
\end{corollary}

\begin{proof}
If $|z-1|\le\zeta$, then
\begin{align}
|z^{r_j}-1| \le r_j|z-1|(1+|z-1|)^{r_j-1} \le(\zeta_j/4)e^{\zeta_j/4}<\zeta_j.
\label{eq:analytic-polydisc-pullback-proof}
\end{align}
Thus $G(z)$ stays inside the joint analytic domain.
On the exterior part,
\begin{align}
\|G(z)\| \le |z|^{\sum_j a_jr_j}.
\end{align}
Differentiating this exterior estimate at $z=1$ along the positive real axis gives $W\preceq A_+I$.
Disk contractivity gives interior exponent zero.
Apply \cref{cor:iir-analytic-two} with $a=h_-=h_+=0$ and $w=A_+$ to obtain \cref{eq:iir-anisotropic-count}.

The exact query counts are then those of \cref{thm:iir-lift}, which proves \cref{eq:iir-anisotropic-count,eq:iir-polydisc-priced-cost}.
\end{proof}

A useful heterogeneous case has the explicit expansion
\begin{align}
F(z_1,z_2)=\sum_{j=0}^{d}f_j(z_2)z_1^j.
\label{eq:mixed-source-expansion}
\end{align}
The first oracle occurs with degree at most $d$; the coefficient functions $f_j$ are analytic in the second delay variable on a common domain.
For instance, the first oracle may prepare the input state and the second may encode a matrix.
The following rule gives a preparation-query bound proportional to $d$ while allowing the matrix-query count to depend logarithmically on precision.

\begin{corollary}[Finite-degree and analytic ports]
\label{thm:iir-mixed}
Let $F(\Bz,w)$ have a unitary realization.
Suppose its polynomial degree in the last coordinate is at most $d\ge1$.
Fix the delays of the other ports, and write $\widehat F(z,w)=F(z^{r_1},\ldots,z^{r_m},w)$.
For every $|w|\le1$, suppose this function is analytic on the same neighborhood in the delay coordinate: $|z-1|\le\zeta$, for $0<\zeta\le1/16$.
Assume the uniform bound
\begin{align}
\sup_{|w|=1}\|\widehat F(z,w)\| \le M_{\rm loc}|z|^{A_+}
\quad (|z|\ge1,\ |z-1|\le\zeta),
\label{eq:analytic-mixed-envelope}
\end{align}
where $A_+\ge0$ and $M_{\rm loc}\ge1$.
Put
\begin{align}
H_{\rm an}&=1+A_++\zeta^{-1}\log(3M_{\rm loc}/\epsilon),
\label{eq:analytic-mixed-H}
\\
r_w&=\max\{1,\lceil H_{\rm an}/d\rceil\}.
\label{eq:analytic-mixed-delay}
\end{align}
Then a clock with normalization two exists with
\begin{align}
N=O(H_{\rm an}+d),
\quad q_w=O(d),
\quad q_j=O((H_{\rm an}+d)/r_j).
\label{eq:analytic-mixed-counts}
\end{align}
If $d=O(H_{\rm an})$, the last bound simplifies to $q_j=O(H_{\rm an}/r_j)$.
All exact counts remain $\lfloor(N-1)/r_j\rfloor$.
\end{corollary}

\begin{proof}
Reversing the coefficients turns growth outside the disk into a bound inside it.
For an operator polynomial of degree at most $d$, define
\begin{align}
p(w)=\sum_{j=0}^d A_jw^j,
\quad q(u)=u^dp(1/u)=\sum_{j=0}^d A_ju^{d-j}.
\end{align}
The reversed function is a polynomial, including at the origin.
On the unit circle it has the same norm bound as the original polynomial.
Apply the maximum-modulus principle to each scalar matrix element and then take the supremum over unit vectors.
Substituting $u=1/w$ gives
\begin{align}
\|p(w)\|\le |w|^d\sup_{|v|=1}\|p(v)\|
\quad(|w|\ge1).
\label{eq:analytic-polynomial-exterior}
\end{align}
Hence $G(z)=\widehat F(z,z^{r_w})$ is analytic on the same local neighborhood and has exterior bound $\|G(z)\|\le M_{\rm loc}|z|^{A_++dr_w}$.
Disk contractivity gives interior exponent zero.
The smoothed-flat construction at constant normalization therefore has $N=O(H_{\rm an}+dr_w)=O(H_{\rm an}+d)$.
The delay obeys $r_w\ge\max\{1,H_{\rm an}/d\}$.
Dividing the horizon by this delay gives $q_w=O(d)$; the other query counts follow from the same horizon.
\end{proof}

The degree bound in $w$ holds throughout a common analytic domain, with a bound uniform on its unit circle.
A cancellation only at $w=1$ gives neither of these hypotheses.
This permits a larger delay for the first oracle while keeping its query count proportional to $d$.

The radial bound itself can be obtained from an application's resolvent, Hermitian-part, or positive-kernel estimate.
A local exponential estimate alone need not imply a radial-power bound: $\|G(z)\|\le\exp(B|z-1|)$ does not imply $ \|G(z)\|\le|z|^A$ for a fixed $A$.
At tangential exterior points, $\log|z|$ can be of order $|z-1|^2$.
It does, however, give the valid constant growth bound $M_{\rm loc}=e^{B\zeta}$ with exponents zero in \cref{lem:iir-flat-smoothing}.
At constant normalization this gives width $O(B+\zeta^{-1}\log(1/\epsilon))$.
A sharper radial estimate is needed to obtain the spread-dependent improvement in \cref{eq:analytic-horizon-bigO}.

\subsubsection{Perturbation stability and corrections}
For finite Schur realizations and truncated recurrences, clock design may use a simpler analytic transfer function whose Toeplitz block approximates the implemented public block.
\cref{prop:iir-comparison-stability} bounds the resulting operator error by the clock norm times the block error, without an additional horizon factor.

\begin{proposition}[Comparison transfer-function stability]
\label{prop:iir-comparison-stability}
Consider a comparison transfer function $\widehat G$.
Suppose a clock matrix $X$ has nuclear norm at most $\alpha_{\rm clk}$ and satisfies
\begin{align}
\left\|\sum_{n<N}c_n(X)\widehat G_{n}-\widehat G\right\| \le\epsilon_{\rm clk}.
\label{eq:comparison-clock-error}
\end{align}
Suppose the implemented finite public block $\mathcal K_N$ and target $U_{\rm task}$ obey
\begin{align}
\|\mathcal K_N-T_N[\widehat G]\|\le\eta_N,
\quad \|\widehat G-U_{\rm task}\|\le\eta_0.
\label{eq:comparison-assumptions}
\end{align}
Then the same clock on the implemented lift has approximation error before normalization at most
\begin{align}
\epsilon_{\rm clk}+\alpha_{\rm clk}\eta_N+\eta_0.
\label{eq:comparison-total-error}
\end{align}
\end{proposition}

\begin{proof}
Apply \cref{lem:clock-extraction} to the difference between the implemented kernel and its comparison Toeplitz block.
It gives $\|\mathcal L_X(\mathcal K_N-T_N[\widehat G])\| \le\|X\|_*\eta_N\le\alpha_{\rm clk}\eta_N$.
Adding the clock and target errors from \cref{eq:comparison-clock-error,eq:comparison-assumptions} by the triangle inequality proves \cref{eq:comparison-total-error}.
\end{proof}

\paragraph{Compilation with an optional correction.}
\label{subsec:mqsp-master}

The clock coefficients and an explicitly implemented correction share a joint normalization bound.
\cref{thm:mqsp-master} implements their sum for a given finite kernel.
In the Toeplitz case, \cref{thm:iir-analytic-clock-selection} gives a feasible clock.

Let $\mathcal K_n$ be an $n$-slot public kernel of a given finite unitary, after specified known input and output operations.
For the unmodified delayed lift, the kernel is the Toeplitz block from \cref{thm:iir-lift}: $\mathcal K_n=\mathcal T_n[G]=[G_{o-i}]_{o,i=0}^{n-1}$.
Its clock horizon is $n$.
More generally the clock horizon is $N(n)$, and the known operations may align clock addresses or restrict the input and output spaces.
Any oracle calls in those operations are counted explicitly.
For a scalar clock matrix $X$, define
\begin{align}
\mathcal L_X(\mathcal K_n)=\sum_{o,i=0}^{n-1}X_{oi}(\mathcal K_n)_{oi},
\quad c_k(X)=\sum_{o-i=k}X_{oi}.
\label{eq:master-extraction}
\end{align}
In particular, $\mathcal L_X(\mathcal T_n[\widehat G])=\sum_{k<n}c_k(X)\widehat G_k$.
The kernels and transfer functions depend on the given oracles.
The known clock coefficients, input and output circuits, and choices of parameters are common to the whole promise; they cannot depend on the unknown instance.

\begin{theorem}[Constructive MQSP compilation]
\label{thm:mqsp-master}
Fix $\alpha_{\rm clk}=1+\delta$, with $0<\delta\le1$.
Suppose $\mathcal K_n$ is implemented as above and a given comparison kernel $\overline{\mathcal K}_n$ satisfies
\begin{align}
\sup_{\rm promise}\|\mathcal K_n-\overline{\mathcal K}_n\|\le\eta_n.
\label{eq:master-comparison}
\end{align}
Optionally, use contractions $B_0,\ldots,B_d$ with block encodings $\Be{\widetilde B_k}$ satisfying $\sup_{\rm promise}\|\widetilde B_k-B_k\|\le\eta_k$.
Assume controlled access to $\sum_{k=0}^d|k\rangle\langle k|\otimes\Be{\widetilde B_k}$ after compatible padding of their ancillas.
This is the SELECT operation for the correction block encodings.
Suppose a clock factorization of $X$ and coefficient preparations for $\boldsymbol{\xi}=(\xi_0,\ldots,\xi_d)$ are explicitly implementable, and
\begin{align}
\|X\|_*+\|\boldsymbol{\xi}\|_1&\le1+\delta,
\label{eq:master-budget}
\\
\sup_{\rm promise}\left\| \mathcal L_X(\overline{\mathcal K}_n)+ \sum_{k=0}^{d}\xi_kB_k-F_{\rm task} \right\|&\le\epsilon_{\rm des}.
\label{eq:master-design}
\end{align}
Use an empty correction sum when the chosen construction has no correction stage.
For a causal Toeplitz comparison, $\overline{\mathcal K}_n=\mathcal T_n[\widehat G]$ and $\mathcal L_X(\overline{\mathcal K}_n)=\sum_{k<n}c_k(X)\widehat G_k$.
Then the finite circuit implements $\Be{\widetilde F/(1+\delta)}$ with
\begin{align}
\|\widetilde F-F_{\rm task}\| \le\epsilon_{\rm des}+\|X\|_*\eta_n+ \sum_k|\xi_k|\eta_k+(1+\delta)\eta_{\rm syn}.
\label{eq:master-error}
\end{align}
Here $\eta_{\rm syn}$ is any additional operator-norm perturbation of the complete assembled unitary, including the clock preparations and known attenuation.
The implementation uses its given common ordered query sequence.
In particular, concatenating the controlled lift and correction query sequences is always valid and gives
\begin{align}
q_j\le \left\lfloor\frac{N(n)-1}{r_j}\right\rfloor +b_j+q_j^{\rm corr},
\label{eq:master-word}
\end{align}
where $b_j$ counts any calls in the input and output operations.
A shorter common query sequence may be used when its execution is given.
Query amplitudes do not discount these counts.

\end{theorem}

\begin{proof}
Factor $X=\sum_s\boldsymbol{h}_s\boldsymbol{g}_s^{\dagger}$ with $\sum_s\|\boldsymbol{h}_s\|\|\boldsymbol{g}_s\|=\|X\|_*$.
\cref{thm:iir-clock-body} constructs normalized input and output clocks from these factors.
Use a further clock label for the correction operators and prepare their magnitudes and phases.
The combined scalar clock matrix is $Y=X\oplus\operatorname{diag}(\boldsymbol{\xi})$ with $\|Y\|_*=\|X\|_*+\|\boldsymbol{\xi}\|_1$.
Apply the given unitary that selects the lift on the first branch and the correction SELECT operation on the other branches.
Orthogonal input and output padding fills unused normalization, or a known rotation attenuates a smaller achieved normalization to $1+\delta$.
The ideal encoded operator before normalization is $\mathcal L_X(\mathcal K_n)+\sum_k\xi_k\widetilde B_k$.
The two directions of the clock characterization already prove that this factorization is a implemented circuit, including its rejection space.

\cref{lem:clock-extraction} bounds the kernel comparison error, and the triangle inequality bounds the errors in the correction block encodings.
A perturbation $\eta_{\rm syn}$ of the whole implemented unitary changes its compressed operator by at most $\eta_{\rm syn}$.
Multiplying by $1+\delta$ proves \cref{eq:master-error}.

\cref{thm:iir-lift}
gives the delayed query count in \cref{eq:master-word}.
A controlled execution of the lift followed by a controlled execution of the correction SELECT operation is a valid circuit for their coherent direct sum.
Every controlled call executed by these circuits is counted, including a call whose control is inactive on some branches.
This proves the resource statement.

\end{proof}

When the input lies in a promised subspace, \cref{cor:iir-task-restricted} applies the analytic theorem to the restriction of the transfer function.
This can reduce both the catalyst and growth bounds while using the same circuit on the full public space.

\begin{corollary}[Restricted-input analytic selection]
\label{cor:iir-task-restricted}
Let $J:\mathcal E\to\mathcal P$ be an isometry and measure task error by $\|(\widetilde G-G)J\|$.
Replace the transfer function by the rectangular analytic function $G(z)J$.
Its delay-weighted catalyst Gram operator is
\begin{align}
W_J=J^\dagger WJ.
\label{eq:analytic-task-energy}
\end{align}
Suppose $aI\preceq W_J\preceq wI$, and suppose the contractivity, analytic continuation, and radial-growth hypotheses of \cref{thm:iir-analytic-clock-selection} hold for $G(z)J$.
Then the same horizon and clock conclusions hold with error $\|(\widetilde G_N-G)J\|\le\epsilon_{\rm clk}$.
\end{corollary}

\begin{proof}
Every step of the analytic proof is an operator-norm estimate on the selected transfer function.
Right multiplication by $J$ preserves analyticity, and the evaluated map $GJ$ is an isometry.
Its energy is $(GJ)^\dagger(GJ)'=J^\dagger WJ$.
The flat-window and contour arguments use contractivity and the norm at $z=1$.
These properties also hold for the stated rectangular operators.
Applying those arguments to $GJ$ proves the same clock bound.
The quditized circuit itself is unchanged; the error estimate is restricted to the specified input space.
\end{proof}

The radial growth bound must hold on the same restricted transfer function as the catalyst estimate.

\subsubsection{Refinements from additional structure}
\label{subsec:additional-structure}

The analytic theorem provides one feasible clock for \cref{eq:iir-joint-frontier}; additional information can improve it.
For instance, suppose the coefficient norms are summable.
A clock with weights in the unit interval and a plateau on $[L,H]$ satisfies
\begin{align}
\sum_n\|G_n\|<\infty,\qquad0\le c_n\le1,
\quad c_n=1
\quad(L\le n\le H).
\end{align}
These conditions give the direct tail estimate
\begin{align}
\left\|\sum_nc_nG_{n}-G\right\| \le\sum_{n<L}\|G_{n}\|+\sum_{n>H}\|G_{n}\|.
\label{eq:clock-design-global-tail}
\end{align}
The plateau can be chosen from the actual tail, using the local smoothing estimate.
For finite impulse support this gives an exact clock.
Similarly, a known scalar factor $z^p$ can be absorbed into a relative translation of the clock states; the analytic estimate then need only control the remaining transfer function.
A known recurrence among the $K_n$ can instead be used directly in \cref{eq:clock-design-residual}.
Each refinement uses information absent from a norm bound at $z=1$ and local neighborhood.

When the fixed lift and its full promise constraints admit a tractable exact representation or a rigorous separation procedure, one can refine the analytic clock by solving \cref{eq:iir-joint-frontier} at the same normalization.
The analytic clock gives a feasible point, so this optimization cannot worsen its error and leaves the ordered query sequence unchanged.
Reducing the horizon requires a new feasible solution at that shorter horizon.

For fixed clock weights, the optimization in \cref{eq:qudit-clock-norm} can reduce normalization while preserving the implemented operator and query schedule.
The analytic clock is a feasible point of that optimization.

After padding clocks to a common support, convex combinations satisfy \begin{align}
\left\|\sum_k\lambda_kX_k\right\|_*\le\sum_k\lambda_k\|X_k\|_*,
\quad\lambda_k\ge0,
\quad\sum_k\lambda_k=1.
\end{align} Computing the combined factors classically does not require one time register per term.
\cref{prop:qudit-rankone} gives a single clock pair for causal scalar coefficients on full equal supports.
The general rank-$R$ formulation remains available for support constraints, terminal private-output cancellation, intermediate query weights, and non-Toeplitz online kernels.
Preparing a generic length-$N$ endpoint requires $O(N)$ known gates in addition to classical factorization.

\providecommand{\vect}[1]{\vec{#1}}

\section{Achievability}
\label{sec:achievable}

Quditization and clock shaping assume that a unitary system matrix is already known.
Algorithm design may instead begin with a desired transformation, in which case we must first determine whether the supplied oracles admit a unitary realization of it.
For univariate QSP, boundedness of an admissible polynomial supplies the complementary amplitudes needed for a unitary completion, which can then be factored into a QSP sequence.
With several oracles, scalar boundedness alone can fail to ensure contractivity after matrix substitution, and commuting variables do not retain the order of oracle calls.

\Cref{sec:ach-polynomial} reviews polynomial completion and the distinction between scalar boundedness and contractivity for commuting matrix tuples.
\Cref{sec:ach-free} retains the order of noncommuting calls and imposes a fixed query schedule.
Every finite circuit with such a schedule has an analytic MQSP representation, so analyticity of the query variables does not itself restrict the finite query model.
The stronger continuation assumptions used for clock shaping provide the additional information needed for efficient compilation.
Finally, \cref{sec:ach-rational,sec:ach-modular} explain how rational transfer functions and modular unitary constructions turn these existence results into useful design rules.

Throughout this section, $\Pi_{\rm in}$ and $\Pi_{\rm out}$ denote orthogonal projectors onto designated public input and output subspaces.
If $V$ is a public unitary or transfer function, the transformation encoded between these subspaces is
\begin{align}
V_{\rm enc}=\Pi_{\rm out}V\Pi_{\rm in}: \operatorname{ran}\Pi_{\rm in}\longrightarrow \operatorname{ran}\Pi_{\rm out}.
\label{eq:ach-pi-block}
\end{align}
When $\Pi_{\rm in}=\Pi_{\rm out}=|0\rangle\!\langle0|\otimes I$, this is the usual block-encoding convention.
The two projectors are kept distinct because some constructions use different public input and output subspaces.

\subsection{Polynomial completion and the Schur--Agler boundary}
\label{sec:ach-polynomial}

Implementing a target block $\Pi_{\rm out}U(t)\Pi_{\rm in}$ requires complementary output amplitudes that make $U(t)$ unitary for every allowed signal.
In one variable, the positivity bound on the target entry also supplies this completion.

Let $p(t)$ be a Laurent polynomial and write $p^{\sharp}(t)=\overline{p(1/\overline t)}$.
On the unit circle, $p^{\sharp}(t)=\overline{p(t)}$.
If $|p(t)|\leq1$ for $|t|=1$, then
\begin{align}
h(t)=1-p^{\sharp}(t)p(t)\geq0,
\quad |t|=1.
\label{eq:ach-univariate-defect}
\end{align}
The Fej\'er--Riesz theorem states that every nonnegative univariate trigonometric polynomial is one Hermitian square.
Hence there is an analytic polynomial $q$ such that
\begin{align}
1-p^{\sharp}(t)p(t)=q^{\sharp}(t)q(t),
\quad |t|=1.
\label{eq:ach-fejer-riesz}
\end{align}
The matrix
\begin{align}
U(t)= \begin{pmatrix} p(t)&-q^{\sharp}(t)
\\
q(t)&p^{\sharp}(t) \end{pmatrix}
\label{eq:ach-paraunitary-completion}
\end{align}
is therefore unitary on the unit circle.
Indeed, its two columns have squared norm one by \cref{eq:ach-fejer-riesz}, and their inner product is $-p^{\sharp}q^{\sharp}+q^{\sharp}p^{\sharp}=0$.
QSP adds the required reciprocity and parity conditions and factors the resulting $2\times2$ paraunitary Laurent polynomial into elementary signal-processing rotations~\cite{Low2016Methodology,Low2016HamSim}.
GQSP enlarges the admissible polynomial class while retaining the same basic completion principle~\cite{Motlagh2024gqsp}.

The corresponding algebraic fact is also special to one variable.
If a real polynomial $h(x)$ is nonnegative on the whole real line, every real root has even multiplicity.
Factoring over the reals gives
\begin{align}
h(x)=c\prod_a(x-a)^{2m_a} \prod_b\bigl[(x-u_b)^2+v_b^2\bigr]^{n_b},
\quad c\geq0,
\label{eq:ach-univariate-real-factorization}
\end{align}
where $v_b>0$.
Each factor is a square or a sum of two squares, and
\begin{align}
(a^2+b^2)(c^2+d^2)=(ac-bd)^2+(ad+bc)^2.
\label{eq:ach-two-square-identity}
\end{align}
Repeated use of \cref{eq:ach-two-square-identity} therefore writes $h$ as a sum of two polynomial squares.
The Fej\'er--Riesz theorem is the unit-circle analogue with the stronger conclusion that one complex polynomial square suffices.

Several variables do not have this elementary positivity-to-polynomial-square implication.
A standard example is the Motzkin polynomial
\begin{align}
M(x,y)=1+x^4y^2+x^2y^4-3x^2y^2.
\label{eq:ach-motzkin}
\end{align}
It is nonnegative because the arithmetic--geometric mean inequality gives
\begin{align}
\frac{1+x^4y^2+x^2y^4}{3} \geq\bigl(x^6y^6\bigr)^{1/3}=x^2y^2.
\label{eq:ach-motzkin-positive}
\end{align}
Nevertheless, $M$ is not a sum of squares of real polynomials.
To see this, suppose $M=\sum_k f_k^2$.
Since $M$ has degree six, every $f_k$ has degree at most three.
Terms $x^3,y^3,x^2,y^2,x,y$ cannot occur in any $f_k$: for example, if an $x^3$ term occurred among the highest $x$-degree terms, the coefficient of $x^6$ in $\sum_k f_k^2$ would be a sum of squares and could not vanish.
Repeating this argument for the absent monomials leaves
\begin{align}
f_k(x,y)=a_k+b_kxy+c_kx^2y+d_kxy^2.
\label{eq:ach-motzkin-support}
\end{align}
The coefficient of $x^2y^2$ in $\sum_k f_k^2$ is then $\sum_k b_k^2\geq0$, whereas \cref{eq:ach-motzkin} has coefficient $-3$, a contradiction.
For univariate QSP, the defect $1-|p|^2$ is precisely the positive object that the spectral factor completes.
In several variables, positivity of the analogous defect on the signal domain does not by itself supply polynomial complementary amplitudes with the required degree and symmetry.
The Motzkin example isolates the underlying algebraic failure of positivity to imply polynomial SOS.
This is one reason that a multivariate polynomial ansatz can leave a separate unitary-completion problem after the polynomial block has already been designed~\cite{HillarNie2008Hilbert17Matrices}.

Strictly positive multivariate trigonometric polynomials admit finite sums of Hermitian squares of analytic polynomials, generally with several factors and increased degrees.
In two variables the corresponding nonnegative finite-dimensional operator-valued result also holds \cite{Dritschel2004Factorization,Dritschel2025TwoVariable}.
A fixed-form quantum circuit, however, requires a completion with controlled degrees and symmetry.
This stronger requirement motivates the architecture-specific realizability conditions in multivariable QSP \cite{Rossi2022MultivariableQSP,Mori2024MultivariableComment,Laneve2025MultivariateQSP,Ito2026ConstructiveDecision}.

A second obstruction appears even before polynomial completion: scalar contractivity need not survive matrix substitution.
For an analytic scalar function $f$ on the polydisk $\mathbb D^m$, call $f$ a Schur function if $|f(\Bz)|\leq1$ throughout $\mathbb D^m$.
It belongs to the Schur--Agler class if
\begin{align}
\|f(T_1,\ldots,T_m)\|\leq1
\label{eq:ach-schur-agler-definition}
\end{align}
for every commuting tuple of strict contractions $T_1,\ldots,T_m$, with the usual radial definition when needed.
The distinction is invisible for one and two variables and becomes real from three variables onward.

\begin{theorem}[Commuting scalar variables]
\label{thm:ach-commuting-hierarchy}
For $m=1$ and $m=2$, the Schur and Schur--Agler classes on $\mathbb D^m$ coincide.
For every $m\geq3$, the Schur--Agler class is a proper subset of the Schur class.
A Schur--Agler function admits positive kernels $K_j$ satisfying
\begin{align}
1-f(\vect{w})^*f(\Bz) =\sum_{j=1}^m(1-\overline w_jz_j)K_j(\vect{w},\Bz),
\quad K_j\succeq0,
\label{eq:ach-agler-decomposition}
\end{align}
and conversely such a decomposition gives a contractive transfer-function realization.
Finite-rank kernels give a finite-dimensional realization.
\end{theorem}

\begin{proof}
For one variable, von Neumann's inequality states that a contraction $T$ obeys $\|p(T)\|\leq\sup_{|z|\leq1}|p(z)|$ for every polynomial $p$; approximation extends the statement to Schur functions.
For two commuting contractions, And\^o's dilation theorem supplies commuting unitaries $U_1,U_2$ on a larger Hilbert space such that
\begin{align}
T_1^{n_1}T_2^{n_2} =P U_1^{n_1}U_2^{n_2}\big|_{\mathcal H}, \quad n_1,n_2\geq0.
\label{eq:ach-ando-dilation}
\end{align}
Hence for every polynomial $p$,
\begin{align}
\|p(T_1,T_2)\| \leq\|p(U_1,U_2)\| \leq\sup_{|z_1|=|z_2|=1}|p(z_1,z_2)|.
\label{eq:ach-ando-von-neumann}
\end{align}
The last inequality follows from the joint spectral theorem for the commuting unitaries.
This proves equality of the scalar Schur and Schur--Agler classes for $m\leq2$~\cite{Ando1963Contraction}.

For $m=3$, the Kaijser--Varopoulos polynomial
\begin{align}
p(z_1,z_2,z_3) =\frac{1}{5}\left( z_1^2+z_2^2+z_3^2 -2z_1z_2-2z_1z_3-2z_2z_3\right)
\label{eq:ach-kv-polynomial}
\end{align}
has scalar supremum norm one on the tridisk, but there are three commuting contractions $T_1,T_2,T_3$ for which $\|p(T_1,T_2,T_3)\|>1$.
Thus it is Schur but not Schur--Agler.
Appending unused variables gives the same separation for every $m>3$~\cite{Agler1990Polydics,Knese2010RationalInner}.

It remains to connect \cref{eq:ach-agler-decomposition} to a unitary realization.
Factor each positive kernel as $K_j(\vect{w},\Bz)=H_j(\vect{w})^*H_j(\Bz)$ and define
\begin{align}
\vect a_{\Bz} =\begin{pmatrix}1
\\
z_1H_1(\Bz)
\\
\vdots
\\
z_mH_m(\Bz)\end{pmatrix},
\quad \vect b_{\Bz} =\begin{pmatrix}f(\Bz)
\\
H_1(\Bz)
\\
\vdots
\\
H_m(\Bz)\end{pmatrix}.
\label{eq:ach-agler-lurking-vectors}
\end{align}
\Cref{eq:ach-agler-decomposition} is exactly $\vect a_{\vect w}^{\dagger}\vect a_{\Bz} =\vect b_{\vect w}^{\dagger}\vect b_{\Bz}$.
This reverses the argument in \cref{thm:iir-energy}: a positive kernel now supplies private amplitudes from which to build the unitary.
The assignment $\vect a_{\Bz}\mapsto\vect b_{\Bz}$ therefore preserves all inner products.
It extends linearly to an isometry between their spans and, after adding orthogonal complements, to a unitary.
Writing that unitary in public/private blocks and eliminating the private vector gives the usual transfer-function realization.
If the kernel factors have finite-dimensional ranges, the construction is finite dimensional.
The converse follows by applying the same inner-product calculation to the steady-state equations of a unitary realization.
This is the standard lurking-isometry proof of the Agler realization theorem~\cite{Agler1990Polydics,Ball1998UnitaryColligations}.
\end{proof}

The theorem identifies a useful boundary.
Two commuting scalar variables are still controlled by scalar boundedness.
Three commuting variables require the stronger Agler positivity.
Noncommuting variables require still more information, because a scalar polynomial has already forgotten the order of its factors.

\subsection{Free functions and causally ordered query constraints}
\label{sec:ach-free}

The delay coordinates can be characterized before any realization theory is used.
Suppose a analytic variable depends only on how many times each oracle type has been queried.
Sequential composition adds these count vectors, so a useful marker should turn addition of counts into multiplication of scalars.

\begin{proposition}[Canonical analytic query-count coordinates]
\label{prop:ach-query-markers}
Let $\chi:\mathbb N_0^m\to\mathbb C$ satisfy
\begin{align}
\chi(\vect 0)=1,
\quad\chi(\vect n+\vect k)=\chi(\vect n)\chi(\vect k)
\quad(\vect n,\vect k\in\mathbb N_0^m).
\label{eq:ach-query-character-law}
\end{align}
If $z_j=\chi(\vect e_j)$ for the $j$th coordinate vector, then
\begin{align}
\chi(\vect n)=\prod_{j=1}^m z_j^{n_j}.
\label{eq:ach-query-character}
\end{align}
Consequently, a causal response graded by its per-oracle query counts has the generating-function form
\begin{align}
F(\Bz)=\sum_{\vect n\in\mathbb N_0^m}F_{\vect n} \prod_{j=1}^m z_j^{n_j}.
\label{eq:ach-count-generating-function}
\end{align}
A finite query history gives a polynomial.
A finite-dimensional unitary feedback realization is analytic for $\max_j|z_j|<1$.
\end{proposition}

\begin{proof}
Every count vector decomposes as $\vect n=\sum_jn_j\vect e_j$.
Repeated use of \cref{eq:ach-query-character-law} therefore gives $\chi(\vect n)=\prod_j\chi(\vect e_j)^{n_j}$, which is \cref{eq:ach-query-character}.
Causality gives $n_j\ge0$, so a finite history contains only finitely many nonnegative monomials and hence gives a polynomial.
For a finite unitary system matrix, its private block satisfies $\|D\|\le1$.
With $\oracle(\Bz)=\bigoplus_jz_jO_j$ and unitary $O_j$,
\begin{align}
\|D\oracle(\Bz)\|\le\max_j|z_j|<1.
\label{eq:ach-query-neumann-bound}
\end{align}
Thus
\begin{align}
[I-D\oracle(\Bz)]^{-1}=\sum_{k\ge0}[D\oracle(\Bz)]^k
\label{eq:ach-query-neumann-series}
\end{align}
converges in operator norm on the open polydisk, proving analyticity of the finite-dimensional feedback response.
\end{proof}

The proposition explains why the relevant hierarchy is not continuous versus smooth versus analytic dependence on an arbitrary real parameter.
Query counts are discrete and additive.
Their composition-compatible scalar generating coordinates are monomials, and causality restricts them to nonnegative powers.
The stronger analytic continuation through $\Bz=\Bone$ used in \cref{sec:iir} is an additional quantitative condition for an efficient finite circuit.

Suppose two independently supplied unitary oracles are denoted by $U_1$ and $U_2$.
The scalar monomials $z_1z_2$ and $z_2z_1$ are equal, whereas the operator products $U_2U_1$ and $U_1U_2$ need not be.
A scalar multivariate polynomial therefore groups distinct query histories.
The simplest example is
\begin{align}
z_1z_2-z_2z_1=0,
\quad XZ-ZX=2XZ
\label{eq:ach-commutator-example}
\end{align}
for anticommuting Pauli matrices $X$ and $Z$.
No scalar test can distinguish the zero polynomial on the left from the nonzero ordered polynomial on the right.

To retain the order of oracle calls, we use a noncommutative power series, also called a \emph{free function}.
For a word $w=(j_1,\ldots,j_k)$, define
\begin{align}
Z_w=Z_{j_k}\cdots Z_{j_1},
\quad \mathfrak F(\vect Z)=\sum_w F_w\otimes Z_w.
\label{eq:ach-free-function}
\end{align}
The product $Z_w$ records the exact order of oracle calls.
Substituting commuting scalars $Z_j=z_jI$ combines words with the same per-oracle query counts into a single monomial; this operation is called abelianization.
A free function is defined at matrices of every size and respects direct sums and simultaneous similarities.

A coefficient-level circuit automatically has these properties.
If its known gates are $W_s\otimes I_n$ and its oracle calls use a tuple of $n\times n$ unitaries, then direct-sum substitution replaces every factor by the direct sum of its two evaluations, so
\begin{align}
F(\vect U\oplus\vect V)=F(\vect U)\oplus F(\vect V).
\label{eq:ach-free-direct-sum}
\end{align}
A simultaneous change of signal basis $T$ conjugates every oracle factor and therefore the complete circuit.
With $T\vect U T^{-1}=(TU_1T^{-1},\ldots,TU_mT^{-1})$, this gives
\begin{align}
F(T\vect U T^{-1}) =(I\otimes T)F(\vect U)(I\otimes T^{-1}).
\label{eq:ach-free-similarity}
\end{align}
Finally, a block obtained from a unitary circuit by the projectors in \cref{eq:ach-pi-block} is a contraction, so
\begin{align}
\|F(\vect U)\|\leq1
\label{eq:ach-free-contractivity}
\end{align}
for every matrix size.
Thus matrix-level contractivity is a necessary condition even when all scalar evaluations look harmless.

Free Schur--Agler theory replaces the commuting contraction tuples in \cref{eq:ach-schur-agler-definition} by the matrix tuples of a free domain.
Contractivity at every matrix size, the corresponding matrix-level positive kernel decomposition, and a transfer-function realization are equivalent descriptions of the resulting class~\cite{Ball2016NoncommutativeSchurAgler}.
The realization theorem therefore applies to the specified ordered free function, preserving the noncommuting oracle words.

We use ordinary block-encoding and state-preparation gates in a fixed causal schedule.
A quantum switch instead supplies coherent control of the order of unknown operations as an additional primitive \cite{Chiribella2013QuantumSwitch,Wechs2021CausalOrder}; it can change query costs for order-sensitive tasks.
For example, a switch uses each oracle once to select $(U_2U_1+U_1U_2)/2$, whereas the ordinary gate model requires the three-slot schedule in \cref{prop:ach-symmetrized-cost}.
More general higher-order models replace the prefix constraints below by their own process-admissibility conditions \cite{Abbott2024IndefiniteQuery,Araujo2014ControlledOrder,RennerBrukner2022QubitOrders}.
All bounds in this paper use the ordinary gate model.

We next give a finite fixed-schedule characterization in a form that uses only Gram matrices.
It is a specialization of the standard quantum-query semidefinite characterization ~\cite{Barnum2007QuerySemidefiniteProgramming} and makes its connection with the Agler identity explicit.

Let $\Omega$ be a promise set of oracle tuples.
A fixed schedule is a word
\begin{align}
\sigma=(j_1,\ldots,j_T),
\label{eq:ach-schedule}
\end{align}
meaning that slot $s$ can apply only oracle $O_{j_s}$.
Coherent bypass is allowed: immediately before the query, the state may be split into an inactive subspace and an active subspace on which the oracle acts.
An inactive slot is still a query slot in the schedule.
All operations between query slots are oracle independent.
For a block-encoding oracle, $\Omega$ includes every unitary completion allowed by the access model, while the prescribed $\Pi_{\rm out}$--$\Pi_{\rm in}$ block may depend only on the encoded matrix.
This prevents a realization from using unspecified completion blocks as additional information.

For an operator-valued kernel $K(\omega,\nu)$ on $\Omega$, write $K\succeq0$ when, for every finite set $\omega_1,\ldots,\omega_r$ and public vectors $\vect u_1,\ldots,\vect u_r$,
\begin{align}
\sum_{a,b=1}^r \langle\vect u_a,K(\omega_a,\omega_b)\vect u_b\rangle\geq0.
\label{eq:ach-positive-kernel}
\end{align}
A finite-rank positive kernel has a finite Gram factorization $K(\omega,\nu)=L(\omega)^\dagger L(\nu)$.
This follows by taking the span of formal vectors whose pairwise inner products are specified by the kernel and quotienting by the zero-norm subspace.

\begin{theorem}[Causal Gram characterization of a fixed schedule]
\label{thm:ach-causal-gram}
Let $\Pi_{\rm in}$ and $\Pi_{\rm out}$ project onto finite-dimensional public input and output subspaces, and let $B(\omega):\operatorname{ran}\Pi_{\rm in}\to \operatorname{ran}\Pi_{\rm out}$ be prescribed.
It is exactly implementable as
\begin{align}
B(\omega)=\Pi_{\rm out}V(\omega)\Pi_{\rm in}
\label{eq:ach-causal-pi-block}
\end{align}
by a finite unitary circuit $V(\omega)$ with the schedule in \cref{eq:ach-schedule} if and only if there are finite-rank kernels $G_0,\ldots,G_T$ and finite maps $\Gamma_s(\omega):\operatorname{ran}\Pi_{\rm in}\to\mathcal A_s$ into copies of the active oracle Hilbert space such that, with $I_{\rm in}$ denoting the identity on $\operatorname{ran}\Pi_{\rm in}$,
\begin{align}
G_0(\omega,\nu)&=I_{\rm in},
\label{eq:ach-causal-initial}
\\
K_s^{(1)}(\omega,\nu) &=\Gamma_s(\omega)^\dagger\Gamma_s(\nu),
\label{eq:ach-causal-active}
\\
K_s^{(0)}(\omega,\nu) &=G_{s-1}(\omega,\nu)-K_s^{(1)}(\omega,\nu)\succeq0,
\label{eq:ach-causal-inactive}
\\
G_s(\omega,\nu) &=K_s^{(0)}(\omega,\nu) +\Gamma_s(\omega)^\dagger O_{j_s}(\omega)^\dagger O_{j_s}(\nu) \Gamma_s(\nu),
\label{eq:ach-causal-update}
\\
G_T(\omega,\nu)-B(\omega)^\dagger B(\nu)&\succeq0.
\label{eq:ach-causal-terminal}
\end{align}
The factors reconstruct the known gates and a finite workspace.
For a finite promise set, the kernel conditions become finite positive-semidefinite constraints after bases are chosen.
\end{theorem}

\begin{proof}
Suppose first that the circuit is given.
Let $\Psi_{s-1}(\omega):\operatorname{ran}\Pi_{\rm in}\to\mathcal W_s$ map the public input to the state immediately before slot $s$.
Split the workspace into the inactive and active query subspaces and write
\begin{align}
\Psi_{s-1}(\omega) =\begin{pmatrix}\Lambda_s(\omega)
\\
\Gamma_s(\omega)\end{pmatrix}.
\label{eq:ach-causal-prequery-split}
\end{align}
Its cross-Gram kernel is
\begin{align}
G_{s-1}(\omega,\nu) =\Lambda_s(\omega)^\dagger\Lambda_s(\nu) +\Gamma_s(\omega)^\dagger\Gamma_s(\nu).
\label{eq:ach-causal-prequery-gram}
\end{align}
Thus \cref{eq:ach-causal-active,eq:ach-causal-inactive} hold.
After the query, the state map is
\begin{align}
\begin{pmatrix} \Lambda_s(\omega)
\\
O_{j_s}(\omega)\Gamma_s(\omega) \end{pmatrix},
\label{eq:ach-causal-postquery}
\end{align}
whose cross-Gram is exactly \cref{eq:ach-causal-update}.
The known unitary between slots preserves this Gram, so it is also the kernel $G_s$ of the next pre-query state.
Initially the prepared state is an isometry independent of $\omega$, giving \cref{eq:ach-causal-initial}.
At the end, restrict the input to $\operatorname{ran}\Pi_{\rm in}$ and split the public output with $\Pi_{\rm out}$ and $I-\Pi_{\rm out}$.
Thus $B(\omega)=\Pi_{\rm out}V(\omega)\Pi_{\rm in}$ and $L(\omega)=(I-\Pi_{\rm out})V(\omega)\Pi_{\rm in}$, so
\begin{align}
G_T(\omega,\nu) =B(\omega)^\dagger B(\nu)+L(\omega)^\dagger L(\nu),
\label{eq:ach-causal-output-split}
\end{align}
which proves \cref{eq:ach-causal-terminal}.

Conversely, factor each positive kernel in \cref{eq:ach-causal-inactive,eq:ach-causal-terminal} as
\begin{align}
K_s^{(0)}(\omega,\nu) &=\Lambda_s(\omega)^\dagger\Lambda_s(\nu),
\label{eq:ach-causal-inactive-factor}
\\
G_T(\omega,\nu)-B(\omega)^\dagger B(\nu) &=L(\omega)^\dagger L(\nu).
\label{eq:ach-causal-terminal-factor}
\end{align}
At slot $s$, the pre-query family $[\Lambda_s(\omega);\Gamma_s(\omega)]$ has Gram $G_{s-1}$, while the post-query family in \cref{eq:ach-causal-postquery} has Gram $G_s$ by \cref{eq:ach-causal-update}.
For $s<T$, factor $G_s$ using \cref{eq:ach-causal-inactive,eq:ach-causal-active} at the next slot.
The two finite vector families with Gram $G_s$ are related by an oracle-independent isometry.
To see this directly, a linear combination has squared norm equal to the same Gram quadratic form on both sides; a zero combination therefore maps to zero, and polarization preserves all inner products.
After finite padding, extend the isometry to a unitary.
This unitary is the known gate between slots $s$ and $s+1$.

The same Gram argument maps the fixed input subspace, whose Gram is $I_{\rm in}$, to the first pre-query family.
At the end it maps the last post-query family to $[B(\omega);L(\omega)]$, whose Gram is $G_T$ by \cref{eq:ach-causal-terminal-factor}.
Concatenating these known unitaries and the prescribed oracle calls gives a finite circuit with exactly the schedule $\sigma$ and $\Pi_{\rm out}V(\omega)\Pi_{\rm in}=B(\omega)$.
\end{proof}

The fixed-schedule characterization has an analytic realization with no loss of query expressivity.
The construction is deliberately simple: it stores the causally ordered stage in the private space.

\begin{theorem}[Analytic completeness of finite causally ordered MQSP]
\label{thm:ach-analytic-completeness}
Let
\begin{align}
V(\omega) =K_TQ_{j_T}(\omega)K_{T-1}\cdots K_1Q_{j_1}(\omega)K_0
\label{eq:ach-finite-circuit}
\end{align}
be any finite coherent query circuit, where the $K_s$ are oracle-independent unitaries and $\sigma=(j_1,\ldots,j_T)$ is its ordered query schedule.
Then there is a finite-dimensional unitary MQSP realization executed with the same ordered query schedule and query counts whose marked transfer function is the polynomial
\begin{align}
F_\sigma(\Bz;\omega) =\left(\prod_{s=1}^Tz_{j_s}\right)V(\omega).
\label{eq:ach-finite-history-transfer}
\end{align}
In particular, $F_\sigma(\Bone;\omega)=V(\omega)$.
Hence restricting the query-count coordinates to analytic functions does not remove any finite causally ordered query circuit.
\end{theorem}

\begin{proof}
Pad the circuit registers to one finite workspace $\mathcal W$.
Use one public copy and $T$ private copies of $\mathcal W$, and define
\begin{align}
S(\vect u,\vect f_1,\ldots,\vect f_T) = \bigl( K_T\vect f_T,\, K_0\vect u,\, K_1\vect f_1,\ldots, K_{T-1}\vect f_{T-1} \bigr).
\label{eq:ach-finite-history-system}
\end{align}
This is a permutation of mutually orthogonal sectors followed by the known unitaries $K_s$, so $S$ is unitary.
Close private sector $s$ through the authorized query $z_{j_s}Q_{j_s}(\omega)$.
The private continuation advances strictly from sector $s$ to sector $s+1$ and is therefore nilpotent.
A public input can reach the public output only after traversing all $T$ private sectors, which gives exactly \cref{eq:ach-finite-history-transfer}.
Directed execution calls the same oracle primitive in the same causally ordered slot as the original circuit, so the original query schedule and counts are unchanged.
\end{proof}

The theorem gives an exact representation of any finite fixed-order query circuit, with the same known gates in its system matrix.
The finite response remains analytic in the query-count variables when those gates and oracles vary with physical time; its input--output kernel then has two time indices, as in \cref{sec:time-dependent-mqsp}.
Compact repeated IIR structure and quantitative continuation through $\Bz=\Bone$ supply the additional efficiency used in our algorithms.

The causal Gram theorem also contains a causally ordered version of the Agler decomposition.
Attach an independent analytic variable $z_s$ to the active branch of slot $s$.
If $\Gamma_s(\Bz;\omega)$ is the active pre-query map, it depends only on the markers of earlier slots.
Telescoping the change in the cross-Gram at each query gives
\begin{align}
I-\Psi_T(\Bz;\omega)^\dagger\Psi_T(\vect w;\nu) =\sum_{s=1}^T \Gamma_s(\Bz;\omega)^\dagger \left[I-\overline z_sw_s O_{j_s}(\omega)^\dagger O_{j_s}(\nu)\right] \Gamma_s(\vect w;\nu).
\label{eq:ach-causal-agler}
\end{align}
This identity follows by subtracting the cross-Gram after slot $s$ from the cross-Gram before it and summing over $s$.
Known gates cancel because they are unitary.
The extra information relative to an unlabelled Agler decomposition is causality: the factor at slot $s$ is a state that can be prepared using only the prefix of the schedule.
Grouping all markers belonging to the same oracle type recovers the port-weight identity of \cref{thm:iir-energy}, but the grouped identity no longer records the order of those visits.
This dependence on query order is already visible for two oracles.
Consider
\begin{align}
P(U_1,U_2)=\frac12\left(U_2U_1+U_1U_2\right).
\label{eq:ach-symmetrized-product}
\end{align}
Its free norm is one: the triangle inequality gives at most one, and $U_1=U_2=I$ attains one.
Its total degree is two and each variable occurs only once in every monomial.
Nevertheless, one call to each independently supplied oracle cannot implement both orders in a common causally ordered schedule.

\begin{proposition}[Order can increase actual query cost]
\label{prop:ach-symmetrized-cost}
In the auxiliary-only free model, where known gates act on a finite auxiliary space and each query slot either bypasses or applies its designated oracle, \cref{eq:ach-symmetrized-product} has exact weighted query cost
\begin{align}
c_1+c_2+\min\{c_1,c_2\}.
\label{eq:ach-symmetrized-cost}
\end{align}
The two minimal schedules are $121$ and $212$.
\end{proposition}

\begin{proof}
A branch of a fixed schedule can only produce a word that is a subsequence of that schedule: bypass deletes slots but cannot reverse them.
The target contains both words $12$ and $21$, so every exact schedule must contain both as subsequences.
A two-letter schedule contains at most one of these orders.
The shortest possibilities are therefore $121$ and $212$, with costs $2c_1+c_2$ and $c_1+2c_2$.

For achievability with schedule $121$, prepare an auxiliary qubit in $(|0\rangle+|1\rangle)/\sqrt2$.
On branch $|0\rangle$, use the first $U_1$ slot and bypass the last; on branch $|1\rangle$, bypass the first and use the last.
Apply $U_2$ on both branches in the middle.
The two branches therefore contain $U_2U_1$ and $U_1U_2$.
With $|+\rangle=(|0\rangle+|1\rangle)/\sqrt2$ and $\Pi_{\rm in}=\Pi_{\rm out}=|+\rangle\!\langle+|\otimes I$, the resulting circuit $V$ satisfies
\begin{align}
\Pi_{\rm out}V\Pi_{\rm in} =|+\rangle\!\langle+|\otimes P(U_1,U_2)
\label{eq:ach-symmetrized-realization}
\end{align}
on $\operatorname{ran}\Pi_{\rm in}$.
The schedule $212$ is analogous.
Taking the cheaper of the two costs proves \cref{eq:ach-symmetrized-cost}.
\end{proof}

More generally, an exact causally ordered schedule for a linear combination of ordered words must contain every word with nonzero coefficient as a subsequence.
For scalar coefficients $a_w$ with $\sum_w|a_w|\le1$, a coherent word label and known attenuation give the converse in this auxiliary-only model.
For arbitrary coefficients, the same construction implements the combination at normalization $\sum_w|a_w|$; it does not establish attainability at a smaller prescribed normalization.
Thus the weighted shortest-common-supersequence problem supplies the schedule for this LCU construction, while the causal Gram conditions also retain the normalization constraint.

\subsection{Rational completion and constructive unitary realizations}
\label{sec:ach-rational}

Allowing rational functions changes the algebraic completion problem.
The classical reason is Artin's solution of Hilbert's seventeenth problem: every real polynomial that is nonnegative on all of $\mathbb R^m$ is a sum of squares of rational functions~\cite{Artin1927Hilbert17}.
The same statement holds directly for a nonnegative rational function.

\begin{proposition}[Rational sums of squares]
\label{prop:ach-rational-sos}
Let $h\in\mathbb R(x_1,\ldots,x_m)$ be nonnegative at every real point where it is defined.
Then there are rational functions $r_k$ such that
\begin{align}
h(\vect x)=\sum_k r_k(\vect x)^2.
\label{eq:ach-rational-sos}
\end{align}
If $H(\vect x)$ is a real symmetric rational matrix that is positive semidefinite wherever it is defined, then there are rational matrices $R_k$ such that
\begin{align}
H(\vect x)=\sum_k R_k(\vect x)^T R_k(\vect x).
\label{eq:ach-rational-matrix-sos}
\end{align}
\end{proposition}

\begin{proof}
Write $h=a/b$ with real polynomials $a,b$.
At every point with $b\neq0$,
\begin{align}
a(\vect x)b(\vect x)=h(\vect x)b(\vect x)^2\geq0.
\label{eq:ach-rational-clear-denominator}
\end{align}
The polynomial $ab$ is therefore nonnegative everywhere: when $b=0$, the product $ab$ vanishes.
Artin's theorem writes $ab$ as a sum of squares of rational functions,
\begin{align}
a(\vect x)b(\vect x)=\sum_k q_k(\vect x)^2.
\label{eq:ach-artin-applied}
\end{align}
Dividing by $b^2$ gives $h=\sum_k(q_k/b)^2$, proving \cref{eq:ach-rational-sos}.

For the matrix statement, choose a nonzero polynomial $d$ and a symmetric polynomial matrix $P$ such that $H=P/d$.
The polynomial matrix $d^2H=dP$ is positive semidefinite wherever $d\neq0$ and is zero when $d=0$, hence is positive semidefinite at every real point.
The matrix extension of Hilbert's seventeenth problem gives rational matrices $Q_k$ such that
\begin{align}
d(\vect x)^2H(\vect x)=\sum_k Q_k(\vect x)^TQ_k(\vect x).
\label{eq:ach-matrix-artin}
\end{align}
Dividing each factor by $d$ proves \cref{eq:ach-rational-matrix-sos}~\cite{HillarNie2008Hilbert17Matrices}.
\end{proof}

Allowing rational factors removes the polynomial sum-of-squares obstruction, but does not specify how to realize those factors using the available oracles.
MQSP also requires each private component to pass through a permitted oracle before being fed back.
We obtain this structure directly from the steady-state equations of a unitary IIR.
Recall from \cref{thm:iir-energy} that the system matrix and feedback operator
\begin{align}
S=\left(\begin{array}{c|c}A&B
\\
\hline
C&D\end{array}\right),
\quad \oracle(\Bz)=\bigoplus_{j=1}^m z_jO_j,
\label{eq:ach-system-recall}
\end{align}
has transfer function and private response
\begin{align}
F(\Bz)=A+B\oracle(\Bz)[I-D\oracle(\Bz)]^{-1}C,
\quad \Gamma(\Bz)=[I-D\oracle(\Bz)]^{-1}C.
\label{eq:ach-transfer-recall}
\end{align}
The steady-state equation is
\begin{align}
S\begin{pmatrix}I
\\
\oracle(\Bz)\Gamma(\Bz)\end{pmatrix} =\begin{pmatrix}F(\Bz)
\\
\Gamma(\Bz)\end{pmatrix}.
\label{eq:ach-steady-state-recall}
\end{align}
Taking inner products at two evaluations gives an identity whose factors are the private responses at the individual ports.
The following theorem also proves the converse: finite-dimensional factors satisfying this identity determine a unitary realization.

\begin{theorem}[Finite oracle-compatible unitary realization]
\label{thm:ach-unitary-characterization}
Let $F(\xi)$ be a square map on a fixed finite public space, where $\xi=(\Bz,\vect O)$ ranges over the prescribed oracle promise and $Q_j(\xi)=z_jO_j$.
Then $F$ has a finite oracle-independent unitary realization with these query ports if and only if there are finite-dimensional maps $H_j(\xi)$ such that, for every two evaluations $\xi,\eta$,
\begin{align}
I-F(\eta)^\dagger F(\xi) =\sum_j H_j(\eta)^\dagger \left[I-Q_j(\eta)^\dagger Q_j(\xi)\right] H_j(\xi).
\label{eq:ach-agler-kernel}
\end{align}
For fixed public projectors $\Pi_{\rm in}$ and $\Pi_{\rm out}$, a prescribed block
\begin{align}
f(\xi)=\Pi_{\rm out}F(\xi)\Pi_{\rm in}
\label{eq:ach-encoded-block}
\end{align}
has the corresponding condition
\begin{align}
\Pi_{\rm in}-f(\eta)^\dagger f(\xi) =L(\eta)^\dagger L(\xi) +\sum_j H_j(\eta)^\dagger \left[I-Q_j(\eta)^\dagger Q_j(\xi)\right] H_j(\xi),
\label{eq:ach-block-agler-kernel}
\end{align}
where $H_j(\xi)=H_j(\xi)\Pi_{\rm in}$ and $L(\xi)=(I-\Pi_{\rm out})F(\xi)\Pi_{\rm in}$ are regarded as maps with input in $\operatorname{ran}\Pi_{\rm in}$.
\end{theorem}

\begin{proof}
For necessity, use \cref{eq:ach-steady-state-recall} at $\xi$ and $\eta$.
Unitarity of $S$ preserves the cross inner product, so
\begin{align}
I+\Gamma(\eta)^\dagger \oracle(\eta)^\dagger \oracle(\xi)\Gamma(\xi) =F(\eta)^\dagger F(\xi)+\Gamma(\eta)^\dagger\Gamma(\xi).
\label{eq:ach-cross-inner-product}
\end{align}
Because the private ports form a direct sum, $\oracle(\eta)^\dagger \oracle(\xi)=\bigoplus_jQ_j(\eta)^\dagger Q_j(\xi)$.
Rearranging gives \cref{eq:ach-agler-kernel} with $H_j=\Gamma_j$.
Restricting the public input with $\Pi_{\rm in}$ and splitting the output with $\Pi_{\rm out}$ and $I-\Pi_{\rm out}$ gives $f$ and $L$ in \cref{eq:ach-encoded-block}; the same calculation gives \cref{eq:ach-block-agler-kernel}.

For sufficiency, stack the factors into $H=(H_1,\ldots,H_m)^T$ and put $Q=\bigoplus_jQ_j$.
Define
\begin{align}
V_{\rm in}(\xi)=\begin{pmatrix}I
\\
Q(\xi)H(\xi)\end{pmatrix},
\quad V_{\rm out}(\xi)=\begin{pmatrix}F(\xi)
\\
H(\xi)\end{pmatrix}.
\label{eq:ach-unitary-columns}
\end{align}
\Cref{eq:ach-agler-kernel} says that every pair of columns from these two families has the same inner product.
Hence the assignment $V_{\rm in}(\xi)\vect u\mapsto V_{\rm out}(\xi)\vect u$ is a well-defined isometry on the span of all such vectors.
Extend it to a unitary $S$ on a finite space.
Writing $S=\begin{psmallmatrix}A&B\\C&D\end{psmallmatrix}$ gives
\begin{align}
H=C+DQH,
\quad F=A+BQH.
\label{eq:ach-unitary-recovered-equations}
\end{align}
For strict loads, $\|DQ\|<1$, so $H=(I-DQ)^{-1}C$ and the second equation is the transfer function in \cref{eq:ach-transfer-recall}.
Analytic continuation gives the regular boundary values.
For the block in \cref{eq:ach-encoded-block}, take $H=H\Pi_{\rm in}$ and replace the two columns in \cref{eq:ach-unitary-columns} by
\begin{align}
\begin{pmatrix}\Pi_{\rm in}
\\
0
\\
QH\end{pmatrix},
\quad \begin{pmatrix}f
\\
L
\\
H\end{pmatrix}.
\label{eq:ach-block-unitary-columns}
\end{align}
Their Gram identity is \cref{eq:ach-block-agler-kernel}, so the same isometry-extension argument constructs the required unitary realization.
\end{proof}

Several necessary conditions now follow without separate assumptions.
At equal interior points with unitary $O_j$,
\begin{align}
I-F(\Bz)^\dagger F(\Bz) =\sum_j(1-|z_j|^2)H_j(\Bz)^\dagger H_j(\Bz)\succeq0.
\label{eq:ach-structured-rational-sos}
\end{align}
Thus the defect is a structured rational sum of Hermitian squares.
At a regular point on the unit torus, the right side vanishes and the full square response is unitary.
In particular, $F(\Bone)^\dagger F(\Bone)=I$.
At zero delay, $F(\vect 0)=A$, which is independent of the unknown oracles.
Expanding the resolvent gives $F(\Bz) =A+\sum_{k\geq0}B\oracle(\Bz)[D\oracle(\Bz)]^kC$, so the coefficient of $z_1^{n_1}\cdots z_m^{n_m}$ is a sum of ordered paths with exactly $n_j$ passages through port $j$.
The zero-delay condition is only the first member of this complete grading by query counts and ordered paths.

For the block in \cref{eq:ach-encoded-block}, \cref{eq:ach-block-agler-kernel} gives instead
\begin{align}
\Pi_{\rm in}-f(\Bz)^\dagger f(\Bz) =L(\Bz)^\dagger L(\Bz) +\sum_j(1-|z_j|^2)H_j(\Bz)^\dagger H_j(\Bz).
\label{eq:ach-block-structured-sos}
\end{align}
On the unit torus, the defect from isometry on $\operatorname{ran}\Pi_{\rm in}$ is exactly the probability transmitted to $\operatorname{ran}(I-\Pi_{\rm out})$.
This is the rational counterpart of the univariate QSP completion: the complementary squares are part of the realization rather than an additional polynomial factorization problem.

For a finite realization written in commuting scalar variables, this rational sum of squares can be cleared to a polynomial identity.
Let $d(\Bz)=\det[I-D\oracle(\Bz)]$.
The adjugate formula shows that $d(\Bz)H_j(\Bz)$ and $d(\Bz)F(\Bz)$ are polynomial in the marked query variables, after including any fixed polynomial denominators already present in the permitted loads.
Multiplying \cref{eq:ach-structured-rational-sos} by $|d(\Bz)|^2$ gives
\begin{align}
|d(\Bz)|^2I-[d(\Bz)F(\Bz)]^\dagger[d(\Bz)F(\Bz)] =\sum_j(1-|z_j|^2) [d(\Bz)H_j(\Bz)]^\dagger[d(\Bz)H_j(\Bz)].
\label{eq:ach-cleared-agler-sos}
\end{align}
For the block in \cref{eq:ach-encoded-block}, multiplying \cref{eq:ach-block-structured-sos} by $|d(\Bz)|^2$ gives
\begin{align}
|d(\Bz)|^2\Pi_{\rm in}-[d(\Bz)f(\Bz)]^\dagger[d(\Bz)f(\Bz)] = [d(\Bz)L(\Bz)]^\dagger[d(\Bz)L(\Bz)] +\sum_j(1-|z_j|^2) [d(\Bz)H_j(\Bz)]^\dagger[d(\Bz)H_j(\Bz)].
\end{align}
Thus every finite rational transfer function with a unitary realization in this commuting rational setting has an exact structured sum-of-squares completion after clearing denominators.
For scalar rational inner Schur--Agler functions, this is the standard rational Agler sum-of-squares decomposition; finite-dimensional transfer realizations and such decompositions are equivalent~\cite{Knese2010RationalInner}.
This is the form relevant to MQSP: the factors are assigned to public complementary outputs and to specific query ports.
General rational sum-of-squares theorems explain why denominators remove the elementary polynomial obstruction; the unitary realization supplies the stronger port structure required by the quantum circuit.

\subsection{From achievability to modular transfer-function design}
\label{sec:ach-modular}

The preceding results characterize achievable transformations while preserving oracle order.
For practical algorithm design, we seek structured IIR modules with compact system matrices, explicit catalyst weights, and analytic neighborhoods that can be bounded before compilation.

A one-state example already separates realization size from clock cost, but the analytic variable must first be attached to an actual oracle.
Let $O(x)=x$ be a one-dimensional phase oracle with $x=e^{\ii\phi}$, and let $z$ mark each call to this oracle.
For a known $0<a<1$, define the first-order all-pass IIR module
\begin{align}
\module{AP1}_a(z;x) =b_a(zx) =\frac{zx-a}{1-azx}.
\label{eq:ach-blaschke}
\end{align}
Thus the familiar Blaschke factor $b_a$ is loaded by the query operation $zx$.
It has the two-dimensional unitary system matrix
\begin{align}
S_{\module{AP1}_a}= \left(\begin{array}{c|c} -a&\sqrt{1-a^2}
\\
\hline
\sqrt{1-a^2}&a \end{array}\right),
\quad \oracle(z;x)=zx.
\label{eq:ach-blaschke-system}
\end{align}
If $\vect u,\vect v$ are the public input and output and $\vect g$ is the private amplitude, the steady-state equations are
\begin{align}
\begin{pmatrix}\vect v
\\
\vect g\end{pmatrix} =S_{\module{AP1}_a} \begin{pmatrix}\vect u
\\
zx\,\vect g\end{pmatrix},
\quad \vect g=\frac{\sqrt{1-a^2}}{1-azx}\,\vect u.
\label{eq:ach-blaschke-elimination}
\end{align}
Eliminating $\vect g$ gives \cref{eq:ach-blaschke}.
The private amplitude is therefore the steady-state catalyst for queries to the phase oracle $O(x)$.
At $z=1$,
\begin{align}
\|\Gamma_a(1;x)\|^2 =\frac{1-a^2}{|1-ax|^2} =\overline{\module{AP1}_a(1;x)} \left.\partial_z\module{AP1}_a(z;x)\right|_{z=1}.
\label{eq:ach-blaschke-weight}
\end{align}
For $x=1$, this reduces to $(1+a)/(1-a)$, and the pole $z=(ax)^{-1}$ approaches the all-ones point as $a\uparrow1$.
The large catalyst weight is then a large query weight to an actual oracle port.

An oracle promise can remove this singular behavior.
Since $|x|=1$, being ``gapped away from $1$'' means that a known $\Delta>0$ satisfies $|1-x|\ge\Delta$.
For every such phase,
\begin{align}
|1-ax|^2 =(1-a)^2+a|1-x|^2,
\quad \|\Gamma_a(1;x)\|^2 \le \frac{2(1-a)}{a\Delta^2}.
\label{eq:ach-ap1-gap-weight}
\end{align}
Hence the catalyst weight tends to zero as $a\uparrow1$, uniformly on the promised set once $a$ is bounded away from zero.
The pole also remains away from the all-ones point: for $a\ge1/2$,
\begin{align}
\left|\frac1{ax}-1\right| =\frac{|1-ax|}{a} \ge\frac{\Delta}{\sqrt a} \ge\Delta,
\label{eq:ach-ap1-gap-pole}
\end{align}
while $a<1/2$ places the pole outside $|z|=2$.
At $a=1$ the system matrix is $\operatorname{diag}(-1,1)$, the private port decouples, and the public response is $-1$; the nonuniform $a\uparrow1$ behavior occurs only when the oracle promise permits $x$ to approach the singular phase $1$.

This is the same query bookkeeping used in the introductory FPAA and Cayley constructions of \cref{sec:intro-simple-proofs}: their private loads are $zO_\theta$ and $zO(x)$, respectively, so their catalyst norms measure weight on Grover-iterate or block-encoding queries.
The example isolates why a useful module specification includes the transfer function, its oracle port, and the oracle promise.
Realization dimension alone omits the information needed by clock shaping in \cref{sec:iir}.

The order-sensitive example in \cref{prop:ach-symmetrized-cost} shows a different separation: free norm and polynomial degree can both be small even when independently supplied oracles require additional causally ordered query slots.
A general positive-kernel factorization can likewise hide a large known unitary whose gate synthesis is expensive.
A useful resource analysis must therefore retain the oracle schedule, the known operations, and the clock parameters.
The following theorem relates these choices to exact clock optimization: optimizing the clock for one realization gives its error--normalization frontier, and optimizing over all realizations with the prescribed schedule recovers the optimal error--normalization tradeoff for the prescribed query schedule.

\begin{theorem}[Clock completeness after optimizing the realization]
\label{thm:ach-clock-completeness}
Fix a prescribed query schedule $\sigma$, an oracle promise $\Omega$, projectors $\Pi_{\rm in},\Pi_{\rm out}$, a target family $P(\omega)$, and a normalization $\alpha\ge1$.
Let
\begin{align}
e_\sigma(P;\alpha) := \inf_{V\ {\rm with\ schedule}\ \sigma} \sup_{\omega\in\Omega} \left\| \alpha\Pi_{\rm out}V(\omega)\Pi_{\rm in}-P(\omega) \right\|.
\label{eq:ach-causal-error-frontier}
\end{align}
For a finite schedule-preserving unitary MQSP realization $K$, write $K_\omega$ for its finite public clock kernel and define its exact error--normalization tradeoff for the chosen lift by
\begin{align}
e_K(P;\alpha) := \min_{\|X\|_*\le\alpha} \sup_{\omega\in\Omega} \left\| \Pi_{\rm out}\mathcal L_X(K_\omega)\Pi_{\rm in}-P(\omega) \right\|.
\label{eq:ach-fixed-realization-frontier}
\end{align}
By \cref{thm:iir-clock-body,lem:clock-extraction}, $\mathcal L_X(K_\omega)/\alpha$ is exactly the implemented block produced by the clock.
If $\mathfrak R_\sigma$ contains all such finite realizations, including finite-history realizations, then
\begin{align}
e_\sigma(P;\alpha) = \inf_{K\in\mathfrak R_\sigma}e_K(P;\alpha).
\label{eq:ach-clock-completeness}
\end{align}
For a finite explicit promise, the left-hand side is the causal Gram semidefinite optimization for the fixed schedule.
\end{theorem}

\begin{proof}
A scheduled realization $K$ is itself a finite circuit with schedule $\sigma$.
Therefore every clock feasible in $e_K(P;\alpha)$ is feasible in \cref{eq:ach-causal-error-frontier}, which gives
\begin{align}
e_\sigma(P;\alpha)\le e_K(P;\alpha)
\label{eq:ach-clock-completeness-forward}
\end{align}
for every $K\in\mathfrak R_\sigma$.

For the converse, take any circuit $V$ with schedule $\sigma$ and relative error $\eta$.
By \cref{thm:ach-analytic-completeness}, its finite-history MQSP realization has only one nonzero total-delay coefficient at delay $T$, namely $V$.
Use horizon $T+1$ and the endpoint clock matrix
\begin{align}
X=\alpha|T\rangle\!\langle0|,
\quad \|X\|_*=\alpha.
\label{eq:ach-endpoint-clock}
\end{align}
The exact clock characterization then extracts the relative numerator $\alpha V$.
The finite public kernel can be implemented with the prescribed schedule $\sigma$: route clock input zero through $V$ to clock output $T$, route all other public clock inputs to orthogonal private outputs, and complete the routing by known swaps.
The oracle slots act only on the branch traversing $V$, with coherent bypass on the other branches.
Thus the only nonzero public block is $V$ at position $(T,0)$, without repeating the schedule for each clock position.
After applying $\Pi_{\rm out}$ and $\Pi_{\rm in}$, its error is exactly the error of the original circuit.
Hence some $K\in\mathfrak R_\sigma$ satisfies $e_K(P;\alpha)\le\eta$.
Taking the infimum over $V$ gives the reverse inequality in \cref{eq:ach-clock-completeness}.
\end{proof}

The theorem makes clock optimization exact for a chosen realization, but selecting that realization remains an algorithm-design problem.
Modular constructions provide transfer functions, private responses, and known operations that can be analyzed once and composed.
The resulting resource bounds retain the contribution of each original oracle.

One useful construction starts from an operator-valued load $M$ with positive semidefinite Hermitian part, $\operatorname{Re}M=(M+M^\dagger)/2\succeq0$.
Such a load is called \emph{accretive}.
Positive sums, known congruences, and regular Schur complements preserve this property, while the Cayley transform produces a contractive transfer function:
\begin{align}
F=(I-M)(I+M)^{-1},
\quad I-F^\dagger F =4(I+M^\dagger)^{-1}(\operatorname{Re}M)(I+M)^{-1}\succeq0.
\label{eq:ach-cayley-defect}
\end{align}
The identity follows by multiplying on the left and right by the two denominators and using
\begin{align}
(I+M^\dagger)(I+M)-(I-M^\dagger)(I-M)=2(M+M^\dagger).
\label{eq:ach-cayley-defect-proof}
\end{align}
A strict contraction $Y$ supplies an accretive load because
\begin{align}
\operatorname{Re}\left[(I+Y)(I-Y)^{-1}\right] =(I-Y^\dagger)^{-1}(I-Y^\dagger Y)(I-Y)^{-1}\succeq0.
\label{eq:ach-oracle-cayley-positive}
\end{align}
This identity is verified by multiplying by $I-Y^\dagger$ and $I-Y$.
It uses no commutativity assumption.

The Cayley module used for Hamiltonian simulation in \cref{sec:intro-simple-proofs} has a different signal model.
There $x\in[-1,1]$ is an eigenvalue encoded in a larger unitary oracle, rather than the phase of the one-dimensional oracle used in \cref{eq:ach-blaschke}.
The scalar quotient
\begin{align}
\frac{z-\ii x}{1+\ii xz}
\label{eq:ach-cayley-scalar-factor}
\end{align}
is all-pass in $z$, but its value at $z=0$ depends on the unknown $x$ and therefore cannot by itself be the MQSP transfer function for that oracle access model.
The query-compatible module is instead
\begin{align}
\module{Cayley}(z;x) =z\frac{z-\ii x}{1+\ii xz},
\label{eq:ach-ap1-cayley}
\end{align}
with the oracle load $zO(x)$ shown in \cref{eq:sp-cayley-warmup-system}.
In contrast, $\module{AP1}_a(0;x)=-a$ in \cref{eq:ach-blaschke} is already independent of the phase oracle, so no additional leading factor of $z$ is required there.
Whether a scalar all-pass formula is query-compatible therefore depends on how its signal variable is supplied.
The multioracle construction of \cref{eq:sp-weighted-cayley} combines the corresponding oracle loads before applying the outer transform below.

To use the same block-encoding notation as the rest of the manuscript, assume that oracle $O_j$ acts on an encoding register and a common signal space, and write its signal embedding as $|0\rangle_j\otimes I_j$.
We abbreviate this map as $|0\rangle_j$ and its adjoint $\langle0|_j\otimes I_j$ as $\langle0|_j$.
This entails no loss of generality for a known isometric signal embedding: after padding if necessary, any known isometry $E_j$ extends to a known unitary $W_j$ satisfying
\begin{align}
E_j=W_j(|0\rangle_j\otimes I_j).
\label{eq:ach-signal-basis-change}
\end{align}
Replacing $O_j$ by $W_j^\dagger O_jW_j$ only adds known gates before and after one call to $O_j$.
We therefore use $|0\rangle_j$ directly below.

\begin{theorem}[Cayley realization with standard signal subspaces]
\label{thm:ach-cayley-realization}
Let $Y_j=z_jO_j$, let $K_0$ be known and Hermitian on the common signal space, and define
\begin{align}
M(\Bz) &=\ii K_0+\sum_j\langle0|_j (I+Y_j)(I-Y_j)^{-1}|0\rangle_j,
\label{eq:ach-cayley-load}
\\
F(\Bz)&=(I-M(\Bz))(I+M(\Bz))^{-1}.
\label{eq:ach-cayley-response}
\end{align}
Let
\begin{align}
L=\begin{pmatrix}|0\rangle_1
\\
\vdots
\\
|0\rangle_m\end{pmatrix},
\quad Y=\bigoplus_jY_j,
\quad R=(I+L^\dagger L+\ii K_0)^{-1} =((m+1)I+\ii K_0)^{-1},
\label{eq:ach-cayley-R}
\end{align}
where $m$ is the number of oracle ports.
Then $F$ has a finite oracle-independent unitary realization using the original query ports.
One such system matrix is
\begin{align}
S= \left(\begin{array}{c|c} 2R-I&-2RL^\dagger
\\
\hline
2LR&I-2LRL^\dagger \end{array}\right).
\label{eq:ach-cayley-system}
\end{align}
\end{theorem}

\begin{proof}
The abbreviation $|0\rangle_j=|0\rangle_j\otimes I_j$ gives
\begin{align}
L^\dagger L=mI,
\quad L^\dagger(I+Y)(I-Y)^{-1}L =\sum_j\langle0|_j(I+Y_j)(I-Y_j)^{-1}|0\rangle_j.
\label{eq:ach-cayley-signal-compression}
\end{align}
The inverse in \cref{eq:ach-cayley-R} exists because the Hermitian part of its denominator is $(m+1)I\succ0$.
From
\begin{align}
R^{-1}+R^{-\dagger}=2(I+L^\dagger L)
\label{eq:ach-cayley-R-inverse}
\end{align}
we obtain
\begin{align}
R+R^\dagger=2R^\dagger(I+L^\dagger L)R.
\label{eq:ach-cayley-R-identity}
\end{align}
For example, the upper-left block of $S^\dagger S$ is
\begin{align}
(2R^\dagger-I)(2R-I)+4R^\dagger L^\dagger LR=I+4R^\dagger(I+L^\dagger L)R-2(R+R^\dagger)=I,
\label{eq:ach-cayley-unitary-11}
\end{align}
and the upper-right block is
\begin{align}
(2R^\dagger-I)(-2RL^\dagger) +2R^\dagger L^\dagger(I-2LRL^\dagger)=2(R+R^\dagger)L^\dagger -4R^\dagger(I+L^\dagger L)RL^\dagger=0.
\label{eq:ach-cayley-unitary-12}
\end{align}
The lower-left block is the adjoint of \cref{eq:ach-cayley-unitary-12}.
The lower-right block is
\begin{align}
4LR^\dagger RL^\dagger +(I-2LR^\dagger L^\dagger)(I-2LRL^\dagger)=I-2L(R+R^\dagger)L^\dagger +4LR^\dagger(I+L^\dagger L)RL^\dagger=I.
\label{eq:ach-cayley-unitary-22}
\end{align}
Hence $S^\dagger S=I$.
Since $S$ is square and finite dimensional, $S$ is unitary.

Write the steady-state relation as
\begin{align}
S\begin{pmatrix}\vect u
\\
Y\vect g\end{pmatrix} =\begin{pmatrix}\vect v
\\
\vect g\end{pmatrix}.
\label{eq:ach-cayley-steady-state}
\end{align}
The lower row and the sum of the upper input and output give
\begin{align}
\vect g-Y\vect g=L(\vect u+\vect v).
\label{eq:ach-cayley-private-equation}
\end{align}
Using $R^{-1}=I+L^\dagger L+\ii K_0$ in the upper row gives
\begin{align}
\vect u-\vect v =\ii K_0(\vect u+\vect v)+L^\dagger(\vect g+Y\vect g).
\label{eq:ach-cayley-public-equation}
\end{align}
Solving \cref{eq:ach-cayley-private-equation} gives
\begin{align}
\vect g=(I-Y)^{-1}L(\vect u+\vect v).
\label{eq:ach-cayley-private-solution}
\end{align}
Substitution into \cref{eq:ach-cayley-public-equation} and \cref{eq:ach-cayley-signal-compression} yields
\begin{align}
\vect u-\vect v=M(\Bz)(\vect u+\vect v).
\label{eq:ach-cayley-impedance-equation}
\end{align}
Therefore $\vect v=(I-M)(I+M)^{-1}\vect u$, which is \cref{eq:ach-cayley-response}.
Every matrix in \cref{eq:ach-cayley-system} is known and oracle independent; all unknown dependence occurs through the original loads $Y_j$.
\end{proof}

The construction is closed under elimination of known internal public coordinates.
Partition an accretive load as
\begin{align}
M= \begin{pmatrix}M_{11}&M_{12}
\\
M_{21}&M_{22}\end{pmatrix},
\quad \widehat M=M_{11}-M_{12}M_{22}^{-1}M_{21},
\label{eq:ach-schur-load}
\end{align}
and assume $M_{22}$ is invertible on the domain of interest.
Define
\begin{align}
V=\begin{pmatrix}I
\\
-M_{22}^{-1}M_{21}\end{pmatrix}.
\label{eq:ach-schur-elimination-map}
\end{align}
The internal rows of $MV$ vanish, so
\begin{align}
MV=\begin{pmatrix}\widehat M
\\
0\end{pmatrix},
\quad \operatorname{Re}\widehat M =V^\dagger(\operatorname{Re}M)V\succeq0.
\label{eq:ach-schur-accretive}
\end{align}
Thus Schur complementation preserves accretivity.
In the unitary realization, identifying the internal public output with its input imposes zero on the corresponding component of $\vect u-\vect v$ in \cref{eq:ach-cayley-impedance-equation}.
Eliminating that component gives exactly the Schur complement in \cref{eq:ach-schur-load}.
Hence
\begin{align}
\widehat F=(I-\widehat M)(I+\widehat M)^{-1}
\label{eq:ach-schur-response}
\end{align}
has a finite unitary realization whenever the internal feedback is regular.
No commutativity of the component operators was used.

Positive sums, known congruences, Cayley transforms, cascades, and regular Schur complements therefore generate a broad class of rational transfer functions whose unitary realizations are known by construction.
This class is only sufficient.
For example, the inner response $Y^2$ is obtained by two successive query sections but is not a single linear-fractional Cayley section.
These closure rules make achievability immediate for transfer functions assembled from the listed operations.

These constructive closure rules motivate the modular construction in \cref{sec:modularity}.
A module is specified by a transfer function together with a proved unitary realization, its original oracle ports, its private response, and a decomposition of the known system operations.
Series composition, direct sums, and regular feedback connect these realizations by known unitary operations, so the composed transfer function is achievable without solving a new global completion problem.
For a public cascade, if two modules have transfer functions $F_1,F_2$ and private responses $\Gamma_1,\Gamma_2$, then the steady-state equations give the composed response
\begin{align}
F_{21}=F_2F_1
\label{eq:ach-cascade-response}
\end{align}
and the private vector
\begin{align}
\Gamma_{21}\vect u =\Gamma_1\vect u\oplus\Gamma_2F_1\vect u.
\label{eq:ach-cascade-private}
\end{align}
Hence the catalyst weight is the sum of the first module's weight and the second module's weight evaluated on the transmitted state.
This is the composition rule already derived in \cref{thm:iir-energy}; it preserves the state-dependent query information before any worst-case norm is taken.
Feedback is more delicate because the inverse arising from feedback elimination can move poles, as \cref{eq:ach-schur-load,eq:ach-schur-accretive} make explicit.

Algorithm design can consequently start from the desired transformation and assemble an IIR from components with known realizations and analytic properties.
The assembled response is analyzed and compiled with one clock, without separately approximating every component.
This gives a sufficient design method within the complete finite query model of \cref{thm:ach-analytic-completeness,thm:ach-clock-completeness}, while leaving query optimality to be established for each construction.
The oracle costs, private responses, analytic neighborhoods, and known-gate decompositions remain available for that analysis.

\section{Modular transfer functions}
\label{sec:modularity}

\cref{sec:achievable} characterizes a broad class of achievable transfer functions and shows that finite online query circuits admit analytic MQSP realizations.
That result answers an existence question.
For algorithm design we need a more constructive question: given the oracle access, which transfer function should we build so that its realization and resource properties remain visible?

We use a small construction language rather than a finite catalogue.
A known unitary system matrix with specified oracle feedback connections gives an achievable transfer function immediately.
Series connections, orthogonal branches, basis changes, and regular feedback then combine such realizations without solving another global unitary-completion problem.
Structured choices of the system matrix give familiar modules such as direct oracle-query modules, all-pass sections, Cayley junctions, reflected walks, and state-preparation queries.
The family of possible unitary junctions is infinite; later algorithms may introduce a new one when its couplings expose a useful response or query balance.

The section proceeds in four steps.
We first state the general unitary-junction construction.
We then give a small set of canonical modules for the supplied oracles for the access models used most often in quantum algorithms.
Next we describe the connection rules and the catalyst, analytic, gate, and clock information that they preserve.
Finally we record several reusable IIR modules.
Application- specific modules are introduced later, where the reader can see the property they were designed to enforce.

The corresponding classical constructions use unitary scattering networks, all-pass filters, and wave-digital filters \cite{Darlington1939NetworkSynthesis,Redheffer1962ScatteringTransfer,Regalia1988AllpassFilters,Fettweis1986WaveDigitalFilters}.
We use their transfer-function algebra but keep the complete quantum Hilbert spaces and all complementary outputs.
For a single Hermitian signal, scalar formulas can be checked on eigenspaces.
For several noncommuting signals or a non-normal encoded matrix, the realization and its bounds are established at the operator level before any scalar specialization.

Each named module denotes a transfer function together with one stated unitary realization and its oracle ports.
Values at $\Bz=\Bone$ suppress the delay arguments.
A different realization of the same function can have different query weights or clock behavior, so resource statements always refer to the realization being used.
The classical and MQSP conventions are summarized in \cref{tab:mod-conventions}.

\begin{table}[H]
\centering\small
\setlength{\tabcolsep}{2.2pt}
\begin{tabularx}{\linewidth}{@{}r|Y|Y@{}}
\hline
\hline
Convention & MQSP IIR & Classical IIR
\\
\hline
Delay variable & $z_j$ & $z_{{\rm DSP},j}^{-1}$
\\
Analytic transfer function & Analytic in $\mathbb D^m$ & Analytic for all $|z_{{\rm DSP},j}|>1$
\\
BIBO-stable & No poles in the closed unit disk & All poles in the open unit disk
\\
System matrix $S$ & $\left(\begin{smallmatrix}A&B\\ C&D\end{smallmatrix}\right)$; public output first & $\left(\begin{smallmatrix}D&C\\ B&A\end{smallmatrix}\right)$; output first
\\
Wires & Amplitudes; direct-sum ports & Scalar or vector samples
\\
Junctions & Unitary mixing & Adders, fan-out, lossless mixing
\\
Realization & Unitary; all ports retained & Passive, lossless, or active
\\
Internal state & Private amplitude; catalyst norm & Filter state; internal energy
\\
\hline
\hline
\end{tabularx}
\caption{Differences between MQSP IIR and classical IIR conventions.
The delay coordinates satisfy $z_j=z_{{\rm DSP},j}^{-1}$.
Open-disk analyticity alone allows boundary poles; the separate stability row states the pole condition for bounded-input bounded-output (BIBO) \cite[Chapter 8]{Smith2007DigitalFilters} of a causal rational function of one variable.
The classical matrix is written in output-first order: $(y,x_{\rm next})^{\mathsf T}=S(u,x)^{\mathsf T}$, with the usual state-space letters.}
\label{tab:mod-conventions}
\end{table}

A first reading need not remember every named IIR.
The reusable facts are simpler: choose a unitary junction that directly queries the supplied oracles, connect it using the rules of \cref{subsec:mod-connect}, analyze the resulting complete transfer function, and compile it with one clock.
\cref{sec:algorithms} follows this order repeatedly and reminds the reader which modules are being used at the start of each algorithm.

\subsection{Constructive realizability}
\label{subsec:mod-junctions}

The general achievability results of \cref{sec:achievable} are most useful for algorithm design when they are turned around.
Instead of first proposing a complete function and then searching for a realization, choose an oracle-independent unitary system matrix and let its closed oracle ports define the transfer function.

Let the public and private spaces be $\cP$ and $\cL=\bigoplus_j\cL_j$, and write
\begin{align}
S=\left(\begin{array}{c|c}A&B
\\
\hline
C&D\end{array}\right),
\quad \oracle(\Bz)=\bigoplus_j z_jO_j,
\label{eq:mod-unitary-junction}
\end{align}
where $S$ is known and unitary and $O_j$ is the declared oracle on port $j$.
Closing the private ports gives
\begin{align}
\Gamma(\Bz)=[I-D\oracle(\Bz)]^{-1}C,
\quad F(\Bz)=A+B\oracle(\Bz)\Gamma(\Bz).
\label{eq:mod-response}
\end{align}
The first equation gives the private response and the second gives the public transfer function.

\begin{proposition}[Unitary-junction construction]
\label{prop:mod-unitary-junction}
For every choice in \cref{eq:mod-unitary-junction}, $F(\Bz)$ is analytic for $\max_j|z_j|<1$.
If the response reachable from the public input is regular at $\Bz=\Bone$, then $F(\Bone)$ is unitary and the steady-state private amplitude $\vect v=\Gamma(\Bone)\ket\psi$ satisfies
\begin{align}
S[I\oplus \oracle(\Bone)](\ket\psi\oplus\vect v) =F(\Bone)\ket\psi\oplus\vect v.
\label{eq:mod-junction-catalyst}
\end{align}
Conversely, every finite-dimensional unitary MQSP realization is a junction of this form after collecting its private oracle ports into the direct sum $\cL$.
Hence arbitrary unitary junctions form a complete low-level family for finite-dimensional MQSP realizations.
\end{proposition}
\begin{proof}
For $\max_j|z_j|<1$, $\|D\oracle(\Bz)\|<1$, so the inverse in \cref{eq:mod-response} is its norm-convergent Neumann series.
This proves analyticity and also shows that every coefficient is a sum of online query paths.
At $\Bz=\Bone$, the feedback operation $\oracle(\Bone)$ is unitary.
Substituting the steady-state private solution into the unitary map $S$ gives \cref{eq:mod-junction-catalyst}; equality of the private input and output leaves a norm-preserving public map, so $F(\Bone)$ is unitary on the specified public input space.
The converse is the block decomposition of the system matrix in the definition of a finite unitary realization.
\end{proof}

The available oracle access guides the choice of a structured junction.
We use examples whose transfer functions, private responses, poles, and known-gate decompositions can be analyzed directly.
They provide reusable constructions within the finite-query class of \cref{thm:ach-analytic-completeness}; later applications may choose different couplings when those give a useful response or query balance.

\begin{table}[H]
\centering\small
\setlength{\tabcolsep}{3pt}
\begin{tabularx}{\linewidth}{c|c|X}
\hline
\hline
Oracle access & Suggested module & Implemented signal
\\
\hline
Unitary $O$ & \module{Query} & Full action $zO$
\\
Phase $x=e^{i\phi}$ & $\module{AP1}_a$ & First-order all-pass phase response
\\
Hermitian block encoding & \CayleyIIR & Hermitian load in a unitary rational response
\\
Hermitian block encodings & \WeightedCayleyIIR & Weighted noncommuting sum with separate oracles
\\
Any block encoding $\Be{A/\lambda}$ & Direct projected query & Matrix block $A/\lambda$ without a normality assumption
\\
\multirow{2}{*}{Singular-value transformation} & Hermitian dilation & \multirow{2}{*}{Hermitian or reflected singular-value signal}
\\
& or \ReflectionWalkIIR &
\\
State-preparation unitary $U$ & \PrepReturnIIR & Bidirectional preparation signal using $U$ and $U^\dagger$
\\
Several full oracle Hilbert spaces & Star junction family & Common hub with application-chosen couplings
\\
\hline
\hline
\end{tabularx}
\caption{A convenient non-exhaustive list of canonical choices for the supplied oracles.
The complete low-level family is the arbitrary unitary junction of \cref{prop:mod-unitary-junction}.}
\label{tab:mod-oracle-interfaces}
\end{table}

\IIRAnchor{Query}
The most direct choice is a direct oracle-query module.
Exchanging a public space with a private copy through
\begin{align}
S_{\module{Query}}=\left(\begin{array}{c|c}0&I
\\
\hline
I&0\end{array}\right),
\quad \oracle(z)=zO
\label{eq:mod-query-system}
\end{align}
gives the transfer function $\module{Query}(z;O)=zO$.
It preserves the full oracle action and is the appropriate starting point when no smaller signal needs to be exposed.

For a one-dimensional phase oracle $O(x)=x$ with $|x|=1$, the first-order all-pass section $\module{AP1}_a(z;x)$ of \cref{eq:ach-blaschke,eq:ach-blaschke-system} is another elementary choice.
Its pole and catalyst weight are explicit, so a phase promise can be translated directly into a clock bound.
This is useful when the oracle already supplies a phase rather than a Hermitian matrix element.

For Hermitian block encodings, a Cayley junction is usually more useful than a direct oracle query because its value at $\Bz=\Bone$ is the Cayley transform of a Hermitian matrix while the complete response remains unitary.
The common multioracle construction below also permits known coupling maps and a known Hermitian load.

\subsubsection{A common unitary Cayley junction}
\label{subsec:common-cayley}
A sum of noncommuting Hermitian signals can be inserted into a single Cayley transfer function.
We design its load to have positive semidefinite Hermitian part throughout the open polydisk; the Cayley transformation then makes it contractive.
At $\Bz=\boldsymbol{1}$, the load becomes the desired Hermitian sum multiplied by $i$.
The wave-digital junction \cite{Fettweis1986WaveDigitalFilters} in \cref{lem:common-cayley} realizes that transfer function by enforcing a shared wave amplitude across the parallel ports.

\begin{lemma}[Cayley junction with a known load]
\label{lem:common-cayley}
Let $K_0=K_0^\dagger$ act on the load space.
Couple it to the complete oracle spaces through $C\boldsymbol{u}=\bigoplus_jC_j\boldsymbol{u}$.
With self-inverse unitaries $V_j$, define the load and target transfer function by
\begin{subequations}
\label{eq:imp-response}
\begin{align}
M(\Bz)&=iK_0+\sum_j C_j^\dagger \frac{(1-z_j^2)I+2iz_jV_j}{1+z_j^2}C_j,
\\
F_0(\Bz)&=(I-M(\Bz))(I+M(\Bz))^{-1}.
\end{align}
\end{subequations}
The load has nonnegative Hermitian part in the open polydisk and evaluates to
\begin{align}
M(\Bone)=i\left(K_0+\sum_jC_j^\dagger V_jC_j\right).
\end{align}
It is realized by the unitary system matrix and private feedback connections
\begin{align}
R=(I+C^\dagger C+iK_0)^{-1},
\quad S_0=\left(\begin{array}{c|c} 2R-I&2RC^\dagger
\\
\hline
2CR&2CRC^\dagger-I \end{array}\right),
\quad \oracle(\Bz)=-i\bigoplus_jz_jV_j.
\label{eq:imp-junction}
\end{align}
The public space is the load space, and the private ports are the full oracle Hilbert spaces.
Equivalently, absorb the phase into the private columns of $S_0$ and use $\oracle(\Bz)=\bigoplus_jz_jV_j$.
\end{lemma}
\begin{proof}
For a self-inverse unitary, spectral decomposition reduces each summand of the load to its two eigenvalues.
Their real parts are
\begin{align}
\operatorname{Re}\frac{1+iz_j}{1-iz_j} =\frac{1-|z_j|^2}{|1-iz_j|^2},
\quad \operatorname{Re}\frac{1-iz_j}{1+iz_j} =\frac{1-|z_j|^2}{|1+iz_j|^2}.
\end{align}
They are nonnegative in the open disk, and congruence by the coupling maps preserves positivity.
The real part of $I+C^\dagger C+iK_0$ is positive definite, so $R$ exists.
Its inverse gives
\begin{align}
R+R^\dagger=2R(I+C^\dagger C)R^\dagger,
\label{eq:imp-unitarity}
\end{align}
which proves the four unitary block identities by multiplication.
For incoming public and private vectors $\boldsymbol{u},\boldsymbol{v}$, set $\boldsymbol{h}=R(\boldsymbol{u}+C^\dagger\boldsymbol{v})$.
The outputs are $2\boldsymbol{h}-\boldsymbol{u}$ and $\boldsymbol{w}=2C\boldsymbol{h}-\boldsymbol{v}$.
Substitute $\boldsymbol{v}=\oracle\boldsymbol{w}$ in these equations.
The resulting load is $iK_0+C^\dagger(I-\oracle)(I+\oracle)^{-1}C$.
Using $V_j^2=I$ yields \cref{eq:imp-response} wherever elimination is regular, with removable singularities defined by analytic continuation.
The private phase change multiplies $S_0$ by the known unitary $\operatorname{diag}(I,-iI)$ and leaves the closed network unchanged.
\end{proof}

A transfer function formula alone gives no short known-gate decomposition.
The applications specify the structure of $C,K_0$ and count their implementation gates.
\cref{lem:closed-system} proves this decomposition for the constraint solver; \cref{lem:multi-sos-known-cascade} handles repeated small known systems, including their boundary and private blocks.

\IIRAnchor{WeightedCayley}
For Hermitian terms $H_j$ with self-inverse encodings $V_j=\Be{H_j/\lambda_j}$, set
\begin{align}
\Lambda=\|\vect{\lambda}\|_1,
\quad C_j=\sqrt{\lambda_j/\Lambda}\,|0\rangle_j.
\label{eq:mod-weighted-cayley-couplings}
\end{align}
The Cayley junction of \cref{lem:common-cayley} with $K_0=0$ is the \WeightedCayleyIIR{} module.
At $\Bz=\Bone$ its load is $i\sum_jH_j/\Lambda$, while every original oracle remains on its own private port.
This is the canonical starting point when the algorithm needs a noncommuting Hermitian sum but wants to optimize the ports separately.

The same construction includes more general weighted parallel Cayley junctions: choosing the coupling maps $C_j$ fixes how strongly each full oracle Hilbert space enters the common load.
A star junction is the corresponding topology in which one public space is coupled to several oracle ports through a known unitary scattering matrix.
The couplings need not be fixed universally.
Later algorithms may choose them to balance query weights or to preserve a useful public symmetry; once the displayed star system matrix is unitary, \cref{prop:mod-unitary-junction} already proves achievability.

A non-normal block-encoded matrix should generally be treated through the complete unitary oracle rather than through its eigenvalues.
If $V=\Be{A/\lambda}$, the direct query already satisfies
\begin{align}
F(z)=zV,
\quad \Pi_{\rm out}F(z)\Pi_{\rm in}=zA/\lambda.
\label{eq:mod-direct-block-query}
\end{align}
This is often enough when the surrounding network only needs the prescribed matrix block.
When a Hermitian signal is preferable, a direction qubit gives the self-inverse Hermitian dilation
\begin{align}
J_V=\begin{pmatrix}0&V^\dagger
\\
V&0\end{pmatrix},
\quad\bigl(|0\rangle\langle0|_{\rm d}\otimes\Pi_{\rm in} +|1\rangle\langle1|_{\rm d}\otimes\Pi_{\rm out}\bigr) J_V \bigl(|0\rangle\langle0|_{\rm d}\otimes\Pi_{\rm in} +|1\rangle\langle1|_{\rm d}\otimes\Pi_{\rm out}\bigr) =\frac1\lambda\begin{pmatrix}0&A^\dagger
\\
A&0\end{pmatrix}.
\end{align}
This converts singular-value information into a Hermitian signal without assuming that $A$ is normal.

\IIRAnchor{Chebyshev2}\IIRAnchor{ReflectionWalk} A closely related choice keeps the original block encoding and reflects its signal subspaces.
With input and output signal projectors $\Pi_{\rm in},\Pi_{\rm out}$, one call to each of $V$ and $V^\dagger$ gives
\begin{align}
R=V^\dagger(2\Pi_{\rm out}-I)V,
\quad (\langle0|_{\rm in}\otimes I)R(|0\rangle_{\rm in}\otimes I) =2A^\dagger A/\lambda^2-I.
\label{eq:mod-chebyshev2}
\end{align}
Thus $R$ block-encodes the degree-two Chebyshev transformation on the right singular vectors, including the nullspace.
The product $(2\Pi_{\rm in}-I)R$ defines \ReflectionWalkIIR.
This is a natural signal obtained from the supplied oracle when a later IIR acts on singular values.

State-preparation access has its own canonical bidirectional query.
If $U|0\rangle=|\psi\rangle$ with all required work erased, then
\begin{align}
V_U=\begin{pmatrix}0&U^\dagger
\\
U&0\end{pmatrix}
\label{eq:mod-preparation-query}
\end{align}
is self-inverse, with $U$ and $U^\dagger$ in its off-diagonal blocks, and thus gives a Hermitian signal on a doubled space.
The weighted state-preparation algorithm in \cref{subsec:wsp} uses this construction for separately charged preparation oracles.

These choices cover the forms of oracle access used later: direct unitary queries, phase oracles, Hermitian block encodings, singular-value signals, and state-preparation unitaries.
They do not restrict the theory.
A custom module is admissible whenever it supplies the same data as \cref{prop:mod-unitary-junction}: a known unitary system matrix, specified oracle feedback connections, and the public subspaces used by the surrounding network.

\subsection{Connections and inherited bounds}
\label{subsec:mod-connect}

Once the modules for the supplied oracles have been chosen, the remaining operations are connections between already unitary realizations.
The following rules combine unitary network connections \cite{Redheffer1962ScatteringTransfer} with their quantum transducer compositions \cite{Belovs2024TamingQuantum}.
They preserve the complete complementary outputs, so achievability is inherited rather than re-established for each composite transfer function.

\IIRAnchor{Wire}
\RuleRow{A known public gate, \EWire, changes the transfer function to $FV$ when applied before a module, or to $VF$ when applied after it.
Preparation is the case $V\ket0=\ket\psi$.}{RulePreparation}

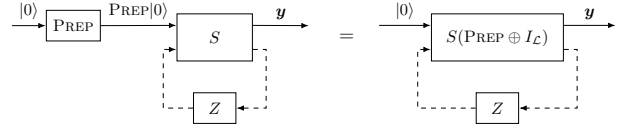
\IIRAnchor{Series}
\RuleRow{For series composition, \ESeries, connect one public output to the next public input.
The cascade transfer function is $F_2F_1$.
It acts on public space $\cP$ and uses the direct-sum private space $\cL_1\oplus\cL_2$.
Extend each known system by identity on the other's private space and multiply in execution order.
This specifies the full unitary, including every complementary output.}{RuleSeries}

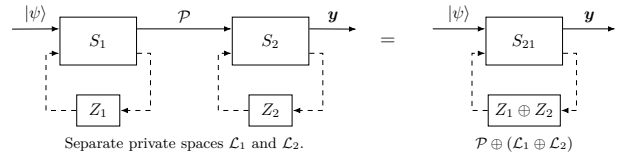
\IIRAnchor{DirectSum}
\RuleRow{For a direct sum, \EDirectSum, use orthogonal branches.
The transfer function is $F_1\oplus F_2$, acting on the direct sum of the public spaces with the direct sum of the private spaces.
A known unitary may prepare and recombine branch amplitudes.
For example, prepare and project the branch flag onto $\ket+$.
Acting between these maps, $F_1\oplus F_2$ block-encodes $(F_1+F_2)/2$; the complementary output remains part of the circuit.}{RuleDirectSum}

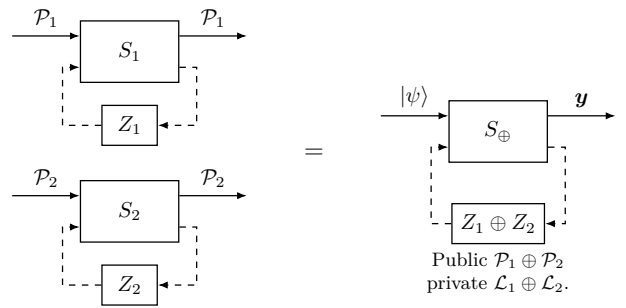
\IIRAnchor{Spectator}
\RuleRow{A register carried through unchanged gives \ESpectator, with transfer function $I_{\mathcal R}\otimes F$.
Tensor the original public and private spaces with that register.
The rule also holds for entangled inputs.
It multiplies Hilbert-space dimensions and preserves the worst-case catalyst weight on normalized inputs.}{RuleSpectator}

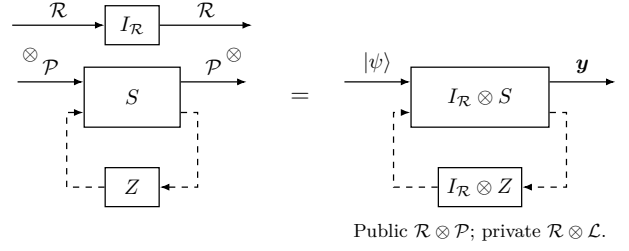
\IIRAnchor{Close}
\RuleRow{To close a feedback loop, \EClose, the output in public sector $\mathcal E$ is fed back to its input through $wV$.
The subspace $\mathcal E$ is moved from the public port into memory.
The resulting spaces are $\text{public space }\cP$ and private space $\mathcal E\oplus\cL$.
If $F$ is partitioned by these public sectors, elimination gives $F_{pp}+F_{pe}wV(I-F_{ee}wV)^{-1}F_{ep}$.
The new inverse must exist on the stated analytic domain.}{RuleClose}

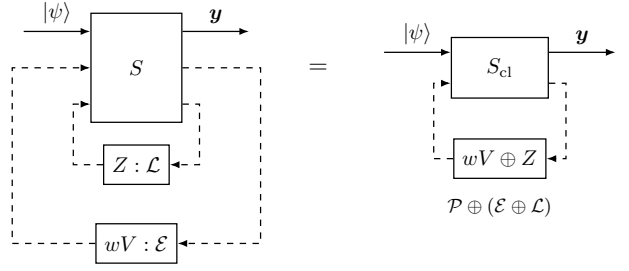
\IIRAnchor{Substitute}
\RuleRow{For module substitution, \ESubstitute, implement an outer feedback connection by an inner transfer function $F_B$.
Its public space $\cP_B$ becomes private relative to the outer module, together with $\cL_B$.
Keeping an analytic variable $w$, the transfer function is $A_A+B_AwF_B(I-D_AwF_B)^{-1}C_A$.
The system is $(S_A\oplus I)(I\oplus S_B)$ in public-first order.}{RuleSubstitute}

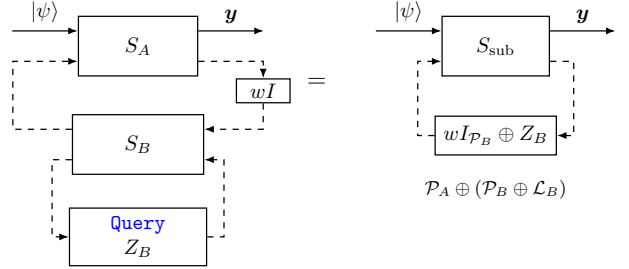
\IIRAnchor{BufferedSeries}
\RuleRow{Storing a connecting wire gives \EBuffered: it adds one copy of its Hilbert space to private space and a marker $w$ to that connection.
Thus buffered series has transfer function $wF_2F_1$ and private space $\cL_1\oplus\cP\oplus\cL_2$.}{RuleBufferedSeries}

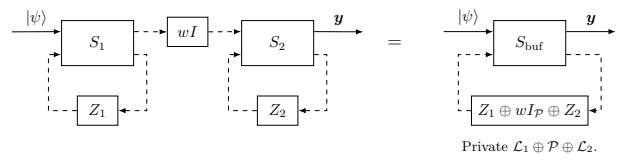
\IIRAnchor{Delay}
\RuleRow{More generally \EDelay\ replaces $zO_j$ by $z^{r_j}O_j$.
It uses $r_j$ storage sectors and one oracle per completed traversal.
At horizon $N$, quditization makes $q_j=\lfloor(N-1)/r_j\rfloor$ oracle calls.}{RuleDelay}

\IIRAnchor{Inverse}
\RuleRow{For the adjoint transformation, \EInverse, reverse the known system and use inverse oracle feedback connections.
At $\Bz=\boldsymbol{1}$ this implements $U^\dagger$ with the same catalyst norm on the corresponding output state.
The new causal transfer function must be derived from the reversed system; $F(z)^{-1}$ may have poles in the disk.}{RuleInverse}

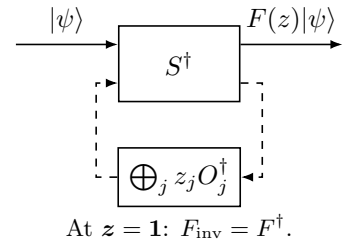
\IIRAnchor{Project}
\RuleRow{For the projection \ESelect, specify the input and output ancilla states: $(\bra0_b\otimes I)V_{\rm out}^\dagger F V_{\rm in}(\ket0_b\otimes I)$.
Known changes of basis cover any given finite-dimensional isometry after padding.
Their gates must be counted.
This gives the corresponding block-encoding, with all signal ancillas included as in \cref{def:resource-accounting}.}{SelectedBlock}

\subsubsection{Catalyst, analytic, and resource bounds}
\label{subsec:mod-certificates}
Write $U_i=F_i$ and $W_i=\Gamma_i^\dagger\Gamma_i$.
For series and branch composition the accounting is exact:
\begin{align}
W_{21}=W_1+U_1^\dagger W_2U_1,
\quad W_{\oplus}=W_1\oplus W_2.
\label{eq:mod-catalyst-compose}
\end{align}
Thus probability $p$ on a branch incurs $p$ times that branch's squared catalyst norm.
This thrifty accounting applies coherently within a single circuit execution.
For feedback, if $\boldsymbol{h}=(I-F_{ee}V)^{-1}F_{ep}\ket\psi$, the exact cost is $\norm{\boldsymbol{h}}^2+\norm{\Gamma(\ket\psi\oplus V\boldsymbol{h})}^2$.
Its input to the old module is generally unnormalized.

The catalyst identity for a cross-product of oracles of \cref{sec:quditization} gives state-conversion adversary constraints; optimizing over feasible conversions yields query lower bounds \cite{Belovs2023LasVegas}.
The pure-delay realization $F(z)=z^kI$ has catalyst weight $k$.
Its value at the evaluation point is the identity, which also admits a circuit with no queries.
This shows why the choice of realization matters.
A chosen realization gives exact network accounting; a query lower bound follows from the optimized state-conversion constraints.

Consider a public cascade on a common neighborhood $|z-1|\le\zeta$.
Suppose each transfer function $F_i$ is contractive in the open unit disk and obeys the two-sided bounds
\begin{align}
\norm{F_i(z)}\le M_i \begin{cases}|z|^{a_i-h_{i,-}},&|z|\le1,
\\
|z|^{w_i+h_{i,+}},&|z|\ge1. \end{cases}
\label{eq:mod-tag}
\end{align}
For a cascade use the smallest neighborhood, multiply the $M_i$, and add $a_i,w_i,h_{i,\pm}$.
The catalyst-weight bounds are $a_i,w_i$.
The radial excesses $h_{i,\pm}$ and prefactor $M_i$ bound growth beyond that interval.
\cref{eq:mod-catalyst-compose} justifies the derivative bounds.
For branches use the maximum norm and the minimum/maximum derivative bounds across the retained sectors.
If two implementations are contractions on the stated domain, $\norm{\widetilde F_2\widetilde F_1-F_2F_1} \le\norm{\widetilde F_2-F_2}+\norm{\widetilde F_1-F_1}$ there.
Outside it, the corresponding norm factors must be retained.

\RuleRow{Feedback can invalidate a large analytic neighborhood even when the constituent transfer functions are entire.
A unitary tap with $0<r<1$ and feedback connection $w=z^2$ has
\begin{align}
F(z)=\frac{z^2-r}{1-rz^2},
\quad W=\frac{2(1+r)}{1-r}.
\end{align}
As $r\to1$, the pole $z=r^{-1/2}$ approaches the evaluation point.
The analytic neighborhood must therefore exclude the zero of $1-rz^2$.}{FeedbackExample}

For one Hermitian $A$, $\norm{p(A)}=\max_{x\in\operatorname{spec}A}|p(x)|$.
For a contraction substituted into a scalar Schur function, the one-variable contraction functional calculus is another valid route.
For several noncommuting matrices, an operator estimate is needed: the scalar polynomial $xy-yx$ is identically zero, while $XY-YX=2iZ$ for Pauli matrices.
Even for one variable, $p(x)=x$ is bounded on $[-1,1]$ but can have arbitrarily large norm on a nonnormal matrix whose eigenvalues all lie in that interval.
The lift therefore uses an invariant-subspace argument or a bound in operator norm.

For Hamiltonian simulation below, the group-delay operator is $W=\lambda|t|I$.
The normalization is $\lambda$ and the evolution time is $t$.
A common clock translation therefore absorbs the whole leading delay.
A one-sided bound discards the lower group-delay bound and increases the leading coefficient.
Analyzing the complete \ExpIIR--\CayleyIIR\ transfer function recovers it.
Finally, optimizing the clock matrix under the nuclear-norm constraint from \cref{sec:quditization} cannot worsen an analytic feasible solution at fixed horizon and normalization.
A numerical optimization can improve this bound; its result must satisfy the same clock-norm and continuum-error constraints.

\label{subsec:mod-ledger}
For each assembled network, \cref{thm:iir-gates} gives the gate and qubit counts from its public and complete private dimensions, gate cost of the known system matrix, oracle costs, and clock preparation.
An unchanged system register can be factored out before counting routing addresses.
The circuit retains all block-encoding ancillas, oracle work registers, and modes of the exponential realization.
The error budget is \cref{eq:qudit-error-budget}: a finite Toeplitz replacement is counted once, while an operation error is multiplied by its actual number of executions.
The examples below specify these construction-dependent inputs.

\subsection{Reusable IIRs for different applications}
\label{subsec:mod-IIR modules}
The preceding constructions give unitary junctions and connection rules.
We now combine them into several IIRs used in the algorithms below.
They are examples rather than an exhaustive library.
In \cref{sec:algorithms}, a custom module is introduced where its design objective first becomes clear and is reduced there to the unitary junctions and connections above.
Its local proof then concentrates on the response, query weights, analytic domain, and circuit cost that are specific to that application.

\begin{table}[H]
\centering\small
\setlength{\tabcolsep}{2.2pt}
\begin{tabularx}{\linewidth}{@{}YccY@{}}
\hline
\hline
Task & MQSP queries & QSP/QSVT comparison & Classical modules
\\
\hline
FPAA / OAA$^*$ \cite{Yoder2014} & $\frac{\log(1/\epsilon)}{2\artanh\lambda_0}+O_{\lambda_0}(\log\log(1/\epsilon))$ & $\frac{\log(1/\epsilon)}{2\artanh\lambda_0}+O_{\lambda_0}(1)$ & All-pass biquad; lossless tap \cite{Regalia1988AllpassFilters}
\\
\hline
Thresholding$^*$ & $\frac{\log(1/\epsilon)}{\artanh\lambda_0}$ & $\frac{\log(1/\epsilon)}{\artanh\lambda_0}$ & Schur lattice; Geronimus
\\
\cite{Low2017USA,Gilyen2018singular,Eremenko2011Sign}&$+O_{\lambda_0}(\log\log(1/\epsilon))$&$+O_{\lambda_0}(\log\log(1/\epsilon))$&\cite{Simon2005OrthogonalPolynomials,Simanek2018GeronimusWeights}
\\
\hline
Hamiltonian simulation$^*$ & $(1+\delta)\lambda|t|$ & $(1+\delta)\lambda|t|$ & All-pass section; bilinear
\\
\cite{Low2016HamSim,Low2016Qubitization}&$+O_\delta(\log(1/\epsilon))$&$+O_\delta(\log(1/\epsilon))$&delay \cite{Fettweis1986WaveDigitalFilters,Smith2010PhysicalAudioSignal}
\\
\hline
Weighted-cost & $(1+\delta)|t|\langle\boldsymbol{C},\boldsymbol{\lambda}\rangle_{1/2}$ & $(1+\delta)|t|\|\boldsymbol{\lambda}\|_1\|\boldsymbol{C}\|_1)$ & Parallel loads; wave digital
\\
simulation$^*$ \cite{Low2016Qubitization}&$+O_\delta(\|\boldsymbol{C}\|_1\log(1/\epsilon))$ &$+O_\delta(\|\boldsymbol{C}\|_1\log(1/\epsilon)$&junction \cite{Fettweis1986WaveDigitalFilters} \\ \hline State preparation$^*$ & $O(\langle\boldsymbol{C},\boldsymbol{\lambda}\rangle_{1/2}/s)$ & \multirow{2}{*}{$O(\|\boldsymbol{\lambda}\|_1\|\boldsymbol{C}\|_1/s)$} & Parallel loads; bilinear\\ \cite{Childs2012LCU,brassard2002quantum}&$+O_\sigma(\|\boldsymbol{C}\|_1\log(1/\epsilon))$&& delay \cite{Fettweis1986WaveDigitalFilters,Smith2010PhysicalAudioSignal} \\ \hline\hline
\end{tabularx}
\caption{Reusable IIR examples and query upper bounds; success conventions and sharper estimates are given in the corresponding subsections.
References in the `Task' column give comparison benchmarks; state preparation uses LCU and amplification.
The initial-amplitude bound or normalized threshold gap is $\lambda_0$; the Hamiltonian normalization is $\lambda$.
Simulation rows use a fixed leading-query allowance $\delta>0$ and an additive logarithmic precision bound.
State preparation assumes a given superposition norm $s$ and fixed simulation normalization $1+\sigma>1$.
Asterisks mark uses in later applications.}
\label{tab:mod-algorithms}
\end{table}

\begin{figure}[H]
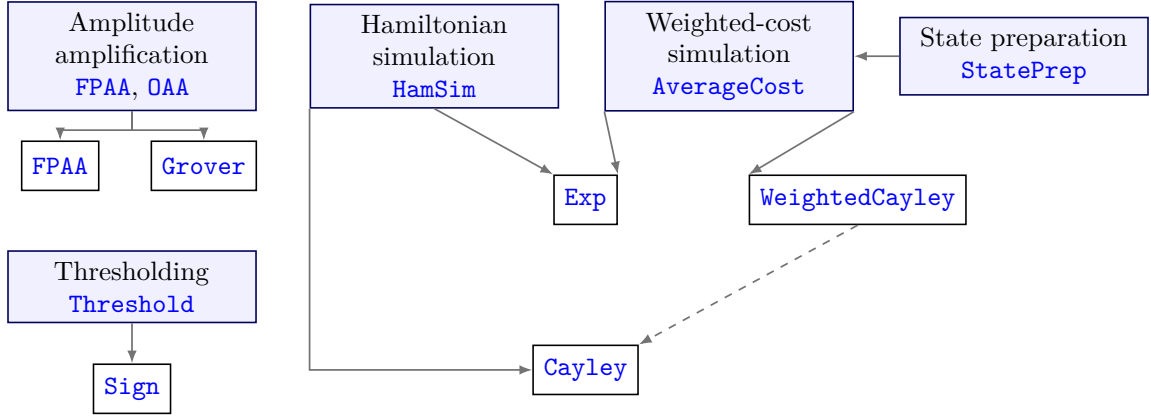

\centering\IIRDiagram{Dependencies}
\caption{Examples of shared processing modules.
An arrow points from a construction to a module it uses.
Shaded boxes name the algorithms and their modules; all module names link to their definitions.
The dashed edge denotes the weighted version of the same Cayley transformation.
Elementary connections are shown in the later IIR diagrams.}
\label{fig:mod-dependencies}
\end{figure}

\subsubsection{Fixed-point amplitude amplification}
\label{subsec:mod-fpaa}\IIRAnchor{FPAA}

The desired response should suppress the bad component for every $\lambda\ge\lambda_0$ while preserving the phase of the good component.
A first-order all-pass tap is a natural module for the preparation oracle because its unitary realization is explicit and its pole controls the damping rate.
Closing and cascading these taps gives the \module{FPAA} response of \cref{eq:sp-fpaa-response,eq:sp-fpaa-system}.
Write $P\ket0=\lambda\ket{\psi_{\rm g}}\ket1_a+ \sqrt{1-\lambda^2}\ket{\psi_{\rm b}}\ket0_a$; a known flag flip identifies this good subspace with the introduction's signal subspace.
The refinement below retains the full promised range and obtains the leading preparation coefficient in \cref{prop:mod-fpaa}.

The damping bound is symmetric in the Grover signal about zero.
We first rescale the success amplitude to $\mu=\lambda/\sqrt{1+\lambda_0^2}$.
Its endpoint squares sum to one, so $2\mu^2-1$ has a symmetric promised interval.
The required known flag rotation is
\begin{align}
W_{\lambda_0}\ket{00}_{ad}=\ket{00}_{ad},
\quad W_{\lambda_0}\ket{10}_{ad} =\frac{\ket{10}_{ad}+\lambda_0\ket{01}_{ad}}{\sqrt{1+\lambda_0^2}}.
\end{align}
Complete the second column orthonormally.
Write $\widehat P=W_{\lambda_0}P$ and $\mu=\lambda/\sqrt{1+\lambda_0^2}$ for the new good amplitude.
The new endpoints satisfy
\begin{align}
\mu_{\min}^2=\frac{\lambda_0^2}{1+\lambda_0^2},
\quad \mu_{\max}^2=\frac1{1+\lambda_0^2},
\quad \mu_{\min}^2+\mu_{\max}^2=1.
\end{align}
Hence the transformed Grover signal has a symmetric range.
This adjustment uses no additional oracle call.
A small offset from critical damping bounds the condition number of the private quadratic metric in \cref{app:mod-fpaa}.
For $0<\epsilon<e^{-2}$, choose
\begin{align}
c=\exp\!\left[-4\artanh\lambda_0 \left(1-\frac1{\log(1/\epsilon)}\right)\right],
\quad \chi=\frac{2\lambda_0(1+c)+(1+\lambda_0^2)(1-c)} {2\lambda_0(1+c)-(1+\lambda_0^2)(1-c)}.
\label{eq:mod-fpaa-choice}
\end{align}
The positive offset $1/\log(1/\epsilon)$ avoids the repeated-pole endpoint of the critically damped example in \cref{sec:intro-simple-proofs}.

\begin{figure}[ht]
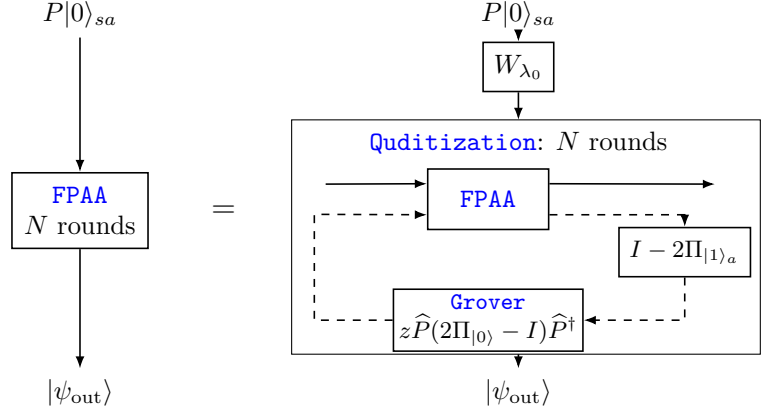

\centering\IIRDiagram[.78\linewidth]{FPAA}
\caption{The same tap and preparation reflection define an IIR and a finite preparation algorithm.
The lift receives the prepared state in its private subspace, initially empty public slots retain every transmitted amplitude, and the final success flag selects their direct sum.}
\label{fig:mod-fpaa}
\end{figure}

\IIRAnchor{TapFPAA}\IIRAnchor{Grover} Use the tap system matrix of \cref{eq:sp-fpaa-system} with the chosen $c$, and connect feedback through the preparation reflection followed by the known signal reflection:
\begin{align}
O_\theta=-\widehat P(2\ket0\bra0-I)\widehat P^\dagger (2\ket1\bra1_a\otimes I-I),
\quad \oracle(z)=zO_\theta.
\label{eq:mod-fpaa-system}
\end{align}
Each traversal uses one $P$ and one $P^\dagger$.
Eliminating the same private amplitudes, now at the reduced amplitude $\mu$, gives
\begin{align}
\module{FPAA}(z;\mu,c)=g(z;\mu)=\frac{c+(1+c)(2\mu^2-1)z+z^2} {1+(1+c)(2\mu^2-1)z+cz^2}.
\label{eq:mod-fpaa-biquad}
\end{align}
On the public target direction the transfer function is $g$; on its orthogonal complement it is $-(z-c)/(1-cz)$.
Thus $F=2|\psi_{\rm g}\rangle\langle\psi_{\rm g}|-I$, with $\module{FPAA}(\mu,c)=1$.
For state preparation, initialize the finite lift with $\widehat P\ket0$ in its private register and collect the transmitted amplitudes in orthogonal clock states.
The finite tap and its retained output slots are shown in \cref{fig:mod-fpaa}.

\begin{proposition}[FPAA with the sharp leading preparation coefficient]
\label{prop:mod-fpaa}
In the parameter range $0<\lambda_0<1$ and $0<\epsilon<e^{-2}$, choose $N=\lceil\log(\chi/\epsilon)/\log(1/c)\rceil$.
The finite construction has failure probability at most $\epsilon$.
Its preparation counts are $Q_P=N+1$ and $Q_{P^\dagger}=N$, with
\begin{align}
Q_P+Q_{P^\dagger}\le \frac{\log(1/\epsilon)+\log\log(1/\epsilon)} {2\artanh\lambda_0}+O_{\lambda_0}(1).
\label{eq:mod-fpaa-sharp}
\end{align}
On success the output is the exact good state tensored with a normalized time record.
The gate and qubit bounds are collected in \cref{eq:mod-fpaa-resources}.
\end{proposition}
\begin{proof}
We bound the probability left in the private register using a positive quadratic form that contracts by $c$ each round.
In bad/good order the private update and transmitted row are
\begin{align}
K=\begin{pmatrix}1-2\mu^2&-2c\mu\sqrt{1-\mu^2}
\\
2\mu\sqrt{1-\mu^2}&c(1-2\mu^2)\end{pmatrix},
\quad L=-\sqrt{1-c^2}(0\ \ 1).
\end{align}
They satisfy $K^\dagger K+L^\dagger L=I$.
The positive metric in \cref{app:mod-fpaa} satisfies $K^\dagger MK=cM$ and $\operatorname{cond}M\le\chi$.
Hence the private residual after $N$ rounds has squared norm at most $\chi c^N$.
Orthogonal output slots retain the rest of the probability.
The chosen damping gives
\begin{align}
\chi=\frac{\sinh[(4-2/\log(1/\epsilon))\artanh\lambda_0]} {\sinh[2\artanh\lambda_0/\log(1/\epsilon)]},
\quad \log\chi=\log\log(1/\epsilon)+O_{\lambda_0}(1).
\end{align}
Each round makes one query in each direction and uses binary routing; the initial preparation contributes the remaining call.
\end{proof}

The same calculation gives analytic bounds: after removing undriven private modes, $\norm{K^j}\le\sqrt\chi c^{j/2}$ implies a resolvent bound on $|z|<c^{-1/2}$.
The public catalyst operator is
\begin{align}
W=\frac{1-c}{(1+c)\mu^2}\ket{\psi_{\rm g}}\bra{\psi_{\rm g}} +\frac{1+c}{1-c}(I-\ket{\psi_{\rm g}}\bra{\psi_{\rm g}}).
\label{eq:mod-fpaa-weight}
\end{align}
The local growth parameters in \cref{app:mod-fpaa} bound the public reflection through \cref{thm:iir-analytic-clock-selection}.
The state-preparation bound in \cref{prop:mod-fpaa} uses the known private input and the full unitary lift directly.
The optimal Chebyshev FPAA sequence \cite{Yoder2014} has $\log(1/\epsilon)/(2\artanh\lambda_0)+O_{\lambda_0}(1)$ preparation calls.
Thus the leading constant agrees, whereas our displayed sufficient bound pays the additional $\log\log(1/\epsilon)$ term.
For small $\lambda_0$, its coefficient is $1/(2\lambda_0)-\lambda_0/6+O(\lambda_0^3)$.

\subsubsection{Oblivious amplitude amplification}
\IIRAnchor{OAA}
When the success amplitude is input independent, \cref{cor:mod-oaa} applies the same amplification to an entire isometry.

\begin{corollary}[A block proportional to an isometry]
\label{cor:mod-oaa}
Suppose a unitary $A$ block-encodes a scalar multiple of an isometry, $\lambda V$, $V^\dagger V=I$, and $\lambda\ge\lambda_0$.
The scalar is the same for every input.
Replace the preparation reflection above by $A(2\Pi_{\rm in}-I)A^\dagger$, and tap the known output signal projector.
Then \cref{prop:mod-fpaa} applies uniformly to every input, including one entangled with a reference.
The successful output is $V\ket\psi$ with a time record independent of $\ket\psi$.
\end{corollary}
The proof is the same two-dimensional calculation on each singular subspace.
This is the oblivious setting of amplitude amplification \cite{Berry2015Truncated,Gilyen2018singular}.
If the singular values vary, their time records generally vary too; tracing them out can decohere the data.
The proportional-isometry hypothesis is therefore part of the module specification.
The gate and qubit counts in \cref{eq:mod-fpaa-resources} apply with $P=A$ and $G_0$ replaced by the cost of reflecting the input signal subspace.
At amplitude $1/2$, the usual oblivious-amplification step uses three forward or inverse calls.
It has constant extra workspace and needs no FPAA record.

\subsubsection{The sign function and a movable threshold}
\label{subsec:mod-sign}\IIRAnchor{Sign}

The sign transformation preserves eigenvectors and places a minus sign on the negative spectrum.
With a Hadamard-controlled use it gives the two spectral projectors, making it a component of energy filtering.
Suppose we can make controlled queries to a self-inverse block-encoding $O=\Be{A}$ of a Hermitian contraction $A$.
The input is supported on eigenspaces with $|A|\ge\lambda_0>0$.

Start with one qubit, $O(x)=\left(\begin{smallmatrix}x&\sqrt{1-x^2}\\ \sqrt{1-x^2}&-x\end{smallmatrix}\right)$.
A scalar coefficient in a larger Hilbert space means $xI$.
Feeding the second output back through a scalar transfer function $w$ gives $(x+w)/(1+xw)$.
Repeating the same unitary section with one delay gives the constant-parameter Schur lattice \cite{Yagle1985SchurAlgorithm,Simon2005OrthogonalPolynomials,Simanek2018GeronimusWeights}:
\begin{align}
f=\frac{x+zf}{1+xzf},
\quad \module{Sign}(z;x)=F_{\rm s}(z;x)=zf =\frac{z-1+\sqrt{(1-z)^2+4x^2z}}{2x}.
\label{eq:mod-sign-response}
\end{align}
At the evaluation point, a finite cascade can be understood by writing each signal as a hyperbolic tangent.
The addition formula gives
\begin{align}
x=\tanh\alpha,
\quad w=\tanh\beta,
\quad \frac{x+w}{1+xw}=\tanh(\alpha+\beta).
\label{eq:mod-sign-hyperbolic}
\end{align}
Thus each section adds the same hyperbolic angle.
Starting from zero and using lattice path length $L$ gives
\begin{align}
r_L(1;x)=\tanh(L\operatorname{artanh}x),
\quad |r_L(1;x)-\operatorname{sgn}x|\le2e^{-2L\lambda_0}
\quad(|x|\ge\lambda_0).
\label{eq:mod-sign-convergence}
\end{align}
The endpoint cases follow by continuity.
This explains the sign limit of the infinite lattice; \cref{app:mod-sign-path} derives the full two-endpoint transfer function.
In \cref{eq:mod-sign-response}, the square root equals one at $z=0$.
The transfer function is contractive in the unit disk.
At the evaluation point, $\module{Sign}(x)=\sgn x$ for $x\ne0$.
This is the classical Geronimus transfer function followed by one delay.

\begin{figure}[ht]
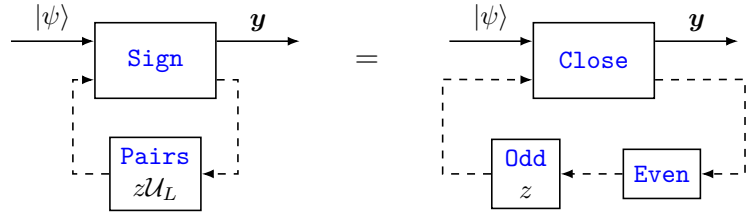

\centering\IIRDiagram[.95\linewidth]{Sign}
\caption{The sign lattice is a path of unitary sections.
The two endpoints are public.
A known boundary swap closes every internal coordinate, and the feedback connection applies the even and odd matchings.
The marker $z$ counts a whole two-query round.}
\label{fig:mod-sign}
\end{figure}

\IIRAnchor{Even}\IIRAnchor{Odd}\IIRAnchor{Pairs} Use a finite path with $L+1$ sites, $L$ even.
Let $\Pi=|0\rangle\langle0|_a\otimes I$ on the full oracle Hilbert space $\mathcal K$.
Even sites carry $\Pi\mathcal K$ and odd sites carry $(I-\Pi)\mathcal K$.
The following maps specify the registers and the oracle action.
For $\boldsymbol{\psi}\in\mathcal K$, define
\begin{subequations}
\label{eq:mod-sign-matchings}
\begin{align}
E_k\boldsymbol{\psi}&=|2k\rangle\Pi\boldsymbol{\psi} +|2k+1\rangle(I-\Pi)\boldsymbol{\psi},
\quad U_{\rm even}E_k\boldsymbol{\psi}=E_kO\boldsymbol{\psi},
\\
D_k\boldsymbol{\psi}&=|2k+2\rangle\Pi\boldsymbol{\psi} +|2k+1\rangle(I-\Pi)\boldsymbol{\psi},
\quad U_{\rm odd}D_k\boldsymbol{\psi}=D_k[-(2\Pi-I)O(2\Pi-I)]\boldsymbol{\psi}.
\end{align}
\end{subequations}
Here $0\le k<L/2$; each matching acts as the identity on its unpaired endpoint.
The isometries $E_k,D_k$ identify disjoint full oracle Hilbert spaces.
A binary pair-address register therefore implements each matching with one controlled $O$, and their product, \PairsIIR, uses two oracle calls.
The address changes are known gates counted by \cref{def:resource-accounting}.
This alternating-reflection quantum path is closely related to the space-efficient purifier of Belovs and Jeffery \cite{Belovs2026SpaceEfficientPurification}; the clock below selects a polynomial window of its classical transfer function.
The alternating lattice and its public endpoints are shown in \cref{fig:mod-sign}.

The public space has two endpoint copies of $\mathcal H$.
Its known address map is
\begin{align}
J\bigl(|0\rangle_{\rm end}|\psi\rangle +|1\rangle_{\rm end}|\phi\rangle\bigr) =|0\rangle_{\rm path}(|0\rangle_a|\psi\rangle) +|L\rangle_{\rm path}(|0\rangle_a|\phi\rangle).
\label{eq:mod-sign-endpoints}
\end{align}
With private path space $\cL_L$, the known boundary system is $S=\left(\begin{smallmatrix}0&J^\dagger\\ J&I-JJ^\dagger\end{smallmatrix}\right)$.
Write $U_{\rm even},U_{\rm odd}$ for these matchings.
The feedback operation is $zU_{\rm odd}U_{\rm even}$.
Take the lattice path length $L=2N$.
Its first $N$ coefficients then agree exactly with $\diag(F_{\rm s},-F_{\rm s})$, because transmission cannot reach the opposite endpoint before that degree.
Thus \cref{prop:iir-comparison-stability} incurs zero error from the finite transfer function at this horizon.
For any $O$ satisfying the block-encoding promise, the invariant plane generated by an eigenvector of $A$ has exactly the displayed one-qubit section.
All unused complementary directions are retained.

On the promised spectral subspace and the neighborhood $|z-1|\le\lambda_0/16$, the classical formula and \cref{app:mod-sign-path} give
\begin{align}
F_{\rm s}^\dagger F'_{\rm s}=\frac{I+|A|^{-1}}2,
\quad \zeta=\lambda_0/16,
\quad M_{\rm loc}=1,
\quad \norm{F_{\rm s}(z;A)}\le |z|^{2/\lambda_0}
\quad(|z|\ge1).
\label{eq:mod-sign-tag}
\end{align}
Together with disk contractivity, these bounds meet the hypotheses of \cref{thm:iir-analytic-clock-selection} with $a=h_-=0$, $w=(1+\lambda_0^{-1})/2$, and $h_+=2/\lambda_0-w$.
The restricted-input result, \cref{cor:iir-task-restricted}, applies on the same spectral subspace.
They give a uniform $O(\lambda_0^{-1}\log(1/\epsilon))$-query result with normalization two.
The fixed-gap polynomial construction in \cref{prop:mod-sign}, proved in \cref{app:improved-polynomials}, improves the count to $[\log(1/\epsilon)+\tfrac12\log\log(1/\epsilon)]/ \artanh\lambda_0+O_{\lambda_0}(1)$ at normalization two.
It matches the leading constant of the QSVT sign approximation \cite{Gilyen2018singular}; the appendix gives the polynomial and its clock-norm bound.

\IIRAnchor{Threshold}
A threshold at $E_*$ is obtained by LCU with the identity:
\begin{align}
\frac{H-E_*I}{\lambda+|E_*|} =\frac{\lambda}{\lambda+|E_*|}\frac H\lambda +\frac{|E_*|}{\lambda+|E_*|}[-\operatorname{sgn}(E_*)I].
\label{eq:mod-threshold}
\end{align}
The standard LCU block-encoding \cite{Childs2012LCU} uses one call to $\Be{H/\lambda}$ and one mixing ancilla.
Applying \SignIIR\ with gap $\lambda_0=\Delta/(\lambda+|E_*|)$ defines \ThresholdIIR; at $E_*=0$ the additional branch is omitted.

\subsubsection{Hamiltonian simulation by an exponential transform}
\label{subsec:mod-ham}\IIRAnchor{HamSim}

Let $O=\Be{H/\lambda}$ be a self-inverse block-encoding with controlled oracle access, and put $T=\lambda|t|$.
We give the construction for positive time.
Conjugating the circuit phases handles negative time; zero time needs no queries.
We use the Cayley system matrix and exponential transform already constructed in \cref{eq:sp-cayley-warmup-transfer,eq:sp-exp-system}.
Their transfer functions \cref{eq:sp-ham-response} compose to
\begin{align}
\module{HamSim}_T(z;x)=F_t(z;x) =\module{Exp}_T(\module{Cayley}(z;x)) =\exp\!\left[-T\frac{1-z^2+2izx}{1+z^2}\right],
\quad F_t(x)=e^{-iTx}.
\label{eq:mod-ham-response}
\end{align}
\IIRAnchor{Cayley}\IIRAnchor{Exp} The scalar system in \cref{eq:sp-cayley-warmup-transfer} lifts to the full oracle Hilbert space by replacing its signal coordinate by $|0\rangle_a\otimes I$:
\begin{align}
S_{\module{Cayley}}=\left(\begin{array}{c|c}0&-i(\bra0_a\otimes I)
\\
\hline
\ket0_a\otimes I&i(I-\Pi)\end{array}\right),
\quad \Pi=\ket0\bra0_a\otimes I,
\quad \oracle=zO.
\label{eq:mod-cayley-system}
\end{align}
The unknown Hamiltonian enters only through this feedback oracle.
The finite realization in \cref{thm:exp-implementation} uses the fused system matrix below.
\cref{prop:mod-ham} sharpens the introduction's clock bound by using the exactly common group delay.

For a finite realization $S_{\module{Exp}}=\left(\begin{smallmatrix}a&b\\ d&D_{\module{Exp}}\end{smallmatrix}\right)$, the explicit fused system is
\begin{align}
S_{\rm ham}=\left(\begin{array}{c|c} aI&-ib\otimes(\bra0_a\otimes I)
\\
\hline
d\otimes(\ket0_a\otimes I)&-iD_{\module{Exp}}\otimes\Pi+iI_D\otimes(I-\Pi) \end{array}\right),
\quad \oracle=z(I_D\otimes O).
\label{eq:mod-ham-system}
\end{align}
It applies the known exponential-transform system to the public port and private signal sectors, and phases to their complement.
These operations act on orthogonal subspaces, so the full matrix is unitary.
The unknown Hamiltonian enters through the block-encoding in the feedback connection.

\begin{proposition}[Analytic simulation with leading time coefficient one]
\label{prop:mod-ham}
For $0<\epsilon\le\delta\le1/4$, the transfer function in \cref{eq:mod-ham-response} has a finite MQSP implementation of $\Be{\widetilde G/(1+\delta)}$.
It satisfies $\norm{\widetilde G-e^{-itH}}\le\epsilon$ and has query bound
\begin{align}
q_O\le T+O\!\left(\delta^{-1/2} [\sqrt{T\log(1/\epsilon)}+\log(1/\epsilon)+1]\right).
\label{eq:mod-ham-seed}
\end{align}
Absorbing the square-root term into the leading term $(1+\delta)T$ gives $q_O\le(1+\delta)T+O(\delta^{-2}\log(1/\epsilon))$.
\end{proposition}
\begin{proof}
We bound the growth of the complete exponential transfer function near $z=1$.
Its radial derivative gives the leading time cost, while its growth on a surrounding neighborhood controls the precision overhead.
Write $z=e^{u+iv}$ and $M(z;x)=(1-z^2+2izx)/(1+z^2)=-\tanh(u+iv)+ix\sech(u+iv)$.
For $|u|,|v|\le s\le1/8$ and real $|x|\le1$, direct calculation gives
\begin{align}
\frac{-\operatorname{Re}M(z;x)}u =\frac{\sinh u}{u}\frac{\cosh u-x\sin v} {\sinh^2u+\cos^2v} \in[1-4s,1+4s].
\label{eq:mod-ham-radial}
\end{align}
On $|z-1|\le s/4$, the two-sided exponents are $T(1\pm4s)$.
The quotient at $u=0$ is defined by continuity.
The group delay is scalar, so \cref{thm:iir-analytic-clock-selection} applies with $F_t^\dagger F_t'=TI$, $a=w=T$, $h_-=h_+=4Ts$, and $M_{\rm loc}=1$.
Choose normalization $1+\delta$ and initial error $\epsilon/16$.
Its horizon is $N\le T+O(\delta^{-1/2}[Ts+s^{-1}\log(1/\epsilon)+1])$.
Choose $s$ of order $\min\{1,\sqrt{\log(2/\epsilon)/T}\}$.
The resulting approximation error before normalization and conditional-state error are below $\epsilon$.
Finite realization is exact on the required prefix, or receives a fixed error allocation as described next.
\end{proof}

For fixed $\delta>0$, the upper bound $(1+\delta)T+O_\delta(\log(1/\epsilon))$ also holds for the QSP approximation benchmark \cite{Low2016HamSim,Low2016Qubitization}; this is the comparison used in \cref{tab:mod-algorithms}.

Since $W=TI$ on every eigenvector, restricting the input to low energy leaves this module's catalyst norm unchanged.
A low-energy improvement requires a different signal or transfer function.

\subsubsection{Finite exponential transforms}
\label{subsec:mod-exp-resources}
Chen et al.\ \cite{Chen2026OptimaTimeDependent} give a transducer construction of Hamiltonian simulation by cascading Cayley transformation.
This can be understood as a first-order Pad\'e approximation, which we generalize in~\cref{app:exp-approx} to obtain better gate and qubit counts.
The higher-order diagonal Pad\'e version uses $R_p(x)=P_p(-x)/P_p(x)$, where $P_p$ is given in \cref{eq:expapp-pade}, and realizes
\begin{align}
\widehat{\module{Exp}}_T(w) =\left[R_p\!\left(\frac{T}{J}\frac{1-w}{1+w}\right)\right]^J.
\label{eq:mod-exp-pade}
\end{align}
Choosing $J=\Theta(\max\{1,T\})$ to be a power of two and
$p=O(\log(J/(\epsilon\delta))/\log\log(J/(\epsilon\delta)))$
gives $Jp$ private modes.
The complete known system can be applied using $O(p+\log(2Jp)\log(2J))$ gates and $\lceil\log_2(Jp+1)\rceil+O(1)$ qubits; these costs exclude the problem oracle and clock registers.
The shared structure of the repeated sections permits this implementation without applying all $J$ sections sequentially.
\Cref{thm:exp-approx} states the precise error allocation and proves these bounds.

Alternatively, one may synthesize the exponential transformation directly.
The following constructions match its finite impulse response or discretize its spectral representation.

\begin{theorem}[Finite exponential transform]\label{thm:exp-implementation}
For $T>0$ and finite horizon $N$, there are two useful finite realizations.
Schur recursion peels off a two-mode unitary and a unit delay at each step, matching successive Taylor coefficients, much as QSP synthesis removes one signal rotation at a time.
Apply $N$ steps of Schur recursion \cite{Schur1917SchurAlgorithm,Yagle1985SchurAlgorithm} to $\module{Exp}_T$, then terminate with a unit phase.
The resulting rational inner function agrees through degree $N-1$.
Because $C(0)=0$, composition preserves this prefix.
Equivalently, the Toeplitz load is a nilpotent contraction, $X=T_N[C]$ and $X^N=0$.
The exact and finite transfer functions therefore agree when applied to this load.
This option has $D=N$ private modes and zero comparison error; a direct unitary lattice implementation gives $G_S=O(N[a+\log(N+2)])$.

The spectral realization gives a smaller proved gate bound.
It implements a finite system with Cauchy weights
\begin{align}
p_k=\frac{2T}{T^2+4\pi^2k^2},
\quad u_k=\frac{T-2\pi ik}{T+2\pi ik},
\quad W_T=\coth(T/2),
\quad |k|\le R,
\label{eq:mod-exp-modes}
\end{align}
with the omitted positive mass assigned either to $u_*=-1$ or to positive quadrature nodes near it.
Its known system is one diagonal phase operation times one reflection.
For implementation error $\xi$, the reflection state has an MPS representation with bond $O(\log(1/\xi))$.
Preparing it on its output qubits and applying a bandwise phase circuit implements the full finite system to isometry error $\xi$, with
\begin{align}
G_{\rm Exp}=O(b_{\module{Exp}}\log^2(1/\xi)),
\quad n_{\rm Exp}\le b_{\module{Exp}}+4,
\label{eq:mod-exp-circuit}
\end{align}
where $b_{\module{Exp}}$ is its padded mode-address width.
The additional four qubits include the public flag.
The phase circuit uses GQSP \cite{Motlagh2024gqsp} on known address-dependent phases; it makes no calls to the problem oracle.
\cref{app:mod-exp} gives the construction and error proof.
The bound uses this explicit preparation and diagonal-phase factorization of the unitary.

For the finite horizon, choose a positive-weight quadrature and cutoff satisfying
\begin{align}
\frac8{\pi(2s)!}\le\frac\epsilon{6(1+\delta)},
\quad R\ge\max\{1,TN/\pi\},
\end{align}
where $s$ is the number of quadrature nodes.
Then $D=O(TN+s)$ modes suffice, with mode-address width $b_{\module{Exp}}=O(\log(2+TN+s))$.
Its substitution into the common resource bound is given in \cref{eq:mod-ham-gates}.

\end{theorem}

The finite realizations are proved in \cref{app:mod-exp}, with the Schur-prefix construction in \cref{app:qudit-schur}.
Allocate the error budget by
\begin{align}
\epsilon_{\rm clk}\le\epsilon/3,
\quad \eta_N\le\frac\epsilon{3(1+\delta)},
\quad \xi\le\frac\epsilon{3(1+\delta)N}.
\end{align}
Further constant splitting covers oracle and endpoint synthesis when needed.
The dimension $D2^a$ includes all encoding-ancilla states.
The MPS bond uses output qubits that have not yet been prepared, giving the stated width with one mode register.

\subsubsection{A rational reciprocal from shifted Cayley transformations}
\label{subsec:mod-reciprocal}

Here we implement a bounded rational transfer function of a Hermitian operator while retaining its original oracle connections.
The desired operation is a block-encoding of that rational function.
A lossless matrix completion can be factored into a cascade of shifted \CayleyIIR\ transformations and known public rotations \cite{Alpay2014RationalUnitaryFunctions}; connecting their public ports then implements the product.
For the reciprocal needed in \cref{cor:app-inverse}, \cref{lem:mod-reciprocal} gives the scalar approximation, its unitary completion, and the required number of Cayley transformations.

\begin{lemma}[Unitary rational reciprocal]
\label{lem:mod-reciprocal}
For an integer $d\ge1$, set
\begin{align}
p_d(u)=u^{2d-1},
\quad h_d(u)=1+u^{2d},
\quad f_d(u)=\frac{p_d(u)}{h_d(u)}.
\label{eq:mod-reciprocal-scalar}
\end{align}
There is a polynomial of degree $2d$ with the spectral-factor identity $q_dq_d^\#=h_d^2-p_d^2$ and $q_d^\#(u)=\overline{q_d(\overline u)}$.
The full transfer function
\begin{align}
\mathcal R_d(u)=\frac1{h_d(u)} \begin{pmatrix}p_d(u)&-q_d^\#(u)
\\
q_d(u)&p_d(u)\end{pmatrix}
\label{eq:mod-reciprocal-completion}
\end{align}
is unitary on the real axis and has entry $f_d(u)$.
It admits a cascade of at most $3d$ shifted Cayley sections and a public rotation.
For $|u|\ge2$,
\begin{align}
|f_d(u)-u^{-1}|\le\frac1{2(1+4^d)}.
\label{eq:mod-reciprocal-error}
\end{align}
For a Hermitian signal, every section has a full unitary realization with both public channels retained.
Factors with opposite signs of the Cayley argument use separate delay variables with nonnegative powers.
\end{lemma}

The completion uses the same spectral-factor idea as QSP unitary completion.
The difference between the squared denominator and squared numerator is strictly positive on the real line: $h_d(u)^2-p_d(u)^2=(1+u^{2d})^2-u^{4d-2}>0 $ for $u\in\mathbb R$.
Its roots occur in conjugate pairs.
Choosing one root from each pair defines a polynomial whose product with its conjugate reflection is this difference.
The resulting two-port matrix is unitary on the real axis.
Multiplying by scalar Blaschke factors removes its upper-half-plane poles; lossless factorization \cite{Alpay2014RationalUnitaryFunctions} then produces rank-one Cayley sections.
The final inverse scalar sections restore the original transfer function.
\cref{app:mod-reciprocal} proves the count of $3d$ sections, including these inverses.

\begin{figure}[ht]
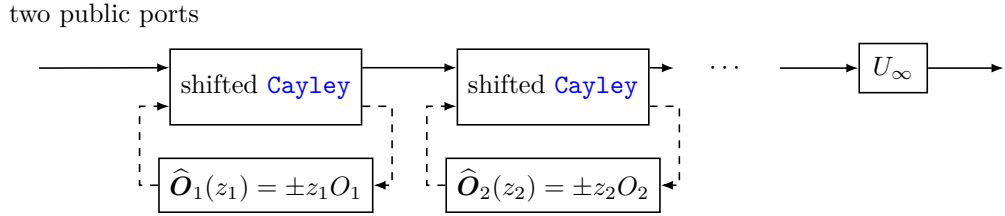

\centering

\IIRDiagram{ReciprocalCascade}

\caption{The reciprocal is block-encoded by a two-port unitary cascade.
Each section retains its feedback connection and its additional public channel.
The final clock acts on the connected network.}
\label{fig:mod-reciprocal}
\end{figure}

To describe the cascade, define the upper-half-plane roots and their Blaschke factors:
\begin{align}
\zeta_j=a_j+ib_j=\exp(i(2j+1)\pi/(2d)),
\quad B_j(u)=\frac{u-\zeta_j}{u-\overline\zeta_j},\qquad0\le j<d.
\end{align}
Each rank-one section is $I_2+(B_j(u)-1)|v\rangle\langle v|$ for a known normalized $|v\rangle$.
It applies a Cayley transfer function to one public combination and transmits the orthogonal combination unchanged.
The scalar sections $B_j(u)^{-1}I_2$ undo the auxiliary all-pass factor on both public ports.
The system matrices and factorization proof appear in \cref{app:mod-reciprocal}.
The unitary reciprocal cascade is shown in \cref{fig:mod-reciprocal}.

For an effective Hamiltonian $H_{\mathrm{eff}}\succeq aI$, substitute $u=2H_{\mathrm{eff}}/a$.
The resulting entry approximates $(a/2)H_{\mathrm{eff}}^{-1}$.
A clock with normalization two then selects $(a/4)H_{\mathrm{eff}}^{-1}$.
The degree is $O(\log(1/\epsilon))$; the oracle cost additionally depends on a common analytic neighborhood of the connected sections.
\cref{cor:app-inverse} proves that neighborhood from the reduced model's response and perturbation bounds.

\subsubsection{A two-term weighted-cost example}
\label{subsec:mod-average}\IIRAnchor{AverageCost}

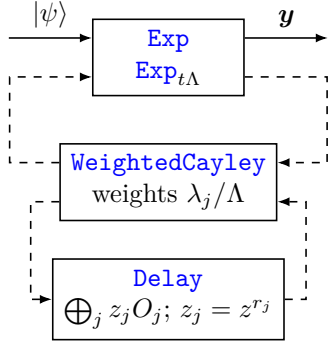
\begin{figure}[ht]
\centering
\begin{minipage}[c]{.43\linewidth}\centering
\IIRDiagram{IntroAverage}
\end{minipage}\hfill
\begin{minipage}[c]{.53\linewidth}
For weights $(1,R^{-1})$ and costs $(1,R)$, choose unequal port delays.
Use one delay on the first feedback connection and $R$ on the second.
With one clock for the complete transfer function, the expensive oracle is queried once every $R$ ticks; the intervening private sectors store its amplitude.
\end{minipage}
\caption{The weighted-cost example uses the same exponential transform, with a weighted Cayley junction and unequal port delays.
Each delay is \EDelay; the weighted preparation uses known gates whose cost is included.}
\label{fig:mod-average}
\end{figure}

The weighted simulation theorem, \cref{thm:multi-hs}, allocates queries according to $\langle\boldsymbol{C},\boldsymbol{\lambda}\rangle_{1/2}$.
We illustrate that allocation with two terms using the same exponential transform as \cref{sec:intro-simple-proofs}.
The classical operation is adding two impedance loads at a lossless wave-digital junction \cite{Fettweis1986WaveDigitalFilters}.
Set $m=2$ and $\Lambda=\lambda_1+\lambda_2$.
The public port is $\mathcal H$ and the private port is the direct sum $\mathcal K_1\oplus\mathcal K_2$ of the full oracle Hilbert spaces.
The known isometry $E\ket\psi=\bigoplus_{j=1}^2 \sqrt{\lambda_j/\Lambda}(\ket0_{a_j}\otimes\ket\psi)$ prepares the two weighted signal sectors.
We replace the single signal sector of \CayleyIIR\ by this weighted combination, leaving its orthogonal complement with phase $i$.
This gives \WeightedCayleyIIR:
\begin{align}
S_{\Sigma}=\left(\begin{array}{c|c}0&-iE^\dagger
\\
\hline
E&i(I-EE^\dagger)\end{array}\right),
\quad \oracle(\Bz)=z_1O_1\oplus z_2O_2.
\label{eq:mod-weighted-system}
\end{align}
Specializing \cref{lem:common-cayley} to $C=E$, $K_0=0$, and the private phase convention gives
\begin{subequations}
\label{eq:mod-average-response}
\begin{align}
M(\Bz)=\sum_{j=1}^2\lambda_j \frac{(1-z_j^2)I+2iz_jH_j/\lambda_j}{1+z_j^2},
\quad C_{\Sigma}(\Bz)= (I-M(\Bz)/\Lambda)(I+M(\Bz)/\Lambda)^{-1},
\\
F_t(\Bz)=\module{Exp}_{t\Lambda}(C_{\Sigma}(\Bz)) =e^{-tM(\Bz)},
\quad F_t=e^{-it(H_1+H_2)}.
\end{align}
\end{subequations}
With $x_j=H_j/\lambda_j$, the module names denote
\begin{align}
\module{WeightedCayley}_{\boldsymbol{\lambda}}(\Bz;\boldsymbol{x})=C_\Sigma(\Bz),
\\
\module{AverageCost}_{t,\boldsymbol{\lambda}}(\Bz;\boldsymbol{x})=F_t(\Bz).
\end{align}
The displayed realization is $S_{\module{WeightedCayley}}=S_\Sigma$ with the stated feedback connections.
Eliminating the private variables adds the two loads inside one matrix exponential.
The displayed identity preserves their order and therefore holds even when $H_1$ and $H_2$ do not commute.
The unequal-delay two-term network is shown in \cref{fig:mod-average}.

Choose positive time and an integer delay with $t>0$, $0<\epsilon\le\delta\le1/4$, and $R\ge1$.
Set the normalizations and costs to $(\lambda_1,\lambda_2)=(1,R^{-1})$ and $(C_1,C_2)=(1,R)$.
With delays $\boldsymbol{r}=(1,R)$, use $G(z)=F_t(z,z^R)$.
Direct differentiation of \cref{eq:mod-average-response} gives $G^\dagger G'=2tI$.
For $|\log z|\le s/R$, the calculation in \cref{eq:mod-ham-radial} bounds the Hermitian part of each load.
The inequality $\norm{e^A}\le e^{\lambda_{\max}((A+A^\dagger)/2)}$ then gives the two-sided growth bound
\begin{align}
a=w=2t,
\quad \zeta=s/(4R),
\quad M_{\rm loc}=1,
\quad h_-=h_+=8ts
\quad(0<s\le1/8).
\label{eq:mod-average-tag}
\end{align}
This operator estimate establishes the noncommuting lift.
For $|z|<1$, each load has positive semidefinite Hermitian part, so the complete transfer function is contractive.
Applying
\cref{thm:iir-analytic-clock-selection}
and optimizing $s$ gives
\begin{subequations}
\label{eq:mod-average-cost}
\begin{align}
N&\le2t+O\!\left(\delta^{-1/2} [\sqrt{tR\log(1/\epsilon)}+R\log(1/\epsilon)+1]\right),
\\
q_1&=N-1,
\quad q_2=\lfloor(N-1)/R\rfloor,
\quad q_1+Rq_2\le4(1+\delta)t+ O\!\left(\delta^{-2}R\log(1/\epsilon)\right).
\end{align}
\end{subequations}
We obtain $\Be{\widetilde G_N/(1+\delta)}$ with approximation error before normalization at most $\epsilon$.
Thus the coefficient of $t$ in the weighted query cost is independent of $R$; the precision term still depends on it.
The aggregate encoding with normalization $1+R^{-1}$ incurs cost $1+R$ per query.
Here $\langle\boldsymbol{C},\boldsymbol{\lambda}\rangle_{1/2}=4$ and $\|\boldsymbol{C}\|_1\|\boldsymbol{\lambda}\|_1=(1+R)(1+R^{-1})$, recovering the comparison in \cref{tab:mod-algorithms}.

For a finite implementation, let the system have $n$ qubits.
The oracle at port $j$ uses $a_j$ encoding ancillas.
Use $D=N$ Schur modes for \ExpIIR.
Since $C_{\Sigma}(0,0)=0$, the delayed transfer function matches all coefficients used by the clock exactly.
Its port and gate counts are
\begin{subequations}
\label{eq:mod-average-gates}
\begin{align}
d_c=N+N(2^{a_1}+R2^{a_2}),
\quad b_c=\lceil\log_2d_c\rceil,
\quad n_{\rm total}\le n+b_c+s_O+O(1),
\\
G\le q_1G_{O_1}+q_2G_{O_2} +O\!\left(N^2b_c+Nb_c\log(2+b_c)\right).
\end{align}
\end{subequations}
The bound counts a direct Schur-lattice circuit, the weighted signal preparation, the delays, and the rank-one input- and output-clock state preparations.
At $R=1$ the two feedback operations can be treated as one aggregate oracle when that access model is cheaper.
The next section optimizes the integer delays for general weights and costs.

\subsubsection{Black-box state preparation from weighted-cost simulation}
\label{subsec:mod-stateprep}\IIRAnchor{StatePrep}

A useful application of \AverageCostIIR\ is to prepare a linear combination of states whose preparation circuits have non-uniform cost.
The classical network is unchanged: parallel Cayley loads feed the same exponential transform.
The Hamiltonian given to it couples one known reference state to the desired combination, so its evolution is a rotation in a two-dimensional subspace.

Suppose $U_j\ket0=\ket{\psi_j}$, $j=1,\ldots,m$, prepares a normalized state with all work erased.
The controlled preparation and its inverse, $U_j,U_j^\dagger$, each have cost $C_j>0$.
For known weights $\lambda_j>0$, set
\begin{align}
\Lambda=\sum_j\lambda_j,
\quad s=\norm{\sum_j\lambda_j\ket{\psi_j}}>0,
\quad \ket\Psi=s^{-1}\sum_j\lambda_j\ket{\psi_j}.
\label{eq:mod-stateprep-target}
\end{align}
The given norm $s$ fixes the simulation time for state preparation in \cref{cor:mod-stateprep}.
Known coefficient phases are absorbed into the preparations, and the states may be nonorthogonal.
The query allocation is deterministic and accounts for non-uniform oracle costs, as in \cref{subsec:mod-average}.

\begin{corollary}[Known-norm state preparation]
\label{cor:mod-stateprep}
The simulated unitary has block-encoding normalization $1+\sigma$; the final state-preparation normalization is two.
For $0<\epsilon\le\sigma\le1/4$, the unnormalized successful output of \StatePrepIIR\ satisfies $\norm{\boldsymbol{b}-\ket\Psi/2}\le\epsilon/4$.
Its total forward and inverse preparation counts $Q_j$ obey, per attempt,
\begin{align}
\sum_jC_jQ_j\le \frac{2(1+\sigma)\arcsin((1+\sigma)/2)}{s} \left(\sum_j\sqrt{\lambda_jC_j}\right)^2 +O\!\left(\sigma^{-2}\sum_jC_j\log(1/\epsilon)\right).
\label{eq:mod-stateprep-cost}
\end{align}
The leading coefficient is $\pi/3+O(\sigma)$.
The success probability is bounded by $(1/2-\epsilon/4)^2\le p_{\rm succ}\le(1/2+\epsilon/4)^2$.
The conditional normalized state has vector error at most $\epsilon$.
\end{corollary}
\begin{proof}
We simulate a Hamiltonian that rotates a known reference state into the desired linear combination.
Adjoin a normalized reference state $\ket{\mathrm{ref}}$ orthogonal to the output space and set
\begin{align}
H_j=\lambda_j(\ket{\psi_j}\!\bra{\mathrm{ref}} +\ket{\mathrm{ref}}\!\bra{\psi_j}),
\quad H=\sum_jH_j,
\quad t=\frac{\arcsin((1+\sigma)/2)}s.
\label{eq:mod-stateprep-star}
\end{align}
On the full state-preparation Hilbert space, define the self-inverse unitary
\begin{align}
V_j=\begin{pmatrix}0&U_j^\dagger
\\
U_j&0\end{pmatrix} =D_j(X\otimes I)D_j^\dagger,
\quad D_j=\ket0\!\bra0\otimes I+\ket1\!\bra1\otimes U_j.
\label{eq:mod-preparation-return}
\end{align}
\IIRAnchor{PrepReturn}
The signal isometry embeds the reference and output sectors as $\ket{\mathrm{ref}}\mapsto\ket0\ket0$ and $\ket v\mapsto\ket1\ket v$ for a normalized state in the output sector.
Sandwiching $V_j$ by this embedding gives $H_j/\lambda_j$.
Known routing and a retained zero-block padding put this in the usual $\Be{H_j/\lambda_j}$ layout, without additional preparation calls.
Each $V_j$ uses one controlled preparation and one controlled inverse.
Its cost is therefore $2C_j$.

On the reference/target plane, the Hamiltonian is $H=sX$.
Here $X$ swaps the two basis states of that plane.
Let $\Pi_{\rm out}$ project onto the output sector.
Then
\begin{align}
e^{-itH}\ket{\mathrm{ref}} =\cos(ts)\ket{\mathrm{ref}}-i\sin(ts)\ket\Psi,
\quad \frac{i\Pi_{\rm out}e^{-itH}\ket{\mathrm{ref}}}{1+\sigma} =\tfrac12\ket\Psi.
\label{eq:mod-stateprep-rotation}
\end{align}
Apply \cref{thm:multi-hs} at normalization $1+\sigma$ with query costs $2C_j$.
Its simulation block encoding has normalization $1+\sigma$; projection onto the output sector and the known phase $i$ give the second identity.
Substitution of $t$ gives \cref{eq:mod-stateprep-cost}.
Allocate $\epsilon/8$ selected-vector error to simulation and $\epsilon/8$ to synthesis.
The reverse triangle inequality gives the success interval, and normalizing $\boldsymbol{b}$ changes its error by at most a factor four.
\end{proof}

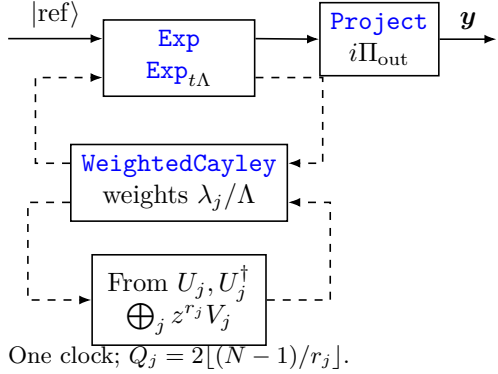
\begin{figure}[H]
\centering
\begin{minipage}[c]{.47\linewidth}\centering
\IIRDiagram{StatePrep}
\end{minipage}\hfill
\begin{minipage}[c]{.49\linewidth}
Replace the feedback operations of \cref{fig:mod-average} by the unitaries $V_j$ above.
Use the weights $\lambda_j/\Lambda$ and the delays given by \cref{thm:multi-hs}.
The transfer function is still $F_t(\Bz)=e^{-tM(\Bz)}$, with the sum in \cref{eq:mod-average-response} running over all terms.
Projection onto the output sector prepares the desired state with the constant success probability above.
\end{minipage}
\caption{\StatePrepIIR\ is Hamiltonian simulation with weighted query cost on the span of the reference and target states.
The output-sector selection is \ESelect; its unnormalized successful output vector is $\boldsymbol{b}\simeq\ket\Psi/2$.}
\label{fig:mod-stateprep}
\end{figure}

Use the \module{AverageCost} transfer function with the star-Hamiltonian signals $x_j=H_j/\lambda_j$ from \cref{eq:mod-stateprep-star}.
This defines $\module{StatePrep}_{t,\boldsymbol{\lambda}}(\Bz;\boldsymbol{x})$; projection onto its output sector gives the prepared vector.
The rotation between the reference and target states is connected as shown in \cref{fig:mod-stateprep}.

To count quditization resources, let each output state have $n$ qubits and use $D=N$ Schur modes.
The complete space of oracle $V_j$, including its sector flag and padding, has dimension at most $2^{n+a_j}$.
The public space has dimension $1+2^n\le2^{n+1}$.
With integer delays $r_j$ from \cref{thm:multi-hs},
\begin{subequations}
\label{eq:mod-stateprep-gates}
\begin{align}
q_j=\lfloor(N-1)/r_j\rfloor,
\quad Q_j=2q_j,
\quad d_c\le2N+N\sum_jr_j2^{a_j},
\quad b_c=\lceil\log_2d_c\rceil,
\\
n_{\rm total}\le n+b_c+s_U+O(1),
\\
G\le\sum_jq_j(G_{U_j}+G_{U_j^\dagger}) +O\!\left(N(N+m)(n+b_c)\log(2+n+b_c)\right).
\end{align}
\end{subequations}
This conservative bound counts the known signal embeddings, weight loading, Schur lattice, delay routing, and input- and output-clock state preparations.
This count precedes rotation synthesis.
The rotation count is $L=O(N(N+m))$.
A total synthesis error budget $O(\epsilon)$ requires an additional $O(L\log(L/\epsilon))$ gates.
Oracle errors are budgeted using their displayed invocation counts, as in \cref{eq:qudit-error-budget}.
The exponential-transform mode label does not multiply the oracle counts.

For comparison, LCU followed by known-angle amplitude amplification uses $O((\Lambda/s)\sum_jC_j)$ queries in this separate-oracle model \cite{Childs2012LCU,brassard2002quantum}.
Its precision logarithm is absent when the preparations and norm are exact.
Cauchy--Schwarz gives $(\sum_j\sqrt{\lambda_jC_j})^2\le\Lambda\sum_jC_j$, so the new leading weighted factor can be smaller.
The additive precision term must also be included when comparing the total costs.
The constant $\pi/3$ is inherited from simulation and the selected amplitude $1/2$.
Establishing an optimal state-preparation constant would require a separate lower bound.
A given multiplexed SELECT changes the access model and the comparison cost.

A nonorthogonal example illustrates the same construction.
Let $U_j\ket0=\ket{\phi_j}$ be training states and suppose their Gram matrix $G_{jk}=\langle\phi_j|\phi_k\rangle$ is known.
For a classically selected coefficient vector $\boldsymbol{a}$ with $\boldsymbol{a}^\dagger G\boldsymbol{a}>0$, \cref{cor:mod-stateprep} prepares
\begin{align}
\ket{\Psi_{\boldsymbol{a}}}=\frac{\sum_ja_j\ket{\phi_j}}{\sqrt{\boldsymbol{a}^\dagger G\boldsymbol{a}}},
\quad \sum_jC_jQ_j= O\!\left(\frac{(\sum_j\sqrt{|a_j|C_j})^2}{\sqrt{\boldsymbol{a}^\dagger G\boldsymbol{a}}} +\sum_jC_j\log(1/\epsilon)\right)
\label{eq:mod-stateprep-ritz}
\end{align}
at fixed $\sigma$, taking $\lambda_j=|a_j|$.
Absorb each coefficient phase into the corresponding preparation unitary.
Terms with $a_j=0$ are omitted.
For example, eigenvector continuation \cite{Frame2018EigenvectorContinuation} chooses the coefficients by solving the small generalized eigenproblem $H^{\rm red}\boldsymbol{a}=EG\boldsymbol{a}$ and $H^{\rm red}_{jk}=\langle\phi_j|H_{\rm phys}|\phi_k\rangle$.
The quantum stage then prepares this Ritz vector without calls to $H_{\rm phys}$.
Obtaining the reduced matrices, solving their eigenproblem, and the subspace approximation error are separate.
If the norm is estimated, a guaranteed relative error $O(\epsilon)$ can be included in the preparation error budget.

The representative IIRs and their reuse are summarized in \cref{tab:mod-algorithms,fig:mod-dependencies}, with resource bounds in \cref{app:resources}.
In \cref{sec:algorithms}, each construction starts from the required transfer function, combines these modules or a known unitary junction, and applies the common connection and clock results.

\section{Multioracle quantum algorithms}\label{sec:algorithms}

The algorithms in this section use the construction language of \cref{sec:modularity}.
The transfer function is chosen before the finite circuit.
For each problem we first identify the transformation required at $\Bz=\Bone$, then choose a module appropriate to the given oracle access.
A reusable IIR or an application-specific unitary junction turns that signal into the desired response.
Only after the complete response has been assembled do we choose oracle delays and a clock and apply quditization.
This order separates three questions that are easy to conflate: achievability of the transfer function, the query weights of the chosen realization, and the cost of its finite implementation.

Any custom module introduced in the following has a known unitary system matrix and specified oracle ports, so achievability follows from~\cref{prop:mod-unitary-junction}.
These will be useful for application-specific reasons, for example, for balancing several oracle costs, making a residual directly accessible, or mapping an eigenvalue to a convenient phase.

\subsection{Hamiltonian simulation and spectral amplification}
\label{sec:multi-hs}

The target is the unitary $e^{-itH}$, while the given access exposes the terms of $H$ through separate oracle ports.
Following \cref{subsec:mod-junctions}, we first choose a Hermitian signal for each port.
For separately block-encoded terms this is the Cayley junction: its load at $\Bz=\Bone$ is $i$ times the desired Hermitian sum, and its complete transfer function has a unitary realization by construction.
We then substitute that response into the reusable exponential IIR.
The resulting analytic function has value $e^{-itH}$ at $\Bz=\Bone$; assigning different delays to its original oracle ports optimizes the weighted query cost before one clock compiles the complete network.
A shared SELECT changes only the realization in terms of the supplied oracles.
The sum-of-squares construction below likewise introduces its specialized phase signal locally and gives its unitary realization before analyzing the clock.

Throughout, $0<\epsilon\le\delta\le1/4$.
The first parameter bounds approximation error before normalization; the second determines the normalization overhead.
The evolution results use $t>0$; the identity handles $t=0$ and the adjoint handles negative time.
The SOS phase amplifier uses an integer walk-step gain $k\ge1$.
All inverse and controlled calls are counted.
We use operator-norm error and state each block-encoding normalization explicitly.

\subsubsection{Separately block-encoded terms}
\label{sec:multi-hs-response}

Let $H=\sum_{j=1}^mH_j$ act on an $n$-qubit system space.
For each term we are given a self-inverse block-encoding with controlled-query cost $C_j>0$,
\begin{align}
V_j=\Be{H_j/\lambda_j},
\quad (\langle0|_j\otimes I)V_j(|0\rangle_j\otimes I)=H_j/\lambda_j,
\quad V_j^2=I,
\label{eq:access31}
\end{align}
where $\lambda_j>0$ is given.
Set $\Lambda=\|\boldsymbol{\lambda}\|_1$.
A combined block encoding that calls every term at each step has a leading cost proportional to $t\|\boldsymbol{\lambda}\|_1\|\boldsymbol{C}\|_1$.
Separate ports permit the smaller leading cost
\begin{align}
t\langle\boldsymbol{C},\boldsymbol{\lambda}\rangle_{1/2} =t\left(\sum_j\sqrt{\lambda_jC_j}\right)^2 \le t\|\boldsymbol{\lambda}\|_1\|\boldsymbol{C}\|_1.
\label{eq:multi-hs-average}
\end{align}
The construction applies to noncommuting terms.
Set $U=e^{-itH}$ for the target.

\begin{theorem}[Simulation with non-uniform query costs]
\label{thm:multi-hs}
Under \cref{eq:access31}, explicitly computable integer delays $r_j\ge1$ and a finite horizon $N$ give
\begin{align}
\mathcal A=\Be{\widetilde G_N/(1+\delta)},
\quad \|\widetilde G_N-U\|\le\epsilon,
\label{eq:multi-hs-guarantee}
\end{align}
with exact controlled-query counts
\begin{align}
q_j=\left\lfloor\frac{N-1}{r_j}\right\rfloor
\label{eq:multi-hs-counts}
\end{align}
and
\begin{align}
\sum_jC_jq_j\le (1+\delta)t\langle\boldsymbol{C},\boldsymbol{\lambda}\rangle_{1/2} +O\!\left(\delta^{-2}\|\boldsymbol{C}\|_1\log\frac1\epsilon\right).
\label{eq:multi-hs-cost}
\end{align}
The implicit constant in the remainder is independent of the number of terms, the Hamiltonians, their strengths, and the query costs.
\end{theorem}

By the block-encoding definition, $(\langle0|\otimes I)\mathcal A(|0\rangle\otimes I) =\widetilde G_N/(1+\delta)$; its success probability is at least $(1-\epsilon)^2/(1+\delta)^2$ on every normalized input.

For every fixed $\delta>0$, the precision contribution is additive and logarithmic in $1/\epsilon$.

The scalar construction first takes a Cayley transform and then inserts it in Exp, whose transfer function satisfies $\module{Exp}_{t\Lambda}(w)=\exp[-t\Lambda(1-w)/(1+w)]$.
Thus a Cayley load $ix$ becomes the phase $e^{-it\Lambda x}$.
The weighted system below replaces that scalar load by $M(\Bz)/\Lambda$.
We keep the full oracle Hilbert spaces so that this substitution is valid for every unitary satisfying the block-encoding promises.

Let $\mathcal K_j$ be the full Hilbert space of $V_j$, including its encoding ancillas and workspace.
On $\mathcal L=\bigoplus_j\mathcal K_j$, define the known coisometry
\begin{align}
E=\begin{pmatrix} \sqrt{\lambda_1/\Lambda}(\langle0|_1\otimes I)&\cdots& \sqrt{\lambda_m/\Lambda}(\langle0|_m\otimes I) \end{pmatrix},
\quad EE^\dagger=I,
\quad \Pi=E^\dagger E.
\label{eq:multi-hs-E}
\end{align}
The weighted Cayley system exchanges the public input with the weighted private signal and retains the complementary private outputs.
In public/private order it is
\begin{align}
S_{\module{Cayley}}=\left(\begin{array}{c|c}0&-iE
\\
\hline
E^\dagger&i(I-\Pi)\end{array}\right),
\quad \oracle(\Bz)=\bigoplus_jz_jV_j.
\label{eq:multi-hs-cayley-system}
\end{align}
This is \cref{lem:common-cayley} with $C=E^\dagger$, $K_0=0$, and the private phase absorbed into the system.
It is unitary on the full Hilbert space: it exchanges the public and selected private coordinates with a phase, and acts as $iI$ on the remaining private coordinates.
Eliminating the private variables gives
\begin{align}
C(\Bz)&=-iE\oracle(\Bz)[I-i(I-\Pi)\oracle(\Bz)]^{-1}E^\dagger =(I-M(\Bz)/\Lambda)(I+M(\Bz)/\Lambda)^{-1},
\label{eq:multi-hs-cayley-response}
\\
M(\Bz)&=\sum_j\left[ \lambda_j\frac{1-z_j^2}{1+z_j^2}I +i\frac{2z_j}{1+z_j^2}H_j\right].
\label{eq:M31}
\end{align}
Indeed, the identity $(I+iz_jV_j)(I-iz_jV_j)^{-1} =[(1-z_j^2)I+2iz_jV_j]/(1+z_j^2)$ follows from $V_j^2=I$.
Its compression with $E$ proves the transfer function formula and shows that $M$ is accretive in the polydisk, independently of the unspecified entries of the block encodings.

Connecting this transfer function to the exponential transform gives
\begin{align}
\module{Exp}_{t\Lambda}(w)=\exp\!\left[-t\Lambda\frac{1-w}{1+w}\right],
\quad F(\Bz)=\module{Exp}_{t\Lambda}(C(\Bz))=e^{-tM(\Bz)},
\quad F=e^{-itH}.
\label{eq:multi-hs-response}
\end{align}
The transfer function exponentiates the sum of the loads, including all orders of the noncommuting terms.
For a finite Exp system $S_{\module{Exp}}=\left(\begin{smallmatrix}A_{\module{Exp}}&B_{\module{Exp}}\\ C_{\module{Exp}}&D_{\module{Exp}}\end{smallmatrix}\right)$ with $D$ private modes, the complete composed system is
\begin{align}
S=\left(\begin{array}{c|c} A_{\module{Exp}} I&-iB_{\module{Exp}}\otimes E
\\
\hline
C_{\module{Exp}}\otimes E^\dagger& iI_D\otimes(I-\Pi)-iD_{\module{Exp}}\otimes\Pi \end{array}\right),
\quad \oracle_{\module{Exp}}(\Bz)=I_D\otimes\bigoplus_jz_jV_j.
\label{eq:multi-hs-fused}
\end{align}
Its selected part is $S_{\module{Exp}}\operatorname{diag}(1,-iI_D)$ and its complement is $iI$, so it is unitary.
The finite transfer function is $\widehat{\module{Exp}}_{t\Lambda}(C(\Bz))$.
The exponential-transform mode label is retained and unchanged by the oracle calls; each scheduled feedback operation uses the relevant oracle once.

\begin{figure}[htbp]
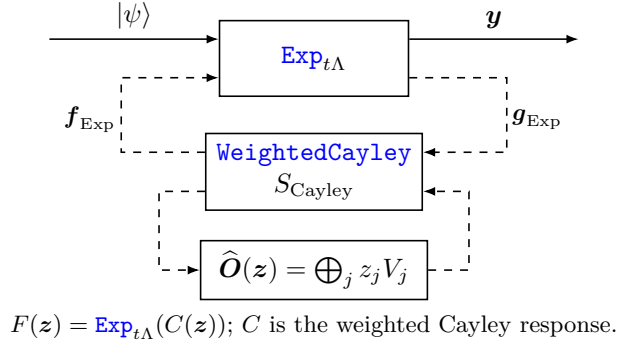

\centering
\IIRDiagram{HamiltonianExp}
\caption{The weighted Cayley transfer function in the feedback connection of the exponential transform.
Solid wires are public and dashed wires are private to the complete system.
The inner feedback connection keeps one separate port $z_jV_j$ per term.
Replacing Exp by its finite realization gives \cref{eq:multi-hs-fused}.}
\label{fig:multi-hs-exp}
\end{figure}

The weighted Cayley load in the exponential transform is shown in \cref{fig:multi-hs-exp}.

At the evaluation point $\Bz=\boldsymbol{1}$, the load and its derivative are $M=iH$ and $\partial_{z_j}M=-\lambda_jI$.
The unitarity identity in \cref{thm:iir-energy} therefore gives
\begin{align}
F^\dagger\partial_{z_j}F=t\lambda_jI.
\label{eq:multi-hs-action}
\end{align}
By the unitarity identity, these operators give the squared norms of the private query amplitudes on each input.
They therefore specify the contribution of each port to the leading propagation scale.
Give port $j$ a positive integer delay $r_j$, and write
\begin{align}
G(z)=F(z^{r_1},\ldots,z^{r_m}) =\sum_{k\ge0}G_{k}z^k,
\quad G=U=e^{-itH},
\quad \Omega=t\sum_j\lambda_jr_j.
\label{eq:multi-hs-delays}
\end{align}
A traversal applies $V_j$ once and contributes a delay $r_j$.
The horizon-$N$ unitary Toeplitz lift in \cref{thm:iir-lift} gives the query counts in \cref{eq:multi-hs-counts}.
For a horizon close to the common delay, minimizing the leading cost over real port delays gives
\begin{align}
t\left(\sum_j\lambda_jr_j\right)\left(\sum_jC_j/r_j\right),
\quad r_j\propto\sqrt{C_j/\lambda_j}.
\end{align}

\begin{proof}[Proof of \cref{thm:multi-hs}]
If $sr_j\le u\le1/2$, the two radial exponents on $|\log z|\le s$ are
\begin{align}
A_-=t\sum_j\frac{\lambda_jr_j}{1+\sin(sr_j)},
\quad A_+=t\sum_j\frac{\lambda_jr_j}{1-\sin(sr_j)}.
\label{eq:multi-hs-radial}
\end{align}
They bound the norm of the delayed transfer function by $|z|^{A_-}$ inside the circle and $|z|^{A_+}$ outside.
The bounds follow from the Hermitian part of the generator, without diagonalizing the terms simultaneously.
Apply the smoothed-window estimate, \cref{lem:iir-flat-smoothing}, using the sine-clock specialization in \cref{app:multi-hs-delays}.
It has amplitude $a=\cos(\pi/(K+1))$ and horizon
\begin{align}
N\le\Omega+\frac{K+\sin u}{\cos^2u}\,st\sum_j\lambda_jr_j^2 +O\!\left(\frac K s\log\frac1{a\epsilon}\right).
\label{eq:multi-hs-moment}
\end{align}
Taking $K=\Theta(\delta^{-1/2})$ gives $a\ge1/(1+\delta)$.
Keeping the second delay moment while clipping and rounding the optimal real rates gives \cref{eq:multi-hs-cost}; \cref{app:multi-hs-delays} records this step.
Finite Exp replacement controls the whole Toeplitz block, as in \cref{sec:multi-hs-implementation}.
Finally, a known attenuation changes $a$ to $1/(1+\delta)$.
Allocate a fixed fraction of $\epsilon$ to each clock, finite-realization, and implementation error.
This gives \cref{eq:multi-hs-guarantee} at the stated normalization.
\end{proof}

\subsubsection{Sum-of-squares phase amplification}
\label{sec:multi-hs-sos}

A sum-of-squares decomposition gives access to the square roots of the energies.
The usual reflection walk turns these square roots into phases, with one forward and one inverse call to each factor per step \cite{Low2016Qubitization,King2025SOSSA}.
We find that the same quadratic amplification can be obtained with non-uniform query rates.
The leading cost then depends on the strengths and costs of the individual factors.

Let $H=\sum_jA_j^\dagger A_j$ act on the public space $\mathcal X$, allowing rectangular factors.
Their squared normalization bounds are positive scalars $\lambda_j$.
Write $\Lambda=\|\boldsymbol{\lambda}\|_1$.
The scalar design is the Cayley transform on the normalized singular value $y=\sqrt{E/\Lambda}$.
For positive integers $k,J_0$, define $\rho=1/J_0$ and $J=kJ_0$.
Then
\begin{align}
\left(\frac{1-i\rho y}{1+i\rho y}\right)^J =e^{-i\phi_k(E)},
\quad \phi_\pm(E)=\pm\phi_k(E),
\quad \phi_k(E)=2kJ_0\arctan\!\left(\frac{\sqrt{E/\Lambda}}{J_0}\right).
\label{eq:multi-sos-phase}
\end{align}
The scalar transfer function has unit magnitude for real $y$.
Its phase gain at zero is the same as that of $k$ ordinary SOS walk steps.
In the centered energy variable $x=2E/\Lambda-1$,
\begin{align}
\frac{d\phi_k}{dx}= \frac{k}{\sqrt{2(x+1)}[1+\rho^2(x+1)/2]},
\quad \phi_k(E)=2k\sqrt{E/\Lambda}+O\!\left(k\rho^2(E/\Lambda)^{3/2}\right).
\label{eq:multi-sos-gradient}
\end{align}
Thus small energies produce phase changes proportional to their square roots.
We choose $J_0$ to control the cost of realizing this phase law; the finite rational transfer function itself is the target.

For electronic structure, explicit SOS representations and rigorous energy lower bounds can supply these factors \cite{Rubin2026LowerBoundCertificates}.
First-quantized plane-wave encodings give a further application \cite{Dutkiewicz2026FirstQuantizedAmplification}.
The oracle counts below apply after the factors and their normalizations have been specified.

We are given $V_j=\Be{A_j/\sqrt{\lambda_j}}$, its inverse, and known input and output signal subspaces in its full Hilbert space $\mathcal K_j$.
With $\Pi_{{\rm out},j}$ the output projector, the access condition is
\begin{align}
\lambda_j(\langle0|_j\otimes I)V_j^\dagger\Pi_{{\rm out},j}V_j (|0\rangle_j\otimes I)=A_j^\dagger A_j.
\label{eq:multi-sos-access}
\end{align}
The cost $c_j=C_{V_j}+C_{V_j^\dagger}>0$ includes controlled calls in both directions.
The factor encodings may be arbitrary unitaries.
Their conjugated projectors and reflections satisfy
\begin{subequations}
\label{eq:multi-hs-reflected-gram}
\begin{align}
P_j&=V_j^\dagger\Pi_{{\rm out},j}V_j,
\quad R_j=2P_j-I=V_j^\dagger(2\Pi_{{\rm out},j}-I)V_j,
\\
(\langle0|_j\otimes I)R_j(|0\rangle_j\otimes I) &=2A_j^\dagger A_j/\lambda_j-I,
\quad P_j^2=P_j,
\quad R_j^2=I.
\end{align}
\end{subequations}
One reflected feedback operation uses exactly one call in each factor-oracle direction.
Because $P_j$ is a projector, the map $\sqrt{\lambda_j}P_j(|0\rangle_j\otimes I)$ has the same Gram matrix as the original factor.
This lets us keep the identity half of $P_j=(I+R_j)/2$ in the known system matrix.

The target phase must vary linearly with the square root of the energy.
A bipartite Hermitian dilation converts singular values of a factor into signed eigenvalues: its square on the original system equals the normalized sum of squares.
Set $\omega_j=\lambda_j/\Lambda$ and use the enlarged public space $\mathcal P=\mathcal X\oplus\bigoplus_j\mathcal K_j$.
Define
\begin{align}
B=\begin{pmatrix} \sqrt{\omega_1}P_1(|0\rangle_1\otimes I)
\\
\vdots
\\
\sqrt{\omega_m}P_m(|0\rangle_m\otimes I) \end{pmatrix},
\quad \mathcal H=\begin{pmatrix}0&B^\dagger
\\
B&0\end{pmatrix},
\quad B^\dagger B=H/\Lambda.
\label{eq:multi-sos-dilation}
\end{align}
For $E>0$, an invariant plane has the orthonormal basis
\begin{align}
|\psi_E\rangle\oplus0,
\quad 0\oplus\frac{B|\psi_E\rangle}{\sqrt{E/\Lambda}}.
\end{align}
On this plane, $\mathcal H=\sqrt{E/\Lambda}\,X$.
Hence
\begin{align}
U=(I-i\rho\mathcal H)^J(I+i\rho\mathcal H)^{-J}
\label{eq:multi-sos-target}
\end{align}
has the phase pair in \cref{eq:multi-sos-phase}.
An input in the original system subspace has equal weight on the two branches; zero-energy inputs are fixed.
All-pass behavior holds on this enlarged public space.
Compressing back to $\mathcal X$ generally loses it.

The leading weighted query cost is
\begin{align}
\frac{k}{\sqrt\Lambda} \left(\sum_j\lambda_j^{1/3}c_j^{2/3}\right)^{3/2}.
\label{eq:multi-sos-leading}
\end{align}
For fixed $\delta$, the construction below has this leading cost plus $O(\|\boldsymbol{c}\|_1\log(1/\epsilon))$.
The sharper statement retains the dependence on the normalization parameter.

\begin{theorem}[Weighted SOS phase amplification]
\label{thm:multi-sos-phase}
Fix $0<\delta\le1/4$ and choose one integer $J_0=\Theta(\delta^{-3/4})$ satisfying \cref{eq:multi-sos-clock-parameters}.
This choice is independent of the amplification gain and approximation error.
For every integer $k\ge1$ and $0<\epsilon<1/4$, integer port delays and a finite horizon can be constructed.
With $r_j\ge1$ and horizon $N$, the circuit is
\begin{align}
\mathcal A=\Be{\widetilde G_N/(1+\delta)},
\quad \|\widetilde G_N-U\|\le\epsilon,
\label{eq:multi-sos-guarantee}
\end{align}
with exact factor-oracle counts
\begin{align}
Q_{V_j}=Q_{V_j^\dagger}=q_j= \left\lfloor\frac{N-1}{r_j}\right\rfloor
\label{eq:multi-sos-counts}
\end{align}
and weighted query cost
\begin{align}
\sum_jc_jq_j\le (1+\delta)\frac{k}{\sqrt\Lambda} \left(\sum_jc_j^{2/3}\lambda_j^{1/3}\right)^{3/2} +O\!\left(\delta^{-3}\|\boldsymbol{c}\|_1\log\frac1\epsilon\right).
\label{eq:multi-sos-cost}
\end{align}
The full enlarged-space phase walk in \cref{eq:multi-sos-target} is fixed by $J_0$ before its implementation precision is chosen.
The implicit constant is uniform in the strengths, costs, number of terms, and $k$.
The error bound holds on the full public space.
\end{theorem}

The compression convention is $(\langle0|\otimes I)\mathcal A(|0\rangle\otimes I) =\widetilde G_N/(1+\delta)$.
Thus the success probability is at least $(1-\epsilon)^2/(1+\delta)^2$, and $\|\widetilde G_N^\dagger\widetilde G_N-I\|\le2\epsilon+\epsilon^2$.
The circuit implements this encoded operator at the stated query cost; subsequent amplification would require additional calls.

The construction has three steps.
We first realize the projector factors by separate feedback connections, choosing their coupling strengths to balance the delays.
We then connect identical Cayley transformations in series.
Finally, we apply the master compilation theorem to the transfer function's radial bounds and implement the resulting clocks and unitary Toeplitz lift.

Splitting $P_j=(I+R_j)/2$ in \cref{eq:multi-sos-dilation} gives
\begin{subequations}
\label{eq:multi-sos-split}
\begin{align}
\mathcal H&=\mathcal H_0+\sum_j\mathcal H_j,
\\
\mathcal H_0&=\frac12\sum_j\sqrt{\omega_j} \begin{pmatrix}0&\langle0|_j\otimes I
\\
|0\rangle_j\otimes I&0\end{pmatrix}_{0,j},
\quad \mathcal H_j=\frac{\sqrt{\omega_j}}2 \begin{pmatrix} 0&(\langle0|_j\otimes I)R_j
\\
R_j(|0\rangle_j\otimes I)&0 \end{pmatrix}_{0,j}.
\end{align}
\end{subequations}
The subscript $0,j$ places the displayed matrix on the data and $j$th output sectors.
The first operator is known, with $\|\mathcal H_0\|=\|\sum_j\mathcal H_j\|=1/2$.

For fixed delays, put $D_{\boldsymbol{r}}=(\sum_j\omega_jr_j^2)^{1/2}$ and $r_* =\max_jr_j$.
Each port delay increases that port's contribution to the catalyst weight.
To equalize the total delay-weighted catalyst contribution across public sectors, we rescale the two sides of each coupling by reciprocal factors.
Their product preserves the Hermitian signal, while their ratio compensates for the delay.
Port $j$ has private space $\mathbb C^2\otimes\mathcal K_j$.
Its feedback operation and coupling map are
\begin{subequations}
\label{eq:multi-sos-couplings}
\begin{align}
O_j&=X_{\rm dir}\otimes R_j,
\\
E_j(\boldsymbol{\xi}\oplus\boldsymbol{\delta}) &=\sqrt{\frac{\omega_jr_j}{2D_{\boldsymbol{r}}}} (\langle0|_j\otimes I)\boldsymbol{\xi} \;\oplus_j\; \sqrt{\frac{D_{\boldsymbol{r}}}{2r_j}}\,\boldsymbol{\delta},
\quad K_j=E_jE_j^\dagger.
\end{align}
\end{subequations}
The product of the two amplitudes is fixed by $E_jO_jE_j^\dagger=\mathcal H_j$.
Their ratio is chosen so that the sum of the delay-weighted loads is the same on every public sector:
\begin{align}
E_jO_jE_j^\dagger=\mathcal H_j,
\quad -K_j\preceq\mathcal H_j\preceq K_j,
\quad \sum_jr_jK_j=\frac{D_{\boldsymbol{r}}}2I_{\mathcal P}.
\label{eq:multi-sos-balance}
\end{align}
The factor oracle acts identically on both direction sectors, followed by the known direction flip.
Each feedback operation therefore uses one forward/inverse pair.

The feedback operation is $\oracle(\Bz)=\bigoplus_jz_jO_j$.
Its load is $i\mathcal H_0+\sum_jE_j(I+iz_jO_j)(I-iz_jO_j)^{-1}E_j^\dagger$.
Expanding with $O_j^2=I$ gives
\begin{subequations}
\label{eq:multi-sos-response}
\begin{align}
M(\Bz)&=i\mathcal H_0+ \sum_j\left[\frac{1-z_j^2}{1+z_j^2}K_j +i\frac{2z_j}{1+z_j^2}\mathcal H_j\right],
\\
F_\rho(\Bz)&=(I-\rho M(\Bz))(I+\rho M(\Bz))^{-1},
\quad M=i\mathcal H.
\end{align}
\end{subequations}
This applies the scalar Cayley design to the factor dilation while keeping the individual oracle feedback connections separate.

To realize the transfer function, concatenate $E=(E_1\ \cdots\ E_m)$ and set $L_\rho=(I+\rho EE^\dagger+i\rho\mathcal H_0)^{-1}$.
Solving for the feedback amplitudes gives the known system matrix
\begin{align}
\oracle(\Bz)=\bigoplus_jz_jO_j,
\quad S_\rho=\left(\begin{array}{c|c} 2L_\rho-I&-2i\sqrt\rho\,L_\rho E
\\
\hline
2\sqrt\rho\,E^\dagger L_\rho&i(I-2\rho E^\dagger L_\rho E) \end{array}\right).
\label{eq:multi-sos-system}
\end{align}
The divider separates $\mathcal P$ from the private space $\bigoplus_j(\mathbb C^2\otimes\mathcal K_j)$, for both inputs and outputs.
\cref{lem:common-cayley} with $C=\sqrt\rho E^\dagger$ and $K_0=\rho\mathcal H_0$ proves unitarity and \cref{eq:multi-sos-response} in this private phase convention.
Each queried load has positive semidefinite Hermitian part for $|z_j|<1$, while $i\mathcal H_0$ has zero Hermitian part.
Hence $F_\rho$ is analytic and contractive throughout the open polydisk.
Its value at $\Bz=\boldsymbol{1}$ is unitary.

The projector transformation and its reflected feedback path are shown in \cref{fig:multi-sos-cell}.

\begin{figure}[htbp]
\centering
\IIRDiagram{SOSProjectorCell}\par\medskip
\IIRDiagram{SOSReflectedReturn}
\caption{The projector Cayley transformation and its reflected feedback operation.
Public wires are solid and private wires are dashed.
The system matrix maps the combined public and private input to the corresponding outputs, as in \cref{eq:mod-response}.
Closing the private connection gives $\boldsymbol{f}=\oracle(\Bz)\boldsymbol{g}$ and $\boldsymbol{y}=F_\rho(\Bz)|\psi\rangle$.
The lower circuit acts identically on the two direction sectors and is read from left to right.
The term $\mathcal H_0$ is included in $S_\rho$.}
\label{fig:multi-sos-cell}
\end{figure}

Connect the public output of each system matrix to the next public input, retaining a separate feedback connection for every system matrix.
The complete system is the unitary series composition $S_\rho^{[J]}$ with feedback operation $I_J\otimes \oracle(\Bz)$, and
\begin{align}
F(\Bz)=F_\rho(\Bz)^J,
\quad G(z)=F(z^{r_1},\ldots,z^{r_m}),
\quad G=U=F=(I-i\rho\mathcal H)^J(I+i\rho\mathcal H)^{-J}.
\label{eq:multi-sos-cascade}
\end{align}
All blocks of $S_\rho^{[J]}$ are given in \cref{eq:multi-sos-cascade-matrix}.
Fixing $J_0$ makes this target the $k$th power of one fixed walk.
The system matrix label is stored coherently and is unchanged by every factor-oracle call.

\begin{proof}[Proof of \cref{thm:multi-sos-phase}]
The balance in \cref{eq:multi-sos-balance} gives the exact delayed catalyst operator
\begin{align}
U^\dagger G'= kD_{\boldsymbol{r}}(I+\rho^2\mathcal H^2)^{-1}.
\label{eq:multi-sos-catalyst}
\end{align}
Its expectation on an input is the squared norm of the private feedback amplitude, so it identifies the leading propagation scale $kD_{\boldsymbol{r}}$.
To control the finite-clock error, we also use \cref{lem:multi-sos-cap}.
On a neighborhood of the evaluation point, its radius and radial exponents are $\Theta(u/r_*)$ and $kD_{\boldsymbol{r}}[1\pm(3\rho^2+5u)]$, respectively.
Substituting these bounds into \cref{thm:iir-analytic-clock-selection}, in the two-sided form \cref{eq:multi-sos-clock-rule}, with $u=\Theta(\delta^{3/2})$ and $\rho^2=O(\delta^{3/2})$, gives
\begin{align}
N\le(1+O(\delta))kD_{\boldsymbol{r}} +O\!\left(\delta^{-2}r_*\log\frac1\epsilon\right).
\label{eq:multi-sos-horizon}
\end{align}
The coefficient of the first term and the success amplitude are allocated explicitly in \cref{app:multi-sos-phase}.
\cref{lem:common-delay-allocation}, with $p=2$, optimizes the leading query cost:
\begin{align}
\inf_{r_j>0}D_{\boldsymbol{r}}\sum_jc_j/r_j =\frac1{\sqrt\Lambda} \left(\sum_jc_j^{2/3}\lambda_j^{1/3}\right)^{3/2},
\quad r_j\propto(c_j/\omega_j)^{1/3}.
\label{eq:multi-sos-delay-optimum}
\end{align}
Clipping and rounding these rates gives the uniform precision term in \cref{eq:multi-sos-cost}.
The finite lift in \cref{thm:iir-lift} gives the exact integer query counts.
Finally, known attenuation sets the normalization to $1+\delta$.
Allocate fixed fractions of $\epsilon$ to clock and known-operation errors.
The finite rational target requires no additional exponential-approximation step.
\end{proof}

For the circuit, the normalized input and output clock states are computed and prepared by known unitaries $P_{\rm in}$ and $P_{\rm out}$.
Their coefficient vectors are denoted $\boldsymbol{c}_{\rm in}$ and $\boldsymbol{c}_{\rm out}$; any auxiliary clock label is included in these preparations.
These states select the required linear combination of the coefficients in $T_N[G]$.
Before attenuation the encoded operator is $aA_N$, with $\|A_N-U\|\le\epsilon/16$ and $a\ge1-\delta/8$.
A known rotation gives amplitude $1/[a(1+\delta)]$.
Its approximating operator at the prescribed normalization is $\widetilde G_N=A_N$.

\begin{figure}[htbp]
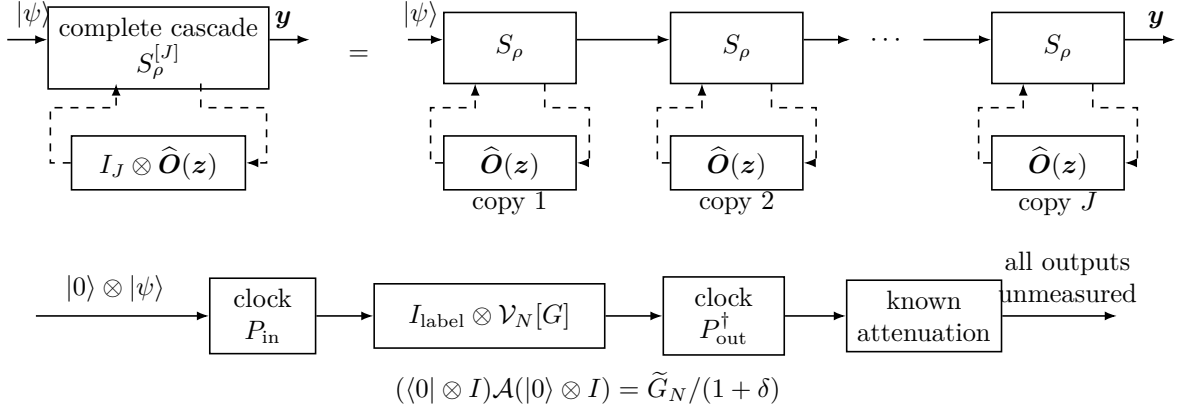

\centering
\IIRDiagram{SOSPhaseCascade}\par\medskip
\IIRDiagram{SOSPhaseCircuit}
\caption{The finite cascade and its circuit.
Each system matrix has a distinct feedback connection, with aggregate feedback operation $I_J\otimes \oracle(\Bz)$.
Clock preparation, the unitary Toeplitz lift, output-clock unpreparation, and known attenuation produce the compression shown below the circuit.
All output coordinates are retained by the unitary implementation.
The fixed $J_0$ depends on $\delta$; increasing precision leaves the target walk unchanged.
\cref{lem:multi-sos-known-cascade} implements the complete known cascade with a cost polynomial in $\log J$.}
\label{fig:multi-sos-cascade}
\end{figure}

The full cascade and finite clock-selected circuit are shown in \cref{fig:multi-sos-cascade}.

Assume the input has support only on energies satisfying $0\le E\le\Gamma\le\Lambda$ and $\Gamma>0$.
For $E>0$, differentiating gives the minimum energy-to-phase sensitivity
\begin{align}
\frac{d\phi_k}{dE}= \frac{k}{\sqrt{\Lambda E}(1+\rho^2E/\Lambda)} \ge t_{\rm eq},
\quad t_{\rm eq}=\frac{k}{\sqrt{\Lambda\Gamma}(1+\rho^2\Gamma/\Lambda)}.
\label{eq:multi-sos-time-equivalent}
\end{align}
The parameter allocation in the proof gives
\begin{align}
\sum_jc_jq_j\le (1+\delta)t_{\rm eq}\sqrt\Gamma\, \left(\sum_jc_j^{2/3}\lambda_j^{1/3}\right)^{3/2} +O\!\left(\delta^{-3}\|\boldsymbol{c}\|_1\log\frac1\epsilon\right).
\label{eq:multi-sos-time-cost}
\end{align}
Here $t_{\rm eq}$ describes the sensitivity of the exact target phase.
The implementation error must be smaller than the phase differences being resolved.
The promise is support below $\Gamma$; a lower bound on the ground energy alone does not provide it.

The exact reflection walk provides a useful comparison at the same endpoint sensitivity.
Define the known isometry
$\mathcal J|\psi\rangle= \bigoplus_j\sqrt{\omega_j}(|0\rangle_j\otimes|\psi\rangle)$
and $R=\bigoplus_jR_j$.
Then
\begin{align}
W=-(2\mathcal J\mathcal J^\dagger-I)R,
\quad \mathcal J^\dagger W^k\mathcal J=T_k(I-2H/\Lambda)
\label{eq:multi-sos-exact-walk}
\end{align}
has phases $\pm2k\arcsin\sqrt{E/\Lambda}$ and uses $q_j=k$ pairs from every term.
Since either phase law is permitted, we may choose the smaller cost of $k\|\boldsymbol{c}\|_1$ and \cref{eq:multi-sos-cost}.
The exact walk may be attenuated to the same selected normalization using known gates.
Its query count is independent of precision; finite-gate-set synthesis still has a precision cost.
For the exact phase-pair guarantee determined only by energy, \cref{prop:multi-sos-exact} proves $q_j\ge k$ for every positive-weight term.
The approximate construction has the stated logarithmic upper bound, with no matching logarithmic lower bound asserted here.

For the known operations, write $d_j=\dim\mathcal K_j$.
The public and private dimensions are \begin{align}
\dim\mathcal X+\sum_jd_j,
\quad \dim\mathcal L_j=2Jd_j
\quad\text{for the }J\text{-stage cascade}.
\end{align} By \cref{thm:iir-lift}, the delayed unitary Toeplitz lift has dimension
\begin{align}
D_c=N\left(\dim\mathcal X+\sum_jd_j\right) +2J\sum_jr_jd_j,
\quad b_c=\lceil\log_2D_c\rceil.
\label{eq:multi-sos-space}
\end{align}
All public output subspaces, private copies, and residual delay-buffer amplitudes are retained.
Let $G_{\rm align}$ count the gates that align the oracle registers and $G_{\rm clk}$ the clock preparations and their inverses.
Substituting \cref{lem:multi-sos-known-cascade} into \cref{thm:iir-gates} gives
\begin{subequations}
\label{eq:multi-sos-gates}
\begin{align}
G_{\rm total}\le{}&\sum_jq_j (G_{V_j}+G_{V_j^\dagger}+G_{2\Pi_{{\rm out},j}-I}) +O(NG_{\rm align})+G_{\rm clk}
\\
&+N\operatorname{poly}\!\left(m,b_c,\log(2J),\log(N/\epsilon)\right) +O\!\left(\left[(m+1)N+\sum_jq_j\right] b_c\log(2+b_c)\right).
\end{align}
\end{subequations}
The polynomial term counts reversible arithmetic on the given small known coefficient matrices, including synthesis and its workspace.
The qubit count is $b_c+O(1)+s_{\rm clk}+s_{\rm work}$.
The last two terms count restored clock-preparation scratch and arithmetic/oracle scratch, respectively.
The full cascade is implemented in each system update with a cost polynomial in $\log J$, giving the horizon dependence displayed in \cref{eq:multi-sos-gates}.
The $1+\delta$ leading coefficient applies to the weighted oracle calls; known gates and their scratch space have the separate costs above.

The optional phase law in \cref{prop:multi-sos-precision-dependent} lets the target depend on the chosen precision; the fixed-target algorithm above does not require that specialization.

\subsubsection{A shared SELECT}
\label{sec:multi-hs-select}

The desired Cayley load is unchanged, but the oracle access model is different.
Rather than treating every $V_j$ as an independent black box, we realize the same transfer function through one shared SELECT and term-dependent known preparations.
This is an alternative unitary junction for the supplied oracles in the sense of \cref{subsec:mod-junctions}; sharing SELECT changes the per-oracle query counts without changing the outer exponential IIR.

Suppose all term encodings act on a common Hilbert space and satisfy the full-unitary identity
\begin{align}
V_j=P_j^\dagger\mathrm{SEL}P_j,
\quad \mathrm{SEL}=\mathrm{SEL}^\dagger=\mathrm{SEL}^{-1}.
\label{eq:multi-hs-select-model}
\end{align}
The cost of SELECT is $C_S>0$, and $c_j=C_{P_j}+C_{P_j^\dagger}\ge0$ includes all coefficient and angle loading.
SELECT acts identically in every term sector and leaves the external term, mode, and delay labels unchanged.

\begin{figure}[htbp]
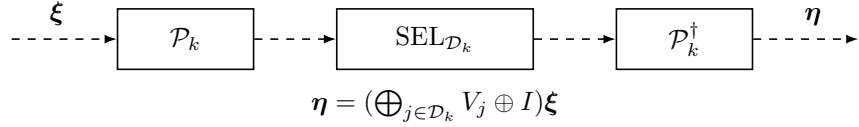

\centering
\IIRDiagram{SharedSelectReturn}
\caption{The feedback connection at tick $k$ with a shared SELECT.
Only the sectors scheduled at that step $\mathcal D_k$ are active.
A preparation and its inverse are counted for each such sector, while SELECT is applied once to their union.
The identity on all inactive sectors is retained.}
\label{fig:multi-hs-select-return}
\end{figure}

At tick $k>0$, let $\mathcal D_k=\{j:r_j\mid k\}$.
Apply the preparations on these sectors, a single SELECT controlled on their union, and the inverse preparations.
On the full Hilbert space,
\begin{align}
\mathcal P_k^\dagger\mathrm{SEL}_{\mathcal D_k}\mathcal P_k =\bigoplus_{j\in\mathcal D_k}V_j\oplus I,
\quad \mathcal P_k=\bigoplus_{j\in\mathcal D_k}P_j\oplus I.
\label{eq:multi-hs-select-batch}
\end{align}
The batched operation is the same feedback operation on the full Hilbert space, so the system matrix and transfer function of the preceding construction apply.
\cref{fig:multi-hs-select-return} shows the shared call.

\cref{thm:multi-hs-select} counts one shared SELECT for all sectors queried at the same tick and optimizes the preparation delays separately.

\begin{theorem}[Simulation with a shared SELECT]
\label{thm:multi-hs-select}
For $0<\epsilon\le\delta\le1/4$, the circuit can satisfy \cref{eq:multi-hs-guarantee} with exact counts
\begin{align}
Q_{P_j}=Q_{P_j^\dagger}=q_j=\left\lfloor\frac{N-1}{r_j}\right\rfloor,
\quad q_S=\left|\bigcup_j\{r_j,2r_j,\ldots\} \cap\{1,\ldots,N-1\}\right|.
\label{eq:multi-hs-select-counts}
\end{align}
There are constructive integer delays for which
\begin{align}
C_Sq_S+\sum_jc_jq_j \le(2-\gamma)t\left[\Lambda C_S+ \left(\sum_j\sqrt{\lambda_jc_j}\right)^2\right] +O_{\boldsymbol{\lambda},\boldsymbol{c},C_S,\delta}\!\left(\log\frac1\epsilon\right).
\label{eq:multi-hs-select-bound}
\end{align}
The displayed $O_\delta(\log(1/\epsilon))$ form is for fixed $\delta>0$.
Here $\gamma\in(0,1]$ is the instance-dependent margin defined in \cref{eq:multi-hs-select-internal-slack}.
The logarithmic coefficient may depend on the strengths and costs.
\end{theorem}
\begin{proof}
Choose an intermediate normalization allowance $0<\nu\le1/4$.
A fixed-delay clock with normalization $1+\nu$ and unnormalized error at most $\min\{\epsilon/16,\nu\}$ gives
\begin{align}
N\le(1+\nu)t\sum_j\lambda_jr_j +O\!\left(\nu^{-2}r_*\log\frac{16}{\min\{\epsilon,\nu\}}\right).
\label{eq:multi-hs-select-horizon}
\end{align}
Since $q_S\le N-1$, multiply this by $C_S+\sum_jc_j/r_j$.
Choose
\begin{align}
\bar r_j=\max\!\left\{1, \left\lceil\sqrt{\frac{\Lambda c_j}{C_S\lambda_j}}\right\rceil\right\}.
\label{eq:multi-hs-select-delays}
\end{align}
Rounding gives
\begin{align}
\left(\sum_j\lambda_j\bar r_j\right) \left(C_S+\sum_j\frac{c_j}{\bar r_j}\right) <2\left[\Lambda C_S+ \left(\sum_j\sqrt{\lambda_jc_j}\right)^2\right].
\label{eq:multi-hs-select-gap}
\end{align}
For a nonzero preparation cost, its left side is strictly below $(\sqrt{\Lambda C_S}+\sum_j\sqrt{\lambda_jc_j})^2$; if all preparations are free it equals $\Lambda C_S$.
The choice of internal allowance and positive margin in \cref{eq:multi-hs-select-internal-slack} bounds that leading product by the coefficient in \cref{eq:multi-hs-select-bound}.
Attenuate the block-encoding by $(1+\nu)/(1+\delta)$.
Its approximation error before normalization is unchanged.
The explicit gap choice and the alternative bound with uniform $\delta$-dependence are in \cref{app:multi-hs-select}.
\end{proof}

Self-inverse SELECT ensures that every scheduled sector uses the same oracle direction.
For a general unitary, implementing direction-controlled access can require both SELECT and SELECT$^\dagger$, and the query count includes both calls.

\subsubsection{Finite realizations and circuit implementation}
\label{sec:multi-hs-implementation}

The separately block-encoded sum and shared-SELECT constructions both substitute a transfer function that vanishes at zero into Exp.
After fixing the delays and horizon, let $C_{\boldsymbol{r}}$ denote that transfer function.
Then $X=T_N[C_{\boldsymbol{r}}]$ is a contraction and $X^{\lceil N/r_{\min}\rceil}=0$.
The finite spectral realization of \cref{thm:exp-implementation}, with parameter $\tau=t\Lambda$, gives
\begin{align}
\|\widehat{\module{Exp}}_{\tau,R}(X)-\module{Exp}_\tau(X)\| \le\frac{\tau^3K_{\rm Exp}(K_{\rm Exp}-1)(2K_{\rm Exp}-1)} {9\pi^4R^3},
\quad K_{\rm Exp}=\lceil N/r_{\min}\rceil.
\label{eq:multi-hs-spectral-error}
\end{align}
Choosing this below a constant fraction of $\epsilon$ controls the entire Toeplitz block.
The full-space identity therefore remains valid after selecting the input and output clocks.
Every original oracle acts identically on all exponential-transform mode labels.
Alternatively, an exact Schur-prefix realization preserves the horizon-$N$ Toeplitz block; its known-gate cost must be counted separately.
The positive boundary quadrature version of the spectral realization reduces its mode count at high precision, with the same oracle schedule.

The SOS phase amplifier instead uses the finite rational cascade in \cref{eq:multi-sos-cascade}.
Its system matrix is \cref{eq:multi-sos-system}; its finite rational target requires only clock and known-operation error budgets.

Let the complete finite system have public space $\mathcal P$ and private oracle spaces $\mathcal L_j$.
Its lift occupies
\begin{align}
(\mathbb C^N\otimes\mathcal P)\oplus \bigoplus_j(\mathbb C^{r_j}\otimes\mathcal L_j),
\quad D_c=N\dim\mathcal P+\sum_jr_j\dim\mathcal L_j.
\label{eq:multi-hs-carrier}
\end{align}
The separately block-encoded and shared-SELECT constructions have $\dim\mathcal P=2^n$ and $\dim\mathcal L_j=D\dim\mathcal K_j$.
All block-encoding ancillas and residual private delay-buffer amplitudes remain part of the circuit.
For SOS phase amplification, use the full Hilbert space in \cref{eq:multi-sos-space}.

Apply \cref{thm:iir-gates} to these full Hilbert spaces, with clock normalization $1+\delta$.
The finite rational SOS transfer function has $\widehat G=G$, and the sine-clock constructions need only a constant number of labels.
Direct clock preparation costs $O(N)$ gates.
The feedback-operation costs are $G_{V_j}$ for separate encodings, $G_{V_j}+G_{V_j^\dagger}+G_{2\Pi_{{\rm out},j}-I}$ for SOS factors.
For shared SELECT, replace the oracle sum by $q_SG_{\rm SEL}+\sum_jq_j(G_{P_j}+G_{P_j^\dagger})$ and use the corresponding routing for the scheduled sectors.
The weighted query bounds include the costs of the known reflections and preparations when these are part of a given oracle; additional known operations contribute the gate costs stated separately.

For private mode count $D$, alignment and one spectral exponential-transform update have gate cost $G_S=O(G_E+\log D\,\log^2(N/\epsilon))$.
The additional scratch width is constant; $G_E$ includes weight preparation and the phase tests on the signal space.
For the SOS phase amplifier, the full known-cascade implementation is given in \cref{app:multi-sos-cascade}, with arithmetic workspace counted in addition to the binary address register.
Clock preparation and all synthesis errors are counted in \cref{thm:iir-gates,eq:qudit-error-budget}.
Allocating a constant fraction of the error to the clock, finite Toeplitz replacement when needed, and known gates yields the stated total error.
The leading coefficient $1+\delta$ applies to the weighted oracle cost; the known operations have the separately displayed gate costs.

\subsubsection{Sparse Hamiltonian simulation}
\label{sec:multi-hs-sparse}
We now simulate a Hamiltonian specified by its nonzero matrix entries.
The construction rotates the two orientations of each edge, combines these rotations through the Sparse2 junction, and applies the exponential transform from \cref{thm:exp-implementation}.
The first compressed moment of the edge matrix recovers the Hamiltonian.
Its second compressed moment bounds the query delay and the analytic growth required for clock shaping.
This permits a simulation cost governed by the largest row Euclidean norm.

Let $H=H^\dagger$ act on $n$ qubits, with at most $d$ nonzero entries per row.
We are given known upper bounds
\begin{align}
s_2\geq\|H\|_{1\to2} =\max_x\left(\sum_y|H_{xy}|^2\right)^{1/2},
\quad h_{\max}\geq\max_{x,y}|H_{xy}|.
\label{eq:s6sp-norms}
\end{align}
The position oracle is an in-place permutation $O_f|x,\ell\rangle=|x,f_x(\ell)\rangle$ whose first $d$ positions contain the row support, padded with distinct zero locations.
The value oracle $O_H$ writes $H_{xy}$ at the address $(x,y)$.
Controlled access to both oracles and their inverses is assumed; in particular, $O_f^\dagger$ performs inverse position lookup.
We use $t>0$, $s_2,h_{\max}>0$, and $0<\epsilon\leq1/4$ below.
The identity handles a zero Hamiltonian or zero time, and negative time is handled by replacing $H$ by $-H$ in the entry arithmetic.

\begin{theorem}[Sparse Hamiltonian simulation]
\label{thm:multi-hs-sparse}
For every $v\geq\sqrt d\,s_2$, there is a finite unitary $\mathcal A=\Be{\widetilde G_N/2}$ such that
\begin{align}
(\langle0|_a\otimes I)\mathcal A(|0\rangle_a\otimes I) =\widetilde G_N/2,
\quad \|\widetilde G_N-e^{-itH}\|\leq\epsilon.
\label{eq:s6sp-guarantee}
\end{align}
Each of $O_f,O_f^\dagger,O_H,O_H^\dagger$ is used $N-1$ times, where
\begin{align}
Q_f=Q_H=2(N-1),
\quad N=O\!\left(tv+ \left(1+\frac{dh_{\max}}v\right)\log\frac1\epsilon\right).
\label{eq:multi-hs-sparse-bound}
\end{align}
Here $Q_f,Q_H$ combine the forward and inverse counts.
Optimizing $v$ gives
\begin{align}
Q_f,Q_H=O\!\left( t\sqrt d\,s_2+ \sqrt{tdh_{\max}\log(1/\epsilon)}+\log\frac1\epsilon \right).
\label{eq:multi-hs-sparse-optimized}
\end{align}
The same asymptotic counts suffice for a unitary $\mathcal A_{\rm sim}$ with
\begin{align}
\|\mathcal A_{\rm sim}(|0\rangle_a\otimes I) -|0\rangle_a\otimes e^{-itH}\|\leq\epsilon.
\label{eq:s6sp-clean}
\end{align}
\end{theorem}

\begin{proof}
We keep the scale $v$ adjustable.
Increasing it reduces the size of each edge rotation and enlarges the analytic neighborhood, at the cost of a larger propagation term $tv$.
Let $P_d|0\rangle_j=d^{-1/2}\sum_{\ell<d}|\ell\rangle_j$, where the $n$-qubit register $j$ stores a column or sparse-list index.
Define
\begin{subequations}
\label{eq:multi-hs-sparse-edge}
\begin{align}
A_v&=\frac d v\sum_{x,y}H_{xy}|x,y\rangle\langle y,x|,
\quad T_v=P_d^\dagger O_f^\dagger A_vO_fP_d,
\\
\mathcal U_v&=(I-iT_v)(I+iT_v)^{-1}.
\end{align}
\end{subequations}
Identities on the row register are implicit next to $P_d$.
The Hermitian matrix $A_v$ decomposes into one-dimensional diagonal edges and two-dimensional unordered edges.
Its Cayley transform is therefore computed locally from one matrix entry, using reversible fixed-point arithmetic.

For $x<y$, write $a=dH_{xy}/v$ and set $\vartheta=2\arctan|a|$, $\phi=\arg a$; set $\phi=0$ when $a=0$.
In the basis $(|x,y\rangle,|y,x\rangle)$, the edge rotation is
\begin{align}
\left.(I-iA_v)(I+iA_v)^{-1}\right|_{\{x,y\}} =\begin{pmatrix} \cos\vartheta&-i e^{i\phi}\sin\vartheta
\\
-i e^{-i\phi}\sin\vartheta&\cos\vartheta \end{pmatrix}.
\label{eq:multi-hs-edge-rotation}
\end{align}
This is the half-angle parametrization $|a|=\tan(\vartheta/2)$.
Reversibly sort the addresses and retain an orientation qubit.
Load $H_{xy}$ at the sorted address, apply \cref{eq:multi-hs-edge-rotation} to the orientation, unload the entry, and undo the sorting.
For the indicated input orientation, the state transformation is
\begin{subequations}
\label{eq:s6sp-edge-state}
\begin{align}
|x,y\rangle|0\rangle_o|0\rangle_b &\xrightarrow{\ \mathrm{sort},\ O_H\ } |x,y\rangle|1\rangle_o|H_{xy}\rangle_b
\\
&\xrightarrow{\ \mathrm{rotation}\ } |x,y\rangle \bigl(\cos\vartheta|1\rangle_o -i e^{-i\phi}\sin\vartheta|0\rangle_o\bigr) |H_{xy}\rangle_b
\\
&\xrightarrow{\ O_H^\dagger,\ \mathrm{unsort}\ } \bigl(\cos\vartheta|x,y\rangle -i e^{-i\phi}\sin\vartheta|y,x\rangle\bigr) |0\rangle_o|0\rangle_b.
\end{align}
\end{subequations}
The sorted address is unchanged during the rotation, so the same inverse value query erases the value on both components.
Reversed inputs use the other column of \cref{eq:multi-hs-edge-rotation}; a diagonal entry uses the phase $(1-ia)/(1+ia)$ with real $a=dH_{xx}/v$.
Conjugating this circuit by $O_fP_d$ implements $\mathcal U_v$ with one call in each direction of each sparse oracle.
We call this fixed sequence a query group.

Compression onto the initial index state now gives
\begin{align}
\langle0|_jT_v|0\rangle_j&=H/v,
\quad D_2:=\langle0|_jT_v^2|0\rangle_j =\frac d{v^2}\operatorname{diag}_x\sum_y|H_{xy}|^2\preceq I,
\label{eq:multi-hs-sparse-moment}
\\
\|T_v\|&\leq dh_{\max}/v.
\label{eq:s6sp-edge-norm}
\end{align}
The first identity contains two uniform amplitudes $1/\sqrt d$, which cancel the edge factor $d$.
Squaring $A_v$ instead produces the diagonal entries $d^2|H_{xy}|^2/v^2$; one uniform average gives the second identity.
In both calculations the power precedes compression.
The bound on $D_2$ is the reason for choosing $v\geq\sqrt d\,s_2$.

To obtain evolution by $H$, use the Sparse2 unitary junction
\begin{align}
\Pi_j=I\otimes(|0\rangle\langle0|)_j,
\quad S_0=\left(\begin{array}{c|c} 0&I\otimes\langle0|_j
\\
\hline
I\otimes|0\rangle_j&-(I-\Pi_j) \end{array}\right),
\quad \oracle_v(z)=z\mathcal U_v.
\label{eq:s6sp-junction}
\end{align}
It swaps the external system space with the internal subspace whose index is $|0\rangle_j$, and applies $-I$ on its orthogonal complement.
Connect the internal output to the input through $z\mathcal U_v$.
The factor $z$ records one query group.
Eliminating these internal amplitudes by \cref{eq:mod-response} gives
\begin{subequations}
\label{eq:multi-hs-sparse-response}
\begin{align}
M_v(z)&=v\langle0|_j (I-z\mathcal U_v)(I+z\mathcal U_v)^{-1}|0\rangle_j,
\\
C_v(z)&=(I-M_v(z)/v)(I+M_v(z)/v)^{-1},
\quad C_v(0)=0,
\\
F_v(z)&=\module{Exp}_{tv}(C_v(z))=e^{-tM_v(z)},
\quad F_v:=F_v(1)=e^{-itH}.
\end{align}
\end{subequations}
Indeed, $(I-\mathcal U_v)(I+\mathcal U_v)^{-1}=iT_v$, so $M_v(1)=iH$.
The exponential transform is applied after compression has recovered the physical generator.
The nested connection is shown in \cref{fig:multi-hs-sparse-exp}.

\begin{figure}[tb]
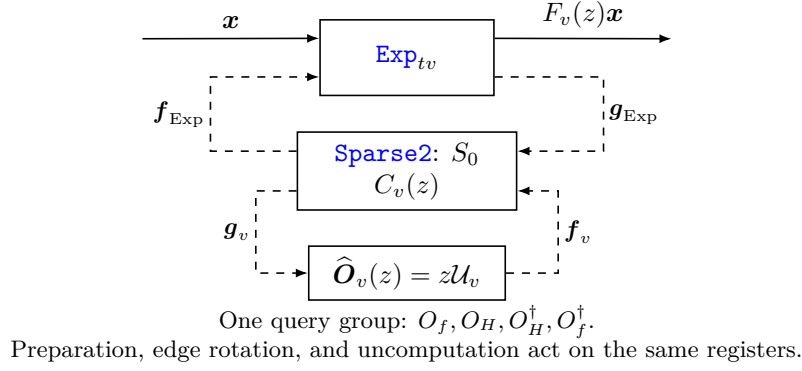

\centering

\IIRDiagram{SparseSimulationV2}

\caption{Sparse Hamiltonian simulation.
Solid arrows are external connections; dashed arrows are internal feedback connections.
The Sparse2 junction implements $C_v(z)$ from the edge query $z\mathcal U_v$.
Connecting this complete transfer function to the exponential transform gives $F_v(z)=e^{-tM_v(z)}$ and $F_v(1)=e^{-itH}$.
Each edge query uses $O_f,O_H,O_H^\dagger,O_f^\dagger$ once.
Quditization is applied to the assembled finite network, with one pair of input and output clocks.}
\label{fig:multi-hs-sparse-exp}
\end{figure}

We next verify the hypotheses of \cref{cor:iir-analytic-two}.
For $|z|<1$, unitarity of $\mathcal U_v$ implies
\begin{align}
\operatorname{Re}M_v(z) =v(1-|z|^2)\langle0|_j (I+z\mathcal U_v)^{-\dagger}(I+z\mathcal U_v)^{-1}|0\rangle_j \succeq0.
\label{eq:s6sp-positive-real}
\end{align}
Hence $F_v$ is analytic and contractive in the unit disk.
The Cayley identity $(I+\mathcal U_v)^{-1}=(I+iT_v)/2$ gives $-M_v'(1)=v(I+D_2)/2$.
Differentiating the exponential therefore yields
\begin{align}
F_v^\dagger F_v'(1) =\frac{tv}{2}\int_0^1 e^{iutH}(I+D_2)e^{-iutH}\,du,
\quad \frac{tv}{2}I\preceq F_v^\dagger F_v'(1)\preceq tvI.
\label{eq:multi-hs-sparse-action}
\end{align}
This calculation permits noncommuting $H$ and $D_2$.

For the analytic neighborhood, put
\begin{align}
\zeta=\frac1{64(1+\max\{1,dh_{\max}/v\})},
\quad D(z)=I+\frac{z-1}{2}(I-iT_v).
\label{eq:s6sp-radius}
\end{align}
On $|z-1|\leq\zeta$, \cref{eq:s6sp-edge-norm} gives $\|D(z)-I\|\leq1/128$, so $D(z)$ is invertible and $\|D(z)^{-1}\|\leq128/127$.
Factoring the two resolvents in \cref{eq:s6sp-positive-real} gives
\begin{align}
-\operatorname{Re}M_v(z) =\frac{v(|z|^2-1)}4\langle0|_j (I+T_v^2)D(z)^{-\dagger}D(z)^{-1}|0\rangle_j.
\label{eq:multi-hs-sparse-radial}
\end{align}
All factors inside the compression commute because they are functions of the same Hermitian matrix $T_v$.
For $|z|\geq1$, the second-moment bound and $(|z|^2-1)/(2\log|z|)\leq|z|^2$ imply
\begin{align}
-\operatorname{Re}M_v(z) \preceq v(128/127)^2|z|^2\log|z|\,I \preceq2v\log|z|\,I,
\quad \|F_v(z)\|\leq |z|^{2tv}.
\label{eq:s6sp-exterior}
\end{align}
The last step uses $\|e^A\|\leq\exp(\lambda_{\max}(\operatorname{Re}A))$.
Inside the disk the bound is one.
Thus the parameters in \cref{cor:iir-analytic-two} may be chosen as $a=tv/2$, $w=tv$, $h_-=tv/2$, $h_+=tv$, and $M_{\rm loc}=1$, with radius \cref{eq:s6sp-radius}.
That corollary supplies the horizon in \cref{eq:multi-hs-sparse-bound} at normalization two.

It remains to realize these coefficients by a finite circuit.
Because $C_v(0)=0$, $T_N[C_v]$ is a strictly lower-triangular contraction and $T_N[C_v]^N=0$.
Apply \cref{thm:exp-implementation} and \cref{eq:multi-hs-spectral-error} with $\tau=tv$ and $r_{\min}=1$.
They supply a finite exponential transform whose error is arbitrarily small on this entire Toeplitz matrix.
If its known unitary has blocks $A_{\module{Exp}},B_{\module{Exp}},C_{\module{Exp}},D_{\module{Exp}}$ and $D$ internal modes, the complete sparse system matrix is
\begin{subequations}
\label{eq:multi-hs-sparse-fused}
\begin{align}
S_v&=\left(\begin{array}{c|c} A_{\module{Exp}}I&B_{\module{Exp}}\otimes(I\otimes\langle0|_j)
\\
\hline
C_{\module{Exp}}\otimes(I\otimes|0\rangle_j)& D_{\module{Exp}}\otimes\Pi_j-I_D\otimes(I-\Pi_j) \end{array}\right),
\\
\oracle_{\module{Exp},v}(z)&=z(I_D\otimes\mathcal U_v).
\end{align}
\end{subequations}
On the subspace coupled to $|0\rangle_j$, $S_v$ acts as the exponential transform's known unitary tensored with the system identity.
It acts as $-I$ on the remaining internal subspace, proving full unitarity.
The mode index is unchanged by the sparse oracle calls.
Consequently, each update needs one query group, independently of $D$.
By \cref{thm:iir-lift}, the horizon-$N$ circuit uses $N-1$ groups, which proves the exact counts.

Allocate a constant fraction of $\epsilon$ to clock shaping, the finite exponential transform, and circuit synthesis.
The whole-kernel comparison bound in \cref{prop:iir-comparison-stability} transfers the finite-transform error to the encoded operator with factor two; unitary telescoping bounds the synthesis error.
These choices give \cref{eq:s6sp-guarantee}.
Optimizing the two competing terms $tv+(dh_{\max}/v)\log(1/\epsilon)$ subject to $v\geq\sqrt d\,s_2$ gives
\begin{align}
v=\max\!\left\{\sqrt d\,s_2, \sqrt{\frac{dh_{\max}\log(1/\epsilon)}t}\right\},
\label{eq:s6sp-optimal-scale}
\end{align}
which proves \cref{eq:multi-hs-sparse-optimized}.
Finally, oblivious amplitude amplification in \cref{lem:iir-cubic} triples the query costs.
Its error is linear in the operator error in \cref{eq:s6sp-guarantee}; rescaling the internal error by a constant gives the claimed simulation with auxiliary qubits returned to their initial state.
\end{proof}

\subsection{Weighted coherent state preparation}
\label{subsec:wsp}

The desired transformation maps one known reference state to a normalized linear combination of states prepared by differently priced oracles.
The oracle access model therefore comes first.
For each preparation $U_j$, use the bidirectional state-preparation query of \cref{eq:mod-preparation-query}; this turns $U_j,U_j^\dagger$ into one Hermitian signal while preserving its original query label.
A sign IIR removes the unknown branch success amplitude without measuring it.
The normalized branches are then coupled by a weighted Cayley junction, and either a weighted Hamiltonian-simulation IIR or a direct Cayley loader transfers amplitude from the reference state to the desired sum.
Fixed-point amplitude amplification preserves the prescribed relative phases across the prepared branches, up to a known common phase, while handling unknown cancellation when the final norm is not given.
We use this phase-preserving form throughout.

For comparison, black-box state preparation from numerical amplitude data can avoid coherent arithmetic~\cite{Sanders2018}.
The present input consists of preparation unitaries for component states, so its resource bound concerns coherent summation with separate query costs.

All of these components are connected before clock selection, so the algorithm has one complete analytic transfer function and one construction of a finite-query circuit.
The application-specific branch normalizers introduced below are unitary junctions of \cref{subsec:mod-junctions}; their local analysis establishes the response and query weights needed here rather than adding new general achievability assumptions.

Assume controlled access to preparations and their inverses, with
\begin{align}
U_j|0\rangle=\lambda_j|0\rangle_f|\psi_j\rangle +\boldsymbol{b}_{{\rm bad},j},
\quad \lambda_j\in[\lambda_{0j},1],
\quad \lambda_{0j}>0,
\label{eq:wsp-input}
\end{align}
where the normalized states share one output space and $(|0\rangle\langle0|_f\otimes I)\boldsymbol{b}_{{\rm bad},j}=0$.
The positive real amplitudes fix the relative phase convention.
All work required to be erased on success is included in the common flag.
The preparation and its inverse, $U_j,U_j^\dagger$, each have cost $C_j>0$.
For known complex coefficients, omit zero coefficients and write
\begin{align}
A=\sum_j|a_j|,
\quad s=\left\|\sum_j a_j|\psi_j\rangle\right\|,
\quad |\Psi\rangle=\frac{\sum_j a_j|\psi_j\rangle}{s}.
\label{eq:wsp-target}
\end{align}
Given a lower bound $s\ge s_0>0$ valid for the possible interference between branch states, \cref{thm:wsp-main} prepares the normalized sum with separate costs for the branch preparations.

\begin{theorem}[State summation with non-uniform preparation costs]
\label{thm:wsp-main}
For $0<\epsilon<1/4$, there is a finite unitary circuit with selected output vector $\boldsymbol{b}$ satisfying
\begin{align}
\left\|\boldsymbol{b}-\tfrac12|\Psi\rangle\right\|\le\epsilon/4.
\label{eq:wsp-output}
\end{align}
All selected work registers have one fixed known value.
Counting every controlled forward and inverse call to the preparations, its query cost is
\begin{align}
\sum_j C_jq_j =O\!\left[ \left\{\frac{\bigl(\sum_j\sqrt{|a_j|C_j/\lambda_{0j}}\bigr)^2}{s_0} +\sum_j\frac{C_j}{\lambda_{0j}}\right\} \log\frac1\epsilon\right].
\label{eq:wsp-unknown-cost}
\end{align}
If $s$ is given exactly, the bound improves to
\begin{align}
\sum_j C_jq_j =O\!\left[ \frac{\bigl(\sum_j\sqrt{|a_j|C_j/\lambda_{0j}}\bigr)^2}{s} +\log\frac1\epsilon\sum_j\frac{C_j}{\lambda_{0j}}\right].
\label{eq:wsp-known-cost}
\end{align}
The constants are universal.
The postselection probability satisfies $(1/2-\epsilon/4)^2\le p_{\rm succ}\le(1/2+\epsilon/4)^2$.
The conditional vector error is at most $\epsilon$.
The counts are per attempt.
\end{theorem}

The circuit approximates the prescribed amplitude $1/2$, so its finite-precision success probability has the error stated in the theorem.
Both unknown branch amplitudes and unknown cancellation are handled by phase-preserving fixed-point amplification.
With known $s$, use its known rotation angle or the corresponding simulation time.
\cref{lem:wsp-certificate} bounds the combined network shown below and determines its integer schedule.

The diagram below shows how four processing layers combine the branch transformations.
The arrows in this first diagram denote substitution of a complete transfer function into its parent before the combined network is compiled.
Original oracle calls remain marked through every layer.
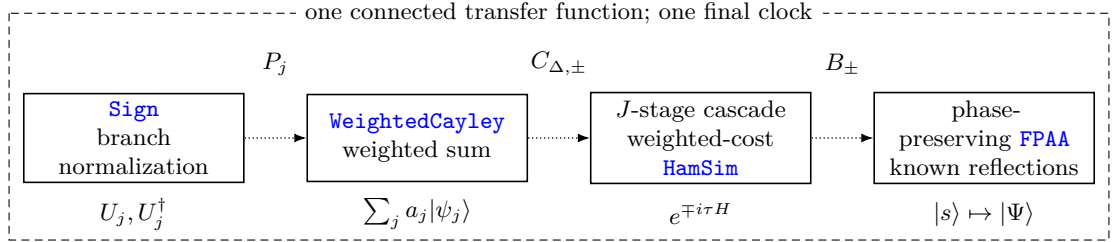
\begin{center}
\begin{adjustbox}{max width=\linewidth}
\IIRDiagram{WeightedLayers}

\end{adjustbox}
\end{center}

Directly combining the successful outputs would multiply each desired coefficient by an unknown branch amplitude.
We remove those amplitudes inside the connected transfer function before combining the branches.
This also allows one clock to approximate the combined network, without assigning a separate compilation error and normalization to every branch.
Let $|\mathrm{in}\rangle$ be the known reference state and $|\mathrm{out},\psi_j\rangle$ the accepted output in an orthogonal sector.
The projected column of $U_j$ gives the Hermitian contraction
\begin{align}
K_j=\lambda_j\bigl( |\mathrm{out},\psi_j\rangle\langle\mathrm{in}| +|\mathrm{in}\rangle\langle\mathrm{out},\psi_j|\bigr),
\quad H_j=\frac{K_j+(\lambda_{0j}/2)I}{1+\lambda_{0j}/2}.
\label{eq:wsp-branch-signal}
\end{align}
Known input and output projections, the Hermitian block-encoding construction, and an identity combination implement an involutory block encoding of $H_j$ with a constant number of calls to $U_j,U_j^\dagger$.
All block-encoding ancillas remain in the circuit.
The spectrum of $H_j$ has absolute value at least $\lambda_{0j}/(2+\lambda_{0j})$, including the original zero eigenspace.
Its sign therefore satisfies
\begin{align}
\operatorname{sgn}(H_j)|\mathrm{in}\rangle =|\mathrm{out},\psi_j\rangle.
\label{eq:wsp-polar}
\end{align}
On the input/output plane, the unshifted operator is the unknown positive amplitude times a Pauli swap.
Its eigenvectors are the symmetric and antisymmetric combinations.
The sign function maps their opposite nonzero eigenvalues to opposite unit phases, giving a unit-amplitude swap independently of the unknown magnitude.
The positive shift makes the orthogonal zero eigenspace gapped as well.
Although the shift changes the Hermitian operator itself, it preserves the signs on the two-dimensional plane and hence the desired column of its sign function.

Use the finite two-endpoint Sign path of \cref{eq:app-sign-finite,lem:finite-sign-cap}, realized by the alternating-reflection purifier \cite{Belovs2026SpaceEfficientPurification}.
Write $P_j(z)=F_{L_j}(z;H_j)$ using spectral functional calculus.
The full endpoint transfer function is retained.
At the evaluation point the transfer function is a Hermitian involution.
Its first input endpoint implements \cref{eq:wsp-polar} to error $\eta$ when $L_j=O(\lambda_{0j}^{-1}\log(2/\eta))$.
Increasing $L_j$ improves this approximation without shrinking its common analytic neighborhood beyond a constant multiple of $\lambda_{0j}$.

The sign path itself is a feedback IIR.
Let $\iota_j$ inject its two public endpoints into the complete path and let $W_j$ be the two matching layers of the alternating-reflection walk.
Its known system and its tagged transfer function are
\begin{align}
S_{{\rm sign},j}=\left(\begin{array}{c|c} 0&\iota_j^\dagger
\\
\hline
\iota_j&I-\iota_j\iota_j^\dagger \end{array}\right),
\quad P_j(z_j)=z_j\iota_j^\dagger W_j [I-z_j(I-\iota_j\iota_j^\dagger)W_j]^{-1}\iota_j.
\label{eq:wsp-sign-system}
\end{align}
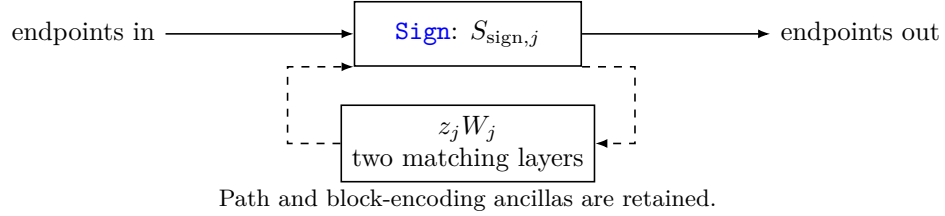
\begin{center}
\begin{adjustbox}{max width=\linewidth}
\IIRDiagram{WeightedSign}

\end{adjustbox}
\end{center}
Each matching layer uses one controlled call to the involutory encoding of $H_j$; expanding that encoding uses a constant number of $U_j,U_j^\dagger$ calls.
Solid wires denote public amplitudes and dashed wires feedback connections.

To combine the normalized branches by a weighted Cayley transformation, pad the endpoint spaces to a common space $\mathcal K$.
Let $|e\rangle$ be the copy of the reference input at the first endpoint.
The public space is $\mathbb C|s\rangle\oplus\mathcal K$.
For each $j$, define the known isometry
\begin{align}
J_j|s\rangle=e^{i\arg a_j}|0\rangle|e\rangle,
\quad J_j|v\rangle=|1\rangle|v\rangle
\quad(v\in\mathcal K),
\quad E=\bigoplus_j\sqrt{|a_j|/A}\,J_j.
\label{eq:wsp-injections}
\end{align}
Then $E^\dagger E=I$.
Form the constituent transfer functions $R_j(z_j)=\sigma_x\otimes P_j(z_j)$.
Their weighted Cayley combination is
\begin{subequations}
\label{eq:wsp-cayley-transfer}
\begin{align}
M_\pm(\Bz) &=\sum_j\frac{|a_j|}{A}J_j^\dagger (I\pm iR_j(z_j))(I\mp iR_j(z_j))^{-1}J_j,
\\
F_\pm(\Bz)&=(I-M_\pm(\Bz))(I+M_\pm(\Bz))^{-1}.
\end{align}
\end{subequations}
A unitary system matrix realizing this transfer function is
\begin{align}
S=\left(\begin{array}{c|c}0&E^\dagger
\\
\hline
E&-I+EE^\dagger\end{array}\right).
\label{eq:wsp-system}
\end{align}
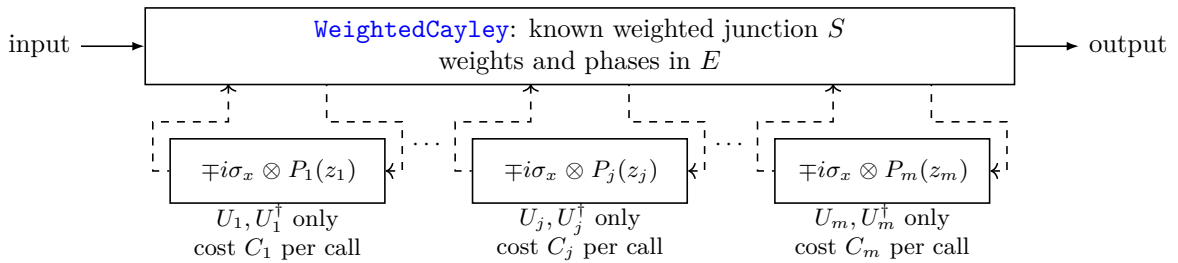
\begin{center}
\begin{adjustbox}{max width=\linewidth}
\IIRDiagram{WeightedJunction}

\end{adjustbox}
\end{center}
Closing its private ports through $\mp iR_j(z_j)$ gives the stated transfer function.
At $\Bz=\boldsymbol{1}$ the branch functions are Hermitian involutions.
Thus $F_-=F_+^\dagger$ even for finite branch paths.
For ideal branch normalization, put
\begin{align}
H=\frac1A\begin{pmatrix} 0&\sum_j\overline{a_j}\langle\psi_j|
\\
\sum_j a_j|\psi_j\rangle&0 \end{pmatrix}.
\label{eq:wsp-sum-signal}
\end{align}
Known output embeddings are suppressed in this display.
Then $F_+=(I-iH)(I+iH)^{-1}$ and
\begin{align}
F_+|s\rangle =\frac{1-(s/A)^2}{1+(s/A)^2}|s\rangle -i\frac{2s/A}{1+(s/A)^2}|\Psi\rangle.
\label{eq:wsp-cayley-column}
\end{align}
The known phase $-i$ is corrected before amplification.

Hamiltonian evolution provides another way to transfer the reference amplitude into the desired sum, because the star Hamiltonian in \cref{eq:wsp-sum-signal} couples these two states.
Approximate its exponential using the small-step Cayley transfer functions and their cascade:
\begin{subequations}
\label{eq:wsp-hamsim-response}
\begin{align}
C_{\Delta,\pm}(\Bz) &=[I-(\Delta/2)M_\pm(\Bz)] [I+(\Delta/2)M_\pm(\Bz)]^{-1},
\\
\mathcal U_{\tau,J,\pm}(\Bz) &=C_{\tau/J,\pm}(\Bz)^J \longrightarrow e^{-\tau M_\pm(\Bz)}.
\end{align}
\end{subequations}
The system matrix for each small step is
\begin{align}
S_\Delta=\left(\begin{array}{c|c} \dfrac{1-\Delta/2}{1+\Delta/2}I& \dfrac{2\sqrt{\Delta/2}}{1+\Delta/2}E^\dagger
\\[3pt]
\hline
\dfrac{2\sqrt{\Delta/2}}{1+\Delta/2}E& -I+\dfrac{\Delta}{1+\Delta/2}EE^\dagger \end{array}\right),
\quad \Delta>0.
\label{eq:wsp-small-cayley-system}
\end{align}
It uses the same separate feedback connections as the weighted Cayley transformation.
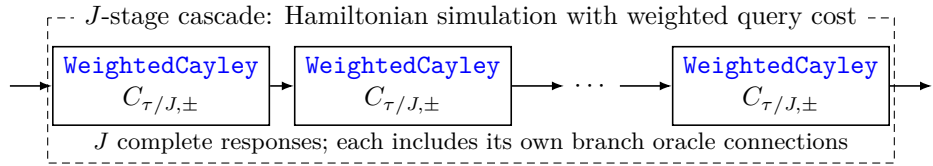
\begin{center}
\begin{adjustbox}{max width=\linewidth}
\IIRDiagram{WeightedCascade}

\end{adjustbox}
\end{center}
In the ideal branch limit, the transfer function at $\Bz=\Bone$ approaches $e^{\mp i\tau H}$.
In particular,
\begin{align}
e^{-i\tau H}|s\rangle =\cos(\tau s/A)|s\rangle-i\sin(\tau s/A)|\Psi\rangle.
\label{eq:wsp-hamsim-column}
\end{align}
When the superposition norm is unknown, choose $\tau=\pi/2$.
The accepted amplitude then satisfies \begin{align}
\sin(\pi s/(2A))\ge s/A\ge s_0/A.
\end{align} For known $s$, choosing $\tau=\pi A/(2s)$ gives unit accepted amplitude and no outer FPAA is needed.
The cascade is finite: taking $J\ge\tau$ and sufficiently large to make $\tau^3/(12J^2)$ smaller than the allocated simulation error is sufficient.
The factor $J$ increases the retained processing dimension while preserving the original query count per clock event.

Let $\Pi_g$ project onto the star Hamiltonian's output sector.
Define the reference projector and phase correction by $\Pi_s=|s\rangle\langle s|$ and $Z_g=\Pi_s+i\Pi_g$.
Define the analytic forward and reverse loaders
\begin{align}
B_+(\Bz)=Z_g\mathcal U_{\tau,J,+}(\Bz),
\quad B_-(\Bz)=\mathcal U_{\tau,J,-}(\Bz)Z_g^\dagger,
\quad B_-=B_+^\dagger.
\label{eq:wsp-hamsim-loaders}
\end{align}
Their ideal accepted amplitudes have the prescribed positive phase.
The phase rotations on the reference state and success subspace are known gates,
\begin{align}
R_s(\phi)=I+(e^{i\phi}-1)\Pi_s,
\quad R_g(\theta)=I+(e^{i\theta}-1)\Pi_g.
\label{eq:wsp-phase-reflections}
\end{align}
One phased Grover step has the full transfer function
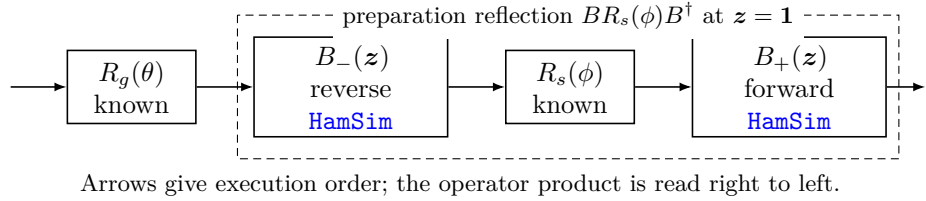
\begin{align}
Q_{\phi,\theta}(\Bz) =B_+(\Bz)R_s(\phi)B_-(\Bz)R_g(\theta).
\label{eq:wsp-grover-response}
\end{align}
\begin{center}
\begin{adjustbox}{max width=\linewidth}
\IIRDiagram{WeightedFpaa}

\end{adjustbox}
\end{center}
Phase-preserving FPAA assembles these forward and reverse calls with the known phase reflections and retains any phase-selection ancilla.
The reverse box has its own analytic delay variables.
Its value at $\Bz=\boldsymbol{1}$ is the adjoint of the forward transformation.

To amplify the combined output before finite-circuit construction from the chosen clock, use a finite phase-preserving amplification circuit, including its known reflections about the reference state and success subspace.
If $s$ is unknown, a real odd singular-value filter gives Euclidean error of the joint output state $O(\epsilon)$ using
\begin{align}
D=O\!\left(\frac{A}{s_0}\log\frac1\epsilon\right)
\label{eq:wsp-outer-degree}
\end{align}
calls to either the direct Cayley loader $F_+$ or the Hamiltonian-simulation loader $B_+$, and the corresponding inverse~\cite{Gilyen2018singular}.
For the latter, $\tau=\pi/2$ is fixed.
The selected-polynomial error is chosen $O(\epsilon^2)$ to control the unselected amplitude as well.
If $s$ is given, known-angle amplitude amplification gives an exact ideal output with $D=O(A/s)$ \cite{brassard2002quantum}.
Alternatively, the known-time Hamiltonian simulation in \cref{eq:wsp-hamsim-column} uses one loader.
Writing the known gates as $V_k$, the direct-Cayley transfer function is a finite product
\begin{align}
\mathcal F(\Bz)= V_D F_{\sigma_D}(\Bz)V_{D-1}\cdots V_1F_{\sigma_1}(\Bz)V_0,
\quad \sigma_k\in\{+,-\}.
\label{eq:wsp-complete-response}
\end{align}
Controlled calls mean direct sums with identity in this formula.
Replacing every $F_\pm$ by $B_\pm$ gives the HamSim--FPAA transfer function.
The branch paths, the Cayley connections, and this product are assembled before any clock is selected.
Take $\eta=\Theta(\epsilon/D)$ for the direct-Cayley branch error.
For HamSim, take branch error $O(\epsilon/(D\tau))$ and per-loader cascade error $O(\epsilon/D)$.

\cref{lem:wsp-certificate} gives one joint analytic neighborhood for the whole preparation network, so its components can share a single clock.

\begin{lemma}[Joint analytic bounds]
\label{lem:wsp-certificate}
Choose integer delays $r_j\ge1$ and form $G(z)=\mathcal F(z^{r_1},\ldots,z^{r_m})$.
This finite transfer function is contractive in the unit disk, unitary at $z=1$, and analytic on
\begin{align}
|z-1|\le c\left(\max_j\frac{r_j}{\lambda_{0j}}\right)^{-1}.
\label{eq:wsp-radius}
\end{align}
On a possibly smaller such neighborhood its exterior growth satisfies
\begin{align}
\log\|G(z)\| \le c'\frac{D}{A}\sum_j\frac{|a_j|r_j}{\lambda_{0j}}\log|z|
\quad(|z|\ge1).
\label{eq:wsp-growth}
\end{align}
Here $c,c'>0$ are universal and independent of the path lengths, once $L_j\lambda_{0j}$ exceeds a universal constant.
\end{lemma}

Applying the analytic clock theorem to these bounds gives normalization two and horizon
\begin{align}
N=O\!\left(\frac{D}{A}\sum_j\frac{|a_j|r_j}{\lambda_{0j}} +\max_j\frac{r_j}{\lambda_{0j}}\log\frac1\epsilon\right).
\label{eq:wsp-horizon}
\end{align}
With normalized input and output clock states, the unitary Toeplitz lift in \cref{thm:iir-lift} then makes $\lfloor(N-1)/r_j\rfloor$ calls to each feedback operation.
Expanding the constant-size block-encoding constructions counts the original $U_j,U_j^\dagger$ calls.

For the HamSim--FPAA route, the same proof uses the product $D\tau$ in place of the outer degree alone.
It is uniform for sufficiently large cascade length $J$.
Hence
\begin{align}
N=O\!\left(\frac{D\tau}{A}\sum_j\frac{|a_j|r_j}{\lambda_{0j}} +\max_j\frac{r_j}{\lambda_{0j}}\log\frac1\epsilon\right).
\label{eq:wsp-hamsim-horizon}
\end{align}
The choices $(\tau,D)=(\pi/2,O((A/s_0)\log(1/\epsilon)))$ and $(\pi A/(2s),1)$ therefore give the two bounds in \cref{thm:wsp-main}.
\cref{app:weighted-state-preparation} proves these bounds, including the finite-cascade error.
There is no intermediate postselection or multiplication of clock normalizations.
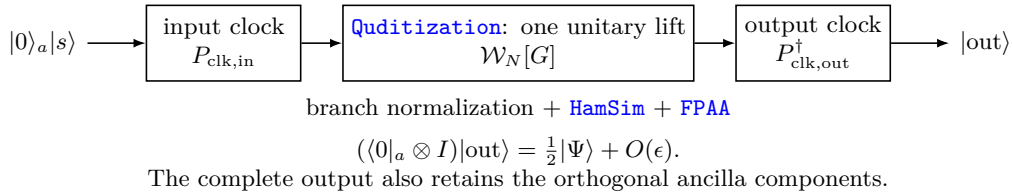
\begin{center}
\begin{adjustbox}{max width=\linewidth}
\IIRDiagram{WeightedClock}

\end{adjustbox}
\end{center}
The last arrow denotes the unnormalized successful output vector.
All unselected and private output amplitudes remain in the unitary circuit until this selection.

The assembled public and private dimensions, including the branch paths and all module occurrences, enter the common resource bound as specified in \cref{eq:wsp-dimensions,eq:wsp-width}.

Optimizing the continuous leading cost by \cref{lem:common-delay-allocation}, with $p=1$, gives
\begin{align}
\inf_{r_j>0} \left(\sum_j\frac{|a_j|r_j}{\lambda_{0j}}\right) \left(\sum_j\frac{C_j}{r_j}\right) =\left(\sum_j\sqrt{\frac{|a_j|C_j}{\lambda_{0j}}}\right)^2.
\label{eq:wsp-sharp-allocation}
\end{align}
For fixed $\lambda_0$, the scalar phase-preserving sign benchmark has degree
\begin{align}
\frac{\log(1/\epsilon)}{\operatorname{atanh}\lambda_0} +O_{\lambda_0}\bigl(\log\log(1/\epsilon)\bigr).
\label{eq:wsp-scalar-benchmark}
\end{align}
This follows from the minimax sign asymptotics \cite{Eremenko2007SignApproximation}.
It does not establish a sharp constant for \cref{thm:wsp-main}: the shift in \cref{eq:wsp-branch-signal}, the oracle constructions, and the joint analytic-clock bound still contribute constant factors.
No optimality over all weighted preparation algorithms is claimed.

An application is physical preparation of a state obtained from a nonorthogonal variational subspace, including eigenvector continuation \cite{Frame2018EigenvectorContinuation}.
Given the Gram matrix $G_{jk}=\langle\psi_j|\psi_k\rangle$ and classically chosen coefficients, $s=\sqrt{a^\dagger Ga}$.
An exact value permits \cref{eq:wsp-known-cost}; a lower bound permits \cref{eq:wsp-unknown-cost}.
Estimating the Gram matrix, finding the coefficients, and controlling subspace approximation error are separate tasks.
The trial-state preparations may already be available with different query costs.

\subsection{Quantum linear systems problems}
\label{wql:sec:qlsp}

The target is the normalized solution of a linear equation, but the transfer function is chosen before an approximation to $1/x$.
We encode the relation $A\vect x=\vect b$ as a unitary constraint network: matrix terms enter through separate Cayley ports, the right-hand-side vectors enter through state-preparation oracles and their inverses, and closing the auxiliary coordinates makes the private amplitudes satisfy the linear equation.
This is an application-specific unitary junction built from \cref{subsec:mod-junctions,subsec:mod-connect}, so its achievability follows from the complete network rather than from a scalar reciprocal ansatz.

A coarse finite clock then leaves the unsatisfied part of the equation as a terminal private vector.
A matrix-only reciprocal IIR corrects this residual, which separates high-precision matrix queries from state preparation.
For the single-matrix case, the exact identity proved in \cref{wql:eq:one-residual} is
\begin{align}
H\boldsymbol{v}_{N_0,\rho} =\frac{2c_{N_0,\rho}}{\rho}|b\rangle +\frac{2i}{\kappa\sqrt{N_0}}L_{\rho,s}\boldsymbol{e}_{N_0,\rho}.
\label{wql:eq:residual-preview}
\end{align}
The first vector is the public output averaged with a uniform clock; the second is the unmeasured amplitude in the final private delay slots.
The terminal map converts that private amplitude into the equation residual.
The coefficient of the prepared input state is a scalar determined by the finite run.
A horizon $N_0=O(\kappa)$ suffices for constant initial accuracy.
Applying a high-precision reciprocal to the terminal map's output then cancels the residual, as derived in \cref{wql:eq:one-readout}.
Accuracy improvements therefore require no additional preparation queries.
The multioracle construction uses the same conservation law and chooses a coarse horizon from the weighted rates in \cref{wql:eq:rates}.

Let $A=\sum_{j=1}^m A_j$ be invertible on a finite-dimensional Hilbert space $\mathcal H$.
For each term, assume controlled access to $U_j=\Be{A_j/\lambda_j}$ and its inverse.
The positive normalization $\lambda_j$ is given.
Thus
\begin{align}
(\langle0|\otimes I)U_j(|0\rangle\otimes I)=A_j/\lambda_j.
\label{wql:eq:access}
\end{align}
Let $B_k|0\rangle=|b_k\rangle$, $k=1,\ldots,n$, be unitary preparations of normalized states in the same output space, with any preparation workspace returned to zero.
Controlled $B_k$ and $B_k^\dagger$ are available.
The states may be nonorthogonal.
For known nonzero complex coefficients $a_k$, define
\begin{align}
\boldsymbol{b}=\sum_{k=1}^n a_k|b_k\rangle,
\quad K\ge\|A^{-1}\|,
\quad 0<R_0\le\|A^{-1}\boldsymbol{b}\|.
\label{wql:eq:promises}
\end{align}
The norm of $\boldsymbol{b}$ and the actual solution norm need not be known beyond the given bounds $K,R_0$.
Let $C_j>0$ and $c_k>0$ be the respective costs of a controlled call in either direction, and write
\begin{align}
\langle\boldsymbol{C},\boldsymbol{\lambda}\rangle_{1/2} =\left(\sum_j\sqrt{C_j\lambda_j}\right)^2,
\quad \langle\boldsymbol{c},|\boldsymbol{a}|\rangle_{1/2} =\left(\sum_k\sqrt{c_k|a_k|}\right)^2.
\label{wql:eq:paired-cost}
\end{align}

\cref{wql:thm:main} bounds the two weighted query costs for the normalized inverse state.
Its realization and error analysis are in \cref{wql:app:proofs}; \cref{prop:qlsp-resources} gives the known-gate and qubit costs of that realization.
Its controlled small matrices and shared-oracle cascade are implemented by \cref{lem:small-matrix,lem:ordered-cascade}.

\begin{theorem}[Linear systems with separately given matrices and states]
\label{wql:thm:main}
Under \cref{wql:eq:access,wql:eq:promises}, for $0<\epsilon<1/4$ a finite MQSP circuit succeeds with probability at least $1/4$.
Its conditional system state has trace distance at most $\epsilon$ from
\begin{align}
|\mathrm{sol}\rangle =\frac{A^{-1}\boldsymbol{b}}{\|A^{-1}\boldsymbol{b}\|}.
\label{wql:eq:target}
\end{align}
Its original-oracle counts satisfy
\begin{align}
\sum_j C_jQ_j &=O\!\left[ \left(K\langle\boldsymbol{C},\boldsymbol{\lambda}\rangle_{1/2} +\|\boldsymbol{C}\|_1\right)\log\frac1\epsilon \right],
\label{wql:eq:matrix-cost}
\\
\sum_k c_kQ_{b,k} &=O\!\left[ \frac K{R_0}\langle\boldsymbol{c},|\boldsymbol{a}|\rangle_{1/2} +\|\boldsymbol{c}\|_1 \right].
\label{wql:eq:state-cost}
\end{align}
One common choice of delays attains both bounds.
Their constants are independent of $m,n$, the dimension, and the given numerical parameters.
All forward, inverse, and unsuccessful-branch calls are included.
No commutativity of the matrix terms is assumed.
\end{theorem}

The effective solution normalization is at most two; the successful amplitude may depend on the instance and need not equal $1/2$.
The output may retain a scale register that factors from the ideal data state.
The query bounds exclude the gates for known coefficient loading, clock preparation, routing, and synthesis.

When the matrix is the square of a given Hermitian factor, \cref{wql:thm:factor} reduces the dependence on its inverse norm.
The factor construction is proved in \cref{wql:subsec:factor-proof}.

\begin{theorem}[A Hermitian factor given as a sum]
\label{wql:thm:factor}
Suppose a Hermitian factor and its components satisfy
\begin{align}
T=\sum_{j=1}^mT_j=T^\dagger,
\quad T_j=T_j^\dagger,
\quad\|T\|\le1.
\end{align}
The positive target and its inverse bound are $H=T^2$, $\|H^{-1}\|\le\kappa$, and $\kappa\ge2$.
Each factor term has a controlled self-inverse encoding $O_{T,j}=\Be{T_j/\lambda_j}$.
Its cost is $C_j$ in either direction.
Use the deterministic preparations and coefficients above, and a given bound $0<R_0\le\|T^{-2}\boldsymbol{b}\|$.
Define
\begin{align}
r_0=\frac{R_0}{\|\boldsymbol{a}\|_1},
\quad D=\frac{\kappa}{r_0}=\frac{\kappa\|\boldsymbol{a}\|_1}{R_0}\ge1.
\label{wql:eq:factor-promises}
\end{align}
There is a finite MQSP circuit with the output guarantee of \cref{wql:thm:main}, for $A=T^2$, and original-oracle counts
\begin{align}
\sum_j C_jQ_{T,j} &=O\!\left[ \left(\sqrt{\kappa D}+\sqrt\kappa\log\frac1\epsilon\right) \langle\boldsymbol{C},\boldsymbol{\lambda}\rangle_{1/2} +\|\boldsymbol{C}\|_1\log\frac1\epsilon\right],
\label{wql:eq:factor-matrix-cost}
\\
\sum_k c_kQ_{b,k} &=O\!\left[ \frac\kappa{R_0}\langle\boldsymbol{c},|\boldsymbol{a}|\rangle_{1/2} +\|\boldsymbol{c}\|_1\right].
\label{wql:eq:factor-state-cost}
\end{align}
The constants are independent of $m,n$ and the given parameters.
The coarse and precision stages may use different matrix delays; both use the original factor oracles.
\end{theorem}

In \cref{wql:thm:factor}, $H=(\sum_j T_j)^2$: the terms are summed before squaring.
The hypotheses do not cover an arbitrary sum $\sum_j T_j^2$ or factor access inferred from a block encoding of $H$.
Cancellation in the right-hand side affects the inverse-amplitude ratio through the lower solution-norm bound.
It can make $r_0$ smaller than one.

\cref{wql:cor:single} specializes both results to one matrix oracle and one preparation, using a lower bound on the inverse-state amplitude.

\begin{corollary}[One matrix and one prepared state]
\label{wql:cor:single}
Let the Hermitian contraction obey $H=H^\dagger$, $\|H\|\le1$, $\|H^{-1}\|\le\kappa$, and $\kappa\ge2$.
Suppose a controlled self-inverse block encoding $O_H=\Be{H}$ and a unitary $B|0\rangle=|b\rangle$ are given.
Given
\begin{align}
\kappa^{-2}\le p\le \frac{\|H^{-1}|b\rangle\|^2}{\kappa^2},
\label{wql:eq:single-promise}
\end{align}
the solution can be prepared with the output guarantee of \cref{wql:thm:main} using
\begin{align}
Q_H=O\!\left(\kappa\log\frac1\epsilon\right),
\quad Q_b=O(p^{-1/2}).
\label{wql:eq:single-cost}
\end{align}
For factor access, assume $H=T^2$ with $T=T^\dagger$.
Given a controlled self-inverse block encoding $O_T=\Be{T}$, the same output guarantee holds with
\begin{align}
Q_T=O\!\left(\sqrt\kappa\,p^{-1/4} +\sqrt\kappa\log\frac1\epsilon\right),
\quad Q_b=O(p^{-1/2}).
\label{wql:eq:single-factor-cost}
\end{align}
\end{corollary}
\begin{proof}
Specialize to one matrix and one preparation, with $m=n=1$, $\lambda_1=a_1=C_1=c_1=1$, and $R_0=\kappa\sqrt p$.
The direct result follows from \cref{wql:thm:main} with $K=\kappa$.
For factor access, \cref{wql:thm:factor} has $D=p^{-1/2}$, and \cref{wql:eq:factor-matrix-cost,wql:eq:factor-state-cost} give \cref{wql:eq:single-factor-cost}.
The extra $\log(1/\epsilon)$ term is absorbed by $\sqrt\kappa\log(1/\epsilon)$.
\end{proof}

The one-matrix construction in \cref{wql:subsec:one,wql:subsec:finite} uses the energy identity to bound the two oracle weights and the unitary Toeplitz lift to retain the finite computation's residual.
The analytic clock theorem determines the horizon of the matrix-only reciprocal circuit applied to that residual.
These steps use the common Cayley junction and compilation results of \cref{lem:common-cayley,thm:iir-lift,thm:iir-analytic-clock-selection}.
Two factor couplings give the $\sqrt\kappa$ dependence in \cref{wql:subsec:factor}; separate couplings for every summand give the weighted bounds in \cref{wql:subsec:sums,wql:subsec:allocation}.
The organization into a coarse computation and a matrix-only correction also occurs in the state-preparation-optimal algorithm of Low and Su~\cite{Low2024QLSP}.
Here the correction follows from the feedback equation itself.

\subsubsection{Nullspace reflection and scale cascade}
\label{wql:subsec:one}

Use the public and private spaces and feedback equations of \cref{eq:intro-response}, with $\oracle(\Bz)$ specifying this construction's private oracle connections.
The finite lift retains its final private amplitudes, as in \cref{thm:iir-lift}.

The desired transfer function is a reflection about the solutions of the constraint.
With invertible $H$, append one scalar reference coordinate and write
\begin{align}
A_\rho=(H,-|b\rangle/\rho),
\quad \ker A_\rho=\operatorname{span}\left\{ \binom{H^{-1}|b\rangle/\rho}{1}\right\}.
\label{wql:eq:nullspace-design}
\end{align}
Reflection about this line sends part of the scalar reference input into the inverse-state direction.
The Hermitian load below realizes that reflection by making the average public wave satisfy the constraint and the difference wave lie in its orthogonal complement.
The public space is $\mathcal H\oplus\mathbb C$.
We initialize its one-dimensional summand and select the output in $\mathcal H$.
Write
\begin{align}
\boldsymbol{u}=(\boldsymbol{u}_H,u_0),
\quad \boldsymbol{v}=(\boldsymbol{v}_H,v_0),
\quad \boldsymbol{x}=\frac{\boldsymbol{u}_H+\boldsymbol{v}_H}{2},
\quad c=\frac{u_0+v_0}{2}.
\label{wql:eq:public}
\end{align}
The average amplitudes are required to satisfy
\begin{align}
H\boldsymbol{x}=\frac c\rho|b\rangle,
\quad \rho>0.
\label{wql:eq:one-constraint}
\end{align}
There is one scalar $c$ for the entire vector equation.
It is determined by feedback and need not be known.

\begin{figure}[tbp]
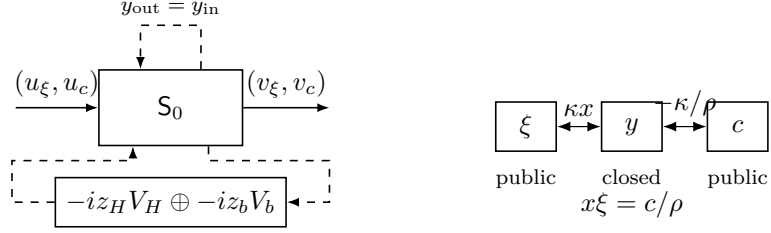

\centering

\IIRDiagram{QLSPConstraint}

\caption{The average of the incoming and outgoing scalar vectors lies on the constraint line (left).
Closing the auxiliary coordinate of the Cayley load enforces that equation (right).
The load connections shown here are realized by the known system and private oracle connections in \cref{wql:fig:returns}.}
\label{wql:fig:constraint}
\end{figure}

For a real scalar signal $x\ne0$, impose $x\xi=c/\rho$ on the average of the incoming and outgoing vectors, and require their difference to be orthogonal to this line.
The outgoing vector is then the reflection of the incoming vector across the line, so its norm is preserved.
\cref{wql:fig:constraint} shows the scalar constraint; one closed coordinate enforces these conditions through the Hermitian load
\begin{align}
\mathcal K_\rho(x) =\kappa\left(\begin{array}{cc|c} 0&0&x
\\
0&0&-1/\rho
\\
\hline
x&-1/\rho&0 \end{array}\right).
\label{wql:eq:scalar-load}
\end{align}
The first two coordinates are public.
The last output is connected to its input.
The Cayley equations give
\begin{align}
\begin{pmatrix}v_1
\\
v_0
\\
y\end{pmatrix} =(I-i\mathcal K_\rho(x))(I+i\mathcal K_\rho(x))^{-1} \begin{pmatrix}u_1
\\
u_0
\\
y\end{pmatrix},
\quad x\frac{u_1+v_1}{2}=\frac{u_0+v_0}{2\rho}.
\label{wql:eq:scalar-closure}
\end{align}
Eliminating $y$ gives
\begin{align}
F_\rho(x) =\frac1{1+\rho^2x^2} \begin{pmatrix} 1-\rho^2x^2&2\rho x
\\
2\rho x&\rho^2x^2-1 \end{pmatrix}.
\label{wql:eq:scalar-response}
\end{align}
The entry $2\rho x/(1+\rho^2x^2)$ depends on the scalar average $c$.
That average itself depends on the signal.
In \cref{wql:eq:one-constraint}, one scalar $c$ multiplies every component of $H^{-1}|b\rangle$, giving the required solution direction.

The same calculation applies to a vector-valued classical network.
Set
\begin{align}
A_\rho=(H,-|b\rangle/\rho),
\quad \mathcal K_\rho =\kappa\left(\begin{array}{cc|c} 0&0&H
\\
0&0&-\langle b|/\rho
\\
\hline
H&-|b\rangle/\rho&0 \end{array}\right).
\label{wql:eq:one-load}
\end{align}
The auxiliary space is a copy $\mathcal Y\simeq\mathcal H$.
Closing it gives $(\boldsymbol{u}+\boldsymbol{v})/2\in\ker A_\rho$ and $(\boldsymbol{u}-\boldsymbol{v})/2\in\ran A_\rho^\dagger$.
These subspaces are orthogonal complements, so the public response at $\Bz=\boldsymbol{1}$ is
\begin{align}
F_\rho(\Bone)=F_\rho=2\Pi_{\ker A_\rho}-I,
\quad \ker A_\rho= \operatorname{span}\left\{ \begin{pmatrix}H^{-1}|b\rangle/\rho
\\
1\end{pmatrix}\right\}.
\label{wql:eq:one-reflection}
\end{align}
For $\boldsymbol{u}=(0,1)$, with $r=\|H^{-1}|b\rangle\|$, this gives
\begin{align}
F_\rho\begin{pmatrix}0
\\
1\end{pmatrix} =\begin{pmatrix} \displaystyle\frac{2H^{-1}|b\rangle}{\rho(1+r^2/\rho^2)}
\\[5pt]
\displaystyle\frac{1-r^2/\rho^2}{1+r^2/\rho^2} \end{pmatrix}.
\label{wql:eq:one-column}
\end{align}
The $\mathcal H$ component is an exact multiple of $H^{-1}|b\rangle$.
In this wave-digital scattering junction~\cite{Fettweis1986WaveDigitalFilters}, the coupling through $|b\rangle$ keeps the second public summand one-dimensional under the matrix substitution.

\label{wql:subsec:finite}

\begin{figure}[tbp]
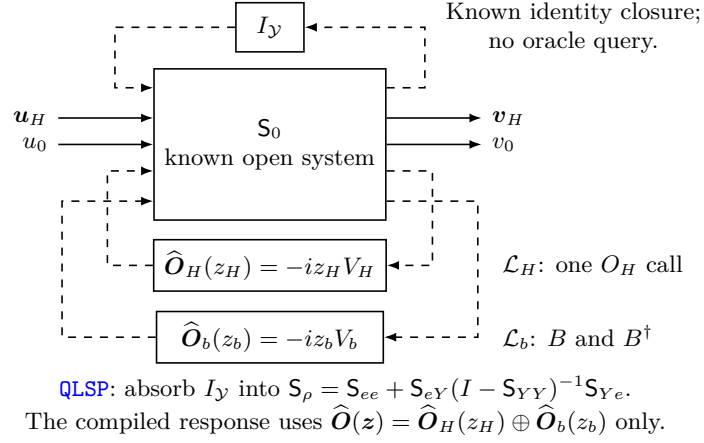

\centering

\IIRDiagram{QLSPReturns}

\caption{The known Cayley system realizes the load through separate matrix and preparation oracles.
Public connections are solid and internal connections are dashed.
The identity connection on $\mathcal Y$ is eliminated in $\mathsf S_\rho$; the full matrix and preparation oracle spaces remain private.}
\label{wql:fig:returns}
\end{figure}

The unknown entries of the load in \cref{wql:eq:one-load} enter through private oracle connections.
On coordinates $(\boldsymbol{x},c,\boldsymbol{y})$, the Cayley junction uses the known maps
\begin{align}
\mathcal C_H(\boldsymbol{x},c,\boldsymbol{y}) &=\sqrt\kappa \begin{pmatrix}(|0\rangle\otimes I)\boldsymbol{x}
\\
(|0\rangle\otimes I)\boldsymbol{y}\end{pmatrix},
\quad V_H=\begin{pmatrix}0&O_H
\\
O_H&0\end{pmatrix},
\label{wql:eq:one-matrix-map}
\\
\mathcal C_b(\boldsymbol{x},c,\boldsymbol{y}) &=\sqrt{\kappa/\rho} \begin{pmatrix}(|0\rangle_{w}\otimes I)\boldsymbol{y}
\\
c|0\rangle\end{pmatrix},
\quad V_b=-\begin{pmatrix}0&B
\\
B^\dagger&0\end{pmatrix}.
\label{wql:eq:one-state-map}
\end{align}
The subscript $w$ denotes the preparation workspace returned to zero; it is omitted when no such workspace is present.
The private spaces contain the full Hilbert spaces of the oracle unitaries, including their ancillas.
With $\mathcal C=\mathcal C_H\oplus\mathcal C_b$, define
\begin{align}
\mathsf R=(I+\mathcal C^\dagger\mathcal C)^{-1},
\quad \mathsf S_0= \left(\begin{array}{c|c} 2\mathsf R-I&2\mathsf R\mathcal C^\dagger
\\
\hline
2\mathcal C\mathsf R&2\mathcal C\mathsf R\mathcal C^\dagger-I \end{array}\right),
\quad \oracle(\Bz)=(-iz_HV_H)\oplus(-iz_bV_b).
\label{wql:eq:one-system}
\end{align}
\cref{wql:fig:returns} shows the oracle connections.
The line in $\mathsf S_0$ separates the open load coordinates from the private oracle spaces.
Closing its $\mathcal Y$ coordinate leaves public space $\mathcal H\oplus\mathbb C$.
The inverse defining $\mathsf R$ is known:
\begin{align}
\mathcal C^\dagger\mathcal C =\diag\bigl(\kappa I,\kappa/\rho, (\kappa+\kappa/\rho)I\bigr).
\label{wql:eq:one-known-gram}
\end{align}
Thus constructing $\mathsf R$ requires only known scalar inverses.
Each matrix-feedback operation uses one call to $O_H$.
Each preparation-feedback operation uses one controlled call to $B$ and one to $B^\dagger$.
\cref{lem:common-cayley} proves the junction identity.

If the solution norm were available, matching the reference scale to it would transmit all probability into the solution component of \cref{wql:eq:one-column}.
This is impedance matching in the scattering interpretation.
With only a lower bound, use a descending sequence of reference scales:
\begin{align}
r_0=\kappa\sqrt p,
\quad D=\kappa/r_0,
\quad \rho_j=2^jr_0,
\quad j=\lceil\log_2D\rceil,\ldots,0.
\label{wql:eq:one-scales}
\end{align}
Here $D$ is the inverse-amplitude ratio.
Connect each scalar output to the next scalar input, and keep the system outputs in orthogonal scale sectors.
They all point in the same solution direction.
A stage whose reference scale greatly exceeds the solution norm reflects most of the amplitude onward, with matrix catalyst weight proportional to $\kappa(r/\rho_j)^2$.
Eventually a scale brackets the solution norm, giving
\begin{align}
\frac12\le\frac r{\rho_j}\le1,
\quad \left[\frac{2(r/\rho_j)}{1+(r/\rho_j)^2}\right]^2\ge\frac{16}{25}.
\end{align}
This lower bound concerns the probability arriving at that stage.
Earlier transmissions are already successful; adding them proves that the total steady-state transmission is at least the same constant.
The cascade starts from one scalar input and requires no independent preparation at each stage.

\begin{figure}[tbp]
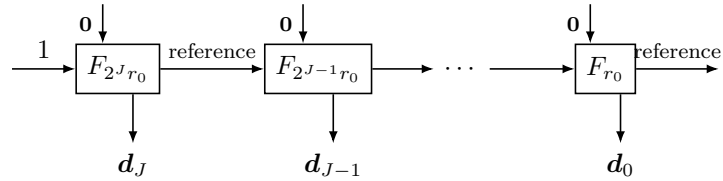

\centering

\IIRDiagram{QLSPCascade}

\caption{The scalar output of each scale system matrix drives the next system matrix.
The vertical outputs retain the scale label and share the solution direction.
Here $J=\lceil\log_2D\rceil$; unused $\mathcal H$ inputs are zero and are omitted.}
\label{wql:fig:cascade}
\end{figure}

\cref{wql:fig:cascade} retains the system output of every scale.
The steady-state internal averages within a single system matrix are
\begin{align}
c=\frac1{1+r^2/\rho^2},
\quad \boldsymbol{x}=\frac c\rho H^{-1}|b\rangle,
\quad \boldsymbol{y}=\frac i\kappa H^{-1}\boldsymbol{x}.
\label{wql:eq:one-steady-state}
\end{align}
By the energy identity, feedback connection $j$ has squared amplitude $2\|\mathcal C_j\boldsymbol{h}\|^2$.
The bound $\|\boldsymbol{y}\|\le\|\boldsymbol{x}\|$ makes the matrix weight at most $\kappa$ times the transmission probability.
Summing over the cascade telescopes these probabilities, while a geometric sum in $1/\rho$ bounds the preparation weights:
\begin{align}
W_H\le\kappa,
\quad W_b<4D.
\label{wql:eq:one-weights}
\end{align}
The factor $\kappa$ in the load balances the two internal vectors in \cref{wql:eq:one-steady-state}; it has no effect on the public reflection.

Choose preparation delay $s=\lceil r_0\rceil$ and matrix delay one.
Use a uniform clock whose horizon is proportional to $\kappa$.
Its delayed steady-state weight is $W\le W_H+sW_b<9\kappa$.
The unitary Toeplitz lift with this clock produces the public vector obtained after clock projection $\boldsymbol{y}_{N_0}$ and a terminal private vector $\boldsymbol{e}_{N_0}$.
Project the averaged public output onto the system output summands, defining $\boldsymbol{v}_{N_0}$.
Reject the last scalar output.
The public and terminal components occupy orthogonal output subspaces.
Summing the finite constraint equations gives
\begin{align}
H\boldsymbol{v}_{N_0,\rho} &=\frac{2c_{N_0,\rho}}\rho|b\rangle +\frac{2i}{\kappa\sqrt{N_0}}L_{\rho,s}\boldsymbol{e}_{N_0,\rho},
\label{wql:eq:one-residual}
\\
L_{\rho,s}L_{\rho,s}^\dagger &=\frac\kappa2(1+s/\rho)I.
\label{wql:eq:one-residual-gram}
\end{align}
\cref{wql:fig:finite} shows the public and terminal outputs of the finite lift.
The terminal-buffer map $L_{\rho,s}$ sums the unmeasured private amplitudes using known coordinate operations and a constant number of original-oracle calls (\cref{wql:lem:terminal}).

A matrix-only block $J_H$ satisfying $\|J_H-H^{-1}/(4\kappa)\|\le\eta$ removes this residual.
If $L$ is the direct sum of the maps in \cref{wql:eq:one-residual} and $\ell\ge\|L\|$, a known row selection implements
\begin{align}
\frac{\boldsymbol{v}_{N_0}-i\beta J_H(L/\ell)\boldsymbol{e}_{N_0}} {\sqrt{1+\beta^2}},
\quad \beta=\frac{8\ell}{\sqrt{N_0}}.
\label{wql:eq:one-readout}
\end{align}
With the exact inverse, every scale component is a scalar multiple of $H^{-1}|b\rangle$.
Choose a sufficiently large constant in $N_0=O(\kappa)$ so that the accepted amplitude exceeds $1/2$.
Taking $\eta=O(\epsilon)$ then gives the required conditional error.
The finite state-preparation circuit uses $O(D)$ calls to $B,B^\dagger$.
\cref{wql:lem:reciprocal} constructs the reciprocal using $O(\kappa\log(1/\epsilon))$ matrix calls and no preparation calls.

\begin{figure}[tbp]
\centering

\IIRDiagram{QLSPFinite}

\caption{The uniform-clock lift places the public output and the equation residual in orthogonal subspaces.
Applying $L/\ell$ and the matrix-only reciprocal to the terminal state corrects the residual.
A known row selects their linear combination; all other outputs remain in failure subspaces.
The resulting accepted amplitude determines the solution normalization, including the reciprocal's normalization.}
\label{wql:fig:finite}
\end{figure}

\Needspace{9\baselineskip}
\subsubsection{Access to a Hermitian factor}
\label{wql:subsec:factor}

For $H=T^2$, two couplings to $T$ split the constraint through an intermediate amplitude.
Scaling that amplitude by the given norm bound reduces the coarse matrix weight from $\kappa$ to $O(\sqrt{\kappa D})$.
The precision correction uses the inverse of $T$, whose norm is at most $\sqrt\kappa$.
In the one-factor, one-state case, $r_0=\kappa\sqrt p$ and $D=\kappa/r_0$.

\begin{figure}[htbp]
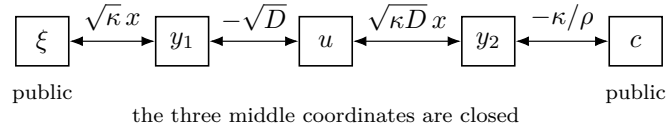

\centering

\IIRDiagram{QLSPFactor}

\caption{The factor-access load has two public end coordinates and three closed auxiliary coordinates.
The bidirectional arrows denote Hermitian couplings: both signal edges use $x$, and closure gives $x^2\xi=c/\rho$.
In the vector-valued realization, the signal edges use the same block encoding of $T$, and the last edge couples to $|b\rangle$.}
\label{wql:fig:factor}
\end{figure}

At the scalar level, we design the rational reflection in \cref{wql:eq:factor-scalar-transfer}, whose average satisfies the quadratic constraint.
Splitting that constraint into two first-order relations gives
\begin{align}
\sqrt\kappa\,x\xi=\sqrt D\,u,
\quad \sqrt{\kappa D}\,xu=\frac\kappa\rho c,
\quad u=\sqrt{r_0}\,x\xi.
\label{wql:eq:factor-design}
\end{align}
The intermediate amplitude balances the two factor couplings.
For the steady-state solution under the stated promises, its norm is at most the solution amplitude's norm, as verified in \cref{wql:eq:factor-steady-state-single}.
The chain in \cref{wql:fig:factor} realizes these relations.
Its open coordinates are $(\xi,c)$, and the three auxiliary coordinates $(y_1,u,y_2)$ are closed.
The Hermitian load is
\begin{align}
\mathcal K^{\rm fac}_\rho(x)= \left(\begin{array}{cc|ccc} 0&0&\sqrt\kappa\,x&0&0
\\
0&0&0&0&-\kappa/\rho
\\
\hline
\sqrt\kappa\,x&0&0&-\sqrt D&0
\\
0&0&-\sqrt D&0&\sqrt{\kappa D}\,x
\\
0&-\kappa/\rho&0&\sqrt{\kappa D}\,x&0 \end{array}\right).
\label{wql:eq:factor-scalar-load}
\end{align}
The transfer function before closing the auxiliary coordinates is $(I-i\mathcal K^{\rm fac}_\rho)(I+i\mathcal K^{\rm fac}_\rho)^{-1}$.
The closed rows impose
\begin{align}
u=\sqrt{r_0}\,x\xi,
\quad y_1=\sqrt\kappa\,xy_2,
\quad x^2\xi=c/\rho.
\label{wql:eq:factor-scalar-closure}
\end{align}
Eliminating them gives the public transfer at $\Bz=\boldsymbol{1}$,
\begin{align}
F^{\rm fac}_\rho(1;x)= \frac1{1+\rho^2x^4} \begin{pmatrix} 1-\rho^2x^4&2\rho x^2
\\
2\rho x^2&\rho^2x^4-1 \end{pmatrix}.
\label{wql:eq:factor-scalar-transfer}
\end{align}
This reflection enforces the scalar constraint.
Substitute the factor operator into the two signal couplings and the preparation into the scalar-to-vector coupling.
Closing the auxiliary coordinates leaves public space $\mathcal H\oplus\mathbb C$ and imposes $T^2\boldsymbol{x}=(c/\rho)|b\rangle$ on its average amplitudes.

In \cref{lem:common-cayley}, the $-\sqrt D$ identity coupling belongs to the known $K_0$.
The two signal couplings occupy separate copies of the oracle space, so one controlled $O_T$ call acts on both at a due event, leaving the edge label unchanged.
\cref{wql:eq:factor-maps,wql:eq:factor-known-gram} give the maps and known system.

The steady-state solution has the same $c,\boldsymbol{x}$ as \cref{wql:eq:one-steady-state}, and
\begin{align}
\boldsymbol{u}=\sqrt{r_0}\,T\boldsymbol{x},
\quad \boldsymbol{y}_1=\frac{i}{\sqrt\kappa}T^{-1}\boldsymbol{x},
\quad \boldsymbol{y}_2=\frac{i}{\kappa}T^{-2}\boldsymbol{x}.
\label{wql:eq:factor-steady-state-single}
\end{align}
All three auxiliary norms are at most $\|\boldsymbol{x}\|$.
For the intermediate vector, use $\|T^{-1}|b\rangle\|^2=\langle b|H^{-1}|b\rangle\le r$ and $r_0\le r$.
The other estimates use the factor inverse bound $\|T^{-1}\|\le\sqrt\kappa$.
The energy identity and the same scale cascade therefore give
\begin{align}
W_T\le\sqrt\kappa+\sqrt{\kappa D},
\quad W_b<4D.
\label{wql:eq:factor-single-weight}
\end{align}
Choose the preparation delay and coarse horizon as $s=\lceil\sqrt{r_0}\rceil$ and $N_0=O(\sqrt{\kappa D})$.
The resulting counts are $O(\sqrt{\kappa D})$ factor queries and $O(D)$ preparation queries.

The finite chain leaves two constraint residuals.
The same buffer-sum identity used in \cref{wql:eq:one-residual} now gives
\begin{align}
H\boldsymbol{v}_{N_0,\rho} =\frac{2c_{N_0,\rho}}\rho|b\rangle +\frac{2i}{\kappa\sqrt{N_0}} (L_{2,\rho}+\sqrt\kappa\,T L_{1,\rho})\boldsymbol{e}_{N_0,\rho}.
\label{wql:eq:factor-residual-single}
\end{align}
The correction applies $T^{-2}L_{2,\rho}$ and $\sqrt\kappa T^{-1}L_{1,\rho}$ to the terminal state.
Apply \cref{wql:lem:reciprocal} with the factor's inverse-norm bound $\sqrt\kappa$.
This gives a block approximating $T^{-1}/(4\sqrt\kappa)$.
Two executions with separate success ancillas give its product, approximating $T^{-2}/(16\kappa)$.
Both branches use $O(\sqrt\kappa\log(1/\epsilon))$ factor queries and no preparation queries; a known row selection accounts for their fixed normalizations.
The resulting success and error bounds are proved in \cref{wql:subsec:factor-proof}.
Thus $p^{-1/4}$ enters only the constant-accuracy constraint computation, while final precision enters only the matrix correction in \cref{wql:eq:single-factor-cost}.
Both stages query $T$ directly, without an intermediate block encoding of $T^2$.

\subsubsection{Multi-oracle linear systems}
\label{wql:subsec:sums}

For Hermitian matrix terms and self-inverse encodings, use \begin{align}
A_j=H_j=H_j^\dagger,
\quad O_j=\Be{H_j/\lambda_j},
\quad H=\sum_jH_j,
\quad\boldsymbol{b}=\sum_ka_k|b_k\rangle.
\end{align} The two sums replace the matrix and preparation couplings in \cref{wql:eq:one-load}:
\begin{align}
\mathcal K_\rho =K\left(\begin{array}{cc|c} 0&0&H
\\
0&0&-\boldsymbol{b}^\dagger/\rho
\\
\hline
H&-\boldsymbol{b}/\rho&0 \end{array}\right).
\label{wql:eq:sum-load}
\end{align}
The public space remains $\mathcal H\oplus\mathbb C$.
Closing the auxiliary space enforces $H\boldsymbol{x}=(c/\rho)\boldsymbol{b}$.
In the known junction, replace the two maps by
\begin{align}
\mathcal C_j(\boldsymbol{x},c,\boldsymbol{y}) &=\sqrt{K\lambda_j} \begin{pmatrix}(|0\rangle_j\otimes I)\boldsymbol{x}
\\
(|0\rangle_j\otimes I)\boldsymbol{y}\end{pmatrix},
\quad V_j=\begin{pmatrix}0&O_j
\\
O_j&0\end{pmatrix},
\label{wql:eq:sum-matrix-maps}
\\
\mathcal D_k(\boldsymbol{x},c,\boldsymbol{y}) &=\sqrt{K|a_k|/\rho} \begin{pmatrix}(|0\rangle_{w_k}\otimes I)\boldsymbol{y}
\\
c|0\rangle\end{pmatrix},
\quad W_k=-\begin{pmatrix} 0&(a_k/|a_k|)B_k
\\
(a_k^*/|a_k|)B_k^\dagger&0 \end{pmatrix}.
\label{wql:eq:sum-state-maps}
\end{align}
Each $W_k$ is self-inverse.
Its coefficient phase is a known controlled phase.
The sum of the compressed couplings is exactly \cref{wql:eq:sum-load}.
Writing $\mathcal C=(\bigoplus_j\mathcal C_j)\oplus (\bigoplus_k\mathcal D_k)$, its known Gram matrix is
\begin{align}
\mathcal C^\dagger\mathcal C =\diag\left( K\|\boldsymbol{\lambda}\|_1 I, \frac K\rho\|\boldsymbol{a}\|_1, K\left[\|\boldsymbol{\lambda}\|_1+\frac{\|\boldsymbol{a}\|_1}\rho\right]I \right).
\label{wql:eq:sum-gram}
\end{align}
The preparation maps have orthogonal private codomains, so this Gram matrix contains no overlaps between the prepared states.
Their compressed outputs contribute to the same vector $\boldsymbol{b}$.

\begin{figure}[tbp]
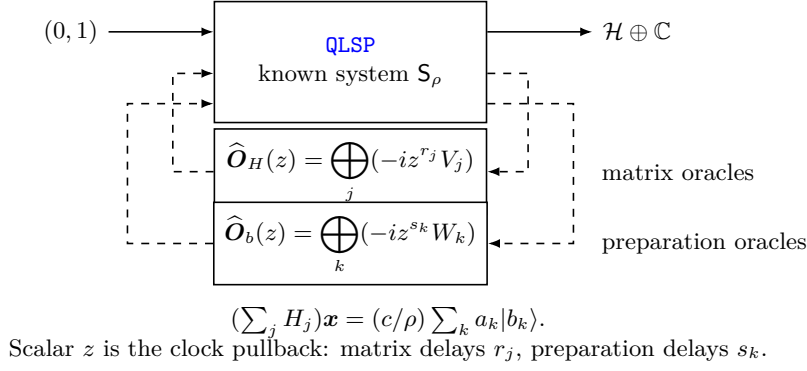

\centering

\IIRDiagram{QLSPSums}

\caption{The summed constraint has one scalar public input and separate private connections for the matrix and preparation terms.
The delays $r_j$ and $s_k$ set their query rates.
Only the matrix connections appear in the precision correction.}
\label{wql:fig:sums}
\end{figure}

Let $R=\|H^{-1}\boldsymbol{b}\|$, used only in the analysis.
The response column is \cref{wql:eq:one-column} with $H^{-1}|b\rangle,r$ replaced by $H^{-1}\boldsymbol{b},R$.
Use the known scales
\begin{align}
\rho_j=2^jR_0,
\quad j=\left\lceil\log_2\frac{K\|\boldsymbol{a}\|_1}{R_0}\right\rceil,\ldots,0.
\label{wql:eq:sum-scales}
\end{align}
The upper endpoint covers $R$ because $R\le K\|\boldsymbol{a}\|_1$.
The steady-state transmission is again at least $16/25$.

The steady-state averages are $\boldsymbol{x}=(c/\rho)H^{-1}\boldsymbol{b}$, $\boldsymbol{y}=(i/K)H^{-1}\boldsymbol{x}$, and $c=(1+R^2/\rho^2)^{-1}$.
\cref{wql:fig:sums} connects the individual matrix and preparation terms.
The matrix-$j$ weight in one system matrix is at most $K\lambda_j$ times its transmission probability.
The preparation-$k$ weight satisfies
\begin{align}
W_{b,k,\rho} =\frac{2K|a_k|}\rho(\|\boldsymbol{y}\|^2+|c|^2) \le\frac{2K|a_k|}{\rho(1+R^2/\rho^2)}.
\label{wql:eq:local-source-weight}
\end{align}
Weighting each system matrix by the squared scalar amplitude reaching it gives
\begin{align}
W_j\le K\lambda_j,
\quad W_{b,k}<\frac{4K|a_k|}{R_0}.
\label{wql:eq:separate-weights}
\end{align}
Transmission probabilities telescope in the matrix bound, and $\sum_j\rho_j^{-1}<2/R_0$ gives the preparation bound.
Neither estimate requires orthogonality among $|b_k\rangle$.

The residual after the coarse finite-clock computation has the same form, with a Gram matrix that records both sets of delays:
\begin{align}
H\boldsymbol{v}_{N_0,\rho} &=\frac{2c_{N_0,\rho}}\rho\boldsymbol{b} +\frac{2i}{K\sqrt{N_0}}L_\rho\boldsymbol{e}_{N_0,\rho},
\label{wql:eq:sum-residual}
\\
L_\rho L_\rho^\dagger &=\frac K2\left( \sum_j\lambda_jr_j+\frac1\rho\sum_k|a_k|s_k \right)I.
\label{wql:eq:sum-residual-gram}
\end{align}
A matrix-only inverse therefore corrects the combined right-hand side without a separate preparation of $\boldsymbol{b}/\|\boldsymbol{b}\|$.

\label{wql:subsec:allocation}

Assign rate parameters
\begin{align}
t_j=1+K\sqrt{\frac{\lambda_j}{C_j}} \sum_i\sqrt{C_i\lambda_i},
\quad u_k=1+\frac K{R_0}\sqrt{\frac{|a_k|}{c_k}} \sum_l\sqrt{c_l|a_l|}.
\label{wql:eq:rates}
\end{align}
Choose a known $L\ge\max\{1,\max_jt_j,\max_ku_k\}$ and integer delays
\begin{align}
r_j=\lceil L/t_j\rceil,
\quad s_k=\lceil L/u_k\rceil.
\label{wql:eq:delays}
\end{align}
The rate definitions imply
\begin{align}
K\sum_j\frac{\lambda_j}{t_j}\le1,
\quad \frac K{R_0}\sum_k\frac{|a_k|}{u_k}\le1.
\label{wql:eq:rate-sums}
\end{align}
Each rate is at most the rate scale $L$, so the unweighted sums satisfy $K\|\boldsymbol{\lambda}\|_1\le L$ and $K\|\boldsymbol{a}\|_1/R_0\le L$.
Rounding therefore gives
\begin{align}
K\sum_j\lambda_jr_j\le2L,
\quad \frac K{R_0}\sum_k|a_k|s_k\le2L,
\quad \max_jr_j,\max_ks_k\le2L.
\label{wql:eq:rounded-sums}
\end{align}
By \cref{wql:eq:separate-weights}, the coarse delayed weight is at most $10L$.
By \cref{wql:eq:sum-residual-gram}, the complete terminal map has squared norm at most $2L$.
A uniform clock with $N_0=O(L)$ therefore gives the same constant-success correction as in the single-state case.

The reciprocal is the only precision-dependent oracle circuit.
Keep the same matrix connections and delays $r_j$.
Then \cref{wql:lem:reciprocal} constructs a block $J$ approximating $H^{-1}/(4K)$, with horizon
\begin{align}
N_1=O\!\left[ \left(K\sum_j\lambda_jr_j+\max_jr_j\right) \log\frac1\epsilon\right] =O\!\left(L\log\frac1\epsilon\right).
\label{wql:eq:inverse-horizon}
\end{align}
Occurrences of the same original matrix oracle across scale system matrices and reciprocal factors occupy a direct sum.
At a due event, one controlled call acts on that entire sum with the occurrence label unchanged.
Calls to different original oracles are counted separately.
For the delayed transfer function $G=F_{\boldsymbol{r},\boldsymbol{s}}$, the finite lift has query counts $\lfloor(N_0-1)/r_j\rfloor$ at matrix port $j$ and $\lfloor(N_0-1)/s_k\rfloor$ at preparation port $k$.
Each preparation feedback operation uses a forward and an inverse call.
The final buffer-to-residual map uses $O(1)$ additional calls of each type.
These exact schedules imply
\begin{align}
Q_j=O\!\left(t_j\log\frac1\epsilon\right),
\quad Q_{b,k}=O(u_k).
\label{wql:eq:individual-counts}
\end{align}
Multiplying by $C_j,c_k$ and summing gives \cref{wql:eq:matrix-cost,wql:eq:state-cost}.
The numbers of scale system matrices and reciprocal sections affect the known gates and private dimensions, but do not multiply the query counts.
\cref{wql:prop:success} proves the success and error bounds for this schedule.

For arbitrary square $A_j$, use the termwise Hermitian dilation
\begin{align}
H_j=\begin{pmatrix}0&A_j
\\
A_j^\dagger&0\end{pmatrix},
\quad O_j=\begin{pmatrix}0&U_j
\\
U_j^\dagger&0\end{pmatrix}.
\label{wql:eq:hermitian-dilation}
\end{align}
The latter is a self-inverse block encoding of $H_j/\lambda_j$, implemented with one controlled original call in each direction and a known swap.
Prepare each state in the first sector.
Then
\begin{align}
\left\|\begin{pmatrix}0&A
\\
A^\dagger&0\end{pmatrix}^{-1}\right\| =\|A^{-1}\|,
\quad \begin{pmatrix}0&A
\\
A^\dagger&0\end{pmatrix}^{-1} \begin{pmatrix}\boldsymbol{b}
\\
0\end{pmatrix} =\begin{pmatrix}0
\\
A^{-1}\boldsymbol{b}\end{pmatrix}.
\label{wql:eq:dilation-norms}
\end{align}
The dilation preserves $K,R_0$, places the ideal solution in a fixed sector, and changes the query counts by constant factors.
It applies to arbitrary square terms, without a commutativity assumption.

\label{wql:subsec:normalization}

The preparation dependence can also be written in the probability notation of \cref{wql:cor:single}.
If the given bounds $s_0>0$ and $p>0$ satisfy
\begin{align}
s_0\le\|\boldsymbol{b}\|,
\quad p\le\frac{\|A^{-1}|b\rangle\|^2}{K^2},
\quad |b\rangle=\boldsymbol{b}/\|\boldsymbol{b}\|,
\label{wql:eq:rhs-p}
\end{align}
then $R_0=Ks_0\sqrt p$ is a valid given lower bound.
Hence
\begin{align}
\sum_kc_kQ_{b,k} =O\!\left[ \frac{\langle\boldsymbol{c},|\boldsymbol{a}|\rangle_{1/2}}{s_0\sqrt p} +\|\boldsymbol{c}\|_1\right].
\label{wql:eq:rhs-p-cost}
\end{align}

For \cref{wql:thm:factor}, take $A=T^2$, $K=\kappa$ in \cref{wql:eq:rhs-p}.
Then $D=\|\boldsymbol{a}\|_1/(s_0\sqrt p)$, and
\begin{align}
\sum_j C_jQ_{T,j} =O\!\left[ \sqrt\kappa\left( \sqrt{\frac{\|\boldsymbol{a}\|_1}{s_0}}\,p^{-1/4} +\log\frac1\epsilon\right) \langle\boldsymbol{C},\boldsymbol{\lambda}\rangle_{1/2} +\|\boldsymbol{C}\|_1\log\frac1\epsilon\right].
\label{wql:eq:factor-p-cost}
\end{align}
The preparation bound is still \cref{wql:eq:rhs-p-cost}.
The norm ratio in the coarse factor term accounts for cancellation in this construction.
For a single normalized state, the known combination norm allows this ratio to be set to one.

Cancellation between the states can reduce $s_0$ or $R_0$ arbitrarily; the coefficient magnitudes alone provide no positive lower bound.
A common rescaling of all $a_k$ and $R_0$ leaves the normalized output and query bounds unchanged.

The state-combination construction in \cref{cor:mod-stateprep} connects these preparation couplings to an \ExpIIR{} transformation.
Here the couplings enter the linear constraint, with feedback determining the normalization and the matrix-only terminal correction controlling final accuracy.

For a flagged preparation with amplitude $\gamma_k$, direct coupling forms $\sum_k a_k\gamma_k|b_k\rangle$.
Known nonzero amplitudes can be absorbed into the coefficients; unknown amplitudes require a separate normalization construction.
\cref{wql:thm:main,wql:thm:factor} assume deterministic preparations and do not claim optimal multi-oracle prefactors.

\subsubsection{Classical certificates beyond the condition number}
\label{cert:sec:main}

\begin{figure}[tbp]
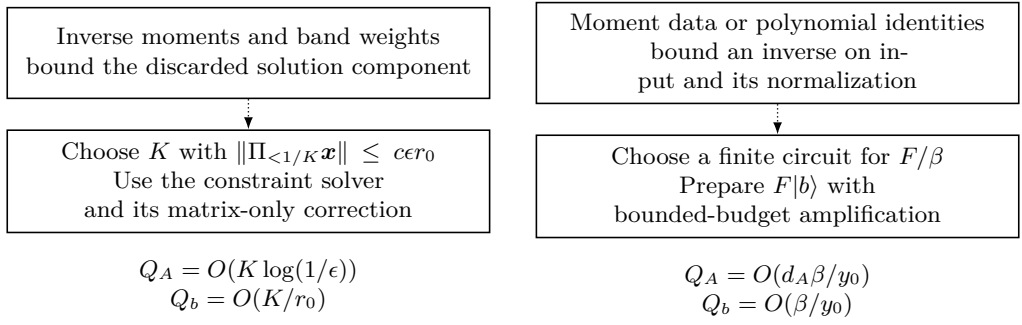

\centering
\IIRDiagram{QLSPClassicalBounds}

\caption{Two ways to use given classical information.
The left branch changes parameters of the existing constraint solver.
The right branch chooses a different finite-query approximation to the inverse on the specified input.
The classical inputs are bounds on the solution tail or on the accuracy and normalization of a finite-query approximation to the inverse on the specified input.}
\label{cert:fig:choices}
\end{figure}

The smallest singular value can have negligible weight in the desired solution.
Classical information about that weight lets the same constraint solver use a smaller inverse scale.
Alternatively, a finite polynomial or circuit may approximate the inverse on the given right-hand-side vector directly.
\cref{cert:fig:choices} separates these two constructions.
State-dependent inverse bounds have been studied by Li \cite{Li2025BeyondConditionNumber}; singular truncation and higher inverse moments are developed by Dalzell, Li, and Su \cite{Dalzell2026BeyondConditionNumber}.
Our extension uses the separate preparation-query bound of \cref{wql:thm:main}, continuing the objective of Low and Su \cite{Low2024QLSP}.

Let the invertible contraction and normalized right-hand-side vector be accessed through
\begin{align}
U_A=\Be{A},
\quad B\ket0=\ket b,
\quad \boldsymbol x=A^{-1}\ket b,
\quad r=\|\boldsymbol x\|,
\quad 1\le r_0\le r.
\label{cert:eq:access}
\end{align}
Controls and adjoints are available, with preparation work restored.
The solution-tail bound in \cref{cert:thm:tail} uses
\begin{align}
|A|=(A^\dagger A)^{1/2},
\quad \tau(K)=\|\Pi_{<1/K}\boldsymbol x\|,
\quad \Pi_{<1/K}=\boldsymbol1_{[0,1/K)}(|A|),
\quad K\ge2.
\label{cert:eq:tail}
\end{align}
The spectral projection defines the classical bound used in the analysis.

\begin{theorem}[QLSP from a solution-tail bound]
\label{cert:thm:tail}
There is a universal constant $c>0$ with the following property.
Under \cref{cert:eq:access,cert:eq:tail}, suppose a classical certificate supplies $K\ge2$ and an upper bound $t_K\ge\tau(K)$ such that
\begin{align}
t_K\le c\epsilon r_0,
\quad 0<\epsilon\le1/10.
\label{cert:eq:tail-condition}
\end{align}
The constraint construction of~\cref{wql:thm:main}, with its inverse scale and clocks chosen from $K$, succeeds with probability at least one quarter.
It prepares the solution with conditional trace-distance error at most $\epsilon$, using
\begin{align}
Q_A=O\!\left(K\log\frac1\epsilon\right),
\quad Q_b=O\!\left(\frac K{r_0}\right).
\label{cert:eq:cutoff-cost}
\end{align}
These counts include forward and inverse calls to the original oracles, with the effective output normalization at most two.
\end{theorem}

For the proof, split the solution at the singular-value threshold and apply the matrix to its discarded part.
This compresses the error in the prepared input:
\begin{align}
\boldsymbol x_< =\Pi_{<1/K}\boldsymbol x,
\quad \boldsymbol b_<=A\boldsymbol x_<,
\quad \|\boldsymbol b_<\|\le\tau(K)/K.
\label{cert:eq:source-compression}
\end{align}
A comparison preparation unitary for the normalized remaining input vector can therefore be chosen within $\sqrt2\tau(K)/K$ of the actual oracle.
\cref{cert:lem:junction-support,cert:lem:support} prove that the complete unitary realization on this input state stays in an invariant space with inverse bound $K$, including arbitrary oracle completion blocks.
Only the preparation calls change in the comparison; the matrix-only precision stage uses identical oracles.
Their hybrid error is
\begin{align}
Q_b\|B_\ge-B\|=O\!\left(\frac{\tau(K)}{r_0}\right),
\quad K\ge\sqrt{r^2-\tau(K)^2}.
\label{cert:eq:necessary-cutoff}
\end{align}
The last inequality gives $r_0=O(K)$ under the small-tail hypothesis.
Thus no precision logarithm multiplies the input-state perturbation.
\cref{cert:proof:tail} includes the conditional-error and success bounds.
A relative tail estimate suffices when an upper bound on the solution norm converts it to the absolute tolerance in \cref{cert:eq:tail-condition}.
The resulting cutoff may depend on accuracy.

For access to a Hermitian factor, \cref{cert:cor:factor} applies the same comparison between input states to the factor construction of \cref{wql:cor:single}.
Its proof is in \cref{cert:app:support}.
\begin{corollary}[One Hermitian factor]
\label{cert:cor:factor}
Suppose the target is the square of a Hermitian contraction: $H=S^2$, $S=S^\dagger$, and $\|S\|\le1$.
The matrix oracle is a controlled self-inverse block encoding of $S$.
Put $\boldsymbol x=H^{-1}|b\rangle$ and $1\le r_0\le\|\boldsymbol x\|$.
If
\begin{align}
\|\mathbf1_{[0,1/K)}(H)\boldsymbol x\|\le c\epsilon r_0,
\label{cert:eq:factor-tail}
\end{align}
then the same output guarantee holds with
\begin{align}
Q_S=O\!\left(\frac K{\sqrt{r_0}}+ \sqrt K\log\frac1\epsilon\right),
\quad Q_b=O\!\left(1+\frac K{r_0}\right).
\label{cert:eq:factor-cost}
\end{align}
This statement uses one factor block encoding and one preparation.
\end{corollary}

Each classical spectral bound bounds the solution-weighted measure
\begin{align}
\nu_x=\sum_i\frac{|\langle u_i|b\rangle|^2}{\sigma_i^2}\delta_{\sigma_i},
\quad \nu_x([0,1])=r^2,
\quad \nu_x([0,1/K))=\tau(K)^2,
\label{cert:eq:solution-measure-main}
\end{align}
where the singular-value decomposition is $A=\sum_i\sigma_i\ket{u_i}\bra{v_i}$.
A global inverse bound makes the tail zero beyond that bound.
An inverse moment instead gives
\begin{align}
\mu_q\ge\||A|^{-q}\boldsymbol x\|
\quad\Longrightarrow
\quad \tau(K)\le\mu_q/K^q
\quad(q>0).
\label{cert:eq:moment-main}
\end{align}
This follows by integrating $\sigma^{-2q}$ against \cref{cert:eq:solution-measure-main}.
If every integer moment obeys the growth bound $\mu_q\le r_+(Lq^a)^q$, with a constant-factor norm estimate, choosing the moment order logarithmically in inverse accuracy yields
\begin{align}
K=O(L\log^a(1/\epsilon)),
\quad Q_A=O(L\log^{a+1}(1/\epsilon)),
\quad Q_b=O(K/r_0).
\label{cert:eq:moment-main-cost}
\end{align}
For orthogonal sectors reducing the absolute value of the matrix, individual solution weights and inverse moments can be combined as
\begin{align}
\tau(K)^2\le\sum_j\min\left\{w_j,\inf_{q\in\mathcal Q_j} \mu_{j,q}^2/K^{2q}\right\}.
\label{cert:eq:sector-main}
\end{align}
The data are defined in \cref{cert:eq:sector-data}; a sector contributes zero once the cutoff reaches its inverse gap.
Singular bands with input-state weights $\pi_j$ and lower endpoints $a_j$ instead give
\begin{align}
\tau(K)^2\le\sum_{j:a_j<1/K}\pi_j/a_j^2,
\label{cert:eq:bands-main}
\end{align}
including a bound on any uncovered spectrum.
The same example in \cref{tab:certificate-resolution} illustrates these different resolutions.

Take 33 singular values $\sigma_j=2^{-j}$, indexed from zero to 32.
Let the normalized solution have amplitudes proportional to $2^{-2j}$ in their right singular vectors.
Choose the normalized right-hand-side vector proportional to $A|\mathrm{sol}\rangle$.
For the relative tail threshold $\tau(K)/r\le10^{-3}$, the sufficient cutoffs are listed in \cref{tab:certificate-resolution}.
\begin{table}[tbp]\centering\small
\begin{tabular}{lc}
\toprule
Classical bound & Sufficient $K$
\\
\midrule
Global inverse norm & $2^{32}\simeq4.29\times10^9$
\\
First inverse moment & $1.119\times10^3$
\\
Second inverse moment & $74.7$
\\
Full solution-weighted tail & $16$
\\
\bottomrule
\end{tabular}
\caption{Sufficient cutoffs for the same input state and spectrum, with relative discarded solution norm at most $10^{-3}$.
The algorithmic tolerance additionally includes the universal constant in \cref{cert:eq:tail-condition}.}\label{tab:certificate-resolution}\end{table} The entries use the same input state and spectrum.
The moment bounds are $\tau(K)/r\le\kappa_q^q/K^q$, where $\kappa_q=(\||A|^{-q}\boldsymbol x\|/r)^{1/q}$.
At $K=16$, the discarded normalized norm is less than $2^{-10}$.
This threshold illustrates the certificate comparison; the algorithmic accuracy must still include the universal error-allocation factor in \cref{cert:eq:tail-condition}.

Positive moments of the input state can also certify a tail or a finite-query approximation to the inverse on the specified input.
For a real polynomial majorant of the threshold indicator, spectral calculus gives
\begin{align}
m_j=\bra b(AA^\dagger)^j\ket b,
\quad \tau(K)^2\le h_0r^2+\sum_{j=1}^d h_jm_{j-1},
\label{cert:eq:positive-main}
\end{align}
with the majorant conditions in \cref{cert:eq:majorant}.
Interval bounds on the moments and solution norm must account for coefficient signs.
For the input-specific inverse approximation defined below, the same moments give its norm and overlap with the solution:
\begin{align}
F=A^\dagger p(AA^\dagger),
\quad p(t)=\sum_{j=0}^d c_jt^j,
\quad \boldsymbol y=F\ket b.
\label{cert:eq:action-main}
\end{align}
\cref{cert:eq:moment-action} converts those two quantities to a normalized state-error bound.
If a circuit block-encodes the action with normalization $\beta$, \cref{cert:prop:short} gives
\begin{align}
Q_A=O(d_A\beta/y_0),
\quad Q_b=O(\beta/y_0),
\quad 0<y_0\le\|\boldsymbol y\|,
\label{cert:eq:action-main-cost}
\end{align}
where $d_A$ is its matrix-query count.
The accuracy premise and bounded-budget amplification are proved in \cref{cert:prop:short,cert:lem:amplification}.
A polynomial annihilating the given input state gives an exact example:
\begin{align}
a(A)\ket b=0,
\quad a(0)\ne0
\quad\Longrightarrow
\quad p(z)=\frac{1-a(z)/a(0)}z,
\quad p(A)\ket b=\boldsymbol x.
\label{cert:eq:annihilator-main}
\end{align}
\cref{cert:prop:annihilator} includes the coefficient normalization and the inverse-weighted residual needed for an approximate identity.
The separation example and normalization lower bound in \cref{cert:app:short} show why degree of the input-specific inverse approximation and an inverse cutoff parameter describe different costs.

The exact-support argument also applies to the separately weighted result under a common invariance condition.
Let $A=\sum_jA_j$ be given through separate block encodings $\Be{A_j/\lambda_j}$.
The right-hand-side vector is $\boldsymbol b=\sum_k a_k|b_k\rangle$.
Suppose there are orthogonal projections $P_R,P_L$ such that
\begin{align}
A_jP_R=P_LA_j
\quad\text{for every }j,
\quad P_L|b_k\rangle=|b_k\rangle
\quad\text{for every }k,
\label{cert:eq:weighted-invariance}
\end{align}
and $A:\ran P_R\to\ran P_L$ is bijective with inverse norm at most $K$.
These are conditions on the given data; no projector oracle is required.
For $0<R_0\le\|A^{-1}\boldsymbol b\|$, the same proof as \cref{cert:lem:support} restricts each feedback connection separately.
\cref{wql:thm:main} then gives
\begin{subequations}
\label{cert:eq:weighted-supported}
\begin{align}
\sum_jC_jQ_j &=O\!\left[ \left(K\langle\boldsymbol C,\boldsymbol\lambda\rangle_{1/2} +\|\boldsymbol C\|_1\right)\log\frac1\epsilon\right],
\\
\sum_k c_kQ_{b,k} &=O\!\left[ \frac K{R_0}\langle\boldsymbol c,|\boldsymbol a|\rangle_{1/2} +\|\boldsymbol c\|_1\right].
\end{align}
\end{subequations}

The paired invariance condition must hold for each oracle term, and each separately prepared vector must lie in the corresponding input space.
These conditions ensure invariance throughout the unequal-delay circuit.

The full spectral identities and proofs for the input-specific inverse approximation are in \cref{cert:app:spectral}; \cref{app:certificate-resources} gives the gate and qubit substitutions for the unchanged solver.

\providecommand{\VDEQLSP}{\cref{wql:thm:main}}
\providecommand{\VDEResources}{\cref{prop:qlsp-resources}}
\providecommand{\VDESynthesis}{\cref{eq:app-synthesis}}
\providecommand{\VDEQLSPTag}{\hyperref[wql:thm:main]{\IIRTag{QLSP}}}

\subsection{Linear differential equations}
\label{sec:alg-volterra}
Consider the inhomogeneous linear differential equation and its equivalent Volterra integral equation,
\begin{align}
\boldsymbol{x}'(t)&=-H(t)\boldsymbol{x}(t)+\boldsymbol{b}(t),
\quad \boldsymbol{x}(0)=\boldsymbol{x}_0,
\label{eq:vde-ode}
\\
\boldsymbol{x}(t)&=\boldsymbol{x}_0+\int_0^t[\boldsymbol{b}(s)-H(s)\boldsymbol{x}(s)]\,ds,
\quad 0\le t\le T.
\label{eq:vde-volterra}
\end{align}
This Volterra integral formulation describes driven evolution for an arbitrary square matrix $H(t)$, including non-Hermitian and nonnormal generators.
We seek the normalized final state $|x_T\rangle=\boldsymbol{x}(T)/\|\boldsymbol{x}(T)\|$ from a block-encoding $\Be{H(t)/\alpha}$ and preparations of the initial vector and forcing term.

We represent the differential equation by its causal Volterra relation.
After discretization, each sampled generator remains an original matrix-oracle port and each initial or forcing term remains an original preparation port.
The resulting lower-triangular linear relation is then passed to the QLSP IIR of \cref{wql:sec:qlsp}.
Thus the new work in this subsection is to choose and condition the Volterra transfer function; its unitary realization and precision refinement are inherited from the modular linear-system construction.

Quantum algorithms for Volterra integro-differential equations with convolution memory kernels were developed by Ameri, Dutt, and Krovi~\cite{Ameri2026NonMarkovian}.
Their formulation models evolution with memory and includes reductions of structured kernels to enlarged ODE systems.
Here the Volterra equation is the integrated form of the original time-local ODE, and its discretization is used to control the QLSP inverse bound.

In a direct finite-difference reduction, the inverse norm grows with the number of time steps \cite{Berry2014Differential}.
Taylor and Dyson reductions control precision through higher-order propagator expansions \cite{Berry2017Differential,Berry2024Dyson}.
We find that normalized Volterra constraints on short intervals retain the Euler endpoint while bounding each local inverse independently of the number of samples.
The global QLSP inverse bound then scales with the number of intervals.
For example, with zero generator the unscaled Euler history has a cumulative-sum inverse whose norm grows linearly with the total sample count.
In the Volterra formulation, the cumulative sum on each interval is divided by that interval's sample count.
Its norm is at most one on every grid.
Choosing a sufficiently short coarse interval therefore gives
\begin{align}
E_j=I+hJ_ND_j,
\quad \|hJ_ND_j\|\le h\alpha\le\tfrac12,
\quad \|E_j^{-1}\|\le2.
\label{eq:vde-local-inverse-preview}
\end{align}
Here $J_N$ is the normalized cumulative-sum matrix defined in \cref{eq:vde-clock}, and $D_j$ stores the sampled generator.
Refining the integration grid improves accuracy without enlarging this local inverse bound.
Row differences still recover the exact Euler trajectory on the fine grid.

The initial preparation gives $\boldsymbol{x}_0/\|\boldsymbol{x}_0\|$ with known norm.
The forcing preparation is a unitary with successful output vector $\boldsymbol{b}(t)/\beta$, including its magnitude and phase.
Controlled oracles and their adjoints are available.
Constant data use time-independent oracles; varying data require coherent control by the sampled time label.
Let $V_H,V_b$ bound the total variations of the matrix and forcing over the time interval.
They vanish for constant data.

\Cref{tab:vde-constructions} summarizes the dependence on time variation and the given preparation oracles.

\begin{theorem}[Volterra reduction with separate oracle costs]
\label{thm:vde-constant}
Assume the parameter and input bounds $T>0$, $0<\epsilon\le1$, $\|H(t)\|\le\alpha$, and $\|\boldsymbol{b}(t)\|\le\beta$.
Suppose known bounds satisfy
\begin{align}
\|U(t,s)\|\le C,
\quad \sup_t\|\boldsymbol{x}(t)\|\le g,
\quad 0<r\le\|\boldsymbol{x}(T)\|,
\quad C\ge1,
\label{eq:vde-promises}
\end{align}
Writing $p=r^2$ matches the introduction\textquoteright s lower bound on $\|\boldsymbol x(T)\|^2$.
Here $U(t,s)=\mathcal T\exp(-\int_s^tH(v)\,dv)$ and $0\le s\le t\le T$.
For constant $H$, the first promise is $\|e^{-Ht}\|\le C$.
Choose the coarse intervals, trajectory bound, and preparation scale by
\begin{align}
m=2^{\lceil\log_2\max\{1,2\alpha T\}\rceil},
\quad h=T/m,
\quad g\ge h\beta,
\quad B=\sqrt{\|\boldsymbol{x}_0\|^2+T^2\beta^2}>0.
\end{align}
For the Euler trajectory on the grid in \cref{eq:vde-time-grid}, assume a classical mean-square history bound $W_2$ satisfying
\begin{subequations}
\label{eq:vde-history-rms}
\begin{align}
W_2^2&\ge\frac1m\left[\sum_{j=0}^{m}\|\boldsymbol x_{\rm e}(jh)\|^2 +\frac1N\sum_{j=0}^{m-1}\sum_{k=0}^{N-1} \|\boldsymbol x_{\rm e}(jh+kh/N)\|^2\right],
\\
W_2&\ge\max\{r,B/m\}.
\end{align}
\end{subequations}
This is classical information about the trajectory; its quantum preparation is unnecessary.
Root-mean-square history norms and classical Lyapunov bounds already enter quantum differential-equation algorithms \cite{Jennings2024CostLinearDE}.
Here the history estimate sets the output scaling parameter in a solver that separates matrix and state-preparation costs.
The peak bound $g$ still determines the discretization mesh.
With only the peak information, $W_2=2g$ is a valid choice.
Then the final state $|x_T\rangle$ can be prepared with success probability at least $2/3$.
Its conditional trace-distance error is at most $\epsilon$, using
\begin{align}
Q_H=O\!\left(\frac{CmW_2}{r}\log\frac1\epsilon\right),
\quad Q_0+Q_b=O\!\left(\frac{CB}{r}\right).
\label{eq:vde-queries}
\end{align}
Zero input terms are omitted; $\alpha=0$ needs no generator calls.
Choose the power-of-two grid size $N$ in \cref{eq:vde-time-grid}.
For $\alpha T\ge1$, the interval count is $m=\Theta(\alpha T)$.
Write $\mathcal K=CmW_2/r$.
Let $n$ be the system width and $a$ the persistent oracle workspace.
For controlled gate costs $G_H,G_0,G_b$, define the total address width $b_V=n+a+O(\log(2mN\mathcal K)+\log\log(4/\epsilon))$.
The gate and qubit counts are
\begin{align}
G_{\rm total} &\le Q_HG_H+Q_0G_0+Q_bG_b +O\!\left(\mathcal K \left[\log\!\left(2+\frac{CB}{r}\right)+\log^2\frac1\epsilon\right] b_V\log(2+b_V)\right),
\label{eq:vde-gates}
\\
q_{\rm total} &\le n+a+O\!\left(\log(2mN\mathcal K)+\log\log\frac4\epsilon\right).
\label{eq:vde-qubits}
\end{align}
Gates are arbitrary one-qubit rotations and CNOTs.
Coherent time evaluation is included in the oracle costs; \VDESynthesis\ gives the additional known-rotation synthesis cost.
\end{theorem}

\begin{table}[H]
\centering
\begin{tabularx}{\linewidth}{@{}l>{\raggedright\arraybackslash}X@{}}
\toprule
Reduction & How precision enters the construction
\\
\midrule
Unscaled Euler history & A finer difference mesh enlarges the inverse norm.
\\
Taylor / Dyson \cite{Berry2017Differential,Berry2024Dyson} & Higher-order propagator expansions control the error.
\\
Volterra (this work) & A finer integration clock leaves the inverse bound unchanged.
\\
\bottomrule
\end{tabularx}
\caption{For $H=0$, the unscaled Euler system has a cumulative-sum inverse of norm $\Theta(mN)$.
Normalized Volterra retains the Euler endpoint with an inverse bound independent of $N$.}
\label{tab:vde-constructions}
\end{table}

\begin{proof}
Choose a power-of-two number of samples per interval satisfying
\begin{align}
N\ge\max\left\{1,16h\alpha,\frac{128Ch}{\epsilon r} [gV_H+V_b+T\alpha(\alpha g+\beta)]\right\},
\label{eq:vde-time-grid}
\end{align}
and put $\delta=h/N$, $t_{j,k}=jh+k\delta$.
The normalized sample clock and the left-quadrature matrix are
\begin{align}
|u\rangle=N^{-1/2}\sum_{k=0}^{N-1}|k\rangle,
\quad J_N=N^{-1}\sum_{0\le l<k<N}|k\rangle\langle l|.
\label{eq:vde-clock}
\end{align}
Hadamards prepare $|u\rangle$.
Every row and column sum of $J_N$ is at most one, so $\|J_N\|\le\sqrt{\|J_N\|_1\|J_N\|_\infty}\le1$.

For constant scalar generator and preparation signals $z_H,z_0,z_b$, initialize $x_0=\|\boldsymbol{x}_0\|z_0$ and impose
\begin{align}
\boldsymbol{y}_j=|u\rangle x_j+hJ_N(\beta|u\rangle z_b-\alpha z_H\boldsymbol{y}_j),
\quad x_{j+1}=x_j+h\langle u|(\beta|u\rangle z_b-\alpha z_H\boldsymbol{y}_j).
\label{eq:vde-scalar-stages}
\end{align}
The first equation imposes the Volterra constraint at each sample; the second gives the endpoint.
Row differences recover Euler with step $h/N$.
For varying operators, make the substitutions in their original multiplication order: \begin{align}
\alpha z_H\mapsto D_j=\operatorname{diag}_kH(t_{j,k}),
\quad \beta|u\rangle z_b\mapsto\boldsymbol{b}_j=N^{-1/2}\sum_k|k\rangle\boldsymbol{b}(t_{j,k}).
\end{align} Preparations give the initial and forcing columns, giving boundaries $\boldsymbol{X}_j$ and stages $\boldsymbol{Y}_j$.

An initial-condition perturbation propagates through every coarse interval.
Scaling the initial row by $m^{-1/2}$ balances its inverse contribution with the interval-forcing contributions, both of order $Cm$.
We then amplify the final coordinate relative to the rest of the history by attaching
\begin{align}
\tau\Bz=\boldsymbol{X}_m,
\quad \tau=\frac{r}{4\sqrt m\,W_2}.
\label{eq:vde-readout-scale}
\end{align}
The lower bound on the final norm makes this coordinate carry more than four fifths of the solution's squared norm; \cref{eq:vde-endpoint-norm} gives the quantitative estimate.
Write $E_j=I+hJ_ND_j$; identities on the system register are implicit.
For two intervals, the coefficient in the interleaved order $(\boldsymbol{X}_0,\boldsymbol{Y}_0,\boldsymbol{X}_1,\boldsymbol{Y}_1,\boldsymbol{X}_2,\Bz)$ is
\begin{align}
\mathcal B= \begin{pmatrix} I/\sqrt2&0&0&0&0&0
\\
-|u\rangle&E_0&0&0&0&0
\\
-I&h\langle u|D_0&I&0&0&0
\\
0&0&-|u\rangle&E_1&0&0
\\
0&0&-I&h\langle u|D_1&I&0
\\
0&0&0&0&-I&\tau I \end{pmatrix}.
\label{eq:vde-system}
\end{align}
For general $m$, repeat the stage and boundary row pair and use $m^{-1/2}I$ in the first row.
Before normalizing the right-hand side, the right-hand side is $(\boldsymbol{x}_0/\sqrt m,\{hJ_N\boldsymbol{b}_j, h\langle u|\boldsymbol{b}_j\}_j,0)^T$.
For $N=4$, the stage block is
\begin{align}
E_j= \begin{pmatrix} I&0&0&0
\\
\delta H_{j,0}&I&0&0
\\
\delta H_{j,0}&\delta H_{j,1}&I&0
\\
\delta H_{j,0}&\delta H_{j,1}&\delta H_{j,2}&I \end{pmatrix},
\quad H_{j,k}=H(t_{j,k}),
\quad\delta=h/4.
\label{eq:vde-local-matrix}
\end{align}
Its $k$th row is $\sqrt N(\boldsymbol{Y}_j)_k=\boldsymbol{X}_j+ \delta\sum_{l<k}[\boldsymbol{b}(t_{j,l})-H_{j,l}\sqrt N(\boldsymbol{Y}_j)_l]$.
The Neumann series gives $\|E_j^{-1}\|\le(1-h\alpha\|J_N\|)^{-1}\le2$, independently of $N$.

The coefficient $\mathcal B$ is the sum of diagonal attenuation, the negative endpoint-coupling term, and generator action, encoded with respective normalizations $1,\sqrt2,h\alpha\sqrt2$.
The endpoint coupling maps each intermediate endpoint into two orthogonal rows of the next interval: $\boldsymbol{X}_j\mapsto(|u\rangle\boldsymbol{X}_j,\boldsymbol{X}_j) $ for $j<m$.
The final endpoint is sent to the final-state output.
The generator-dependent column is
\begin{align}
h\begin{pmatrix}J_N
\\
\langle u|\end{pmatrix}D_j.
\label{eq:vde-shared-query}
\end{align}
To encode $J_N$, prepare a uniform auxiliary index beside the input index.
Flag the event $k\le l$ as failure, swap the indices, and unprepare the auxiliary index.
Zero-flag compression gives $N^{-1}\mathbb 1_{k>l}$.
For $N=1$ the encoded matrix is zero.
A branch rotation combines $J_N$ and $\langle u|$.
The normalization is $\sqrt2$; continuation uses the same construction.
Three-term LCU gives $\Be{\mathcal B/4}$, since $1+\sqrt2+h\alpha\sqrt2<4$.
It uses one coherent generator query and $O(n+a+\log(2mN))$ comparison, shift, Hadamard, and signal-test gates.

Prepare initial and forcing branches with amplitudes $\|\boldsymbol{x}_0\|/B$ and $T\beta/B$, whose squared magnitudes sum to one.
The forcing branch contains uniform $m$- and $N$-clocks and the full preparation column.
A known rectangular encoding sends the initial branch to the first row and forcing interval $j$ through $(J_N,\langle u|)^T$ with the projection onto its forcing signal.
The comparator circuit implements this map with normalization $\sqrt2$.
The full prepared state $|c\rangle$ has component on the coordinates representing the discretized equation
\begin{align}
\frac1{\sqrt2 B} \left(\boldsymbol{x}_0,\ \{\sqrt m\,hJ_N\boldsymbol{b}_j,\ \sqrt m\,h\langle u|\boldsymbol{b}_j\}_j,\ 0\right)^T.
\label{eq:vde-source-clock}
\end{align}
Retain all other amplitudes and extend $\mathcal B$ by the identity there, giving $\mathcal A=(\mathcal B\oplus I)/4$.
The algorithm prepares $|c\rangle$, applies QLSP to $\mathcal A$, and selects the final-state output coordinates (\cref{fig:vde-assembly}).

\begin{figure}[tbp]
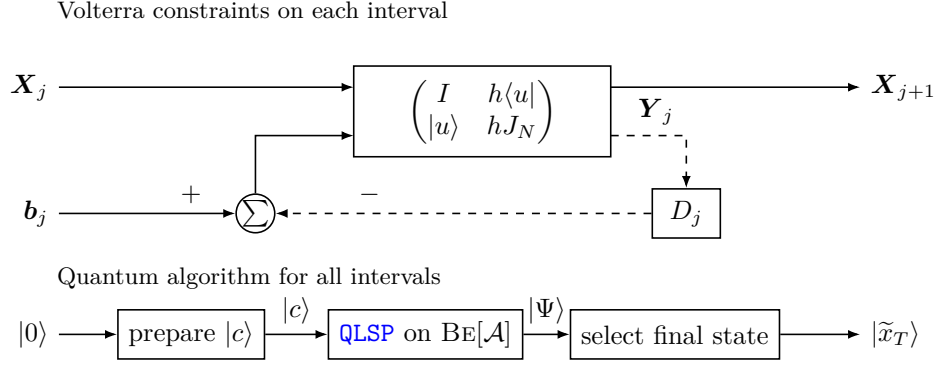

\centering
\IIRDiagram{VolterraAssemblyCurrent}
\caption{Top: the common residual $\boldsymbol{b}_j-D_j\boldsymbol{Y}_j$ determines both interval outputs, with the dashed feedback path imposing the Volterra stage constraint.
Arrows denote linear relations; the sum node is a row of the encoded coefficient matrix.
Bottom: conditioned on its success flag, QLSP solves all interval constraints and prepares $|\Psi\rangle=\mathcal A^{-1}|c\rangle/\|\mathcal A^{-1}|c\rangle\|$.
Selecting the final-state output gives the final state with constant probability.}
\label{fig:vde-assembly}
\end{figure}

Subtracting consecutive stage rows recovers $\widetilde{\boldsymbol{x}}_{l+1}=(I-\delta H(t_l))\widetilde{\boldsymbol{x}}_l+ \delta\boldsymbol{b}(t_l)$, with the boundary row giving the final fine step.
Consequently, the solution on the coordinates representing the discretized equation is
\begin{align}
(\boldsymbol{X}_j,\boldsymbol{Y}_j,\Bz)=\frac{\sqrt m}{\sqrt2 B} \left(\widetilde{\boldsymbol{x}}(jh),\ N^{-1/2}\sum_k|k\rangle\widetilde{\boldsymbol{x}}(t_{j,k}),\ \widetilde{\boldsymbol{x}}(T)/\tau\right).
\label{eq:vde-exact-output}
\end{align}
For the discretization error, set $\mathcal E=gV_H+V_b+T\alpha(\alpha g+\beta)$.
The total local propagator defect is at most $\delta(V_H+T\alpha^2)$: on a fine interval its norm is bounded by $\delta\operatorname{Var}(H)+ \alpha^2\delta^2e^{\alpha\delta}/2$.
Write $p_b$ for the product norm from a fixed starting sample and $e_l$ for each local propagator defect norm.
Ordered telescoping against the exact propagators gives $p_b\le C+C\sum_{l<b}e_lp_l$; induction bounds this by $C\prod_l(1+Ce_l)$.
The chosen mesh gives $C\delta\mathcal E\le\epsilon r/128$, and $\mathcal E\ge g(V_H+T\alpha^2)$.
Using $1+x\le e^x$ therefore gives, on every consecutive subinterval,
\begin{align}
\|(I-\delta H(t_{b-1}))\cdots(I-\delta H(t_a))\| \le C e^{C\delta(V_H+T\alpha^2)}\le2C.
\label{eq:vde-time-stability}
\end{align}
The exact state is Lipschitz with constant $\alpha g+\beta$.
Splitting interval residuals into variations of $H,\boldsymbol{b},\boldsymbol{x}$ gives total residual at most $\delta\mathcal E$, including jumps.
Propagating this residual using \cref{eq:vde-time-stability} gives grid error at most $\epsilon r/64$.
The case $\mathcal E=0$ is exact.
In particular, grid-state norms are at most $65g/64$ and the final norm is at least $63r/64$.
Constant data use the same proof with $V_H=V_b=0$.

Let $L$ be \cref{eq:vde-system} with its final-state output row and column removed, and set $G=L^{-1}$.
Let $P$ project onto the final endpoint $\boldsymbol{X}_m$.
For residuals $(\boldsymbol{e}_0,\{\boldsymbol{d}_j,\boldsymbol{e}_{j+1}\}_j)$, eliminating the stages gives
\begin{subequations}
\label{eq:vde-eliminate}
\begin{align}
\boldsymbol{Y}_j&=E_j^{-1}(|u\rangle\boldsymbol{X}_j+\boldsymbol{d}_j),
\quad \boldsymbol{X}_0=\sqrt m\,\boldsymbol{e}_0,
\\
\boldsymbol{X}_{j+1}&=\Phi_j\boldsymbol{X}_j+\boldsymbol{e}_{j+1} -h\langle u|D_jE_j^{-1}\boldsymbol{d}_j,
\\
\Phi_j&=I-h\langle u|D_jE_j^{-1}|u\rangle =(I-\delta H_{j,N-1})\cdots(I-\delta H_{j,0}).
\end{align}
\end{subequations}
Row differences give the product identity in \cref{eq:vde-eliminate}.
For $m=4$, the Schur complement on boundaries, $L_{\partial}$, has inverse
\begin{align}
L_{\partial}^{-1} =\begin{pmatrix} I/2&0&0&0&0
\\
-\Phi_0&I&0&0&0
\\
0&-\Phi_1&I&0&0
\\
0&0&-\Phi_2&I&0
\\
0&0&0&-\Phi_3&I \end{pmatrix}^{-1}=\begin{pmatrix} 2I&0&0&0&0
\\
2\Phi_0&I&0&0&0
\\
2\Phi_1\Phi_0&\Phi_1&I&0&0
\\
2\Phi_2\Phi_1\Phi_0&\Phi_2\Phi_1&\Phi_2&I&0
\\
2\Phi_3\Phi_2\Phi_1\Phi_0&\Phi_3\Phi_2\Phi_1&\Phi_3\Phi_2&\Phi_3&I \end{pmatrix}.
\label{eq:vde-expanded-history}
\end{align}
Multiplication verifies the inverse.
For a general number of intervals, replace the first-column factor by $\sqrt m$.
Each nonzero block is an ordered propagator, bounded by $2C$ before this first-column factor.
These ordered products depend on both interval indices, so the causal inverse is generally not Toeplitz.
The unitary Toeplitz lift implements the QLSP iteration; stage elimination bounds the Volterra inverse.

Since $\|h\langle u|D_jE_j^{-1}\|\le1$, each boundary forcing receives two residuals with coefficients of norm at most one.
The scalar majorant of the boundary solution map therefore has squared Frobenius norm at most $4C^2[m(m+1)+2m(m+1)/2]\le16C^2m^2$; its final row has squared norm at most $12C^2m$.
Let $\boldsymbol{e}$ collect all the residuals.
The boundary and stage norms then obey
\begin{subequations}
\label{eq:vde-residual-norms}
\begin{align}
\|\boldsymbol{X}\|&\le4Cm\|\boldsymbol{e}\|,
\\
\|\boldsymbol{Y}\|&\le2(\|\boldsymbol{X}\|+\|\boldsymbol{d}\|) \le10Cm\|\boldsymbol{e}\|,
\\
\|G\|&\le11Cm,
\quad \|PG\|\le4C\sqrt m.
\end{align}
\end{subequations}
On the history coordinates, final-state output coordinates, and coordinates containing the orthogonal completion of the state preparation,
\begin{align}
\mathcal A=\frac14 \begin{pmatrix}L&0&0
\\
-P&\tau I&0
\\
0&0&I\end{pmatrix},
\quad \mathcal A^{-1}=4 \begin{pmatrix}G&0&0
\\
\tau^{-1}PG&\tau^{-1}I&0
\\
0&0&I\end{pmatrix}.
\label{eq:vde-border-inverse}
\end{align}
Since $W_2\ge r$, the block inverse gives $\|\mathcal B^{-1}\|\le31CmW_2/r$.
Together with $\|\mathcal A\|\le1$, this implies
\begin{align}
\kappa_2(\mathcal A)=\|\mathcal A\|\,\|\mathcal A^{-1}\| \le\|\mathcal A^{-1}\|\le K:=128CmW_2/r.
\label{eq:vde-condition}
\end{align}
The inverse bound depends on coarse-interval propagation and the output scaling parameter; $N$ enters the clock-register size and known-gate count.
For a zero generator, the boundary inverse satisfies $\|\text{first column of }L_{\partial}^{-1}\|=\sqrt{m(m+1)}$ and $\|L_{\partial}^{-1}\|=\Theta(m)$ for every fine-grid size $N$.

Write $\boldsymbol x_{\rm hist}$ for the boundary values and rescaled stage values appearing in \cref{eq:vde-history-rms}.
Before the common factor four in the inverse, the norms of the remaining coordinates and final-state output obey
\begin{subequations}
\label{eq:vde-endpoint-norm}
\begin{align}
\frac{m}{2B^2}\|\boldsymbol x_{\rm hist}\|^2+\|c_\perp\|^2 &\le\frac32(mW_2/B)^2,
\\
\|\boldsymbol z\|^2 &=\frac{m}{2B^2\tau^2}\|\boldsymbol x_{\rm e}(T)\|^2 \ge8(63/64)^2(mW_2/B)^2.
\end{align}
\end{subequations}
The first bound includes the full preparation completion, whose norm is at most one, and uses the floor $W_2\ge B/m$.
The probability of selecting the final-state output therefore exceeds $4/5$.
In particular,
\begin{align}
\|\mathcal A^{-1}|c\rangle\|\ge R_0:=4mW_2/B\ge4, \quad K/R_0=32CB/r.
\label{eq:vde-certificates}
\end{align}
Apply \VDEQLSP\ with inverse-state error $\min\{10^{-3},\epsilon/256\}$.
It uses $O(K\log(1/\epsilon))$ coefficient calls and $O(K/R_0)$ state preparations, counting both directions.
Conditioning on selection of the final-state output increases the inverse-state error by at most a factor eight, while normalizing the discretized endpoint contributes at most $\epsilon/32$.
Their sum is below $\epsilon$.
A constant number of attempts raises the overall success probability to at least $2/3$.

Each coefficient call uses one generator query; each preparation of $|c\rangle$ uses one initial and one forcing query.
Binary sector and time labels require $O(\log(2mN))$ qubits in addition to the padded oracle registers.
Apply \VDEResources\ with $J=O(\log(2+K/R_0))$, $N_0=O(K)$, and $N_1=O(K\log(1/\epsilon))$ and register width $n+a+O(\log(2mN))$.
This gives \cref{eq:vde-gates,eq:vde-qubits}.
\end{proof}

Suppose the analytic extensions are bounded by $M_H,M_b$ in a complex neighborhood of the real interval.
Cauchy's estimate on circles of radius $\sigma/2$ gives $\operatorname{dist}(z,[0,T])<\sigma$, $V_H\le2TM_H/\sigma$, and $V_b\le2TM_b/\sigma$.
Bounded variation also permits jumps: \cref{eq:vde-time-grid} refines the sampled grid, while the resource bounds depend on its size through $\log N$ and the cost of the coherently controlled oracles.
State-dependent propagation bounds can reduce $g$ through variation of constants.

\providecommand{\VDEMultiQLSP}{\cref{wql:thm:main}}
\providecommand{\VDEMultiRates}{\cref{wql:subsec:allocation}}
\providecommand{\VDEMultiJunction}{\cref{lem:common-cayley}}
\providecommand{\VDEMultiReciprocal}{\cref{wql:lem:reciprocal}}

\subsubsection{Multiple matrix and preparation oracles}
\label{sec:vde-multioracle}
The Volterra transfer function is the same as above.
We now replace each aggregate matrix and preparation port by the corresponding direct sum of modules querying the original oracles, so the QLSP realization inherits separate query labels without changing the causal discretization.
Suppose the generator, initial vector, and forcing are given as
\begin{align}
H(t)=\sum_{\ell=1}^{p}H_\ell(t),
\quad \boldsymbol{x}_0=\sum_{k=1}^{s}\xi_k|v_k\rangle,
\quad \boldsymbol{b}(t)=\sum_{k=1}^{q}\beta_k\boldsymbol{v}_k(t),
\label{eq:vde-multi-input}
\end{align}
with matrix encodings $\Be{H_\ell(t)/\alpha_\ell}$ and normalized-state preparations of $|v_k\rangle$.
The forcing preparations must provide complete columns $\boldsymbol{v}_k(t)$ with $\|\boldsymbol{v}_k(t)\|\le1$.
The coefficients $\xi_k$ are known complex numbers and $\alpha_\ell,\beta_k>0$ are known.
Controls and adjoints are available, with coherent sampled-time control for varying data, as in \cref{thm:vde-constant}.
Zero terms are omitted.

Assume the bounds in \cref{eq:vde-promises} for the combined evolution.
Sum the component normalizations,
\begin{align}
\alpha=\sum_\ell\alpha_\ell,
\quad\beta=\sum_k\beta_k.
\end{align}
Use the interval count, interval length, fine-grid size, and output scaling parameter from \cref{thm:vde-constant}, with $g\ge h\beta$.
Use the history bound in \cref{eq:vde-history-rms}, with the floor replaced by $W_2\ge\max\{r,h\beta\}$ for the combined unnormalized right-hand side.
Bounds on the total variations can be obtained by summing those of the terms.
Define the known weights
\begin{align}
\lambda_0=\frac{1+\sqrt2}{4},
\quad \lambda_\ell=\frac{\sqrt2 h\alpha_\ell}{4}
\quad(1\le\ell\le p),
\quad \boldsymbol{\omega}=\left(\frac{\xi_1}{\sqrt2},\ldots, \frac{\xi_s}{\sqrt2},T\beta_1,\ldots,T\beta_q\right).
\label{eq:vde-multi-weights}
\end{align}
Index $0$ denotes the known initial, continuation, and final-state output coefficient.
Let $C_\ell,c_k>0$ be the matrix and preparation query prices.
They index $p$ matrix terms and $s+q$ preparations.
For nonnegative vectors, write $\langle\boldsymbol{C},\boldsymbol{\alpha}\rangle_{1/2} =(\sum_{\ell=1}^{p}\sqrt{C_\ell\alpha_\ell})^2$.

\cref{thm:vde-multi} assigns separate delays to each generator and prepared input term, using the weighted solver in \cref{wql:thm:main}.

\begin{theorem}[Separate matrix and preparation costs]
\label{thm:vde-multi}
For $0<\epsilon\le1$, the final state in \cref{thm:vde-constant} can be prepared with conditional trace-distance error at most $\epsilon$.
The success probability is at least $2/3$.
With $K=128CmW_2/r$, one common schedule gives
\begin{align}
\sum_{\ell=1}^{p}C_\ell Q_\ell &=O\!\left[\left(\frac{CTW_2}{r} \langle\boldsymbol{C},\boldsymbol{\alpha}\rangle_{1/2} +\|\boldsymbol{C}\|_1\right)\log\frac1\epsilon\right],
\label{eq:vde-multi-matrix-cost}
\\
\sum_{k=1}^{s+q}c_k Q_{{\rm prep},k} &=O\!\left[\frac{C}{r} \langle\boldsymbol{c},|\boldsymbol{\omega}|\rangle_{1/2} +\|\boldsymbol{c}\|_1\right].
\label{eq:vde-multi-source-cost}
\end{align}
Counts include controlled original calls in both directions.
Set $t_0=1+2K\lambda_0$ for the known coefficient and, for $\ell\ge1$, put
\begin{align}
t_\ell=1+2K\sqrt{\lambda_\ell/C_\ell} \sum_{i=1}^{p}\sqrt{C_i\lambda_i},
\quad u_k=1+\frac{32C}{r}\sqrt{|\omega_k|/c_k} \sum_{i=1}^{s+q}\sqrt{c_i|\omega_i|}.
\label{eq:vde-multi-rates}
\end{align}
The simultaneous query and coefficient-operation bounds are $Q_\ell=O(t_\ell\log(1/\epsilon))$ and $Q_{{\rm prep},k}=O(u_k)$, together with $O(t_0\log(1/\epsilon))$ coefficient operations.
\end{theorem}
\begin{proof}
Decompose the coefficient as $\mathcal A=\mathcal A_0+\sum_{\ell=1}^{p}\mathcal A_\ell$.
Setting $H=0$ in \cref{eq:vde-system,eq:vde-border-inverse}, with all scales fixed, gives $\mathcal A_0$.
For each interval, the only nonzero column of $\mathcal A_\ell$ acts as
\begin{align}
\boldsymbol{Y}_j\mapsto\frac h4 \begin{pmatrix}J_N
\\
\langle u|\end{pmatrix} D_{j,\ell}\boldsymbol{Y}_j,
\quad D_{j,\ell}=\operatorname{diag}_k H_\ell(t_{j,k}).
\label{eq:vde-multi-column}
\end{align}
The construction of \cref{eq:vde-shared-query} gives $\Be{\mathcal A_\ell/\lambda_\ell}$ with one $H_\ell$ query.
The diagonal and continuation terms give $\|\mathcal A_0\|\le(1+\sqrt2)/4=\lambda_0$, including its identity completion block.
Its block-encoding uses only known gates.

For an initial component, prepare $|c_k\rangle$ by placing $|v_k\rangle$ in the initial-row coordinates.
For forcing component $k$, prepare uniform interval and sample clocks, call its full oracle, and apply the map $(J_N,\langle u|)^T/\sqrt2$ to the selected column.
Routing each forcing completion to its own orthogonal sector produces a normalized $|c_{s+k}\rangle$ with one forcing query.
The component on the coordinates representing the discretized equation of $\boldsymbol{c}_{\rm sum}=\sum_{k=1}^{s+q}\omega_k|c_k\rangle$ is
\begin{align}
\frac1{\sqrt2} \left(\boldsymbol{x}_0,\{\sqrt m\,hJ_N\boldsymbol{b}_j, \sqrt m\,h\langle u|\boldsymbol{b}_j\}_j,0\right)^T.
\label{eq:vde-multi-source}
\end{align}
Retaining oracle workspace in the states and uncomputing temporary work gives the full unitary preparations required by \VDEMultiQLSP.

The forcing completions are mutually orthogonal, and their total squared norm is at most $T^2\sum_k\beta_k^2\le m^2W_2^2$.
The inverse applied to the right-hand side on the coordinates representing the discretized equation is $\sqrt m/\sqrt2$ times the solution history and final-state output.
Hence
\begin{align}
\|\text{other output coordinates}\|^2\le\tfrac32m^2W_2^2,
\quad \|\boldsymbol z\|^2
\quad \ge8(63/64)^2m^2W_2^2.
\label{eq:vde-multi-output}
\end{align}
The probability of selecting the final-state output exceeds $4/5$, and the complete inverse applied to the sum of prepared input vectors has norm at least $4mW_2$.
Apply \VDEMultiQLSP\ with $(\boldsymbol{\lambda},\boldsymbol{\omega},K,4mW_2)$.
The rates satisfy the requirements of \VDEMultiRates:
\begin{align}
K\lambda_0/t_0\le1/2,
\quad K\sum_{\ell=1}^p\lambda_\ell/t_\ell\le1/2,
\quad \frac K{4mW_2}\sum_k|\omega_k|/u_k\le1.
\end{align}
Multiplying each rate by its query cost and summing, with $Kh=128CTW_2/r$ and $K/(4mW_2)=32C/r$, gives \cref{eq:vde-multi-matrix-cost,eq:vde-multi-source-cost}.
Hermitian dilation preserves the norm bounds.
Use inverse-state error $\min\{10^{-3},\epsilon/256\}$ and the final-state output and repetition argument of \cref{thm:vde-constant}.
\end{proof}

For the known-gate implementation, put
\begin{align}
L_*&=\left\lceil\max_{\ell,k}\{t_\ell,u_k\}\right\rceil,
\quad J=1+\left\lceil\log_2\frac{32C\|\boldsymbol{\omega}\|_1}{r}\right\rceil,
\label{eq:vde-multi-horizons}
\\
b_*&=n+a+O\!\left(\log[2mN(p+s+q+1)JL_*] +\log\log\frac4\epsilon\right).
\label{eq:vde-multi-width}
\end{align}
The maximum defining $L_*$ includes the coefficient term with index zero.
The workspace width $a$ includes the persistent registers of the padded oracles.
If $G_{H,\ell}$ and $G_{{\rm prep},k}$ count gates in controlled original calls, the implementation below uses
\begin{subequations}
\label{eq:vde-multi-resources}
\begin{align}
G_{\rm total} &\le\sum_{\ell=1}^{p}Q_\ell G_{H,\ell} +\sum_{k=1}^{s+q}Q_{{\rm prep},k}G_{{\rm prep},k}+O\!\left(L_*\left[(p+s+q+1)J +(p+1)\log^2\frac1\epsilon\right] b_*\log(2+b_*)\right),
\\
q_{\rm total}&\le b_*.
\end{align}
\end{subequations}

Use \VDEMultiRates\ with matrix and preparation delays $\lceil L_*/t_\ell\rceil$ and $\lceil L_*/u_k\rceil$.
The corresponding horizons are $N_0=O(L_*)$ and $N_1=O(L_*\log(1/\epsilon))$.
In \VDEMultiJunction, the known Gram matrix is diagonal on three coordinate types of the Cayley load.
Each coupling is a known weighted coordinate embedding, so $O(p+s+q+1)$ two-level rotations align its coupling vector with one coordinate.
The remaining unitary acts on a fixed-dimensional subspace of each type, including after the constraint coordinate is closed.
Each system matrix therefore costs $O((p+s+q+1)b_*)$ gates.
Cascading $J$ system matrices and routing their binary addresses costs $O((p+s+q+1)Jb_*\log(2+b_*))$ per coarse step.

The construction of \VDEMultiReciprocal\ uses $O(\log(1/\epsilon))$ Cayley sections with $p+1$ matrix queries each.
Traversing their known systems costs $O((p+1)\log(1/\epsilon)b_*\log(2+b_*))$ per precision step.
A due oracle acts once on the direct sum of its occurrences, preserving \cref{eq:vde-multi-matrix-cost}.
The two clocks cost $O(N_0+N_1)$ gates by direct state preparation.
Buffer-index vectors of length $O(L_*)$ and their alignment add $O((p+s+q+1)JL_*b_*\log(2+b_*))$ gates at the final output stage.
Summing gives \cref{eq:vde-multi-resources}.

The direct sum contains $O((p+s+q+1)JL_*)$ padded oracle spaces; the precision stage enlarges it by $O((p+1)L_*\log^2(1/\epsilon))$.
Binary addresses and the reusable scratch of \VDEResources\ give \cref{eq:vde-multi-width}.
Apply \VDESynthesis\ for a finite gate set.

\subsubsection{Mean-square histories and classical certificates}
\label{sec:vde-certificates}

The output scaling parameter in \cref{eq:vde-readout-scale} depends on the norm of the whole sampled history.
A short transient can therefore give a much smaller cost than its peak amplitude suggests.
A second improvement uses a bound on the discarded solution component for the already balanced coefficient matrix.
These bounds apply after the output scaling parameter has been chosen.

Write the coefficient in \cref{eq:vde-border-inverse} as $A_\tau$ and set
\begin{align}
A_\tau=\mathcal A(\tau),
\quad \boldsymbol z_\tau=A_\tau^{-1}\ket c,
\quad R_0=4mW_2/B,
\label{vdc:eq:balanced}
\end{align}
where the full right-hand-side vector includes the preparation-completion coordinates.
\cref{vdc:thm:combined} applies the QLSP tail theorem to this inverse state.

\begin{theorem}[Mean-square histories with a classical bound on the discarded solution component]
\label{vdc:thm:combined}
Let $0<\epsilon\le1/10$.
Under the hypotheses of \cref{thm:vde-constant}, suppose a classical bound gives $K_\epsilon\ge2$ with
\begin{align}
\norm{\mathbf1_{[0,1/K_\epsilon)}(|A_\tau|)\boldsymbol z_\tau} \le c\epsilon R_0,
\label{vdc:eq:tail}
\end{align}
for a sufficiently small universal constant $c$.
MQSP prepares the normalized final state with conditional trace-distance error at most $\epsilon$ and success at least $2/3$, using
\begin{align}
Q_H=O\!\left(K_\epsilon\log\frac1\epsilon\right),
\quad Q_0+Q_b=O\!\left(\frac{B K_\epsilon}{mW_2}\right).
\label{vdc:eq:combined-cost}
\end{align}
Zero input terms are omitted.
The sufficient fallback
\begin{align}
K_0=\frac{128CmW_2}{r}
\label{vdc:eq:baseline}
\end{align}
has zero discarded tail.
Thus choosing the better certified scale never worsens the existing bounds
\begin{align}
Q_H=O\!\left(\frac{CmW_2}{r}\log\frac1\epsilon\right),
\quad Q_0+Q_b=O(CB/r).
\label{vdc:eq:rms-baseline}
\end{align}
The original coefficient and state-preparation oracles implement the circuit.
\end{theorem}

\begin{proof}
\cref{eq:vde-endpoint-norm,eq:vde-certificates} give probability of selecting the final-state output greater than $4/5$ and the inverse-state norm lower bound.
Apply \cref{cert:thm:tail} to the normalized right-hand-side vector with an internal accuracy that is a sufficiently small constant multiple of the target.
Conditioning on selection of the final-state output changes trace-distance error by only a constant factor.
The endpoint discretization error is bounded by \cref{eq:vde-time-grid}, and a constant number of bounded attempts gives the stated success probability.
One coefficient call uses a constant number of original generator calls; one state preparation uses at most one call to each nonzero preparation oracle.
Substituting the norm lower bound gives \cref{vdc:eq:combined-cost}.
\cref{eq:vde-condition} proves the global inverse-norm bound, for which no singular component is discarded.
\end{proof}

If a further certificate gives $K_\epsilon\le\theta K_0$, both query bounds improve by the factor $\theta$, up to the unavoidable constant preparation budget.
The mean-square improvement alone cancels from that preparation count.
A certified cutoff is always of order at least the inverse-state norm lower bound, by \cref{cert:eq:necessary-cutoff}.

A continuous energy bound can supply the sampled-history estimate:
\begin{align}
E_2^2\ge T^{-1}\int_0^T\|\boldsymbol x(t)\|^2\,dt,
\quad W_2=O\!\left(\max\{r,E_2,h\beta\}\right).
\label{vdc:eq:energy-main}
\end{align}
\cref{vdc:prop:continuous} proves this sampling bound, including Euler error and the lower bound $W_2\ge B/m$ accounting for orthogonal preparation-completion amplitudes.
Positive classical Lyapunov data give one way to establish the energy bound:
\begin{align}
H(t)^\dagger\mathsf P+\mathsf P H(t)\succeq2\eta I,
\quad \mathsf P\succ0,
\quad\eta>0.
\label{vdc:eq:lyapunov-main}
\end{align}
For homogeneous evolution, \cref{vdc:prop:lyapunov} gives an energy upper bound from the initial quadratic form.
For a driven equation it adds the integrated squared norm of the weighted forcing.

The inverse norm can also be bounded through propagation between different times.
Let $G=L^{-1}$ be the history inverse before the final-state output is appended, and let $P$ select its final boundary.
\cref{vdc:thm:green} distinguishes the norms of the full history and of this last row.

\begin{theorem}[Two-time Green-function bounds]
\label{vdc:thm:green}
Suppose classical bounds supply $k\ge\norm G$ and $\gamma\ge\norm{PG}$.
A sufficient inverse-norm upper bound that discards no singular component in \cref{vdc:thm:combined} is
\begin{align}
K_{\rm G}=\max\{2,4[k+(\gamma+1)/\tau+1]\}.
\label{vdc:eq:green-scale}
\end{align}
In particular,
\begin{subequations}
\label{vdc:eq:green-cost}
\begin{align}
Q_H&=O\!\left(\left[k+1+(\gamma+1)\frac{\sqrt m W_2}{r}\right] \log\frac1\epsilon\right),
\\
Q_0+Q_b&=O\!\left(1+\frac{B(k+1)}{mW_2} +\frac{B(\gamma+1)}{r\sqrt m}\right).
\end{align}
\end{subequations}
These are full inverse bounds and therefore also support the weighted multi-oracle theorem, without a singular-support assumption.
\end{theorem}
\begin{proof}
The block inverse in \cref{eq:vde-border-inverse} has norm at most the sum of the displayed three contributions and the retained identity.
This is bounded by \cref{vdc:eq:green-scale}.
Substitute it and $R_0$ into \cref{vdc:eq:combined-cost}.
\end{proof}

Scalar bounds on the ordered coarse propagators give both quantities by \cref{vdc:prop:envelope}.
For an envelope decaying geometrically with coarse-time separation, let its effective memory length be
\begin{align}
\ell=\min\{m+1,(1-q)^{-1}\},
\quad 0\le q<1.
\label{vdc:eq:memory-main}
\end{align}
Then the history and final-row bounds have orders
\begin{align}
k=O(\ell+\sqrt{m\ell}),
\quad\gamma=O(\sqrt\ell).
\label{vdc:eq:memory-norms}
\end{align}
\cref{vdc:eq:decay} gives the resulting query costs.
The envelope must bound the numerical propagators used by the circuit; a continuous-time bound includes the corresponding discretization allowance.

Inverse moments and spectral-band information from \cref{cert:app:spectral} apply to the balanced matrix in \cref{vdc:eq:balanced}.
For example,
\begin{align}
\mu_q\ge\||A_\tau|^{-q}\boldsymbol z_\tau\|
\quad\Longrightarrow
\quad \|\boldsymbol1_{[0,1/K)}(|A_\tau|)\boldsymbol z_\tau\| \le\mu_q/K^q.
\label{vdc:eq:moment-main}
\end{align}
The first such moment includes a backward response as well as the forward history; \cref{vdc:prop:adjoint} gives its exact expression.
Changing the output scaling parameter changes that moment and the singular spectrum.
The scale and scalar-profile choices in \cref{vdc:app:certificates} therefore recompute the inverse-moment or solution-tail bound after each rescaling.

For separate generator and preparation oracles, a common reducing subspace of the original system space supports the weighted result in \cref{vdc:thm:sectors}.
Its proof applies \cref{cert:lem:junction-support} to each oracle term.
\begin{theorem}[Reducing system subspaces and multiple oracles]
\label{vdc:thm:sectors}
Suppose $H(t)=\sum_jH_j(t)$, with block normalizations $\alpha_j$, and each initial or forcing vector is given separately.
Let an orthogonal projector on the original system space $\Pi$ obey
\begin{align}
H_j(t)\Pi=\Pi H_j(t)
\quad\text{for every }j,t,
\quad \Pi\boldsymbol x_{0,k}=\boldsymbol x_{0,k},
\quad \Pi\boldsymbol b_k(t)=\boldsymbol b_k(t).
\label{vdc:eq:common-sector}
\end{align}
No projector oracle is required.
Bounds on propagation, history norms, and the Green inverse may be restricted to $\ran\Pi$.
In particular, let $C_\Pi$ bound the restricted propagators and let the corresponding $W_2$ and mesh satisfy \cref{thm:vde-constant}.
Writing $\boldsymbol x_0=\sum_k\xi_k\ket{v_k}$ and $\boldsymbol b(t)=\sum_k\beta_k\boldsymbol v_k(t)$, the weighted bounds are
\begin{subequations}
\label{vdc:eq:weighted}
\begin{align}
\sum_j C_jQ_j &=O\!\left[\left(\frac{C_\Pi T W_2}{r} \langle\boldsymbol C,\boldsymbol\alpha\rangle_{1/2}+\norm{\boldsymbol C}_1\right) \log\frac1\epsilon\right],
\\
\sum_k c_kQ_{{\rm prep},k} &=O\!\left[\frac{C_\Pi}{r}\langle\boldsymbol c,|\boldsymbol\omega|\rangle_{1/2} +\norm{\boldsymbol c}_1\right],
\quad \boldsymbol\omega=(\xi_1/\sqrt2,\ldots,T\beta_1,\ldots).
\end{align}
\end{subequations}
Here $\langle\boldsymbol C,\boldsymbol\alpha\rangle_{1/2}=(\sum_j\sqrt{C_j\alpha_j})^2$.
For the version with an unnormalized combined right-hand side it suffices to impose the history bound and $W_2\ge\max\{r,h\sum_k\beta_k\}$.
\end{theorem}

The full proof is in \cref{vdc:app:certificates}.
A bound on the complete history inverse, such as \cref{vdc:eq:green-scale}, can be used in the weighted theorem directly.
An inverse cutoff parameter with unequal delays instead requires the paired invariance and assumptions on the individual input vectors in \cref{cert:eq:weighted-invariance}; a general cutoff remains available for a bundled coefficient encoding.

A classical polynomial identity can provide a finite-query approximation to the inverse on the specified input in place of an inverse cutoff parameter.
\cref{vdc:prop:finite} keeps the circuit's block-encoding normalization in the query count; its proof and a space-efficient implementation of matrix powers are in \cref{vdc:app:certificates}.
\begin{proposition}[Preparation from a finite-query inverse approximation]
\label{vdc:prop:finite}
Suppose a specified circuit using $d_H$ generator queries block-encodes $F/\beta_F$.
Let $\boldsymbol w=F\ket c$, with a given norm lower bound $0<y_0\le\norm{\boldsymbol w}$.
Suppose its normalized direction is within $O(\epsilon)$ of the normalized inverse history, up to global phase.
Then the normalized final solution $\boldsymbol x(T)/\|\boldsymbol x(T)\|$ can be prepared with constant success using
\begin{align}
Q_H=O(d_H\beta_F/y_0),
\quad Q_0+Q_b=O(1+\beta_F/y_0).
\label{vdc:eq:finite-action}
\end{align}
There is no additional inverse-approximation logarithm; $F,d_H,\beta_F,y_0$ may themselves depend on $\epsilon$.
\end{proposition}

Three examples separate the improvements.
For the dissipative shift generator in \cref{vdc:eq:shift}, the endpoint and mean-square trajectory norms both scale as $T^{-1/4}$.
The mean-square bound reduces the generator estimate from $O(T^{5/4}\log(1/\epsilon))$ to $O(T\log(1/\epsilon))$.
Preparation still costs $O(T^{1/4})$ queries.
For a dissipative sector coupled by direct sum to an unused unstable sector, \cref{vdc:thm:sectors} removes the latter's exponential propagator factor when every input vector and oracle term respects the sector.
Finally, an annihilating identity can give a constant-degree action:
\begin{align}
H=\lambda I+N_0,
\quad N_0^2\ket\psi=0
\quad\Longrightarrow
\quad e^{-HT}\ket\psi=e^{-\lambda T}(I-TN_0)\ket\psi.
\label{vdc:eq:nilpotent-main}
\end{align}
With the orthogonality and block-normalization assumptions in \cref{vdc:eq:nilpotent}, a constant number of queries prepares its normalized state.
\cref{vdc:prop:finite,vdc:lem:powers} give the general finite-query inverse construction and a logarithmic-space implementation of matrix powers.
The three example calculations are in \cref{vdc:app:certificates}.
Although a smaller cutoff reduces the query counts, the fine-grid address width still depends on $\log N$; improving that cost also requires a sharper discretization bound on the supported sector or trajectory.
\cref{vdc:sec:resources} gives the resulting gate and qubit counts.

\subsubsection{Time-dependent classical bounds}
\label{vde:sec:time-envelopes}
The Volterra solver can use the integrated forcing normalization and time-resolved propagation and trajectory bounds.
Dependence on the $L^1$ norm of the forcing normalization already occurs in LCHS \cite{an2023lchs,Low2025NonunitaryDynamics}; we obtain this dependence for the Volterra reduction by changing the clock, as in norm-rescaled Hamiltonian simulation \cite{Berry2020TimeDependentL1}.
Assume classical envelopes $\|U(t,s)\|\le f(t,s)$, $\|\boldsymbol x(t)\|\le g(t)$, and $\|\boldsymbol b(t)\|\le\beta(t)$.
In addition to the constant-normalization access of \cref{thm:vde-constant}, suppose the matrix oracle block-encodes $H(t)/\alpha(t)$ and the preparation oracle has successful column $\boldsymbol b(t)/\beta(t)$, with coherent sampled-time control and adjoints.
When $\beta(t)=\|\boldsymbol b(t)\|$, this is preparation of normalized forcing states.
Preparing only the direction requires separate access to the actual magnitude.
For a fixed-normalization matrix oracle, take $\alpha(t)=\alpha$.

Let $\rho>0$ be a probability density on $[0,T]$.
Assume reversible evaluation of $t(u)=F_\rho^{-1}(u)$, where $F_\rho(t)=\int_0^t\rho(s)ds$, and of the scalar attenuation factors below.
Their implementation cost is included in the gate bound.

\begin{theorem}[Classical time envelopes]
\label{vde:thm:time-envelopes}
Suppose $H(t(u))/\rho(t(u))$ and $\boldsymbol b(t(u))/\rho(t(u))$ have bounded variation.
Let $C\ge\max\{1,\sup_{s\le t}f(t,s)\}$ and $0<r\le\|\boldsymbol x(T)\|$.
Set
\begin{subequations}
\label{vde:env:history}
\begin{align}
\alpha_\rho&=\sup_t\frac{\alpha(t)}{\rho(t)},
\quad \beta_\rho=\sup_t\frac{\beta(t)}{\rho(t)},
\quad B_\rho=\sqrt{\|\boldsymbol x_0\|^2+\beta_\rho^2},
\\
m&=2^{\lceil\log_2\max\{1,2\alpha_\rho\}\rceil},
\quad W_\rho=8\max\left\{r,\left(\int_0^T\rho(t)g(t)^2dt\right)^{1/2}, B_\rho/m\right\}.
\end{align}
\end{subequations}
For $0<\epsilon\le1$, MQSP prepares the normalized endpoint with success probability at least $2/3$ and conditional trace-distance error $\epsilon$ using
\begin{align}
Q_H=O\!\left(\frac{CmW_\rho}{r}\log\frac1\epsilon\right),
\quad Q_0+Q_b=O(CB_\rho/r).
\label{vde:env:queries}
\end{align}
More generally, use the numerical ordered-propagator envelope $F_{jk}\ge f(t(j/m),t(k/m))+\eta$ for $j>k$, with $F_{jj}=1$, $F_{jk}=0$ for $j<k$, and certified propagator error at most $\eta$.
Compute $k\ge\|G\|$ and $\gamma\ge\|PG\|$ from \cref{vdc:prop:envelope}.
With $\tau=r/(4\sqrt m W_\rho)$ and $K=\max\{2,4[k+(\gamma+1)/\tau+1]\}$, \cref{vdc:thm:green}\ gives
\begin{align}
Q_H=O(K\log(1/\epsilon)),
\quad Q_0+Q_b=O(1+KB_\rho/(mW_\rho)).
\label{vde:env:green-queries}
\end{align}
The smaller certified inverse bound may be used.
Zero input ports are omitted.
The gate and qubit bounds of \cref{thm:vde-constant}\ hold for the transformed data, with the clock circuit's gates charged per transformed oracle call and its reusable workspace added.
\end{theorem}

\begin{proof}
The substitution $\boldsymbol y(u)=\boldsymbol x(t(u))$ gives
\begin{align}
\boldsymbol y'=-\frac{H(t(u))}{\rho(t(u))}\boldsymbol y +\frac{\boldsymbol b(t(u))}{\rho(t(u))},
\quad \widehat U(u,v)=U(t(u),t(v)).
\label{vde:env:transformed}
\end{align}
After computing $t(u)$, attenuate the original matrix block and preparation column by $\alpha(t)/(\alpha_\rho\rho(t))$ and $\beta(t)/(\beta_\rho\rho(t))$, respectively.
Each factor is at most one, so each transformed oracle uses one original call and known rotations.
Uncompute the arithmetic workspace and retain the completion flags.
The uniform $u$ register uses Hadamard gates.

Apply \cref{thm:vde-constant}\ with horizon one, normalizations $\alpha_\rho,\beta_\rho$, and preparation scale $B_\rho$.
The energy identity $\int_0^1\|\boldsymbol y(u)\|^2du =\int_0^T\rho(t)\|\boldsymbol x(t)\|^2dt$ and \cref{vdc:prop:continuous}\ give the stated history bound, including $W_\rho\ge8B_\rho/m$ to account for orthogonal preparation-completion amplitudes.
The balanced coefficient has inverse bound $128CmW_\rho/r$, and the solution for the full prepared right-hand-side vector has norm at least $4mW_\rho/B_\rho$.
Substitution into \cref{wql:thm:main}\ proves \eqref{vde:env:queries}; the bound on the ordered Green matrix gives \eqref{vde:env:green-queries}.
The mesh and circuit estimates are those of \cref{thm:vde-constant}\ evaluated on \eqref{vde:env:transformed}.
\end{proof}

\begin{corollary}[$L^1$ scaling of the forcing normalization]
\label{vde:cor:source-integral}
For $\alpha(t)=\alpha$ and $I_\beta=\int_0^T\beta(t)dt>0$, choose
\begin{align}
Z=\int_0^T\max\{1/T,\beta(t)/I_\beta\}\,dt,
\quad \rho(t)=Z^{-1}\max\{1/T,\beta(t)/I_\beta\}.
\label{vde:env:max-clock}
\end{align}
Then $1\le Z\le2$, $\alpha_\rho\le Z\alpha T$, $\beta_\rho\le ZI_\beta$, and
\begin{align}
Q_H=O\!\left(\frac{C(1+\alpha T)W_\rho}{r} \log\frac1\epsilon\right),
\quad Q_0+Q_b=O\!\left(\frac{C}{r} \sqrt{\|\boldsymbol x_0\|^2+I_\beta^2}\right).
\label{vde:env:integral-queries}
\end{align}
\end{corollary}
\begin{proof}
Bound the maximum in \eqref{vde:env:max-clock} between $1/T$ and $1/T+\beta/I_\beta$, then use \cref{vde:thm:time-envelopes}.
This clock also minimizes $\max\{\sup_t1/(T\rho(t)),\sup_t\beta(t)/(I_\beta\rho(t))\}$: any upper bound $c$ must satisfy $c\rho(t)\ge\max\{1/T,\beta(t)/I_\beta\}$, hence $c\ge Z$.
\end{proof}

For constant $\beta$, the clock is uniform and $I_\beta=T\beta$.
If $I_\beta=0$, omit the forcing term and use $\rho=1/T$.
A peak trajectory bound remains valid for either clock; an existing mean-square bound must be converted to the weighted integral in \eqref{vde:env:history}.
A bound $C(t)$ on every $U(t,s)$ may fill the corresponding row of the scalar Green envelope, whereas a bound on $U(t,0)$ alone applies only to its first column.
State-dependent bounds $\|U(t,0)\boldsymbol x_0\|\le g_0(t)$ and $\|U(t,s)\boldsymbol b(s)\|\le g_b(t,s)\beta(s)$ give $g(t)=g_0(t)+\int_0^t g_b(t,s)\beta(s)ds$ by variation of constants.
With only constant envelopes, $\rho=1/T$ recovers \cref{thm:vde-constant}.
The original clock remains a candidate when minimizing any chosen weighted query or gate cost.

Efficient scalar evaluation does not follow from preparation of normalized forcing states alone.
The inverse cumulative time map $t(u)=F_\rho^{-1}(u)$ may be evaluated by reversible bisection when its cumulative distribution has an efficient evaluator.
A clock circuit with polylogarithmic gate cost and logarithmic workspace preserves the corresponding Volterra resource bounds.
If the forcing normalization or clock map is instead supplied by a black-box oracle, count its calls separately from $Q_b$.

\subsection{Generalized eigenvalue estimation}
\label{subsec:app-generalized-eigenvalues}

The desired output is a classical estimate of a generalized eigenvalue, so we first choose the phase response that should be attached to a generalized eigenstate.
For scalar signals $a$ and $b>0$, the Cayley ratio $(b+ia/R)/(b-ia/R)$ has exactly the phase needed below.
Its powers form the phase-estimation history.
The matrix algorithm therefore realizes this rational history from the original $A$ and $B$ ports, using one affine linear system and the QLSP module, instead of first compiling the Hermitian reduction $B^{-1/2}AB^{-1/2}$ as a separate oracle.
This is the same modular pattern as \cref{sec:modularity}: specify the response, realize it from the given ports, then choose the finite clock used for readout.

For a Hermitian matrix and a positive definite metric, the generalized eigenvalue problem is $A\boldsymbol{x}=\lambda B\boldsymbol{x}$, $\boldsymbol{x}\ne0$, $A=A^\dagger$, and $B\succ0$.
For example, in a nonorthogonal basis, $A$ represents a Hermitian operator and $B$ is the Gram matrix of the basis.
We seek a classical eigenvalue estimate from separate encodings of $A,B$ and a prepared quantum state.
The usual Hermitian reduction, $B^{-1/2}AB^{-1/2}$, places a synthesized function of $B$ inside each matrix application.
Keeping the original oracles separate can reduce the cost when their normalizations or query prices differ.

Assume known positive bounds
\begin{align}
\beta I\preceq B\preceq b_{\max}I,
\quad \kappa_B=b_{\max}/\beta,
\quad H=B^{-1/2}AB^{-1/2},
\quad \operatorname{spec}H\subseteq[-R,R].
\label{eq:gep-promises}
\end{align}
The oracle access and positive normalizations are $\Be{A/\alpha_A}$, $\Be{B/\alpha_B}$, $U_\psi|0\rangle=|\psi\rangle$, $\alpha_A\ge\|A\|$, and $\alpha_B\ge b_{\max}$.
Controlled forward and inverse calls are available.
Their costs, including unsuccessful attempts, follow \cref{def:resource-accounting}.
The preparation is a unitary with its workspace returned to zero.
We assume coherent access to the preparation and its inverse, $U_\psi,U_\psi^\dagger$.
This access supports operations beyond those available from independent copies of the prepared state.

\cref{thm:gep-eigenstate} preserves separate costs for the two matrices and the eigenstate preparation.
Its proof uses the inverse-state guarantee of \cref{wql:thm:main}.

\begin{theorem}[Generalized eigenvalue estimation]
\label{thm:gep-eigenstate}
Assume the input is a generalized eigenstate satisfying $A|\psi\rangle=\lambda B|\psi\rangle$.
For $0<\epsilon<R$ and $0<\delta<1/2$, an algorithm returns a classical eigenvalue estimate.
Denoting it by $\widehat\lambda$, its success guarantee is $\Pr(|\widehat\lambda-\lambda|\le\epsilon)\ge1-\delta$, using
\begin{align}
Q_A=O\!\left[\left(\frac{\alpha_A}{\beta\epsilon}\right) \log\frac1\delta\right],
\quad Q_B=O\!\left[\left(\frac{R\alpha_B}{\beta\epsilon}\right) \log\frac1\delta\right],
\quad Q_\psi=O\!\left(\sqrt{\kappa_B}\log\frac1\delta\right).
\label{eq:gep-eigenstate-cost}
\end{align}
No eigenvalue gap is required.
The same orders hold for an input within Euclidean distance $\xi/\sqrt{\kappa_B}$ of the target eigenstate, for a sufficiently small universal constant $\xi>0$.
\end{theorem}

\cref{tab:gep-comparison} compares QEP \cite{Low2024Eigenvalue,Low2024QLSP}, Hermitian reduction \cite{Gilyen2018singular}, and ordinary LCU followed by state-preparation-optimal QLSP \cite{Low2024QLSP}, under the same eigenstate-input promise.
The complete logarithms and conversion costs are derived in \cref{app:gep-comparison}.

Under the change of coordinates below, an input satisfying $A|\psi\rangle=\lambda B|\psi\rangle$, obeys
\begin{align}
|\chi\rangle=\frac{B^{1/2}|\psi\rangle} {\sqrt{\langle\psi|B|\psi\rangle}},
\quad H|\chi\rangle=\lambda|\chi\rangle,
\quad H=B^{-1/2}AB^{-1/2}.
\label{eq:gep-hermitian-reduction}
\end{align}
Thus the generalized eigenvalue is an ordinary eigenvalue of a Hermitian matrix.
Given $|\chi\rangle$ and controlled evolution $e^{iH/R}$, standard quantum phase estimation estimates the eigenvalue divided by the reference scale.
A direct implementation converts the prepared input to $|\chi\rangle$.
It also constructs an encoding of the Hermitian reduction using matrix functions of $B$.
Both costs are included in the Hermitian-reduction row of \cref{tab:gep-comparison}.

Our construction uses this Hermitian problem to identify the desired phase, then prepares the Kaiser-window phase-estimation state from the original three oracles.
We first prepare the known Kaiser amplitudes in a phase register.
A single QLSP solve encodes all the controlled Cayley steps that attach the eigenphase to those amplitudes.
Fourier readout then gives an eigenvalue estimate.
The same construction applies to eigenstates and arbitrary trial states.

\begin{figure}[ht]
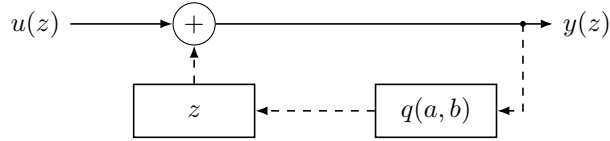

\centering

\IIRDiagram{GeneralizedEigenvalue}

\caption{The scalar IIR generating the phases $1,q,q^2,\ldots$; the dashed path is feedback.
The Kaiser amplitudes are prepared in the phase register.
Controlled queries attach the phase $q^k$ to the corresponding clock amplitude.
\cref{eq:gep-affine} encodes these operations in one linear system.}
\label{fig:gep-junction}
\end{figure}

Start with scalar signals $a\in\mathbb R$ and $b>0$.
A classical feedback loop \cite{Fettweis1986WaveDigitalFilters} satisfies $y=u+zq(a,b)y$, where $z$ denotes one delay.
Its infinite impulse response (IIR) transfer function is
\begin{align}
q(a,b)=\frac{b+ia/R}{b-ia/R},
\quad \mathcal H(z;a,b)=\frac1{1-zq(a,b)} =\sum_{k\ge0}z^k q(a,b)^k,
\quad |z|<1.
\label{eq:gep-scalar-history}
\end{align}
Its impulse coefficients are $e^{ik\theta}$, where $\theta=2\arctan(a/(Rb))$.
Weighting these coefficients by known Kaiser amplitudes $c_k$ gives the response used for phase estimation.
The variable $z$ records the history index; it is separate from the scalar matrix signals $a,b$.
\cref{fig:gep-junction} shows the recurrence.

For the matrix problem, define the ordered Cayley transforms
\begin{align}
V=(I-iH/R)^{-1}(I+iH/R),
\quad W=(B-iA/R)^{-1}(B+iA/R)=B^{-1/2}VB^{1/2}.
\label{eq:gep-cayley}
\end{align}
On the transformed eigenstate, the unitary $V$ has eigenphase $\theta=2\arctan(\lambda/R)\in[-\pi/2,\pi/2]$.
The similarity relation gives $W|\psi\rangle=e^{i\theta}|\psi\rangle$.
Prepare the known normalized Kaiser state $|c\rangle=\sum_{k=0}^{n-1}c_k|k\rangle$.
The desired output is
\begin{align}
\sum_{k=0}^{n-1}c_k|k\rangle W^k|\psi\rangle =\left(\sum_{k=0}^{n-1}c_ke^{ik\theta}|k\rangle\right) \otimes|\psi\rangle.
\label{eq:gep-phase-history}
\end{align}
Its phase register has the Kaiser-window QPE distribution.
The known window and shifted Fourier readout can be chosen using \cite[Sections II.B.3 and IV.A]{Berry2024Rapid} so that
\begin{align}
\Pr(\text{circular phase error}>\epsilon/R)\le\eta,
\quad n=O\!\left(\frac R\epsilon\log\frac1\eta\right)
\label{eq:gep-kaiser}
\end{align}
Take $n$ to be a power of two.
The readout includes a known randomized phase-grid offset, corrected in the reported phase, making the guarantee uniform in $\theta$.
Window preparation and this readout use only known operations.
We use $\eta=1/10$ for \cref{thm:gep-eigenstate}.

To prepare \cref{eq:gep-phase-history}, apply the walk at each step under the control $t=1,\ldots,n-1$ and $k\ge t$.
After step $t$, the component with phase label $k$ is $c_kW^{\min\{k,t\}}|\psi\rangle$.
All components are complete after step $n-1$.
Append $n$ idle steps so that at least half the computation history contains this completed output.
There are $N=2n$ history times.
The phase label $k$ and computation time $t$ occupy separate registers.

The projectors controlling these walk applications are \begin{align}
E_t=\begin{cases} \sum_{k=t}^{n-1}|k\rangle\langle k|,&1\le t<n,
\\
0,&n\le t<N. \end{cases}
\end{align} Clearing each controlled Cayley denominator gives
\begin{align}
s\Bz_0=|c,\psi\rangle,
\quad (I\otimes B-iE_t\otimes A/R)\Bz_t =(I\otimes B+iE_t\otimes A/R)\Bz_{t-1},
\quad 1\le t<N,
\label{eq:gep-circuit-history}
\end{align}
The boundary scale balances the inverse contribution of the initial row against the triangular history blocks:
\begin{align}
s=\beta\sqrt{\kappa_B/N},
\quad \frac{\sqrt{N\kappa_B}}s=\frac N\beta.
\label{eq:gep-boundary-balance}
\end{align}
Although this scale cancels from the normalized output, it controls the inverse bound.
\cref{eq:gep-inverse-bound} bounds the full inverse by twice the balanced scale.
For an eigenstate, the solution norm is $N/(\beta\sqrt{\kappa_B})$.
The algorithm uses half that value as its lower bound, so the ratio controlling preparation queries is $K/\rho_0=4\sqrt{\kappa_B}$, independently of history length.
On the time register, define the initial projector, shift, and controlled diagonal and shift operators:
\begin{align}
P_0&=|0\rangle\langle0|,
\quad S_N=\sum_{t=1}^{N-1}|t\rangle\langle t-1|,
\\
D_E&=\sum_{t=1}^{N-1}|t\rangle\langle t|\otimes E_t,
\quad T_E=\sum_{t=1}^{N-1}|t\rangle\langle t-1|\otimes E_t.
\end{align}
The full system is
\begin{subequations}
\label{eq:gep-affine}
\begin{align}
M_N\Bz&=|0,c,\psi\rangle,
\\
M_N&=sP_0\otimes I\otimes I +(I-P_0-S_N)\otimes I\otimes B -\frac iR(D_E+T_E)\otimes A.
\end{align}
\end{subequations}
Thus the Kaiser state is part of the prepared right-hand side.
The two matrix coefficients have norms at most two.
The multi-oracle QLSP theorem (\cref{wql:thm:main}) therefore applies with strengths $2\alpha_A/R$ and $2\alpha_B$, and one known boundary term.
There are three given oracle types: $\Be{A/\alpha_A}$, $\Be{B/\alpha_B}$, and state preparation.
One preparation of $|0,c,\psi\rangle$ uses one call to $U_\psi$.

Use the inverse bound, solution-norm lower bound, and conditional trace-distance error
\begin{align}
K=\frac{2N}{\beta},
\quad \rho_0=\frac{N}{2\beta\sqrt{\kappa_B}},
\quad \tau=\frac1{100}.
\label{eq:gep-solver-parameters}
\end{align}
Each estimation attempt runs at most eight fresh QLSP executions, stopping at the first success.
Since each succeeds with probability at least $1/4$, inverse success is at least $1-(3/4)^8>2/3$.
Measure the time register and retain the output only when $t\ge n-1$.
Apply the shifted inverse Fourier transform to the phase register, clip the signed phase to $[-\pi/2,\pi/2]$, and return $R\tan(\widehat\theta/2)$.
After $\lceil50\log(1/\delta)\rceil$ independent attempts, take the median of accepted estimates, returning zero if there are none.
The query counts include every attempt.

\begin{proof}[Proof of \cref{thm:gep-eigenstate}]
If $\epsilon\ge\alpha_A/\beta$, return zero, since $|\lambda|\le\alpha_A/\beta$.
Otherwise each matrix term in the stated query bounds is bounded below by a positive constant.
For an eigenstate, \cref{eq:gep-circuit-history} gives
\begin{align}
M_N^{-1}|0,c,\psi\rangle =s^{-1}\sum_{t=0}^{N-1}\sum_{k=0}^{n-1} c_ke^{i\min\{k,t\}\theta}|t,k,\psi\rangle,
\quad \rho:=\|M_N^{-1}|0,c,\psi\rangle\| &=\frac N{\beta\sqrt{\kappa_B}}.
\label{eq:gep-source-response}
\end{align}
For arbitrary input, each partial circuit is similar to a unitary through $I\otimes B^{1/2}$ and has norm at most $\sqrt{\kappa_B}$.
The block column of the inverse corresponding to the initial time coordinate consequently has norm at most $\sqrt{N\kappa_B}/s=N/\beta$.
Every other inverse block is
\begin{align}
(I\otimes B^{-1/2})\mathcal U_{t,u} (I-iE_u\otimes H/R)^{-1}(I\otimes B^{-1/2}),
\quad\text{of norm at most }\beta^{-1},
\label{eq:gep-impulse}
\end{align}
where $\mathcal U_{t,u}$ is the product of the intervening controlled $V$ steps.
It is unitary, and the middle inverse is a contraction.
The remaining triangular blocks have row and column norm sums at most $N/\beta$, proving
\begin{align}
\|M_N^{-1}\|\le K:=2N/\beta.
\label{eq:gep-inverse-bound}
\end{align}
To obtain the separate query bounds, set
\begin{align}
\mu_0=s,
\quad \mu_A=2\alpha_A/R,
\quad \mu_B=2\alpha_B.
\label{eq:gep-term-strengths}
\end{align}
Tensor the known clock contractions with the given matrix encodings.
In the allocation \cref{wql:eq:rates}, choose auxiliary matrix weights $1/\mu_j$ and one prepared state with coefficient and weight one.
This gives simultaneous rates $t_j=1+3K\mu_j$ and $u=1+K/\rho_0$.
The individual counts \cref{wql:eq:individual-counts} are
\begin{align}
Q_A=O\!\left[K\mu_A\log\frac1\tau\right],
\quad Q_B=O\!\left[K\mu_B\log\frac1\tau\right],
\quad Q_\psi=O(K/\rho_0).
\label{eq:gep-solver-cost}
\end{align}
These auxiliary weights define one schedule; actual query prices may be arbitrary.
The known boundary encoding contributes only known operations.
With horizon $N=O(R/\epsilon)$ and constant solver tolerance, these are the claimed per-attempt matrix costs.
The state-preparation count is $O(\sqrt{\kappa_B})$.

The exact eigenstate history has completed-time probability $h=(n+1)/(2n)\ge1/2$.
Its accepted phase estimate is correct with probability at least $1-\eta=9/10$ by \cref{eq:gep-kaiser}.
After clipping the phase estimate to its allowed interval, use $R\tan(\widehat\theta/2)$ as the eigenvalue estimate.
The conversion has derivative at most $R$ on $[-\pi/2,\pi/2]$, so the phase accuracy gives the required eigenvalue accuracy.
Score each attempt by \begin{align}
X=\begin{cases} 1,&\text{correct accepted estimate},
\\
-1,&\text{incorrect accepted estimate},
\\
0,&\text{discarded attempt}. \end{cases}
\end{align} Including inverse failures and trace-distance error gives $\mathbb E X\ge(2/3)(9h/10-\tau)-(h/10+\tau) =h/2-5\tau/3>1/5$.
Hoeffding's inequality bounds failure of the median of accepted estimates after $r$ attempts by $\Pr(\sum X\le0)\le e^{-r/50}$.
The input-tolerance argument is given in \cref{app:gep-trials}.
\end{proof}

With query prices $C_A,C_B,C_\psi$, the cost in the regime of \cref{tab:gep-comparison}, at fixed confidence, is
\begin{align}
O\!\left(\frac{C_A\alpha_A+C_BR\alpha_B}{\beta\epsilon} +C_\psi\sqrt{\kappa_B}\right).
\label{eq:gep-priced-cost}
\end{align}
For arbitrary trial states, \cref{prop:gep-trials} gives an overlap-dependent list guarantee using this same construction.
Identifying a named low-weight mode requires information such as an isolating interval; a median of arbitrary trial-state readouts does not identify it.

\subsubsection{Amplification for a known spectral interval}
\label{sec:gep-amplification}
For a trial state, overlap is measured after the Hermitian change of coordinates.
Let $P_I$ be the spectral projector of $H$ onto an interval $I\subseteq[-R,R]$, and define
\begin{align}
|\chi\rangle=\frac{B^{1/2}|\psi\rangle} {\sqrt{\langle\psi|B|\psi\rangle}},
\quad p_I=\langle\chi|P_I|\chi\rangle.
\label{eq:gepamp-overlap}
\end{align}
For an isolated eigenvalue this is its squared overlap in the orthonormal eigenbasis of $H$.
It need not equal the Euclidean overlap with a normalized generalized eigenvector.

Amplitude amplification~\cite{brassard2002quantum} improves the dependence on this overlap when a reversible test identifies the desired output.
A known spectral interval supplies such a test: mark a successful readout exactly when its estimated eigenvalue lies in the interval.
An interval containing a cluster is sufficient; its internal eigenvalues need not be separated.

\begin{theorem}[Amplified generalized eigenvalue estimation]
\label{thm:gepamp-main}
Assume \cref{eq:gep-promises} and coherent access to the separate matrix encodings and the trial-state preparation $U_\psi$, whose work registers are restored to $|0\rangle$.
Let $0<p\le p_I$, $0<\delta<1/2$, and $0<\epsilon<R$ be given.
Suppose every eigenvalue of $H$ has distance at least $g>0$ from the endpoints of the known interval $I$.
Put $h=\min\{\epsilon,g/2\}$.
There is an algorithm returning a classical estimate such that
\begin{align}
\Pr\!\left(\operatorname{dist} (\widehat\lambda,\operatorname{spec}(H)\cap I)\le\epsilon\right) \ge1-\delta.
\end{align}
With the rational metric reweighting, its query bounds are
\begin{subequations}
\label{eq:gepamp-cost}
\begin{align}
Q_A&=O\!\left[\left(1+\frac{\alpha_A}{\beta h}\right) \frac1{\sqrt p}\log^2\!\frac{1}{p\delta} \log\frac1\delta\right],
\\
Q_B&=O\!\left[\left(1+\frac{R\alpha_B}{\beta h}\right) \frac1{\sqrt p}\log^2\!\frac{1}{p\delta} \log\frac1\delta\right],
\\
Q_\psi&=O\!\left(\sqrt{\frac{\kappa_B}{p}} \log\frac1\delta\right).
\end{align}
\end{subequations}
The simpler direct output selection instead gives
\begin{align}
Q_\psi=O\!\left(\frac{\kappa_B}{\sqrt p}\log\frac1\delta\right).
\label{eq:gepamp-direct-prep}
\end{align}
For this version, multiply each matrix bound in \cref{eq:gepamp-cost} by $\sqrt{\kappa_B}$ and replace $1$ inside its squared logarithm by $\kappa_B$.
All forward, inverse, controlled, and unsuccessful calls are counted.
\end{theorem}

The direct version needs initial and final idle segments in the trial-state history.
An initial idle segment as long as the phase-estimation computation retains copies of the input.
These copies increase the solution norm and reduce one attempt's preparation cost to $O(\sqrt{\kappa_B})$.
We then amplify the joint event of successful inversion, completed readout, and an estimate in $I$.
Its probability is $\Omega(p/\kappa_B)$, giving \cref{eq:gepamp-direct-prep}.
The rational metric reweighting changes this probability to $\Omega(p)$ without increasing the preparation cost per attempt, proving the stronger bound in \cref{eq:gepamp-cost}.
\Cref{app:gepamp} gives the construction and proof using the same multi-oracle QLSP construction.

Without a recognizable target event, \cref{prop:gep-trials} still gives an overlap-dependent list guarantee.
The interval promise supplies the acceptance test needed for amplitude amplification; it need not identify an individual eigenvalue within the accepted cluster.
This use of amplitude amplification follows the usual filtering and search principle \cite{brassard2002quantum,Berry2024Rapid}.
The bounds above retain the separate matrix and preparation costs of the affine GEVP construction.

\subsection{Ground-state preparation and energy estimation}
\label{mqspgsp:sec:ground-state}

Let $H$ be a finite-dimensional Hermitian matrix with a unique ground state $|g\rangle$ and energies $E_0<E_1\le\cdots$.
We are given $\Be{H/\lambda}$ with $\lambda>0$ and $\lambda\ge\|H\|$, a guiding-state oracle $B|0\rangle=|\psi\rangle$, and a known lower bound $p_0>0$ on $p=|\langle g|\psi\rangle|^2$.
The aim is to prepare a state close to $|g\rangle$ while accounting separately for calls to $B$ and to a block-encoding of $H$.
The target conditional trace-distance error is $0<\epsilon\le1/4$.
The problem and its relation to previous algorithms are summarized in \cref{mqspgsp:sec:intro}.

With a known energy threshold, earlier eigenstate-preparation methods already reduce repeated preparation of the guiding state \cite{Berry2018FermionicEigenstates}.
We retain separate costs for that preparation and for the spectral filter, as in the refined filtering and linear-system analyses \cite{Berry2024Rapid,Low2024QLSP}.

\begin{figure}[!htbp]
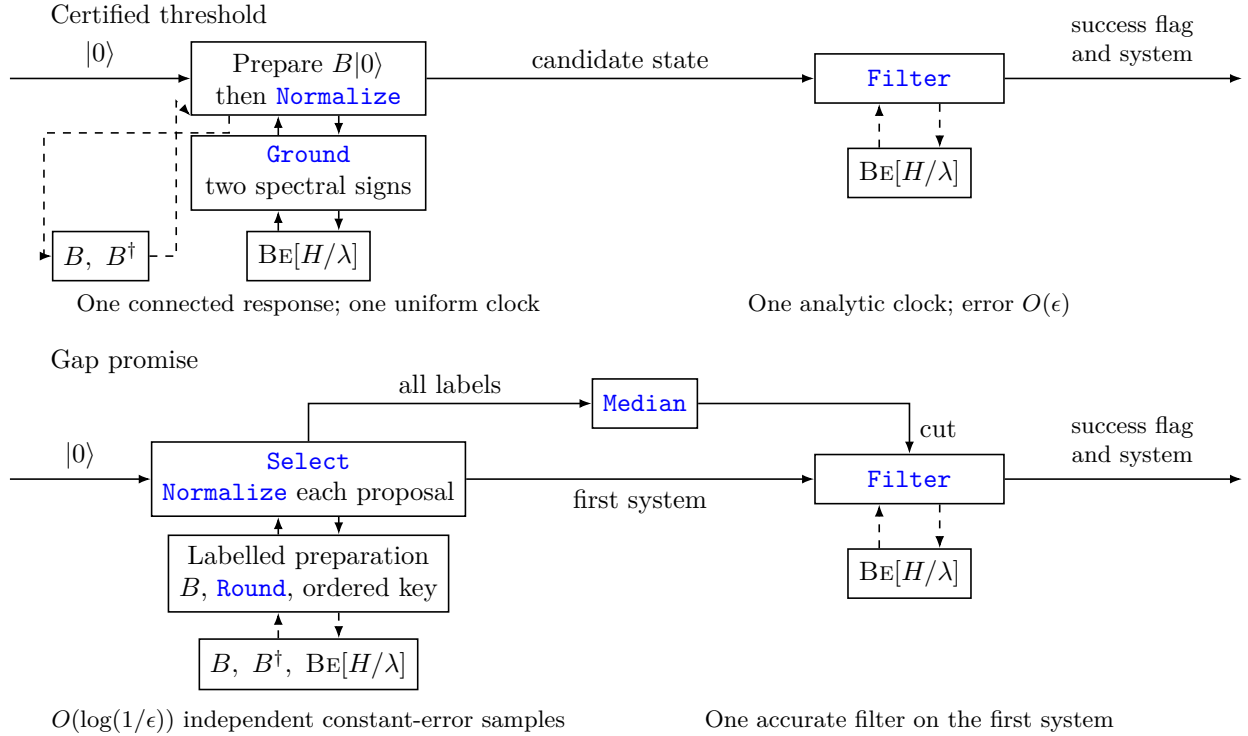

\centering
\IIRDiagram{GroundOverview}
\caption{The two preparation algorithms.
Each dashed pair connects an outer module to an inserted module and back.
Expanding the inserted modules into their unitary realizations gives the original oracle feedback connections.
The connected coarse stage is compiled once per sample.
Solid arrows between the coarse stage and Filter show the actual finite-circuit state flow.
The gap algorithm retains every sample's labels and histories; the dotted arrow controls the filter by their reversibly computed median.
The final filter uses the Hamiltonian oracle alone.}
\label{mqspgsp:fig:overview}
\end{figure}

The two networks in \cref{mqspgsp:fig:overview} give the proof strategy.
With a known threshold between the ground and excited energies, we combine the sign response of \cref{subsec:mod-sign} with the reflection about $|\psi\rangle$ to normalize its ground component.
A uniform clock compiles this coarse stage at constant error.
An accurate spectral filter then uses only the Hamiltonian oracle.

With only a gap bound, coarse energy labels and coherent selection supply candidate states.
The median of independent labels chooses a threshold for filtering the first candidate.
We derive the responses and their steady-state weights below, then apply the clock theorems of \cref{sec:iir}.

\FloatBarrier

We assume controlled access to an exact self-inverse $\Be{H/\lambda}$, where $\lambda\ge\|H\|$ and its signal state is $|0\rangle_H$, and to $B,B^\dagger$.
The counts $q_H,q_B$ include controlled and inverse calls, with the resource convention of \cref{def:resource-accounting}.
A certified threshold consists of numbers $-1\le u<v\le1$ satisfying $E_0/\lambda\le u<v\le E_1/\lambda$.
Its angular separation is $d=\arccos u-\arccos v>0$.
Alternatively, we are supplied a gap bound $E_1-E_0\ge\gamma_0>0$.
We may also be supplied the endpoint bound $E_0/\lambda\ge-1+\beta$, with $0\le\beta\le2-\gamma_0/\lambda$; take $\beta=0$ when this information is unavailable.
The guaranteed angular gap is
\begin{align}
d_*=\min_{-1+\beta\le x\le1-\gamma_0/\lambda} [\arccos x-\arccos(x+\gamma_0/\lambda)].
\label{mqspgsp:eq:gsp-angular-gap}
\end{align}
These angles are eigenphases of the qubitized Hamiltonian walk.
We call $|0\rangle_H\otimes\mathcal H_{\rm sys}$ the encoding signal subspace.
With $\beta=0$, $d_*=2\arcsin(\gamma_0/(2\lambda)) =\Theta(\gamma_0/\lambda)$.
The exact angular parameters retain the improvement available when the supplied energy bands lie near a spectral endpoint.

\begin{theorem}[Ground-state preparation with separate oracle costs]
\label{mqspgsp:thm:gsp-main}
Under the preceding promises, let $0<\epsilon\le1/4$ be the target conditional trace-distance error.
A finite unitary circuit reports success with probability $\Omega(1)$.
Conditioned on success, its system state $\rho$ satisfies $\tfrac12\|\rho-|g\rangle\langle g|\|_1\le\epsilon$.
Its worst-case query counts are
\begin{align}
\begin{array}{c|cc} \text{promise}&q_B&q_H
\\
\hline
\text{certified threshold}&O(p_0^{-1/2})& O\!\left(d^{-1}[p_0^{-1/2}+\log(1/\epsilon)]\right)
\\[3pt]
\text{gap lower bound}&O(p_0^{-1/2}\log(1/\epsilon))& O(d_*^{-1}p_0^{-1/2}\log(1/\epsilon)). \end{array}
\label{mqspgsp:eq:gsp-counts}
\end{align}
The query schedule is fixed by the supplied promises.
The actual success probability may be unknown, and auxiliary records may remain after success.
\end{theorem}

The flag and the conditional error have different roles: the theorem allows a constant probability of reporting failure, while the system state conditioned on success obeys the stated error bound.
Repeating until success has constant expected overhead.
The gate and space counts are given alongside the two constructions below; all retained histories are included in those counts.

The certified-threshold algorithm uses one coarse preparation and one accurate filter.
Only the filter depends on $\epsilon$, and it uses only the Hamiltonian oracle.
With a gap promise, a coarse sampler first finds a near-ground label with probability exceeding one half by a fixed margin.
A median of $O(\log(1/\epsilon))$ such labels supplies the threshold for filtering the first sample.
All coarse computations use a fixed constant accuracy.
The constants in the bounds are universal.

\subsubsection{Normalizing a good component}
\label{mqspgsp:sec:gsp-responses}

The same normalization step is used for ground-state overlap and for acceptable energy proposals.
We build it from the sign response $F_{\rm s}(z;A)$ in \cref{eq:mod-sign-response}, with the known unitary system and oracle matchings of \cref{eq:mod-sign-matchings}.
Its value at $z=1$ is $\operatorname{sgn}(A)$.
By \cref{eq:mod-sign-tag,thm:iir-energy}, its derivative gives the steady-state query weight.
Each sign round has two matching layers, so its block-encoding weight on an unnormalized vector $\boldsymbol{v}$ is
\begin{align}
L_{\Be{A}}(\boldsymbol{v}) \le\langle\boldsymbol{v},(I+|A|^{-1})\boldsymbol{v}\rangle.
\label{mqspgsp:eq:gsp-sign-weight}
\end{align}
We use this inequality only on the prepared invariant subspace, where $A$ is invertible.
The response may have a kernel on other subspaces.

Suppose $|\omega\rangle=V|0\rangle$, and let $\Pi$ project onto the good subspace we wish to retain.
Let $a=\langle\omega|\Pi|\omega\rangle>0$ be its probability.
For $0<a<1$, the normalized good and bad components are $|\omega_{\rm good}\rangle=\Pi|\omega\rangle/\sqrt a$ and $|\omega_{\rm bad}\rangle=(I-\Pi)|\omega\rangle/\sqrt{1-a}$.
In this ordered basis, the two projectors are
\begin{align}
\Pi=\begin{pmatrix}1&0
\\
0&0\end{pmatrix},
\quad |\omega\rangle\langle\omega| =\begin{pmatrix}a&\sqrt{a(1-a)}
\\
\sqrt{a(1-a)}&1-a\end{pmatrix}.
\end{align}
Average their reflections $R_\omega=2|\omega\rangle\langle\omega|-I$ and $R_\Pi=2\Pi-I$.
Their average $A=(R_\omega+R_\Pi)/2$ consequently obeys
\begin{align}
A=\begin{pmatrix}a&\sqrt{a(1-a)}
\\
\sqrt{a(1-a)}&-a\end{pmatrix},
\quad A^2=aI,
\quad A|\omega\rangle=\Pi|\omega\rangle.
\label{mqspgsp:eq:gsp-normalization-matrix}
\end{align}
Both projectors preserve this span, and $\operatorname{sgn}(A)=A/\sqrt a$ on it.
Thus the response we call \textsf{Normalize} satisfies
\begin{align}
F_{\rm Normalize}(z)=F_{\rm s}(z;A),
\quad F_{\rm Normalize}(1)|\omega\rangle =\frac{\Pi|\omega\rangle}{\sqrt a}.
\label{mqspgsp:eq:gsp-normalization}
\end{align}
The case $a=1$ is immediate.
The circuit uses the two reflections; the probability $a$ is needed only for the analysis.

The required $\Be{A}$ is an equal-weight LCU of those reflections.
Let $H_c$ be a Hadamard on a choice qubit.
The known system matrix, private feedback operators, and complete response are
\begin{align}
S_{\rm choice}=\left(\begin{array}{c|c}0&H_c
\\
\hline
H_c&0\end{array}\right),
\quad Z=R_\omega\oplus R_\Pi,
\quad F_{\rm choice}=H_cZH_c=\Be{A}.
\label{mqspgsp:eq:gsp-choice}
\end{align}
Here $R_\omega=V(2|0\rangle\langle0|-I)V^\dagger$.
Substituting this LCU into the sign module, as drawn in \cref{mqspgsp:fig:normalize}, specifies the complete realization by the connection rules of \cref{subsec:mod-connect}.
\begin{figure}[!htbp]
\centering
\IIRDiagram{GroundNormalize}
\caption{Normalize is Sign applied to the average of two reflections.
The choice qubit $c$ selects $R_\omega$ or $R_\Pi$ and is recombined by Hadamards, as specified by \cref{mqspgsp:eq:gsp-choice}.
The left diagram expands each matching layer of Sign into these reflections.
On the prepared span, $A^2=aI$, so its boundary action is $\Pi|\omega\rangle/\sqrt a$.
The reflection about the prepared state is implemented by $V^\dagger$, the known zero-state reflection, and $V$, in execution order.}
\label{mqspgsp:fig:normalize}
\end{figure}

The sign gap on this span is $\sqrt a$.
By \cref{mqspgsp:eq:gsp-sign-weight}, the choice weight is $O(a^{-1/2})$.
Each reflection about the prepared state uses $V$ and $V^\dagger$, so the catalyst weights for state preparation, the projector reflection, and the other private connections are all $O(a^{-1/2})$, including the initial preparation.
The same bounds hold with spectator registers.

We substitute exact responses into the connected network and add their original-oracle weights using \cref{eq:mod-catalyst-compose}.
A single application of \cref{thm:iir-uniform} then compiles the network at constant error.
The proof of its bound \cref{eq:uniform-clean-error} applies separately to each vector in the prepared span, so it covers the prepared span here even if $A$ has a kernel elsewhere.

Quditization turns the finite response into a unitary circuit using a time register and known input and output clock states (\cref{thm:iir-lift,thm:iir-clock-body}).
For horizon $N$ and integer port delays $r_j$, oracle $j$ is used $O(N/r_j)$ times.
The coarse stages use uniform clocks; the accurate filter uses the analytic clock of \cref{thm:iir-analytic-clock-selection}.
Both have a constant number of labels and $O(N)$ known preparation gates.
The gate and space bounds in \cref{eq:qudit-resource-summary,eq:qudit-main-cost} apply to the full register space, including every retained proposal record.

\FloatBarrier
\subsubsection{Preparation with a certified threshold}
\label{mqspgsp:sec:gsp-certified}

The certified-threshold construction in \cref{mqspgsp:fig:overview} normalizes the ground component at constant accuracy, then filters out the remaining excited component.
We first give the exact reflection response needed by both stages.

We first construct the reflection required by \textsf{Normalize}.
Let $\Pi_g=|g\rangle\langle g|$ and choose the angular cut $\theta_c=(\arccos u+\arccos v)/2$.
On an eigenvalue $x$ of $H/\lambda$, the desired reflection is $\operatorname{sgn}(\cos\theta_c-x)$: it is positive on the ground band and negative on the excited band.
For an angle $\phi$ with $x=\cos2\phi$, the identity
\begin{align}
\operatorname{sgn}\sin(\phi+\theta_c/2)\, \operatorname{sgn}\sin(\phi-\theta_c/2) =\operatorname{sgn}(\cos\theta_c-\cos2\phi)
\label{mqspgsp:eq:gsp-cut-identity}
\end{align}
expresses this reflection as a product of two signs.
We realize these factors with \textsf{Sign}.

To obtain the angle from the block-encoding, reflect its signal with $R_H=2|0\rangle\langle0|_H-I$ and form the qubitized walk
\cite{Low2016Qubitization}
\begin{align}
W=R_H\Be{H/\lambda}.
\label{mqspgsp:eq:gsp-walk}
\end{align}
This walk has phases $\pm\theta$ on the invariant subspace associated with an eigenvalue $x=\cos\theta$ of $H/\lambda$.
Ground energy corresponds to the largest folded walk angle $\theta=\arccos(E/\lambda)$.
Introduce a qubit $t$ to take the two square roots of the walk, and rotate their phases to the two sign cuts:
\begin{align}
U_{1/2}=\begin{pmatrix}0&W
\\
I&0\end{pmatrix}_t,
\quad V_\pm=-ie^{\pm i\theta_c/2}U_{1/2}.
\label{mqspgsp:eq:gsp-cut}
\end{align}
The qubit $t$ gives $U_{1/2}^2=I_t\otimes W$.
Thus, if $e^{i\phi}$ is an eigenvalue of $U_{1/2}$, then $\cos 2\phi=x$ on the walk invariant subspace associated with the system eigenvalue.
The half-angle construction treats the two walk branches together, and $\operatorname{Re}V_\pm$ has eigenvalue $\sin(\phi\pm\theta_c/2)$ on that phase.

For a unitary $V$, let $H_b,Z_b$ be Hadamard and Pauli-$Z$ gates on a direction qubit.
To supply its real part to \textsf{Sign}, choose between $V$ and $V^\dagger$ and interfere the two choices:
\begin{subequations}
\label{mqspgsp:eq:gsp-real-system}
\begin{align}
S_{\rm real}=\left(\begin{array}{c|c}0&Z_bH_b
\\
\hline
H_b&0\end{array}\right),
\quad Z=V\oplus V^\dagger,
\\
F_{\rm real}=Z_bH_b\operatorname{diag}(V,V^\dagger)H_b =\Be{\operatorname{Re}V}
\end{align}
\end{subequations}
is a self-inverse block-encoding of the real part.
To derive the response, take a public input $|v\rangle$ that includes the direction qubit.
The private-port equations of this system are $\boldsymbol{g}=H_b|v\rangle$, $\boldsymbol{f}=(V\oplus V^\dagger)\boldsymbol{g}$, and $\boldsymbol{y}=Z_bH_b\boldsymbol{f}$.
Substituting the first two equations into the third eliminates both private vectors and gives the stated $F_{\rm real}$.
Multiplying the two Hadamards makes its blocks explicit:
\begin{align}
F_{\rm real}=\frac12 \begin{pmatrix} V+V^\dagger&V-V^\dagger
\\
V^\dagger-V&-V-V^\dagger \end{pmatrix}.
\end{align}
This matrix is Hermitian.
Its square is the identity: the off-diagonal blocks cancel because $V$ and $V^\dagger$ commute, and each diagonal block is $[(V+V^\dagger)^2-(V-V^\dagger)^2]/4=I$.
The zero-direction block is $(V+V^\dagger)/2=\operatorname{Re}V$.
One direction-selected invocation of $V_\pm$ uses one controlled Hamiltonian query: $U_{1/2}$ is a known swap followed by a controlled $W$, and $W$ itself uses one query.

We define \textsf{Ground} by connecting the two sign systems for $\operatorname{Re}V_-$ and $\operatorname{Re}V_+$ in series, using the known system-matrix composition of \cref{subsec:mod-connect}.
Before substituting their realizations in terms of the original oracles, the response is
\begin{align}
F_{\rm Ground}(z_-,z_+) =F_{\rm s}(z_+;\operatorname{Re}V_+)\,F_{\rm s}(z_-;\operatorname{Re}V_-).
\label{mqspgsp:eq:gsp-ground-response}
\end{align}
\begin{figure}[!htbp]
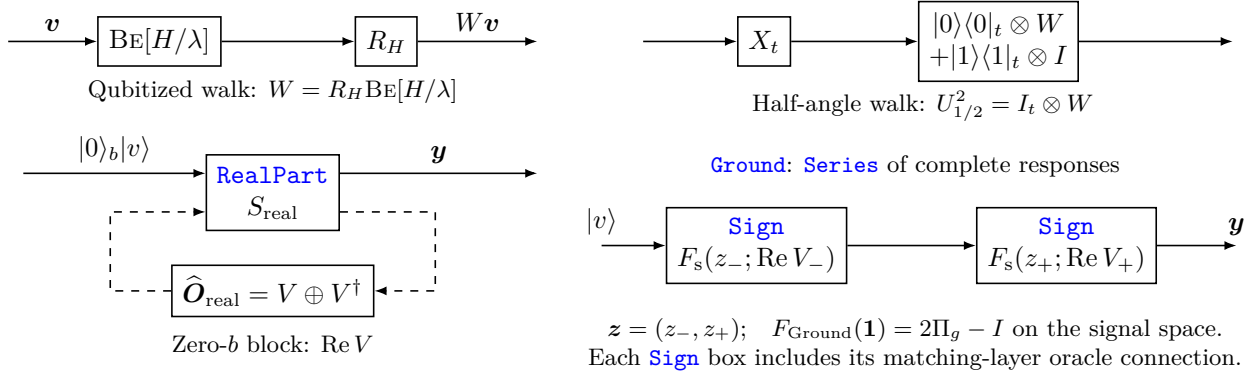

\centering
\IIRDiagram{GroundSpectral}
\caption{From the Hamiltonian oracle to the ground reflection.
The upper row shows the walk and its half-angle construction.
The lower left uses $S_{\rm real}$ to form a self-inverse block-encoding of $\operatorname{Re}V$.
Substituting $V=V_-$ and $V=V_+$ into two Sign responses and connecting them in the indicated order gives Ground.
The product identity \cref{mqspgsp:eq:gsp-cut-identity} makes its boundary action equal on both walk eigenphase branches.}
\label{mqspgsp:fig:spectral}
\end{figure}

\FloatBarrier The walk, real-part construction, and two-sign connection are shown in \cref{mqspgsp:fig:spectral}.
The two matrix arguments commute because both are functions of $U_{1/2}$.
Their product at the all-ones vector is given by \cref{mqspgsp:eq:gsp-cut-identity}.
Thus $F_{\rm Ground}(1,1)$ restricts to $2\Pi_g-I$ on the encoding signal subspace and acts identically on the two square-root branches.
The encoding signal subspace is invariant.

Each folded walk angle is at least $d/2$ from the cut.
Each argument of the sign consequently has magnitude at least $\sin(d/4)$.
The unused walk-completion phases are $0$ or $\pi$ and obey the same bound, because $d/2\le\theta_c\le\pi-d/2$.
Thus the sign gap holds on the full register space of every component invocation.
Applying \cref{mqspgsp:eq:gsp-sign-weight} to both factors gives uniform catalyst weight on the Hamiltonian-oracle and other private connections $O(d^{-1})$ per unit squared norm of the input to the module.

We can now prepare the coarse candidate.
Set $V=B$ and $\Pi=\Pi_g$ in \textsf{Normalize}, and supply its projector reflection by \textsf{Ground}.
Let $F$ be the response obtained by this substitution.
At the all-ones vector, the inner response is $F_{\rm Ground}(1,1)=2\Pi_g-I$ on the encoding signal subspace.
The matrix supplied to the outer sign response is consequently
\begin{align}
A_g=\frac{(2|\psi\rangle\langle\psi|-I)+F_{\rm Ground}(1,1)}2 =|\psi\rangle\langle\psi|+\Pi_g-I.
\end{align}
The normalization calculation in \cref{mqspgsp:eq:gsp-normalization-matrix} gives $A_g^2=pI$ on the span of $|\psi\rangle$ and its ground component, and $A_g|\psi\rangle=\Pi_g|\psi\rangle$.
Using the sign response at one therefore gives
\begin{align}
F(\mathbf1)|\psi\rangle =F_{\rm s}(1;A_g)|\psi\rangle =\frac{A_g|\psi\rangle}{\sqrt p} =\frac{\Pi_g|\psi\rangle}{\sqrt p}.
\label{mqspgsp:eq:gsp-target}
\end{align}
Up to the phase already present in the overlap, this is the ground state.
Normalization contributes $O(p_0^{-1/2})$ catalyst weight for state preparation and $O(p_0^{-1/2})$ catalyst weight on the reflection feedback connection.
Substituting the uniform reflection bound therefore gives total catalyst weights on the Hamiltonian-oracle and other private connections $O(d^{-1}p_0^{-1/2})$.

To compile this connected response, set $r_B=r_{\bar B}=\Theta(d^{-1})$ and $r_H=1$, with integer preparation-oracle delays and unit delays for known operations.
If $L_j$ is the squared steady-state amplitude at original port $j$, as in \cref{eq:qudit-lv-weight}, then $\sum_jr_jL_j=O(d^{-1}p_0^{-1/2})$.
Choose a horizon of this order with a sufficiently large constant.
\Cref{thm:iir-uniform,eq:uniform-delayed-energy,eq:uniform-clean-error}
give a candidate circuit $C_0$ at any fixed small vector error, with
\begin{align}
q_B(C_0)=O(p_0^{-1/2}),
\quad q_H(C_0)=O(d^{-1}p_0^{-1/2}).
\label{mqspgsp:eq:gsp-candidate-cost}
\end{align}
The ideal output is a ground state with the ancillary work registers in $|0\rangle$, so this joint event has probability $\Omega(1)$ in the candidate.
The longer delay between preparation queries gives the smaller preparation-query count.
MQSP controls the error of the entire connected response; each internal ground-reflection call is charged through its steady-state weight.
For a finite circuit, end each sign path beyond the sites reachable in the chosen matching layers and synthesize known gates to a sufficiently small total constant error.

It remains to suppress the candidate's excited component.
Prepare a flag with a Hadamard $H_f$, apply the exact ground-reflection response controlled on the flag, and undo the Hadamard.
Write $A_\pm=\operatorname{Re}V_\pm$.
The known closing system and its full response are
\begin{subequations}
\label{mqspgsp:eq:gsp-simple-filter-response}
\begin{align}
S_{\rm filter}=\left(\begin{array}{c|c}0&H_f
\\
\hline
H_f&0\end{array}\right),
\quad \oracle(z)=I\oplus F_{\rm s}(z;A_+)F_{\rm s}(z;A_-),
\\
F_{\rm filter}(z)=H_f\oracle(z)H_f.
\end{align}
\end{subequations}
At $z=1$ the full response is unitary, while its zero-flag block on the encoding signal subspace is $(I+2\Pi_g-I)/2=\Pi_g$.
We compile this connected response once at operator error $\eta$, using a constant clock normalization with corresponding known signal amplitude $c>0$.
\begin{figure}[!htbp]
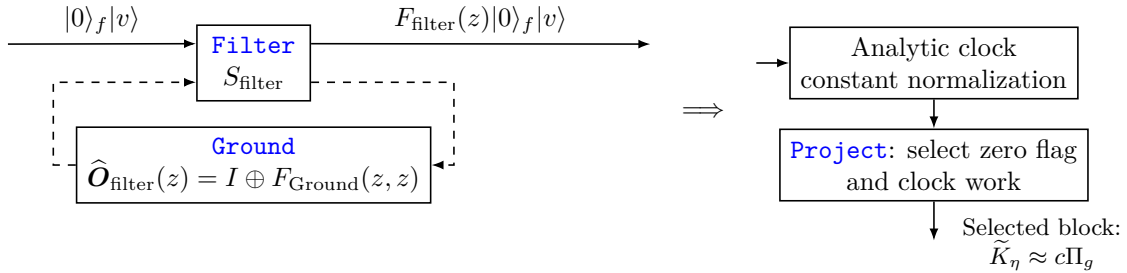

\centering
\IIRDiagram{GroundFilter}
\caption{The accurate filter.
The flag Hadamards in $S_{\rm filter}$ combine the identity and Ground feedback branches, giving selected response $(I+F_{\rm Ground})/2$.
Its boundary value is $\Pi_g$ on the encoding signal subspace.
A single analytic clock compiles the whole left network; selecting its work and the filter flag gives the contraction in \cref{mqspgsp:eq:gsp-simple-filter}.
Both private sign paths use only $\Be{H/\lambda}$.}
\label{mqspgsp:fig:filter}
\end{figure}

\FloatBarrier

Let $h>0$ be a common lower bound on $|A_\pm|$.
The two sign factors commute, so their derivative weights add and their exterior growth bounds multiply.
On a disc $|z-1|\le\zeta$ of radius $\zeta=\Theta(h)$ with a sufficiently small coefficient, \cref{eq:mod-sign-tag} gives
\begin{align}
0\preceq F_{\rm filter}(1)^\dagger F_{\rm filter}'(1) \preceq O(h^{-1})I,
\quad \|F_{\rm filter}(z)\|\le |z|^{O(h^{-1})}
\quad (|z|\ge1,\ |z-1|\le\zeta).
\label{mqspgsp:eq:gsp-simple-filter-growth}
\end{align}
The response is analytic there and contractive in the unit disc.
In \cref{thm:iir-analytic-clock-selection}, take $a=h_-=0$, $w,h_+=O(h^{-1})$, $M_{\rm loc}=1$, and $\zeta=\Theta(h)$.
Fixing the theorem's normalization slack at a constant and choosing clock error of order $\eta$ give horizon $O(h^{-1}\log(1/\eta))$.
The amplitude $c$ is independent of the problem parameters.
Selecting zero filter flag and clock work gives a contraction $\widetilde K_\eta$ with
\begin{align}
\|\widetilde K_\eta-c\Pi_g\|\le\eta,
\quad q_B(\widetilde K_\eta)=0,
\quad q_H(\widetilde K_\eta)=O(h^{-1}\log(1/\eta)).
\label{mqspgsp:eq:gsp-simple-filter}
\end{align}
The selected work includes the fresh walk signal and square-root qubit.
The clock and gate errors are each chosen of order $\eta$.
The theorem supplies the clock states and the finite filter directly from the analytic bounds.
Take $h=\sin(d/4)=\Theta(d)$ and accept when the candidate work and selected filter registers are zero.
Let $\boldsymbol{v}$ be the candidate vector after selecting zero work.
It has norm at most one and ground norm $\|\Pi_g\boldsymbol{v}\|=\Omega(1)$.
The operator approximation in \cref{mqspgsp:eq:gsp-simple-filter} bounds the success probability $s$ and conditional system state $\rho$ by
\begin{align}
\sqrt s=\|\widetilde K_\eta\boldsymbol{v}\| \ge c\|\Pi_g\boldsymbol{v}\|-\eta=\Omega(1),
\quad \frac12\|\rho-|g\rangle\langle g|\|_1 \le\frac{\eta}{\sqrt s}=O(\eta),
\label{mqspgsp:eq:gsp-simple-certified-success}
\end{align}
The second inequality uses $\|(I-\Pi_g)\widetilde K_\eta\boldsymbol{v}\|\le\eta$; it also holds with retained records by purification and partial trace.
Choose $\eta=\Theta(\epsilon)$ with a sufficiently small constant.
Adding the coarse and accurate stages proves
\begin{align}
q_B=O(p_0^{-1/2}),
\quad q_H=O\!\left(d^{-1}[p_0^{-1/2}+\log(1/\epsilon)]\right),
\label{mqspgsp:eq:gsp-simple-certified-cost}
\end{align}
which is the certified-threshold claim of \cref{mqspgsp:thm:gsp-main}.

If the guiding state were filtered directly, its ground norm would be only $\sqrt p$, and suppressing the excited part relative to it would require a filter error proportional to $\epsilon\sqrt p$.
Here the candidate already has constant ground norm.
An $O(\epsilon)$ filter therefore suffices, independently of $p$, giving the additive Hamiltonian precision term in the theorem.

For gate and qubit counts, the candidate uses one connected lift with horizon $O(d^{-1}p_0^{-1/2})$.
The entire final filter uses one analytic lift with horizon $O(d^{-1}\log(1/\epsilon))$.
Apply \cref{eq:qudit-dimensions,eq:qudit-main-cost} to their finite register spaces.
If the complete candidate and filter have gate counts $G_{\rm cand},G_{\rm filt}$ and widths $w_{\rm cand},w_{\rm filt}$, respectively, the sequential construction uses at most
\begin{subequations}
\label{mqspgsp:eq:gsp-certified-gates-space}
\begin{align}
G_{\rm total}&\le G_{\rm cand}+G_{\rm filt} +O\!\left((w_{\rm cand}+w_{\rm filt}) \log(2+w_{\rm cand}+w_{\rm filt})\right),
\\
n_{\rm total}&\le w_{\rm cand}+w_{\rm filt}+O(1).
\end{align}
\end{subequations}
The extra gate term computes the success flag from the tests that the work registers are in $|0\rangle$ with constant extra workspace.
The space bound retains the candidate's work during filtering; the common system register is counted twice only to simplify the upper bound.

\FloatBarrier
\subsubsection{Preparation with a gap promise}
\label{mqspgsp:sec:gsp-simple-rounding}

Without a certified threshold, we first attach coarse energy labels and favor the smallest ones.
An incorrectly small label could itself be amplified by this selection, so we design an exact rounding response supported only on neighboring phase bins.
We connect it to selection before compiling a finite approximation.
This lets the MQSP uniform-clock theorem control the complete sampler at once.

Choose a power of two $M=\Theta(d_*^{-1})$ large enough that $2\pi/M\le d_*/8$, and use the walk $W$ in \cref{mqspgsp:eq:gsp-walk}.
For an eigenphase $\vartheta$, write $M\vartheta/(2\pi)=k+t$, where $k$ is an integer and $0\le t<1$.
On the cyclic bin register, let $T|y\rangle=|y+1\bmod M\rangle$.
A cosine-squared profile joins neighboring bins smoothly.
Its Fourier coefficients decay as $|\ell|^{-3}$, giving a finite weighted sum $\sum_\ell(1+|\ell|)|b_{\varphi,\ell}|$ and bounded mean walk cost.
We use the rounding contraction
\begin{align}
R(\vartheta)=T^k\left[\cos^2(\pi t/2)I+\sin^2(\pi t/2)T\right].
\label{mqspgsp:eq:gsp-simple-round}
\end{align}
Its column at $|0\rangle$ has two nonnegative amplitudes, supported on bins $k,k+1$, and squared norm between $1/2$ and $1$.
We implement $R/C$ as a selected block of a unitary, for a fixed sufficiently large normalization $C$, and retain all its outputs.
A valid label means that this unitary's preparation work is zero.
Validity will be part of the ordered label supplied coherently to selection.

The Fourier formula for this contraction is elementary.
In a bin-momentum block $r\in\{-M/2,\ldots,M/2-1\}$, put $\varphi=2\pi r/M$, so $T$ has eigenvalue $e^{i\varphi}$.
Define
\begin{align}
f_\varphi(t)=e^{-i\varphi t} [\cos^2(\pi t/2)+e^{i\varphi}\sin^2(\pi t/2)].
\label{mqspgsp:eq:gsp-simple-round-fourier}
\end{align}
Then the eigenvalue of $R(\vartheta)$ is $e^{ir\vartheta}f_\varphi(t)$.
The function $f_\varphi$ and its first derivative match at zero and one.
Its Fourier coefficients are
\begin{align}
b_{\varphi,\ell} =\int_0^1f_\varphi(t)e^{-2\pi i\ell t}\,dt =\frac{-\pi^2\sin\varphi} {(\varphi+2\pi\ell)[(\varphi+2\pi\ell)^2-\pi^2]},
\quad \ell\in\mathbb Z,
\label{mqspgsp:eq:gsp-simple-round-coefficients}
\end{align}
with removable values defined by the integral.
This follows by writing $\cos^2(\pi t/2)=(1+\cos\pi t)/2$ and integrating the three exponentials.
Since $|f_\varphi(t)|\le1$, all coefficients have magnitude at most one.
For $|\ell|\ge2$, the displayed denominator also gives $|b_{\varphi,\ell}|\le|\ell|^{-3}$.

\begin{figure}[!htbp]
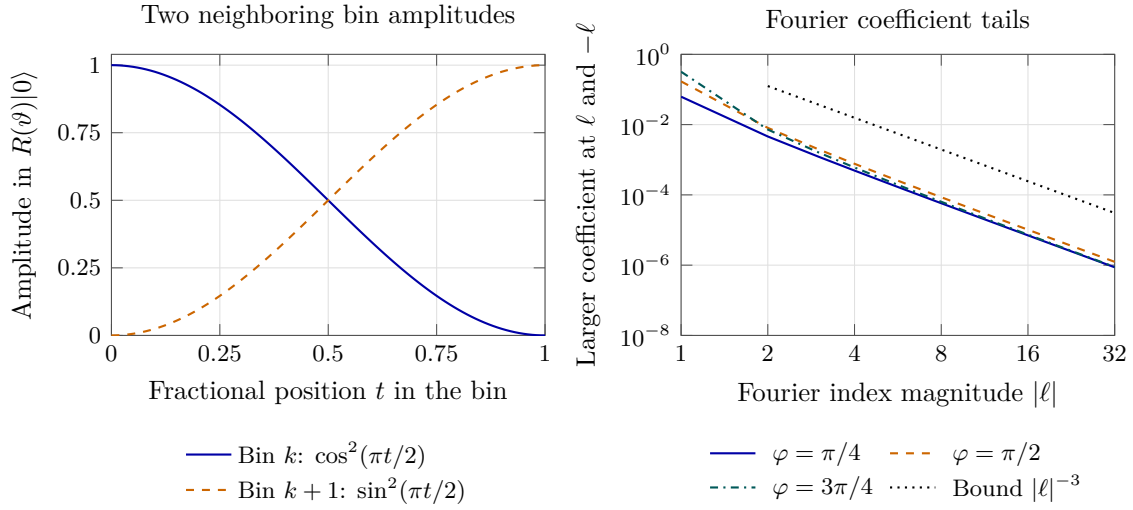

\centering
\IIRDiagram{GroundRoundingPlots}
\caption{The rounding response in \cref{mqspgsp:eq:gsp-simple-round} and the coefficients in \cref{mqspgsp:eq:gsp-simple-round-coefficients}.
Left: the only nonzero amplitudes of $R(\vartheta)|0\rangle$.
The implemented selected block is $R/C$, so its bin probabilities are $\cos^4(\pi t/2)/C^2$ and $\sin^4(\pi t/2)/C^2$, with total probability $\Theta(1)$.
Right: for each displayed momentum, $\max\{|b_{\varphi,\ell}|,|b_{\varphi,-\ell}|\}$ at positive integer $\ell$, with curves connecting the values.
Both tails obey the $|\ell|^{-3}$ bound for $|\ell|\ge2$.}
\label{mqspgsp:fig:rounding-plots}
\end{figure}

\Cref{mqspgsp:fig:rounding-plots} shows both parts of this construction.
The two neighboring amplitudes vary continuously as the phase crosses a bin.
Their Fourier coefficients decay cubically, which makes both the total branch weight and its first moment finite.

The cubic decay gives $\sum_\ell(1+|\ell|)|b_{\varphi,\ell}|=O(1)$ uniformly in $\varphi$.
Choose the universal constant $C$ larger than this bound, so that the valid-label probability is bounded away from one.
The coefficients are real.
Introduce two orthogonal padding labels $|\mathrm{pad}_+\rangle,|\mathrm{pad}_-\rangle$ and prepare the known normalized branch state
\begin{align}
|\chi_\varphi\rangle =\sum_\ell\sqrt{\frac{|b_{\varphi,\ell}|}{C}}\,|\ell\rangle +\sqrt{\frac{C-\sum_\ell|b_{\varphi,\ell}|}{2C}}\, (|\mathrm{pad}_+\rangle+|\mathrm{pad}_-\rangle).
\label{mqspgsp:eq:gsp-simple-round-preparation}
\end{align}
Let $U_\chi|0\rangle=|\chi_\varphi\rangle$, controlled by the known momentum label $r$.
In that block, the system and response are
\begin{subequations}
\label{mqspgsp:eq:gsp-simple-round-system}
\begin{align}
S_{\rm branch}=\left(\begin{array}{c|c}0&U_\chi^\dagger
\\
\hline
U_\chi&0\end{array}\right),
\quad \oracle_{\rm branch}=\left[\bigoplus_\ell \operatorname{sgn}(b_{\varphi,\ell})W^{M\ell+r}\right] \oplus I\oplus(-I),
\\
F_{\rm branch}=U_\chi^\dagger \oracle_{\rm branch}U_\chi.
\end{align}
\end{subequations}
The sign at a zero coefficient can be chosen as $+1$.
The padding terms cancel in the selected block.
Selecting the branch preparation work to be zero gives $R(\vartheta)/C$, because $e^{iM\ell\vartheta}=e^{2\pi i\ell t}$.
Known Fourier basis changes on the bin register complete the construction.
Each signed walk word has length at most $M(|\ell|+1/2)$, so the first-moment bound gives $O(M)$ catalyst weight on the Hamiltonian-oracle and other private connections on every input with work registers in $|0\rangle$, for either the forward or inverse unitary.
MQSP uses this first moment to bound the steady-state query weight.
The infinite response is used only to derive this bound; we give its finite realization after analyzing the connected sampler.

\begin{figure}[!htbp]
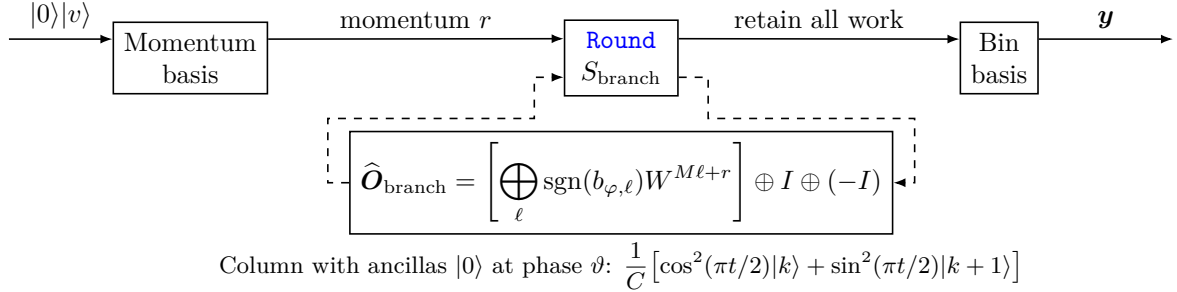

\centering
\IIRDiagram{GroundRounding}
\caption{The positive two-bin rounder.
In each known momentum block, $S_{\rm branch}$ prepares $|\chi_\varphi\rangle$, routes through a signed walk word, and undoes that preparation.
The padding terms cancel in the block with work registers initialized and projected onto $|0\rangle$.
Returning to the bin basis gives $R(\vartheta)/C$ and its two nonnegative amplitudes.
All completion outputs are retained and receive the failure key.
The weighted word length is controlled by $\sum_\ell(1+|\ell|)|b_{\varphi,\ell}|$, giving the $O(M)$ weight.}
\label{mqspgsp:fig:rounding}
\end{figure}

The Fourier branches and their selected two-bin output are shown in \cref{mqspgsp:fig:rounding}.

For a successful label, use the folded angle $\widehat\theta_y=\arccos\cos(2\pi y/M)$ and order keys by increasing $-\widehat\theta_y$.
Assign every failed label the same additional largest key.
Folding is $1$-Lipschitz in circular distance, so a valid label differs from its folded walk angle by at most $2\pi/M\le d_*/8$.
Every valid ground key therefore precedes every valid excited key: their angular separation is at least $d_*-4\pi/M>0$.
The selected column has squared norm $\Theta(1)$.
Thus the valid ground prefix has probability $\Omega(p)$, and all valid labels together have probability at most $1-\Omega(1)$.

Rounding also entangles the two walk branches with their labels.
We must retain ground-state weight after projecting the original signal register onto $|0\rangle_H$.
Away from energies $\pm\lambda$, a system eigenstate with encoding ancilla $|0\rangle_H$ is an equal superposition of the two walk eigenvectors.
At a fixed valid key, denote their unnormalized label vectors by $\boldsymbol{a}_+,\boldsymbol{a}_-$.
Both have real nonnegative entries by \cref{mqspgsp:eq:gsp-simple-round}.
The key's probability and its joint probability with the encoding ancilla in $|0\rangle_H$ are therefore
\begin{align}
\Pr(x)=\frac{\|\boldsymbol{a}_+\|^2+\|\boldsymbol{a}_-\|^2}{2},
\quad \Pr(x,0_H)=\frac{\|\boldsymbol{a}_++\boldsymbol{a}_-\|^2}{4} \ge\frac12\Pr(x).
\label{mqspgsp:eq:gsp-signal-recovery}
\end{align}
At endpoint energies the walk block already lies in the encoding signal subspace.
Selection will reweight whole keys and retain their raw labels, so this bound survives selection and its orthogonal histories.
Conditional on any selected valid ground key, at least half the weight is on the ground state with encoding ancilla $|0\rangle_H$ state.

\label{mqspgsp:sec:gsp-selection}

We now bias these labels toward the ground energy.
Let $V$ denote the state-preparation unitary $B$ followed by the complete rounding unitary and the known computation of its ordered key, including the failure key.
Let $m$ be the number of distinct keys, and let $\pi_x$ be the probability of key $x$.
Let $|\varphi_x\rangle$ be the normalized conditional state of the system, raw-bin, and rounding-work registers at key $x$, choosing it arbitrarily when $\pi_x=0$.
On zero work, write
\begin{align}
V|0\rangle=\sum_{x=1}^m\sqrt{\pi_x}\,|x\rangle|\varphi_x\rangle,
\quad P_x=\sum_{y\le x}\pi_y,
\quad P_0=0.
\label{mqspgsp:eq:gsp-labelled-source}
\end{align}
Here $x$ indexes the valid keys in increasing energy order with the failure key last, and $P_x$ is the cumulative probability.
The system register in $|\varphi_x\rangle$ includes the walk's signal register.
The probabilities enter the analysis.
The circuit uses the known comparison of two keys to implement the acceptance projector.

The \textsf{Select} response begins with one sample of $V$.
Given current key $x$, it proposes a fresh sample or a stop symbol.
Choose a known stop weight $\eta_0=\Theta(p_0)$ with a sufficiently small fixed coefficient, and let $U_\Omega$ prepare
\begin{align}
|\Omega\rangle= \frac{\sqrt{\eta_0}|\mathrm{stop}\rangle+V|0\rangle} {\sqrt{1+\eta_0}}.
\label{mqspgsp:eq:gsp-proposal}
\end{align}
The stop state is orthogonal to all preparation labels.
A known rotation followed by a controlled $V$ implements $U_\Omega$.
For current key $x$, the known projector $\Pi_x$ accepts the stop symbol or a proposal key $y<x$.
We apply \textsf{Normalize} with preparation unitary $U_\Omega$ and projector $\Pi_x$.
Denote this response by $\mathcal N_x$.
It prepares the normalized acceptable proposal coherently, without estimating its probability.

A known reversible permutation $R_{\rm route}$ then acts on the current state, the proposal, and a history register.
On stop it transmits the current system and retained label registers to the public output.
Otherwise it archives these registers and keeps the strictly smaller proposal as the next current state.
The archived records make different histories orthogonal.
Registers are enlarged as necessary to complete this routing to a permutation.

The known system for one proposal and routing step is
\begin{align}
S_{\rm step}=\left(\begin{array}{c|cc} 0&0&R_{\rm route}
\\
\hline
I&0&0
\\
0&I&0 \end{array}\right),
\quad Z=U_\Omega\oplus\mathcal N_x,
\quad F_{\rm step}=R_{\rm route}\mathcal N_xU_\Omega.
\label{mqspgsp:eq:gsp-selector-system}
\end{align}
The current key controls the known projector inside $\mathcal N_x$.
Connecting the continuation output to a new proposal step and leaving the stop output public defines the selector.
This connection is a known permutation of ports; substituting the choice and sign systems for each $\mathcal N_x$ specifies its unitary system matrix entirely in terms of known routing and the original oracle feedback connections.

To express its response, let $K_{\rm next}$ be the continuation block of the routed step on the history space, and $D_{\rm out}$ its public-output block.
The initial history is $|\emptyset\rangle$.
With an analytic variable $z$ for each continuation, the public response is
\begin{align}
F_{\rm Select}(z)=D_{\rm out}(I-zK_{\rm next})^{-1} (|\emptyset\rangle\otimes V).
\label{mqspgsp:eq:gsp-selection-response}
\end{align}
Expanding the inverse sums the amplitudes for stopping immediately, after one decrease, after two decreases, and so on.
The blocks $D_{\rm out}$ and $K_{\rm next}$ are obtained from the complete step response in \cref{mqspgsp:eq:gsp-selector-system}.
On the subspace reachable from the prepared input this series has at most $m$ terms, because the key strictly decreases before every continuation.
Fresh proposal registers are retained together with all previous records; inverse preparation queries inside normalization act on the same registers as their corresponding forward calls.

\begin{figure}[!htbp]
\centering
\IIRDiagram{GroundSelector}
\caption{One selector step and its closed response.
The left IIR has the system matrix in \cref{mqspgsp:eq:gsp-selector-system}; its two feedback operations prepare and normalize a proposal.
The right diagram draws the linear equation for the history amplitudes: the initial prepared vector plus the continuation $zK_{\rm next}$, followed by the stop block $D_{\rm out}$.
The plus sign denotes this sum in the response equation.
The unitary implementation is the known routing and child systems on the left.
Eliminating the continuation gives \cref{mqspgsp:eq:gsp-selection-response}.}
\label{mqspgsp:fig:selector}
\end{figure}

The two views in \cref{mqspgsp:fig:selector} show one proposal step and the continuation that turns it into Select.
The retained histories make the branches in that continuation orthogonal.

We can compute the output distribution directly.
At key $x$, the acceptable proposal probability is $a_x=(\eta_0+P_{x-1})/(1+\eta_0)$.
Normalization gives
\begin{align}
\Pr(\mathrm{stop}\mid x)=\frac{\eta_0}{\eta_0+P_{x-1}},
\quad \Pr(y\mid x)=\frac{\pi_y}{\eta_0+P_{x-1}}
\quad(y<x).
\label{mqspgsp:eq:gsp-transitions}
\end{align}
Although the response is coherent, its orthogonal histories allow these probabilities to be added just as in a decreasing random process.
Let $r_x$ be the probability of visiting $x$ and $w_x$ the probability of stopping there.
Each key can be visited at most once, so
\begin{align}
r_x=\pi_x+\sum_{z>x}r_z\frac{\pi_x}{\eta_0+P_{z-1}}.
\label{mqspgsp:eq:gsp-visitation-recursion}
\end{align}
The solution and its cumulative output law are
\begin{align}
r_x=\frac{(1+\eta_0)\pi_x}{\eta_0+P_x},
\quad w_x=\frac{r_x\eta_0}{\eta_0+P_{x-1}},
\quad \sum_{y\le x}w_y=\frac{(1+\eta_0)P_x}{\eta_0+P_x}.
\label{mqspgsp:eq:selector-law}
\end{align}
For the first identity, substitute into the sum in \cref{mqspgsp:eq:gsp-visitation-recursion} and use
\begin{align}
\frac{\pi_z}{(\eta_0+P_{z-1})(\eta_0+P_z)} =\frac1{\eta_0+P_{z-1}}-\frac1{\eta_0+P_z}.
\label{mqspgsp:eq:gsp-probability-telescope}
\end{align}
The sum telescopes and $P_m=1$.
The same identity telescopes the stopping probabilities to give the last formula in \cref{mqspgsp:eq:selector-law}.
In particular the total output probability is one.
The valid ground labels form a prefix of initial weight $\Omega(p_0)$.
Choosing the coefficient in $\eta_0=\Theta(p_0)$ small enough makes its output probability exceed one half by a fixed margin.
This uses only the supplied lower bound on $p$.

The same output law bounds the steady-state query weight.
At a visit to $x$, \textsf{Normalize} uses catalyst weight for state preparation $O(a_x^{-1/2})$.
Orthogonal histories add their weights.
Since $\eta_0=\Theta(p_0)$,
\begin{align}
\sum_x\frac{r_x}{\sqrt{a_x}} =O\!\left(\sum_x \frac{\pi_x}{(\eta_0+P_x)\sqrt{\eta_0+P_{x-1}}}\right) =O(p_0^{-1/2}).
\label{mqspgsp:eq:occupation}
\end{align}
To prove the last bound, for $0<A\le B$ use $(B-A)/(B\sqrt A)\le2(A^{-1/2}-B^{-1/2})$.
After canceling $\sqrt B-\sqrt A$, this is just $\sqrt A+\sqrt B\le2\sqrt B$.
Substituting $A=\eta_0+P_{x-1}$ and $B=\eta_0+P_x$ makes the sum telescope to $O(\eta_0^{-1/2})$.
Since $a_x\le1$, the same bound covers the routing and preparation work per visit.

To substitute the original oracles for each proposal call, we must also bound the rounder's cost on the vectors reached during normalization.
Its $O(M)$ bound was proved for zero rounding work; failed proposals may have nonzero work.
At current key $x$, let $|\gamma_x\rangle=\Pi_x|\Omega\rangle/\sqrt{a_x}$ be the normalized acceptable proposal.
It contains only stop or valid smaller keys, so its rounding work is zero.
The valid-label probability is bounded away from one, and therefore
\begin{align}
a_x=\frac{\eta_0+P_{x-1}}{1+\eta_0}\le1-\Omega(1).
\label{mqspgsp:eq:gsp-simple-proposal-bound}
\end{align}
The inverse proposal circuit has catalyst weights on the Hamiltonian-oracle and other private connections $O(M)$ on $|\gamma_x\rangle$ by the bound for work registers initialized to $|0\rangle$, and on $|\Omega\rangle$ by reversing its preparation.
Their overlap is $\sqrt{a_x}$, so their Gram matrix has smallest eigenvalue $\Omega(1)$.
Thus a normalized vector $c|\Omega\rangle+d|\gamma_x\rangle$ has $|c|^2+|d|^2=O(1)$.
Each steady-state weight is the squared norm of a linear response; its value on this combination is at most twice the sum of the two individual weights multiplied by $|c|^2,|d|^2$.
It is consequently $O(M)$ on the whole span used by \textsf{Normalize}.
The reflection about the prepared state preserves this span, and reversing the circuit gives the same bound for its forward invocations.
The bound is uniform in the key; orthogonal retained histories contribute additively.

Including the initial labelled preparation and using \cref{eq:mod-catalyst-compose}, we obtain original-oracle weights $O(p_0^{-1/2})$ for $B,B^\dagger$ and $O(Mp_0^{-1/2})$ for the Hamiltonian-oracle and other private feedback connections.
Compile the connected sampler with integer delays $r_B=r_{\bar B}=M$ and $r_H=1$, and unit delays for known operations.
A horizon $N=O(Mp_0^{-1/2})$ gives any fixed small vector error by \cref{eq:uniform-clean-error}.

The exact output has a valid ground label with probability $1/2+\Omega(1)$.
Its probability of having a valid ground label and encoding ancilla $|0\rangle_H$ is $\Omega(1)$ by \cref{mqspgsp:eq:gsp-signal-recovery}.
A sufficiently small constant compilation error preserves both bounds, including the ancillary work registers being in $|0\rangle$ in the latter event.
Thus one finite sample satisfies
\begin{align}
q_B=O(p_0^{-1/2}),
\quad q_H=O(Mp_0^{-1/2})=O(d_*^{-1}p_0^{-1/2}).
\label{mqspgsp:eq:gsp-sampler-cost}
\end{align}
MQSP converts the sum of steady-state catalyst weights into these fixed worst-case counts.
The preparation-oracle delay $M$ accounts for the difference between the two oracle costs.
The maximum history length determines storage, while the sum of steady-state catalyst weights determines queries.

\label{mqspgsp:sec:gsp-finite}
For completeness, the steady-state response has a finite implementation.
There are $m$ coarse keys, so at most $m$ visits and their full proposal records need be stored.
Choose the horizon $N$ from the preceding weight bound, then truncate the Fourier branches to $|\ell|\le J$.
Their omitted probability is $O(J^{-2})$ by \cref{mqspgsp:eq:gsp-simple-round-coefficients}.
Moving this mass into the two padding labels changes the normalized branch state by $O(J^{-1})$.

To control both the preparation and its inverse, implement $U_\chi$ as the reflection
\begin{align}
U_\chi=I-2|d_\varphi\rangle\langle d_\varphi|,
\quad |d_\varphi\rangle= \frac{|0\rangle-|\chi_\varphi\rangle} {\sqrt{2(1-\langle0|\chi_\varphi\rangle)}}.
\end{align}
Choose $|0\rangle$ to be the $\ell=0$ branch label.
Its overlap with $|\chi_\varphi\rangle$ is real and at most $C^{-1/2}$, so the denominator is bounded away from zero.
Substitution gives $U_\chi|0\rangle=|\chi_\varphi\rangle$.
Truncation changes this reflection by $O(J^{-1})$ in operator norm: normalized difference states change by that amount, and the difference of their rank-one projectors is bounded by twice their vector distance.
The fixed finite clock invokes these known reflections finitely often, so choosing $J$ sufficiently large makes their total hybrid error a small constant.
The oracle schedule was fixed before this choice.

\begin{figure}[!htbp]
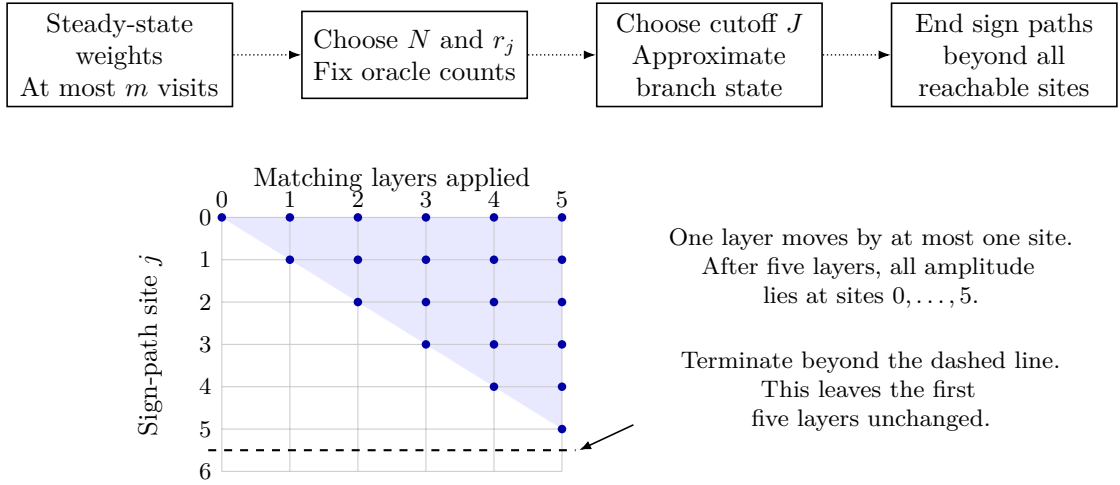

\centering
\IIRDiagram{GroundFinite}
\caption{Finite implementation after the sampler weights have been bounded.
The clock horizon and delays determine the oracle counts.
The Fourier cutoff then bounds the accumulated change in the known branch-preparation reflections.
Finally, sign paths are truncated beyond the sites reachable in the executed matching layers.
The five-layer example shades a containing region for every possible path, including later injections at site zero.
Its points indicate allowed sites, independently of their actual amplitudes.
Finite history storage, Fourier truncation, and path truncation provide finite register-space dimensions for the resource bound in \cref{eq:qudit-main-cost}.}
\label{mqspgsp:fig:finite}
\end{figure}

Finally, end every sign path beyond the sites reachable in the chosen matching layers.
Each layer moves by at most one site, so this change preserves all executed amplitudes.
Synthesizing the remaining known gates to a small total constant error completes the finite sampler.
\Cref{mqspgsp:fig:finite} illustrates the order of these choices.
Its finite history, Fourier, and sign-path dimensions enter the gate and space bounds below.

\FloatBarrier

\label{mqspgsp:sec:gsp-gap-completion}
We finish the gap-promise algorithm using independent copies of the coarse sampler.
Let $\theta_0=\arccos(E_0/\lambda)$ be the ground angle and $\theta_1=\arccos(E_1/\lambda)$ the first excited angle.
Each sample's folded label lies within $d_*/8$ of $\theta_0$ with probability $1/2+\Omega(1)$.
Assign angle zero to the failure key.
Take an odd number $n=\Theta(\log(1/\epsilon))$ of samples, with a sufficiently large coefficient, and keep the first sample's system for the final filter.
All sample records remain in separate registers.

The \textsf{Median} operation is a known reversible computation.
On folded label registers $y_1,\ldots,y_n$ and a zero output register it acts as
\begin{align}
U_{\rm med}|y_1,\ldots,y_n\rangle|0\rangle =|y_1,\ldots,y_n\rangle |\operatorname{median}(y_1,\ldots,y_n)\rangle.
\label{mqspgsp:eq:gsp-median-system}
\end{align}
On an arbitrary output register, we XOR the median label into that register; this extends the displayed action to a permutation.
Reversible arithmetic then computes the threshold.
All registers in this module are public, with $S_{\rm Median}=F_{\rm Median}=U_{\rm med}$ and zero oracle cost.
Keeping the labels makes the operation unitary and allows the final analysis to condition on classical records without inserting intermediate measurements.

Let $\widehat\theta$ be the median folded angle.
If it is more than $d_*/8$ from $\theta_0$, at least half of the samples have bad labels.
Independence and the constant bias above one half give, by Hoeffding's inequality,
\begin{align}
\tau:=\Pr(|\widehat\theta-\theta_0|>d_*/8) \le e^{-\Omega(n)}=O(\epsilon).
\label{mqspgsp:eq:gsp-median}
\end{align}
Let clipping mean restricting a real number to the indicated interval.
The filter is controlled by the angular cut and bands
\begin{subequations}
\label{mqspgsp:eq:gsp-median-cut}
\begin{align}
\theta_c&=\operatorname{clip}_{[d_*/4,\,\pi-d_*/4]} (\widehat\theta-d_*/2),
\\
u&=\cos(\theta_c+d_*/4),
\quad v=\cos(\theta_c-d_*/4).
\end{align}
\end{subequations}
\begin{figure}[!htbp]
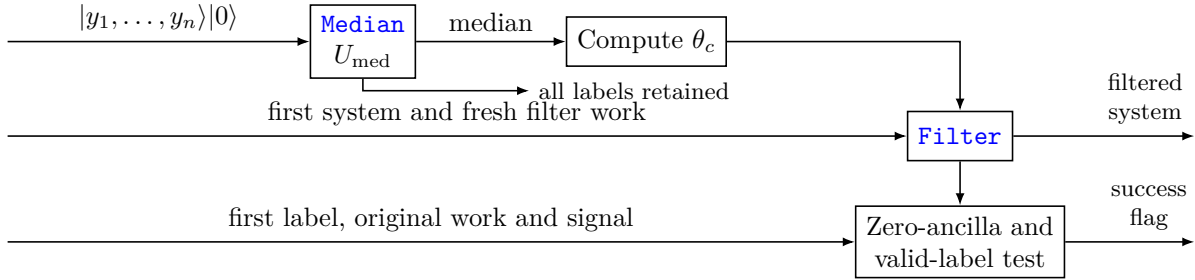

\centering
\IIRDiagram{GroundMedian}
\caption{Median control and the final success test.
Median is a known unitary on public registers, with $S=F=U_{\rm med}$ and no private feedback.
The cut computation in \cref{mqspgsp:eq:gsp-median-cut} controls the Hamiltonian-only filter.
The first system, its original records, and fresh filter work remain distinct.
The final known reversible test checks the registers stated in the proof; its value is the observable success flag.}
\label{mqspgsp:fig:median}
\end{figure}

\FloatBarrier The public computations and the final success test are shown in \cref{mqspgsp:fig:median}.
The bands are therefore defined for every record.
On a good record, $\theta_0-\theta_1\ge d_*$ implies
\begin{align}
\theta_1+3d_*/8\le\theta_c\le\theta_0-3d_*/8.
\label{mqspgsp:eq:gsp-good-cut}
\end{align}
These inequalities also place the un-clipped cut inside the clipping interval, so clipping is inactive.
The bands therefore satisfy $E_0/\lambda\le u<v\le E_1/\lambda$ and have angular separation $d_*/2$.

Apply the filter in \cref{mqspgsp:eq:gsp-simple-filter} to the retained first system with fresh signal and work registers.
Use $h=\sin(d_*/8)=\Theta(d_*)$ and $\eta=\Theta(\epsilon)$.
The clock and horizon depend only on these promised parameters, so every record has the same query schedule.
On good records the accepted block obeys $\|\widetilde K_\eta-c\Pi_g\|\le\eta$.
On every other record the circuit remains unitary.

Accept when the first label is valid and the first sample's ancillary work registers, original signal register, and new filter flags and work are all zero.
The first state and the median are correlated.
Nevertheless, removing the bad median records loses at most $\tau$ from the first sample's constant probability of being in the ground state with its work registers in $|0\rangle$.
On the remaining records, the operator approximation changes the accepted vector from $c$ times its ground component by at most $\eta$.
The same triangle inequality as in the certified case therefore gives
\begin{align}
s=\Omega(1).
\label{mqspgsp:eq:gsp-gap-success}
\end{align}

On good records the unnormalized accepted excited norm is at most $\eta$.
Bad records have total accepted probability at most $\tau$.
After conditioning on success and tracing out the orthogonal records, convexity of trace distance gives
\begin{align}
\frac12\|\rho-|g\rangle\langle g|\|_1 \le\frac{\eta}{\sqrt s}+\frac\tau s =O(\eta+\tau).
\label{mqspgsp:eq:gsp-gap-error}
\end{align}
The first term bounds the excited component on good records as in the certified case.
The second is the conditional probability of bad records; tracing them out produces a mixture.
Taking the constants in $\eta=\Theta(\epsilon)$ and $n=\Theta(\log(1/\epsilon))$ sufficiently small and large, respectively, makes this error at most $\epsilon$.
The cost is $n$ coarse samples and one accurate filter:
\begin{align}
q_B=O(p_0^{-1/2}\log(1/\epsilon)),
\quad q_H=O(d_*^{-1}p_0^{-1/2}\log(1/\epsilon)).
\label{mqspgsp:eq:gsp-simple-gap-cost}
\end{align}
This proves the gap-promise claim of \cref{mqspgsp:thm:gsp-main}.
The uniform-clock theorem supplied the constant-error samples from their steady-state catalyst weights; analytic clock selection supplied the final filter from its growth bound.
The median and conditional-state arguments connect these two implementations.

The same decomposition gives a gate and space bound.
Let one complete coarse sampler have gate count $G_{\rm samp}$ and width $w_{\rm samp}$, including its retained proposal histories, and let the controlled final filter have counts $G_{\rm filt},w_{\rm filt}$, excluding the stored sample records.
These quantities are obtained from \cref{eq:qudit-main-cost}: a coarse lift has horizon $O(d_*^{-1}p_0^{-1/2})$, while the final filter lift has horizon $O(d_*^{-1}\log(1/\epsilon))$.
Here the dimensions of the coarse circuit registers include the finite history, Fourier, and sign-path registers described in \cref{mqspgsp:sec:gsp-finite}.

The $n$ labels have $O(\log M)$ bits each.
Copying them to work registers and reversibly sorting them with a fixed sequence of $O(n^2)$ comparisons computes the median using $O(n^2\log M)$ gates and $O(n\log M+n^2)$ workspace qubits.
Keep the comparison bits, copy the median, and reverse the sorting computation.
Including this computation and the success test gives
\begin{subequations}
\label{mqspgsp:eq:gsp-gap-gates-space}
\begin{align}
G_{\rm total} &\le nG_{\rm samp}+G_{\rm filt}+O\!\left(n^2\log M +(w_{\rm samp}+w_{\rm filt})\log(2+w_{\rm samp}+w_{\rm filt})\right),
\\
n_{\rm total} &\le nw_{\rm samp}+w_{\rm filt}+O(n\log M+n^2).
\end{align}
\end{subequations}
These are explicit resource bounds for the construction with retained records.
They include all $n=O(\log(1/\epsilon))$ sampled quantum states and their histories, in addition to the clock and routing registers.
The logarithmic address cost in quditization applies to the dimension of the full register space in \cref{eq:qudit-dimensions}.

We close by evaluating the angular parameter in the input promise.
Write $\Delta=\gamma_0/\lambda$.
If $x=E_0/\lambda$, the physical angular gap is at least $\arccos x-\arccos(x+\Delta)$.
Differentiating this expression shows that it decreases up to $x=-\Delta/2$ and increases afterward.
Its minimum over the promised interval is attained at $x=\max\{-1+\beta,-\Delta/2\}$, which proves \cref{mqspgsp:eq:gsp-angular-gap}.
Without an endpoint bound,
\begin{align}
d_*=2\arcsin\frac{\gamma_0}{2\lambda} =\Theta(\gamma_0/\lambda).
\label{mqspgsp:eq:gsp-unrestricted-gap}
\end{align}
If $2\gamma_0/\lambda>2-\beta$, the same promises already certify $u=1-\gamma_0/\lambda<v=-1+\beta+\gamma_0/\lambda$.
The certified-threshold algorithm then applies directly.
If $p_0=1$, one call to $B$ already prepares the ground state.

\FloatBarrier
\subsubsection{Ground-state energy estimation}
\label{mqspgsp:sec:gsp-energy-estimation}

Ground-state preparation can be followed by phase estimation at an additive cost.
Let $0<\epsilon_E\le\lambda/8$ be the desired additive energy error and let $0<q\le1/4$ be the allowed failure probability of the estimate.
These have separate roles: $\epsilon_E$ fixes the phase resolution, while $q$ fixes the required ground-state weight and the confidence of the phase measurement.

After a successful preparation, run phase estimation on the qubitized walk $W$ in \cref{mqspgsp:eq:gsp-walk}.
On a ground state its two walk eigenphases are $\pm\theta_0$, where $E_0=\lambda\cos\theta_0$.
A phase estimate $\widehat\theta$ gives the energy estimate $\widehat E=\lambda\cos\widehat\theta$.
The cosine is $1$-Lipschitz in circular distance, so angular error at most $\epsilon_E/\lambda$ suffices.

\begin{figure}[!htbp]
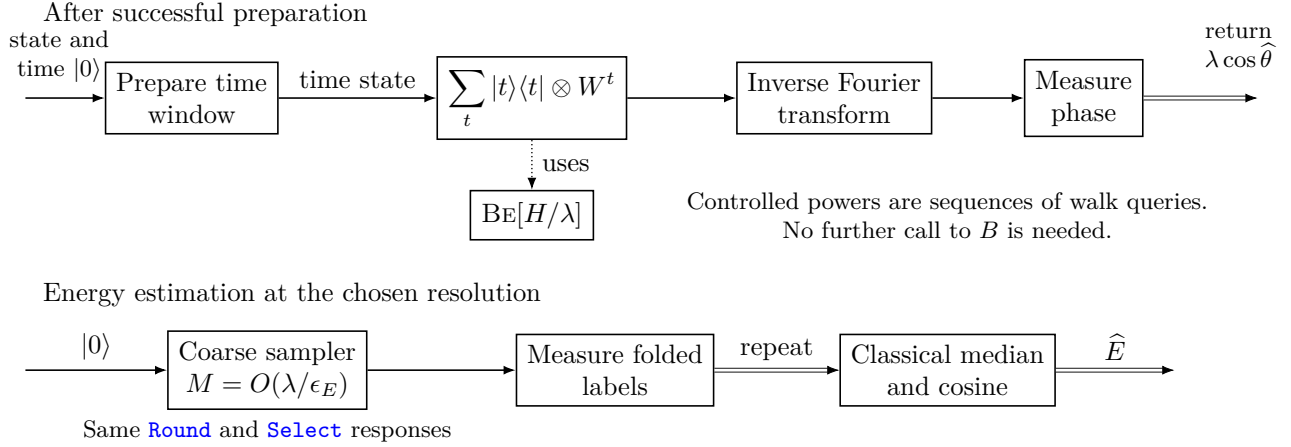

\centering
\IIRDiagram{GroundEnergy}
\caption{The two energy-estimation routes.
In the upper circuit, $t$ is the integer time label of the phase-estimation window.
The public Hilbert space throughout the upper row contains both time and the already prepared system registers; the dashed pair opens the controlled walk powers into Hamiltonian queries.
This finite sequence adds the phase-estimation cost after preparation succeeds.
The lower route reads the coarse sampler's labels at the desired energy resolution and takes their median, as proved in \cref{mqspgsp:prop:gsp-gap-free-energy}.
It stops before the accurate state filter.}
\label{mqspgsp:fig:energy}
\end{figure}

Apply the Kaiser-window phase measurement of \cref{eq:gep-phase-history,eq:gep-kaiser} to $W$, with angular accuracy $\epsilon_E/\lambda$ and failure probability $\Theta(q)$, chosen with a sufficiently small coefficient.
Each walk call uses one Hamiltonian query, giving
\begin{align}
q_H^{\rm phase}=O\!\left(\frac\lambda{\epsilon_E}\log\frac1q\right),
\quad q_B^{\rm phase}=0.
\label{mqspgsp:eq:gsp-phase-cost}
\end{align}
Both ground-state walk branches give the same energy, so the measurement has this guarantee on the physical ground state.
The upper circuit in \cref{mqspgsp:fig:energy} shows this additive phase measurement; its lower circuit previews the energy-only construction proved below.

\FloatBarrier

\Needspace{10\baselineskip}
\begin{proposition}[Energy estimation after successful preparation]
\label{mqspgsp:prop:gsp-energy-after-preparation}
Run the preparation theorem with trace-distance error $\epsilon=\Theta(q)$, then apply the phase measurement when its success flag is observed, choosing sufficiently small error coefficients.
The flag occurs with probability $\Omega(1)$.
Conditioned on it, $\Pr(|\widehat E-E_0|>\epsilon_E)\le q$.
The query bounds are
\begin{align}
\begin{array}{c|cc} \text{promise}&q_H&q_B
\\
\hline
\text{threshold}& O\!\left(d^{-1}[p_0^{-1/2}+\log(1/q)] +\dfrac\lambda{\epsilon_E}\log(1/q)\right)&O(p_0^{-1/2})
\\[3pt]
\text{gap}& O\!\left([d_*^{-1}p_0^{-1/2}+\lambda/\epsilon_E]\log(1/q)\right)& O(p_0^{-1/2}\log(1/q)). \end{array}
\label{mqspgsp:eq:gsp-energy-after-preparation}
\end{align}
\end{proposition}

\begin{proof}
Trace distance bounds the change in the probability of every measurement event.
Thus the conditional failure probability is at most the preparation trace-distance error plus the phase-estimation failure probability.
Choosing each to be a sufficiently small multiple of $q$ gives the claim.
The phase measurement uses the prepared state throughout, so its cost in \cref{mqspgsp:eq:gsp-phase-cost} is additive.
\end{proof}

Repeating preparation until the success flag appears has constant expected overhead; the phase measurement is then performed once.
Thus the same bounds hold in expectation for an algorithm that always returns an estimate.
With a fixed cutoff of $O(\log(1/q))$ preparation attempts, this repetition factor multiplies the preparation costs; exhaustion can be included in the failure budget.
The heralded rows of \cref{mqspgsp:tab:comparison} state the per-attempt bounds.

The phase step also has additive gate and space costs.
If its window length is $N=O((\lambda/\epsilon_E)\log(1/q))$, direct preparation of the known window state costs $O(N)$ gates and its Fourier transform costs $O(\log^2N)$ gates.
It adds $O(\log N)$ time-register qubits and a fresh block-encoding signal register.
The controlled walk calls include their usual oracle and signal-reflection gate costs.

The substitution $\gamma_0=\epsilon_E$ in a ground-state preparation theorem is valid under the promise $E_1-E_0\ge\epsilon_E$.
Keeping a larger supplied gap can reduce the preparation cost.
For energy estimation alone, eigenstates within the target energy interval are also acceptable.
The coarse-label construction provides a direct way to use this weaker requirement.

\Needspace{9\baselineskip}
\begin{proposition}[Energy estimation at a chosen resolution]
\label{mqspgsp:prop:gsp-gap-free-energy}
Given the Hamiltonian and guiding-state oracles with ground-state probability at least $p_0$, choose $\epsilon_E,q$ as above.
For arbitrary spectral gaps, a fixed-schedule algorithm outputs an estimate satisfying $\Pr(|\widehat E-E_0|>\epsilon_E)\le q$ with
\begin{align}
q_H=O\!\left(\frac\lambda{\epsilon_E\sqrt{p_0}}\log\frac1q\right),
\quad q_B=O\!\left(\frac1{\sqrt{p_0}}\log\frac1q\right).
\label{mqspgsp:eq:gsp-gap-free-energy-cost}
\end{align}
The same statement holds for a degenerate ground space when $p_0$ lower-bounds its total probability.
\end{proposition}

\begin{proof}
Use the rounder and selector already constructed, taking a power of two $M=\Theta(\lambda/\epsilon_E)$ and writing $h=2\pi/M$.
Choose its coefficient large enough to leave constant slack in the energy accuracy.
Every valid folded label is within $h$ of the folded walk angle that produced it.
Since $\theta_0$ is the largest folded walk angle $\theta=\arccos(E/\lambda)$, every valid label is at most $\theta_0+h$.
The valid labels at least $\theta_0-h$ form a prefix in the selection order.
This prefix contains all valid labels contributed by the ground space, so its initial probability is $\Omega(p_0)$.

Keep $\eta_0=\Theta(p_0)$ with a sufficiently small coefficient.
By the output law \cref{mqspgsp:eq:selector-law}, the ideal sampler has \begin{align}
\Pr(|\widehat\theta-\theta_0|\le h)\ge\frac12+\Omega(1).
\label{mqspgsp:eq:gsp-energy-prefix}
\end{align} The occupation bounds and the bounds on the span of vectors supplied to the modules use only the stop weight, valid-label probabilities, and Fourier first moment.
They therefore give the same $O(p_0^{-1/2})$ catalyst weight for state preparation and $O(Mp_0^{-1/2})$ Hamiltonian weight for this choice of $M$.
The uniform-clock theorem compiles the entire sampler at a sufficiently small constant vector error, preserving the fixed bias above one half.

Measure the labels of independent samples and take the median of the folded angles, assigning angle zero to a failed label.
For an odd $n=\Theta(\log(1/q))$ with a sufficiently large coefficient, \cref{mqspgsp:eq:gsp-median} bounds an inaccurate median by $e^{-\Omega(n)}\le q$.
On the complementary event, its energy differs from $E_0$ by at most $\lambda h=O(\epsilon_E)$.
The slack in $M$ leaves enough accuracy to evaluate the final cosine, so the total error is at most $\epsilon_E$.
There are $n$ samples, each costing $O(p_0^{-1/2})$ state-preparation queries and $O(Mp_0^{-1/2})$ Hamiltonian queries, proving the bounds.
\end{proof}

This energy algorithm stops after reading the coarse labels.
The accurate state filter and the subsequent phase measurement can both be omitted.
Its choice $h=O(\epsilon_E/\lambda)$ specifies a numerical resolution and is valid for arbitrarily small gaps.
The proof needs only a prefix of labels close to the ground energy; states contributing to that prefix may occupy several nearby eigenspaces.
Sequential measurement permits reuse of one sampler's quantum workspace, with $O(n\log M)$ classical bits to retain the labels.
The MQSP role is the same as in preparation: steady-state oracle weights and one constant-error compilation give the sampler's query bounds.

\section{Quantum simulation of downfolded Hamiltonians}\label{sec:downfolding}

The target transformation is evolution under a Schur-complement Hamiltonian.
The modular design starts from that transfer function rather than from a separately compiled inverse.
Hermitian matrix components enter through the Cayley junctions of \cref{eq:sp-cayley-warmup-system}; closing the internal public coordinates performs the Schur complement by the feedback rule of \cref{subsec:mod-connect}.
An exponential IIR then turns the resulting effective Hermitian load into the desired evolution.
The original matrix ports remain separately labelled, so their internal response determines the query allocation.

Eliminating internal degrees of freedom gives a reduced model, but evaluating its response can still require solving the full internal linear system.
The MQSP network includes this solve in its infinite-impulse-response dynamics.
Closing the internal ports makes the feedback amplitudes satisfy the coupled equations.
Their influence on the active-space response determines how often each component is queried and how accurately its matrix data must be represented.
This allows us to retain small corrections even when accessing the corresponding components at full resolution would be expensive.

Let $P,Q_1,\ldots,Q_h$ be mutually orthogonal projectors whose sum is the identity.
The active space and its orthogonal complement are $\mathcal P=\operatorname{ran}P$, $Q=I-P$, and $Q_0=P$.
For a Hermitian matrix, define its active block, coupling, internal block, and effective Hamiltonian by
\begin{align}
\mathcal H=\left(\begin{array}{c|c}A&B
\\
\hline
B^\dagger&K\end{array}\right),
\quad H_{\mathrm{eff}}=A-BK^{-1}B^\dagger,
\quad X=-K^{-1}B^\dagger.
\label{eq:app-schur}
\end{align}
We assume that the internal block is invertible and that a positive gap bound $\Delta$ is given, with $\|K^{-1}\|\le\Delta^{-1}$.
For an active-space input $|\psi\rangle$, the internal amplitude satisfies $KX|\psi\rangle=-B^\dagger|\psi\rangle$.
Substitution into the active-space equation gives $H_{\mathrm{eff}}|\psi\rangle$.
This is the elimination underlying the Feshbach--Schur map \cite{Dusson2021SchurPerturbation} and Kron reduction \cite{Dorfler2013KronReduction}.
In the MQSP construction, the unknown internal amplitude is an unnormalized catalyst fixed by the feedback equations~\cite{Belovs2024TamingQuantum}.
We compile the complete network using a unitary Toeplitz lift and known clock states.
Neither preparation of the virtual wavefunction nor a separately compiled quantum linear-system algorithm is required.
The simulation below applies to indefinite $K$ as well as positive $K$.

Decompose the model into Hermitian components $\mathcal H=\sum_{j=1}^mH_j$.
Each component has one of two support patterns: $H_j=Q_iH_jQ_i $ or $ H_j=Q_iH_jQ_k+Q_kH_jQ_i $ for $i\ne k$.
Its support projector is $\Pi_j=Q_i$ or $\Pi_j=Q_i+Q_k$, respectively.
We omit known zero components and assume $m\ge1$; if no components remain, the target is the identity.
Access each component through a controlled encoding $\Be{H_j/\lambda_j}$.
Its normalization and query cost satisfy $\lambda_j\ge\|H_j\|$, $\lambda_j>0$, and $C_j>0$.
In a known support basis, its signal input is $(|0\rangle_j\otimes I)\Pi_j$, where $|0\rangle_j$ is the all-zero encoding register.
We require the given full unitary to be an involution:
\begin{subequations}
\label{eq:app-access}
\begin{align}
\Pi_j(\langle0|_j\otimes I)\Be{H_j/\lambda_j} (|0\rangle_j\otimes I)\Pi_j&=H_j/\lambda_j,
\\
\Be{H_j/\lambda_j}^{\dagger}=\Be{H_j/\lambda_j},
\quad \Be{H_j/\lambda_j}^{2}&=I.
\end{align}
\end{subequations}
The oracle acts on its full encoding space, including all block-encoding ancillas and work registers; $\Pi_j$ restricts only the signal input with ancillas initialized to zero.
We count a coupling pair as one component call and count the known support-basis changes in the circuit bounds.
If an encoding must first be converted to an involution, its forward and inverse queries, controls, and additional qubits are included in $C_j$ and the corresponding gate and space counts.

Assume positive bounds on the internal sector responses: $\|Q_iX\|\le\varrho_i$, $\varrho_i>0\ (i\ge1)$, and $\varrho_0=1$.
The bounds are uniform over active-space inputs; no entries of $X$ are given.
Define the sensitivity weights
\begin{align}
\chi_j=\begin{cases} \varrho_i^2,&\Pi_j=Q_i,
\\
2\varrho_i\varrho_k,&\Pi_j=Q_i+Q_k. \end{cases}
\label{eq:app-weights}
\end{align}
We will show that $\lambda_j\chi_j$ is the effective normalization governing the simulation cost.
The two adjoint blocks of a coupling give the factor two in $\chi_j$ while sharing one oracle call.

\Needspace{7\baselineskip} The internal inverse also limits how coarsely we may specify the component data.
We describe this tolerance by component perturbation-sensitivity factors $\kappa_j\ge1$.
A sufficient inequality is
\begin{align}
\sum_{j:\Pi_j=Q_i}\frac{\lambda_j}{\kappa_j} +\sum_{k\ne i}\ \sum_{j:\Pi_j=Q_i+Q_k} \frac{\lambda_j}{\kappa_j}\frac{\varrho_k}{\varrho_i} \le\Delta
\quad (i=1,\ldots,h),
\label{eq:app-kappa}
\end{align}
where $k$ ranges over $\{0,\ldots,h\}\setminus\{i\}$ and an empty sum is zero.
An active-space-only component can use $\kappa_j=1$.
To interpret the inequality, perturb each component while preserving its Hermitian diagonal or coupling-pair support.
Suppose the normalized errors satisfy $\|\widehat H_j-H_j\|\le\lambda_j\eta_j$ and $\eta_j\ge0$.
For the resulting internal block $\widehat K$, we prove
\begin{align}
\max_j\kappa_j\eta_j<1
\quad\Longrightarrow
\quad \|\widehat K^{-1}\|\le \frac{1}{\Delta(1-\max_j\kappa_j\eta_j)}.
\label{eq:app-coarse-stability}
\end{align}
Thus $1/\kappa_j$ gives a sufficient normalized data-error scale for preserving the internal solve.
The weights $\chi_j$ will separately bound the change in the active-space output, provided the response bounds remain valid under the perturbation.
These two roles also appear in the simulation cost.

\Needspace{18\baselineskip} \cref{thm:app-main} simulates the reduced Hamiltonian while weighting each oracle by its bound on its contribution to the internal response.

\begin{theorem}[Sensitivity-weighted simulation]
\label{thm:app-main}
Assume \cref{eq:app-schur}--\cref{eq:app-kappa}.
Fix real $t$ and error $0<\epsilon<1/4$.
A finite unitary circuit $U$ can be chosen to satisfy
\begin{align}
\left\|U(|0\rangle\otimes I_{\mathcal P}) -|0\rangle\otimes e^{itH_{\mathrm{eff}}}\right\|\le\epsilon.
\label{eq:app-clean}
\end{align}
Let $q_j$ count all calls to $\Be{H_j/\lambda_j}$, including the final oblivious amplitude amplification.
The total query cost is
\begin{align}
\quad \sum_{j=1}^m C_jq_j =O\!\left[ |t|\left(\sum_{j=1}^m\sqrt{C_j\lambda_j \chi_j}\right)^2 +\log\frac1\epsilon\sum_{j=1}^m C_j\kappa_j \right].
\quad
\label{eq:app-maincost}
\end{align}
The constants are universal, and the only matrix oracles are the given $\Be{H_j/\lambda_j}$.
\cref{thm:app-gates} gives the gate and workspace counts.
For $t=0$, the identity circuit suffices.
\end{theorem}

The leading term weights the cost $C_j$ by the component's effective normalization $\lambda_j\chi_j$.
A small internal response therefore reduces the queries assigned to that component.
The second term accounts for the precision needed to resolve the coupled feedback through $\kappa_j$.
We obtain the two terms together by optimizing the delays between component queries.
\cref{lem:app-allocation} proves constant-factor optimality of the allocation within this construction and its given bounds.
All internal sectors share the same lift and clock normalization, so repeated excursions between levels of a hierarchy are included in this single cost bound.

A separate spectral bound can improve the response estimate even when the given encoding is unchanged.
For one internal sector, write the three components as
\begin{align}
H_A=\left(\begin{array}{c|c}A&0
\\
\hline
0&0\end{array}\right),
\quad H_B=\left(\begin{array}{c|c}0&B
\\
\hline
B^\dagger&0\end{array}\right),
\quad H_K=\left(\begin{array}{c|c}0&0
\\
\hline
0&K\end{array}\right).
\label{eq:app-three-components}
\end{align}
Let their encodings have normalizations $\lambda_A,\lambda_B,\lambda_K$ and costs $C_A,C_B,C_K$.
Suppose a known upper bound $0<h_B\le\lambda_B$ satisfies $\|B\|\le h_B$.
A larger given bound may first be replaced by its minimum with $\lambda_B$.
In \cref{cor:app-spectral}, $h_B$ enters the response bound through the external resolvent, while $\lambda_B$ remains the oracle normalization.

\Needspace{18\baselineskip}

\begin{corollary}[Spectral-norm response bound]
\label{cor:app-spectral}
For the components in \cref{eq:app-three-components}, assume $\|K^{-1}\|\le\Delta^{-1}$.
Use
\begin{align}
\varrho=\frac{h_B}{\Delta},
\quad \chi_A=1,
\quad \chi_B=\frac{2h_B}{\Delta},
\quad \chi_K=\frac{h_B^2}{\Delta^2},
\quad \kappa_A=1,
\quad \kappa_B=\frac{2\lambda_B}{h_B},
\quad \kappa_K=\frac{2\lambda_K}{\Delta}.
\label{eq:app-spectral-parameters}
\end{align}
Then Hamiltonian simulation of $e^{itH_{\mathrm{eff}}}$ to error $\epsilon$ has query cost
\begin{align}
\sum_{j\in\{A,B,K\}}C_jq_j =O\!\Bigg[&|t|\left( \sqrt{C_A\lambda_A} +\sqrt{\frac{2C_B\lambda_Bh_B}{\Delta}} +\frac{h_B}{\Delta}\sqrt{C_K\lambda_K}\right)^2
\\
&+\log\frac2\epsilon\left( C_A+\frac{C_B\lambda_B}{h_B} +\frac{C_K\lambda_K}{\Delta}\right)\Bigg].
\label{eq:app-spectral-cost}
\end{align}
The circuit uses the original three encodings.
Its gate and workspace counts are those of \cref{thm:app-gates}, with these weights, conditioning factors, and the corresponding delays.
The proof appears below after the general finite-circuit construction.
\end{corollary}

The physical coupling is bounded by $h_B$.
Its encoding retains the original normalization $\lambda_B$.
Relative to the estimate using the encoding normalization, the coupling and internal contributions to the catalyst-weight bound are reduced by $h_B/\lambda_B $ and $ (h_B/\lambda_B)^2$, respectively.
These are reductions of the individual contributions to the catalyst-weight bound before square-root allocation.
The conditioning term is retained: the displayed balanced construction pays $C_B\lambda_B/h_B$ in its precision coefficient.
A tighter response bound therefore improves the time-dependent part without necessarily improving the full cost at every $t$ and $\epsilon$.
If the coupling bound is zero, then $B=0$ and $H_{\mathrm{eff}}=A$.
The coupling and internal queries can be omitted without using the singular formula in \cref{eq:app-spectral-cost}.

The same weights determine the data accuracy required for the active-space transformation.
Let $\widehat H_{\mathrm{eff}}$ be formed from the components $\widehat H_j$ above.
When the internal inverse exists and the bounds $\varrho_i$ hold along the linear interpolation between the two models, \cref{prop:app-data} gives
\begin{align}
\|\widehat H_{\mathrm{eff}}-H_{\mathrm{eff}}\|\le\sum_j\lambda_j \chi_j\eta_j,
\quad \|e^{it\widehat H_{\mathrm{eff}}}-e^{itH_{\mathrm{eff}}}\| \le |t|\sum_j\lambda_j \chi_j\eta_j.
\label{eq:app-motivation-precision}
\end{align}
Consequently, a component with small $\chi_j$ may be represented more coarsely while preserving the requested simulation accuracy.
If the coarse representation admits a cheaper circuit with common valid bounds, its cost $C_j(\eta_j)$ enters \cref{eq:app-maincost}.
This couples the allocation of queries to the choice of data resolution.
Sensitivity-based refinement and coarsening have a classical antecedent in goal-oriented error estimation~\cite{Becker2001ErrorEstimation}; here the sensitivities also bound the quantum query cost.

For a hierarchy of increasingly detailed components, the leading cost is controlled by $\sum_j\sqrt{C_j\lambda_j\chi_j}$.
Refinement can be inexpensive when the response weights decrease rapidly enough to offset the growing access costs, provided the conditioning and known-operation costs remain controlled.
If $K\succ0$, the identity $BK^{-1}B^\dagger=X^\dagger KX$ further relates these weights to the energy, or quadratic-objective change, produced by internal relaxation.
\cref{prop:app-positive} and \cref{cor:app-relaxation} quantify this relation.
They provide the connection to correlation contributions in electronic structure and to eliminated variables in graph-based learning.

Related quantum algorithms exploit structure in the inverse or the requested output.
Fast inversion and preconditioning use structured inverse access~\cite{Tong2021FastInversion}, while other linear-system algorithms separate state-preparation costs or use instance-dependent conditioning~\cite{Low2024QLSP,Dalzell2026BeyondConditionNumber}.
A Schur-complement construction based on quantum singular value transformation~\cite{Gilyen2018singular} first forms an inverse encoding, composes the reduced matrix, and then applies a second inversion~\cite[Algorithm 2]{BinandehDehaghani2025OptimalControl}.
Our construction closes the internal equations before construction of a finite-query circuit, giving an operator transformation whose cost is expressed directly in the original component queries.

Li's multilevel observable-estimation algorithm uses Schur-complement factorization to encode small interlevel corrections without preparing the full solution state~\cite{Li2026MultilevelCorrection}.
Its implementation assumes efficient Ritz-complement block encodings with normalizations that decrease under refinement.
\cref{thm:app-main} instead starts from the component encodings and given uniform response and conditioning bounds.
It produces active-space time evolution; the inverse result below produces a block encoding with known normalization.
State preparation and observable estimation are then counted according to the application.

In \cref{subsec:app-theory}, we prove the theorem by constructing a unitary junction whose closed-port equations give \cref{eq:app-schur}.
Rescaling the internal sectors balances the two ends of each coupling and yields $\lambda_j\chi_j$ without changing the queried matrices.
The perturbation estimate controls the analytic continuation needed by the analytic clock, and optimizing the component delays gives \cref{eq:app-maincost}.
To obtain a finite circuit, we approximate the entire clock Toeplitz block by a Cayley cascade and factor its full system matrix.
This makes the cascade length enter the gate count through a logarithmic address width.
We then differentiate the elimination equations to prove the data-error bounds, relate the positive response to internal relaxation, and construct a reciprocal block with the same component access.
These results give the common theory for the subsequent electronic-structure and Kron-learning subsections, where the response bounds, input vectors, and measured outputs are specified.

\subsection{Sensitivity-weighted simulation}
\label{subsec:app-theory}

We first design the effective transfer function so that its value at the evaluation point generates the desired Schur complement.
Rescaling the internal equations preserves this target but changes the transfer function away from the evaluation point, and hence its group delay.
Choose positive sector scales and define
\begin{align}
T_Q=\sum_{i\ge1}s_iQ_i,
\quad s_0=1.
\end{align}
The identity $BT_Q(T_QKT_Q)^{-1}T_QB^\dagger=BK^{-1}B^\dagger$ shows that this congruence preserves the effective Hamiltonian.
\cref{lem:app-loads} proves that taking each sector scale equal to its response bound minimizes the stated componentwise upper bounds on the delay-weighted catalyst bound.
This balances the two ends of every coupling; it makes no claim to minimize the actual group-delay operator over all realizations.
The congruent matrix and its elimination map are
\begin{align}
\widetilde{\mathcal H}= \left(\begin{array}{c|c}A&BT_Q
\\
\hline
T_QB^\dagger&T_QKT_Q\end{array}\right),
\quad V=\begin{pmatrix}I_{\mathcal P}
\\
T_Q^{-1}X\end{pmatrix}.
\label{eq:app-congruence}
\end{align}
The identity $(T_QKT_Q)^{-1}=T_Q^{-1}K^{-1}T_Q^{-1}$ shows that the effective Hamiltonian remains $H_{\mathrm{eff}}$.
A diagonal component acquires normalization $\widetilde\lambda_j=s_i^2\lambda_j$, whereas a coupling acquires $\widetilde\lambda_j=s_is_k\lambda_j$.
In both cases the rescaled component $\widetilde H_j$ satisfies $\widetilde H_j/\widetilde\lambda_j=H_j/\lambda_j$ on its support.
We can therefore implement the rescaling by changing known junction coefficients while using the original $\Be{H_j/\lambda_j}$ and signal input with ancillas initialized to zero.
\cref{lem:app-native} realizes these rescaled elimination equations by closing the internal public coordinates of a Cayley transformation.

\Needspace{8\baselineskip}

\begin{lemma}[Schur complement from internal feedback]
\label{lem:app-native}
For complex markers $\Bz=(z_1,\ldots,z_m)$ in the open unit polydisk, define the supported load
\begin{align}
\mathsf M(\Bz)=\sum_j\left[ \widetilde\lambda_j\frac{1-z_j^2}{1+z_j^2}\Pi_j -i\frac{2z_j}{1+z_j^2}\widetilde H_j\right].
\label{eq:app-load}
\end{align}
Let $\mathscr S(\Bz)$ be its Schur complement on $\mathcal P$.
For every $u>0$, we can connect the component oracles to a known unitary system matrix and feed the public output of each internal sector $Q_i$ back to its input.
The active-space response is
\begin{align}
F_u(\Bz)=(I-u\mathscr S(\Bz)) (I+u\mathscr S(\Bz))^{-1}.
\label{eq:app-cell}
\end{align}
The load $\mathscr S$ is analytic and accretive in the polydisk: $\appRe\mathscr S=(\mathscr S+\mathscr S^\dagger)/2\succeq0$.
It continues analytically to $\Bz=\boldsymbol{1}$, where $\mathscr S=-iH_{\mathrm{eff}}$.
At $\Bz=\boldsymbol{1}$, the active-space input and output are $\boldsymbol{x},\boldsymbol{y}$.
Together with the internal feedback amplitude $\boldsymbol{g}$, they obey
\begin{align}
\boldsymbol{g}=T_Q^{-1}X\frac{\boldsymbol{x}+\boldsymbol{y}}{2},
\quad \boldsymbol{y}=(I+iuH_{\mathrm{eff}})(I-iuH_{\mathrm{eff}})^{-1}\boldsymbol{x}.
\label{eq:app-selfconsistency}
\end{align}
\end{lemma}
\begin{proof}
We couple the signal inputs with ancillas initialized to zero to their supported public sectors through the known map $L_u=(\sqrt{u\widetilde\lambda_j}\,\Pi_j(\langle0|_j\otimes I))_j$.
Its domain is the direct sum of the full oracle spaces.
Define $R_u=(I+L_uL_u^\dagger)^{-1}$ and the system matrix
\begin{align}
S_u=\left(\begin{array}{c|c} 2R_u-I&-2iR_uL_u
\\
\hline
2L_u^\dagger R_u&i(I-2L_u^\dagger R_uL_u) \end{array}\right).
\label{eq:app-junction}
\end{align}
Block multiplication with $R_u+R_uL_uL_u^\dagger=I$ verifies unitarity.
Connect private port $j$ through $-z_j\Be{H_j/\lambda_j}$.
The involution property gives
\begin{align}
&\bigl(I-iz_j\Be{H_j/\lambda_j}\bigr) \bigl(I+iz_j\Be{H_j/\lambda_j}\bigr)^{-1}
\\
&
\quad=\frac{1-z_j^2}{1+z_j^2}I -i\frac{2z_j}{1+z_j^2}\Be{H_j/\lambda_j}.
\end{align}
Compressing by $(|0\rangle_j\otimes I)\Pi_j$ and summing yields \cref{eq:app-load}, so the response before closing the internal sectors is $(I-u\mathsf M)(I+u\mathsf M)^{-1}$.
Each compressed Cayley transform has positive definite real part on its support for $|z_j|<1$.
Invertibility of $K$ implies that these supports cover the internal space.
Hence $\mathsf M_{QQ}$ is invertible throughout the polydisk.
Equating the input and output on each internal sector and eliminating them gives \cref{eq:app-cell}.
To bound the resulting load, define $V(\Bz)=\binom{I}{-\mathsf M_{QQ}^{-1}\mathsf M_{QP}}$.
The internal rows of $\mathsf M(\Bz)V(\Bz)$ vanish, which implies
\begin{align}
\appRe\mathscr S(\Bz) =V(\Bz)^\dagger\appRe\mathsf M(\Bz)V(\Bz) \succeq0.
\label{eq:app-energy-identity}
\end{align}
An operator is accretive when its Hermitian part is positive semidefinite.
The identity above proves this property for the effective load; the internal inverse also proves analyticity.
Accretivity explains why the Cayley transfer function is contractive: for any vector $\boldsymbol{v}$ and $u>0$, $\|(I+u\mathscr S)\boldsymbol{v}\|^2 -\|(I-u\mathscr S)\boldsymbol{v}\|^2 =4u\boldsymbol{v}^\dagger(\operatorname{Re}\mathscr S)\boldsymbol{v}\ge0$.
The inverse exists since the real part of the denominator is bounded below by the identity.
This proves contractivity without diagonalizing the noncommuting component operators.
At $\boldsymbol{1}$, the open response is $(I+iu\widetilde{\mathcal H})(I-iu\widetilde{\mathcal H})^{-1}$.
Write the closed internal input and output as the same vector $\boldsymbol{g}$.
Multiplying by $I-iu\widetilde{\mathcal H}$ gives $T_QKT_Q\boldsymbol{g}=-T_QB^\dagger(\boldsymbol{x}+\boldsymbol{y})/2$.
Solving this equation and substituting into the active-space row proves \cref{eq:app-selfconsistency}.
\end{proof}

\cref{eq:app-selfconsistency} identifies the internal solve performed by the IIR: its feedback amplitude is the full coupled response to the mean active-space input and output.
Closing all sectors uses only invertibility of $K$.
Successive Schur complements give the same answer when their intermediate pivots exist, since they eliminate the same equations.
Simultaneous closure therefore includes excursions between all internal sectors in a single active-space response.

\cref{lem:app-loads} balances the component loads so that the catalyst weight follows the response bound.

\begin{lemma}[Balanced response-weighted normalization]
\label{lem:app-loads}
For every component $j$,
\begin{align}
-\left.\partial_{z_j}\mathscr S\right|_{\Bz=\boldsymbol1} =\widetilde\lambda_j V^\dagger\Pi_jV\succeq0.
\label{eq:app-port-action}
\end{align}
The choice $s_i=\varrho_i$ minimizes, simultaneously for all couplings, the upper bounds obtained from the given $\varrho_i$.
For positive integer delays $r_1,\ldots,r_m$, define $D_{\boldsymbol{r}}=\sum_j r_j\widetilde\lambda_j\Pi_j$.
Then
\begin{align}
0\preceq V^\dagger D_{\boldsymbol{r}}V =-\left.\frac{d}{dz}\mathscr S(z^{r_1},\ldots,z^{r_m})\right|_{z=1} \preceq\left(\sum_jr_j\lambda_j \chi_j\right)I_{\mathcal P}.
\label{eq:app-delayed-action}
\end{align}
\end{lemma}
\begin{proof}
Differentiate the load in \cref{eq:app-load}.
At $z_j=1$, the identity coefficient has derivative $-1$ and the matrix coefficient has derivative zero.
Differentiating the Schur complement sandwiches $-\widetilde\lambda_j\Pi_j$ between the elimination map and its adjoint, proving \cref{eq:app-port-action}.
For the scale optimization, write the sector response maps as $X_i=Q_iX $ for $i\ge1$ and $X_0=I_{\mathcal P}$.
A diagonal component contributes $\lambda_jX_i^\dagger X_i$ for every $s_i$.
A coupling between $i$ and $k$ contributes
\begin{align}
\lambda_j\left[\frac{s_k}{s_i}X_i^\dagger X_i +\frac{s_i}{s_k}X_k^\dagger X_k\right] \preceq\lambda_j\left[\frac{s_k}{s_i}\varrho_i^2 +\frac{s_i}{s_k}\varrho_k^2\right]I.
\label{eq:app-balancing}
\end{align}
The bracket is minimized at $s_k/s_i=\varrho_k/\varrho_i$, where its value is $2\varrho_i\varrho_k$ by the arithmetic--geometric mean inequality.
Setting $s_i=\varrho_i$ attains this minimum on every edge, including those incident to $P$.
The chain rule for $z_j=z^{r_j}$ and summation with coefficients $r_j$ give \cref{eq:app-delayed-action}.
The argument uses no commutativity assumption on the component matrices.
\end{proof}

We use the balanced choice $T_Q=\sum_{i\ge1}\varrho_iQ_i$ from now on.
The junction normalization is then $\widetilde\lambda_j=\lambda_j\chi_j$ for a diagonal component and $\widetilde\lambda_j=\lambda_j\chi_j/2$ for a coupling.
The latter contributes twice its junction normalization to the bound on the active-space response in \cref{eq:app-delayed-action}.

The derivative bound controls how the active-space response changes with the markers.
Finite-circuit construction also needs a neighborhood in which the internal inverse exists.
For a chosen schedule, let $r_*=\max_jr_j$, set $D_{\boldsymbol{r},Q}=QD_{\boldsymbol{r}}Q$, and define the weighted inverse norm
\begin{align}
\tau_{\boldsymbol{r}}=\left\| D_{\boldsymbol{r},Q}^{1/2}(T_QKT_Q)^{-1}D_{\boldsymbol{r},Q}^{1/2} \right\|.
\label{eq:app-tau}
\end{align}
This quantity measures the stability of the internal equations in the same weighted norm used by the load.
Bounding it by the given $\kappa_j$ gives the analytic neighborhood in \cref{lem:app-cap}, which controls dispersion and truncation in the clock theorem.

\begin{lemma}[Local analyticity with explicit conditioning]
\label{lem:app-cap}
Let $t\ge0$ and let $\widehat\tau_{\boldsymbol{r}}$ be a known uniform upper bound on \cref{eq:app-tau}.
After substituting $z_j=z^{r_j}$, write $\mathscr S(z)=\mathscr S(z^{r_1},\ldots,z^{r_m})$.
The response $e^{-t\mathscr S(z)}$ has norm at most one for $|z|<1$.
It is analytic on an open neighborhood of the closed disk $|z-1|\le\zeta$, where
\begin{align}
\zeta= \frac{1}{64(r_*+\widehat\tau_{\boldsymbol{r}})},
\quad \|e^{-t\mathscr S(z)}\| \le |z|^{3t\sum_j r_j\lambda_j \chi_j}
\quad (|z|\ge1,\ |z-1|\le\zeta).
\label{eq:app-cap}
\end{align}
Condition \cref{eq:app-kappa} implies $\tau_{\boldsymbol{r}}\le\max_j\kappa_jr_j$.
In particular, the smaller radius $[128\max_j\kappa_jr_j]^{-1}$ is valid.
\end{lemma}
\begin{proof}
We estimate changes in the load in the logarithmic coordinate $z=e^w$.
Restrict to its active support, on whose complement the active-space transformation is the identity.
Within this proof write $D=D_{\boldsymbol{r}}$ and regard the load as a function of $w$:
\begin{align}
\mathsf M(w)=-\sum_j\widetilde\lambda_j\tanh(r_jw)\Pi_j -i\sum_j\operatorname{sech}(r_jw)\widetilde H_j.
\label{eq:app-log-load}
\end{align}
For $|w|\le1/(16r_*)$, we have $\|D^{-1/2}\mathsf M'(w)D^{-1/2}\|\le2$.
The bound is independent of the number of components: the maps $\sqrt{r_j\widetilde\lambda_j}\Pi_jD^{-1/2}$ form an isometry when stacked, so each weighted sum is a compression of a direct-sum contraction.
The two scalar derivative coefficients have combined magnitude less than $2$ on this disk.
Integrating from zero therefore gives
\begin{align}
\|D^{-1/2}[\mathsf M(w)+i\widetilde{\mathcal H}]D^{-1/2}\| \le2|w|.
\label{eq:app-load-perturbation}
\end{align}
Since the load is anti-Hermitian on the imaginary axis, integration in the real direction also gives $-\appRe\mathsf M(u+i\theta)\preceq2uD$ for $u\ge0$ in the disk.

A Neumann series in the weighted load perturbation shows that the internal inverse is analytic when $2\tau_{\boldsymbol{r}}|w|<1$, with weighted norm at most $\tau_{\boldsymbol{r}}/(1-2\tau_{\boldsymbol{r}}|w|)$.
Let $V(w)=\binom{I}{-\mathsf M_{QQ}(w)^{-1}\mathsf M_{QP}(w)}$.
Subtract the internal equations at $w$ and at zero, where $V(0)=V$.
Applying \cref{eq:app-load-perturbation} to the residual and \cref{eq:app-delayed-action} to $V$ gives
\begin{align}
\|D^{1/2}[V(w)-V]\| \le\frac{2\tau_{\boldsymbol{r}}|w|}{1-2\tau_{\boldsymbol{r}}|w|} \sqrt{\sum_jr_j\lambda_j \chi_j}.
\label{eq:app-return-perturbation}
\end{align}
On the neighborhood in \cref{eq:app-cap}, $|w|\le2\zeta$, so this change is small enough that $\|D^{1/2}V(w)\|\le(8/7)\sqrt{\sum_jr_j\lambda_j \chi_j}$.
Combining this estimate with \cref{eq:app-energy-identity} yields
\begin{align}
-\appRe\mathscr S(e^{u+i\theta}) \preceq\frac{128}{49}u\left(\sum_jr_j\lambda_j \chi_j\right)I \preceq3u\left(\sum_jr_j\lambda_j \chi_j\right)I.
\end{align}
The bound $\|e^Y\|\le e^{\lambda_{\max}(\appRe Y)}$ proves \cref{eq:app-cap} outside the disk.
Inside the disk, accretivity gives norm at most one.
The strict Neumann-series inequality gives an open set containing the stated closed disk on which the function is analytic.

It remains to express its radius in given parameters.
The delay-weighted internal matrix commutes with the sector scaling.
Using the inverse bound $\|K^{-1}\|\le\Delta^{-1}$ gives
\begin{align}
\tau_{\boldsymbol{r}}\le\frac1\Delta\max_{i\ge1} \left[\sum_{j:\Pi_j=Q_i}r_j\lambda_j +\sum_{k\ne i}\ \sum_{j:\Pi_j=Q_i+Q_k} r_j\lambda_j\frac{\varrho_k}{\varrho_i}\right] \le\max_j\kappa_jr_j,
\label{eq:app-scalar-tau}
\end{align}
where \cref{eq:app-kappa} gives the last inequality.
The use of the full coupled inverse is what permits the same bound when different internal sectors interact.
\end{proof}

For $K\succ0$, the operator order can give a sharper sufficient perturbation bound than the scalar bound \cref{eq:app-kappa}:
\begin{align}
\sum_j\frac{\widetilde\lambda_j}{\kappa_j}Q\Pi_jQ \preceq T_QKT_Q,
\quad \kappa_j\ge1.
\label{eq:app-positive-kappa}
\end{align}
Multiplying this inequality by $\max_j\kappa_jr_j$ bounds $D_{\boldsymbol{r},Q}$ term by term.
Congruence by $(T_QKT_Q)^{-1/2}$ then gives $\tau_{\boldsymbol{r}}\le\max_j\kappa_jr_j$ as before.
The subsequent results therefore also apply with \cref{eq:app-positive-kappa}.
This inequality is assumed or proved from the structure of the application before applying the quantum algorithm.

\begin{lemma}[Allocation with upper bounds on the scaled delays]
\label{lem:app-allocation}
Assume nonnegative catalyst-weight bounds and positive query costs.
For parameters satisfying $a_j\ge0$, $C_j>0$, $b_0>0$, and $\kappa_j\ge1$, one has
\begin{subequations}
\label{eq:app-capped-optimum}
\begin{align}
\left(\sum_j\sqrt{C_ja_j}\right)^2+b_0\sum_jC_j\kappa_j&\le\inf_{r_j\in\mathbb N_{>0}} \left(\sum_ja_jr_j+b_0\max_j\kappa_jr_j\right) \sum_j\frac{C_j}{r_j}
\\
&\le4\left[\left(\sum_j\sqrt{C_ja_j}\right)^2 +b_0\sum_jC_j\kappa_j\right].
\end{align}
\end{subequations}
A constructive schedule is obtained from the real numbers
\begin{align}
x_j=\min\left\{\frac1{\kappa_j},\, \sqrt{\frac{b_0C_j}{a_j\sum_kC_k\kappa_k}}\right\},
\quad r_j=\left\lceil\frac{x_j}{\min_kx_k}\right\rceil.
\label{eq:app-capped-delays}
\end{align}
At $a_j=0$, take $x_j=1/\kappa_j$.
\end{lemma}
\begin{proof}
We bound the time and conditioning contributions separately.
Cauchy--Schwarz gives $(\sum_j\sqrt{C_ja_j})^2$ for the first.
For the conditioning contribution, use
\begin{align}
\frac{\max_k\kappa_kr_k}{r_j}\ge\kappa_j
\quad\text{for every }j.
\end{align}
This gives $b_0\sum_jC_j\kappa_j$.
Adding proves the lower bound.
For the constructive upper bound, write $A_0=\sum_j\sqrt{C_ja_j}$ and $B_0=\sum_jC_j\kappa_j$ within this proof.
The real delays in \cref{eq:app-capped-delays} obey
\begin{align}
\sum_ja_jx_j\le\sqrt{b_0/B_0}\,A_0,
\quad \max_j\kappa_jx_j\le1,
\quad \sum_jC_j/x_j\le B_0+\sqrt{B_0/b_0}\,A_0.
\end{align}
Their objective is at most $A_0^2+b_0B_0+2A_0\sqrt{b_0B_0}\le2(A_0^2+b_0B_0)$.
Rescale all delays so that the smallest is one; the objective is unchanged.
Upward integer rounding multiplies the first factor by at most two and can only decrease the second.
This proves the factor-four bound and gives the stated schedule.
\end{proof}

We now apply the unitary Toeplitz lift in \cref{thm:iir-lift} and analytic clock shaping in \cref{thm:iir-analytic-clock-selection} to the response bound.
For fixed delays, let $q_j^{\rm lift}$ count the calls to component $j$ in one forward lift.
\cref{lem:app-cap} gives the analytic radius and exterior growth required by \cref{thm:iir-analytic-clock-selection}, which gives
\begin{align}
N=O\!\left[|t|\sum_jr_j\lambda_j \chi_j +\max_j\kappa_jr_j\log\frac1\epsilon\right],
\quad q_j^{\rm lift}=\left\lfloor\frac{N-1}{r_j}\right\rfloor
\label{eq:app-horizon}
\end{align}
The clock has nuclear norm at most two.
The circuit obtains $\Be{\widetilde G_N/2}$, approximating the desired evolution at normalization two.
Its coefficients depend only on the given bounds, without requiring $X$ or its operator derivative.

For an explicit sufficient horizon, put
\begin{align}
\zeta=[128\max_j\kappa_jr_j]^{-1},
\quad M_{\rm sm}=\left\lceil\frac{2e(1+\zeta)}\zeta\right\rceil,
\quad q_{\rm sm}=\left\lceil\log\frac{49152}{\epsilon}\right\rceil,
\quad d_{\rm sm}=q_{\rm sm}(M_{\rm sm}-1).
\end{align}
Apply \cref{lem:iir-optimal-flat,lem:iir-flat-smoothing} with two coarse blocks, interior exponent zero, and exterior exponent $3|t|\sum_jr_j\lambda_j\chi_j$.
The horizon
\begin{align}
N=2\max\left\{1,\left\lceil3|t|\sum_jr_j\lambda_j\chi_j\right\rceil\right\} +3d_{\rm sm}
\label{eq:app-explicit-N}
\end{align}
follows from \cref{eq:analytic-horizon-exact}.
The flat clock has normalization at most $\sqrt2$; positive smoothing cannot increase it, and known attenuation sets the normalization to two.
Since $\beta\le e^{-1}$ in \cref{eq:analytic-smoothing-parameters-a}, the clock approximation satisfies $\|\widetilde G_N-e^{itH_{\mathrm{eff}}}\|\le3e^{-q_{\rm sm}}\le\epsilon/16384$ before block-encoding normalization.
The encoded-operator error is half this bound at normalization two.
This leaves the error budget needed for finite realization and oblivious amplitude amplification.
The endpoint vectors are computed from the two-block singular vectors and the finite displacement distribution of \cref{eq:analytic-smoothing-random-variable}.

The analytic exponential need not have a finite-dimensional realization.
\cref{lem:app-finite-cascade} constructs a finite Cayley cascade with the required length-$N$ Toeplitz compression.
Approximating this whole block controls the impulse coefficients sampled by the clock; agreement only at $z=1$ would not suffice.

Oblivious amplitude amplification in \cref{lem:iir-cubic} gives time evolution with bounded isometry error.
The total component counts are
\begin{align}
q_j=3\left\lfloor\frac{N-1}{r_j}\right\rfloor.
\label{eq:app-exact-calls}
\end{align}
\cref{lem:iir-cubic} controls both the block-encoding error and leakage for this unitary target.
The reserved error fractions therefore give \cref{eq:app-clean}.
For negative time we reverse the entire positive-time circuit, preserving these counts.

The total query cost is bounded by multiplying \cref{eq:app-horizon} by $\sum_jC_j/r_j$ and the constant cost of oblivious amplitude amplification.
Apply \cref{lem:app-allocation} with $a_j=|t|\lambda_j\chi_j$ and $b_0=\log(2/\epsilon)$.
Since $\max_j\kappa_jr_j\ge1$, the constant in the horizon is absorbed into the precision term.
This proves \cref{eq:app-maincost}, subject to the finite realization established next, and the claimed constant-factor bound for the allocation.

We obtain a finite unitary network by cascading the Cayley transformations of \cref{lem:app-native}.
Sum the rescaled normalizations incident to each sector: \begin{align}
\Lambda_i=\sum_{j:\,Q_i\Pi_j=Q_i}\widetilde\lambda_j,
\quad D_{\boldsymbol{1}}=\sum_i\Lambda_iQ_i.
\end{align} These scalars are computed from the given normalizations.
If $\Lambda_0=0$, the active-space generator vanishes and the identity circuit suffices.
Otherwise the active incident normalization is positive.
Invertibility of the internal block also ensures $\Lambda_i>0$ in every internal sector.

\begin{lemma}[A finite realization accurate on the whole clock]
\label{lem:app-finite-cascade}
Let $t>0$ and let $N$ be chosen as above.
Choose $J$ as the smallest power of two satisfying
\begin{align}
J\ge\max\left\{2,\ 4t\Lambda_0N,\ 64\sqrt{\frac{\mathrm e}{\epsilon}}(4t\Lambda_0N)^{3/2}\right\}.
\label{eq:app-J}
\end{align}
The cascade of $J$ system matrices from \cref{lem:app-native}, with $u=t/(2J)$, has response
\begin{align}
F_J(z)=\left[(I-t\mathscr S(z)/(2J)) (I+t\mathscr S(z)/(2J))^{-1}\right]^J.
\label{eq:app-cascade}
\end{align}
Let $\appT_N[F]$ denote the causal block Toeplitz matrix of an analytic transfer function $F$.
It contains the first $N$ Taylor coefficients in an array with that many block rows and columns.
Then
\begin{align}
\|\appT_N[F_J]-\appT_N[e^{-t\mathscr S}]\| \le\epsilon/4096.
\label{eq:app-toeplitz-approx}
\end{align}
Moreover $\log J=O(\log(1+t\Lambda_0N)+\log(1/\epsilon))$.
\end{lemma}
\begin{proof}
We control the impulse coefficients by bounding the response on a circle inside the unit disk.
Since $\mathsf M(0)=D_{\boldsymbol{1}}$, $\mathscr S(0)=\Lambda_0I_{\mathcal P}$.
Define the analytic contraction $C(z)=(\mathscr S(z)-\Lambda_0I)(\mathscr S(z)+\Lambda_0I)^{-1}$ and $C(0)=0$.
The quotient by the delay variable has a removable singularity at zero.
Applying the maximum modulus principle on circles of radius approaching one gives $\|C(z)\|\le |z|$, the operator Schwarz lemma.
Inverting the Cayley transform then gives
\begin{align}
\|\mathscr S(z)\|\le\Lambda_0\frac{1+|z|}{1-|z|}.
\label{eq:app-carath}
\end{align}
On the chosen interior circle, $|z|=e^{-1/N} $ implies $ \|t\mathscr S(z)\|\le4t\Lambda_0N$.
To approximate the exponential there, use the power-series estimate $\|(I-Z/2)(I+Z/2)^{-1}-e^{-Z}\|\le\|Z\|^3$, valid for accretive $Z$ with $\|Z\|\le1$.
Both the Cayley transform and the exponential are contractions.
Set $Z=t\mathscr S(z)/J$.
Telescoping the cascade factors bounds the error on the circle by $(4t\Lambda_0N)^3/J^2$.
To pass from this transfer-function error to the clock block, use a diagonal rescaling of the clock basis.
For an analytic matrix function, $R_r=\operatorname{diag}(1,r,\ldots,r^{N-1})$, $E_r(z)=E(rz)$, and $R_r\appT_N[E]R_r^{-1}=\appT_N[E_r]$.
The rescaled Toeplitz matrix is a compression of multiplication by the rescaled transfer function on the unit circle.
Its operator norm is therefore bounded by the maximum there, so $\|\appT_N[E]\| \le\|R_r^{-1}\|\,\|R_r\|\sup_{|z|=r}\|E(z)\| =r^{-(N-1)}\sup_{|z|=r}\|E(z)\| $ for $0<r<1$.
At $r=e^{-1/N}$ the prefactor is at most $\mathrm e$.
The choice \cref{eq:app-J} consequently gives \cref{eq:app-toeplitz-approx}.
The endpoint clock maps are contractions, so the block-encoding error is no larger.
Rounding $J$ to a power of two also gives the stated logarithmic bound.
\end{proof}

The number of Cayley transformations can be large, so executing them serially would lose the query advantage.
We instead factor their known system matrix while retaining the full private space.
Let $n$ count qubits in a common physical system register.
Each oracle uses $a_j$ persistent encoding qubits and has controlled gate cost $G_j$, including its representation changes.
We pad these encoding registers to $a_{\max}=\max_ja_j$, with the extra qubits as spectators.
Let $G_{\rm align}$ count a controlled change of basis from the signal inputs with ancillas initialized to zero to the weighted superpositions defined in \cref{eq:app-bright}, together with its inverse, sector tests, and support-basis changes.
Let $s_{\rm supplied}$ be the maximum number of reusable ancillas initialized to $|0\rangle$ used by an oracle or this alignment.
Known changes of basis must be implemented with standard quantum gates, as in \cref{def:resource-accounting}.

\begin{theorem}[Gate, oracle, and workspace counts]
\label{thm:app-gates}
For $N,J,r_j$ as above, define the address qubit count
\begin{align}
b=\max\left\{8,\left\lceil \log_2\left(N+J2^{a_{\max}}\sum_jr_j\right)\right\rceil\right\}.
\label{eq:app-address}
\end{align}
The amplified circuit of \cref{thm:app-main} makes exactly the involution calls in \cref{eq:app-exact-calls}.
With reusable prefix-computation ancillas initialized to $|0\rangle$ it has elementary gate count and total qubit count
\begin{align}
G_{\rm total} &=O\!\left[\sum_jq_jG_j+ N\bigl(G_{\rm align}+(m+1)b\bigr)\right],
\label{eq:app-gates}
\\
n_{\rm total} &=n+b+O\!\left(\log(N+1)+\max\{b,s_{\rm supplied}\}\right)
\\
&=n+O\!\left[a_{\max}+s_{\rm supplied} +\log(2+N+Jmr_*)\right].
\label{eq:app-space}
\end{align}
Elementary gates here include arbitrary one-qubit rotations; their finite-gate-set synthesis is counted in \cref{eq:app-synthesis}.
The cascade length $J$ enters the gate count through the logarithmic width $b$.
A serial execution has depth at most the corresponding fully decomposed gate count.
\end{theorem}

The factorization and circuit proof are in \cref{app:downfolding-circuit}.

When the sectors have explicit labels and their signal inputs with ancillas initialized to zero share a known basis, a direct alignment circuit gives
\begin{align}
G_{\rm align}= O\!\left(m\,[a_{\max}+\log(m+1)+\log(h+1)]\right).
\label{eq:app-explicit-alignment}
\end{align}
There are at most $2m$ sector--component incidences.
Their weighted amplitudes can be prepared by two-level rotations controlled on the sector label and zero encoding ancillas.
Computing and erasing these predicates takes a number of gates linear in the displayed widths.
Given changes of system basis add their actual costs and are the only parts of this alignment that act nontrivially on the physical system register.

The $n$-qubit physical register contains every qubit acted on by the component oracles.
Independent physical subsystems still require their tensor-product space.
The direct-sum packing in \cref{eq:app-carrier} applies to alternative sectors and histories, which accounts for the logarithmic additional address space.

Delaying a component reduces its query frequency but increases the number of known-operation passes through the horizon.
To account for this tradeoff, fix a conservative address-width bound and let $g_{\rm step}$ bound $G_{\rm align}+(m+1)b$.
The leading gate cost is obtained by minimizing
\begin{align}
\left(\sum_j\lambda_j \chi_jr_j\right) \left(g_{\rm step}+\sum_jG_j/r_j\right),
\quad r_j\ge1.
\label{eq:app-gate-objective}
\end{align}
The minimum is within a universal constant of
\begin{align}
g_{\rm step}\sum_j\lambda_j \chi_j+ \left(\sum_j\sqrt{G_j\lambda_j \chi_j}\right)^2
\label{eq:app-gate-leading}
\end{align}
using
\begin{align}
r_j=\left\lceil\max\left\{1, \sqrt{\frac{G_j\sum_k\lambda_k \chi_k}{g_{\rm step}\lambda_j \chi_j}}\right\}\right\rceil.
\label{eq:app-gate-delays}
\end{align}
The constraint $r_j\ge1$ gives the first lower-bound term, and Cauchy--Schwarz gives the second.
For the constructive upper bound, apply $\lceil\max(1,x)\rceil\le2(1+x)$ in the first factor and $r_j\ge\max(1,x)$ in the second.
Their product is at most four times \cref{eq:app-gate-leading}.
To optimize the precision term as well, introduce $R\ge\kappa_jr_j$ for all $j$ and minimize the full objective
\begin{align}
\left(1+|t|\sum_j\lambda_j \chi_jr_j+R\log\frac2\epsilon\right) \left(g_{\rm step}+\sum_jG_j/r_j\right),
\quad r_j\ge1,
\quad R\ge\kappa_jr_j.
\label{eq:app-gate-GP}
\end{align}
With fixed width bounds, this is a geometric program: expanding the objective gives a sum of positive monomials.
Introduce logarithmic coordinates for the delays and their common upper bound, $r_j=e^{y_j}$ and $R=e^Y$.
Each monomial becomes an exponential of an affine function, and the constraints become affine inequalities.
The continuous relaxation can therefore be minimized by convex optimization to a chosen accuracy.
Upward integer rounding changes the cost by at most a universal factor.
The width chosen for $g_{\rm step}$ must cover the resulting $N,J,r_j$, so the optimization accounts for the known operations between delayed oracle calls.

Finite-gate-set synthesis follows \cref{def:resource-accounting}.
A full-unitary implementation error $\xi_j$ in controlled $\Be{H_j/\lambda_j}$ is counted at every occurrence.
Its circuit-error budget is
\begin{align}
\sum_jq_j\xi_j=O(\epsilon).
\label{eq:app-oracle-error}
\end{align}
The hybrid argument bounds the complete-circuit error by this sum.
The costs $G_j$ and scratch widths are evaluated at the corresponding precision.
An error in matrix data encoded by an otherwise exact oracle has a different effect: it changes the internal response and can be bounded using the weights $\chi_j$, as we now show.

\subsubsection{Spectral bounds, correlation weights, and finite resources}
\label{subsubsec:app-spectral}

We now derive the response bound in \cref{cor:app-spectral} and its conditioning and circuit consequences.
Squaring the Hermitian coupling component gives
\begin{align}
H_B^2=\left(\begin{array}{c|c}BB^\dagger&0
\\
\hline
0&B^\dagger B\end{array}\right),
\quad \|H_B\|=\|B\|,
\quad \|X\|=\|K^{-1}B^\dagger\|\le\frac{h_B}{\Delta}.
\label{eq:app-spectral-response-proof}
\end{align}
Thus the bound concerns every active-space input and requires only the given bounds $h_B$ and $\Delta$.

\begin{proof}[Proof of \cref{cor:app-spectral}]
Use the internal scale $T_Q=\varrho I_Q$, with $\varrho=h_B/\Delta$.
The congruent matrix and elimination map are
\begin{align}
\widetilde{\mathcal H} =\left(\begin{array}{c|c}A&\varrho B
\\
\hline
\varrho B^\dagger&\varrho^2K\end{array}\right),
\quad V=\begin{pmatrix}I_P
\\
X/\varrho\end{pmatrix},
\quad \widetilde\lambda_A=\lambda_A,
\quad \widetilde\lambda_B=\varrho\lambda_B,
\quad \widetilde\lambda_K=\varrho^2\lambda_K.
\label{eq:app-spectral-congruence}
\end{align}
Indeed, $(\varrho^2K)^{-1}=\varrho^{-2}K^{-1}$, so eliminating the internal block gives $A-BK^{-1}B^\dagger=H_{\mathrm{eff}}$.
On each support, $\widetilde H_j/\widetilde\lambda_j=H_j/\lambda_j$.
The known junction coefficients change, but no additional or renormalized matrix oracle is synthesized.

For arbitrary positive delays $r_A,r_B,r_K$, the positive action operator in \cref{eq:app-delayed-action} is exactly
\begin{subequations}
\label{eq:app-spectral-action}
\begin{align}
V^\dagger D_{\boldsymbol{r}}V &=\bigl(r_A\lambda_A+r_B\lambda_B\varrho\bigr)I_P +\left(\frac{r_B\lambda_B}{\varrho}+r_K\lambda_K\right)X^\dagger X
\\
&\preceq\left(r_A\lambda_A+2r_B\lambda_B\varrho +r_K\lambda_K\varrho^2\right)I_P.
\end{align}
\end{subequations}
This proves the three weights in \cref{eq:app-spectral-parameters}.
For the inverse bound, $D_{\boldsymbol{r},Q} =(r_B\varrho\lambda_B+r_K\varrho^2\lambda_K)I_Q$, and hence
\begin{align}
\tau_{\boldsymbol{r}} =\left(\frac{r_B\lambda_B}{\varrho}+r_K\lambda_K\right)\|K^{-1}\| \le \frac{r_B\lambda_B}{h_B} +\frac{r_K\lambda_K}{\Delta} \le\max\{\kappa_Br_B,\kappa_Kr_K\}.
\label{eq:app-spectral-tau}
\end{align}
The last step uses $u+v\le2\max\{u,v\}$.
The normalizations satisfy $\lambda_K\ge\|K\|\ge\Delta$ and $\lambda_B\ge h_B$.
Thus all three conditioning parameters are at least one.
Direct substitution verifies the scalar sufficient feedback bound:
\begin{align}
\frac{\lambda_K}{\kappa_K} +\frac{\lambda_B}{\kappa_B\varrho}=\frac\Delta2+\frac\Delta2=\Delta.
\label{eq:app-spectral-kappa-check}
\end{align}
Consequently, \cref{lem:app-cap} provides a common analytic neighborhood.
Its exponent is at most $3|t|(r_A\lambda_A+2r_B\lambda_B\varrho+r_K\lambda_K\varrho^2)$.
The signed-clock construction and the finite Toeplitz approximation in \cref{lem:app-finite-cascade} give
\begin{align}
N=O\!\left[|t|\left(r_A\lambda_A+ \frac{2r_B\lambda_Bh_B}{\Delta} +\frac{r_K\lambda_Kh_B^2}{\Delta^2}\right) +\max\left\{r_A,\frac{2r_B\lambda_B}{h_B}, \frac{2r_K\lambda_K}{\Delta}\right\} \log\frac1\epsilon\right].
\label{eq:app-spectral-horizon}
\end{align}
Apply the integer allocation in \cref{lem:app-allocation} with $a_A=|t|\lambda_A$, $a_B=2|t|\lambda_Bh_B/\Delta$, $a_K=|t|\lambda_Kh_B^2/\Delta^2$, and $b_0=\log(2/\epsilon)$.
Multiplication by $\sum_j C_j/r_j$ and the three executions used for oblivious amplitude amplification proves \cref{eq:app-spectral-cost}.
The counts in this construction are $q_j=3\lfloor(N-1)/r_j\rfloor$, including controlled inactive occurrences.
No preparation of $X|\psi\rangle$ is part of these circuits.

The active-space incident normalization that fixes the finite Cayley length is
\begin{align}
\Lambda_0=\lambda_A+\lambda_Bh_B/\Delta.
\label{eq:app-spectral-Lambda0}
\end{align}
Use this value in \cref{eq:app-J}, and use the resulting $J,N,r_j$ in \cref{eq:app-address}.
In this three-component case, the actual Hilbert space is
\begin{align}
(\mathbb C^N\otimes\mathcal P)\ \oplus\! \bigoplus_{j\in\{A,B,K\}} (\mathbb C^{r_j}\otimes\mathbb C^J\otimes\mathcal L_j),
\label{eq:app-spectral-carrier}
\end{align}
where $\mathcal L_j$ is the full Hilbert space of $\Be{H_j/\lambda_j}$.
Its stage label is a spectator during a component query.
The exact full-space factorization in \cref{eq:app-dyadic} therefore applies without introducing an oracle for the internal response.
With the gate and scratch definitions of \cref{thm:app-gates}, the resulting bounds are
\begin{align}
G_{\rm total} &=O\!\left[\sum_{j\in\{A,B,K\}}q_jG_j+ N(G_{\rm align}+4b)\right],
\label{eq:app-spectral-gates}
\\
n_{\rm total} &=n+O\!\left[a_{\max}+s_{\rm supplied} +\log(2+N+3Jr_*)\right].
\label{eq:app-spectral-space}
\end{align}
Here the $4N$ routing term counts the public stream and three private streams; the remaining query-routing events are at most $3N$ per lift.
Clock preparation uses $O(N)$ elementary gates and is included.
Synthesis and oracle errors use \cref{eq:app-synthesis,eq:app-oracle-error} under the cost model of \cref{def:resource-accounting}.
\end{proof}

For a scalar coupling, the response and energy correction distinguish the physical coupling strength from its oracle normalization: $B=g$, $K=\Delta$, $X=-\overline g/\Delta$, and $BK^{-1}B^\dagger=|g|^2/\Delta$.
With $h_B=|g|$, the internal and coupling action bounds are $\lambda_K|g|^2/\Delta^2 $ and $ 2\lambda_B|g|/\Delta$, respectively.
The latter becomes $2|g|^2/\Delta$ only when the coupling encoding is tight.

The choice $T_Q=\varrho I_Q$ minimizes the displayed time-dependent coupling contribution to the catalyst-weight bound.
It need not minimize time and precision together.
One can retain the same response bound $\|X\|\le\varrho$ and use any internal scale $s>0$.
Here $V_s=\binom{I_P}{X/s}$ is the elimination map.
Form the diagonal action operator $D_{\boldsymbol{r},s}$ from the rescaled normalizations $\lambda_A$, $s\lambda_B$, and $s^2\lambda_K$.
Define $\tau_{\boldsymbol{r},s}=\|D_{\boldsymbol{r},s,Q}^{1/2}(s^2K)^{-1} D_{\boldsymbol{r},s,Q}^{1/2}\|$, with $D_{\boldsymbol{r},s,Q}=QD_{\boldsymbol{r},s}Q$.
The exact calculation above gives
\begin{subequations}
\label{eq:app-spectral-unbalanced}
\begin{align}
V_s^\dagger D_{\boldsymbol{r},s}V_s &\preceq\left[r_A\lambda_A+ r_B\lambda_B\left(s+\frac{\varrho^2}{s}\right) +r_K\lambda_K\varrho^2\right]I_P,
\\
\tau_{\boldsymbol{r},s} &\le\frac{r_B\lambda_B}{s\Delta} +\frac{r_K\lambda_K}{\Delta},
\quad V_s=\begin{pmatrix}I_P
\\
X/s\end{pmatrix}.
\end{align}
\end{subequations}
These follow by multiplying the diagonal support weights exactly as in \cref{eq:app-spectral-action,eq:app-spectral-tau}; the internal scale cancels from the $K$ contribution.
The proof of \cref{lem:app-cap} uses only the action and weighted inverse bounds, so it applies with these expressions.
The same capped allocation gives
\begin{align}
O\!\Bigg[&|t|\left(\sqrt{C_A\lambda_A} +\sqrt{C_B\lambda_B(s+\varrho^2/s)} +\varrho\sqrt{C_K\lambda_K}\right)^2
\\
&+\log\frac2\epsilon\left(C_A+ C_B\max\left\{1,\frac{\lambda_B}{s\Delta}\right\} +\frac{C_K\lambda_K}{\Delta}\right)\Bigg].
\label{eq:app-spectral-unbalanced-cost}
\end{align}
Choose the conditioning parameters as $\kappa_B=2\max\{1,\lambda_B/(s\Delta)\}$ and $\kappa_K=2\lambda_K/\Delta$.
Each contribution to $\tau_{\boldsymbol{r},s}$ is then at most half the maximum delayed conditioning.
Since $s+\varrho^2/s\ge2\varrho$ with equality at $s=\varrho$, this family recovers \cref{eq:app-spectral-cost} and permits a different balance when precision dominates.
This bounds the cost within the stated family of realizations.

\subsubsection{Response bounds for coupled internal sectors}
\label{subsubsec:app-comparison}

Separate spectral bounds also give computable response estimates for an interacting hierarchy.
For internal sectors, write $K_{ik}=Q_iKQ_k$ and $B_i=BQ_i$ for $i,k\ge1$.
Suppose given diagonal gaps and coupling bounds satisfy $\Delta_i>0$, $h_i\ge0$, and $h_{ik}=h_{ki}\ge0$, and
\begin{align}
\|K_{ii}^{-1}\|\le\Delta_i^{-1},
\quad \|B_i\|\le h_i,
\quad \|K_{ik}\|\le h_{ik}
\quad(i\ne k).
\label{eq:app-block-spectral-data}
\end{align}
The diagonal blocks may be indefinite.
Form the real symmetric comparison matrix $M$ with entries
\begin{align}
M_{ii}=\Delta_i,
\quad M_{ik}=-h_{ik}
\quad(i\ne k).
\label{eq:app-comparison-matrix}
\end{align}
These are physical norm bounds.
They may be tighter than sums of the normalizations of the component encodings that implement the blocks.

\cref{prop:app-comparison} replaces a large operator inequality by a sufficient bound from a scalar comparison matrix.

\begin{proposition}[Response bounds from a comparison matrix]
\label{prop:app-comparison}
If $M\succ0$, the internal block $K$ is invertible.
The comparison inverse $M^{-1}$ is entrywise nonnegative, and
\begin{align}
\|K^{-1}\|\le\lambda_{\min}(M)^{-1},
\quad \|Q_iK^{-1}Q_k\|\le(M^{-1})_{ik},
\quad \|Q_iX\|\le\sum_k(M^{-1})_{ik}h_k.
\label{eq:app-comparison-response}
\end{align}
The self-energy also satisfies
\begin{align}
\|BK^{-1}B^\dagger\|\le\sum_{i,k}h_i(M^{-1})_{ik}h_k.
\label{eq:app-comparison-energy}
\end{align}
Consequently, one may take $\Delta=\lambda_{\min}(M)$ and any positive $\varrho_i$ at least the final expression in \cref{eq:app-comparison-response} in the simulation, inverse, and sensitivity results.
Their query factors remain $\lambda_j\chi_j$, with the original component normalizations.
\end{proposition}
\begin{proof}
Write the scalar comparison matrix as a diagonal part minus its nonnegative coupling matrix: $D=\operatorname{diag}(\Delta_i)$, $M=D-H$, $H_{ii}=0$, and $H_{ik}=h_{ik}\ (i\ne k)$.
Congruence by the inverse square root of the positive diagonal matrix gives $D^{-1/2}MD^{-1/2}=I-D^{-1/2}HD^{-1/2}\succ0$.
Consequently, the largest eigenvalue of the rescaled coupling matrix is strictly less than one.
Since this matrix has nonnegative entries, its operator norm equals its largest eigenvalue: for every real vector $\boldsymbol{v}$, the absolute value of its quadratic form is bounded by the quadratic form on the entrywise absolute vector $|\boldsymbol{v}|$.
Thus $\|D^{-1/2}HD^{-1/2}\|<1$, and a Neumann series gives
\begin{align}
M^{-1}=D^{-1/2}\sum_{n=0}^{\infty} (D^{-1/2}HD^{-1/2})^nD^{-1/2}\ge0
\quad\text{entrywise}.
\label{eq:app-comparison-positive-inverse}
\end{align}
To prove invertibility, suppose $K\boldsymbol{v}=0$.
Collect the sector norms in $\boldsymbol{u}=(\|Q_i\boldsymbol{v}\|)_i$.
Multiply each block row by its diagonal-block inverse and take norms.
The triangle inequality gives
\begin{align}
\Delta_i u_i\le\sum_{k\ne i}h_{ik}u_k.
\end{align}
Thus $M\boldsymbol{u}\le0$ entrywise.
Multiplication by the nonnegative matrix $M^{-1}$ implies $\boldsymbol{u}\le0$.
Since $\boldsymbol{u}\ge0$, $\boldsymbol{v}=0$.
In the finite-dimensional oracle spaces, trivial kernel proves invertibility.

For $K\boldsymbol{v}=\boldsymbol{f}$, define $\boldsymbol{f}_{\rm blk}=(\|Q_i\boldsymbol{f}\|)_i$.
The same calculation gives $\Delta_i u_i\le (f_{\rm blk})_i+\sum_{k\ne i}h_{ik}u_k$.
Thus $M\boldsymbol{u}\le\boldsymbol{f}_{\rm blk}$ and $\boldsymbol{u}\le M^{-1}\boldsymbol{f}_{\rm blk}$, so
\begin{align}
\|\boldsymbol{v}\|=\|\boldsymbol{u}\|_2 \le\|M^{-1}\boldsymbol{f}_{\rm blk}\|_2 \le\lambda_{\min}(M)^{-1}\|\boldsymbol{f}\|.
\label{eq:app-comparison-inverse-proof}
\end{align}
For a forcing vector supported on sector $k$, the same inequality gives $\|Q_iK^{-1}\boldsymbol{f}\|\le(M^{-1})_{ik}\|\boldsymbol{f}\|$.
This proves the blockwise inverse estimate.
Set $\boldsymbol{f}=-B^\dagger|\psi\rangle$ for an arbitrary normalized active-space vector.
Then $(f_{\rm blk})_k\le h_k$, and $\|Q_iX|\psi\rangle\|\le\sum_k(M^{-1})_{ik}h_k$.
Taking the supremum proves \cref{eq:app-comparison-response}.
Finally, $BK^{-1}B^\dagger=-\sum_iB_iQ_iX$ has norm at most $\sum_i h_i\|Q_iX\|$.
Substituting the response bounds gives \cref{eq:app-comparison-energy}.
No positivity of $K$ is needed for this norm bound; positivity is used later to infer response bounds from a separately known self-energy bound.
\end{proof}

For two internal sectors with $\Delta_1\Delta_2>h_{12}^2$, the response bounds are explicitly
\begin{align}
\varrho_1=\frac{\Delta_2h_1+h_{12}h_2}{\Delta_1\Delta_2-h_{12}^2},
\quad \varrho_2=\frac{h_{12}h_1+\Delta_1h_2}{\Delta_1\Delta_2-h_{12}^2}.
\label{eq:app-two-sector-response}
\end{align}
This is direct inversion of \cref{eq:app-comparison-matrix}; the positive determinant and diagonal entries ensure positivity.
It includes coupling between the internal sectors in both the numerators and denominator.
Bounds from isolated diagonal inverses alone would omit these terms.
A zero right-hand side in \cref{eq:app-comparison-response} may be replaced by any positive upper bound for use in the junction, or by exact removal of a structurally disconnected sector when that removal is known.

Once the response and gap bounds have been chosen, the scalar conditioning inequality still uses the original component normalizations $\lambda_j$.
For example, assigning every component the common value
\begin{align}
\kappa_j=\max\!\left\{1,\frac1\Delta\max_{i\ge1} \left[\sum_{\ell:\Pi_\ell=Q_i}\lambda_\ell+ \sum_{k\ne i}\sum_{\ell:\Pi_\ell=Q_i+Q_k} \lambda_\ell\frac{\varrho_k}{\varrho_i}\right]\right\}
\label{eq:app-comparison-kappa}
\end{align}
verifies \cref{eq:app-kappa} by division of each row by this common value.
Separate choices can be smaller.
Constructing $M$ and solving its scalar linear system uses the given block bounds, avoiding explicit entries of the exponentially larger internal response.
Obtaining valid block bounds is a separate classical or separately costed computational task.

\subsubsection{Uniform spectral bounds under coarse data}
\label{subsubsec:app-spectral-data}

The response bound used in a cost statement must also apply to the matrix actually encoded.
The following finite perturbation calculation provides this link without a given trial wavefunction.
Perturb the three blocks as $A\mapsto A+\delta A$, $B\mapsto B+\delta B$, and $K\mapsto K+\delta K$, where
\begin{align}
\|\delta A\|\le\lambda_A\eta_A,
\quad \|\delta B\|\le\lambda_B\eta_B,
\quad \|\delta K\|\le\lambda_K\eta_K<\Delta.
\label{eq:app-spectral-data-errors}
\end{align}
For the entire straight interpolation, a uniform response bound is
\begin{align}
\|K(u)^{-1}\|\le\frac1{\Delta-\lambda_K\eta_K},
\quad \|X(u)\|\le \frac{h_B+\lambda_B\eta_B}{\Delta-\lambda_K\eta_K}
\quad(0\le u\le1).
\label{eq:app-spectral-path}
\end{align}
Indeed, the inverse Neumann series converges since $\|K^{-1}u\delta K\|\le\lambda_K\eta_K/\Delta<1$, and $\|B+u\delta B\|\le h_B+\lambda_B\eta_B$.
Multiplication gives \cref{eq:app-spectral-path}.

Let $H_{\mathrm{eff}}'$ be the perturbed Schur complement and put $K'=K+\delta K$.
The resolvent identity and an exact expansion give
\begin{subequations}
\label{eq:app-spectral-perturbation-expansion}
\begin{align}
K'^{-1}-K^{-1}&=-K'^{-1}\delta K K^{-1},
\\
(B+\delta B)K'^{-1}(B+\delta B)^\dagger-BK^{-1}B^\dagger &=\delta B K'^{-1}(B+\delta B)^\dagger +B K'^{-1}\delta B^\dagger +B(K'^{-1}-K^{-1})B^\dagger.
\end{align}
\end{subequations}
Taking norms term by term proves the explicit output bound
\begin{align}
\|H_{\mathrm{eff}}'-H_{\mathrm{eff}}\|\le\lambda_A\eta_A +\frac{2h_B\lambda_B\eta_B+(\lambda_B\eta_B)^2} {\Delta-\lambda_K\eta_K} +\frac{h_B^2\lambda_K\eta_K} {\Delta(\Delta-\lambda_K\eta_K)}.
\label{eq:app-spectral-finite-error}
\end{align}
The first-order coefficients are $\lambda_A$, $2\lambda_Bh_B/\Delta$, and $\lambda_Kh_B^2/\Delta^2$.
The quadratic coupling term and inverse stability denominator quantify the limitation of that first-order interpretation.
Alternatively, \cref{prop:app-data} with the uniform bound in \cref{eq:app-spectral-path} gives a common sensitivity bound for all three perturbations.
Duhamel's identity multiplies either matrix-error bound by $|t|$ for time evolution.
If both active-space matrices are at least $aI$, the inverse resolvent identity multiplies it by $a^{-2}$ for the inverse task.

To compile the coarse model, use valid bounds for that model: the internal singular-value floor is at least $\Delta-\lambda_K\eta_K$ and the coupling norm is at most $h_B+\lambda_B\eta_B$.
When both endpoint oracles have the same normalization $\lambda_B$, this upper bound can be clamped to $\lambda_B$.
The original and perturbed encodings must both obey their stated normalizations; otherwise the common normalization must be updated too.
The cheaper costs $C_j(\eta_j)$ then enter the cost formula with these updated bounds.
For several sectors, \cref{prop:app-comparison} applies uniformly to any comparison matrix formed from lower diagonal bounds and upper coupling bounds valid throughout the data interpolation.

\Needspace{10\baselineskip}
\begin{proposition}[Sensitivity to coarse approximations]
\label{prop:app-data}
Perturb each component while preserving its Hermitian diagonal or coupling-pair support: $\widehat H_j=H_j+\delta H_j$ and $\|\delta H_j\|\le\lambda_j\eta_j$.
Interpolate the component matrices along
\begin{align}
\mathcal H(s)=\sum_j(H_j+s\delta H_j),\qquad0\le s\le1.
\end{align}
Denote its internal block by $K(s)$ and its Schur complement by $H_{\mathrm{eff}}(s)$.
Either sufficient feedback bound \cref{eq:app-kappa} or \cref{eq:app-positive-kappa}, under its stated assumptions, implies
\begin{align}
\max_j\kappa_j\eta_j<1
\quad\Longrightarrow
\quad \|K(s)^{-1}\|\le \frac{1}{\Delta(1-s\max_j\kappa_j\eta_j)}
\quad(0\le s\le1).
\label{eq:app-coarse-path}
\end{align}
Alternatively, invertibility along the path may be established separately.
In either case, suppose in addition that the response satisfies $\|Q_iX(s)\|\le\varrho_i$ throughout the path.
Then
\begin{align}
\|H_{\mathrm{eff}}(1)-H_{\mathrm{eff}}(0)\|\le\sum_j\lambda_j \chi_j\eta_j.
\label{eq:app-data-bound}
\end{align}
\cref{eq:app-motivation-precision} follows.
If in addition both endpoint Schur complements are at least $aI$ for a known $a>0$, then
\begin{align}
\|H_{\mathrm{eff}}(1)^{-1}-H_{\mathrm{eff}}(0)^{-1}\| \le\frac1{a^2}\sum_j\lambda_j \chi_j\eta_j.
\label{eq:app-inverse-data}
\end{align}
\end{proposition}
\begin{proof}
We first preserve invertibility along the interpolation.
Under the balanced internal congruence, a supported perturbation obeys $-\widetilde\lambda_j\eta_j\Pi_j \preceq(P+T_Q)\delta H_j(P+T_Q) \preceq\widetilde\lambda_j\eta_j\Pi_j$.
For a coupling pair both off-diagonal blocks acquire the same scalar, so the same bound applies.
Use $\eta_j\le(\max_k\kappa_k\eta_k)/\kappa_j$, sum on the internal space, and undo the congruence.
The scalar inequality \cref{eq:app-kappa} gives
\begin{align}
\left\|\sum_jQ\delta H_jQ\right\| \le\Delta\max_j\kappa_j\eta_j.
\end{align}
A Neumann series around $K^{-1}$ proves \cref{eq:app-coarse-path}.
Under the positive-operator inequality \cref{eq:app-positive-kappa}, the rescaled perturbation bound gives instead $K(s)\succeq(1-s\max_j\kappa_j\eta_j)K$.
Using $K\succeq\Delta I$ proves the same inverse estimate.
This establishes \cref{eq:app-coarse-stability}; the separate pathwise assumption on $\varrho_i$ is needed for output sensitivity.

Differentiate the elimination equations along the path.
Since $X=-K^{-1}B^\dagger$, the derivative of the Schur complement is
\begin{align}
\dot H_{\mathrm{eff}}=\dot A+\dot B X+X^\dagger\dot B^\dagger +X^\dagger\dot KX =\begin{pmatrix}I
\\
X\end{pmatrix}^{\!\dagger} \dot{\mathcal H}\begin{pmatrix}I
\\
X\end{pmatrix}.
\label{eq:app-sensitivity-identity}
\end{align}
At each point on the path, a diagonal perturbation is weighted by the squared response of its sector.
A coupling and its adjoint contribute the product of the two sector responses.
Their combined norm is bounded by $\lambda_j\chi_j\eta_j$, so integrating the derivative proves \cref{eq:app-data-bound}.
Duhamel's identity then gives the evolution bound in \cref{eq:app-motivation-precision}: the two surrounding propagators are unitary and the integration interval has length $|t|$.
For inverses, use the exact identity $H_{\mathrm{eff}}(1)^{-1}-H_{\mathrm{eff}}(0)^{-1} =H_{\mathrm{eff}}(1)^{-1} [H_{\mathrm{eff}}(0)-H_{\mathrm{eff}}(1)]H_{\mathrm{eff}}(0)^{-1}$.
The two endpoint inverse norms are at most $1/a$.
Taking norms therefore proves \cref{eq:app-inverse-data}; no inverse bound at intermediate points is needed for this last step.
\end{proof}

When only the internal block changes, a direct resolvent estimate avoids a separate pathwise response bound.
Perturb only the internal block, with error $\eta=\|\delta K\|<\Delta$.
Denote the new effective Hamiltonian by $H_{\mathrm{eff}}'$, and define $K'=K+\delta K$ and $X'=-(K+\delta K)^{-1}B^\dagger$.
Then
\begin{align}
H_{\mathrm{eff}}'-H_{\mathrm{eff}}=X^\dagger\delta KX',
\quad \|X'\|\le\frac{\|X\|}{1-\eta/\Delta},
\quad \ \|H_{\mathrm{eff}}'-H_{\mathrm{eff}}\|\le \frac{\|X\|^2\eta}{1-\eta/\Delta}.\
\label{eq:app-finite-K-error}
\end{align}
The denominator quantifies the loss of internal stability; the squared response norm suppresses its effect on the effective Hamiltonian.
Sectorwise coarsening uses the uniform bounds of \cref{prop:app-data}, which may themselves be established by corresponding resolvent estimates.

These estimates allow the data representation and query schedule to be chosen together.
Under common valid normalizations, supports, response bounds, and perturbation bounds, use the coarse-data costs $C_j(\eta_j)$ in \cref{eq:app-maincost} and require $|t|\sum_j\lambda_j\chi_j\eta_j$ to fit within the model-error budget.
If coarsening changes one of these bounds, update it before applying the cost theorem.
The saving is determined by the resulting access circuit: a lower bit precision is useful to the extent that it reduces its actual cost.

For a positive internal block, the response has an additional interpretation as the decrease in a quadratic form under relaxation.
It also bounds the inverse of the full coupled matrix, which will be used for the reciprocal transformation.

\begin{proposition}[Positive response and inverse bounds]
\label{prop:app-positive}
If $K\succ0$, define $\Sigma=BK^{-1}B^\dagger$.
Then
\begin{align}
\Sigma=X^\dagger KX.
\label{eq:app-positive-energy}
\end{align}
Spectral bounds on the internal block imply $\Delta I\preceq K\preceq dI $ implies $ \Delta X^\dagger X\preceq\Sigma\preceq dX^\dagger X$ and $\|X\|^2\le\|\Sigma\|/\Delta$.
If also $\mathcal H\succ0$ and $H_{\mathrm{eff}}\succeq aI$ for a given positive gap bound, then
\begin{align}
\widetilde{\mathcal H}^{-1} &=VH_{\mathrm{eff}}^{-1}V^\dagger+ \operatorname{diag}(0,(T_QKT_Q)^{-1}),
\label{eq:app-inverse-decomposition}
\\
\left\|D_{\boldsymbol{r}}^{1/2}\widetilde{\mathcal H}^{-1} D_{\boldsymbol{r}}^{1/2}\right\| &\le \frac1a\sum_jr_j\lambda_j \chi_j+\max_j\kappa_jr_j.
\label{eq:app-positive-load}
\end{align}
Either perturbation bound \cref{eq:app-kappa} or \cref{eq:app-positive-kappa} may be used.
\end{proposition}
\begin{proof}
Substitute the internal response into its quadratic form: $X=-K^{-1}B^\dagger$ and $X^\dagger KX=BK^{-1}B^\dagger$.
Sandwiching the spectral bounds on the internal block gives \cref{eq:app-positive-energy} and its inequalities.
Block inversion gives \cref{eq:app-inverse-decomposition}.
After congruence by $D_{\boldsymbol{r}}^{1/2}$, the first term has norm at most $\|H_{\mathrm{eff}}^{-1}\|\,\|V^\dagger D_{\boldsymbol{r}}V\|$, bounded by \cref{lem:app-loads}.
The second term is bounded by $\tau_{\boldsymbol{r}}\le\max_j\kappa_jr_j$ from \cref{lem:app-cap}.
Adding proves \cref{eq:app-positive-load}.
\end{proof}

\Needspace{18\baselineskip}
\begin{corollary}[Relaxation energy with separate spectral bounds]
\label{cor:app-relaxation}
Suppose the internal block is one encoded component with normalization $\lambda_K$.
Its positive spectral bounds satisfy $\Delta I\preceq K\preceq h_KI$ and $h_K\le\lambda_K$.
Assume positive bounds on the relaxation and coupling norms: $\|BK^{-1}B^\dagger\|\le\mathcal E$, $\mathcal E>0$, $\|B\|\le h_B\le\lambda_B$, and $h_B>0$.
The choice
\begin{align}
\varrho=\min\left\{\frac{h_B}{\Delta},\sqrt{\frac{\mathcal E}{\Delta}}\right\},
\quad \chi_K=\varrho^2,
\quad \chi_B=2\varrho
\label{eq:app-combined-response}
\end{align}
is valid and gives
\begin{subequations}
\label{eq:app-relaxation-weights}
\begin{align}
\lambda_K\chi_K &\le\lambda_K\min\left\{\frac{h_B^2}{\Delta^2}, \frac{\mathcal E}{\Delta}\right\},
\\
\lambda_B\chi_B &\le2\lambda_B\min\left\{\frac{h_B}{\Delta}, \sqrt{\frac{\mathcal E}{\Delta}}\right\}.
\end{align}
\end{subequations}
If also $\lambda_B\le\beta\|B\|$ for a known $\beta\ge1$, then
\begin{align}
\lambda_B\chi_B\le2\beta\sqrt{\frac{h_K}{\Delta}}\,\mathcal E,
\quad \lambda_K\chi_K\le\frac{\lambda_K}{\Delta}\mathcal E.
\label{eq:app-relaxation-spectral}
\end{align}
The simulation and reciprocal bounds use these same weights and, for one internal sector, the valid conditioning choices
\begin{align}
\kappa_A=1,
\quad \kappa_B=2\max\left\{1,\frac{\lambda_B}{\Delta\varrho}\right\},
\quad \kappa_K=\frac{2\lambda_K}{\Delta}.
\label{eq:app-relaxation-kappa}
\end{align}
\end{corollary}
\begin{proof}
By \cref{prop:app-positive}, $\Delta X^\dagger X\preceq X^\dagger KX=\Sigma$.
It follows that $\|X\|^2\le\mathcal E/\Delta$.
Combining this with \cref{eq:app-spectral-response-proof} proves the minimum in \cref{eq:app-combined-response}, and substitution into $\lambda_K\varrho^2$ and $2\lambda_B\varrho$ proves \cref{eq:app-relaxation-weights}.
The internal block is positive with eigenvalues at most $h_K$.
Spectral decomposition therefore gives $K^2\preceq h_KK$.
Therefore
\begin{align}
BB^\dagger=X^\dagger K^2X\preceq h_KX^\dagger KX=h_K\Sigma,
\quad \|B\|\le\sqrt{h_K\mathcal E}.
\label{eq:app-B-spectral-energy}
\end{align}
Under the additional assumption on the encoding normalization, $\lambda_B\le\beta\sqrt{h_K\mathcal E}$ and $\varrho\le\sqrt{\mathcal E/\Delta}$, whose product proves the coupling estimate in \cref{eq:app-relaxation-spectral}.
The internal estimate follows directly from $\varrho^2\le\mathcal E/\Delta$.
Finally $\lambda_K/\kappa_K=\Delta/2$ and $\lambda_B/(\varrho\kappa_B)\le\Delta/2$, proving \cref{eq:app-kappa} and the stated conditioning choices.
\end{proof}

The physical spectral width $h_K/\Delta$ improves the coupling estimate.
The physical norm bound may be much smaller than the encoding normalization $\lambda_K$.
The internal contribution still contains $\lambda_K/\Delta$.
Replacing this factor by $h_K/\Delta$ would require a different encoding or an additional implementation theorem.
If the physical spectral width, encoded condition bound, and normalization factor are moderate, both effective normalizations scale with the relaxation contribution $\mathcal E$.
This is a uniform statement on the active space; a single-state energy correction need not bound $\|\Sigma\|$.
Inferring the response bound from the relaxation operator requires positivity of the internal block.
For example, an indefinite block can cancel the relaxation while leaving a nonzero response: $K=\operatorname{diag}(1,-1)$, $B=(g,g)$, $BK^{-1}B^\dagger=0$, and $\|K^{-1}B^\dagger\|^2=2|g|^2$.
The norm-based bound in \cref{cor:app-spectral} still applies to this indefinite example.

The same elimination can be used inside another matrix transformation.
For a positive effective Hamiltonian, substitute the closed internal network into the unitary reciprocal construction of \cref{lem:mod-reciprocal}.
Applying a single clock to the completed response gives a block encoding of $H_{\mathrm{eff}}^{-1}$ with known normalization and costs in the original component oracles.

\Needspace{23\baselineskip}
\begin{corollary}[Positive inverse block encoding with known normalization]
\label{cor:app-inverse}
Under the positive-definite assumptions of \cref{prop:app-positive}, the Reciprocal module and one shared analytic clock give a unitary circuit which obtains $\Be{B_\epsilon}$, where $B_\epsilon$ satisfies
\begin{align}
\|B_\epsilon-(a/4)H_{\mathrm{eff}}^{-1}\|\le\epsilon,
\label{eq:app-inverse-block}
\end{align}
with total component-oracle cost
\begin{align}
O\!\left[\left\{ \frac1a\left(\sum_j\sqrt{C_j\lambda_j \chi_j}\right)^2 +\sum_jC_j\kappa_j\right\}\log\frac1\epsilon\right].
\label{eq:app-inverse-cost}
\end{align}
For specified delays its horizon is
\begin{align}
N=O\!\left[\left(\frac1a\sum_jr_j\lambda_j \chi_j+ \max_j\kappa_jr_j\right)\log\frac1\epsilon\right],
\quad q_j=O(N/r_j).
\label{eq:app-inverse-horizon}
\end{align}
This is a block-encoding guarantee with normalization $4/a$.
A solution-state or observable-estimation task additionally counts its state preparation, measurement, and amplification.
\end{corollary}
\begin{proof}
We approximate the reciprocal with $f_d(u)=u^{2d-1}/(1+u^{2d})$.
The Reciprocal construction completes this function to a unitary response using at most $3d$ shifted Cayley factors.
For real $u\ge2$,
\begin{align}
|f_d(u)-u^{-1}|=\frac1{u(1+u^{2d})} \le\frac1{2(1+4^d)}.
\end{align}
Thus a degree $d=O(\log(1/\epsilon))$ suffices on the spectrum of $2H_{\mathrm{eff}}/a$.
Substitute the closed system matrices of \cref{lem:app-native} into the full unitary completion before applying the clock.

The main step is to bound every factor on one analytic neighborhood, independently of its pole width.
Within this proof put $D=D_{\boldsymbol{r}}$ and $\tau=\|D^{1/2}\widetilde{\mathcal H}^{-1}D^{1/2}\|$.
Each factor has real shift and positive width satisfying $|c|\le a/2$ and $0<\gamma\le a/2$.
Up to its phase and sign of the Cayley factor, its full resolvent is $R=(\widetilde{\mathcal H}-cP+i\gamma P)^{-1}$; a known unitary termination implements the shift.
For fixed active-space amplitude $\boldsymbol{x}$, minimizing over the internal amplitudes gives the quadratic form $\boldsymbol{x}^\dagger H_{\mathrm{eff}}\boldsymbol{x}$.
Since $H_{\mathrm{eff}}\succeq aI$, this proves $\widetilde{\mathcal H}\succeq aP$.
Congruence by $\widetilde{\mathcal H}^{-1/2}$ then gives
\begin{subequations}
\label{eq:app-inverse-poles}
\begin{align}
\|D^{1/2}RD^{1/2}\|&\le2\tau,
\\
RP&=V(H_{\mathrm{eff}}-c+i\gamma)^{-1},
\\
\|\widetilde{\mathcal H}^{1/2}RP\|^2 &=\max_{\sigma\in\operatorname{spec}H_{\mathrm{eff}}} \frac{\sigma}{(\sigma-c)^2+\gamma^2}\le4/a,
\quad \|D^{1/2}RP\|^2\le4\tau/a.
\end{align}
\end{subequations}
The first estimate, together with \cref{eq:app-load-perturbation}, gives a Neumann-series inverse on a neighborhood of radius a universal constant times $(r_*+\tau)^{-1}$.
On this neighborhood the last weighted estimate in \cref{eq:app-inverse-poles} changes by at most a constant factor.
Let $F(z)$ be the shifted Cayley transfer function.
For a normalized active-space input, its feedback amplitude is $\boldsymbol{v}=(\mathsf M(z)+icP+\gamma P)^{-1}P|\psi\rangle$.
The energy identity for this response is $\|F(z)|\psi\rangle\|^2-1 =-4\gamma\,\boldsymbol{v}^\dagger\appRe\mathsf M(z)\boldsymbol{v}$.
The radial load bound from \cref{lem:app-cap} and $\gamma\le a/2$ bound this norm increase by $O(\tau\log|z|)$ on the exterior part of the neighborhood.
Each factor consequently has exterior exponent $O(\tau)$ and norm at most one inside the disk, with no dependence on $\gamma^{-1}$ in these estimates.
Separate system matrices with nonnegative powers of the delay variables implement opposite sign choices for the Cayley factors and preserve the disk bound.

Composing the $O(d)$ factors gives exterior exponent $O(d\tau)$ on this same neighborhood.
Known rotations and active public channels preserve the bounds.
One analytic clock therefore gives \cref{eq:app-inverse-horizon}, using the estimate for $\tau$ in \cref{prop:app-positive}.
Its encoded operator is $f_d(2H_{\mathrm{eff}}/a)/2+O(\epsilon)=(a/4)H_{\mathrm{eff}}^{-1}+O(\epsilon)$; choosing the internal error fractions proves \cref{eq:app-inverse-block}.
Occurrences of the same component oracle are grouped coherently, with a constant overhead for the sign choices for the Cayley factors.
Apply \cref{lem:app-allocation} with $a_j=\lambda_j\chi_j/a$ and $b_0=1$ to obtain \cref{eq:app-inverse-cost}.
\end{proof}

The reciprocal factors have different shifts and widths, so we count their known operations directly.
Let $p_0,p_j$ be the public and component-private multiplicities in the completed realization.
These are relative to a common space of $n+a_{\max}$ qubits.
These multiplicities include every reciprocal stage and active public channel.
Let $G_{\rm work}$ count a full pass of known operations, and let $s_{\rm work}$ be the maximum reusable scratch required by a given oracle, the known operations, or routing.
Direct-sum packing gives the address width and circuit counts
\begin{align}
b_{\rm inv}&=\left\lceil\log_2(Np_0+\sum_jr_jp_j)\right\rceil,
\\
G_{\rm inv}&=O\!\left[\sum_jq_jG_j+NG_{\rm work} +b_{\rm inv}\bigl((m+1)N+\sum_jq_j\bigr)+N\right],
\label{eq:app-inverse-gates}
\\
n_{\rm inv}&=n+a_{\max} +O\!\left[b_{\rm inv}+\log(N+1)+s_{\rm work}\right].
\label{eq:app-inverse-space}
\end{align}
The proof uses the coherent arithmetic and clock preparation of \cref{thm:app-gates}, with the actual multiplicities in place of the equal-stage cascade.
State preparation and readout, including scalar-estimation registers, are additional to these block-encoding counts.

For an explicit bound, take the reciprocal degree $d=O(\log(1/\epsilon))$ used above.
Let $G_{\rm sector}$ count a reversible sector-label circuit, its inverse, and the needed basis changes.
Direct synthesis of the $O(d)$-factor realization gives
\begin{align}
G_{\rm work}=O\!\left[ G_{\rm sector}+d(h+1)(m+2)^2 \bigl(a_{\max}+d+\log(2(h+1)(m+2))\bigr)\right].
\label{eq:app-inverse-work}
\end{align}
There are $O(d(h+1))$ factor--sector cases.
Each is a known unitary on $O(m+2)$ port coordinates and admits two-level synthesis quadratic in that dimension.
The control widths include the signal, ancilla, and channel labels; predicates and their inverses cost linearly in those widths.
Structure in the given operations may reduce this bound.
The dependence on the physical space enters through the given registers and circuits.
Finite-gate synthesis and occurrence errors are counted by \cref{eq:app-synthesis}--\cref{eq:app-oracle-error}.

Applications give the target, response bounds, and access circuits.
Input-state preparation and readout complete the cost of the requested computation.

\subsection{Electronic-structure downfolding}
\label{subsec:app-electronic}

Electronic-structure calculations often seek an accurate description of a small set of chemically important configurations, although their energies and interactions depend on many external excitations.
An active space captures competing configurations responsible for static correlation.
Dynamic correlation describes the additional adjustments by which electrons avoid one another, including for a time-independent Hamiltonian.
These adjustments can involve a large virtual orbital space even when individual amplitudes are small.
Multireference perturbation theory \cite{Andersson1992ActiveSpacePerturbation} and coupled-cluster methods \cite{Bartlett2007coupledcluster}, including active-space downfolding \cite{Bauman2019Downfolding}, incorporate this external response.

Two complementary choices reduce different quantum resources.
Orbital-defined partitions \cite{Bauman2019Downfolding,Bauman2019ExcitedStateDownfolding} expose blocks that can use smaller integral tables or factorizations, reducing their block-encoding gate counts, in particular Toffoli counts.
Occupation-defined restrictions, as used in the multireference linearized coupled-cluster constructions of Larsson, Zhai, Gunst, and Chan \cite{Larsson2022LargeSites}, limit the number of particles and holes that must be represented, reducing data qubits when implemented in a compact encoding.
These gains are not automatic consequences of drawing fewer matrix blocks: both the access circuits and any excitation truncation must be justified.

The preceding sensitivity-weighted simulation theorem retains the remaining virtual response in the MQSP self-consistent equations, without supplying its wavefunction or calling a separate linear-system solver.
Small known sensitivity bounds reduce the cost of querying expensive orbital blocks and can justify removing high-occupation sectors altogether.
A coarse calculation proposes the electronic representation; gap and operator-residual estimates control its approximation error.
We distinguish exact downfolding of the given Hamiltonian from approximation of that target with a proved error bound by a smaller occupation space.

\subsubsection{Orbital partitions and compact particle--hole spaces}
\label{subsec:df-partitions}
We specify the electronic Hamiltonian and the active-space target.
Fix the nuclei, an orthonormal basis of $n_{\rm orb}$ spin orbitals, and an $n_{\rm e}$-electron sector.
In atomic units the given Hamiltonian is
\begin{align}
H=E_{\rm nuc}I+\sum_{p,q=1}^{n_{\rm orb}}h_{pq}a_p^\dagger a_q +\frac14\sum_{p,q,r,s=1}^{n_{\rm orb}} v_{pqrs}a_p^\dagger a_q^\dagger a_s a_r.
\label{eq:df-electronic-H}
\end{align}
The scalar $E_{\rm nuc}$ is the nuclear repulsion energy.
The coefficients $h_{pq}$ contain electronic kinetic energy and nuclear attraction, while $v_{pqrs}$ are antisymmetrized two-electron Coulomb integrals.
The operators $a_p^\dagger,a_p$ create and annihilate an electron in the indicated spin orbital.
The identity $I$ acts on the fixed-electron-number sector.
The finite basis and integrals define the model whose downfolding is exact.

Partition the spin orbitals into inactive core, active, and external virtual orbitals.
Every reference determinant occupies all core orbitals and leaves all virtual orbitals empty, while the active occupations vary subject to the fixed total electron number.
These determinants span the complete active space $\mathcal P$, with projector $P$.
Their competing occupations describe static correlation; excitations into the complementary space describe the additional external response commonly called dynamic correlation.
Write $Q=I-P$ for this external space's projector.
If a determinant has $h$ core holes and $p$ virtual particles, its substitution distance from the complete active space is $d=\max\{h,p\}$.
\cref{prop:df-compact-space} proves this counting rule.
Since the Hamiltonian contains at most two-body operators, a nonzero matrix element changes this distance by at most two.
Pairing adjacent distances into excitation bands therefore yields the exact block-tridiagonal form in \cref{eq:df-tridiagonal-matrix}, before any truncation.
Choose the energy parameter $E$ so that the external excitation operator $K=Q(H-EI)Q$ is invertible on the external space.
Using the common applications notation, put
\begin{subequations}
\label{eq:df-target}
\begin{align}
A&=PHP,
\quad B=PHQ,
\quad K=Q(H-EI)Q,
\\
H_{\rm eff}(E)&=A-BK^{-1}B^\dagger,
\quad X=-K^{-1}B^\dagger.
\end{align}
\end{subequations}
The target is $e^{itH_{\rm eff}(E)}$ on arbitrary active-space inputs, to operator error $\epsilon$.
Time is measured with $\hbar=1$.
The map $X$ gives the virtual response to an active-space vector.
The two block equations prove the Feshbach--L\"owdin identity \cite{Lowdin1951PerturbationTheory,Feshbach1958NuclearReactions}.
An eigenvector at the chosen energy has external component $Xu$ and active component satisfying $H_{\rm eff}(E)u=Eu$.
This energy-dependent Feshbach--L\"owdin elimination is algebraic at the chosen energy.
The simulated effective evolution is defined by that Hamiltonian, whereas compressing the original time evolution generally gives a different operator.
An energy-independent Schrieffer--Wolff transformation \cite{Bravyi2011SchriefferWolff} instead decouples a chosen invariant subspace and can generate interactions of higher particle rank.
Energy self-consistency is a separate task, addressed below.
The common simulation theorem applies to $H-EQ$, whose Schur complement is exactly \cref{eq:df-target}.

Orbital-defined blocks reduce query cost when their smaller integral tensors have cheaper access circuits.
Use nested orbital active spaces with projectors $P=P_0\le P_1\le\cdots\le P_L=I$ on the specified electron sector.
Each enlargement releases selected occupied orbitals or admits additional virtual orbitals; all use the same fixed orbital basis.
The mutually orthogonal sectors $P_\ell-P_{\ell-1}$ therefore partition the external configurations without double counting configurations that excite several orbital groups.
Suppose $P_\ell$ permits variable occupations in only $m_\ell$ orbitals.
The compressed Hamiltonian then acts on those modes, with additional scalar and one-body contributions from the fixed occupied orbitals.
This follows by contracting the spectator creation and annihilation operators against their fixed occupation state.
For $i,j\le\ell$, a block between sectors $i,j$ can consequently be encoded from this smaller Hamiltonian and the two sector tests.

Thus orbital partitioning can replace repeated access to a full $n_{\rm orb}$-orbital tensor by access to differently sized tensors or factorizations.
A dense $m_\ell$-orbital two-electron tensor has $O(m_\ell^4)$ entries.
The query's Toffoli cost depends additionally on how those entries are accessed and used.
Low-rank molecular-orbital encodings \cite{Berry2019CholeskyQubitization,vonBurg2020carbon}, tensor hypercontraction \cite{Lee2020hypercontraction}, and first-quantized encodings \cite{Su2021FirstQuantization} offer different gate, normalization, and workspace tradeoffs.
Achieving this saving requires a block-encoding circuit that loads the smaller orbital tensor or factorization.
Integral preprocessing, spectator contractions, controls, and basis changes remain included in the cost.
Static core Coulomb and exchange contributions belong in $PHP$ even when core holes are rare; only the additional virtual response receives a small-response weight.

Restricting occupations reduces the system register when the encoding stores only the allowed particles and holes.
Relative to an occupied reference determinant, a hole records an emptied occupied orbital and a particle records a filled virtual orbital.
Merely changing to particle--hole operators leaves the Hilbert-space dimension unchanged.
The space saving requires both a restriction justified by an error bound on these occupations and an encoding that stores their indices.
Compact configuration-interaction and occupied-mode simulation provide established access constructions \cite{Babbush2016Fermions,Kirby2021CompactEncoding}.

For a reference determinant with $n_{\rm e}$ occupied spin orbitals, the number of determinants with at most $k$ particle--hole pairs is
\begin{align}
D_{\le k}=\sum_{r=0}^{k} \binom{n_{\rm e}}r\binom{n_{\rm orb}-n_{\rm e}}r,
\quad n_{\rm data}^{\rm min}=\lceil\log_2 D_{\le k}\rceil =O\!\left(k\log n_{\rm orb}+\log(k+1)\right).
\label{eq:df-single-reference-dimension}
\end{align}
Choose the $r$ holes and $r$ particles independently to obtain the count.
Here $k$ is the maximum occupation distance admitted by the encoding.
For fixed $k$, this restriction makes the represented dimension polynomial in the orbital count.
The dimension-minimal address requires reversible ranking and unranking.
The padded encoding in \cref{prop:df-compact-space} instead stores particle and hole indices directly; the proposition also counts the admissible determinants.

\begin{proposition}[Compact particle--hole space around an active space]
\label{prop:df-compact-space}
Partition the orbitals into occupied inactive core, empty external virtual, and variable-occupation active sets.
Allow every active occupation compatible with the fixed electron number.
The orbital counts obey $n_{\rm core}+n_{\rm act}+n_{\rm virt}=n_{\rm orb}$ and $0\le n_{\rm e}-n_{\rm core}\le n_{\rm act}$.
Let $\Pi_k$ contain determinants in the fixed-electron-number sector at substitution distance at most $k$ from the active space.
Its dimension is
\begin{align}
\dim\Pi_k=\sum_{h=0}^{k}\sum_{p=0}^{k} \binom{n_{\rm core}}h\binom{n_{\rm virt}}p \binom{n_{\rm act}}{n_{\rm e}-n_{\rm core}+h-p},
\label{eq:df-CAS-dimension}
\end{align}
where an invalid binomial coefficient is zero.
It has an explicit encoding using at most
\begin{subequations}
\label{eq:df-compact-qubits}
\begin{align}
n_{\rm data}\le{}&n_{\rm act} +k\lceil\log_2(n_{\rm core}+1)\rceil +k\lceil\log_2(n_{\rm virt}+1)\rceil
\\
&+2\lceil\log_2(k+1)\rceil+O(1) =n_{\rm act}+O(k\log n_{\rm orb}+\log(k+1))
\end{align}
\end{subequations}
system qubits, before block-encoding ancillas and temporary work registers.
\end{proposition}
\begin{proof}
A determinant with $h$ inactive holes and $p$ external particles has active occupation $n_{\rm e}-n_{\rm core}+h-p$.
At least $\max(h,p)$ substitutions are necessary to restore the inactive occupations.
They suffice: pair as many external particles with inactive holes as possible, then use the excess occupied or empty active orbitals.
Thus its distance from the complete active space is $\max(h,p)$.
Counting the hole set, particle set, and active determinant proves \cref{eq:df-CAS-dimension}.
Store the active occupations directly and store two sorted lists of at most $k$ orbital indices, padded by a sentinel, with their lengths $h,p$.
This is injective and has the width in \cref{eq:df-compact-qubits}.
Invalid list strings and electron counts are excluded from the encoded system subspace.
\end{proof}

For fixed active-orbital count and occupation cutoff, the compact system register uses $O(\log n_{\rm orb})$ qubits.
An unrestricted occupation encoding uses $n_{\rm orb}$ qubits.
This saving assumes a fixed active-orbital count and a certified cutoff $k$.
Many weakly correlated fragments can require high excitation ranks, and exponentiating a low-rank coupled-cluster operator can populate those ranks \cite{Bartlett2007coupledcluster}.
The error test below decides whether a proposed occupation cutoff is sufficient uniformly for the active-space problem.

\subsubsection{Excitation bands and tridiagonal response bounds}
\label{subsec:df-bands}
The electronic Hamiltonian contains at most two-body terms.
By the Slater--Condon rule, one application changes the minimum substitution distance from the active space by at most two.
Pairing adjacent distances therefore gives an exact block-tridiagonal Hamiltonian.
For a determinant $D$, let $d(D)$ be the minimum number of single-orbital substitutions needed to reach any determinant in the active space.
Set the active projector to $Q_0=P$.
The external bands contain the paired substitution distances $Q_i: d(D)\in\{2i-1,2i\} $ for $i\ge1$.
All bands are retained up to the largest possible distance; $L$ is their number.
If $P$ allows all configurations in the active orbitals and leaves all other orbitals empty, $d(D)$ is the number of electrons outside the active orbitals.
The general definition also includes holes in inactive occupied orbitals.
For a standard complete active space its tests are implemented by counting orbital occupations.

Write the diagonal and nearest-band couplings as $H_i=Q_iHQ_i$, $T_i=Q_iHQ_{i+1}$, and $T_0=PHQ_1$.
The many-electron Hamiltonian is exactly
\begin{align}
H=\begin{pmatrix} H_{\rm act}&T_0&0&0&\cdots
\\
T_0^\dagger&H_1&T_1&0&\cdots
\\
0&T_1^\dagger&H_2&T_2&\cdots
\\
0&0&T_2^\dagger&H_3&\ddots
\\
\vdots&\vdots&\vdots&\ddots&\ddots \end{pmatrix},
\quad H_{\rm act}=A.
\label{eq:df-tridiagonal-matrix}
\end{align}
Bands $1,2,3,\ldots$ group the excitation distances $(1,2),(3,4),(5,6),\ldots$.

To prove the zeros, the Slater--Condon rules \cite{Scemama2013SlaterCondon} imply that a matrix element of \cref{eq:df-electronic-H} vanishes between determinants differing by more than two substitutions.
Distance to a fixed set changes by at most the distance between its arguments.
Thus a nonzero matrix element has $|d(D)-d(D')|\le2$, whereas non-neighbouring bands are separated by at least three.
No excitation-rank truncation is made.
Only neighboring bands are coupled, giving $L$ off-diagonal block pairs.
Eliminating from the end updates only the next diagonal block, provided the pivots are invertible.
Higher-body terms produced by a prior effective-model transformation require a wider band construction; each band still needs a certified gap and its own integral-access cost.
Even the first band can involve every external orbital.
The useful joint partition intersects these occupation bands with the orbital-defined sectors above.
The outer selection-rule zeros survive; within each surviving block, the orbital subsets determine the data access and its cost.
Coarser sectors are preferable whenever finer resolution costs more than it saves.
Exact downfolding keeps all bands; the optional projector $\Pi_k$ removes high bands only with the tail bound below.

The sector response bounds determine the query cost of electronic simulation.
For any chosen orthogonal partition, write $Q_0=P$ and denote the external sectors by $Q_i$.
They may be orbital sectors, excitation bands, or their intersections.
Resolve $H-EQ=\sum_jV_j$ into components supported on one sector or one Hermitian pair of sectors.
If occupations are truncated, apply the same construction to $\Pi_kH\Pi_k$ on $\operatorname{ran}\Pi_k$.
Several components may share a support.
The controlled component encodings are $\Be{V_j/\lambda_j}$.
They have positive normalization and query cost $C_j\ge1$.
For a Toffoli budget, $C_j$ bounds the Toffoli count of a controlled call, including lookup, arithmetic, fermionic signs, sector tests, and representation changes.
Non-Clifford rotation synthesis is counted separately if not already included in the chosen gate cost.

Assume uniform response bounds on the active space, $\|Q_iX\|\le\varrho_i$, $\varrho_i>0$, and $\varrho_0=1$.
The dimensionless sensitivity is
\begin{align}
\chi_j=\begin{cases} \varrho_i^2,&V_j\text{ is supported in }Q_i,
\\
2\varrho_i\varrho_\ell,&V_j\text{ couples }Q_i\text{ and }Q_\ell, \end{cases}.
\label{eq:df-weights}
\end{align}
The factor two accounts for the two adjoint blocks at the stipulated oracle cost.
With the feedback-conditioning factors $\kappa_j$ required by \cref{thm:app-main}, the complete weighted query bound is
\begin{align}
\quad \sum_j C_jq_j=O\!\left[ |t|\left(\sum_j\sqrt{C_j\lambda_j \chi_j}\right)^2 +\log\frac1\epsilon\sum_j C_j\kappa_j \right].
\quad
\label{eq:df-maincost}
\end{align}
Here $q_j$ counts controlled oracle calls, including oblivious amplitude amplification.
This directly instantiates the preceding theorem for the chosen model and its proved response bounds.
The query delays have already been optimized up to a constant for the given bounds.
All bands are closed in one self-consistent network before one unitary Toeplitz lift; recursive Schur identities do not imply separately compiled solvers.
The conditioning sum remains included in the cost when correlation weights are small.
Taking $C_j$ to be a Toffoli cost bounds the oracle contribution.
The total also includes the known alignment, clock, and routing work in \cref{thm:app-gates}.
Optimize with those terms included if they are significant.
The cost of each actual integral call, together with data preparation and certification of the response bounds, enters this comparison.

A bound on the external relaxation operator controls the response uniformly over the active space.
For the correlation interpretation assume the full coupled external operator satisfies $K\succeq\Delta I_Q$ with a known lower bound $\Delta>0$.
Set $\Sigma=BK^{-1}B^\dagger$.
Completing the square at fixed active-space amplitude gives
\begin{subequations}
\label{eq:df-relaxation}
\begin{align}
u^\dagger(A-EI_P)u &-\min_{v\in\operatorname{ran}Q}(u+v)^\dagger(H-EI)(u+v) =u^\dagger\Sigma u,
\\
\Sigma&=X^\dagger KX,
\quad \Delta X^\dagger X\preceq\Sigma.
\end{align}
\end{subequations}
Thus $u^\dagger\Sigma u$ is the external relaxation contribution to this quadratic form.
The identity concerns the displayed quadratic form at fixed $u$ and $E$.
A uniform bound $0\preceq\Sigma\preceq\mathcal E I_P$ implies $\|X\|^2\le\mathcal E/\Delta$.
The quantity $\mathcal E$ bounds the largest contribution over all normalized active-space vectors.

Suppose the internal block $K$ is encoded as one component with normalization $\lambda_K$.
If the coupling normalization satisfies $\lambda_{\rm cpl}\le\beta\|B\|$, then \cref{cor:app-relaxation}\ gives
\begin{align}
\lambda_K \chi_K\le\frac{\lambda_K}{\Delta}\mathcal E,
\quad \lambda_{\rm cpl}\chi_{\rm cpl} \le2\beta\sqrt{\frac{\lambda_K}{\Delta}}\,\mathcal E.
\label{eq:df-energy-weights}
\end{align}
Here $\beta\ge1$ is the coupling normalization factor.
A large external scale can therefore have a small simulation weight when the normalization-to-gap ratio and $\beta$ are controlled.
Here $\mathcal E$ must bound the operator uniformly on the active space.
For an indefinite internal block, cancellations in the relaxation operator can coexist with a large response norm.
The simulation theorem still applies when the response bounds are given directly.

\cref{prop:df-chain-certificate} bounds the response across coupled excitation bands using a scalar tridiagonal comparison matrix.

\begin{proposition}[Scalar bounds for the coupled excitation chain]
\label{prop:df-chain-certificate}
In this proposition $Q_i,H_i,T_i$ refer to the excitation bands in \cref{eq:df-tridiagonal-matrix}.
Assume diagonal gap and coupling bounds $H_i-EI_{Q_i}\succeq d_iI_{Q_i}$, $d_i>0$, $\|T_i\|\le b_i$, and $b_i\ge0$.
The initial bound $b_0$ controls the active-to-first-band coupling.
Form the small $L\times L$ comparison matrix
\begin{align}
\mathsf K=\begin{pmatrix} d_1&-b_1&0&\cdots
\\
-b_1&d_2&-b_2&\cdots
\\
0&-b_2&d_3&\ddots
\\
\vdots&\vdots&\ddots&\ddots \end{pmatrix}.
\label{eq:df-comparison}
\end{align}
If $\mathsf K\succ0$, valid bounds are
\begin{align}
\Delta=\lambda_{\min}(\mathsf K),
\quad \|Q_iX\|\le b_0(\mathsf K^{-1})_{i1},
\quad \|\Sigma\|\le b_0^2(\mathsf K^{-1})_{11}.
\label{eq:df-chain-bounds}
\end{align}
An exactly disconnected tail can be removed; otherwise positive upper bounds on the responses are used in clock shaping.
\end{proposition}
\begin{proof}
Decompose a vector into its sector components and apply blockwise Cauchy--Schwarz:
\begin{align}
v=\sum_iv_i,
\quad z_i=\|v_i\|,
\quad v^\dagger Kv\ge z^\top\mathsf Kz.
\end{align}
This proves $K\succeq\Delta I_Q$.
Decompose the comparison matrix as $\mathsf K=D-W$ and $D=\operatorname{diag}(d_i)$.
The matrix $W$ has nonnegative nearest-neighbor entries $b_i$.
Positivity implies $\rho(D^{-1/2}WD^{-1/2})<1$ by Perron--Frobenius.
Expand the block inverse of $K$ about its block diagonal.
Bounding each path in its convergent Neumann expansion by the corresponding scalar path gives $\|Q_iK^{-1}Q_k\|\le(\mathsf K^{-1})_{ik}$.
Only $Q_1B^\dagger$ is nonzero, so multiplication by $B^\dagger$ proves the response bound.
Sandwiching the $(1,1)$ inverse block by $T_0$ proves the self-energy bound.
\end{proof}

Including the interband couplings gives the scalar pivots
\begin{align}
g_L=d_L,
\quad g_i=d_i-b_i^2/g_{i+1}
\quad(i=L-1,\ldots,1)
\label{eq:df-dressed-pivots}
\end{align}
which test $\mathsf K\succ0$ and give
\begin{align}
\mathcal E=\frac{b_0^2}{g_1},
\quad \|Q_iX\|\le\frac{b_0}{g_1} \prod_{k=2}^{i}\frac{b_{k-1}}{g_k}.
\label{eq:df-chain-product}
\end{align}
The product follows by back-substitution in the first column of $\mathsf K^{-1}$; the empty product for $i=1$ is one.
Ratios $b_{k-1}/g_k<1$ prove decreasing response with excitation distance, without calculating the virtual wavefunction.
The scalar pivots and first column take $O(L)$ classical arithmetic operations once the block bounds are given.
The $g_i$ must lower-bound the \emph{dressed} Schur pivots, including the effects of eliminated couplings.
A verified sum-of-squares identity on a band, $Q_i(H-EI)Q_i-d_iQ_i=\sum_\alpha F_{i\alpha}^\dagger F_{i\alpha}$, is one way to prove its lower bound.
Sum-of-squares methods yield rigorous energy lower bounds, although finding useful sum-of-squares bounds need not be inexpensive \cite{Hastings2024PerturbationTheory}.
Symmetry-aware electronic-structure constructions provide concrete examples~\cite{Rubin2026LowerBoundCertificates}.
Each certificate must still be applied to the relevant band and propagated through the eliminated couplings to bound a dressed Schur pivot.
Isolated-fragment bounds must include the remaining interactions before they bound these many-electron blocks.
The comparison test is sufficient; other certificates may establish invertibility when this test is inconclusive.

\cref{prop:df-pilot} converts a given coarse response and an operator-norm residual bound into rigorous response and correlation bounds.

\begin{proposition}[Response and correlation bounds from a coarse approximation]
\label{prop:df-pilot}
Let $\widehat X:P\to Q$ be a cheaply represented trial response and $R=K\widehat X+B^\dagger$ its operator residual.
If $K\succeq\Delta I_Q$, then
\begin{align}
\|Q_iX\|\le\|Q_i\widehat X\|+\frac{\|R\|}{\Delta}.
\label{eq:df-residual-response}
\end{align}
Define the Hermitian trial self-energy $\widehat\Sigma=-B\widehat X-\widehat X^\dagger B^\dagger -\widehat X^\dagger K\widehat X$.
Then
\begin{align}
\Sigma=\widehat\Sigma+R^\dagger K^{-1}R,
\quad \widehat\Sigma\preceq\Sigma\preceq \widehat\Sigma+\frac{\|R\|^2}{\Delta}I_P.
\label{eq:df-residual-energy}
\end{align}
Thus a known upper bound on $\lambda_{\max}(\widehat\Sigma)+\|R\|^2/\Delta$ gives a valid $\mathcal E$.
\end{proposition}
\begin{proof}
The response error solves the residual equation, $X-\widehat X=-K^{-1}R$ and $(X-\widehat X)^\dagger K(X-\widehat X)=R^\dagger K^{-1}R$.
Taking the norm of its component in each sector gives \cref{eq:df-residual-response}.
Expanding the quadratic expression yields \cref{eq:df-residual-energy}.
Its two inequalities follow from $0\preceq K^{-1}\preceq\Delta^{-1}I_Q$.
\end{proof}

The preliminary classical calculation need only provide useful operator bounds at the accuracy required by the simulation theorem.
For coarse internal data, assume $\bar K\succeq\bar\Delta I_Q$, $\|K-\bar K\|\le\eta_K<\bar\Delta$, and $\|B-\bar B\|\le\eta_B$.
Uniformly throughout the interpolation to the true data, use
\begin{align}
\Delta=\bar\Delta-\eta_K,
\quad \|R\|\le\|\bar K\widehat X+\bar B^\dagger\| +\eta_B+\eta_K\|\widehat X\|.
\label{eq:df-coarse-residual}
\end{align}
Using the coarse blocks in the trial self-energy adds at most $2\eta_B\|\widehat X\|+\eta_K\|\widehat X\|^2$ to its upper eigenvalue bound.
If the comparison matrix is available, a sharper sectorwise residual radius is $\sum_k(\mathsf K^{-1})_{ik}\|Q_kR\|$ instead of $\|R\|/\Delta$.
These bounds must hold on the full active space.
Establishing them, including any simultaneous statistical confidence guarantees, contributes to the classical preprocessing cost.
The trial $\widehat X=0$ recovers $\mathcal E\le\|B\|^2/\Delta$, so an initial bound needs no detailed virtual solution.

A variational upper energy and a proved ground-energy lower bound give $0\le E_{\rm var}-E_0\le E_{\rm var}-E_{\rm lb}$.
This bounds the ground-energy error.
Bounding $\|\Sigma\|$ on a multidimensional active space additionally requires an operator inequality.
An equivalent sufficient condition is
\begin{align}
\left(\begin{array}{c|c}\mathcal E I_P&B
\\
\hline
B^\dagger&K\end{array}\right)\succeq0,
\label{eq:df-selfenergy-PSD}
\end{align}
by the positive Schur-complement criterion.
\cref{prop:df-chain-certificate,prop:df-pilot} give sufficient ways to establish it, without assuming universally cheap norm estimation.

\subsubsection{Excitation tails and coarse-data optimization}
\label{subsec:df-coarse}
Removing high-excitation sectors requires an operator error bound on the resulting active-space evolution.
Small amplitude in one approximate ground state does not justify deleting a sector for arbitrary active-space inputs.
The following operator test controls the actual downfolded evolution and, unlike an exact Schur elimination, permits the discarded sector's quantum registers to disappear.

\cref{prop:df-tail} bounds the error caused by removing high-excitation sectors from the electronic model.

\begin{proposition}[Excitation-tail error]
\label{prop:df-tail}
Assume the full external operator satisfies $K\succeq\Delta I_Q$.
Let $H_{\rm eff}^{(k)}(E)$ be the exact downfolding of $\Pi_kH\Pi_k$ onto the original active space.
Extend its external response $\widehat X$ by zero on the omitted sectors.
The residual is $R=K\widehat X+B^\dagger$.
Then
\begin{subequations}
\label{eq:df-tail-error}
\begin{align}
H_{\rm eff}^{(k)}(E)-H_{\rm eff}(E) &=R^\dagger K^{-1}R,
\\
0\preceq H_{\rm eff}^{(k)}(E)-H_{\rm eff}(E) &\preceq\frac{\|R\|^2}{\Delta}I_P,
\\
\|e^{itH_{\rm eff}^{(k)}(E)}-e^{itH_{\rm eff}(E)}\| &\le |t|\,\frac{\|R\|^2}{\Delta}.
\end{align}
\end{subequations}
The residual vanishes on the included external space $\Pi_k-P$.
\end{proposition}
\begin{proof}
The compression of positive $K$ to $\Pi_k-P$ is invertible.
Its defining equation gives $(\Pi_k-P)(K\widehat X+B^\dagger)=0$ and $\widehat X^\dagger K\widehat X=-\widehat X^\dagger B^\dagger$.
Consequently, the trial self-energy in \cref{eq:df-residual-energy} is exactly $-B\widehat X=A-H_{\rm eff}^{(k)}(E)$.
That identity proves the first two lines of \cref{eq:df-tail-error}; the integral identity for the difference of two unitary evolutions proves the last line.
\end{proof}

The test need not construct $\widehat X$.
Place the cutoff $k=2\ell$ after excitation band $\ell<L$.
Only the coupling $T_\ell$ crosses it, so
\begin{align}
\|R\|\le b_\ell\|Q_\ell\widehat X\| \le b_\ell b_0 (\mathsf K_{\le\ell}^{-1})_{\ell1},
\label{eq:df-tail-comparison}
\end{align}
where $\mathsf K_{\le\ell}$ is the leading principal submatrix of the full comparison matrix in \cref{eq:df-comparison}.
The second inequality follows from \cref{prop:df-chain-certificate} applied to the included chain.
The full comparison matrix gives $\Delta$; its principal submatrix gives the truncated response bound.
Thus classical scalar bounds can justify a smaller quantum system space without preparing the omitted wavefunction.
Their construction still requires valid bounds on the original Hamiltonian.
For a cutoff at an unpaired rank, only the last two included ranks can connect to the tail, and \cref{prop:df-tail} applies unchanged.

A smaller quantum register requires removing the tail from the implemented model using the error bound above, or encoding its self-energy with a separately certified cost.
If the tail is solved inside the IIR, its oracle Hilbert spaces remain part of the circuit.

Spatial localization changes which orbital integrals a component query must access.
Spatial locality organizes orbital subsets and their integral access.
The occupation cutoff separately specifies the admissible excitation distances.
Localized orbitals separate short-range chemistry from distant electrostatics and polarization.
Nearsightedness supports controlled local approximations for suitable electronic observables, while distant interactions retain their screened or unscreened contributions \cite{Prodan2005Nearsightedness}.
A high-rank band can contain local or delocalized excitations, whose rank counts occupied-orbital changes rather than spatial separation.

For well-separated neutral fragments with fixed local charge, the leading fluctuating dipole interaction scales as $\ell^{-3}$, with $\ell$ their separation.
For gapped fragments its second-order nonretarded dispersion contribution scales as $\ell^{-6}$; self-consistent many-body polarization can modify a pairwise description \cite{Tkatchenko2012Dispersion}.
For a gapped, weakly coupled external channel, $\|B_\ell\|=O(\ell^{-3})$ and $K_\ell\succeq\Delta_\ell I$ imply
\begin{align}
\|K_\ell^{-1}B_\ell^\dagger\|=O(\ell^{-3}/\Delta_\ell),
\quad \|B_\ell K_\ell^{-1}B_\ell^\dagger\|=O(\ell^{-6}/\Delta_\ell).
\label{eq:df-spatial-response}
\end{align}
This can motivate a small-response estimate; the full environment must still satisfy the stated coupled response bound.
Collective modes, charge transfer, and the number of distant contributions must still be included.
Total dynamic correlation can be extensive even in neutral matter.
Classical static mean fields alone do not bound the remaining quantum response.

A useful active space includes the chemically active configurations and nearby or soft modes with large response.
Distant interactions can be grouped through multipole or other structured factorizations instead of assigning one controlled oracle to every microscopic pair.
Quantum fast-multipole constructions illustrate both the opportunity and the cost of coherent data access \cite{Berry2026FastMultipole}.
Approximate multipole, integral-factorization, and polarizable-embedding models enter with an explicit model-error bound \cite{Olsen2010PolarizableEmbedding,Lee2020hypercontraction}.
A screened interaction and an explicit polarization sector must be derived consistently to avoid double counting.

Orbital and spatial refinements may establish smaller response bounds than one worst-case bound for an excitation band, but introduce more component oracles and sector tests.
Configurations exciting several regions must be included exactly once, with their interregion couplings.
Use the weights for each actual sector pair, including pairs inside the same excitation band.
A useful partition must lower the total implementation cost, including the operations used within each block.

The response weights also determine how accurately each component must be represented.
At fixed $E$, differentiation gives
\begin{align}
dH_{\rm eff}= \begin{pmatrix}I_P
\\
X\end{pmatrix}^{\!\dagger} dH \begin{pmatrix}I_P
\\
X\end{pmatrix}.
\label{eq:df-sensitivity}
\end{align}
A diagonal perturbation is multiplied by two copies of its response; an intersector coupling by the responses on its two ends.
These are the weights in \cref{eq:df-weights}.
Suppose cheaper data $\widehat V_j$ have the same component supports and $\|\widehat V_j-V_j\|\le\lambda_j\eta_j$.
If the gap and response bounds hold throughout the interpolation, \cref{prop:app-data}\ gives
\begin{align}
\|\widehat H_{\rm eff}-H_{\rm eff}\| \le\sum_j\lambda_j\chi_j\eta_j,
\quad \|e^{it\widehat H_{\rm eff}}-e^{itH_{\rm eff}}\| \le |t|\sum_j\lambda_j\chi_j\eta_j.
\label{eq:df-coarsening}
\end{align}
This bounds the full finite change from the pilot to the proposed model.
\cref{eq:df-coarse-residual} gives a conservative bound along the whole interpolation path.
A derivative that vanishes at the preliminary model does not bound the error of removing a finite coupling.

Optimize the active space, occupation cutoff, and integral representation using their gap, residual, and sensitivity bounds to compare the resulting Toffoli and qubit costs.
Apply \cref{eq:df-coarsening} on the chosen occupation space and add its previously bounded tail error separately.
Exact regrouping has no model bias but requires a new cost calculation.
An approximate replacement is accepted only if its cumulative bound, including earlier changes, satisfies
\begin{align}
|t|\left(\frac{\|R\|^2}{\Delta} +\sum_j\lambda_j\chi_j\eta_j\right) +\epsilon_{\rm circ}\le\epsilon
\label{eq:df-error-budget}
\end{align}
and the total implementation cost improves in the resource being optimized.
Set the tail residual $R=0$ when no occupation truncation is made.
Here $\epsilon_{\rm circ}$ is the circuit error for the chosen coarse model.
Use its actual costs $C_j(\eta_j)$ in \cref{eq:df-maincost} with precision $\epsilon_{\rm circ}$; update its normalizations, response bounds, conditioning factors, and known gate work when they change.
Include derivation of the bounds and rebuilding of the data in this comparison.
Changing $P$ changes the target and requires a specified embedding and a new error bound for the approximation.

Model coarsening is counted once through \cref{eq:df-coarsening}.
Inaccurate implementation of a chosen oracle is a different error: its unitary error obeys the per-call accumulation budget of \cref{thm:app-gates}.
Small Schur sensitivity does not assign each circuit invocation the larger model-error tolerance.

Implementing the compact encoding requires reversible occupation tests and access to the component integrals.
\cref{eq:df-compact-qubits} bounds the system register.
With core, active, and virtual labels grouped contiguously, a constant-length fermionic monomial can be evaluated on the sorted particle--hole lists by membership tests, insertions or deletions, and parity counts for its fermionic sign.
Scanning the lists and active occupations gives an implementation with $O((n_{\rm act}+k+1)\log n_{\rm orb})$ reversible elementary gates for these index operations, with temporary copies of comparable width; the coefficient lookup and superposition over monomials are additional.
The signed partial permutation is completed reversibly on its valid domain.
During a two-body monomial allow up to $k+2$ holes and particles in scratch, and test the final cutoff: projecting after every ladder operator could incorrectly delete a transition whose final state is valid.
No $n_{\rm orb}$-bit occupation string is needed for this implementation.
Alternative configuration-interaction oracle constructions provide another route \cite{Babbush2016Fermions,Kirby2021CompactEncoding}.

A factorized oracle based on full orbital rotations may instead require a full occupation register or leave the low-excitation subspace during the rotation.
Its low Toffoli count cannot be combined with \cref{eq:df-compact-qubits} without a compatible compact implementation.
Block-encoding ancillas, reversible data lookup, and workspace are part of the choice of oracle.
More compact addressing can cost more gates; more parallel lookup can cost more scratch.
These gate and workspace costs are included in addition to the system-register dimension bound.

Applying \cref{thm:app-gates}, the sorted-list implementation has total workspace
\begin{align}
n_{\rm total}=n_{\rm act} +O\!\left(k\log n_{\rm orb}+\log(k+1)+b+s_{\rm supplied}\right).
\label{eq:df-total-qubits}
\end{align}
Here $b$ is the address qubit count of that theorem, including the clock, scheduled query positions, and oracle ancillas; $s_{\rm supplied}$ is the largest temporary oracle or alignment scratch.
The width $b$ is logarithmic in the represented address dimensions; their dependence on the physical orbital count follows from the chosen encoding.
A logarithmic total qubit count therefore requires a fixed active-orbital count and occupation cutoff.
The precision and scratch registers must also have logarithmic width.
Supplying an arbitrary dense state or oracle is not implicitly efficient.
Overlapping supports share one system register; independent simultaneously present subsystems still require their joint state space.
State preparation and requested readout are separate.

For fixed oracles and bounds, query complexity is linear in evolution time with the stated precision term, and gate complexity is linear up to the displayed routing and synthesis logarithms.
Any increase of the cutoff or data precision needed as $|t|$ grows must be substituted into these resource bounds.
The resource optimization is to choose orbital blocks and access circuits for their actual gate costs, and the smallest occupation space whose tail and data errors satisfy \cref{eq:df-error-budget}, subject to the available qubit budget.
The cost of establishing the bounds is included in preprocessing; the chosen partition and cutoff may depend on the system.

The energy-dependent effective Hamiltonian also gives a residual bound for an approximate eigenvalue.
If the eventual objective is a ground energy, choose an interval containing $E_0$ and below the spectrum of $QHQ$, with uniform operator bounds.
Differentiation gives
\begin{align}
\frac{d}{dE}\bigl(H_{\rm eff}(E)-EI_P\bigr) =-I_P-X(E)^\dagger X(E)\preceq-I_P.
\label{eq:df-energy-monotonicity}
\end{align}
Its lowest eigenvalue crosses zero at $E_0$, by Schur-complement inertia.
Suppose the coarse effective Hamiltonian has operator error at most $\eta$ throughout the interval.
If a verified solve produces an energy $\widehat E$ in that interval with residual $|\lambda_{\min}(\widehat H_{\rm eff}(\widehat E)-\widehat E I_P)| \le\epsilon_{\rm solve}$, then
\begin{align}
|\widehat E-E_0|\le\eta+\epsilon_{\rm solve}.
\label{eq:df-ground-energy}
\end{align}
Weyl's inequality bounds the true scalar residual by the right-hand side, while \cref{eq:df-energy-monotonicity} makes its decrease at least one per unit energy.
Thus \cref{eq:df-ground-energy} bounds the error of the computed energy; convergence of an iteration used to find it requires a separate argument.
Bracketing, eigenvalue estimation, and basis-model errors remain separately counted.

The virtual amplitudes and induced many-body Hamiltonian need not be stored explicitly.
Orbital structure reduces the cost of accessing their surviving interactions; occupation restrictions justified by operator bounds reduce the space in which those interactions are evaluated.
Both reductions are measured against the same active-space target.

\subsection{Kron reduction and graph learning}\label{subsec:kron-graph-learning}

A graph Laplacian penalizes disagreement between values on connected vertices.
For nonnegative edge weights its quadratic form is
\begin{align}
\boldsymbol{f}^\dagger L\boldsymbol{f} =\sum_{(u,v)}w_{uv}|f_u-f_v|^2.
\end{align}
Kron reduction \cite{Dorfler2013KronReduction} eliminates internal vertex values while preserving their effect on a chosen set of active, or boundary, vertices.
Solving the internal equations gives their optimal continuation from the active values; substitution produces the Schur complement on the active space.
We use $M_{\mathrm{eff}}$ for this effective learning matrix.
In MQSP this is exactly the feedback-elimination module already used for downfolding: graph-edge components supply the Hermitian loads, and closing the internal-vertex coordinates constructs $M_{\mathrm{eff}}$ before the final readout is chosen.
The quantum task is to estimate a linear prediction on that space for a fixed coarse graph model.
Classical error and response bounds are assumed given; they determine the query allocation and the permitted model bias.

Consider a graph on $N$ vertices, with positive semidefinite multiscale Laplacians and nonnegative coefficients: $L_\ell\succeq0$ and $t_\ell\ge0$.
Let $R$ map vertex values to observations, and let $\boldsymbol{y}$ contain the observed labels.
The observation weights satisfy $W\succeq0$, and the ridge parameter is $\mu>0$.
The regularized learning problem is
\begin{subequations}
\label{eq:kg-learning}
\begin{align}
\min_{\boldsymbol{f}}\;\mathcal J(\boldsymbol{f}) &=\frac12\sum_\ell t_\ell \boldsymbol{f}^\dagger L_\ell \boldsymbol{f} +\frac12\kgNorm{W^{1/2}(R \boldsymbol{f}-\boldsymbol{y})}^2+\frac\mu2\kgNorm{\boldsymbol{f}}^2,
\\
M \boldsymbol{f}&=\boldsymbol{b},
\quad M=\sum_\ell t_\ell L_\ell+R^\dagger WR+\mu I,
\quad \boldsymbol{b}=R^\dagger W \boldsymbol{y}.
\end{align}
\end{subequations}
For ordinary edge and vertex observations this can be written recordwise as
\begin{align}
M=\mu I+\sum_{d\in\mathcal D}A_d,
\quad \boldsymbol{b}=\sum_{d\in\mathcal D}\boldsymbol{b}_d,
\quad A_d=\omega_d \boldsymbol{a}_d\boldsymbol{a}_d^\dagger,
\quad \boldsymbol{b}_d=\omega_d y_d\boldsymbol{a}_d,
\quad\omega_d\ge0.
\label{eq:kg-records}
\end{align}
The edge and labelled-vertex records are \begin{align}
\begin{array}{lll} \text{edge }(u,v):&\boldsymbol{a}_d=\boldsymbol{e}_u-\boldsymbol{e}_v,&y_d=0,
\\
\text{labelled vertex }u:&\boldsymbol{a}_d=\boldsymbol{e}_u,&y_d=\text{observed label}. \end{array}
\end{align} More general positive data terms can also be used.
The graph has $N$ vertices and $|\mathcal D|$ records; these are different size parameters.

Fix a data partition $\Pi$.
Assign each group its representation level $h_G$.
Write
\begin{align}
\bar M=\mu I+\sum_{G\in\Pi}\bar A_G(h_G)\succeq\mu I,
\quad \bar{\boldsymbol{b}}=\sum_{G\in\Pi}\bar{\boldsymbol{b}}_G(h_G),
\quad \bar{\boldsymbol{f}}=\bar M^{-1}\bar{\boldsymbol{b}}.
\label{eq:kg-coarse}
\end{align}
Exact regrouping leaves $M,\boldsymbol{b}$ unchanged.
Rounding, replacing, or dropping records can change both $M$ and $\boldsymbol{b}$; the given estimates bound these changes.
The ridge is retained.
A choice with positive semidefinite representatives automatically satisfies the displayed lower bound.

Let $P,Q_1,\ldots,Q_h$ be orthogonal vertex or hierarchy projectors summing to the identity.
Write $Q_0=P$ and $Q=I-P$.
A data group need not coincide with a sector.
Choose $P$ to contain the original and coarse right-hand-side vectors and the normalized test vector $\Bz$.
Fix a known $\beta>0$ with $\kgNorm{\boldsymbol{b}},\kgNorm{\bar{\boldsymbol{b}}}\le\beta$ for the allowed choices.
Our main output is the bounded linear prediction
\begin{align}
s=\frac\mu\beta\kgRe(\Bz^\dagger \boldsymbol{f}),
\quad |s|\le1.
\label{eq:kg-score}
\end{align}
The scale $\mu/\beta$ is fixed across proposals.
A coordinate prediction uses $\Bz=\boldsymbol{e}_v$; an unscaled prediction is recovered by multiplying by $\beta/\mu$.
The distinct normalized overlap $r=\kgRe(\Bz^\dagger \boldsymbol{f}/\kgNorm{\boldsymbol{f}})$ is treated below.

Partitioning $M$ into active and internal variables gives
\begin{align}
M=\left(\begin{array}{c|c}A&B
\\
\hline
B^\dagger&K\end{array}\right),
\quad M_{\mathrm{eff}}=A-BK^{-1}B^\dagger,
\quad X=-K^{-1}B^\dagger,
\quad \boldsymbol{f}=\begin{pmatrix}I
\\
X\end{pmatrix}M_{\mathrm{eff}}^{-1}\boldsymbol{b}.
\label{eq:kg-schur}
\end{align}
For active vertices, $M_{\mathrm{eff}}$ is the regularized Kron reduction \cite{Dorfler2013KronReduction}.
No coupling between eliminated sectors has been discarded.
The coarse matrix has analogous blocks $\bar A,\bar B,\bar K$ and maps $\bar M_{\mathrm{eff}},\bar X$.
Since $M,\bar M\succeq\mu I$, their Schur complements are at least $\mu I$.

The correlation quantity
\begin{align}
\Sigma=BK^{-1}B^\dagger=X^\dagger KX
\label{eq:kg-correlation}
\end{align}
has a direct learning interpretation.
At fixed active-vertex labels $\boldsymbol{u}$, the decrease in the homogeneous quadratic objective from relaxing the internal labels is $\boldsymbol{u}^\dagger\Sigma \boldsymbol{u}/2$.
If $K\succeq\Delta I$, then $\kgNorm X^2\le\kgNorm\Sigma/\Delta$.
Large internal interaction strength need not imply large influence on the active-vertex prediction.

For the implemented coarse model, decompose $\bar M=\sum_e\bar M_e$ into supported diagonal blocks and Hermitian coupling pairs, including the ridge.
Each actual component has a given controlled block encoding
\begin{align}
O_e=\Be{\bar M_e/\lambda_e},
\quad \lambda_e>0,
\quad \text{query cost }C_e>0.
\label{eq:kg-oracles}
\end{align}
We use the cost model of \cref{def:resource-accounting}.
A group spanning several sectors is resolved into components $(G,i,j)$ before its normalization and circuit cost are specified.

Choose positive response bounds satisfying $\kgNorm{Q_i\bar X}\le\varrho_i$, $\varrho_i>0$, and $\varrho_0=1$.
If the support of component $e$ is one sector or one pair, respectively, define
\begin{align}
\chi_e=\begin{cases}\varrho_i^2,&\Pi_e=Q_i,
\\
2\varrho_i\varrho_j,&\Pi_e=Q_i+Q_j.\end{cases}
\label{eq:kg-weights}
\end{align}
The coupling is queried once; the factor two accounts for its adjoint blocks.
Both $C_e$ and $\lambda_e$ depend on the chosen partition and coarseness.
The $\chi_e$ bound the operator response used by clock shaping.

We apply the inverse block-encoding construction with known normalization of \cref{cor:app-inverse}.
It provides $\Be{(a/4)\bar M_{\mathrm{eff}}^{-1}}$ with known normalization, so a scalar matrix-element estimate recovers the prescribed linear score.
The application gives the component access, response bounds, and sufficient feedback bound required by that corollary.
\cref{thm:kg-main} translates these bounds into a cost for the classical prediction.

\begin{theorem}[Correlation-weighted coarse Kron prediction]
\label{thm:kg-main}
Assume the access above and a known bound $\bar M_{\mathrm{eff}}\succeq aI$ with $a\ge\mu$.
Scale the internal sectors and component normalizations by \begin{align}
T=P+\sum_{i\ge1}\varrho_iQ_i,
\quad T_Q=QTQ,
\\
\widetilde\lambda_e= \begin{cases} \lambda_e\varrho_i^2,&\text{diagonal support},
\\
\lambda_e\varrho_i\varrho_j,&\text{coupling support}. \end{cases}
\end{align} Suppose given $\kappa_e\ge1$ satisfy the sufficient feedback bound
\begin{align}
\sum_e\frac{\widetilde\lambda_e}{\kappa_e}Q\Pi_eQ \preceq T_Q\bar K T_Q.
\label{eq:kg-feedback}
\end{align}
Suppose the chosen coarse prediction $\bar s=(\mu/\beta) \kgRe(\Bz^\dagger\bar{\boldsymbol{f}})$ has bias bounded by $|s-\bar s|\le\epsilon/2$.
Define
\begin{subequations}
\label{eq:kg-cost}
\begin{align}
\kgSens(\Pi,h)&=\frac1a \left(\sum_e\sqrt{C_e\lambda_e \chi_e}\right)^2,
\\
\kgFb(\Pi,h)&=\sum_e C_e\kappa_e.
\end{align}
\end{subequations}
All quantities on the right are evaluated for the chosen $(\Pi,h)$.
Under these assumptions, the MQSP inverse block-encoding circuit and scalar amplitude estimation return an estimate $\widehat s$ satisfying $|\widehat s-s|\le\epsilon$.
The readout fails with probability at most $\vartheta$, at total cost
\begin{align}
C_{\rm total}=C_{\rm build}(\Pi,h) +\widetilde O\!\left[ \frac{\kgSens+\kgFb+C_b+C_z+C_{\rm known}}{\epsilon} \log\frac2\vartheta\right].
\label{eq:kg-total}
\end{align}
Here $C_b$ is the cost of preparing the normalized coarse right-hand-side vector, whose norm is given.
The cost $C_z$ prepares the normalized test vector.
The cost $C_{\rm known}$ is actual nonoracle work per reciprocal evaluation.
If the input-vector norm is estimated, its cost and propagated scalar error are included in $C_b$ and the readout error budget.
The quantum algorithm assumes the classical response, inverse-norm, and bias bounds; establishing them is an additional cost.
\end{theorem}

For a nonzero coarse right-hand-side vector, prepare $\bar{\boldsymbol{b}}/\kgNorm{\bar{\boldsymbol{b}}}$.
A zero right-hand side gives linear score zero without a solver call.
Response weights reduce the query counts, while each data-access circuit still has cost $C_e$ and the feedback contribution remains in the bound.
The constant-factor optimality concerns allocation within the stated response-and-feedback upper bound, with the algorithm and partition fixed.

For reference, if absolute component errors are at most $\eta_e$, the input-vector error is at most $\eta_b$, and the response bounds hold along the entire interpolation between models, a sufficient bias bound is
\begin{align}
|s-\bar s|\le\frac1\mu\sum_e \chi_e\eta_e+\frac{\eta_b}{\beta}.
\label{eq:kg-uniform-bias}
\end{align}
This permits coarser access to weakly coupled sectors when the stated classical bound is available.

\begin{proof}[Proof of \cref{thm:kg-main}]
Apply \cref{prop:app-positive} to the given coarse matrix and the rescaling $T$.
With $D_{\boldsymbol{r}}=\sum_e r_e\widetilde\lambda_e\Pi_e$, it gives
\begin{align}
\|D_{\boldsymbol{r}}^{1/2}(T\bar M T)^{-1}D_{\boldsymbol{r}}^{1/2}\| \le a^{-1}\sum_e r_e\lambda_e\chi_e+\max_e\kappa_e r_e.
\label{eq:kg-resolvent-bound}
\end{align}
The reciprocal construction in \cref{cor:app-inverse} therefore obtains $\Be{(a/4)\bar M_{\mathrm{eff}}^{-1}}$ to operator error $\eta$ with horizon
\begin{align}
L=O\!\left[\left(a^{-1}\sum_e r_e\lambda_e\chi_e+ \max_e\kappa_e r_e\right)\log\frac2\eta\right],
\quad q_e=O(L/r_e).
\label{eq:kg-horizon}
\end{align}
The capped square-root allocation in \cref{lem:app-allocation} gives
\begin{align}
\kgSens+\kgFb\le \inf_{r_e\in\mathbb N_{>0}} \left(a^{-1}\sum_e r_e\lambda_e\chi_e+\max_e\kappa_e r_e\right) \sum_e\frac{C_e}{r_e} \le4(\kgSens+\kgFb).
\label{eq:kg-delay-optimum}
\end{align}
This proves the weighted reciprocal cost.

For $\beta_h=\|\bar{\boldsymbol{b}}\|>0$, the matrix element of this encoding is
\begin{align}
A_h=\operatorname{Re}\left\langle\Bz, \frac a4\bar M_{\mathrm{eff}}^{-1}\frac{\bar{\boldsymbol{b}}}{\beta_h}\right\rangle,
\quad \bar s=\frac{4\mu\beta_h}{a\beta}A_h.
\end{align}
The known prefactor is at most four.
Amplitude estimation \cite{brassard2002quantum} on a Hadamard test determines the score $A_h$ to error $O(\epsilon)$.
It uses $O(\epsilon^{-1})$ controlled reciprocal calls.
Allocate constant fractions of the error to the inverse, preparations, and scalar estimation, and use $O(\log(2/\vartheta))$ repetitions for confidence.
Adding the given bias proves the claim.
The gate and qubit costs follow from \cref{thm:iir-gates} and are recorded in \cref{tab:resources-applications}.

The uniform bias bound follows directly from the Schur derivative in \cref{eq:app-sensitivity-identity} and the resolvent identity:
\begin{align}
\|M_{\mathrm{eff}}-\bar M_{\mathrm{eff}}\|\le\sum_e\chi_e\eta_e,
\quad \|M_{\mathrm{eff}}^{-1}\boldsymbol{b}-\bar M_{\mathrm{eff}}^{-1}\bar{\boldsymbol{b}}\| \le\mu^{-2}\beta\sum_e\chi_e\eta_e+\mu^{-1}\eta_b.
\end{align}
Multiplication by $\mu/\beta$ gives \cref{eq:kg-uniform-bias}.
Completing the internal quadratic form gives the interpretation of $\Sigma$ in \cref{eq:kg-correlation}.
\end{proof}

A normalized full-state overlap additionally requires an implementation of the extension $\bar V=(I,\bar X)^\top$.
Suppose the continuation map has a block encoding $\Be{\bar V/G_V}$ at cost $C_V$, with $G_V\ge\|\bar V\|$.
Assume a positive lower solution-norm bound $n_-\le\|\bar{\boldsymbol{f}}\|$.
The successful amplitude after state preparation, inversion, and extension is at least
\begin{align}
t_- =\frac{a n_-}{4\|\bar{\boldsymbol{b}}\|G_V}.
\label{eq:kg-success}
\end{align}
Estimating the successful-branch overlap and its amplitude to $O(t_-\epsilon)$, then dividing, determines $\operatorname{Re}\langle\Bz,\bar{\boldsymbol{f}}/\|\bar{\boldsymbol{f}}\|\rangle$ with evaluation cost
\begin{align}
\widetilde O\!\left[ \frac{\kgSens+\kgFb+C_V+C_b+C_z+C_{\rm known}} {t_-\epsilon}\log\frac2\vartheta\right].
\label{eq:kg-normalized-cost}
\end{align}
This statement assumes a separate given bias bound for the normalized target.
Its extra lifting and success-amplitude factors specify the resources beyond the active-vertex prediction theorem.

\section{Online multivariate quantum signal processing}
\label{sec:time-dependent-mqsp}

Online MQSP keeps the query-counting structure of MQSP while allowing the known system matrix and the available oracles to change with physical time.
At event $t$, a classical controller chooses the known unitary $S_t$ from the record already available and the quantum state then passes through the current oracle calls.
Past oracles need not remain available.
The resulting public response is a causal two-time kernel $H_{t,s}$, generally not a Toeplitz matrix.
The analytic variables introduced in \cref{sec:achievable} still count oracle occurrences and remain analytic coordinates; physical time $t$ is a separate index.
Thus time independence is a useful specialization of MQSP rather than part of its definition.

The exact clock factorization also survives this change.
For a general causal kernel the full input--output clock matrix is used, whereas a time-independent kernel reduces it to the diagonal sums of a Toeplitz lift.
This distinction lets the same framework process coherent observations as they arrive.
We use online quantum sensor fusion as an example, with a Kalman filter whose high-dimensional transport is queried quantum mechanically while a low-dimensional anisotropic covariance and gain are updated classically.
The filtered mean can then enter a later quantum simulation or overlap measurement.

Quantum Kalman algorithms based on block-encoded matrix arithmetic \cite{Shi2024QuantumKalman,Dehaghani2025QuantumLQG} provide related constructions for prediction, gain, and covariance updates.
Continuous-measurement Kalman filtering in atomic magnetometry \cite{Geremia2003QuantumMagnetometry} motivates the physical setting, while online Hamiltonian learning \cite{Granade2012HamiltonianLearning} can provide a classical calibration record.
Our output is a quantum state proportional to the prescribed final conditional mean; the access model and the low-dimensional covariance structure used for the explicit online algorithm are stated below.

At each event $j$, assume controlled access to the current oracle $O_{a,j}$ for every type scheduled at that event.
A self-inverse oracle uses the same operation in either direction; for a general oracle, inverse access must be given explicitly if a construction uses it.
After its event the oracle is unavailable.
In particular, the inverse of an entire past sweep and an outer amplitude-amplification loop require access beyond this online model.
The Kalman construction uses only current forward calls.

The controller chooses the known unitary $S_j$ from the external classical record available before event $j$.
A horizon-$N$ sweep has $N$ online layers.
Port delays $r_a$ reduce individual oracle counts as in \cref{eq:intro-delays}.
Query each oracle at positive multiples of its delay, and let the subsequent system matrix read private residue $j\bmod r_a$.
The feedback amplitude used at $j\ge r_a$ contains the oracle from event
\begin{align}
e_a(j)=r_a\lfloor j/r_a\rfloor,
\quad q_a=\lfloor(N-1)/r_a\rfloor.
\label{eq:iir-td-timestamp}
\end{align}
An absent call is the identity on that sector.
More general schedules count their actual calls.
Several oracle types scheduled in one layer still contribute their individual calls and their prescribed order.

\begin{figure}[htbp]
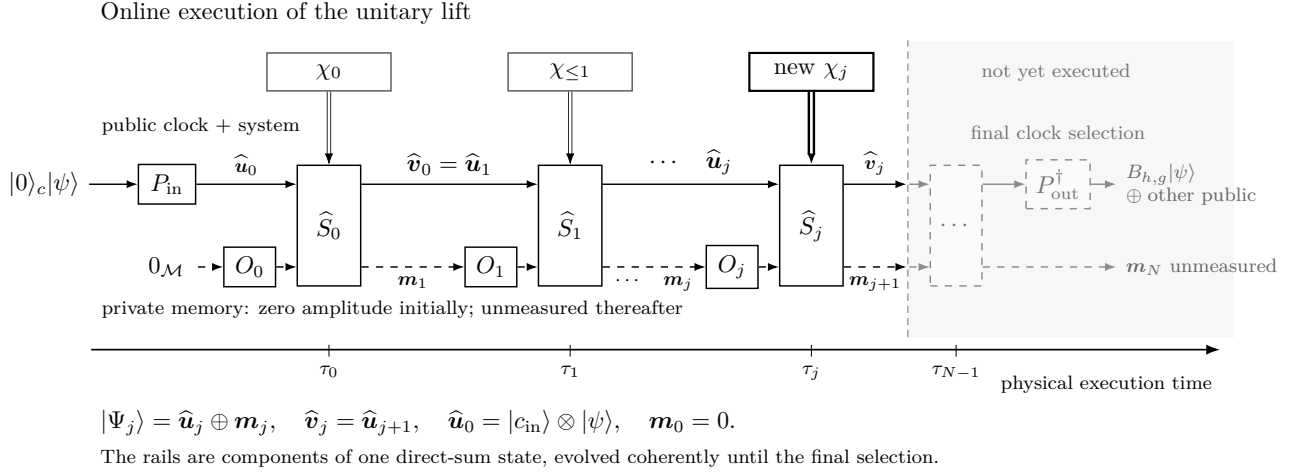

\centering\TDfit{\TDChronologyDiagram}
\caption{Online state flow.
The upper and lower branches are public and private direct-sum components of one state.
Each event applies its scheduled controlled oracles, followed by the embedded system matrix $\widehat S_j$.
The private input is initially $0_{\mathcal M}$.
Classical controls arrive before their system operation.
Clock preparation and final readout enclose the whole sweep.}
\label{fig:td-chronology}
\end{figure}

The public and private spaces $\mathcal P,\mathcal M$ include the registers required up to the chosen horizon, following \cref{def:resource-accounting}.
The initial private input is the zero vector $0_{\mathcal M}$, as in the feedback decomposition \cref{eq:intro-response}.
The input clock is prepared before the stream; the output clock is unprepared before the output projection.
The directed state flow in \cref{fig:td-chronology} preserves the entire quantum state between events.

The general lift in \cref{prop:iir-td-lift} and the clock theorem in \cref{cor:iir-td-sdp} apply before specializing to Kalman filtering.
In detail, \cref{prop:iir-td-lift} identifies the exact online kernel; \cref{cor:iir-td-sdp} extends clock factorization and perturbation bounds to that kernel; \cref{thm:qkf-online} gives an explicit clock over input times for the structured Kalman application.
Stability and output conditioning determine how much old history can be omitted and how accurately the circuit must be executed.

\subsection{Online unitary lift and clock shaping}
\label{subsec:td-lift}
Let $Q_j$ be the ordered sequence of queries scheduled at event $j$, extended by the identity on inactive memory sectors.
In the query-before-work convention, the complete system matrix is
\begin{align}
W_j=S_j(I_{\cP}\oplus Q_j) =\left(\begin{array}{c|c}A_j&B_j
\\
\hline
C_j&D_j\end{array}\right),
\quad \begin{pmatrix}\boldsymbol{v}_j
\\
\boldsymbol{m}_{j+1}\end{pmatrix} =W_j\begin{pmatrix}\boldsymbol{u}_j
\\
\boldsymbol{m}_j\end{pmatrix}.
\label{eq:iir-td-cell}
\end{align}
The divider separates public and private coordinates.
The blocks $A_j,B_j,C_j,D_j$ include the queries; $S_j$ is the known system matrix.
Additional interleaved known gates and queries are represented by refining a tick into ordered substeps.

To obtain the unitary lift for a fixed external record, embed $W_j$ on public slot $j$ and the common memory, acting identically on the other public slots.
Write the embedded operation as $\widehat W_j$ and define $\cP_N=\mathbb C^N\otimes\cP$ and $\mathcal W_N^{\rm td}=\widehat W_{N-1}\cdots\widehat W_0$.

The product specifies the actual circuit schedule:
\begin{subequations}
\label{eq:td-online-execution}
\begin{align}
\mathcal W_N^{\rm td} &=\bigl[\widehat S_{N-1}(I_{\cP_N}\oplus Q_{N-1})\bigr] \cdots\bigl[\widehat S_0(I_{\cP_N}\oplus Q_0)\bigr],
\\
\text{physical order:}
\quad &Q_0\longrightarrow\widehat S_0\longrightarrow Q_1\longrightarrow\widehat S_1\longrightarrow\cdots \longrightarrow Q_{N-1}\longrightarrow\widehat S_{N-1}.
\end{align}
\end{subequations}
The rightmost factor acts first, and an empty $Q_j$ is the identity.
Each system matrix need only be given before its own work tick.
Consequently, the circuit can continue when a new classical update arrives, using the state produced by the preceding operations.

Use the clock states $|c_{\rm in}\rangle,|c_{\rm out}\rangle$ and preparation unitaries $P_{\rm in},P_{\rm out}$ from \cref{eq:iir-clock-amplitudes}.
With amplitudes $g_s$ and $h_t$, the initial state and the state flow are
\begin{align}
|\Psi_0\rangle=(|c_{\rm in}\rangle\otimes|\psi\rangle)\oplus0_{\mathcal M},
\quad \binom{\widehat{\boldsymbol{v}}_j}{\boldsymbol{m}_{j+1}} =\widehat W_j\binom{\widehat{\boldsymbol{u}}_j}{\boldsymbol{m}_j},
\quad\widehat{\boldsymbol{u}}_{j+1}=\widehat{\boldsymbol{v}}_j.
\label{eq:td-zero-private-input}
\end{align}
Here $\boldsymbol{m}_0=0$, and each public slot receives $\boldsymbol{u}_s=g_s|\psi\rangle$.
Forward substitution gives the kernel in \cref{prop:iir-td-lift}, which determines exactly what the final clock measurement block-encodes.

\begin{proposition}[Causal lift and prefix consistency]
\label{prop:iir-td-lift}
The map $\mathcal W_N^{\rm td}$ is unitary on $(\mathbb C^N\otimes\cP)\oplus\cM$.
At each step it executes $Q_j$ followed by $\widehat S_j$.
The steps occur in increasing physical time as in \cref{eq:td-online-execution}.
Its public block, at zero initial private input, is
\begin{align}
\cK_N=[H_{t,s}]_{t,s=0}^{N-1},
\quad H_{t,s}=\begin{cases} 0,&t<s,
\\
A_t,&t=s,
\\
B_tD_{t-1}\cdots D_{s+1}C_s,&t>s. \end{cases}
\label{eq:iir-td-kernel}
\end{align}
If $E_N$ is the private-output block, then
\begin{align}
\cK_N^\dagger\cK_N+E_N^\dagger E_N=I,
\quad\norm{\cK_N}\le1.
\label{eq:td-boundary-energy}
\end{align}
Each shorter online kernel is exactly the corresponding leading principal block of the longer one.
In symbols, $\cK_N[0{:}n-1,0{:}n-1]=\cK_n $ for $n<N$.
Appending a system matrix therefore leaves every earlier public output slot unchanged.
No continuity or analyticity in $j$ is required.
\end{proposition}
\begin{proof}
Execute the factors in \cref{eq:td-online-execution} from right to left.
Each factor uses only the queries and known gates scheduled at that event, so lifting preserves their online order.
Starting with $\boldsymbol{m}_0=0$, forward substitution gives $\boldsymbol{m}_t=\sum_{s<t}D_{t-1}\cdots D_{s+1}C_s\boldsymbol{u}_s$.
The output equation in \cref{eq:iir-td-cell} gives the claimed blocks, with their online order unchanged.
Each embedding is unitary; restricting its product's isometry identity to public inputs gives \cref{eq:td-boundary-energy}.
The formula for $H_{t,s}$ uses only system matrices $s,\ldots,t$, proving prefix consistency.
\end{proof}

For example, the first four slots give the operator-valued matrix
\begin{align}
\cK_4=\begin{pmatrix} A_0&0&0&0
\\
B_1C_0&A_1&0&0
\\
B_2D_1C_0&B_2C_1&A_2&0
\\
B_3D_2D_1C_0&B_3D_2C_1&B_3C_2&A_3 \end{pmatrix}.
\label{eq:td-matrix}
\end{align}
The public block is lower triangular.
Its unitary completion includes the final memory.
For time-independent system matrices, the blocks depend only on $t-s$ and recover the Toeplitz lift.
More generally, the time-independent buffered schedule recovers the delayed Toeplitz kernel from \cref{thm:iir-lift}.

After the final tick, the clock unpreparation gives
\begin{align}
P_{\rm out}^\dagger\mathcal W_N^{\rm td}P_{\rm in} =\Be{B_{h,g}},
\quad B_{h,g}=\sum_{t,s}\overline h_tg_sH_{t,s}.
\label{eq:td-clock-readout}
\end{align}
This is \cref{eq:qudit-general-clock} applied to $\mathcal K_N$.
The continuation and input injections required for a recursive filter are included in its specified system matrices; \cref{thm:qkf-online} specializes this causal lift to the online Kalman construction.

The gate and qubit bounds are those of \cref{thm:iir-gates}.
The only change is that $NG_S$ and $q_aG_{O_a}$ become sums over the actual, possibly nonuniform, event costs; see \cref{tab:resources-applications}.
All statements condition on the external classical record.
A controller that measures the processed quantum state requires a measurement dilation with its records included in the resource count of \cref{def:resource-accounting}.

Use the clocks with two time registers from \cref{eq:qudit-general-clock}.
For an external record and oracle instance $\omega$, write $\mathcal K_N(\omega)=[H_{t,s}(\omega)]$ and
\begin{align}
\mathcal L_X[\mathcal K_N(\omega)] =\sum_{t,s}X_{t,s}H_{t,s}(\omega).
\label{eq:td-clock-contraction}
\end{align}
Because $H_{t,s}$ depends on the input and output times separately, \cref{cor:iir-td-sdp} applies the clock factorization to this two-time kernel and bounds the effect of approximating its entries.

\begin{theorem}[Input- and output-time clock selection and perturbation]
\label{cor:iir-td-sdp}
For a given causal lift, the normalized clock states just described realize exactly the relative overlap matrices $\nuc X\le\alpha_{\rm clk}$, selecting $\mathcal L_X[\mathcal K_N(\omega)]/\alpha_{\rm clk}$.
For fixed clocks common to a promise set, the least normalization for uniform error $\eta$ to $T(\omega)$ is
\begin{align}
\inf\left\{\alpha\ge1:\ \exists X,\ \nuc X\le\alpha, \quad\sup_\omega\norm{\mathcal L_X[\mathcal K_N(\omega)]-T(\omega)}\le\eta\right\}.
\label{eq:iir-td-frontier}
\end{align}
For any two block kernels $\cK,\widetilde{\cK}$,
\begin{align}
\norm{\cL_X(\cK)-\cL_X(\widetilde{\cK})} \le\nuc X\,\norm{\cK-\widetilde{\cK}}.
\label{eq:td-kernel-perturbation}
\end{align}
In particular, two full unitary system matrix sequences satisfy
\begin{align}
\norm{\mathcal L_X[\mathcal K_N(\omega)]-\mathcal L_X[\widetilde{\mathcal K}_N(\omega)]} \le\nuc X\min\left\{2,\sum_{j<N}\norm{W_j-\widetilde W_j}\right\}.
\label{eq:iir-td-drift}
\end{align}
An efficient algorithm additionally requires an explicit clock factorization and preparation circuit.
\end{theorem}
\begin{proof}
The clock characterization follows from \cref{thm:iir-clock-body,lem:clock-extraction}, whose factorization applies to rectangular time kernels.
Apply \cref{lem:clock-extraction} to $\cK-\widetilde{\cK}$ for \cref{eq:td-kernel-perturbation}.
Telescope the two products of unitary embeddings; the intervening norms are one.
Compression to the public block and the bound two for a difference of unitaries prove \cref{eq:iir-td-drift}.
\end{proof}

Upper-triangular entries of $X$ must remain available for nuclear-norm completion, even though they have zero causal response.
For a finite explicit promise, \cref{eq:iir-td-frontier} has the semidefinite form already given in the clock-norm section.
The frontier optimizes scalar clocks of the declared online lift, with its schedule and known operations fixed.

For online execution, each clock preparation can depend only on information available when it is executed.
One sufficient choice is a factorization computed before the first event from known envelopes, with uniform error and success bounds over all allowed records.
The output clock may use the record available at readout while preserving the already prepared input factor.
This causal requirement is additional to the nuclear-norm constraint.

For a given approximation $\overline{\cK}_N$ and known $X$, suppose
\begin{align}
\sup_\omega\norm{\cL_X(\overline{\cK}_N)-T(\omega)}\le\epsilon_{\rm des},
\quad \sup_\omega\norm{\cK_N-\overline{\cK}_N}\le\eta_N,
\quad\nuc X\le\alpha.
\label{eq:td-compilation-data}
\end{align}
The finite-contraction proof of \cref{thm:mqsp-master} then gives
\begin{align}
\Be{\widetilde T/\alpha},
\quad \norm{\widetilde T-T}\le \epsilon_{\rm des}+\nuc X\,\eta_N+\alpha\eta_{\rm syn},
\label{eq:td-master-error}
\end{align}
where $\eta_{\rm syn}$ bounds the additional full-circuit synthesis error.
Corrections may be included under the joint normalization bound in
\cref{thm:mqsp-master}
only when their common ordered query sequence respects the same online access schedule.
Any retrospective sweep requires replayable access and is counted.

To use the time-independent analytic clause, one must additionally establish a uniform comparison $\norm{\cK_N-T_N[G]}\le\eta_N$, or provide a two-time approximation bound.
This comparison concerns the ordered history, including its timestamps; estimates for an instantaneous time-independent system matrix alone leave it unresolved.

\subsection{Stability, drift, and regularity}
\label{subsec:td-stability}

The exact lift permits abrupt updates.
To approximate an update we need a bound on its error; to discard old inputs we need a bound on their propagation; to interpolate sampled data we need quantitative temporal regularity.

For drift or synthesis errors, the complete system matrix estimate follows from
\begin{align}
\norm{W_j-\widetilde W_j}\le \norm{S_j-\widetilde S_j} +\sum_{a:\,j\in\mathcal D_a}\norm{O_{a,j}-\widetilde O_{a,j}}.
\label{eq:td-cell-error}
\end{align}
For example, $\norm{W_{j+1}-W_j}\le v$ gives a worst-case error budget $vN(N-1)/2$ when every system matrix is replaced by the last one.
Smoothness of a signal block alone need not bound the drift of its unitary completion; \cref{eq:td-cell-error} concerns the actual operations.

\cref{prop:td-storage} bounds ordered memory products in a positive-definite metric.
Summing these bounds controls the tail omitted from the online kernel $\mathcal K_N$ and hence its truncation error.

\begin{proposition}[A bound on decay of the memory contribution]
\label{prop:td-storage}
Suppose positive-definite Hermitian matrices $P_j$ define the quadratic memory norm $\|\boldsymbol{m}\|_{P_j}^2=\boldsymbol{m}^\dagger P_j\boldsymbol{m}$, and satisfy
\begin{align}
D_j^\dagger P_{j+1}D_j\preceq\rho^2P_j,
\quad p_-I\preceq P_j\preceq p_+I, \qquad0<\rho<1.
\label{eq:td-storage}
\end{align}
Then, with $C_0=\sqrt{p_+/p_-}$,
\begin{align}
\norm{D_{t-1}\cdots D_s}\le C_0\rho^{t-s},
\quad \norm{H_{t,s}}\le C_0\rho^{t-s-1}
\quad(t>s).
\label{eq:td-product-stability}
\end{align}
Define the memory age operators and their generating function by
\begin{align}
K_\ell=\sum_{t\ge\ell}\ket t\bra{t-\ell}\otimes H_{t,t-\ell},
\quad \cK_N(z)=\sum_\ell z^\ell K_\ell.
\end{align}
Uniformly in $N$, one has
\begin{align}
\sup_{|z|=R}\norm{\cK_N(z)}\le M_R:=1+\frac{C_0R}{1-\rho R},\qquad1<R<\rho^{-1},
\quad \norm{\sum_{\ell\ge L}K_\ell}\le \frac{M_RR^{-L}}{1-R^{-1}}.
\label{eq:iir-td-tail}
\end{align}
\end{proposition}
\begin{proof}
Iterate the storage inequality and use the two bounds on $P_j$.
The blocks of a unitary have norm at most one, so \cref{eq:iir-td-kernel} gives the second estimate.
The different blocks on a single memory age diagonal have disjoint input and output slots, giving $\norm{K_\ell}\le C_0\rho^{\ell-1}$ for $\ell\ge1$.
Sum the geometric series for the circle bound, then use Cauchy's coefficient estimate and sum its tail.
\end{proof}

The frozen spectral radius is the largest modulus of an eigenvalue of one fixed $D_j$.
It describes repeated use of that fixed matrix, whereas memory follows the ordered product of different $D_j$.
For example, alternating the nilpotent contractions $\ket0\bra1$ and $\ket1\bra0$ gives ordered products of norm one.
The two-time product estimate controls the cumulative influence of old observations in an online filter.
The strict storage inequality is a sufficient condition on the memory dynamics to which it is applied.
It cannot hold on an unchanged coherently retained failure subspace, where the transition is the identity.
The retained failure subspace can be omitted from a transfer-function estimate when subsequent operations have zero matrix elements from that subspace to the successful output.
Its quantum state is still retained in the unitary circuit.
A terminal window must begin when its first retained input arrives; evaluating the same window retrospectively requires replay.

The variable $z$ in \cref{eq:iir-td-tail} counts memory age.
Analyticity in external time instead controls interpolation.
Rescale a physical-time interval to $[-1,1]$, and suppose the kernel coefficients have analytic continuations satisfying $H_{t,t-\ell}=B_\ell(\tau_t)$ and $\sup_{E_\chi}\norm{B_\ell}\le MR^{-\ell}$ on a common Bernstein ellipse $E_\chi$.
This ellipse is the image of a complex circle under the Joukowski map; \cref{tdreg:eq:ellipse-fit} explains how it fits inside a strip of analyticity in physical time.
Truncation at memory age $L-1$ and polynomial degree $p$ gives
\begin{align}
\norm{\cK_N-\widetilde{\cK}_N}\le \frac{MR^{-L}}{1-R^{-1}} +\frac{2M\chi^{-p}}{(\chi-1)(1-R^{-1})} =:E_{L,p}.
\label{eq:iir-td-analytic-compression}
\end{align}
The operator-valued Chebyshev coefficients are bounded by $2MR^{-\ell}\chi^{-k}$.
Summing over polynomial degree and then memory age proves the claim.
The additional operator-norm error before normalization satisfies $\|\mathcal L_X[\mathcal K_N]-\mathcal L_X[\widetilde{\mathcal K}_N]\| \le\nuc X E_{L,p}$.
For the block encoding in \cref{eq:td-master-error}, this contribution is divided by $\alpha$.
The required moments are $\sum_{t\ge\ell}X_{t,t-\ell}T_k(\tau_t)$; controlling their number does not automatically give a low-rank or cheaply preparable clock.

The actual finite row response is $F_t(z)=\sum_{s=0}^t z^{t-s}H_{t,s}$.
It is generally different from the frozen expression $A_t+zB_t(I-zD_t)^{-1}C_t$.
Also, memory decay does not imply a clock norm near one: any selected target $T$ obeys $\alpha\ge\norm T-\epsilon$, and a sum of many input contributions can have norm greater than one.
A feasible clock and its normalization remain separate hypotheses of \cref{eq:td-master-error}.

The additional certificates in \cref{app:td-proofs} give sufficient conditions for ordered memory stability and rank-two moment-corrected clocks with normalization close to one.
Their temporal windows must fit the same analytic domain.
The Kalman algorithm instead uses the weighted controllability Gramian in \cref{eq:qkf-gramian-recursion}; it requires no analyticity in physical time.

\subsection{Online quantum Kalman filtering}
\label{sec:online-qkf}

The desired object is the causal map from the initial mean and arriving observations to the current conditional mean.
Because the gains and transport change with the classical record, this response is generally not Toeplitz.
The modular principle of \cref{sec:modularity} nevertheless applies at each event: the current transport oracle is a direct query module, while the small classical gain and covariance data determine a known unitary junction that mixes the previous memory with the new observation.
The online lift then composes these event modules in physical order.
The clock over input times is chosen from the resulting causal kernel, so no time-independent transfer function or replay of earlier oracles is assumed.

\subsubsection{Model, access, and classical prior}
\label{subsec:qkf-model}

Consider a finite-dimensional linear model
\begin{align}
\boldsymbol{x}_t=A_t\boldsymbol{x}_{t-1}+\boldsymbol{\xi}_t,
\quad \boldsymbol{y}_t=C_t\boldsymbol{x}_t+\boldsymbol{\zeta}_t,
\quad 1\le t\le T,
\label{eq:qkf-model}
\end{align}
with independent zero-mean process and observation noises having covariances \(Q_t\) and \(R_t\), independent also of the initial state.
Complex notation is convenient for wave amplitudes; take proper complex Gaussian noises, or equivalently use the real and imaginary quadratures.
With a Gaussian initial prior, let \(\boldsymbol{\mu}_t,P_t\) be the conditional mean and error covariance after observation \(\boldsymbol{y}_t\).
The standard Kalman equations~\cite{Kalman1960LinearFiltering} are
\begin{align}
P_t^-&=A_tP_{t-1}A_t^\dagger+Q_t,
\quad V_t=C_tP_t^-C_t^\dagger+R_t,
\label{eq:qkf-covariance}
\\
K_t&=P_t^-C_t^\dagger V_t^{-1},
\quad \boldsymbol{\mu}_t=F_t\boldsymbol{\mu}_{t-1}+K_t\boldsymbol{y}_t,
\quad F_t=(I-K_tC_t)A_t,
\label{eq:qkf-mean}
\\
P_t&=(I-K_tC_t)P_t^-(I-K_tC_t)^\dagger+K_tR_tK_t^\dagger.
\label{eq:qkf-joseph}
\end{align}
The matrices $P_t,K_t$ are the covariance and Kalman gain, respectively.
The observation matrix is $C_t$.
A known affine control can be included as an additional input in \cref{eq:qkf-mean}.

An outer classical controller gives $Q_t,R_t,C_t$ and the prior before each update, using external calibration and past observations.
The theorem implements the resulting prescribed mean recurrence.
Its Kalman optimality interpretation assumes that these choices specify the correct conditional Gaussian model.
The quantum circuit does not estimate the noise covariances.

The evolution access consists of controlled short-time unitaries
\begin{align}
O_{j,t}=e^{-i\tau_jH_{j,t}},
\quad H_{j,t}=H_{j,t}^\dagger,
\label{eq:qkf-oracles}
\end{align}
at specified events \(t\in\mathcal D_j\).
The propagator \(U_t\) is a prescribed ordered product of the calls scheduled at event \(t\) and known controls.
No commutation among Hamiltonians, or across times, is assumed.
Query each oracle at positive multiples of its port delay.
Over ticks $1,\ldots,T$, the count is $q_j=\lfloor T/r_j\rfloor$.
This is a model of the scheduled dynamics.
Replacing continuous evolution by such a product or using a stale calibration is a separate approximation, bounded below.
In particular, the timestamp convention of \cref{eq:iir-td-timestamp} is retained.
The target conditions on these given propagators.
Marginalizing an unknown Hamiltonian-parameter distribution is generally a different, non-Gaussian filtering problem.

At its arrival time, an observation is given through a controlled preparation with a known phase convention and bound \(Y_t\):
\begin{align}
D_t|0\rangle =|0\rangle_{\rm ok}\frac{\boldsymbol{y}_t}{Y_t} +\boldsymbol{b}_{{\rm bad},t},
\quad (\langle0|_{\rm ok}\otimes I)\boldsymbol{b}_{{\rm bad},t}=0.
\label{eq:qkf-source}
\end{align}
We similarly have \(\boldsymbol{\mu}_0/M_0\), with known \(M_0>0\).
These are controlled unitary preparation oracles in the sense of \cref{def:resource-accounting}; their loading costs are included.
The circuit calls $D_t$ coherently on its arrival-time branches.
This access specifies the controlled action and relative phase in \cref{eq:qkf-source}, in addition to the prepared state.
Gaussian observations are unbounded; a fixed \(Y_t\) is therefore a promise on the processed records, or a bound on a chosen high-probability event whose failure probability must be reported separately.

\subsubsection{Structured anisotropic filtering with high-dimensional unitary transport}
\label{subsec:qkf-structured}

A fully dense covariance on the quantum mode space would itself be a large matrix that must be represented and updated.
We instead separate a small classical channel space from a high-dimensional quantum mode space.
Let the state space be $\mathbb C^d\otimes\mathbb C^N$ and the observation space be $\mathbb C^{d_y}\otimes\mathbb C^N$, with
\begin{align}
A_t=a_t\otimes U_t,
\quad C_t=c_t\otimes I_N,
\quad Q_t=q_t\otimes I_N,
\quad R_t=v_t\otimes I_N,
\quad P_0=p_0\otimes I_N .
\label{eq:qkf-structured-model}
\end{align}
The matrices $a_t,p_t,q_t\in\mathbb C^{d\times d}$ and $c_t\in\mathbb C^{d_y\times d}$ are classically represented, while $U_t$ acts on the $N$-dimensional mode register and is available only at event $t$.
The observation covariance $v_t\succ0$ is a $d_y\times d_y$ matrix.
No commutation is assumed among the $U_t$.

The Kalman covariance recursion stays inside this small matrix algebra.
If $P_{t-1}=p_{t-1}\otimes I_N$, define
\begin{subequations}
\label{eq:qkf-small-riccati}
\begin{align}
s_t&=a_tp_{t-1}a_t^\dagger+q_t,
\\
k_t&=s_tc_t^\dagger(c_ts_tc_t^\dagger+v_t)^{-1},
\quad f_t=(I-k_tc_t)a_t,
\\
p_t&=(I-k_tc_t)s_t(I-k_tc_t)^\dagger+k_tv_tk_t^\dagger .
\end{align}
\end{subequations}
Then $P_t=p_t\otimes I_N$, $K_t=k_t\otimes I_N$, and the physical mean obeys
\begin{align}
\vect\mu_t=(f_t\otimes U_t)\vect\mu_{t-1} +(k_t\otimes I_N)\vect y_t .
\label{eq:qkf-structured-mean}
\end{align}
Thus anisotropy is retained among the $d$ classical channels without explicitly storing the $(dN)\times(dN)$ covariance.
The persistent covariance $p_t$ and the Gramian $g_t$ below require $O(d^2)$ classical numbers.
Including the current dense model matrices and working storage requires $O(d^2+dd_y+d_y^2)$ numbers, and a conventional dense update uses $O(d^3+d^2d_y+dd_y^2+d_y^3)$ classical arithmetic operations per event.
These bounds exclude storage of the classical record and clock weights.
They count scalar values and arithmetic operations; bit complexity also depends on numerical precision and conditioning.
When $d_y=O(d)$, the current-event storage and arithmetic bounds reduce to $O(d^2)$ and $O(d^3)$.

To turn \cref{eq:qkf-structured-mean} into an online unitary circuit, assign fixed positive probabilities $\pi_0,\ldots,\pi_T$ to the input times before the stream begins, with $\sum_{s=0}^T\pi_s=1$.
Let $M_0\ge\|\vect\mu_0\|$ and $Y_t\ge\|\vect y_t\|$ be known preparation bounds.
The previous memory and new observation have different gains, so we rescale the memory to make their combined update a coisometry.
The weighted controllability Gramian gives this rescaling and a small unitary completion; a singular Gramian uses its support as detailed below.
Define the $d\times d$ matrices
\begin{align}
g_0=\frac{M_0^2}{\pi_0}I_d,
\quad g_t=f_tg_{t-1}f_t^\dagger +\frac{Y_t^2}{\pi_t}k_tk_t^\dagger .
\label{eq:qkf-gramian-recursion}
\end{align}
We call $g_t$ the weighted controllability Gramian.
The matrices may be computed from the actual small gains as they arrive; only the probabilities $\pi_s$ are fixed in advance.
Here $g_t^{-1/2}$ denotes the inverse square root on the known support of $g_t$, extended by zero on its orthogonal complement.
The proof includes the additional zero-input directions needed when this support is proper.

\begin{theorem}[Forward-only structured anisotropic Kalman filtering]
\label{thm:qkf-online}
Under \cref{eq:qkf-source,eq:qkf-structured-model}, let the small covariance and gain be updated by \cref{eq:qkf-small-riccati}, and let $g_t$ satisfy \cref{eq:qkf-gramian-recursion}.
If $g_T\ne0$, there is an online unitary circuit whose selected output is
\begin{align}
(\langle0|_{\rm ok}\otimes I)U_{\rm alg}|0\rangle =\frac{\vect\mu_T}{Z_T},
\quad Z_T=\sqrt{\lambda_{\max}(g_T)} .
\label{eq:qkf-output}
\end{align}
For $\vect\mu_T\ne0$, the conditional state and success probability are
\begin{align}
|\mu_T\rangle=\frac{\vect\mu_T}{\|\vect\mu_T\|},
\quad p_{\rm succ}=\frac{\|\vect\mu_T\|^2}{\lambda_{\max}(g_T)} =\chi_T^{-2},
\quad \chi_T=\frac{Z_T}{\|\vect\mu_T\|}.
\label{eq:qkf-success}
\end{align}
If $g_T=0$, the target mean is zero for every allowed input and the circuit instead reports failure; no normalized mean state is requested in this case.
One execution uses each scheduled $U_t$ once and each nonzero-weight observation preparation once.
It uses no historical inverse and no intermediate postselection.
Let $w$ count the mode register and all supplied oracle and synthesis work registers, but exclude channel and event-address registers.
The quantum space is $w+O(\log(d+d_y)+\log(T+1))$ qubits.
A completion chosen from the required rows uses $O(d(d+d_y))$ known two-level rotations per event, with $O(d^2)$ additional rotations for the final readout.
Synthesizing an arbitrary dense completion instead gives the sufficient bound $O((d+d_y)^2)$ per event.
Clock loading, addressing, routing, the given oracle costs, and synthesis into the chosen elementary gate set are additional.
No relation between $d_y$ and $d$ is assumed.
\end{theorem}

The normalization has a useful closed form.
With $Y_0=M_0$, define
\begin{align}
h_{T0}=f_T\cdots f_1,
\quad h_{Ts}=f_T\cdots f_{s+1}k_s
\quad(1\le s\le T).
\label{eq:qkf-source-propagators}
\end{align}
Expanding \cref{eq:qkf-gramian-recursion} gives
\begin{align}
g_T=\sum_{s=0}^T\frac{Y_s^2}{\pi_s}h_{Ts}h_{Ts}^\dagger .
\label{eq:qkf-terminal-gramian}
\end{align}
When the small model is known before execution, the infimum of the squared preparation scale is the convex optimization value
\begin{align}
\inf_{\pi_s>0,\ \sum_s\pi_s=1} \lambda_{\max}\!\left( \sum_{s=0}^T\frac{Y_s^2}{\pi_s}h_{Ts}h_{Ts}^\dagger \right).
\label{eq:qkf-gramian-clock}
\end{align}
The infimum need not be attained if a response block vanishes.
Every implemented clock uses a chosen strictly positive feasible probability vector.
If future gains are revealed only online, choose the $\pi_s$ from known envelopes.
In particular, known positive bounds $\beta_s\ge Y_s\|h_{Ts}\|$ give the feasible choice $\pi_s=\beta_s/\sum_u\beta_u$ and
\begin{align}
Z_T\le\sum_{s=0}^T\beta_s .
\label{eq:qkf-gramian-envelope}
\end{align}
For this feasible choice the Gramian scale is no larger than the scalar envelope.
Optimizing over the probabilities can further exploit anisotropy among the small channel directions.

The isotropic full-state model is the special case $d=d_y=1$, $a_t=c_t=1$, $q_t\ge0$, and $v_t>0$.
Then
\begin{align}
s_t=p_{t-1}+q_t,
\quad k_t=\frac{s_t}{s_t+v_t},
\quad p_t=\frac{s_tv_t}{s_t+v_t},
\quad \vect\mu_t=(1-k_t)U_t\vect\mu_{t-1}+k_t\vect y_t .
\label{eq:qkf-scalar-prior}
\end{align}
Thus the earlier scalar-gain filter is recovered without changing the online query model.

\subsubsection{Implementation and readout}
\begin{proof}[Proof of \cref{thm:qkf-online}]
The tensor structure in \cref{eq:qkf-structured-model} gives $P_t=p_t\otimes I_N$ and $K_t=k_t\otimes I_N$ by direct substitution in \cref{eq:qkf-covariance,eq:qkf-joseph}; hence the mean recursion is \cref{eq:qkf-structured-mean}.

Allocate one persistent memory sector and separate sectors for future observations and retained failure outputs.
Before the stream, prepare the input-time amplitudes $\sqrt{\pi_s}$.
The initial preparation places $\sqrt{\pi_0}\,\vect\mu_0/M_0$ in the memory signal subspace.
At event $t$, the current observation preparation places $\sqrt{\pi_t}\,\vect y_t/Y_t$ in that event's observation signal subspace.
Preparation failures remain in orthogonal subspaces on which subsequent filter updates act trivially.

From \cref{eq:qkf-gramian-recursion}, the $d\times(d+d_y)$ row
\begin{align}
\mathcal R_t^{(g)}= \left[ g_t^{-1/2}f_tg_{t-1}^{1/2}\ ,\ \frac{Y_t}{\sqrt{\pi_t}}g_t^{-1/2}k_t \right]
\label{eq:qkf-gramian-row}
\end{align}
satisfies $\mathcal R_t^{(g)}\mathcal R_t^{(g)\dagger}=I_d$ when $g_t\succ0$.
Extend it to the first $d$ rows of a known $(d+d_y)\times(d+d_y)$ unitary $\mathcal S_t$.
If $g_t$ is singular, let $\Pi_t$ be its support projector.
Then $\mathcal R_t^{(g)}\mathcal R_t^{(g)\dagger}=\Pi_t$; append the known columns $I_d-\Pi_t$ on a fresh auxiliary input subspace with zero amplitude.
The enlarged row has product with its adjoint equal to $I_d$ and has a unitary completion of dimension $2d+d_y$.
All such auxiliary inputs and unused output subspaces are included in the preallocated direct sum.

Apply $U_t$ to the mode register of the persistent memory before applying $\mathcal S_t\otimes I_N$ to the memory, current observation, and any auxiliary input.
The observation and auxiliary inputs bypass $U_t$.
The first $dN$ outputs become the persistent memory for the next event.
Retain every other output in an orthogonal subspace and do not feed it back into that memory.
All input-time contributions interfere coherently in this fixed $dN$-dimensional memory sector through the event unitaries.

Inductively, the memory amplitude after event $t$ is
\begin{align}
(g_t^{-1/2}\otimes I_N)\vect\mu_t .
\label{eq:qkf-gramian-memory}
\end{align}
The base case follows from $g_0=M_0^2I_d/\pi_0$.
The mean at each stage lies in the support of $g_t\otimes I_N$: this follows inductively from the positive-semidefinite sum in \cref{eq:qkf-gramian-recursion}.
Consequently, the memory contribution is $(g_t^{-1/2}f_t\otimes U_t)\vect\mu_{t-1}$ even with the stated pseudoinverse convention.
The observation contribution is $(g_t^{-1/2}k_t\otimes I_N)\vect y_t$, since its amplitude $\sqrt{\pi_t}/Y_t$ cancels the corresponding coefficient in \cref{eq:qkf-gramian-row}.
This proves the induction.

If $g_T\ne0$, apply a unitary extension of the known contraction $g_T^{1/2}/Z_T$, with $Z_T=\sqrt{\lambda_{\max}(g_T)}$, to the persistent memory.
Select only its successful output in that fixed memory sector.
Its channel and mode amplitudes are $\vect\mu_T/Z_T$, proving \cref{eq:qkf-output,eq:qkf-success}.
Any necessary fixed output-sector relabeling is a known routing operation.
The event unitaries have already combined the input-time contributions, so the final extraction acts only on the persistent memory.
If $g_T=0$, the support property implies $\vect\mu_T=0$, and a classical control sets the output flag to failure.

Each event uses one current transport call and one current nonzero-weight observation preparation, with all failure amplitudes retained.
No past oracle is replayed.
Binary addresses for the persistent memory, future inputs, and retained failure sectors require $O(\log(d+d_y)+\log(T+1))$ qubits in addition to the $w$ mode and work qubits specified in the theorem.

The prescribed row has $d$ orthonormal rows and at most $2d+d_y$ columns.
Successive right-acting Givens rotations eliminate its undesired entries row by row without changing rows already reduced.
There are at most $d(2d+d_y)$ entries to eliminate.
Reversing this sequence gives a completion with $O(d(d+d_y))$ two-level rotations, together with $O(d)$ phases.
The final contraction has a $2d$-dimensional unitary extension and requires $O(d^2)$ additional two-level rotations.
These operations act on the small channel subspaces and are tensored with $I_N$; they do not synthesize a generic unitary on all channel and mode coordinates.
\end{proof}

Iterating \cref{eq:qkf-gramian-recursion} proves \cref{eq:qkf-terminal-gramian}.
The objective in \cref{eq:qkf-gramian-clock} is convex because $\pi\mapsto A/\pi$ is matrix convex for $A\succeq0$ and $\lambda_{\max}$ is convex and monotone in the positive-semidefinite order.
For the envelope choice in \cref{eq:qkf-gramian-envelope},
\begin{align}
\frac{Y_s^2}{\pi_s}h_{Ts}h_{Ts}^\dagger \preceq \left(\sum_u\beta_u\right)\beta_s I_d ,
\label{eq:qkf-envelope-proof}
\end{align}
and summing over $s$ gives $g_T\preceq(\sum_s\beta_s)^2I_d$.
Thus the Gramian normalization is never worse than the corresponding scalar bound on the input contributions.

The probabilities assigned to input times are part of the input clock and are fixed before execution.
The small matrices $f_t,k_t,g_t$, by contrast, can be computed from the classical record available at event $t$.
If the complete small model is known in advance, \cref{eq:qkf-gramian-clock} optimizes the clock.
If only envelopes are known, \cref{eq:qkf-gramian-envelope} gives a uniform feasible choice.
In both cases the clock itself has unit norm; the separate scale $Z_T$ determines the selected mean amplitude.

\subsubsection{General conditioning and stability}
\label{subsec:qkf-general}

Once the gains are fixed, \cref{eq:qkf-mean} is a time-dependent IIR.
Its impulse kernel is
\begin{align}
\mathcal U(t,s)=F_tF_{t-1}\cdots F_{s+1},
\quad \mathcal U(s,s)=I,
\quad \boldsymbol{\mu}_T=\mathcal U(T,0)\boldsymbol{\mu}_0+\sum_{s=1}^T\mathcal U(T,s)K_s\boldsymbol{y}_s.
\label{eq:qkf-variation-constants}
\end{align}
There are three distinct conditioning parameters: the inverse innovation scale, the decay and transient growth of \(\mathcal U\), and the norm of the final mean relative to its preparation normalization.

\begin{lemma}[Innovation conditioning]
\label{lem:qkf-gain}
Compare two gain constructions.
Their predicted covariance norms are at most $p$, and their observation-matrix norms are at most $c$.
Each innovation matrix is bounded below by $\lambda I$, for $\lambda>0$.
Writing their differences as \(\delta P^-,\delta C,\delta R\) in operator norm, their gains satisfy
\begin{align}
\delta K\le \frac{c\,\delta P^-+p\,\delta C}{\lambda} +\frac{pc}{\lambda^2} \bigl(c^2\delta P^-+2pc\,\delta C+\delta R\bigr).
\label{eq:qkf-gain-perturbation}
\end{align}
\end{lemma}
\begin{proof}
The numerator and innovation perturbations are bounded by $\delta(P^-C^\dagger)\le c\delta P^-+p\delta C$ and $\delta V\le c^2\delta P^-+2pc\delta C+\delta R$.
Apply
\(V^{-1}-\widetilde V^{-1}=V^{-1}(\widetilde V-V)\widetilde V^{-1}\)
and the triangle inequality to the gain product.
\end{proof}

An upper innovation bound $V_t\preceq\Lambda I$ gives condition number at most $\Lambda/\lambda$.
Absolute perturbation bounds also depend on the lower spectral scale itself.
Likewise, if both dynamics and covariances are bounded by \(a,p\), prediction gives \(\delta P_t^-\le a^2\delta P_{t-1}+2pa\,\delta A_t+\delta Q_t\).
These estimates require no commutation.
These bounds control local sensitivity once the matrix descriptions are available; obtaining those descriptions has its own cost.

Stability of the Kalman filter does not require each mean update to contract the Euclidean norm: the covariance may rotate or contract at different rates in different directions.
Expressing the mean in covariance-whitened coordinates uses the covariance as a time-dependent metric.
For the whitened blocks defined in \cref{eq:qkf-whitened}, the covariance identity gives $\mathsf D_t\mathsf D_t^\dagger+\mathsf B_t\mathsf B_t^\dagger =I-P_t^{-1/2}(I-K_tC_t)Q_t(I-K_tC_t)^\dagger P_t^{-1/2} \preceq I$.
The term subtracted on the right is the whitened process-noise covariance.
Including a factor of that term as an additional input block would give an equality with the identity.
For the observation-input block alone, the inequality suffices for the unitary completion in \cref{prop:qkf-storage}.

\begin{proposition}[Covariance storage and a unitary realization]
\label{prop:qkf-storage}
For \(P_t\succ0\) and \(R_t\succ0\), define
\begin{align}
N_t=(I-K_tC_t)Q_t(I-K_tC_t)^\dagger+K_tR_tK_t^\dagger,
\quad \mathsf D_t=P_t^{-1/2}F_tP_{t-1}^{1/2}, \mathsf B_t=P_t^{-1/2}K_tR_t^{1/2}.
\label{eq:qkf-whitened}
\end{align}
Then \(P_t=F_tP_{t-1}F_t^\dagger+N_t\) and the row \(\mathcal R_t=[\mathsf D_t\ \mathsf B_t]\) is contractive.
It therefore has the unitary completion
\begin{align}
\mathcal J_t= \left(\begin{array}{c|c} \mathcal R_t&(I-\mathcal R_t\mathcal R_t^\dagger)^{1/2}
\\
\hline
(I-\mathcal R_t^\dagger\mathcal R_t)^{1/2}&-\mathcal R_t^\dagger \end{array}\right).
\label{eq:qkf-julia}
\end{align}
The first output block maps the whitened memory and observation to the next whitened mean:
\begin{align}
\left(P_{t-1}^{-1/2}\boldsymbol{\mu}_{t-1},\ R_t^{-1/2}\boldsymbol{y}_t,\ 0\right) \longmapsto P_t^{-1/2}\boldsymbol{\mu}_t.
\end{align}
The zero denotes the additional input subspace of the unitary completion.
If, uniformly,
\begin{align}
p_-I\preceq P_t\preceq p_+I,
\quad N_t\succeq\eta P_t,
\quad 0<\eta<1,
\label{eq:qkf-storage-certificate}
\end{align}
then
\begin{align}
\|\mathcal U(t,s)\|\le C\rho^{t-s},
\quad C=\sqrt{p_+/p_-},
\quad \rho=\sqrt{1-\eta}<1.
\label{eq:qkf-stability}
\end{align}
\end{proposition}
\begin{proof}
Expand \cref{eq:qkf-joseph}.
Whitening the resulting covariance identity shows \(\mathsf D_t\mathsf D_t^\dagger+ \mathsf B_t\mathsf B_t^\dagger\preceq I\), proving row contractivity.
A singular-value decomposition verifies \cref{eq:qkf-julia}.
Under \cref{eq:qkf-storage-certificate}, \(\mathsf D_t\mathsf D_t^\dagger\preceq(1-\eta)I\).
The covariance square roots telescope through a product of the \(\mathsf D_t\), giving \cref{eq:qkf-stability}.
\end{proof}

This gives a unitary realization of a general Kalman update once the covariance factors and their inverses can be implemented.
For a dense high-dimensional covariance, constructing those factors can be as difficult as representing the covariance itself.
The tensor model in \cref{subsec:qkf-structured} avoids this problem: the covariance remains $p_t\otimes I_N$, so the whitening and the weighted controllability Gramian are small classical matrices while the high-dimensional transport remains the queried unitary $U_t$.

\begin{theorem}[Stability-weighted model error]
\label{thm:qkf-robust}
Compare \cref{eq:qkf-mean} with \(\widetilde{\boldsymbol{\mu}}_t=\widetilde F_t\widetilde{\boldsymbol{\mu}}_{t-1} +\widetilde K_t\widetilde{\boldsymbol{y}}_t\).
Suppose the implemented propagators obey \(\|\widetilde{\mathcal U}(t,s)\|\le C\rho^{t-s}\).
With \(e_0=\|\widetilde{\boldsymbol{\mu}}_0-\boldsymbol{\mu}_0\|\),
\begin{align}
\|\widetilde{\boldsymbol{\mu}}_T-\boldsymbol{\mu}_T\| \le C\rho^Te_0+C\sum_{t=1}^T\rho^{T-t} \bigl(&\delta F_t\|\boldsymbol{\mu}_{t-1}\|+\delta K_t\|\boldsymbol{y}_t\| +\|\widetilde K_t\|\delta_{y,t}\bigr).
\label{eq:qkf-general-error}
\end{align}
The same estimate holds with individual product-norm bounds in place of \(C\rho^{T-t}\).
\end{theorem}
\begin{proof}
Subtract the two updates.
Propagate the error using \(\widetilde F_t\); the forcing is \((\widetilde F_t-F_t)\boldsymbol{\mu}_{t-1} +(\widetilde K_t-K_t)\boldsymbol{y}_t+\widetilde K_t(\widetilde{\boldsymbol{y}}_t-\boldsymbol{y}_t)\).
Iterating and taking norms proves the claim.
\end{proof}

For example,
\(\delta F_t\le(1+\widetilde k\widetilde c)\delta A_t +a(c\delta K_t+\widetilde k\delta C_t)\)
under the corresponding norm bounds.
Innovation conditioning and ordered stability thus enter separately.
Stable eigenvalues of each frozen update do not alone imply \cref{eq:qkf-stability}, as the switching example in \cref{sec:time-dependent-mqsp} shows.

\subsubsection{Online access and independent online copies}
\label{subsec:qkf-live-copies}

To raise success probability under the online access model \cref{prop:qkf-live-copies} initializes independent online circuit copies before the input stream begins and queries each at every scheduled event.
Replay-based amplification would require access to past oracles.
Each copy is a separate tensor factor, with its own input-time preparation, state-preparation calls, and evolution calls.
At the final time, measure each online circuit copy's success flag for the final contraction in the fixed memory sector, and return the channel and mode registers of the first success.
No additional projection onto the input-time clock is performed.
Classical priors and controls may be shared across online circuit copies.

\begin{proposition}[Success from independent online copies]
\label{prop:qkf-live-copies}
Fix a classical record.
Suppose each online circuit succeeds with probability at least $p_*>0$ and its conditional output is within trace distance $\epsilon$ of the prescribed target state.
Denote that target by $\tau_T$.
Suppose the online circuit implementations factor across copies, conditional on that record.
Then \(m\) copies, with selection of the first successful register, satisfy
\begin{align}
\Pr[\text{all fail}]\le(1-p_*)^m\le e^{-mp_*},
\quad D(\rho_{\rm selected},\tau_T)\le\epsilon.
\label{eq:qkf-copy-guarantee}
\end{align}
Here \(D(\rho,\sigma)=\tfrac12\|\rho-\sigma\|_1\).
Thus failure probability at most \(\vartheta\in(0,1/2]\) is achieved by
\begin{align}
m=\left\lceil p_*^{-1}\log(1/\vartheta)\right\rceil.
\label{eq:qkf-copy-count}
\end{align}
The state-error statement is conditioned on at least one success.
\end{proposition}
\begin{proof}
Let \(p_\ell\ge p_*\) be the success probability of copy \(\ell\).
Conditional independence gives \(\Pr[\text{all fail}]=\prod_{\ell=1}^m(1-p_\ell)\).
The inequalities in \cref{eq:qkf-copy-guarantee} follow from \(1-p_*\le e^{-p_*}\).
Given that the first success is copy \(\ell\), the failures of preceding copies do not change that copy's conditional successful state.
The returned state is therefore a convex combination of states each within \(\epsilon\) of \(\tau_T\); convexity of trace distance proves the remaining claim.
\end{proof}

The bound \(p_*\) must be known before the online circuit copies are initialized and hold uniformly over allowed future records.
An expectation of success over records is insufficient: all copies experience the same record.
Shared classical drift is compatible with the proposition when the conditional quantum operations factor; correlated hardware failures require their own joint bound.
Replication reduces the probability of obtaining no successful output.
It does not reduce common model error or inaccurate priors.

This construction requires \(m\) independent uses of every nonzero-weight state-preparation oracle and every scheduled evolution oracle at its original event, as well as \(m\) preparations of the initial state if it is retained.
The event must support those calls while the specified oracle remains available, or give simultaneous uses.
If serial calls encounter different Hamiltonians, their mismatch is included in the drift bound below.
Independent initialization and oracle calls give the copies; no cloning of an unknown state is assumed.
If only one use per event is permitted, only the one-pass guarantee is established.
Starting another online circuit after learning the final failure generally encounters a different stream.
Likewise, a succession of intermediate output states requires additional online circuit copies or replay access.

\subsubsection{Drift and final-state guarantees}
\label{subsec:qkf-drift}

The requested online target includes Hamiltonian drift through the actual \(U_t\) queries.
Stale dynamics, inaccurate priors or observations, and circuit implementation introduce approximation error relative to that target.
Duhamel's formula gives the dimension-independent bound
\begin{align}
\|e^{-i\tau H}-e^{-i\tau\widetilde H}\| \le\min\{2,|\tau|\|H-\widetilde H\|\}.
\label{eq:qkf-duhamel}
\end{align}
For a time-ordered propagator, replace the right-hand side by the integral of the Hamiltonian difference.
For a product of scheduled oracles, telescope to obtain
\begin{align}
d_t:=\|U_t-\widetilde G_t\| \le\sum_{j\text{ scheduled at }t}|\tau_j| \|H_{j,t}-\widetilde H_{j,t}\|+d_t^{\rm control}.
\label{eq:qkf-oracle-drift}
\end{align}
A calibration of age $\ell$ gives the error bound $\|H_{j,t+1}-H_{j,t}\|\le\nu_j$ implies an oracle error at most $|\tau_j|\ell\nu_j$.
The estimate applies to the declared query events, with the original availability schedule preserved.

For the scalar specialization $d=1$, \cref{cor:qkf-isotropic-drift} substitutes the scalar Kalman gains into the stability estimate to quantify stale-propagator and observation errors.

\begin{corollary}[Explicit drift bound for the online filter]
\label{cor:qkf-isotropic-drift}
Suppose the reference and implemented gains lie in $[\gamma,k_+]$, where $0<\gamma\le k_+\le1$.
Set the memory contraction factor and input bounds to $\rho=1-\gamma$, $\|\boldsymbol{\mu}_0\|\le M_0$, and $\|\boldsymbol{y}_t\|\le Y$.
Define the mean bound and the gain and observation errors by $M=\max\{M_0,Y\}$, $h_t=|k_t-\widetilde k_t|$, and $o_t=\|\boldsymbol{y}_t-\widetilde{\boldsymbol{y}}_t\|$.
Then
\begin{align}
E_T^{\rm model}:=\|\boldsymbol{\mu}_T-\widetilde{\boldsymbol{\mu}}_T\| \le\rho^Te_0+\sum_{t=1}^T\rho^{T-t} [\rho M d_t+(M+Y)h_t+k_+o_t].
\label{eq:qkf-isotropic-error}
\end{align}
For uniform bounds \(d,h,o\), the sum is at most \([\rho Md+(M+Y)h+k_+o]/\gamma\).
\end{corollary}
\begin{proof}
The convex update gives \(\|\boldsymbol{\mu}_t\|\le M\).
Also \(\|\widetilde a_t\widetilde G_t-a_tU_t\| \le\rho d_t+h_t\), and every implemented product contracts by \(\rho\) per step.
Substitute in \cref{thm:qkf-robust}.
\end{proof}

Old local errors are exponentially suppressed; the error floor is set by recent drift divided by the assimilation rate \(\gamma\).
For a comparison with a terminal frozen propagator satisfying \(\|U_{t+1}-U_t\|\le\nu_U\), with the same gains and input vectors, \cref{eq:qkf-isotropic-error} bounds the resulting model error by \(E_T^{\rm model}\le M\nu_U\rho^2/\gamma^2\), apart from any initial mismatch.
This follows from \(d_t\le(T-t)\nu_U\) and \(\sum_{n\ge0}n\rho^n=\rho/(1-\rho)^2\).
Exact online execution uses the actual oracle at each scheduled event and permits the drift specified by the model.

\cref{prop:qkf-prior-drift} bounds how errors in the scalar covariance prior perturb the gains used by the quantum circuit.

\begin{proposition}[Sensitivity of the scalar covariance prior]
\label{prop:qkf-prior-drift}
Compare two recursions \cref{eq:qkf-scalar-prior} whose gains lie in \([\gamma,k_+]\), with $0<\gamma\le k_+\le1$ and $\rho=1-\gamma$, and whose innovation variances are at least \(\sigma>0\).
With differences denoted by \(\delta\),
\begin{align}
\delta p_t&\le\rho^2\delta p_{t-1} +\rho^2\delta q_t+k_+^2\delta v_t,
\label{eq:qkf-prior-cov-bound}
\\
h_t&\le\frac{\delta p_{t-1}+\delta q_t+\delta v_t}{\sigma}.
\label{eq:qkf-prior-gain-bound}
\end{align}
In particular, uniform variance errors \(\Delta_q,\Delta_v\) give
\begin{align}
\delta p_t\le\rho^{2t}\delta p_0+ \frac{\rho^2\Delta_q+k_+^2\Delta_v}{1-\rho^2}.
\label{eq:qkf-prior-cov-floor}
\end{align}
\end{proposition}
\begin{proof}
Write the scalar covariance update and gain as
\begin{align}
f(s,v)=\frac{sv}{s+v},
\quad k=\frac{s}{s+v},
\quad \partial_sf=(1-k)^2\le\rho^2,
\quad \partial_vf=k^2\le k_+^2.
\end{align}
Along the segment joining the two variance pairs, the gain lies between its endpoint values and the innovation remains at least $\sigma$.
Integrating the derivatives along this segment proves \cref{eq:qkf-prior-cov-bound}.
The two gain derivatives have absolute value at most $1/\sigma$, which proves \cref{eq:qkf-prior-gain-bound}.
Iterating the covariance inequality gives \cref{eq:qkf-prior-cov-floor}.
\end{proof}

These formulas distinguish a deliberately changed prior from an error relative to a reference prior.
If the former is the desired model, there is no gain-mismatch term.
They also show why a small innovation variance makes inaccurate prior parameters consequential, even when its scalar matrix condition number is one.

\begin{theorem}[Conditional final-state guarantee]
\label{thm:qkf-final-state}
Let \(\beta=\|\boldsymbol{\mu}_T\|>0\).
Suppose the implemented vector $\widetilde{\boldsymbol{\mu}}_T$ uses a given preparation scale $Z$.
Assume that the implemented circuit differs from its ideal unitary by operator norm at most $\eta_{\rm circ}$.
Let \(E_{\rm trunc}\) bound omitted history.
Define
\begin{align}
E=E_T^{\rm model}+E_{\rm trunc}+Z\eta_{\rm circ}<\beta.
\label{eq:qkf-total-error}
\end{align}
Conditioned on the selected output, the state \(\widehat\rho_T\) satisfies
\begin{align}
D\bigl(\widehat\rho_T,|\mu_T\rangle\langle\mu_T|\bigr) \le\frac{2E}{\beta},
\quad p_{\rm succ}\ge\frac{(\beta-E)^2}{Z^2},
\label{eq:qkf-final-state-bound}
\end{align}
where \(D(\rho,\sigma)=\frac12\|\rho-\sigma\|_1\).
For $\epsilon\le1$, it suffices to require $E\le\epsilon\beta/2$ and $\chi=Z/\beta$.
The conditional trace-distance error is then at most $\epsilon$, and success is at least $1/(4\chi^2)$.
\end{theorem}
\begin{proof}
Unitary error bounds the selected-vector error by \(\eta_{\rm circ}\).
Multiply the encoded output by its preparation scale $Z$.
The model and truncation bounds then place the unnormalized output within $E$ of the desired mean.
Apply \cref{lem:output-conditioning} for the success and conditional-error bounds.
A classical mixture of synthesis choices obeys the same bound when it holds uniformly for each choice.
\end{proof}

If \(\chi\le\bar\chi\) is known before execution and the error budget holds uniformly, take \(p_*=1/(4\bar\chi^2)\) in \cref{prop:qkf-live-copies}.
Use the independent parallel executions from \cref{prop:qkf-live-copies}, with $m=\lceil4\bar\chi^2\log(1/\delta)\rceil$.
They give failure at most $\delta$ and preserve the conditional trace-distance error bound $\epsilon$.
For exact preparation of the target vector, the factor four can be omitted.
Additional physical noise must be included in the per-copy success and error bounds before applying the proposition.

For incoherent hardware noise, suppose the actual channel differs from this circuit by diamond distance at most \(2\nu\).
Suppose the noise-free success probability has a positive lower bound $p\ge p_*>0$.
For noise strength $\nu<p_*$, the noisy circuit satisfies $\text{success}\ge p_*-\nu$ and additional conditional trace-distance error at most $2\nu/p_*$.
Indeed, successful subnormalized outputs differ by trace norm at most \(2\nu\), their probabilities differ by at most \(\nu\), and normalization gives the stated bound.
Thus hardware noise is amplified by the inverse success probability; it is separate from the classical process and measurement uncertainties.

\subsubsection{Finite memory, precision, and costs}
\label{subsec:qkf-resources}

Assume the stability certificate in \cref{eq:qkf-storage-certificate,eq:qkf-stability}, and suppose \(\|\vect\mu_t\|\le M\) on the promised records.
Initialize the mean to zero at tick $T-L$ and execute only the final memory window.
Then
\begin{align}
E_{\rm trunc}\le C M\rho^L,
\quad L=\min\left\{T,\left\lceil \frac{\log(4CM/(\epsilon\beta_*))}{-\log\rho} \right\rceil\right\}
\label{eq:qkf-window-length}
\end{align}
suffices when \(\beta=\|\vect\mu_T\|\ge\beta_*>0\); take at least one step.
If $L=T$, retain the given initial mean and incur no truncation error.
For fixed $C$ and a gap $1-\rho$ bounded below, the quantum history is logarithmic in $1/\epsilon$.
The small covariance recursion is still continued classically, because resetting $p_t$ would define a different filter.
For an online stream, start the quantum online circuit when the retained window begins; retrospective window evaluation requires replay access.

If \(\|k_t\|\le k_+\), the mean kernel satisfies \(\|\mathcal U(t,s)k_s\|\le Ck_+\rho^{t-s}\).
Its memory age generating function therefore obeys
\begin{subequations}
\label{eq:qkf-analytic-bound}
\begin{align}
\mathcal G_T(z)&=\sum_{n=0}^{T-1}z^n \sum_{t=n+1}^T|t\rangle\langle t-n|\otimes \mathcal U(t,t-n)k_{t-n},
\\
\sup_T\sup_{|z|=R}\|\mathcal G_T(z)\| &\le\frac{Ck_+}{1-\rho R},
\quad 1<R<\rho^{-1}.
\end{align}
\end{subequations}
The coefficient of $z^n$ has norm at most $Ck_+\rho^n$.
This is analyticity in the memory-age variable, which controls omitted causal history.
It does not assume analytic dependence on physical time: the small matrices and transport oracles may change abruptly from one event to the next.
The full unitary circuit also retains orthogonal completion sectors, so the weighted controllability Gramian and the final output norm continue to determine success separately from this tail bound.

For one window execution, let \(c_{j,t}\) and \(c_{y,t}\) be the costs of a scheduled controlled evolution and a state preparation.
Then
\begin{align}
Q_j=|\mathcal D_j\cap\{T-L+1,\ldots,T\}| \le\lceil L/r_j\rceil,
\quad C_{\rm pass}=c_{\rm init}+\sum_j\sum_{t\in\mathcal D_j\cap\mathrm{window}}c_{j,t} +\sum_{t\in\mathrm{window}}c_{y,t}+C_{\rm known}.
\label{eq:qkf-cost}
\end{align}
Here \(c_{\rm init}\) is the initial-state preparation cost if retained, and zero otherwise.
Oracle accuracy is part of these given costs.
For \(J\) known gates whose individual operator errors are at most \(\eta_{\rm circ}/J\), the telescoping bound \cref{eq:iir-td-drift} ensures the circuit error budget.
Ordinary rotation synthesis therefore needs precision logarithmic in \(J\chi/\epsilon\), with the historical-Hamiltonian query count unchanged.
The upper bound also includes oracle implementations, classical prior computation, and the success factor \(\chi\).

With \(m\) copies as in \cref{eq:qkf-copy-count}, the high-success resources are
\begin{align}
Q_j^{\rm total}&=mQ_j\le m\lceil L/r_j\rceil,
\quad Q_y^{\rm total}=mQ_y,
\label{eq:qkf-copy-queries}
\\
C_{\rm total}&=mC_{\rm pass}+C_{\rm select},
\quad S_{\rm total}=m\bigl[w+O(\log((d+d_y)(L+1)))\bigr].
\label{eq:qkf-copy-resources}
\end{align}
Here \(Q_y\) counts nonzero-weight observation-preparation calls in one pass, and \(C_{\rm pass}\) includes an initial-state preparation call when needed.
The register width $w$ is defined in \cref{thm:qkf-online}.
The final flag measurements and routing use at most $C_{\rm select}=O(m[w+\log((d+d_y)(L+1))])$ additional operations.
The per-copy known-gate cost includes $O(Ld(d+d_y)+d^2)$ small-channel two-level rotations before addressing and elementary-gate synthesis.
The persistent small covariance and weighted controllability Gramian require $O(d^2)$ classical numbers.
Including current dense model matrices and working storage requires $O(d^2+dd_y+d_y^2)$ numbers, with $O(d^3+d^2d_y+dd_y^2+d_y^3)$ conventional dense arithmetic operations per update.
Clock weights and classical records have their own storage costs.
The covariance recursion is maintained for all events needed by the model, including those preceding a truncated quantum window; the Gramian and channel-unitary construction need only the retained window.
The resulting classical controls are shared by all copies.
In particular, a known bound \(\chi\le\bar\chi\) gives
\begin{align}
Q_j^{\rm total} =O\!\left(\bar\chi^2\log(1/\delta) \left\lceil L/r_j\right\rceil\right).
\label{eq:qkf-high-success-cost}
\end{align}
The logarithmic window length does not eliminate this quadratic dependence on \(\bar\chi\).
The number of elapsed stream ticks is unchanged if the required parallel event access is available; the total query count and quantum memory both increase by \(m\).

The online construction illustrates the hierarchy developed earlier in the paper.
Query-marker analyticity remains exact even though the filter is time-varying and its public kernel is not Toeplitz.
The full two-time clock SDP characterizes a fixed online lift, while the weighted controllability Gramian gives a small constructive clock for the structured Kalman recurrence.
The tensor assumption in \cref{eq:qkf-structured-model} is what keeps the classical covariance update independent of the large mode dimension.
More general anisotropic models fit the causal lift formally, but an efficient live implementation requires a comparably succinct representation of their covariance and gain.

\settocdepth{1}
\appendix
\section{Quditization, clocks, and circuit resources}
\label{app:qudit-tradeoffs}
This appendix collects domain conventions, finite-output error conversion, routing constructions, clock optimization, and resource substitutions.
\subsection{Domain and finite-precision conventions}
\label{app:conditions}
These conventions collect routine qualifications used in the statements and proofs; any stronger hypothesis is given with the result that uses it.

Asymptotic resource bounds truncate logarithms below at one and include a constant-size regime when a displayed scale is below one.
A bound written with $\alpha T$ uses $\max\{1,\alpha T\}$ for short nonzero evolutions; an identically zero generator needs no generator calls.
Exact integer horizons, floors, and finite-error inequalities retain their stated constants.

Circuit choices and numerical certificates are fixed uniformly over the stated oracle promise.
Norms are operator norms unless a vector or another norm is specified.
All finite circuit spaces include the unitary completion of each oracle and the private output registers.
Known zero-cost identity connections still occupy space.
Padding, support restrictions, and routing are understood on those complete spaces as in \cref{def:resource-accounting}.

Continuation through a closed local disk means analyticity on an open neighborhood of that disk.
Contour arguments use radial limits on any remaining part of the unit circle.
Inverses are taken where defined; removable transfer-function singularities use their analytic continuation.
A logarithmic-coordinate energy estimate uses every horizontal segment from its unitary boundary to the point being bounded.

The smoothed-window estimate is valid for rectangular operator-valued responses.
Its physical realization additionally requires a finite unitary circuit with the stated public Toeplitz block; this circuit is supplied by quditization or by a specified compression of its lift.
The estimate alone assigns no gate cost to an arbitrary analytic function.
Oblivious amplitude amplification uses a unitary target on the full reflected public space, its uniform catalyst bound, and access to the complete circuit inverse.

Clock accuracy and normalization slack are independent positive parameters.
The explicit smoothing formulas use $\epsilon_{\rm clk}<1$; a requested endpoint tolerance of one can use any smaller internal tolerance.
Preparing clock vectors or known rotations to finite precision contributes through the operator-error budgets in \cref{eq:qudit-error-budget,eq:app-synthesis}.
For normalized state outputs, the target-vector lower bound is applied before conditioning, as in \cref{lem:output-conditioning}.

\subsection{Resource convention, normalization, and output error}
\label{subsec:qudit-overhead}

The following conventions make query bounds comparable and specify the additional work needed to turn them into circuits.
We use \cref{def:resource-accounting} throughout unless a result explicitly changes the access model.
\cref{tab:resources-modules,tab:resources-algorithms,tab:resources-applications} tabulate the resulting oracle, gate, and qubit bounds; \cref{app:resources} gives the substitutions and conventions.

\begin{definition}[Resource accounting]
\label{def:resource-accounting}
A given block-encoding $O=\Be{A/\lambda}$ is a unitary on its specified signal and system registers satisfying $(\langle0|_{\rm sig}\otimes I)O(|0\rangle_{\rm sig}\otimes I) =A/\lambda$.
Its other matrix blocks complete this contraction to a unitary; we call their choice the \emph{unitary completion}.
The entire unitary acts in the circuit.
Amplitudes outside the signal subspace remain in their registers until the stated measurement or uncomputation.
A theorem uniform over these choices applies to every unitary satisfying the stated oracle promises; its other blocks are unrestricted subject to those promises.

One oracle query means one invocation of a stipulated controlled unitary.
Forward and inverse invocations are counted separately; self-inverse encodings use the same unitary for both.
Controlled access and inverse access are promises whenever used.
A state-preparation oracle is a unitary preparation with its specified inverse access.
Independent trials, rejected branches, and amplitude amplification contribute all of their executed queries.
A gate cost for an oracle includes its controls, signal tests, arithmetic, lookup, and required changes of basis.
Nonuniform costs are permitted.
Classical construction of these circuits and classical parameter selection are separate preprocessing costs.

Known operations, including system matrices, changes of basis, address permutations, and clock preparations, must be implemented by quantum gates.
Our gate counts use arbitrary one-qubit rotations and two-qubit gates.
For a finite gate set, divide the synthesis error budget equally among the $L$ rotations.
Approximating each to operator error at most $\eta/L$ gives total synthesis error at most $\eta$ by the hybrid bound.
For Clifford+$T$, the single-qubit synthesis bound \cite{Ross2016Optimal} gives
\begin{align}
O\!\left(L\log\frac{2L}{\eta}\right)
\quad\text{$T$ gates}.
\label{eq:app-synthesis}
\end{align}
The efficient classical synthesis search has the number-theoretic qualifications of the cited construction.
Exact reversible classical operations use the corresponding standard Toffoli and Clifford implementations.

Qubit counts include the system, signal ancillas, coherent labels, delay positions, and work or measurement records that coexist.
Workspace may be reused after it is restored.
Direct-sum positions are encoded by one binary address register, with the routing cost in \cref{thm:iir-gates}; tensor-product records require their full joint space.
A preparation scratch register and classical read-only storage are reported separately when needed.
For time-dependent circuits, the oracle and system costs may vary at every event.
\end{definition}

For a unitary target $G$, we measure the circuit's isometry error by
\begin{align}
\bigl\|U_{\rm circ}(|0\rangle_{\rm aux}\otimes I) -|0\rangle_{\rm aux}\otimes G\bigr\|.
\label{eq:implementation-isometry-error}
\end{align}
Known input and output ancillary states can be absorbed into their preparation unitaries.
This operator norm includes ancillary leakage and remains valid when the input is entangled with a reference.
A fixed-input state-preparation bound instead uses the Euclidean norm of the corresponding joint output vectors.

The clock and routing bounds below give the common implementation overhead.
Later results specify their horizon, query multiplicities, and any additional operations; the resource table substitutes these parameters into the same bounds.

To compare circuits with different normalizations, \cref{lem:normalization-error} tracks the induced approximation and conditional-state errors.

\begin{lemma}[Normalization and approximation error]
\label{lem:normalization-error}
Let $G$ be unitary and suppose a circuit implements $\Be{A/(1+\eta)}$ with $\|A-G\|\le\xi<1$.
For any $0<\delta\le\eta$, the same circuit is $\Be{\widetilde G_N/(1+\delta)}$, with
\begin{align}
\widetilde G_N=\frac{1+\delta}{1+\eta}A,
\quad \|\widetilde G_N-G\| \le\frac{\eta-\delta+(1+\delta)\xi}{1+\eta} \le\eta-\delta+\xi.
\label{eq:normalization-rescale}
\end{align}
It has success probability at least $(1-\xi)^2/(1+\eta)^2$ and, conditioned on success, Euclidean state error at most $2\xi$ on every normalized input, including an entangled reference.
Rescale the clock matrix according to
\begin{align}
X_\delta=\frac{1+\delta}{1+\eta}X,
\quad\|X_\delta\|_*\le1+\delta.
\end{align}
This changes the approximating operator's normalization while preserving the query schedule.
\end{lemma}
\begin{proof}
Writing $c=(1+\delta)/(1+\eta)\le1$ gives $cA-G=c(A-G)-(1-c)G$, proving the operator bound.
For $|\psi\rangle$, put $s=\|A|\psi\rangle\|$.
Then $s\ge1-\xi$, and the success probability is $s^2/(1+\eta)^2$.
Add and subtract $A|\psi\rangle$ to obtain $\|A|\psi\rangle/s-G|\psi\rangle\| \le |1-s|+\xi\le2\xi$.
Tensoring the inequalities with the identity leaves their operator norms unchanged.
The clock claim follows from homogeneity of the nuclear norm.
No amplification or inverse oracle calls are added.
\end{proof}

For a nonunitary target $A_0$, the same operator estimate holds after replacing the attenuation contribution by $(1-c)\|A_0\|$.
A conditional-state claim additionally requires a lower bound on $\|A_0|\psi\rangle\|$ and a corresponding allocation of approximation error before normalization.
\cref{lem:output-conditioning} gives the resulting success probability and conditional-state accuracy.
For a linear system, the required lower bound concerns the norm of the inverse applied to the given input; it is an additional assumption to the choice of block-encoding normalization.

\begin{lemma}[Conditioning an approximate output vector]
\label{lem:output-conditioning}
Let the nonzero target vector and its approximation satisfy $\beta=\|\boldsymbol{x}\|>0$ and $\|\widetilde{\boldsymbol{x}}-\boldsymbol{x}\|\le E<\beta$.
Suppose the unnormalized successful output is $\widetilde{\boldsymbol{x}}/Z$.
Then its probability is at least $(\beta-E)^2/Z^2$, and
\begin{align}
\left\|\frac{\widetilde{\boldsymbol{x}}}{\|\widetilde{\boldsymbol{x}}\|} -\frac{\boldsymbol{x}}{\beta}\right\|\le\frac{2E}{\beta}.
\end{align}
The same bound controls pure-state trace distance and survives tracing out retained records.
\end{lemma}
\begin{proof}
The reverse triangle inequality gives the probability bound.
Add and subtract $\widetilde{\boldsymbol{x}}/\beta$; the two resulting norms are at most $E/\beta$ each.
Pure-state trace distance is bounded by vector distance, and partial trace is contractive.
\end{proof}

Oblivious amplitude amplification implements the desired unitary transformation from a block encoding with normalization two, returning the auxiliary registers close to their selected state.
We use the robust oblivious amplitude amplification introduced by Berry et al.\ in their truncated Taylor-series algorithm for Hamiltonian simulation \cite{Berry2015Truncated}; \cref{lem:iir-cubic} gives the error bound needed for \cref{cor:iir-analytic-two}.

\begin{lemma}[Oblivious amplitude amplification {\cite{Berry2015Truncated}}]
\label{lem:iir-cubic}
Suppose a complete unitary $\mathcal A$ block-encodes $\widetilde G/2$.
Let the target $G$ be unitary and satisfy
\begin{align}
\|\widetilde G-G\|\le\delta\le1.
\end{align}
Let $\mathcal R_{\rm clk}$ reflect about the selected input/output clock sector.
Then
\begin{align}
\mathcal A_{\rm OAA} =-\mathcal A\mathcal R_{\rm clk} \mathcal A^\dagger\mathcal R_{\rm clk}\mathcal A
\label{eq:iir-oaa}
\end{align}
has isometry error at most
\begin{align}
\sqrt{3\delta^2+\delta^3}\le2\delta.
\label{eq:iir-oaa-clean}
\end{align}
Its encoded operator is
\begin{align}
\frac{3\widetilde G- \widetilde G\widetilde G^\dagger\widetilde G}{2}.
\label{eq:iir-oaa-selected}
\end{align}
The circuit uses two forward executions and one inverse execution.
\end{lemma}

\begin{proof}
The standard two-reflection expansion gives \cref{eq:iir-oaa-selected}.
To bound the isometry error, write $\widetilde G=G(I+E)$ with $\|E\|\le\delta$.
Let
\begin{align}
C=\frac{3(I+E)-(I+E)(I+E)^\dagger(I+E)}2.
\end{align}
Then the encoded operator in the target frame is $C$.
Expanding the product,
\begin{align}
C =I+\frac{E-E^\dagger}{2} -\frac{E^2+E^\dagger E+EE^\dagger+EE^\dagger E}{2}.
\label{eq:iir-oaa-expanded}
\end{align}
For the input isometry $J$, both the amplified circuit and target maps are isometries: $\mathcal A_{\rm OAA}J$ and $JG$.
Hence their squared difference has Gram matrix
\begin{align}
2I-C-C^\dagger.
\end{align}
The first-order skew-Hermitian term in \cref{eq:iir-oaa-expanded} cancels.
The norm of the remaining Hermitian matrix is at most $3\delta^2+\delta^3$.
Taking the square root proves \cref{eq:iir-oaa-clean}.
The circuit expression \cref{eq:iir-oaa} gives the forward and inverse query counts directly.
\end{proof}

To include approximation of the circuit, let $\eta_S,\eta_j$ be uniform operator errors for the given operations with their work registers restored, and let $\eta_{\rm in},\eta_{\rm out}$ be the clock preparation errors.
If a finite system approximates an ideal transfer function, define $\eta_{\rm rep}=\|T_N[G_{\rm finite}]-T_N[G]\|$.
With ideal clock error $\epsilon_{\rm clk}=\|\widetilde G_N-G\|$, a hybrid argument gives
\begin{align}
\epsilon_{\rm final}\le\epsilon_{\rm clk} +\alpha_{\rm clk}\left(\eta_{\rm rep}+N\eta_S+\sum_jq_j\eta_j +\eta_{\rm in}+\eta_{\rm out}\right).
\label{eq:qudit-error-budget}
\end{align}
Thus a finite approximation to an infinite system is accounted for by its explicit Toeplitz-matrix error.
The operation bounds are uniform on the full retained Hilbert spaces, including entangled reference systems.

\subsection{Routing and its space--time tradeoffs}
\label{app:qudit-arithmetic}
For the arithmetic bounds, set $b=\lceil\log_2\dim\mathcal H_{N,\boldsymbol r}\rceil$ and $\mathcal R=(m+1)N+\sum_jq_j$.

Known clock states can be prepared by standard uniformly controlled rotations \cite{shende2006synthesis,Mottonen2005StatePreparation}.
On $M$ padded basis states the cost is $O(M)$ gates with no preparation scratch.
For real clocks with at most two labels, a conservative bound for both preparations together is
\begin{align}
\#R_y\le4\,2^{\lceil\log_2N\rceil}-2,
\quad \#\mathrm{CNOT}\le4\,2^{\lceil\log_2N\rceil}-4.
\label{eq:qudit-clock-preparation}
\end{align}
Thus the explicit clocks cost $O(N)$ gates and $O(1)$ label qubits.
This is subdominant whenever the cost of the system and oracle operations grows faster than $N$; it need not be negligible for constant-cost oracles.

When dirty ancillary qubits are available, known-state preparation and unitary synthesis admit a different tradeoff between workspace and $T$ gates~\cite{Low2024TradingTGates}.
Those implementations can be substituted when evaluating a particular clock circuit.
The bounds above use arbitrary rotations and require no preparation scratch.

\subsubsection{Binary encoding of the direct sum}
\label{subsec:iir-gates}

The direct sum in \cref{eq:iir-buffer-space} is represented by one register addressing all its coordinates:
\begin{align}
\dim\mathcal H_{N,\boldsymbol{r}}=N\dim\mathcal P+\sum_jr_j\dim\mathcal L_j.
\label{eq:qudit-dimensions}
\end{align}
These dimensions include block-encoding ancillas and workspaces; there need not be a common tensor factor across ports.
Power-of-two padding increases the Hilbert-space dimension by less than a factor four.
Let $G_{\rm clk}$ count both clock preparations and let $s_{\rm clk}$ be their maximum additional scratch, restored afterwards.
The scratch excludes the binary address register and the $r_{\rm clk}$ clock-label values.

\cref{thm:iir-gates} bounds the complete binary-address implementation, including clock preparation and restored routing workspace.

\begin{theorem}[Quantum gate and qubit overhead]
\label{thm:iir-gates}
Given the system matrix, horizon, delays, and circuit of \cref{thm:iir-lift}, give normalized clocks realizing $X/\alpha_{\rm clk}$ and preparation circuits with costs $G_{\rm clk},s_{\rm clk},r_{\rm clk}$ as above.
Suppose the embedded, controlled system and oracle operations have gate costs $G_S,G_{O_j}$, including local encoding conversions, and their work registers are counted in the Hilbert-space dimensions.
Then quditization implements
\begin{subequations}
\label{eq:iir-complete-be}
\begin{align}
\mathcal A &=P_{\rm out}^\dagger(I_{\rm label}\otimes\mathcal W_N[G])P_{\rm in} =\Be{\widetilde G_N/\alpha_{\rm clk}},
\\
\widetilde G_N&=\sum_{n=0}^{N-1}c_n(X)G_{n}.
\end{align}
\end{subequations}
It uses exactly $N$ system operations.
Each port receives $q_j=\lfloor(N-1)/r_j\rfloor$ controlled oracle calls.
A coherent implementation has
\begin{subequations}
\label{eq:qudit-main-cost}
\begin{align}
n_{\rm total}&\le\lceil\log_2\dim\mathcal H_{N,\boldsymbol{r}}\rceil +\lceil\log_2r_{\rm clk}\rceil+s_{\rm clk}+O(1),
\\
G_{\rm total}^{(1,2)}&\le NG_S+\sum_jq_jG_{O_j}+G_{\rm clk}
\\
&
\quad+O\!\left(\left[(m+1)N+\sum_jq_j\right] \log(2\dim\mathcal H_{N,\boldsymbol{r}}) \log\log(2\dim\mathcal H_{N,\boldsymbol{r}})\right).
\end{align}
\end{subequations}
For a constant number of clock labels and direct preparation, $G_{\rm clk}=O(N)$ and $s_{\rm clk}=0$.
The total width is then $\lceil\log_2\dim\mathcal H_{N,\boldsymbol{r}}\rceil+O(1)$.
The bounds apply separately to one- and two-qubit counts in the arbitrary-rotation model.
\end{theorem}
\begin{proof}
The lift and clock characterization give \cref{eq:iir-complete-be}.
To implement them on the binary address register, assign each public position and private residue a contiguous interval of addresses.
At each system operation, move the $m+1$ active intervals to a contiguous prefix in the order expected by $S$.
Apply that system matrix and reverse the permutation.
At a query, move the entire type-$j$ interval into position, apply its controlled oracle with the residue and occurrence labels unchanged, and undo the permutation.

For a basis coordinate, the routing and scratch use can be written as
\begin{subequations}
\label{eq:qudit-routing-registers}
\begin{align}
&|x\rangle_a|0\rangle_b \xrightarrow{\text{route, test}} |\pi_k(x)\rangle_a|\mathrm{active}_k(x)\rangle_b
\\
&\xrightarrow{\text{controlled }S\text{ or }O_j} \sum_y v_y|y\rangle_a|\mathrm{active}_k(x)\rangle_b \xrightarrow{\text{untest, unroute}} \sum_y v_y|\pi_k^{-1}(y)\rangle_a|0\rangle_b.
\end{align}
\end{subequations}
The local operation preserves the active interval, so its flag can be uncomputed; $v_y$ are the resulting amplitudes.
Linearity gives the same sequence on superpositions.
The register $b$ is reused.

Extracting an interval exchanges two adjacent blocks.
There are at most $2[(m+1)N+\sum_jq_j]$ such exchanges and $O(N+\sum_jq_j)$ activation comparisons.
The constant-adder construction in \cref{lem:iir-routing} implements each exchange and comparison in $O(\log(2\dim\mathcal H_{N,\boldsymbol{r}})\log\log(2\dim\mathcal H_{N,\boldsymbol{r}}))$ gates with constant additional workspace.
The address permutations and their proofs are given in \cref{app:qudit-arithmetic}.
Reusing this workspace, and adding the clock label, endpoint scratch, and constant padding, gives the qubit bound in \cref{eq:qudit-main-cost}.

Counting the $N$ system operations, $q_j$ queries, and routing operations, then adding the two clock preparations once, gives the gate bound.
\cref{eq:qudit-clock-preparation} gives the stated direct-preparation specialization.
\end{proof}

The number of additional address qubits is logarithmic in the horizon and private storage.
For fixed port count the routing cost is nearly linear in $N$, up to logarithms of the Hilbert-space dimension.
If the given system and oracle operations dominate this routing cost, they dominate the total gate count.
No such dominance is assumed for a trivial oracle.

The binary encoding in \cref{thm:iir-gates} exchanges adjacent intervals of addresses.
We implement each exchange using constant addition, comparison, and controlled complement.

For an offset $d$ known classically, use the in-place constant-adder construction with one restored borrowed bit~\cite{Haner2016factoring}.
Its controlled and uncontrolled variants on $k$ address bits use
\begin{align}
A(k)=A(\lceil(k+1)/2\rceil)+A(\lfloor(k-1)/2\rfloor)+O(k) =O(k\log(2+k))
\label{eq:qudit-adder-recurrence}
\end{align}
one- and two-qubit gates.
The carry subroutines temporarily borrow the opposite half of the address and restore it.
The borrowed bit may be entangled with the other registers.
The carry-cancellation identity is proved below.

Only constant addition, comparison, and complement are needed for packing.
In execution order, the two-addition circuit
\begin{align}
\operatorname{Add}_{b+1,2^b-c}\text{ on }(f,x),
\quad \operatorname{Add}_{b,c}\text{ on }x,
\quad X_f
\label{eq:iir-comparator-construction}
\end{align}
implements $|f,x\rangle\mapsto|f\oplus[x<c],x\rangle$.
The first addition uses the existing predicate bit as its most significant bit; the second restores $x$.
This identity holds for either initial value of $f$ and for the endpoint thresholds.

Compute $[a\le x<c]$ into one control flag qubit initialized to $|0\rangle$ with two threshold comparisons.
Conditional complement followed by addition of $a+c$ implements
\begin{align}
\operatorname{Rev}_{[a,c)}(x)= \begin{cases}a+c-1-x,&a\le x<c,
\\
x,&\text{otherwise}. \end{cases}
\label{eq:iir-interval-reversal}
\end{align}
The reversal preserves interval membership, so recomputing the predicate clears the enable.
Reverse the adjacent intervals $[a,m)$ and $[m,c)$ separately.
Then reverse their union $[a,c)$.
This exchanges the two blocks.
Each exchange therefore uses a constant number of comparisons, controlled additions, and $O(b)$ CNOTs.
By \cref{eq:qudit-adder-recurrence}, it costs $O(b\log(2+b))$ gates and constant extra workspace.

There is one routing flag qubit initialized to $|0\rangle$ and one borrowed bit.
A clock-label qubit can give the latter because the arithmetic restores it exactly; otherwise add a single constant-size wire.
Registers are reused between routing, work, and oracle operations.
Local padding costs $O(1)$ qubits.
The label and endpoint-preparation registers use, respectively, $\lceil\log_2r_{\rm clk}\rceil $ and at most $ s_{\rm clk}$ additional qubits.
Attenuation uses a constant-size flag.
These costs give the qubit bound in \cref{eq:qudit-main-cost}, while retaining all block-encoding ancillas.

\begin{lemma}[Modular constant addition using a borrowed ancilla]
\label{lem:iir-routing}
Adding a classical constant modulo $2^b$, including one external control, has an exact circuit with $O(b\log(2+b))$ gates.
It uses one restored borrowed bit and a constant number of other flags.
No length-$b$ offset register is required.
The action on the borrowed bit is identity, including when it is entangled with the input.
\end{lemma}
\begin{proof}
Use the in-place constant-adder recursion of H\"aner, Roetteler, and Svore~\cite{Haner2016factoring}.
We recall the cancellation that permits the stated space.
Split the address into a low block of $\ell=\lceil(b+1)/2\rceil$ bits and a high block of $k=\lfloor(b-1)/2\rfloor$ bits.
A forward and backward carry sweep XORs the low carry into an arbitrary target using $O(\ell)$ Toffolis and dirty wires from the high block.
A controlled high-block increment uses $O(k)$ Toffolis, borrowing and restoring low-block wires; sufficiently small blocks use direct circuits.
The offset bits are classical gate choices.
These are the carry and increment primitives of the cited constant-adder construction; an external control adds linear overhead.

Let the borrowed bit start at $v$, and let $c$ be the computed low carry.
Write $I^v$ for a high-block increment controlled on this bit and $J^v$ for controlled complement.
Denote carry XOR by $C$.
Execute
\begin{align}
I^v,\ J^v,\ C,\ I^{v\oplus c},\ C,\ J^v,
\quad\text{followed by the recursive low and high constant additions}.
\label{eq:iir-borrowed-schedule}
\end{align}
On the high block the net preliminary action is $J^vI^{v\oplus c}J^vI^v=I^c$, since $JIJ=I^{-1}$.
The two carry XORs restore the borrowed bit.
Linearity proves the same identity for a superposed or entangled borrowed state.
The opposite-half borrow requirements fit this split, and the carry and increment work at a node is $O(b)$.
This gives \cref{eq:qudit-adder-recurrence}.
Each level has $O(b)$ gates and there are $O(\log b)$ levels.
Reversing the circuit gives subtraction at the same cost.
\end{proof}

For a ripple-carry implementation, allocate an additional $b+O(1)$-qubit register, initialize it with the classical offset by $O(b)$ NOTs, and use a reversible ripple-carry adder~\cite{Cuccaro2004Adder,Takahashi2010Adder}.
It preserves the offset operand and uses $O(b)$ Toffolis and CNOTs with constant carry scratch.
Controlling each constant-size primitive adds only a constant factor.
Erase the offset with the same NOTs.
The comparator and reversal operations in the main proof each cost $O(b)$ gates.
The total cost is $O(\mathcal Rb)$ gates with $O(b)$ additional workspace, including the adder's second operand.

\subsection{Semidefinite representation of the error--normalization tradeoff for the chosen lift}
\label{app:clock-frontier-sdp}
For the finite-promise case of \cref{cor:iir-clock-frontier}, the two norm constraints can be expressed by positive semidefinite blocks.
For a finite promise set, introduce Hermitian matrices $L,R$ and $\eta\ge0$.
Minimize $\eta$ subject to
\begin{subequations}
\label{eq:qudit-sdp}
\begin{align}
\begin{pmatrix}L&X
\\
X^\dagger&R\end{pmatrix}\succeq0,
\quad\frac{\operatorname{tr}L+\operatorname{tr}R}{2}\le\alpha_{\rm clk},
\\
\begin{pmatrix}\eta I&E_{\boldsymbol{O}}(X)
\\
E_{\boldsymbol{O}}(X)^\dagger&\eta I\end{pmatrix}\succeq0
\quad(\boldsymbol{O}\in\mathfrak O),
\quad E_{\boldsymbol{O}}(X)=\widetilde G_N-G.
\end{align}
\end{subequations}
The first constraint represents the nuclear norm \cite{Recht2010NuclearNorm}.
For the singular-value decomposition of the clock matrix, choose
\begin{align}
X=\sum_\ell s_\ell a_\ell b_\ell^\dagger,
\quad L=\sum_\ell s_\ell a_\ell a_\ell^\dagger,
\quad R=\sum_\ell s_\ell b_\ell b_\ell^\dagger.
\end{align}
Then the block matrix is a sum of positive rank-one matrices, and its half-trace equals the nuclear norm:
\begin{align}
\begin{pmatrix}L&X
\\
X^\dagger&R\end{pmatrix} =\sum_\ell s_\ell\binom{a_\ell}{b_\ell} \binom{a_\ell}{b_\ell}^{\!\dagger},
\quad \tfrac12(\operatorname{tr}L+\operatorname{tr}R)=\|X\|_*.
\end{align}
Conversely, positivity evaluated on each vector with opposite singular-vector components gives $a_\ell^\dagger L a_\ell+b_\ell^\dagger R b_\ell \ge2\operatorname{Re}(a_\ell^\dagger Xb_\ell)=2s_\ell$.
Summing and using positivity of the unused diagonal components bounds the nuclear norm by the half-trace.
The second block constraint in \cref{eq:qudit-sdp} similarly bounds each operator error by $\eta$.

\subsection{A single-pair normal form for causal scalar clocks}
\label{app:qudit-rankone}

The spectator formulation is convenient for explicit analytic clocks.
For full equal supports, spectators do not improve the optimum of the causal coefficient problem, although eliminating them can complicate the classical preparation data.

\begin{proposition}[Causal rank-one replacement]
\label{prop:qudit-rankone}
Suppose the causal coefficients have a finite factorization with norm bound \begin{align}
c_n=\sum_\ell\langle h_\ell,S_N^ng_\ell\rangle
\quad(0\le n<N),
\quad A_0=\sum_\ell\|h_\ell\|\|g_\ell\|.
\end{align} There exist length-$N$ vectors with the same nonnegative-clock weights and norm product $\|g\|\|h\|\le A_0$.
In particular, an optimal scalar Toeplitz clock has a rank-one replacement with no larger normalization.
\end{proposition}
\begin{proof}
Form the scalar polynomials
\begin{align}
a_\ell(z)=\sum_{t=0}^{N-1}\overline{h_{\ell,t}}z^{N-1-t},
\quad b_\ell(z)=\sum_{s=0}^{N-1}g_{\ell,s}z^s,
\quad P(z)=\sum_\ell a_\ell(z)b_\ell(z).
\end{align}
The coefficient at degree $N-1-n$ is the required clock weight.
To control the cost of factoring this polynomial, define the Hardy norms using normalized integration on the unit circle:
\begin{subequations}
\label{eq:qudit-hardy-norms}
\begin{align}
\|P\|_{H^1}&=\int_{-\pi}^{\pi}|P(e^{i\theta})|\frac{d\theta}{2\pi},
\\
\|a\|_{H^2}&=\left(\int_{-\pi}^{\pi}|a(e^{i\theta})|^2 \frac{d\theta}{2\pi}\right)^{1/2} =\left(\sum_{k\ge0}|a_k|^2\right)^{1/2}.
\end{align}
\end{subequations}
The second equality is Parseval's identity.
Cauchy--Schwarz applied to each product in the defining sum gives
\begin{align}
\|P\|_{H^1}\le\sum_\ell\|a_\ell\|_{H^2}\|b_\ell\|_{H^2}=A_0.
\end{align}
One-variable Riesz factorization \cite{Jury2019WeakFactorization} splits the polynomial into analytic factors of balanced boundary modulus.
For this polynomial case, remove its zeros inside the disk with their Blaschke product.
The remaining function has no zeros in the open disk and therefore has an analytic square root.
Multiplying that square root by the Blaschke product restores the removed zeros in the other factor.
Since the product has unit modulus on the circle, the two factors satisfy $P=ab$, $|a(e^{i\theta})|=|b(e^{i\theta})| =|P(e^{i\theta})|^{1/2}$, and $\|a\|_{H^2}\|b\|_{H^2}=\|P\|_{H^1}$.
Boundary equalities hold almost everywhere.
These analytic factors can have infinite Taylor series.
Truncate both after their first $N$ coefficients: their product still agrees through degree $N-1$, while neither coefficient norm increases.
Reverse and conjugate the first vector to obtain the output clock, and use the second as the input clock.
This preserves every required causal correlation and proves the norm bound.
For $P=0$ take one vector zero.
Normalized clocks are obtained by the same orthogonal padding as in \cref{thm:iir-clock-body}.
\end{proof}

The replacement preserves only the causal coefficient extraction.
It need not preserve private garbage, exact terminal-catalyst cancellation, or Las Vegas query weights.
It can populate addresses forbidden by an additional support restriction and does not apply to a general time-varying causal kernel.
Constructing its exact factors can require nontrivial classical work; no polylogarithmic classical algorithm follows.
Once the resulting two finite vectors are given, their direct loading cost is $O(N)$.
The constant-label analytic clocks do not need this factorization at all.

\subsection{Resources for the algorithms and applications}
\label{app:resources}

All rows use \cref{def:resource-accounting}.
Let a finite system have public dimension $P$ and complete private dimensions $L_j$.
At horizon $N$, define its Hilbert-space dimension and binary width by
\begin{align}
D_c=NP+\sum_jr_jL_j,
\quad b=\lceil\log_2D_c\rceil.
\label{eq:resource-carrier}
\end{align}
For the explicit constant-label clocks, \cref{thm:iir-gates} gives
\begin{subequations}
\label{eq:resource-common}
\begin{align}
G_{\rm total}&\le \sum_jq_jG_j+NG_S +O\!\left(N+[(m+1)N+\sum_jq_j]b\log(2+b)\right),
\\
n_{\rm total}&\le b+s_{\rm clk}+O(1),
\quad q_j=\lfloor(N-1)/r_j\rfloor.
\end{align}
\end{subequations}
The oracle Hilbert spaces in $L_j$ include their persistent workspace; additional restored scratch is added if it was excluded from those spaces.
$G_S$ includes the implemented known system matrix.
These bounds retain that cost explicitly when an arbitrary known matrix has no given efficient circuit.
They do not infer gate efficiency from query complexity alone.
Finite-gate-set synthesis is counted once by \cref{eq:app-synthesis}.
Sequential repetitions multiply gate and query counts; their quantum workspace is reused.
Simultaneous online copies multiply all simultaneously occupied registers.

\begin{table}[htbp]
\centering\small
\begin{tabularx}{\linewidth}{@{}lXX@{}}
\toprule
Module or task & Oracle calls & Gates and qubits
\\
\midrule
\IIRTag{Chebyshev2} & One $V$ and one $V^\dagger$ & $G_V+G_{V^\dagger}+G_{2\Pi-I}$; original oracle registers and reflection scratch
\\
FPAA / OAA & $Q_P=N+1$, $Q_{P^\dagger}=N$; \cref{prop:mod-fpaa,cor:mod-oaa} & \cref{eq:mod-fpaa-resources}; the time record adds logarithmic width
\\
Sign / threshold & $4m+2$; choose $m$ by \cref{prop:mod-sign} & \cref{eq:mod-sign-gates}; threshold adds one LCU flag
\\
Hamiltonian simulation & $N-1$; \cref{prop:mod-ham} & \cref{eq:mod-ham-gates}; finite spectral or Schur realization
\\
Two-oracle average example & $N-1$, $\lfloor(N-1)/R\rfloor$ & \cref{eq:resource-common}, with the dimensions in \cref{eq:mod-average-gates}
\\
Known-norm state preparation & \cref{eq:mod-stateprep-cost} & Common formula for the star Hamiltonian, with two preparation calls per feedback operation
\\
Pad\'e exponential transform & \cref{eq:expapp-queries} & \cref{eq:expapp-resources,eq:expapp-packed}; degree $p$ controls the additional approximation
\\
Rational reciprocal & Clock schedule for at most $3d$ Cayley factors & \cref{eq:resource-common}; include all factors in $P,L_j,G_S$
\\
\bottomrule
\end{tabularx}
\caption{Simple modules.
Query bounds include the oracle directions specified in each construction.
Additional known gates are included explicitly or through the common finite-system formula.}
\label{tab:resources-modules}
\end{table}

\begin{table}[htbp]
\centering\small
\begin{tabularx}{\linewidth}{@{}lXX@{}}
\toprule
Algorithm & Oracle bound & Circuit resources
\\
\midrule
Separate-term simulation & \cref{eq:multi-hs-cost}, with $q_j=\lfloor(N-1)/r_j\rfloor$ & \cref{eq:resource-common}; $P=2^n$, $L_j=D\dim\mathcal K_j$
\\
Shared SELECT simulation & \cref{eq:multi-hs-select-counts,eq:multi-hs-select-bound} & Same registers; replace query gates by $q_SG_{\rm SEL}+\sum_jq_j(G_{P_j}+G_{P_j^\dagger})$
\\
Sparse-access simulation & \cref{eq:multi-hs-sparse-bound,eq:multi-hs-sparse-optimized} & Common formula with $m=r_1=1$, $P=2^n$, $L_1=D2^{2n}$ plus retained value and arithmetic registers; each feedback operation includes both directions of the position and value oracles
\\
SOS phase amplification & \cref{eq:multi-sos-cost} & Common formula with \cref{eq:multi-sos-space}; known cascade cost in \cref{app:multi-sos-cascade}
\\
Weighted state preparation & \cref{eq:wsp-unknown-cost,eq:wsp-known-cost} & Common formula with \cref{eq:wsp-dimensions,eq:wsp-width}
\\
Weighted direct QLSP & \cref{wql:eq:matrix-cost,wql:eq:state-cost} & \cref{eq:imp-gates,eq:imp-qubits}; matrix-only precision stage
\\
Weighted factor QLSP & \cref{wql:eq:factor-matrix-cost,wql:eq:factor-state-cost} & Same formula with the two factor-reciprocal invocations
\\
QLSP with classical certificates & \cref{cert:eq:cutoff-cost,cert:eq:factor-cost,cert:eq:action-main-cost} & \cref{cert:eq:gates,cert:eq:width}; a finite-query approximation to the inverse on the specified input includes its coefficient-loading circuit
\\
Linear DE, constant or varying coefficients & \cref{eq:vde-queries} & \cref{eq:vde-gates,eq:vde-qubits}; coherent time evaluation belongs to the given oracles
\\
Linear DE with several forcing terms & \cref{eq:vde-multi-matrix-cost,eq:vde-multi-source-cost} & \cref{eq:vde-multi-resources}; individual generator and preparation delays
\\
Volterra with classical certificates & \cref{vdc:eq:combined-cost,vdc:eq:weighted} & \cref{vdc:eq:resources}; the mesh contributes to the address width
\\
Linear DE with time envelopes & \cref{vde:env:queries,vde:env:green-queries} & Transformed-data Volterra bounds; coherent evaluation of $t(u)=F_\rho^{-1}(u)$ and the forcing normalization charged per call
\\
Amplified generalized eigenvalues & \cref{eq:gepamp-cost} & QLSP resources for \cref{eq:gepamp-full}, finite Fourier readout, and coherent amplitude amplification
\\
Generalized eigenvalues & \cref{eq:gep-eigenstate-cost} & Common formula for the three-port system, plus inverse QFT and confidence repetitions, as detailed below
\\
\bottomrule
\end{tabularx}
\caption{Multioracle algorithms.
Spectral simulation uses $G_S=O(G_E+\log D\,\log^2(N/\epsilon))$, where $G_E$ implements the weighted signal alignment.
The Schur-prefix alternative has its own gate cost of the known system matrix from \cref{app:qudit-schur}.}
\label{tab:resources-algorithms}
\end{table}

\begin{table}[htbp]
\centering\small
\begin{tabularx}{\linewidth}{@{}lXX@{}}
\toprule
Application & Oracle or weighted cost & Gates and qubits
\\
\midrule
Downfolded simulation & \cref{eq:app-exact-calls,thm:app-main} & \cref{eq:app-gates,eq:app-space}; cascade length enters logarithmically
\\
Electronic downfolding & Same queries with the given $\lambda_j,\chi_j,\kappa_j$ & \cref{eq:df-total-qubits}; compact orbital access costs are specified in \cref{subsec:app-electronic}
\\
Positive reduced inverse & \cref{eq:app-inverse-cost} & \cref{eq:app-inverse-gates,eq:app-inverse-space}; unequal rational factors use their actual gate cost of the known system matrix
\\
Kron score & \cref{thm:kg-main} & Reduced-inverse resources, multiplied by estimation repetitions; input-state preparation, test-state preparation, and readout circuits included
\\
Online MQSP & Actual event counts; periodic case \cref{eq:iir-td-timestamp} & Common routing and clock bound, with $NG_S$ replaced by $\sum_tG_{S_t}$ and time-specific oracle costs
\\
Online Kalman mean & One call per nonzero-weight observation preparation and scheduled transport event & Mode/work registers: $w$ as in \cref{thm:qkf-online}; additional addresses: $O(\log(d+d_y)+\log(T+1))$ qubits. Small-channel two-level rotations: $O(Td(d+d_y)+d^2)$, plus clock loading, routing, oracle gates, and elementary-gate synthesis.
\\
\bottomrule
\end{tabularx}
\caption{Applications.
Kron reduction assumes the given classical error and response bounds.
Using $m$ simultaneous online copies multiplies all three resource counts by that number.
The required number is given in \cref{eq:qkf-copy-count}; all calls occur while their event is available.}
\label{tab:resources-applications}
\end{table}

For generalized eigenvalue estimation, the linear system in \cref{eq:gep-affine} has three matrix terms, including one known boundary term, and one prepared right-hand side.
Use \cref{prop:qlsp-resources} with the rates in \cref{eq:gep-solver-cost}.
Both the $n$-position phase register and the $N=2n$ computation-time register belong to its oracle Hilbert space.
Here $n=O(R/\epsilon)$ at constant window error.
Direct Kaiser state preparation uses $O(n)$ known gates per right-hand-side preparation, and the inverse quantum Fourier transform uses $O(\log^2 n)$ gates per accepted attempt.
More efficient structured preparation may replace the direct bound if given.
Sequential confidence repetition multiplies the gates and queries by $O(\log(2/\delta))$ while reusing quantum registers; stored outcomes use $O(\log n\log(2/\delta))$ classical bits.

For the amplified interval construction, use the padded history and metric-reweighted output in \cref{eq:gepamp-full}.
The metric branch count is $1+\lceil\log_2\kappa_B\rceil$, and the finite Fourier register has $O(\log(n/\eta))$ qubits with the error budget in \cref{app:gepamp}.
Apply the same QLSP resource theorem to these enlarged coefficient registers, counting their known branch rotations and arithmetic.
The eight coherently retained inversion attempts cost only a constant factor.
Amplitude amplification then multiplies the per-attempt gate count by $O(p^{-1/2}\log(2/\delta))$ for metric-reweighted output, while reusing its quantum registers.
Each reflection also tests those registers; its known-gate cost is included.
No metric branch multiplies the number of calls to the original $B$ oracle.

\subsubsection{FPAA and OAA}
Let $w_P$ be the preparation width, and let $G_P,G_{P^\dagger}$ count its forward and inverse gates.
Write $G_0$ for the cost of a controlled reflection about the all-zero state.
One implementation uses
\begin{align}
w_P+\lceil\log_2(N+2)\rceil+O(1)\text{ qubits},
\quad (N+1)G_P+NG_{P^\dagger}+NG_0+O(N\log(N+2))\text{ gates}.
\label{eq:mod-fpaa-resources}
\end{align}

\subsubsection{Exponential-transform simulation}
Here $a$ is the oracle signal width and $s_O$ its reusable scratch.
Let $T$ be the exponential-transform parameter and $s$ the quadrature-node count from \cref{thm:exp-implementation}.
With $D$ finite modes and mode-address width $b_{\module{Exp}}$,
\begin{subequations}
\label{eq:mod-ham-gates}
\begin{align}
b_{\module{Exp}}=O(\log(2+TN+s)),
\quad d_c=N+D2^a,
\quad b_c=\lceil\log_2d_c\rceil,
\quad q_O=N-1,
\\
n_{\rm total}\le n+b_c+s_O+O(1),
\\
G\le q_OG_O+ O\!\left(N\left[(a+b_{\module{Exp}})\log^2\frac{(1+\delta)N}{\epsilon} +b_c\log(2+b_c)\right]\right).
\end{align}
\end{subequations}
\subsubsection{Sign and threshold}
For an $n$-qubit matrix and $a$ encoding ancillas, its direct-sum port dimension and resources satisfy
\begin{subequations}
\label{eq:mod-sign-gates}
\begin{align}
d_c&=2N+1+N2^a,
\\
n_{\rm total}&\le n+\lceil\log_2d_c\rceil+s_O+O(1),
\\
G&\le q_OG_O+O\!\left(N[a+\log(N+2)] \log(2+a+\log(N+2))\right).
\end{align}
\end{subequations}

\subsubsection{Volterra differential equations}
The one-generator construction has resources \cref{eq:vde-gates,eq:vde-qubits}; the ordered reciprocal gives the $\log^2(1/\epsilon)$ known-gate term in that bound.
For several generators and forcing terms, \cref{eq:vde-multi-resources} includes all port preparations, scale operations, and routing.
Coherent time evaluation belongs to the given-oracle gate costs in both cases.

\paragraph{Classical-certificate Volterra bounds}
\label{vdc:sec:resources}
For the single-encoding certificate solver, let $w_V$ include the data and full oracle workspaces and the $O(\log(mN))$ Volterra labels.
Let $G_H,G_0,G_b$ bound controlled forward or inverse oracle costs.
With
\begin{align}
b_V=w_V+O\!\left(\log K_\epsilon +\log\log(1/\epsilon) +\log\log(K_\epsilon/R_0)\right),
\label{vdc:eq:width}
\end{align}
the unchanged QLSP construction yields
\begin{subequations}
\label{vdc:eq:resources}
\begin{align}
G_{\rm gates}&\le Q_HG_H+Q_0G_0+Q_bG_b
\\
&
\quad+O\!\left(K_\epsilon\left[ \log(K_\epsilon/R_0)+\log^2(1/\epsilon)\right] b_V\log(2+b_V)\right),
\\
q_{\rm qubits}&\le b_V+O(1),
\end{align}
\end{subequations}
in addition to explicitly given reusable arithmetic scratch.
To obtain this bound, use the circuit parameters $J=O(\log(K_\epsilon/R_0))$, $N_0=O(K_\epsilon)$, $d=O(\log(1/\epsilon))$, and $N_1=O(K_\epsilon\log(1/\epsilon))$.
The known work is $O((JN_0+dN_1)b_V\log(2+b_V))$; the labels for those dimensions use the space in \cref{vdc:eq:width}.
The extra precision logarithm in known work is not an extra matrix-query factor.
Coefficient and clock operations for the combined preparation are absorbed in this bound or in the given oracle costs.
The synthesis and oracle-error convention is \cref{def:resource-accounting}.
The mesh contributes $\log N$ to the width; a certificate that requires exponentially many samples can therefore still have a large gate overhead.

\subsubsection{Weighted state preparation}

Let $P$ be the public dimension of the assembled finite system.
Denote its complete private dimensions by $L_j^{\rm full}$, indexed by oracle type.
In particular, the latter include the branch paths and all $D$ occurrences.
With
\begin{align}
D_c=NP+\sum_jr_jL_j^{\rm full},
\quad b_{\rm pack}=\lceil\log_2D_c\rceil,
\label{eq:wsp-dimensions}
\end{align}
a binary-address implementation requires $b_{\rm pack}+O(1)$ qubits for the constant-label analytic clocks, including direct clock preparation.
If every preparation acts on at most $n+b$ qubits, direct-sum storage gives
\begin{align}
b_{\rm pack}=n+b+ O\!\left(\log\left[N+D\sum_jr_jL_j\right]\right).
\label{eq:wsp-width}
\end{align}
For the explicit cascade with $J$ HamSim stages, replace the private mode count by $DJ$ in this dimension bound.
Include the cascade's arithmetic and rotations in the system-matrix gate cost $G_S$ below.
Thus its additional spatial depth costs a logarithmic number of qubits, and its query horizon remains \cref{eq:wsp-hamsim-horizon}.
The gate count follows from \cref{thm:iir-gates} using this Hilbert space, the assembled system cost $G_S$, and the costs of the specified feedback operations $G_j$.
Here $q_j^{\rm ret}$ counts those feedback operations; any constant multiplicity of ports for the forward and adjoint oracles is absorbed into this bound.
Coefficient loading, reflections about the initial state, and phase-table evaluation are included in $G_S$.
Efficient indexed circuits or grouped delays may improve this gate count; their cost is not implied by the query bound.
Rotation synthesis adds its usual logarithmic precision cost.

\subsubsection{Classical-certificate linear-system solvers}\label{app:certificate-resources}
For the single-encoding cutoff construction, \cref{prop:qlsp-resources} applies to the unchanged circuit with the following sizes:
\begin{subequations}
\label{cert:eq:resources-parameters}
\begin{align}
J&=O\!\left(1+\log\left(2+K/r_0\right)\right),
\quad N_0=O(K),
\quad s=O(1+r_0),
\\
d&=O\!\left(\log\frac1\epsilon\right),
\quad N_1=O\!\left(K\log\frac1\epsilon\right).
\end{align}
\end{subequations}
The symbols denote, respectively, the number of scale stages, coarse horizon, preparation delay, reciprocal section count, and precision horizon.
Let $w$ count the full original oracle registers, including block-encoding ancillas and any unmeasured work, and put
\begin{align}
b=O\!\left(w+\log\bigl[2+(J+1)(N_0+s+1)(d+1)(N_1+1)\bigr]\right).
\label{cert:eq:width}
\end{align}
A conservative implementation has $b+O(1)$ logical qubits, in addition to any separately specified reusable arithmetic scratch, and
\begin{align}
G\le Q_AG_A+Q_bG_b+ O\!\left[(JN_0+dN_1)b\log(2+b)\right]
\label{cert:eq:gates}
\end{align}
known one- and two-qubit gates before discrete-gate-set synthesis.
The costs $G_A,G_b$ bound the gate cost of a controlled forward or inverse call to the corresponding original oracle.
The bound includes arbitrary clock-state loading linear in the horizons.
The coarse labels coexist with the precision clock.
Reciprocal-section processing contributes the factor $d$ to the gate count while the matrix-query count remains as stated in \cref{cert:eq:cutoff-cost}.

For \cref{cert:prop:short}, the gate count is multiplied by the bounded amplification budget.
One attempt reuses the given circuit's workspace; coherently retaining both bounded attempts requires only a constant number of such workspaces and the selection records.
For an LCU of powers, \cref{vdc:lem:powers} reuses one signal register and stores the first failure in a logarithmic-size register.
Loading an arbitrary coefficient list still costs $O(d)$ gates.

\section{Module realizations and approximation bounds}
\label{app:mod-details}

\subsection{Primitive module realizations}

\subsubsection{The FPAA metric and analytic bounds}
\label{app:mod-fpaa}
In the bad/good basis of \cref{prop:mod-fpaa}, put $s_\mu=2\mu\sqrt{1-\mu^2}$ and $t_\mu=2\mu^2-1$.
Direct multiplication gives
\begin{align}
M=\begin{pmatrix}s_\mu/\sqrt c&(1-c)t_\mu/(2\sqrt c)
\\
(1-c)t_\mu/(2\sqrt c)&s_\mu\sqrt c\end{pmatrix},
\quad K^\dagger MK=cM.
\label{eq:app-fpaa-metric}
\end{align}
The two eigenvalues of $M$ are $[(1+c)/(2\sqrt c)]\,[s_\mu\pm(1-c)/(1+c)]$.
Since
\begin{align}
\frac{\lambda_0}{\sqrt{1+\lambda_0^2}}\le\mu \le\frac1{\sqrt{1+\lambda_0^2}},
\quad s_\mu\ge\frac{2\lambda_0}{1+\lambda_0^2} >\frac{1-c}{1+c},
\end{align}
the metric is positive and its condition number is at most $\chi$.
For every vector $\boldsymbol{v}$, $\norm{K^j\boldsymbol{v}}^2\le\chi c^j\norm{\boldsymbol{v}}^2$.
The transmitted amplitudes are $LK^j\boldsymbol{v}$; the identity $K^\dagger K+L^\dagger L=I$ telescopes to
\begin{align}
\sum_{j=0}^{N-1}|LK^j\boldsymbol{v}|^2+\norm{K^N\boldsymbol{v}}^2 =\norm{\boldsymbol{v}}^2.
\label{eq:app-fpaa-conservation}
\end{align}
On the prepared plane every transmitted amplitude is proportional to the same good state, proving the factorization of the successful record.
The flag rotation ensures that this metric bound covers the entire original range $\lambda_0\le\lambda\le1$.

For $1<R<c^{-1/2}$, the convergent geometric series gives
\begin{align}
\sup_{|z|\le R}\norm{(I-zK)^{-1}} \le\frac{\sqrt\chi}{1-R\sqrt c},
\quad M_R=c+\frac{(1-c^2)R\sqrt\chi}{1-R\sqrt c}.
\label{eq:app-fpaa-resolvent}
\end{align}
Here $M_R$ bounds the scalar target transfer function and the orthogonal first-order all-pass transfer function.
Disk contractivity gives the interior bound.
On the exterior side, the actual nonnegative weight in \cref{eq:mod-fpaa-weight} allows $M_R\le M_R|z|^w$.
Thus \cref{thm:iir-analytic-clock-selection} applies with $\zeta\le\min\{1/16,(R-1)/2\}$, $M_{\rm loc}=M_R$, and $a=h_-=h_+=0$.
These conservative analytic bounds already ensure logarithmic precision cost.
The finite preparation bound is sharper because it uses the private contraction directly.

For the QSP comparison, the Chebyshev failure amplitude is $T_q(\sqrt{1-\lambda^2}/\sqrt{1-\lambda_0^2})/ T_q(1/\sqrt{1-\lambda_0^2})$ for odd $q$.
Its numerator is bounded by one on the promise, and its denominator is $\cosh(q\artanh\lambda_0)$.
The smallest eligible odd degree is $\log(1/\epsilon)/(2\artanh\lambda_0)+O_{\lambda_0}(1)$ for failure probability $\epsilon$.
Counting both directions of preparation and the initial call gives the convention used in the main table.
The physical damping picture is also related to critically damped quantum search \cite{Mizel2009DampedSearch}; the metric, record, and query convention above specify the construction being compared.

\subsubsection{The finite sign lattice}
\label{app:mod-sign-path}
For a scalar $x\in[-1,1]$, define
\begin{align}
T(z;x)=\begin{pmatrix}z&x
\\
xz&1\end{pmatrix},
\quad r_L=\frac{[T(z;x)^L]_{12}}{[T(z;x)^L]_{22}},
\quad u_L=\frac{(1-x^2)^{L/2}z^{L/2}}{[T(z;x)^L]_{22}}.
\end{align}
Eliminating the internal coordinates of the alternating path gives
\begin{align}
F_L(z;x)=\begin{pmatrix}zr_L&zu_L
\\
u_L&-zr_L\end{pmatrix}.
\label{eq:app-sign-finite}
\end{align}
Its public spaces are the two endpoint copies of $\mathcal H$ and its private space is the full path.
Equivalently it is $zJ^\dagger W(I-z(I-JJ^\dagger)W)^{-1}J$, where $W=U_{\rm odd}U_{\rm even}$.
At the evaluation point, the two coefficients are $r_L(1;x)=\tanh(L\artanh x)$ and $u_L(1;x)=\sech(L\artanh x)$.
These formulas follow by taking powers of the two-by-two transfer matrix.

The inverse Schur map $\mathcal S_{x,z}(v)=(x+zv)/(1+xzv)$ satisfies
\begin{align}
\mathcal S_{x,z}(v)-\mathcal S_{x,z}(w)= \frac{z(1-x^2)(v-w)}{(1+xzv)(1+xzw)}.
\end{align}
Consequently, $r_L-f=O(z^L)$.
The transmission $u_L$ begins at degree $L/2$.
Choose $L=2N$.
Then \cref{eq:app-sign-finite} agrees with $\diag(zf,-zf)$ in every coefficient below degree $N$.
Agreement of these coefficients gives exact equality of the finite Toeplitz blocks used at horizon $N$.

For the analytic bound, take $\lambda_0\le|x|\le1$ and put $\Delta(z;x)=(1-z)^2+4x^2z$.
Its branch points are distance $2|x|$ from $1$.
The square-root argument stays separated from zero on the local disk: $|z-1|\le\lambda_0/16 $ implies $ |\Delta/(4x^2)-1|<1/8$ and $|\sqrt\Delta|\ge\sqrt{7/2}|x|$.
Differentiating the quadratic equation for $F_{\rm s}$ gives
\begin{align}
\frac{zF'_{\rm s}}{F_{\rm s}} =\frac12+\frac{z+1}{2\sqrt{\Delta(z;x)}},
\quad \left|\frac{zF'_{\rm s}}{F_{\rm s}}\right|\le\frac{2}{\lambda_0}.
\label{eq:app-sign-derivative}
\end{align}
On the unit-circle arc through the neighborhood, $|F_{\rm s}|=1$.
Integration along exterior radial segments proves \cref{eq:mod-sign-tag}.
Evaluating the exact derivative at $z=1$ gives $(1+|x|^{-1})/2$.
The functional calculus transfers these bounds uniformly to the gapped reducing subspace of $A$.

\begin{lemma}[Finite Sign error and analytic neighborhood]
\label{lem:finite-sign-cap}
Take an even lattice length and a gapped signal satisfying $g\le|x|\le1$ and $Lg\ge c_0$, where $c_0$ is a sufficiently large universal constant.
The complete endpoint transfer function in \cref{eq:app-sign-finite} is bounded by a universal constant on $|z-1|\le cg$, for a sufficiently small universal $c>0$.
On a smaller such disk its derivative is $O(1/g)$.
Its physical error is at most $2e^{-L\operatorname{artanh}|x|}$.
\end{lemma}
\begin{proof}
At $z=1$,
\begin{align}
r_L(1;x)=\tanh\bigl(L\operatorname{atanh}x\bigr),
\quad u_L(1;x)=\operatorname{sech}\bigl(L\operatorname{atanh}x\bigr).
\label{eq:wsp-path-physical}
\end{align}
Therefore $F_L(1;x)$ is a Hermitian involution and
\begin{align}
\left\|F_L(1;x)- \begin{pmatrix}\operatorname{sgn}x&0
\\
0&-\operatorname{sgn}x\end{pmatrix} \right\| =\sqrt{2\bigl(1-\tanh(L\operatorname{atanh}|x|)\bigr)} \le2e^{-L\operatorname{atanh}|x|}.
\label{eq:wsp-path-error}
\end{align}
The norm estimate includes the contribution of the other endpoint.

To obtain a bound uniform in the lattice path length, restrict the signal and delay to $g\le|x|<1$ and $|z-1|\le cg$, for a sufficiently small universal constant $c$.
The eigenvalues of the transfer matrix are
\begin{align}
\mu_\pm(z;x)=\frac{1+z\pm\sqrt{(1-z)^2+4x^2z}}2,
\quad \mu_\pm(1;x)=1\pm|x|.
\label{eq:wsp-transfer-roots}
\end{align}
The square root is the analytic branch positive at $z=1$.
It is nonzero on this disk.
Its expansion at $z=1$ gives, uniformly in $x$,
\begin{align}
\left|\frac{\mu_-}{\mu_+}\right|\le e^{-c_1|x|},
\quad \frac{1-\mu_-}{\mu_+-\mu_-}=\frac12+O(c),
\quad \frac{\mu_+-1}{\mu_+-\mu_-}=\frac12+O(c)
\label{eq:wsp-root-bounds}
\end{align}
for a universal $c_1>0$.
These inequalities also follow directly by factoring $4x^2z$ out of the square root; the remaining relative correction has size $O(c^2)$.
In particular,
\begin{align}
[T^L]_{22} =\mu_+^L\left[ \frac{1-\mu_-}{\mu_+-\mu_-} +\frac{\mu_+-1}{\mu_+-\mu_-} \left(\frac{\mu_-}{\mu_+}\right)^L\right]
\label{eq:wsp-path-denominator}
\end{align}
has no zero if $Lg$ exceeds a universal constant.
Also $[T^L]_{12}=x(\mu_+^L-\mu_-^L)/(\mu_+-\mu_-)$ and $\det T=\mu_+\mu_-=(1-x^2)z$.
Substitution in \cref{eq:app-sign-finite} bounds the complete $2\times2$ transfer function by a universal constant on this disk, independently of $L$.
On a smaller disk, Cauchy's estimate consequently gives $\|\partial_zF_L(z;x)\|=O(1/g)$.
The cases $x=\pm1$ follow from $F_L(z;\pm1)=\operatorname{diag}(\pm z,\mp z)$.

\end{proof}

\subsubsection{Rational reciprocal completion and its unitary sections}
\label{app:mod-reciprocal}

\begin{proof}[Proof of \cref{lem:mod-reciprocal}]
For real $u$, $|p_d(u)|<h_d(u)$.
Inside the unit interval, use $|p_d(u)|\le1$ and strict inequality at the possible endpoints.
Outside it, use $|p_d(u)|\le|u|^{2d}$.
Thus the real monic polynomial $h_d^2-p_d^2$ is strictly positive on the real axis.
Its roots come in conjugate pairs.
Selecting one root from each pair constructs a monic $q_d$ of degree $2d$ with the required product.
Matrix multiplication proves unitarity of \cref{eq:mod-reciprocal-completion}.
The identity
\begin{align}
u^{-1}-f_d(u)=\frac1{u(1+u^{2d})}
\end{align}
gives \cref{eq:mod-reciprocal-error}.

Let $B(u)=\prod_{j=0}^{d-1}B_j(u)$.
Multiplication by $B$ cancels the upper-half-plane poles of $\mathcal R_d$, leaving a rational matrix analytic in that half-plane, unitary on its boundary, and constant at infinity.
Its determinant is $B^2$, since $\det\mathcal R_d=1$.
Lossless factorization gives
\begin{align}
B(u)\mathcal R_d(u)=U_\infty \prod_{k=1}^{2d}\left[I_2+(B_{j_k}(u)-1) |v_k\rangle\langle v_k|\right].
\label{eq:mod-reciprocal-factors}
\end{align}
One may obtain the factors by choosing a kernel vector at an upper zero and dividing out the corresponding rank-one all-pass factor.
That step reduces the zero multiplicity by one and preserves boundary unitarity and analyticity; repeating removes the $2d$ zeros of the determinant.
The remaining rational inner matrix is constant.
This is the square lossless factorization used in \cite{Alpay2014RationalUnitaryFunctions}.
Append the $d$ scalar inverse factors $B_j^{-1}I_2$.
This recovers $\mathcal R_d$ and proves the section count.

For a Hermitian contraction accessed through a self-inverse encoding, use the scaled signal $H=H^\dagger$, $\|H\|\le1$, $O_H=\Be{H}$, and $u=2\kappa H$.
Set
\begin{align}
\ell_j=1+\frac{|a_j|}{2\kappa},
\quad H_j=\frac{H-a_jI/(2\kappa)}{\ell_j},
\quad t_j=\frac{2\kappa\ell_j}{b_j},
\quad c_j=\frac{1-t_j}{1+t_j},
\quad s_j=\frac{2\sqrt{t_j}}{1+t_j}.
\end{align}
A two-term LCU of $O_H$ and $-\operatorname{sgn}(a_j)I$ gives a self-inverse encoding of the shifted normalized signal.
Write this encoding as $O_j=\Be{H_j}$; it uses one controlled call to the original oracle.
Its signal embedding is $E_j=|0\rangle_{\rm shift}|0\rangle_H\otimes I$ and $\Pi_j=E_jE_j^\dagger$.
A zero shift omits the identity branch.
For the factor associated with the known public state $|v\rangle$, use the system matrix in public/private order
\begin{align}
S_{j,v}=\left(\begin{array}{c|c} [I_2-(1+c_j)|v\rangle\langle v|]\otimes I &is_j|v\rangle\otimes E_j^\dagger
\\
\hline
s_j\langle v|\otimes E_j &i[I-(1-c_j)\Pi_j] \end{array}\right),
\quad \oracle_j(z)=-zO_j.
\label{eq:mod-reciprocal-system}
\end{align}
On the active signal and private signal it is the unitary matrix $\left(\begin{smallmatrix}-c_j&is_j\\ s_j&ic_j\end{smallmatrix}\right)$; it is the identity on the other public combination and $iI$ on the remaining private coordinates.
Eliminating the private vector, using $O_j^2=I$, gives at $z=1$ the active transfer function $-(I+it_jH_j)(I-it_jH_j)^{-1}=B_j(2\kappa H)$.
For $B_j^{-1}I_2$, apply this active system on both public ports and use feedback $+zO_j$.
The result at $z=1$ is $-(I-it_jH_j)(I+it_jH_j)^{-1}$ on each port.
Each system is lossless and contractive for $|z|<1$.
The cascade rule retains all private copies and inserts the known rotations determining $|v_k\rangle$ and $U_\infty$.
It therefore realizes the full unitary extension with the specified delay variables.
\end{proof}

For this factorization, define $\module{Reciprocal}_{d,\kappa}(\Bz;x)$ by the ordered cascade $A_k+B_k\oracle_k(z_k)(I-D_k\oracle_k(z_k))^{-1}C_k$ and the final public rotation $U_\infty$.
The operator substitution is $x=H$.
Its value at $z=1$ is $\mathcal R_d(2\kappa x)$.
The known cascade system is $S_{\module{Reciprocal}}$; the retained section oracle connections, including their signs, are part of this realization.

For an effective Hamiltonian obtained by closing the internal feedback connections, the same sections use $u=2H_{\mathrm{eff}}/a$, $c=(a/2)a_j$, and $\gamma=(a/2)b_j$.
A known termination adds $icP$ to the load of \cref{lem:app-native}; choosing system matrix scale $1/\gamma$ and a known minus phase gives the required all-pass factor.
Reversing the feedback sign and the shift gives its inverse at one while preserving nonnegative powers of the delay variables.
Hence the factorization does not require an oracle for $H_{\mathrm{eff}}$ itself.
The common analytic estimate for these original component feedback connections is proved in \cref{eq:app-inverse-poles}.

\subsection{Improved polynomials for QSP/QSVT}
\label{app:improved-polynomials}
\subsubsection{Sign polynomials and their clock realization}
The coefficient window below gives the sharp fixed-gap leading sign count used in \cref{subsec:mod-sign}.
Its clock realization connects this polynomial to the finite IIR lattice.

\begin{proposition}[A feasible clock with sharp fixed-gap rate]
\label{prop:mod-sign}
Fix $0<\lambda_0<1$ and choose $0<\epsilon<e^{-2}$.
Then \SignIIR\ implements $\Be{\widetilde S/2}$ with error $\norm{\widetilde S-\sgn A}\le\epsilon$ on the gapped subspace, using
\begin{align}
q_O\le\frac{\log(1/\epsilon)+\tfrac12\log\log(1/\epsilon)} {\artanh\lambda_0}+O_{\lambda_0}(1).
\label{eq:mod-sign-count}
\end{align}
Choose an integer $m$ satisfying the error and clock-norm tests in the proof.
One finite construction uses $N=2m+2$, $L=2N$, and $q_O=2(N-1)=4m+2$.
Its gate and qubit counts are \cref{eq:mod-sign-gates}.
\end{proposition}
\begin{proof}
We construct a bounded sign polynomial by integrating a nonnegative kernel, then bound the clock norm that selects it from the lattice.
Use
\begin{align}
K_m(x)=\left[\frac{T_m((1+\lambda_0^2-2x^2)/(1-\lambda_0^2))} {T_m((1+\lambda_0^2)/(1-\lambda_0^2))}\right]^2,
\quad p_m(x)=\frac{\int_0^xK_m(u)\,du}{\int_0^1K_m(u)\,du}.
\label{eq:mod-sign-kernel}
\end{align}
It is a monotone odd polynomial bounded by one on $[-1,1]$.
For $m\ge2/\lambda_0$, its error on the gap is at most $4e\sqrt{8m/(3\lambda_0)}e^{-4m\artanh \lambda_0}$.
\cref{app:mod-sign-clock} constructs an actual clock selecting $p_m$ and proves that its sufficient nuclear-norm bound tends to $1-2\erf(1)+12/(e\sqrt\pi)<2$.
We choose $m$ satisfying both the error inequality and the finite clock-norm test.
The norm limit proves termination at each fixed $\lambda_0$, and inversion of the error inequality gives \cref{eq:mod-sign-count}.
The full-support causal factorization of \cref{sec:quditization} replaces this clock by one pair of endpoint vectors at no greater normalization, with $O(N)$ direct loading gates.
The path has private dimension $2^n(1+N2^a)$ and two public copies of $\mathcal H$; matchings and endpoint routing give \cref{eq:mod-sign-gates}.
\end{proof}

The classical minimax error in odd degree $2k+1$ is asymptotic to $(1-\lambda_0)e^{-2k\artanh \lambda_0}/\sqrt{\pi\lambda_0 k}$ \cite{Eremenko2007SignApproximation}.
To obtain a QSVT implementation, let $E$ be this minimax error.
Alternation places all $2k$ derivative zeros in the two retained intervals, so the odd polynomial is monotone inside the gap and has global norm at most $1+E$.
Dividing by $1+E$ gives a globally bounded odd polynomial with gap error at most $2E$.
Consequently, QSVT achieves degree $(\log(1/\epsilon)-\tfrac12\log\log(1/\epsilon))/\artanh\lambda_0 +O_{\lambda_0}(1)$ \cite{Gilyen2018singular}.
Our clock has the same leading coefficient and the larger logarithmic correction displayed in \cref{eq:mod-sign-count}.
For a full-space gapped input, oblivious amplification at normalization two uses three copies of the selected circuit and gives a sign operation with isometry error $O(\epsilon)$.
A controlled Hadamard construction then selects $(I\pm\sgn A)/2$.
Its extra control flag is constant-sized.

\subsubsection{The sign clock: boundedness and nuclear norm}
\label{app:mod-sign-clock}
Write $f(z;x)=\sum_{j\ge0}b_j(x)z^j$.
The Legendre generating function yields $b_0=x$ and
\begin{align}
b_j(x)=\frac{P_{j-1}(1-2x^2)-P_{j+1}(1-2x^2)}{2x(2j+1)}
\quad(j\ge1).
\label{eq:sc-basis}
\end{align}
These form a triangular basis of the odd polynomials.
For an even nonnegative polynomial of degree $2d$, define its integral normalization and the corresponding odd polynomial by
\begin{align}
Z=\int_0^1K,
\quad p(x)=Z^{-1}\int_0^xK=\sum_{j=0}^da_jb_j(x).
\end{align}
The coefficients $a_j$ describe this polynomial in the transfer function basis.
Set $N=d+2$ and extend the coefficient sequence $a_j$ by zero.
Define the clock weights $\boldsymbol{c}$ and their matrix $X$ by
\begin{subequations}
\label{eq:sc-weights}
\begin{align}
c_0=2a_0-a_1,
\quad c_k=a_{k-1}\ (1\le k<N),
\quad c_N=c_{N+1}=0,
\\
d_n=n(c_{n-1}-2c_n+c_{n+1}),
\quad X=\sum_{n=1}^Nd_n\ket{v_n}\bra{v_n},
\quad \ket{v_n}=n^{-1/2}\sum_{j=0}^{n-1}\ket j.
\end{align}
\end{subequations}
Each uniform prefix state gives a triangular clock-weight profile, also called a Fej\'er profile.
Its second difference has only one nonzero entry:
\begin{align}
f_n(k)=(1-k/n)_+,
\quad f_n(k)-2f_n(k+1)+f_n(k+2)=\frac1n\boldsymbol{1}_{\{k=n-1\}}.
\end{align}
The second differences of the desired weights therefore determine the coefficients of these rank-one projectors.
With the two zero endpoint weights in \cref{eq:sc-weights}, summing twice reconstructs the sequence,
\begin{align}
\sum_{n=1}^N d_n(1-k/n)_+=c_k.
\end{align}
Thus the nonnegative-lag diagonal sums of the clock matrix are exactly the desired weights.
Since every prefix projector has nuclear norm one, the triangle inequality bounds the clock norm by the absolute sum of its coefficients.
The factor $zf$ has zero constant coefficient, so the zeroth clock weight is free.
The displayed choice sets $d_1=0$.
The selected transfer function is $p$ and $\norm X_*\le A(K):=\sum_n|d_n|$.
This norm bound proves that the polynomial can be selected with normalization at most $A(K)$.
Since these are full equal time supports, \cref{app:qudit-rankone} also permits a single clock pair at the same normalization once its classical factors have been computed.

To bound the clock norm, let $\ell_j$ be the Legendre coefficients of $(1-t)K(\sqrt{(1-t)/2})$.
Then, for $1\le j\le d+1$,
\begin{align}
d_{j+1}=-\frac{j+1}{Z}\ell_j =-\frac{4(j+1)(2j+1)}Z\int_0^1x^3K(x)P_j(1-2x^2)\,dx.
\label{eq:sc-legendre}
\end{align}
To prove it, differentiate \cref{eq:sc-basis}: $(xb_j)'=2xP_j(1-2x^2)$, including $j=0$.
Legendre orthogonality therefore gives directly
\begin{align}
a_j&=(2j+1)\int_0^1[p(x)+xp'(x)]P_j(1-2x^2)\,dx =Z^{-1}\int_0^1 K(u)C_j(u)\,du,
\\
C_j(u)&=(2j+1)\left[\int_u^1P_j(1-2x^2)\,dx+uP_j(1-2u^2)\right].
\end{align}
The Legendre recurrence and $P'_{j+1}-P'_{j-1}=(2j+1)P_j$ imply $C_{j-1}(u)-2C_j(u)+C_{j+1}(u) =-4(2j+1)u^3P_j(1-2u^2)$.
Substituting this into \cref{eq:sc-weights}, followed by $t=1-2u^2$, proves both forms of \cref{eq:sc-legendre} for $j\ge1$.
The free constant coefficient sets $d_1=0$ separately.

Choose the derivative kernel
\begin{align}
K_m(x)=\left[ \frac{T_m((1+\lambda_0^2-2x^2)/(1-\lambda_0^2))}{T_m((1+\lambda_0^2)/(1-\lambda_0^2))}\right]^2,
\quad Z_m=\int_0^1K_m(x)\,dx,
\quad p_m(x)=Z_m^{-1}\int_0^xK_m(y)\,dy.
\label{eq:sc-kernel}
\end{align}
The affine map sends $[\lambda_0^2,1]$ onto the interval with endpoints $1,-1$.
On that interval the Chebyshev polynomial has modulus at most one.
Squaring makes a nonnegative derivative kernel; integrating gives a globally bounded monotone odd polynomial.
Thus sign approximation reduces to the kernel mass outside the gap.
The derivative kernel's normalization denominator cancels from the integrated polynomial and clock weights.
A rational gap $\lambda_0$ therefore permits exact rational arithmetic throughout.

For $m\ge2/\lambda_0$ one has
\begin{align}
\delta_m:=1-p_m(\lambda_0) =\sup_{\lambda_0\le|x|\le1}|p_m(x)-\sgn x| \le4e\sqrt{\frac{8m}{3\lambda_0}}\,e^{-4m\artanh \lambda_0}.
\label{eq:sc-error}
\end{align}
Indeed, the Chebyshev value outside the interval gives $T_m((1+\lambda_0^2)/(1-\lambda_0^2))=\cosh(2m\artanh\lambda_0)$, so $K_m\le4e^{-4m\artanh\lambda_0}$ on $[\lambda_0,1]$.
For $0\le x\le \lambda_0/2$, put $v(x)=\operatorname{arcosh}[(1+\lambda_0^2-2x^2)/(1-\lambda_0^2)]$.
Then
\begin{align}
v(0)-v(x)=\int_0^x\frac{2u\,du}{\sqrt{(\lambda_0^2-u^2)(1-u^2)}} \le\frac{4x^2}{3\lambda_0},
\quad v(0)=2\artanh \lambda_0.
\end{align}
The ratio $\cosh(A-B)/\cosh A\ge e^{-B}$ gives $K_m(x)\ge e^{-8mx^2/(3\lambda_0)}$.
Integrating to $\sqrt{3\lambda_0/(8m)}<\lambda_0/2$ yields $Z_m\ge e^{-1}\sqrt{3\lambda_0/(8m)}$, which proves \cref{eq:sc-error}.
Monotonicity proves the exact supremum.

\paragraph{The clock-norm limit and its domination bound}
For every fixed $\lambda_0\in(0,1)$,
\begin{align}
A_m:=A(K_m)\longrightarrow A_*:=\frac8{\sqrt\pi}\int_0^\infty s^2|1-s^2|e^{-s^2}\,ds =1-2\erf(1)+\frac{12}{e\sqrt\pi}<2.
\label{eq:sc-limit}
\end{align}
We prove this limit by bounding the absolute sum uniformly in $m$.
Set $\sigma=\sqrt{2m/\lambda_0}$ and $k_m=K_m/Z_m$.
Since $v(x)=2\artanh \lambda_0-x^2/\lambda_0+O_{\lambda_0}(x^4)$, Laplace scaling gives $\sigma^{-1}k_m(y/\sigma)\to g(y)=2e^{-y^2}/\sqrt\pi$.

On $[0,\lambda_0/2]$, this convergence has Gaussian majorants with polynomial factors for every fixed number of derivatives: use $v(0)-v(x)\ge x^2/\lambda_0$, the smooth even exponent, and its derivatives.
On $[\lambda_0/2,1]$, the kernel and each fixed derivative are exponentially small: $O_{\lambda_0}(m^C e^{-c_{\lambda_0}m})$ and $c_{\lambda_0}>0$.
To avoid differentiating the square root at the gap edge, bound the polynomial by its value at $\lambda_0/2$, followed by the oscillatory bound beyond the gap.
Then apply the polynomial Markov derivative bound on that fixed interval.
Division by $Z_m=\Theta_{\lambda_0}(m^{-1/2})$ preserves exponential decay.
For $r=0,3$, these estimates imply
\begin{align}
\int_0^1x\left|\mathcal D^r[x^2k_m(x)]\right|\,dx \le C_{\lambda_0,r}\sigma^{2r-3},
\quad \mathcal D=-\frac1{4x}\frac d{dx}\left[x(1-x^2)\frac d{dx}\right].
\label{eq:sc-domination}
\end{align}
Under the rescaling $x=y/\sigma$, the factors contribute $x^2k_m:\ \sigma^{-1}$, $x\,dx:\ \sigma^{-2}$, and $\mathcal D:\ \text{at most }\sigma^2$.
The apparent singularity at zero cancels on even smooth functions.
This proves the stated powers using the majorants.

Use the integral representation
\begin{align}
P_j(\cos\theta)=\frac1\pi\int_0^\pi (\cos\theta+i\sin\theta\cos\phi)^j\,d\phi.
\end{align}
When $j/\sigma\to s$, it gives $P_j(1-2y^2/\sigma^2)\to J_0(2sy)$ uniformly on compact sets of the two scaled variables.
Its integrand has modulus at most one.
Applying \cref{eq:sc-legendre} and the Gaussian majorants gives
\begin{align}
\sigma d_{j+1}\longrightarrow -8s^2\int_0^\infty y^3g(y)J_0(2sy)\,dy =-\frac8{\sqrt\pi}s^2(1-s^2)e^{-s^2}.
\label{eq:sc-profile}
\end{align}
The last equality follows by differentiating the Gaussian--Bessel identity,
\begin{align}
\int_0^\infty ye^{-ay^2}J_0(2sy)\,dy=\frac{e^{-s^2/a}}{2a},
\end{align}
with respect to $a$, then setting $a=1$.

For an integrable bound on the coefficient sum, observe $\mathcal DP_j(1-2x^2)=j(j+1)P_j(1-2x^2)$.
The differential operator is self-adjoint for measure $x\,dx$; three integrations by parts have zero endpoint terms, since $x(1-x^2)$ vanishes at the endpoints and all functions are even regular polynomials at zero.
With $|P_j|\le1$ and \cref{eq:sc-domination},
\begin{align}
|d_{j+1}|\le\frac{C_{\lambda_0}}{\sigma} \begin{cases}1,&1\le j\le\sigma,
\\
(j/\sigma)^{-4},&j>\sigma.\end{cases}
\end{align}
This dominates the Riemann sums by an integrable function; coefficients past $2m+2$ are exactly zero by polynomial orthogonality.
Hence \cref{eq:sc-profile} can be integrated in absolute value.
Splitting the Gaussian integral at $s=1$ proves \cref{eq:sc-limit}.

Choose the least integer $m\ge2/\lambda_0$ for which the analytic error bound is at most $\epsilon/2$ and $A(K_m)\le2$.
For rational $\lambda_0$, the polynomial coefficients and their integrals admit exact rational arithmetic; otherwise an interval implementation with rigorous rounding bounds or a smaller rational gap suffices.
The fixed-gap limit proves eventual feasibility for each $\lambda_0$.
For a bound uniform in $\lambda_0$, the local growth estimate and \cref{thm:iir-analytic-clock-selection} give the $O(\lambda_0^{-1}\log(1/\epsilon))$ alternative.
Classical coefficient computation is separate from the quantum gate count.
Approximate endpoint preparations are allocated their physical error through \cref{eq:qudit-error-budget}.

\subsection{Repeated known systems and their full cascade}
\label{app:multi-sos-cascade}
We give a common implementation for a cascade of repeated system matrices.
The SOS application uses the biased system matrix in \cref{eq:multi-sos-system}.
The construction uses reversible arithmetic on the small known coefficient matrices.
Their classical descriptions, and the required alignments of the oracle input subspaces, are part of the implementation data.
The coefficients at binary addresses are computed from these matrices by reversible arithmetic.

After factoring out the unchanged data and aligning each input subspace $|0\rangle_j\otimes\mathcal X$, the nontrivial coefficient dimension of one system matrix is $d=O(m)$, together with scalar actions repeated on the orthogonal complements of the signal subspaces, indexed by the binary addresses.
In this subsection, $A,B,C,D$ denote the four known blocks of one system matrix.
The argument applies to the unitary colligation
\begin{align}
S=\left(\begin{array}{c|c}A&B
\\
\hline
C&D\end{array}\right).
\label{eq:multi-sos-general-cell}
\end{align}
It acts on all public and private inputs, with every output retained.
The $J$-stage cascade has full matrix
\begin{align}
S^{[J]}=\left(\begin{array}{c|c} A^J&B_J
\\
\hline
C_J&D_J\end{array}\right),
\quad B_J=(A^{J-1}B\ \cdots\ B),
\quad C_J=\begin{pmatrix}C
\\
CA
\\
\vdots
\\
CA^{J-1}\end{pmatrix},
\label{eq:multi-sos-cascade-matrix}
\end{align}
where $D_J$ is block lower triangular.
Its diagonal block is $D$.
At subdiagonal displacement $n\ge1$, the block is $CA^{n-1}B$.
Each of the four blocks in \cref{eq:multi-sos-cascade-matrix} is a contraction.
The public output of system matrix $\ell$ is the public input of system matrix $\ell+1$.
All $J$ private copies remain present.
Regrouping them by oracle label is a known permutation, and changes the feedback operation from $I_J\otimes \oracle(\Bz)$ to $\bigoplus_j[I_J\otimes z_jO_j]$.

\begin{lemma}[Known-cascade implementation]
\label{lem:multi-sos-known-cascade}
Suppose arithmetic on the specified $d$-dimensional known coefficients is available to the required precision.
The complete $S^{[J]}$ can be implemented to isometry error $\eta$ with
\begin{align}
\operatorname{poly}(d,\log(2J),\log(1/\eta))
\label{eq:multi-sos-poly-cascade}
\end{align}
known gates and arithmetic workspace, including the coefficient arithmetic, plus the gate cost of aligning the oracle registers.
\end{lemma}
\begin{proof}
The block $A^J$ is computed by repeated squaring.
The two boundary blocks have telescoping Gramians
\begin{align}
B_JB_J^\dagger=I-A^J(A^\dagger)^J,
\quad C_J^\dagger C_J=I-(A^\dagger)^JA^J.
\label{eq:multi-sos-boundary-gram}
\end{align}
Their normalized isometries have a binary-tree preparation.
At a node, compute the positive Gram matrices $Q_0,Q_1$ of its two subintervals and implement the isometry $\boldsymbol{x}\mapsto |0\rangle Q_0^{1/2}(Q_0+Q_1)^{-1/2}\boldsymbol{x} +|1\rangle Q_1^{1/2}(Q_0+Q_1)^{-1/2}\boldsymbol{x}$ on the support of $Q_0+Q_1$.
Each interval Gram is a difference of powers, by the same telescoping as \cref{eq:multi-sos-boundary-gram}.
At the leaves include the polar isometry of the corresponding $A^hB$ or $CA^h$ block.
A root contraction gives the unnormalized Gram square root.
This block-encodes either off-diagonal block, $B_J$ or $C_J$, at constant normalization.
It requires $O(\log J)$ levels of small-matrix arithmetic.
Node indices are binary addresses from which the matrices are computed coherently by the stated arithmetic.
Numerically negligible Gram singular values can be discarded below the desired squared-error threshold; supported inverses then have condition polynomial in the inverse target error.
Allocating accuracy by the number of levels gives a logarithmic bit-precision requirement.
Arbitrary unitary extensions on nullspaces do not change the specified boundary blocks.

For the interior block define the known Schur function
\begin{align}
W(z)=D+zC(I-zA)^{-1}B.
\label{eq:multi-sos-swapped-response}
\end{align}
Unitarity of the system matrix makes $W$ contractive for $|z|<1$.
Its Toeplitz compression is $D_J=\mathcal T_J[W]$.
Put $q=e^{-1/J}$.
Choose a power-of-two Fourier length $M=O(J\log(2J/\eta))$ and use the block circulant whose Fourier blocks are $W(qe^{2\pi i\ell/M})$.
These are contractions and can be block-encoded by their known unitary dilations.
Their small matrices are computable coherently by inversion, with $\|(I-qe^{i\theta}A)^{-1}\|\le(1-q)^{-1}\le J+1$.
A Fourier transform and its inverse therefore implement the whole circulant block encoding using only logarithmic address size and small-matrix arithmetic.

To recover the finite Toeplitz block from the circular embedding, we bound the aliasing error and undo the radial damping.
The Schur coefficient bound is $\|W_n\|\le1$.
In the upper square block of dimension $J$, each entry's alias error is at most $q^{M-J+1}/(1-q^M)$.
Multiplying this bound by the block dimension bounds its operator norm.
Choosing the constant in $M$ sufficiently large makes it $O(\eta)$.
With $L_q=\operatorname{diag}(q^{-h})_{h=0}^{J-1}$,
\begin{align}
\mathcal T_J[W]=L_q\mathcal T_J[W(qz)]L_q^{-1},
\quad \|L_q\|<e,
\quad\|L_q^{-1}\|=1.
\label{eq:multi-sos-radial-similarity}
\end{align}
Apply the diagonal contractions $L_q/e$ and $L_q^{-1}$.
This block-encodes $D_J$ at normalization $e$, with the stated alias and arithmetic errors.

Embed the four blocks of \cref{eq:multi-sos-cascade-matrix} in the full direct sum of public and history subspaces and combine their block encodings by LCU.
Their total normalization is at most $3+e<6$; known attenuation can set it to eight.
Since $S^{[J]}$ itself is unitary, constant-length phase-matched oblivious amplification implements it with controlled isometry error, with no original-oracle queries.
Performing the intermediate block encodings to $O(\eta^2)$ accuracy suffices for $O(\eta)$ isometry error.
All arithmetic accuracies and the Fourier register length have logarithmic bit size in $J$ and $1/\eta$, proving \cref{eq:multi-sos-poly-cascade}.
\end{proof}

The scalar rotation specialization also admits the exact dyadic factorization used in \cref{eq:app-dyadic}, following \cite{Chen2026GateEfficientSimulation}.
That factorization applies after the stated sector alignments.
The matrix proof above retains nonnormal public blocks and all boundary maps.
Unequal constraint system matrices instead use the ordered implementation in \cref{lem:ordered-cascade}; their $J$-stage known-gate cost is counted explicitly.
Online oracle calls retain their physical execution order under either known-gate construction.

\subsection{Spectral Exp and its finite-horizon error}
\label{app:mod-exp}
The spectral system is obtained by expanding the positive-real transform in \cref{eq:sp-exp-positive-real}.
In the impedance variable, the paired partial-fraction identity is
\begin{align}
\coth(Tq/2)=\frac{2}{Tq} +\sum_{k=1}^{\infty}\frac{4Tq}{T^2q^2+4\pi^2k^2}.
\label{eq:app-exp-coth}
\end{align}
Substitute the disk-to-half-plane change of variable and pair the opposite mode indices.
The resulting representation is
\begin{align}
q=\frac{1-w}{1+w},
\quad \frac{1+\module{Exp}_T(w)}{1-\module{Exp}_T(w)} =\sum_{k\in\mathbb Z}p_k\frac{1+u_kw}{1-u_kw},
\label{eq:app-exp-spectral-representation}
\end{align}
with the Cauchy weights and phases in \cref{eq:mod-exp-modes}.
For example, the zero mode gives the first term of \cref{eq:app-exp-coth}; each opposite-index pair gives one summand.
The weights are positive and sum to $\coth(T/2)$ by evaluating at the origin.
Normalizing these weights gives the reflection state in \cref{eq:sp-exp-system}.
Eliminating its private coordinates recovers the displayed positive-real function and hence Exp.
For the finite weights in \cref{eq:mod-exp-modes}, append the tail $p_*=W_T-\sum_{|k|\le R}p_k\ge0$ at $u_*=-1$ and set
\begin{subequations}
\label{eq:app-exp-factor}
\begin{align}
\ket v=\sum_k\sqrt{p_k/W_T}\ket k,
\quad c=e^{-T},
\quad \ket\chi=\sqrt{\frac{1-c}{2}}\ket{\mathrm{pub}} -\sqrt{\frac{1+c}{2}}\ket v,
\\
S_{T,R}=(1\oplus\diag(u_k))(I-2\ket\chi\bra\chi).
\end{align}
\end{subequations}
This is a unitary on the public mode and all private modes.
Padding unused public addresses by identity fixes its unitary completion.

For a nilpotent contraction, define $X^K=0$, $g_u(X)=(I+uX)(I-uX)^{-1}$, and $|u|=1$.
Its Hermitian part is $\operatorname{Re}g_u(X)= (I-\bar uX^\dagger)^{-1}(I-X^\dagger X)(I-uX)^{-1}\succeq0$.
The exact and finite positive-real transfer functions are therefore accretive.
Their Schur transforms differ by at most twice their operator difference.
The omitted pair of modes has phases $u_{\pm k}=-e^{\pm i\vartheta_k}$ and $\vartheta_k=2\arctan(T/(2\pi k))\le T/(\pi k)$.
Moving the pair's mass to $-1$ changes its moment of order $n$ by at most $p_kn^2\vartheta_k^2$.
Use the coefficient and tail bounds in the finite expansion: \begin{align}
p_k\le\frac{T}{2\pi^2k^2},
\quad \sum_{k>R}k^{-4}\le\frac1{3R^3},
\quad g_u(X)=I+2\sum_{n=1}^{K-1}u^nX^n.
\end{align} It follows that
\begin{align}
\norm{\module{Exp}_{T,R}(X)-\module{Exp}_T(X)} \le\frac{T^3K(K-1)(2K-1)}{9\pi^4R^3}.
\label{eq:app-exp-tail}
\end{align}
The constant term cancels because the mass is preserved.
This estimates the two transfer functions evaluated on the same finite contraction, as required by \cref{prop:iir-comparison-stability}.

The approximation used in the main text replaces the omitted positive symmetric spectral measure by its $s$-node Gaussian quadrature, exact through degree $2s-1$.
All phases remain on the unit circle.
Let $a_R=2\arctan(T/[2\pi(R+1)])$ and $M_{\rm tail}$ be the omitted mass.
Taylor's real-axis remainder bounds each paired moment error by $2M_{\rm tail}(na_R)^{2s}/(2s)!$.
Thus
\begin{align}
\norm{\module{Exp}_{T,R,s}(X)-\module{Exp}_T(X)} \le\frac{8M_{\rm tail}(K-1)[(K-1)a_R]^{2s}}{(2s)!} \le\frac8{\pi(2s)!}
\label{eq:app-exp-quadrature}
\end{align}
if $R\ge\max\{1,TK/\pi\}$.
Indeed, $(K-1)a_R\le1$ and $M_{\rm tail}(K-1)\le1/\pi$.
Taking $K=N$ proves the cutoff used in \cref{eq:mod-ham-gates}.
Known classical moment and quadrature preprocessing is required; positivity and moment-error tolerances must be certified.

\subsubsection{MPS preparation for the finite exponential transform}
The reflection amplitudes have a narrow central region and a slowly varying tail.
A polynomial over the whole address range would have to resolve both scales.
Dyadic bands avoid this problem: the leading nonzero address bit identifies an interval whose endpoints differ by a factor of two.
On a positive band define
\begin{align}
k=L(3/2+y/2),
\quad L=2^j,
\quad a=\frac{T}{2\pi L},
\quad -1\le y<1.
\label{eq:app-exp-dyadic-band}
\end{align}
The amplitude relative to its value at the lower endpoint is
\begin{align}
f_a(y)=\sqrt{\frac{a^2+1}{a^2+(3/2+y/2)^2}}.
\label{eq:app-exp-band-amplitude}
\end{align}
Its nearest possible complex singularities lie at $-3\pm2ia$, uniformly away from the approximation interval.
They are square-root branch points when distinct.
On the disk of radius $5/4$, the function is analytic and has modulus at most $8/7$.
On the real interval it is at least $1/2$.
Cauchy's coefficient estimate therefore gives a polynomial of degree $d=O(\log(1/\xi))$ with uniform relative error of order $\xi$.
The state error is weighted by the original probabilities, so summing over addresses introduces no dimension factor.

A matrix product state (MPS) describes the amplitude by passing a small auxiliary vector, called the bond, from one address bit to the next.
Here the bond initially has one state marking that no nonzero bit has appeared.
The first nonzero bit identifies the dyadic band and loads its polynomial coefficients.
Each later bit shifts the polynomial argument by its binary place value.
Such shifts act linearly on a degree-$d$ coefficient vector:
\begin{align}
p(y)=\sum_{k=0}^d a_ky^k
\quad\Longrightarrow
\quad [y^\ell]p(y+\eta)=\sum_{k=\ell}^d\binom{k}{\ell}a_k\eta^{k-\ell}.
\label{eq:app-exp-polynomial-shift}
\end{align}
Thus the bond needs only $d+1$ coefficient states and the initial state.
The terminal constant coefficient gives the desired amplitude.
Negative addresses, zero, and the exceptional tail add a constant number of states; the quadrature nodes add a number proportional to their count.
For the precision allocation above, the total bond is $O(\log(1/\xi))$, independently of the number of binary address qubits.

Sequential MPS preparation \cite{Schon2005MPS} uses unitary synthesis \cite{shende2006synthesis} on the output qubits.
For a normalized state on $b_{\module{Exp}}$ qubits with bond dimension $\chi$, it uses $O(b_{\module{Exp}}\chi^2)$ arbitrary one- and two-qubit gates.
One way to see the absence of a second bond register is to disentangle one qubit at a time using a window of $1+\lceil\log_2\chi\rceil$ output qubits, then reverse the circuit.
All virtual information occupies output qubits not yet prepared.
Conjugating a known zero-state reflection by this preparation implements the reflection about $\ket v$ on every input.
A normalized state error $\eta$ gives reflection error at most $2\eta$.

\subsubsection{The diagonal phases with constant extra workspace}
Rescale the positive dyadic band by $a=T/(2\pi)$, $k=Lx$, $1\le x<2$, and $A=a/L$.
The phase is $u_A(x)=(A-ix)/(A+ix)$.
Define
\begin{align}
\theta=\frac\pi2(x-3/2),
\quad y=\sin\theta,
\quad h(y)=\frac2\pi\arcsin y,
\quad x=3/2+h(y).
\end{align}
The coefficient one-norm of $h$ is one.
The geometric expansion in the rescaled band variable gives
\begin{align}
\frac{A-i(3/2+h)}{A+i(3/2+h)}
\quad\text{as a series in }h/(3/2).
\end{align}
Its coefficient one-norm is at most three, uniformly in $A$.
The Taylor truncation of degree $d$ is bounded by three on the entire unit disk.
Its error is at most $3\,2^{-(d+1)/2}$ on $|y|\le1/\sqrt2$.
The bound on the entire unit disk permits unitary synthesis while preserving the approximation on the band.

Normalize by $C=1+\sqrt5=1/\sin(\pi/10)$.
Substituting $y=(z-z^{-1})/(2i)$ gives a Laurent polynomial bounded by $3/C<1$ on the whole unit circle.
GQSP implements it with one signal qubit and $O(d)$ controlled phases \cite{Motlagh2024gqsp}.
Those phases are known bitwise operations:
\begin{align}
e^{i\theta(k/L)}=\exp\!\left[-\frac{i\pi}4 +i\sum_{j<\log_2L}k_j\frac\pi2\frac{2^j}{L}\right].
\end{align}
No numerical value is computed into a precision register.
Two steps of oblivious amplitude amplification use five copies and raise the constant amplitude $1/C$ to one.
Suppose the encoded operator differs from $U/C$ by at most $\eta\le.01$.
Its isometry error is then at most $24\eta$.
Polar decomposition controls the unitary part, while the amplification polynomial satisfies $f_5(s)=5s-20s^3+16s^5$, $f_5(1/C)=1$, and $f'_5(1/C)=0$.
The vanishing derivative makes the leakage linear in the input error as well.

Only $O(\log(1/\xi))$ bands need this processing.
On a band beginning at $L$, the following sufficient conditions place the phase within $\eta$ of a constant: $4L/a\le\eta $ for $ +1$ and $2a/L\le\eta $ for $ -1$.
The remaining band endpoints lie between $a\eta/4$ and $2a/\eta$.
With $d=O(\log(1/\xi))$, a band uses $O(b_{\module{Exp}} d)$ gates and one band flag.
The disjoint bands preserve their addresses, so their errors combine by a maximum.
Known prefix tests use constant restored scratch \cite{Khattar2025ConditionallyCleanAncillae}.
The zero and tail phases are exact; a quadrature boundary has an $O(s)$-gate known diagonal table.

Together with reflection-state preparation this proves $O(b_{\module{Exp}}\log^2(1/\xi))$ gates on the mode address, public flag, one signal qubit, one band flag, and one restored work qubit.
The isometry error includes ancillary leakage.
It can therefore be inserted $N$ times using the hybrid allocation in \cref{eq:qudit-error-budget}.
The gate bound follows from the diagonal-times-reflection form of \ExpIIR.

\subsubsection{Exp from a finite Schur prefix}
\label{app:qudit-schur}

Let $\varphi_{\lambda t}(q) =\exp[-\lambda t(1-q)/(1+q)]$ and $\module{Exp}_{\lambda t}(z;w)=\varphi_{\lambda t}(zw)$.
This scalar function is analytic and contractive on the disk.
Schur recursion strips off one lossless section at a time, much as QSP synthesis strips its first signal-processing rotation.
Each step extracts the value at zero and leaves a contractive remainder after one delay.
Set $f_0=\varphi_{\lambda t}$ and use the classical recursion
\begin{align}
\gamma_j=f_j(0),
\quad f_{j+1}(q)=\frac{f_j(q)-\gamma_j} {q[1-\overline{\gamma_j}f_j(q)]},
\quad f_j(q)=\frac{\gamma_j+qf_{j+1}(q)} {1+\overline{\gamma_j}qf_{j+1}(q)}.
\label{eq:qudit-schur-recursion}
\end{align}
Each inverse step is a known two-mode unitary with reflection parameter $\gamma_j$ and transmission $\sqrt{1-|\gamma_j|^2}$.
Terminate $f_K$ by a known unit phase.
Subtracting two inverse steps shows that each adds one order of agreement at zero, so the finite inner transfer function agrees with $\varphi_{\lambda t}$ through degree $K-1$~\cite{Garcia2017BlaschkeProducts}.
For a proper inserted load with minimum delay $r_{\min}$, the preserved coefficient count is multiplied by that delay.
Thus the first $N$ clock slots agree whenever $Kr_{\min}\ge N$.
It does not make the finite rational transfer function equal to the nonrational exponential everywhere.

The finite unitary has $K$ private modes.
Ordering them in the Cantero--Moral--Vel\'azquez (CMV) basis \cite{Simon2006CMVMatrices} gives two staggered layers of adjacent two-mode rotations, as in a one-dimensional quantum walk.
Each layer acts on disjoint mode pairs.
This is a depth-two factorization in mode coordinates; implementing those rotations on a binary address register still requires the angle-controlled gates counted below.
One layer consists of pair-controlled rotations on a target bit.
A generic multiplexor over $O(K)$ known angles has $O(K)$ gates~\cite{Bergholm2005ControlledGates}.
A known shift of the mode address converts the staggered layer into the same pairing; Fourier constant addition costs $O(\log^2(2K))$ gates with constant scratch~\cite{Draper2000QuantumAddition}.
This proves the stated exact work bound.
No polylogarithmic exact synthesis of the general Schur-angle table is implied by the two-layer description.

\subsection{Pad\'e implementation of the exponential transform}
\label{app:exp-approx}

The diagonal Pad\'e approximants give a finite implementation of Exp for the bounded Cayley loads used in Hamiltonian simulation.
This is the classical scaling-and-squaring approximation to the matrix exponential \cite[Sec.~2, Eq.~(2.1), and Theorem~2.1]{Higham2005MatrixExponential}: scale the argument, approximate the small exponential, and take a power of two.
Diagonal Pad\'e approximants are suitable here because they are contractive in the appropriate half-plane and exactly unitary on skew-Hermitian inputs \cite{Ehle1973PadeApproximations,Higham2005MatrixExponential}.
Thus the approximation itself admits a unitary realization.
The additional issue is to implement its complete system unitary inexpensively, without applying every section separately.

We retain the convention
\begin{align}
\ExpIIR_\tau(w)=\exp\!\left[-\tau\frac{1-w}{1+w}\right],
\quad P_p(x)=\sum_{k=0}^{p}\frac{(2p-k)!\,p!}{(2p)!\,k!\,(p-k)!}x^k,
\quad R_p(x)=\frac{P_p(-x)}{P_p(x)}.
\label{eq:expapp-pade}
\end{align}
The rational function is the standard diagonal Pad\'e approximant of order $[p/p]$ to $e^{-x}$.
In particular, $R_1(x)=(1-x/2)/(1+x/2)$ is the Cayley step.

\cref{thm:exp-approx} bounds the approximation error and implementation of the full known system matrix of the Pad\'e cascade.
\begin{theorem}[Finite implementation of Exp]
\label{thm:exp-approx}
Let $\tau>0$ and $0<\epsilon\le\delta\le1/4$.
Choose
\begin{align}
J&=2^{\lceil\log_2\max\{1,16\tau\}\rceil},
\quad \sigma=\tau/J,
\quad \beta_p=\frac{4(p!)^2}{(2p)!(2p+1)!\,4^{2p+1}},
\label{eq:expapp-parameters}
\\
J\beta_p&\le c\epsilon\delta^2,
\quad \widehat{\ExpIIR}_\tau(w) =\left[R_p\!\left(\sigma\frac{1-w}{1+w}\right)\right]^J,
\label{eq:expapp-response}
\end{align}
where $c>0$ is a sufficiently small absolute constant and $p\ge1$ is the least integer satisfying the inequality.
Then $\widehat{\ExpIIR}_\tau$ is rational inner.
It has a unitary realization $S_{\rm Exp}$ with private mode count $D=Jp$.
One application, including its controlled or inverse form, requires
\begin{align}
q_{\rm Exp}&\le\lceil\log_2(D+1)\rceil+O(1),
\\
G_{\rm Exp}&=O\!\left(p+\log(2Jp)\log(2J)\right),
\label{eq:expapp-resources}
\\
p&=O\!\left( \frac{\log(J/(\epsilon\delta))} {\log\log(J/(\epsilon\delta))}\right).
\label{eq:expapp-order}
\end{align}
The gate count is in arbitrary one- and two-qubit gates, with constant additional workspace.
These are the resources of the known scalar system unitary, excluding oracle, system, delay, and clock registers.
For a Hermitian Hamiltonian, assume $H=H^\dagger$, $\|H\|\le\Lambda$, $\tau=t\Lambda $ for $t>0$.
The simulation error then obeys
\begin{align}
\left\| \widehat{\ExpIIR}_\tau\!\left((I+iH/\Lambda)(I-iH/\Lambda)^{-1}\right) -e^{itH}\right\| =\left\|R_p(-itH/J)^J-e^{itH}\right\| \le J\beta_p.
\label{eq:expapp-physical}
\end{align}
The Cayley transform in this statement is implemented by the Cayley construction from the given Hamiltonian block encodings.
\end{theorem}

\begin{proof}
The denominator roots of $R_p(x)$ lie in the open left half-plane \cite{Ehle1973PadeApproximations}.
On the imaginary axis its numerator and denominator are complex conjugates, so its modulus is one.
The maximum-modulus principle, including the finite limit at infinity, therefore gives
\begin{align}
|R_p(x)|\le1
\quad(\operatorname{Re}x\ge0),
\quad |R_p(iy)|=1
\quad(y\in\mathbb R).
\label{eq:expapp-half-plane}
\end{align}
The change of variable maps the disk to this half-plane.
Consequently, $R_p(\sigma(1-w)/(1+w))$ is a degree-$p$ rational inner function.
The Pad\'e remainder, obtained by coefficient matching, is
\begin{align}
e^{-x}P_p(x)-P_p(-x) =\frac{(-1)^{p+1}x^{2p+1}}{(2p)!} \int_0^1 e^{-sx}s^p(1-s)^p\,ds.
\label{eq:expapp-remainder}
\end{align}
The coefficients of the denominator polynomial satisfy $[x^k]P_p(x)\le2^{-k}/k!$ and $\|P_p(X)-I\|\le e^{1/8}-1 $ for $\|X\|\le1/4$.
The remainder and the convergent logarithm series, as in the backward-error argument of \cite[Theorem~2.1]{Higham2005MatrixExponential}, give
\begin{subequations}
\label{eq:expapp-log}
\begin{align}
E_p(X)&:=\log(e^XR_p(X)),
\quad \|E_p(X)\|\le\beta_p
\quad(\|X\|\le1/4),
\\
\|DE_p(X)[Y]\|&\le16\beta_p\|Y\|
\quad(\|X\|\le1/8).
\end{align}
\end{subequations}
Apply Cauchy's derivative estimate in a unit matrix direction on the circle $z\mapsto E_p(X+zY)$ with $\|Y\|=1$ and $|z|=1/8$.
The perturbed argument has norm at most $1/4$ there.
It gives $8\beta_p$; the displayed constant $16$ is a convenient upper bound.
Scaling $Y$ proves the claim for arbitrary, possibly noncommuting directions.
For skew-Hermitian arguments in this norm range, both factors inside the logarithm are unitary.
Their product lies in the neighborhood of the identity where the principal logarithm is defined; the logarithm is therefore skew-Hermitian.
It also commutes with the argument, since all factors are functions of that one matrix.
Thus
\begin{align}
X=-itH/J,
\quad R_p(X)^J=e^{itH}\exp[J E_p(X)].
\label{eq:expapp-backward-evolution}
\end{align}
The last exponential is unitary, and its distance from the identity is at most $J\|E_p(X)\|$.
This proves \cref{eq:expapp-physical}; Stirling's estimates give \cref{eq:expapp-order}.

For the implementation, choose a minimal unitary realization $S_0$ of one section in a finite CMV basis.
Its two staggered layers of two-dimensional blocks have an $O(p)$-gate implementation \cite{Simon2006CMVMatrices,shende2006synthesis}.
Its public entry is the real scalar $a=R_p(\sigma)\in(0,1)$.
The scalar-public cosine--sine decomposition gives
\begin{align}
S_0=(1\oplus U_{\rm out}) [\mathcal C(a)\oplus I_{p-1}](1\oplus U_{\rm in}^\dagger),
\quad \mathcal C(a)=\left(\begin{array}{c|c}a&\sqrt{1-a^2}
\\
\hline
\sqrt{1-a^2}&-a\end{array}\right).
\label{eq:expapp-local}
\end{align}
Choose $U_{\rm in}$ by preparing the normalized input-coupling vector, at cost $O(p)$ \cite{shende2006synthesis}.
The identity $1\oplus U_{\rm out}=S_0(1\oplus U_{\rm in}) [\mathcal C(a)\oplus I_{p-1}]$ implements the other basis change at the same cost, using a temporary local public coordinate whose final amplitude is zero for every private input.

In the cascade, the sections share a public coordinate but have disjoint private spaces.
Their private basis changes commute past the other sections, giving
\begin{align}
S_{0,J-1}\cdots S_{0,0} =(1\oplus I_J\otimes U_{\rm out}) \left[\mathcal C_{J-1}(a)\cdots\mathcal C_0(a) \oplus I_{J(p-1)}\right] (1\oplus I_J\otimes U_{\rm in}^\dagger).
\label{eq:expapp-cascade}
\end{align}
Thus each local basis change is applied once, with the stage label unchanged.
The middle product has the scalar dyadic factorization of \cite[Lemma~10]{Chen2026GateEfficientSimulation}, including the known phases on the remaining coordinates.
There are $O(\log J)$ layers with identical rotations within a layer.
Recomputing their control predicates costs $O(\log(2Jp))$ gates per layer with constant restored workspace \cite{Khattar2025ConditionallyCleanAncillae}.
This proves \cref{eq:expapp-resources} for the complete unitary on every input state.
The degree-one case recovers the scalar Cayley/dyadic construction of \cite{Chen2026GateEfficientSimulation}.
\end{proof}

\cref{cor:exp-approx-hamsim} gives the analytic bounds for composing this approximation with the weighted Cayley transformation.
\begin{corollary}[Replacement in Hamiltonian simulation]
\label{cor:exp-approx-hamsim}
Suppose a sum of Hermitian terms is accessed through controlled self-inverse encodings,
\begin{align}
H=\sum_jH_j,
\quad O_j=\Be{H_j/\lambda_j},
\quad\lambda_j>0.
\end{align}
The system uses $n$ qubits; the encoding at each port uses $a_j$ additional qubits.
Its query cost is $C_j>0$.
Put $\Lambda=\|\boldsymbol{\lambda}\|_1$ and $\tau=t\Lambda$.
Compose the realization of \cref{thm:exp-approx} with weighted Cayley for $-H_j$, using the known sign change of $O_j$.
The complete transfer function is
\begin{subequations}
\label{eq:expapp-loaded}
\begin{align}
M(\Bz)&=\sum_j\left[ \lambda_j\frac{1-z_j^2}{1+z_j^2}I -\frac{2iz_j}{1+z_j^2}H_j\right],
\quad M(\Bone)=-iH,
\\
\widehat F(\Bz)&=\left[R_p\!\left(\frac{t}{J}M(\Bz)\right)\right]^J.
\end{align}
\end{subequations}
With clocks chosen for this transfer function, quditization produces
\begin{align}
(\langle0|\otimes I)\Be{\widetilde G_N/(1+\delta)} (|0\rangle\otimes I)&=\frac{\widetilde G_N}{1+\delta},
\quad \|\widetilde G_N-e^{itH}\|\le\epsilon,
\label{eq:expapp-selected}
\\
q_j&=\left\lfloor\frac{N-1}{r_j}\right\rfloor,
\\
\sum_j C_jq_j &\le(1+\delta)t\langle\boldsymbol{C},\boldsymbol{\lambda}\rangle_{1/2} +O\!\left(\delta^{-2}\|\boldsymbol{C}\|_1\log(1/\epsilon)\right).
\label{eq:expapp-queries}
\end{align}
Here $r_j$ are constructive positive integer delays and $\langle\boldsymbol{C},\boldsymbol{\lambda}\rangle_{1/2} =(\sum_j\sqrt{C_j\lambda_j})^2$.
Neither $J$ nor $p$ multiplies a scheduled oracle count.
The finite lift retains all private modes and all oracle output subspaces.
Its packed coordinate register has dimension
\begin{align}
d_c=N+Jp\sum_jr_j2^{a_j},
\quad n_{\rm total}\le n+\lceil\log_2d_c\rceil+s_{\rm work}+O(1),
\label{eq:expapp-packed}
\end{align}
where $s_{\rm work}$ is the maximum restored scratch required by the oracle, signal-alignment, and endpoint circuits, reused between those operations.
The Exp contribution to known gates is $O(NG_{\rm Exp})$; oracle, alignment, routing, and endpoint costs remain separate.
\end{corollary}

\begin{proof}
The existing zero-delay Cayley--Exp fusion applies to the full $S_{\rm Exp}$; the mode address is unchanged during each feedback operation.
Eliminating that connection gives \cref{eq:expapp-loaded}, a rational function of the whole sum, with no product formula over the $H_j$.
Since $\partial_{z_j}M(\Bone)=-\lambda_jI$, logarithmic differentiation gives
\begin{align}
\widehat F(\Bone)^\dagger\partial_{z_j}\widehat F(\Bone) =t\lambda_j[I-E_p'(-itH/J)],
\quad (1-16\beta_p)I\preceq I-E_p'(-itH/J)\preceq I.
\label{eq:expapp-port}
\end{align}
For the upper inequality, clearing denominators in the Pad\'e agreement through order $2p$ gives
\begin{align}
E_p'(iy)= \left(\frac{p!}{(2p)!}\right)^2 \frac{y^{2p}}{|P_p(iy)|^2}\ge0
\quad(y\in\mathbb R,\ |y|\le1/8).
\label{eq:expapp-positive}
\end{align}
The lower inequality follows from \cref{eq:expapp-log}.
Thus all port energies have the same operator multiplier and their relative weights are unchanged.

To bound the clock horizon, restrict to the delayed slice and define its total delay weight: \begin{align}
\widehat G=\widehat F_r,
\quad \widehat G(z)=\widehat F(z^{r_1},\ldots,z^{r_m}),
\quad \Omega=t\sum_j\lambda_jr_j.
\end{align} For $|\log z|\le s\le1/(8\max_jr_j)$,
\begin{align}
\|\widehat G(z)\|\le \begin{cases} |z|^{\displaystyle t\sum_j\frac{\lambda_jr_j}{1+\sin(sr_j)}-32\beta_p\Omega}, &|z|\le1,
\\[3pt]
|z|^{\displaystyle t\sum_j\frac{\lambda_jr_j}{1-\sin(sr_j)}+32\beta_p\Omega}, &|z|\ge1. \end{cases}
\label{eq:expapp-radial}
\end{align}
Indeed, the delayed load is $M_{\rm del}(z)=\sum_j[-\lambda_j\tanh(r_j\log z)I -iH_j\operatorname{sech}(r_j\log z)]$.
It obeys $\|M_{\rm del}(z)\|\le2\Lambda$ and $\|\partial_uM_{\rm del}(e^{u+iv})\|\le2\sum_j\lambda_jr_j$ on this neighborhood.
Use the exact backward-error identity
\begin{align}
\widehat G(z)=\exp[-tM_{\rm del}(z)+J E_p(tM_{\rm del}(z)/J)].
\end{align}
The additional Hermitian part vanishes on $u=0$.
Integrating its derivative in the radial variable, bounded by \cref{eq:expapp-log}, gives the bound $32\beta_p\Omega|u|$.
Combining this with the original Hermitian-part estimate for weighted Cayley proves \cref{eq:expapp-radial}, without assuming commuting terms.

Apply the main text's clock-shaping and integer-delay construction with a fixed fraction of the available error and normalization slack.
In its smoothed-window horizon formula, changing each radial exponent by $O(\beta_p\Omega)$ adds at most $O(\delta^{-1/2}\beta_p\Omega)$ clock steps.
Since the original horizon is at least $\Omega$, \cref{eq:expapp-response} makes this a sufficiently small relative change to absorb in the remaining slack.
The rational error is at most $J\beta_p$; allocate the other errors to the clock and gate synthesis.
This gives \cref{eq:expapp-selected,eq:expapp-queries}.
The standard packed lift gives \cref{eq:expapp-packed}.
\end{proof}

For a discrete gate set, synthesize each application of the full known system matrix to error $O(\epsilon/((1+\delta)N))$.
Standard ancilla-free Clifford+$T$ synthesis then gives $O(G_{\rm Exp}\log((1+\delta)NG_{\rm Exp}/\epsilon))$ gates per application \cite{Ross2016Optimal}; oracle, routing, and endpoint costs remain separate.
The circuit obtains $\Be{\widetilde G_N/(1+\delta)}$ with $\|\widetilde G_N-e^{itH}\|\le\epsilon$ before normalization.
For a deterministic unitary approximation, apply oblivious amplitude amplification as analyzed in \cref{sec:iir}.
The clock is chosen for $\widehat G$ using \cref{eq:expapp-radial}.
This comparison applies to bounded Cayley loads; the finite Toeplitz comparison for a general contraction remains the spectral construction of \cref{app:mod-exp}.

\section{Hamiltonian algorithms and weighted state preparation}

\subsection{Finite-circuit construction from the clock and structured Hamiltonian transformations}
\label{app:multi-hs}

\subsubsection{Clocks with normalization close to one and weighted delays}
\label{app:multi-hs-delays}

We retain the portwise analytic estimates before choosing the delays.
Write the logarithmic delay coordinate as $r_j\log z=v+iw$.
An eigenvalue $x\in[-1,1]$ of the normalized component contributes the following scalar to the Hermitian part of $-tM$:
\begin{align}
t\lambda_j\sinh v\left[ \frac{(1+x)/2}{\cosh v+\sin w} +\frac{(1-x)/2}{\cosh v-\sin w}\right].
\end{align}
On $v^2+w^2\le y^2\le1/4$,
\begin{align}
\frac1{1+\sin y}\le \frac{\sinh v/v}{\cosh v\pm\sin w} \le\frac1{1-\sin y}.
\end{align}
Summing these Hermitian-part bounds and using $\|e^T\|\le e^{\lambda_{\max}((T+T^\dagger)/2)}$ proves \cref{eq:multi-hs-radial}.
For the lower inequality use $\cosh|v|+\sin|w|\le1+\sin y$ on the disk; for the upper one use $\sinh v/v\le\cosh v$.
Continuity gives the values at zero.

For flat-clock block count $K\ge2$, define $\theta=\pi/(K+1)$ and $a=\cos\theta$.
For a plateau half-width $L_0\ge1$, form the normalized clocks
\begin{align}
g_\pm(rL_0+s)=\frac{(\pm1)^r}{\sqrt{L_0}} \sqrt{\frac2{K+1}}\sin((r+1)\theta),
\quad 0\le r<K,
\quad 0\le s<L_0.
\label{eq:multi-hs-sine-clock}
\end{align}
Let $k_\pm$ be their autocorrelations.
The signed correlation
\begin{align}
W(n)=\frac{(1+a)k_+(n)-(1-a)k_-(n)}{2a}
\label{eq:multi-hs-sine-window}
\end{align}
is exactly one on $|n|\le L_0$, is nonnegative and symmetric, and is nonincreasing for $n\ge0$.
Its total variation is two.
To check this, the coarse autocorrelation at displacement $bL_0$ is
\begin{align}
k_+(bL_0)= \frac{(K+1-b)\cos(b\theta)+\cot\theta\sin(b\theta)}{K+1}.
\end{align}
The fine correlations linearly interpolate these values, and $a k_+(bL_0)-k_+((b+1)L_0) =(K-b)\sin\theta\sin(b\theta)/(K+1)\ge0$.
The branch probabilities $(1\pm a)/2$ and a minus sign on the output minus branch implement $aW$ with normalized clock states.

To apply \cref{lem:iir-flat-smoothing}, translate the base window to $c_n^0=W(n-L_0)$, which is one on $[0,2L_0]$.
Round the radial exponents and choose the smoothing width by $\ell=\lfloor A_-\rfloor$, $u_0=\lceil A_+\rceil$, and $d=O(s^{-1}\log(1/(a\epsilon)))$.
The neighborhood $|z-1|\le s/4$ is contained in $|\log z|\le s$ for the range used here.
Choose $L_0=\max\{1,\lceil(u_0-\ell+d)/2\rceil\}$, smooth the output by the positive box average, and translate it by $L_0+\ell-d$.
The input support is $[0,KL_0-1]$ and the output support is contained in $[L_0+\ell-d,KL_0-1+L_0+\ell]$.
A common translation therefore gives
\begin{align}
N=KL_0+\max\{L_0+\ell,d\} \le\frac{K+1}{2}A_+-\frac{K-1}{2}A_-+O(Kd+K).
\end{align}
Positive smoothing does not increase the nuclear norm.
Thus the selected amplitude is $a$, with known attenuation if the endpoint norms shrink.
Substitution of \cref{eq:multi-hs-radial} gives
\begin{align}
N\le\Omega+\kappa s\sum_jt\lambda_jr_j^2+\frac Zs,
\quad \kappa=\frac{K+\sin u}{\cos^2u},
\quad Z=O\!\left(K\log\frac1{a\epsilon}\right),
\label{eq:multi-hs-app-horizon}
\end{align}
provided $sr_j\le u\le1/2$.
The exact propagation term before this bound is
\begin{align}
\frac{K+1}{2}A_+-\frac{K-1}{2}A_- =\sum_jt\lambda_jr_j\frac{1+K\sin(sr_j)}{\cos^2(sr_j)}.
\end{align}
Subtracting $\Omega$ and using $\sin x\le x$ gives the second moment in \cref{eq:multi-hs-app-horizon}.
The box width is $O(1/s)$ and its repetition count is $O(\log(1/(a\epsilon)))$.
The contour bound uses both the analytic neighborhood of one and contractivity on the remaining disk boundary, so it introduces no logarithm of a strength ratio or delay ratio.
For integer clocks the ceilings are absorbed by a fixed constant in the logarithm.

The displacement average admits a constant-rank overlap matrix.
Split displacements of each sign according to whether the input or output clock is shifted away from zero.
In each set one clock is fixed and only the other is averaged, giving one outer product.
There are at most four such products, and their nuclear norm is at most one.
By \cref{thm:iir-clock-body}, at most six labels suffice after normalization padding.
All discarded private amplitudes are retained in the lift; averaging displacements does not repeat that lift.

The weighted step uses the following inequality.
Collect the time weights, total oracle cost, and square-root allocation cost in
\begin{align}
\mu_j=t\lambda_j,
\quad C_\Sigma=\sum_jC_j,
\quad X=\left(\sum_j\sqrt{\mu_jC_j}\right)^2.
\end{align}
For $\eta,u,\kappa>0$ with $\kappa u\ge1$, set
\begin{align}
x_j=\min\{u,\eta\sqrt{C_j/\mu_j}\},
\quad A=\sum_j\mu_jx_j,
\quad J_2=\sum_j\mu_jx_j^2,
\quad R=\sum_jC_j/x_j.
\end{align}
Then
\begin{align}
(A+\kappa J_2)R\le X+\kappa\eta C_\Sigma\sqrt X +\frac{\eta^2C_\Sigma^2}{4u^2},
\quad R\le\frac{\sqrt X}{\eta}+\frac{C_\Sigma}{u}.
\label{eq:multi-hs-clipped}
\end{align}
For the proof, use a scalar random variable with probabilities and values $\Pr(j)=C_j/C_\Sigma$ and $y_j=(u/\eta)\sqrt{\mu_j/C_j}$.
Write the four moments as $U=\mathbb E\min(y^2,y)$, $V=\mathbb E\max(1,y)$, $W_0=\mathbb Ey$, and $J_0=\mathbb E\min(y^2,1)$.
The first inequality is equivalent to $UV-W_0^2+\kappa u(J_0V-W_0)\le1/4$.
First $J_0V\le W_0$.
At $\kappa u=1$, separate the two clipping regions using \begin{align}
p&=\Pr(y\ge1),
\quad b=\mathbb E[y\boldsymbol{1}_{y<1}],
\\
c&=\mathbb E[y^2\boldsymbol{1}_{y<1}]\le b,
\quad v_0=\mathbb E(y-1)_+.
\end{align} Expansion bounds the expression by $(p+b)(1-p-b)\le1/4$.
Increasing $\kappa u$ only decreases it.
The second inequality follows from $V\le W_0+1$.

Take
\begin{align}
K=\left\lceil\frac\pi{\sqrt\delta}\right\rceil-1,
\quad u=1/K,
\quad \beta=\delta^2/36,
\quad d_0=\frac{1+\delta}{1+\beta}-1>0.
\label{eq:multi-hs-physical-parameters}
\end{align}
These choices give $a\ge1-\delta/2\ge1/(1+\delta)$, $\kappa=\Theta(K)$, and $d_0=\Theta(\delta)$.
Choose $\eta=\sqrt{Z/(\kappa C_\Sigma)}$ and form the clipped $x_j$ above.
For the integer implementation, use
\begin{align}
P=\left\lceil\frac1{\beta\min_jx_j}\right\rceil,
\quad r_j=\lceil Px_j\rceil,
\quad s=\frac1{(1+\beta)P}.
\label{eq:multi-hs-integer-rates}
\end{align}
This gives $x_j/(1+\beta)\le sr_j\le x_j$ and $s\le\beta u/(1+\beta)$.
Choose $Z$ using this last upper bound on $s$, so that it is fixed before computing the rates.
Combining \cref{eq:multi-hs-app-horizon} with the exact integer query counts gives
\begin{subequations}
\label{eq:multi-hs-physical-certificate}
\begin{align}
\sum_jC_jq_j &\le(1+\beta)(A+\kappa J_2+Z)R
\\
&\le(1+\beta)\left[ X+2\sqrt{\kappa ZC_\Sigma X} +ZC_\Sigma\left(\frac1u+\frac1{4\kappa u^2}\right)\right].
\end{align}
\end{subequations}
Apply $2\sqrt{\kappa ZC_\Sigma X}\le d_0X+ \kappa ZC_\Sigma/d_0$.
Since $(1+\beta)(1+d_0)=1+\delta$, this proves the actual query-cost bound with additive term $O(K^2 C_\Sigma\delta^{-1}\log(1/\epsilon)) =O(\delta^{-2}C_\Sigma\log(1/\epsilon))$.

This allocation is for actual oracle calls.
Dividing the weighted query cost by the selected amplitude gives $\sum_jC_jq_j/a$, which also divides the leading coefficient in \cref{eq:multi-hs-physical-certificate} by the selected amplitude $a$, or equivalently multiplies it by the block-encoding normalization $1/a$.
That objective requires a different choice of $d_0$.
The theorem bounds actual oracle calls with a universal big-$O$ remainder.

\subsubsection{Projector SOS phase amplification}
\label{app:multi-sos-phase}

To prove \cref{thm:multi-sos-phase}, we establish the SOS spectrum, unitarity, and two-sided analytic bounds for the finite rational transfer function in \cref{eq:multi-sos-cascade}, then apply clock shaping.

Since $P_j^2=P_j$, the signal condition gives
\begin{align}
\lambda_j(\langle0|_j\otimes I)P_j^\dagger P_j (|0\rangle_j\otimes I)=A_j^\dagger A_j.
\label{eq:multi-sos-factor-check}
\end{align}
Thus $B^\dagger B=H/\Lambda$.
At positive energy, the two vectors in the main text are orthonormal.
On their span the dilated Hamiltonian acts as $\sqrt{E/\Lambda}\,X$.
The Cayley transform on this plane is
\begin{align}
\frac1{1+\rho^2E/\Lambda} \begin{pmatrix} 1-\rho^2E/\Lambda&-2i\rho\sqrt{E/\Lambda}
\\
-2i\rho\sqrt{E/\Lambda}&1-\rho^2E/\Lambda \end{pmatrix}.
\label{eq:multi-sos-plane}
\end{align}
Taking its $J$th power proves the phase formula.
The original input is an equal superposition of its two eigenvectors.
At zero energy, $B|\psi_0\rangle=0$, so $|\psi_0\rangle\oplus0$ is fixed.

The coupling in \cref{eq:multi-sos-couplings} satisfies
\begin{align}
K_j=\frac{\omega_jr_j}{2D_{\boldsymbol{r}}}I_{\mathcal X} \oplus_j\frac{D_{\boldsymbol{r}}}{2r_j}I_{\mathcal K_j}.
\label{eq:multi-sos-load-blocks}
\end{align}
All other leaf blocks are zero.
Summing $r_jK_j$ gives $(D_{\boldsymbol r}/2)I$ on the public space.
Compressing the involution with its coupling map gives $E_jO_jE_j^\dagger=\mathcal H_j$ and $-K_j\preceq\mathcal H_j\preceq K_j$.

For $L=L_\rho$, we have
\begin{align}
L+L^\dagger=2L^\dagger(I+\rho EE^\dagger)L =2L(I+\rho EE^\dagger)L^\dagger.
\label{eq:multi-sos-unitarity-identity}
\end{align}
Substitution in $S_\rho^\dagger S_\rho$ and $S_\rho S_\rho^\dagger$ proves that the entire system is unitary, including private complements.
Eliminating its private input using $\boldsymbol{f}=Z\boldsymbol{g}$ gives the outer Cayley transform of
\begin{align}
i\mathcal H_0+E(I+iZ)(I-iZ)^{-1}E^\dagger.
\label{eq:multi-sos-elimination}
\end{align}
Expansion of $(I+iz_jO_j)(I-iz_jO_j)^{-1}$ with $O_j^2=I$ gives \cref{eq:multi-sos-response}.
The formula is analytic in the open polydisk; at $\Bz=\boldsymbol{1}$ it is defined by regular continuation.
The inverse there is regular because $\mathcal H$ is Hermitian.

The following estimate controls the full public space near $\Bz=\boldsymbol{1}$.
Its two exponents determine the clock's propagation and the additional support needed for precision.
\begin{lemma}[Two-sided bound for the projector system matrix]
\label{lem:multi-sos-cap}
Let $0<u\le1/4$ and $0<\rho\le1/4$.
Write the delayed load in logarithmic coordinates as $M(w)=M(e^{r_1w},\ldots,e^{r_mw})$.
On $|w|\le u/r_*$,
\begin{align}
\|M'(w)+D_{\boldsymbol{r}}I/2\|\le uD_{\boldsymbol{r}},
\quad \|M(w)\|\le11/8,
\label{eq:multi-sos-load-cap}
\end{align}
and the single-stage transfer function satisfies
\begin{align}
\|F_\rho'(w)F_\rho(w)^{-1}-\rho D_{\boldsymbol{r}}I\| \le \rho D_{\boldsymbol{r}}(3\rho^2+5u).
\label{eq:multi-sos-generator}
\end{align}
If $3\rho^2+5u<1$, the cascade has radial exponents
\begin{align}
A_\pm=kD_{\boldsymbol{r}}[1\pm(3\rho^2+5u)]
\label{eq:multi-sos-radial}
\end{align}
on a neighborhood of the evaluation point with radius $1-e^{-u/r_*}$.
The norm bounds are $\|G(z)\|\le|z|^{A_-} $ for $|z|\le1$ and $\|G(z)\|\le|z|^{A_+} $ for $|z|\ge1$.
\end{lemma}
\begin{proof}
The logarithmic load is exactly
\begin{align}
M(w)=i\mathcal H_0+ \sum_j[-\tanh(r_jw)K_j+i\operatorname{sech}(r_jw)\mathcal H_j].
\label{eq:multi-sos-log-load}
\end{align}
By \cref{eq:multi-sos-balance}, the row concatenation of $\sqrt{2r_j/D_{\boldsymbol{r}}}E_j$ is a coisometry.
For either delayed family, scalar coefficients bounded in modulus by $b$ give a sum of norm at most $bD_{\boldsymbol{r}}/2$.
The two families are $(r_jK_j)_j$ and $(r_j\mathcal H_j)_j$.
On the local disk, the scalar functions obey $|z|\le u $ implies $ |\operatorname{sech}z|\le1/\cos u$ and $|\tanh z|\le\sinh u/\cos u$.
Differentiating the load gives the first bound in \cref{eq:multi-sos-load-cap}; here $\tanh^2u$ and $\operatorname{sech}u\tanh u$ are bounded by their corresponding complex absolute-value bounds.
Integrate from $M(0)=i\mathcal H$ and use $D_{\boldsymbol{r}}\le r_*$.
This gives $\|M(w)\|\le1+u/2+u^2\le11/8$.

Differentiate in the indicated order, without commuting $M$ and $M'$:
\begin{align}
F_\rho'F_\rho^{-1} =\rho D_{\boldsymbol{r}}(I-\rho^2M^2)^{-1}
\quad
\quad-2\rho(I+\rho M)^{-1} (M'+D_{\boldsymbol{r}}I/2)(I-\rho M)^{-1}.
\label{eq:multi-sos-log-derivative}
\end{align}
The first inverse differs from $I$ by at most $3\rho^2$.
After division by $\rho D_{\boldsymbol{r}}$, the second term is at most $2u/(1-11/32)^2<5u$.
This proves \cref{eq:multi-sos-generator}.
On the imaginary axis the transfer function is unitary.
\cref{lem:iir-radial-energy} gives the two radial exponents for one Cayley transfer function; submultiplicativity gives their $J$th powers.
This argument bounds the entire public space and requires no invariance of the low-energy subspace away from $\Bone$.
\end{proof}

For the catalyst, $M'(0)=-D_{\boldsymbol{r}}I/2$ in logarithmic coordinates.
At $\Bz=\boldsymbol{1}$, each Cayley transformation consequently has group-delay operator $\rho D_{\boldsymbol{r}}(I+\rho^2\mathcal H^2)^{-1}$.
It commutes with the evaluated transfer function, so composition adds the $J$ equal actions.
This proves \cref{eq:multi-sos-catalyst} with $J\rho=k$.

Applying \cref{thm:iir-analytic-clock-selection} with these radial exponents gives the following clock bounds.
Consider a disk-contractive transfer function with unitary value at $z=1$.
Let its radial exponents be $A_-,A_+$ on a neighborhood of radius $\zeta$.
A smoothed flat clock with integer block count $K\ge2$ attains amplitude at least $a=\cos(\pi/(2K))$.
For unnormalized error $\xi>0$, a sufficient horizon is
\begin{align}
N\le K A_+-(K-1)A_- +O\!\left(K\zeta^{-1}\log\frac{16}{\xi}\right).
\label{eq:multi-sos-clock-rule}
\end{align}
It may be attenuated to the stated $a$.
Integer endpoints, displacement, and clock synthesis are included in the remainder.
This is \cref{eq:analytic-horizon-exact} with rounded exponents and smoothing width $\ell=\lfloor A_-\rfloor$, $u=\lceil A_+\rceil$, and $d=O(\zeta^{-1}\log(16/\xi))$.
Since $\ell\ge0$, its remaining displacement term is $O(d+1)$.

For the fixed positive parameter $\delta$ in \cref{thm:multi-sos-phase}, choose
\begin{align}
K=\left\lceil\frac\pi{\sqrt\delta}\right\rceil,
\quad J_0=\left\lceil\sqrt{\frac{128K}{\delta}}\right\rceil,
\quad \rho=J_0^{-1},
\quad J=kJ_0,
\quad u=\frac{\delta}{256K},
\quad a=\cos\frac\pi{2K}.
\label{eq:multi-sos-clock-parameters}
\end{align}
These parameters fix $J_0=\Theta(\delta^{-3/4})$ independently of the gain and implementation precision.
Allocate a total approximation error before normalization $\xi=\epsilon/16$ to clock approximation and known-operation synthesis.
Then $a\ge1-\delta/8$ and $(2K-1)(3\rho^2+5u)\le11\delta/128$.
Applying \cref{eq:multi-sos-clock-rule} gives
\begin{align}
N\le(1+11\delta/128)kD_{\boldsymbol{r}} +O\!\left(\delta^{-2}r_*\log\frac{16}{\xi}\right).
\label{eq:multi-sos-horizon-explicit}
\end{align}
Since $(1+11\delta/128)/(1-\delta/8)\le1+\delta/4$, the exact integer query counts give the stronger fixed-delay statement
\begin{align}
\frac1a\sum_jc_jq_j\le (1+\delta/4)kD_{\boldsymbol{r}}\sum_j\frac{c_j}{r_j} +O\!\left(\frac{r_*}{\delta^2}\sum_j\frac{c_j}{r_j} \log\frac{16}{\xi}\right).
\label{eq:multi-sos-fixed-normalized}
\end{align}
Here $a$ is the preattenuation success amplitude, with ideal success probability $a^2$.
The coefficient $1+\delta/4$ includes the division by $a$.

Write $S=\sum_j\omega_j^{1/3}c_j^{2/3}$ and $C_\Sigma=\sum_jc_j$ in the following allocation proof.
Let $p=1+\delta/4$ and let $B_\delta=O(\delta^{-2}\log(16/\xi))$ be a sufficient coefficient of the fixed-delay precision term.
Set
\begin{align}
h=\sqrt{\frac{B_\delta\sqrt S}{pkC_\Sigma}},
\quad x_j=\min\{1,h(c_j/\omega_j)^{1/3}\}.
\label{eq:multi-sos-clipping}
\end{align}
The inequalities $\sqrt{\sum_j\omega_jx_j^2}\le h\sqrt S$ and $\sum_jc_j/x_j\le C_\Sigma+S/h$ imply
\begin{align}
\left(pk\sqrt{\sum_j\omega_jx_j^2}+B_\delta\right) \sum_j\frac{c_j}{x_j} \le\left(\sqrt{pkS^{3/2}}+\sqrt{B_\delta C_\Sigma}\right)^2.
\label{eq:multi-sos-clipped-cost}
\end{align}
This also covers the case in which every scaled delay is clipped.
Choose the clipping parameter and integer delays by $\beta=\delta/8$, $P=\lceil1/(\beta\min_jx_j)\rceil$, and $r_j=\lceil Px_j\rceil$.
Then $Px_j\le r_j\le(1+\beta)Px_j$ and $r_*\le(1+\beta)P$.
The fixed-delay objective is consequently at most $1+\beta$ times \cref{eq:multi-sos-clipped-cost}.
The inequality $2\sqrt{XY}\le(\delta/4)X+(4/\delta)Y$, and $(1+\delta/8)(1+\delta/4)^2\le1+3\delta/4$, give
\begin{align}
\frac1a\sum_jc_jq_j\le (1+3\delta/4)\frac{k}{\sqrt\Lambda} \left(\sum_j\lambda_j^{1/3}c_j^{2/3}\right)^{3/2} +O\!\left(\delta^{-3}\sum_jc_j\log\frac{16}{\xi}\right).
\label{eq:multi-sos-weighted-normalized}
\end{align}
This proves \cref{eq:multi-sos-cost} for the physical cost as well.
This clipping estimate gives the $\delta^{-3}$ dependence for SOS amplification.
Its implicit constant is specific to this transfer function.

The clock selects $a\widetilde G_N$, with $\|\widetilde G_N-U\|\le\xi$.
Since $a\ge1/(1+\delta)$, a one-qubit attenuation of amplitude $1/[a(1+\delta)]$ is feasible.
It yields the encoded operator $\widetilde G_N/(1+\delta)$.
This proves \cref{eq:multi-sos-guarantee,eq:multi-sos-cost}; by \cref{lem:normalization-error}, its conditional error is at most $2\xi/(1-\xi)<\epsilon$.
For the energy-sensitivity bound, substitute $k=t_{\rm eq}\sqrt{\Lambda\Gamma}(1+\rho^2\Gamma/\Lambda)$ in \cref{eq:multi-sos-weighted-normalized}.
The remaining factor is absorbed because $(1+3\delta/4)(1+\delta/(128K))\le1+\delta$.
This proves \cref{eq:multi-sos-time-cost}.

At every query event the original factor oracle acts identically on every stage and direction sector, leaving these labels unchanged.
Each traversal therefore uses one $V_j$ and one $V_j^\dagger$.
Known system and routing errors can be allocated to $O(\xi/N)$ per update, and the clock error to a fixed fraction of $\xi$.
Absorbing these constant error allocations in the logarithms preserves all displayed leading coefficients.

\subsubsection{The exact reflection-walk alternative}
\label{app:multi-sos-exact}

Let $\mathcal J$ and $R$ be as in \cref{eq:multi-sos-exact-walk}.
Then $R^2=I$ and $\mathcal J^\dagger R\mathcal J=2H/\Lambda-I$.
For a system eigenvector with centered signal $x=2E/\Lambda-1$, the two reflections preserve the span $\operatorname{span}\{\mathcal J|\psi_E\rangle,R\mathcal J|\psi_E\rangle\}$.
Their product has the usual two-dimensional rotation representation.
The additional minus sign in $W$ gives phases $\pm2\arcsin\sqrt{E/\Lambda}$ and therefore
\begin{align}
\mathcal J^\dagger W^k\mathcal J=T_k(I-2H/\Lambda).
\label{eq:multi-sos-exact-compression}
\end{align}
The endpoint cases follow by continuity.
One $R$ is implemented by addressed uses of every $R_j$; thus each step uses one pair from every factor.
This proves the exact alternative and its cost $k\sum_jc_j$.
Its known-gate cost is $k$ times the single-step weight-preparation, reflection, and routing costs.

\begin{proposition}[Query cost of an exact phase pair determined by energy]
\label{prop:multi-sos-exact}
Consider a finite coherent circuit for every admissible factor tuple and unitary completion.
Suppose a reference embedding of every energy eigenstate is independent of the unknown oracles.
Its spectral support must lie entirely on $e^{\pm i\phi(E)}$ for $E\in[0,\Gamma]$, with a positive upper energy bound.
Assume the phase law depends only on $E$ and the given parameters and satisfies $\phi(E)=2k\sqrt{E/\Lambda}+o(\sqrt{E/\Lambda})$ at zero.
Let $n_j$ bound the total forward and inverse calls to factor oracle $V_j$.
Then $\lfloor n_j/2\rfloor\ge k $ for every $ j$.
If $q_j$ counts pairs consisting of one forward and one inverse query, $q_j\ge\lceil k\rceil$.
\end{proposition}
\begin{proof}
We apply the polynomial method \cite{Beals2001polynomial} to scalar test oracles, with independent signs $s_j,t_j\in\{-1,1\}$,
\begin{align}
V_j(h_j;s_j,t_j)=\begin{pmatrix} s_j\sqrt{h_j}&t_j\sqrt{1-h_j}
\\
t_j\sqrt{1-h_j}&-s_j\sqrt{h_j} \end{pmatrix},
\quad h_j\in[0,1].
\label{eq:multi-sos-scalar-test}
\end{align}
They have factors $A_j=s_j\sqrt{\lambda_jh_j}$ and energy $E=\sum_j\lambda_jh_j$.
Circuit amplitudes are polynomials in the oracle entries, of degree at most $n_j$ in the entries of oracle $j$.
Average the real part of the reference-state matrix element over the signs.
The resulting polynomial satisfies $P=P(h_1,\ldots,h_m)$, $\deg_{h_j}P\le\lfloor n_j/2\rfloor$, and $|P|\le1$ on the cube.
The phase-pair guarantee implies $P(h)=\cos\phi(\sum_j\lambda_jh_j)$ on the promised simplex, regardless of the relative weights of the two eigenbranches.

Restrict to one coordinate near zero.
This defines a univariate polynomial $p(E)$ agreeing with $\cos\phi(E)$ on a nonempty interval.
Polynomial identity then gives $P(h)=p(\sum_j\lambda_jh_j)$ everywhere.
All normalization weights are positive.
Therefore the degree in every individual variable equals the degree of the resulting univariate polynomial: $d:=\deg p=\deg_{h_j}P $ for every $ j$.
Setting all signals to $h_j=E/\Lambda$ also gives $|p(E)|\le1$ throughout the energy interval.
Its endpoints are zero and $\Lambda$.
The endpoint Markov inequality yields $|p'(0)|\le2d^2/\Lambda$, whereas the assumed phase expansion gives $p'(0)=-2k^2/\Lambda$.
Hence $d\ge k$.
\end{proof}

The endpoint inequality follows directly by interpolation at the Chebyshev extrema: the absolute values of the Lagrange-basis derivatives at the endpoint sum to $T_d'(1)=d^2$.
Rescaling $[-1,1]$ gives the factor $2/\Lambda$.
The result therefore applies even when exact spectral correctness is required only on a low-energy interval.
It permits arbitrary finite workspace.
It does not concern oracle-dependent reference preparations, decomposition-dependent phase laws, or expected-cost algorithms with unbounded stopping time.

For an exact encoded operator $aU$ with given $a>0$, the same argument applies to the attenuated phase: $a\cos\phi(E)$ and $q_j\ge\lceil k\sqrt a\rceil$.
Thus the lower bound depends on the normalization of the implemented block-encoding.
A quantitative $\log(1/\epsilon)$ lower bound for the approximate construction remains open here.
The minimum-cost choice in the main text uses two different permitted phase laws, and does not assert that they approximate one fixed target.

\begin{proposition}[Precision-dependent phases determined by energy]
\label{prop:multi-sos-precision-dependent}
Use the finite coherent separate-factor model and known reference embedding of \cref{prop:multi-sos-exact}.
Fix positive strengths and query upper bounds $n_j$.
Suppose the real part of the reference-state matrix element approximates $\cos\phi_\epsilon(E)$ uniformly to error $\epsilon$.
This must hold for arbitrarily small positive precision, on the promised instances and all sign choices for the factors and unspecified unitary blocks.
Assume continuous unwrapped phases with $2k_-\sqrt{E/\Lambda}\le\phi_\epsilon(E) \le2k_+\sqrt{E/\Lambda} $ for $0\le E\le E_0$, where $E_0>0$ and $0<k_-\le k_+<\infty$ are fixed, and the interval is contained in the promise.
Then $\lfloor n_j/2\rfloor\ge k_-$ for every $j$.
Finite workspace and phase laws may vary with precision.
\end{proposition}
\begin{proof}
Sign-average the real part of the reference-state matrix element as in \cref{prop:multi-sos-exact}.
The resulting polynomials $P_\epsilon$ have coordinate degrees at most $\lfloor n_j/2\rfloor$ and absolute value at most one on the cube.
This is a bounded subset of a finite-dimensional polynomial space, so a sequence with errors tending to zero has a coefficient-convergent subsequence.
Its limit $P$ is constant on each promised energy level, since two such instances differ in this matrix element by at most $2\epsilon$.
The polynomial-identity argument in \cref{prop:multi-sos-exact} gives
\begin{align}
P(h)=p\left(\sum_j\lambda_jh_j\right),
\quad |p|\le1\text{ on }[0,\Lambda],
\quad d=\deg p\le\min_j\lfloor n_j/2\rfloor.
\end{align}
For sufficiently small fixed positive $E$, both bounding phases lie in $[0,\pi]$.
Monotonicity of cosine gives the endpoint and slope bounds $p(0)=1$, $p(E)\le\cos(2k_-\sqrt{E/\Lambda})$, and $p'(0)\le-2k_-^2/\Lambda$.
The endpoint Markov bound $|p'(0)|\le2d^2/\Lambda$ proves $d\ge k_-$.
A fixed selected amplitude $a>0$ similarly gives $d\ge k_-\sqrt a$.
\end{proof}
The interval and query budgets in this proposition stay fixed as error tends to zero.
It gives a qualitative zero-error-limit obstruction; it leaves the precision logarithm and exponent of $1/\delta$ unrestricted.
Decomposition-dependent phases and useful intervals shrinking with precision require a different argument.

\subsubsection{Resources for the SOS cascade}
\label{app:sos-cascade-resources}
Use \cref{lem:multi-sos-known-cascade} with isometry error $O(\epsilon/N)$ per known update.
For the projector system matrix it gives the gate-count bound
\begin{subequations}
\label{eq:multi-sos-gates-poly}
\begin{align}
G_{\rm total}\le{}&\sum_jq_j (G_{V_j}+G_{V_j^\dagger}+G_{2\Pi_{{\rm out},j}-I}) +O(NG_{\rm align})
\\
&+N\operatorname{poly}(m,b,\log(2J),\log(N/\epsilon)) +G_{\rm clk}.
\end{align}
\end{subequations}
The arithmetic workspace in this implementation must be added to the address qubit count and the largest oracle scratch space.
This bound keeps every history mode and block-encoding ancilla.
Each known update has polynomial dependence on $\log J$.
The displayed arithmetic and gate counts are sufficient upper bounds; only the oracle-cost theorem carries a leading query prefactor $1+\delta$.

\subsubsection{The shared-SELECT rounding gap}
\label{app:multi-hs-select}

Let $\bar r_j$ be as in \cref{eq:multi-hs-select-delays}, and put
\begin{align}
L_0=\left(\sum_j\lambda_j\bar r_j\right) \left(C_S+\sum_jc_j/\bar r_j\right),
\quad D_0=\Lambda C_S+\left(\sum_j\sqrt{\lambda_jc_j}\right)^2.
\end{align}
If some $c_j>0$, the strict inequality $\bar r_j<1+\sqrt{\Lambda c_j/(C_S\lambda_j)}$ on every positive-cost term gives
\begin{align}
\sum_j\lambda_j\bar r_j<\Lambda+ \sqrt{\Lambda/C_S}\sum_j\sqrt{\lambda_jc_j},
\quad \sum_jc_j/\bar r_j\le\sqrt{C_S/\Lambda} \sum_j\sqrt{\lambda_jc_j}.
\end{align}
Thus $L_0<(\sqrt{\Lambda C_S}+\sum_j\sqrt{\lambda_jc_j})^2 \le2D_0$.
If all $c_j=0$, then $L_0=D_0<2D_0$.
Define the positive relative rounding margin and the internal normalization allowance by
\begin{align}
\gamma=1-\frac{L_0}{2D_0}\in(0,1],
\quad \nu=\min\!\left\{\delta,\frac{2D_0-L_0}{2L_0}\right\}.
\label{eq:multi-hs-select-internal-slack}
\end{align}
These choices give the non-strict bound
\begin{align}
(1+\nu)L_0\le\frac{2D_0+L_0}{2}=(2-\gamma)D_0.
\end{align}
Compile at normalization $1+\nu$ with unnormalized error at most $\min\{\epsilon/16,\nu\}$.
This also satisfies the clock theorem's precision hypothesis when the rounding margin is small.
Attenuation by $(1+\nu)/(1+\delta)$ sets the requested normalization without changing the approximating operator before normalization.
The logarithmic term can be bounded by
\begin{align}
O\!\left(\nu^{-2}\bar r_* \left(C_S+\sum_jc_j/\bar r_j\right)\log\frac1\epsilon\right),
\quad \bar r_*=\max_j\bar r_j.
\end{align}
At fixed positive $\delta$, the coefficient is independent of time and accuracy and depends on the instance through the rounding gap.
Choosing $\nu=\delta$ gives the alternative with the prescribed $\delta$
\begin{align}
C_Sq_S+\sum_jc_jq_j \le2(1+\delta)tD_0+ O\!\left(\delta^{-2}\bar r_* \left(C_S+\sum_jc_j\right)\log\frac1\epsilon\right).
\label{eq:multi-hs-select-uniform}
\end{align}
For finite precision one may optimize the complete fixed-delay objective in \cref{eq:multi-hs-select-horizon}, including its remainder.
Because $C_S>0$, its precision part lower bounds a positive multiple of $r_*$, so comparison with the all-unit-delay choice restricts any improvement to a finite integer search.
The exact SELECT count is the size of the union of scheduled query times in \cref{eq:multi-hs-select-counts}; simultaneous preparations share the SELECT call.

\subsection{Weighted state preparation}
\label{app:weighted-state-preparation}

On the input/output plane, the operator is an amplitude-weighted swap: $K_j\big|_{\operatorname{span}\{|\mathrm{in}\rangle,|\mathrm{out},\psi_j\rangle\}} =\lambda_j\sigma_x$.
Its two eigenvalues, shifted by $\lambda_{0j}/2$, have opposite signs.
On the kernel, the shifted operator is positive.
This proves the gap and \cref{eq:wsp-polar}.
The finite Sign bounds in \cref{lem:finite-sign-cap} now apply.
The spectral theorem makes these full bounds on the public transfer function valid for $H_j$ and all unitaries satisfying the oracle promises.
Private modes uncoupled from the input are interpreted through the regular continuation of the transfer function reachable from the public input, as in the Toeplitz-lift construction.
In particular, for $|z_j-1|\le c\lambda_{0j}$,
\begin{align}
\|P_j(z_j)-P_j\|\le C\frac{|z_j-1|}{\lambda_{0j}},
\quad \|\partial_{z_j}P_j(z_j)\|\le C/\lambda_{0j},
\label{eq:wsp-child-perturbation}
\end{align}
with universal constants independent of $L_j$.
The finite unitary realization is contractive in the disk and unitary on the part of the unit circle in this neighborhood.

Since $E^\dagger E=I$, the system matrix in \cref{eq:wsp-system} swaps the public space with the image of the coupling map.
It acts as $-I$ on the orthogonal complement, proving unitarity.
Eliminating its private variables gives \cref{eq:wsp-cayley-transfer}.
At $\Bz=\boldsymbol{1}$, $R_j$ is a Hermitian involution, so
\begin{align}
(I\pm iR_j)(I\mp iR_j)^{-1}=\pm iR_j,
\quad \|(I\mp iR_j)^{-1}\|=1/\sqrt2.
\label{eq:wsp-return-inverse}
\end{align}
Thus $M_\pm$ is skew-Hermitian and $\|(I+M_\pm)^{-1}\|\le1$.
Using \cref{eq:wsp-child-perturbation}, first in each private inverse and then in the remaining public inverse, the Neumann-series criterion gives uniform bounded inverses whenever $\max_j|z_j-1|/\lambda_{0j}$ is sufficiently small.
On a slightly smaller polydisk the resolvent identity gives
\begin{align}
\|\partial_{z_j}F_\pm(\Bz)\| \le C\frac{|a_j|}{A\lambda_{0j}}.
\label{eq:wsp-weighted-derivative}
\end{align}
For example, $(I+iR)(I-iR)^{-1}=2(I-iR)^{-1}-I$ and $(I-M)(I+M)^{-1}=2(I+M)^{-1}-I$ exhibit the bounded inverse factors in both differentiations.
This bounds the full connected module on the analytic domain, including its new feedback inverses.

At $\Bz=\boldsymbol{1}$, the compressed branch operator is a Hermitian star Hamiltonian.
Its first column is $e^{i\arg a_j}P_j|e\rangle$.
Replacing these columns by their ideal limits gives \cref{eq:wsp-sum-signal}.
Its action is $(s/A)\sigma_x$ on $\operatorname{span}\{|s\rangle,|\Psi\rangle\}$ and zero on the orthogonal complement.
This proves \cref{eq:wsp-cayley-column} and the correct relative phases.

\begin{proof}[Proof of \cref{lem:wsp-certificate}]
For $|z-1|$ bounded as in \cref{eq:wsp-radius}, $|z^{r_j}-1|\le2r_j|z-1|$, so every argument lies in the common polydisk just proved.
The derivative of one delayed Cayley transfer function is bounded by $C\sum_j |a_j|r_j/(A\lambda_{0j})$.
At $z=e^{i\theta}$ in this neighborhood all of its constituent unitary transfer functions are unitary, hence the complete transfer function is unitary.
Integrating the derivative radially from $e^{i\theta}$ to $\rho e^{i\theta}$ gives
\begin{align}
\|F_\pm(z^{\boldsymbol{r}})\| \le1+C\sum_j\frac{|a_j|r_j}{A\lambda_{0j}}(|z|-1) \le\exp\left(C'\sum_j\frac{|a_j|r_j}{A\lambda_{0j}}\log|z|\right)
\label{eq:wsp-cell-growth}
\end{align}
for $|z|\ge1$ on a smaller neighborhood.
Contractivity inside the disk follows from the finite unitary realizations.
The finite product has $D$ factors, so submultiplicativity proves \cref{eq:wsp-growth}.
Known gates and controlled direct sums do not change the bound.
Its common domain is independent of $D$ and all $L_j$.
\end{proof}

The ideal accepted amplitude in \cref{eq:wsp-cayley-column} is at least $s_0/A$ and at most one.
A real odd bounded polynomial of degree $O((A/s_0)\log(1/\epsilon))$ approximates the sign function on this interval to error $O(\epsilon^2)$.
Singular-value transformation preserves the accepted singular vectors and implements a finite sequence of unitary operations of that degree.
Its selected amplitude is real and within $O(\epsilon^2)$ of one; unitarity bounds the norm of its failure component by $O(\epsilon)$.
This proves the Euclidean error of the joint output state bound with the prescribed relative phase used in \cref{eq:wsp-outer-degree}.
For a given superposition norm, exact amplitude amplification uses \begin{align}
\text{angle }\arcsin\!\left(\frac{2(s/A)}{1+(s/A)^2}\right),
\quad O(A/s)\text{ calls}.
\end{align} A known final phase correction fixes its phase.

Choose even path lengths with $L_j\lambda_{0j}$ above the constant required above and $2e^{-L_j\operatorname{atanh}(\lambda_{0j}/(2+\lambda_{0j}))} \le\eta$.
By \cref{eq:wsp-path-error}, each implemented branch module differs from its ideal limiting involution by at most $\eta$.
Their compressed star Hamiltonians and convex combination also differ by at most $\eta$.
The Cayley identity bounds the system-matrix error by $2\eta$.
Summing over all $D$ occurrences bounds the product error by $2D\eta$.
Allocate the approximation errors by $\eta\le\epsilon/(128D)$ and ideal outer joint-output error at most $\epsilon/16$.
Reserve a further $\epsilon/16$ each for clock approximation and synthesis.
These bounds imply \cref{eq:wsp-output}.

Apply the local analytic-clock theorem with interior exponent zero, the exterior exponent of \cref{eq:wsp-growth}, the radius in \cref{eq:wsp-radius}, and normalization two.
It gives a clock matrix with horizon \cref{eq:wsp-horizon}.
The finite network has already been chosen, so no infinite-register implementation or interchange of limits is required.
The same clock works for all state-preparation unitaries satisfying the promises.
\cref{sec:quditization}\textquoteright s exact clock characterization and unitary Toeplitz lift implement it.

For the Hamiltonian-simulation route in \cref{eq:wsp-hamsim-response,eq:wsp-grover-response}, the matrix acts separately on the coupled and uncoupled subspaces.
On the public space and the image of the coupling map, the matrix $S_\Delta$ is a two-mode reflection.
Its entries are \begin{align}
\text{diagonal: }\pm\frac{1-\Delta/2}{1+\Delta/2},
\quad \text{off diagonal: }\frac{2\sqrt{\Delta/2}}{1+\Delta/2}.
\end{align} On the remaining private space it is $-I$.
It is therefore unitary.
Eliminating the same separate feedback connections $\mp iR_j$ gives $C_{\Delta,\pm}$ in \cref{eq:wsp-hamsim-response}.
Thus each finite cascade has a full unitary realization, including its complementary output coordinates.

At $\Bz=\boldsymbol{1}$, write
\begin{align}
M_\pm=\pm i\widetilde H,
\quad \widetilde H=\sum_j\frac{|a_j|}{A}J_j^\dagger R_jJ_j,
\quad \|\widetilde H\|\le1.
\label{eq:wsp-finite-star}
\end{align}
This is an exactly Hermitian star Hamiltonian, even before taking the branch limit.
If every branch column has error at most $\eta$, the resulting Hamiltonian differs from $H$ by at most the same amount.
For an eigenvalue $h\in[-1,1]$, the forward cascade has eigenvalue
\begin{align}
\left(\frac{1-i\tau h/(2J)}{1+i\tau h/(2J)}\right)^J =\exp\!\left[-2iJ\arctan\!\left(\frac{\tau h}{2J}\right)\right].
\label{eq:wsp-cascade-eigenvalue}
\end{align}
The inequality $|\arctan y-y|\le |y|^3/3$ and the spectral theorem give
\begin{align}
\left\|\mathcal U_{\tau,J,\pm} -e^{\mp i\tau\widetilde H}\right\| \le\frac{\tau^3}{12J^2},
\quad \left\|e^{\mp i\tau\widetilde H}-e^{\mp i\tau H}\right\| \le\tau\eta.
\label{eq:wsp-cascade-errors}
\end{align}
The second inequality follows from Duhamel's integral and unitarity.
The action of $H$ on its span of the reference and target states proves \cref{eq:wsp-hamsim-column}.
Concavity of sine on $[0,\pi/2]$ gives $\sin(\pi s/(2A))\ge s/A$.
Thus the same real, phase-preserving filter used above has degree $D=O((A/s_0)\log(1/\epsilon))$ at $\tau=\pi/2$.
For a given superposition norm, choose $\tau=\pi A/(2s)$.
This transfers the ideal reference state to the target with a phase corrected by $Z_g$, so outer degree $D=1$ suffices.
Both signs of the Cayley argument in \cref{eq:wsp-hamsim-response} are analytic transfer functions, and their values at $\Bz=\boldsymbol{1}$ are mutual inverses.
Consequently, \begin{align}
B_+R_s(\phi)B_- =I+(e^{i\phi}-1)B_+|s\rangle\langle s|B_+^\dagger.
\label{eq:wsp-preparation-reflection}
\end{align} This justifies the expanded preparation-reflection diagram, including the order and the phase convention of its inverse call.

The cascade horizon is independent of $J$ because the private-inverse bounds in \cref{eq:wsp-return-inverse} give
\begin{align}
\|\partial_{z_j}M_\pm(\Bz)\| \le C\frac{|a_j|}{A\lambda_{0j}}
\label{eq:wsp-load-derivative}
\end{align}
on the common polydisk.
Take $J\ge\tau$, so $0<\Delta=\tau/J\le1$.
Since $M_\pm$ is skew-Hermitian, $\|[I+(\Delta/2)M_\pm]^{-1}\|\le1$.
Shrinking the same polydisk by a universal factor bounds this inverse by two throughout, uniformly in $\Delta$.
Differentiating $C_{\Delta,\pm}=2[I+(\Delta/2)M_\pm]^{-1}-I$ yields
\begin{align}
\|\partial_{z_j}C_{\Delta,\pm}(\Bz)\| \le C\Delta\frac{|a_j|}{A\lambda_{0j}}.
\label{eq:wsp-small-cell-derivative}
\end{align}
Each small Cayley transfer function is unitary on the local unit-circle arc.
Applying the radial integration argument of \cref{eq:wsp-cell-growth}, multiplying the $J$ transfer functions, and then multiplying the $D$ loader occurrences gives
\begin{align}
\log\|G(z)\| \le C\frac{D\tau}{A}\sum_j\frac{|a_j|r_j}{\lambda_{0j}}\log|z|
\quad(|z|\ge1),
\label{eq:wsp-hamsim-growth}
\end{align}
on the radius in \cref{eq:wsp-radius}.
The radius and the constant are independent of $J$, $D$, and the path lengths.
Interior contractivity follows from the finite unitary realization.
The local analytic-clock theorem and \cref{sec:quditization}\textquoteright s exact unitary lift therefore give \cref{eq:wsp-hamsim-horizon}.

For explicit finite choices, take
\begin{align}
\eta\le\frac{\epsilon}{256D\tau},
\quad J\ge\max\!\left\{\tau, \sqrt{\frac{64D\tau^3}{3\epsilon}}\right\},
\quad J\in\mathbb N.
\label{eq:wsp-finite-hamsim-choices}
\end{align}
Then \cref{eq:wsp-cascade-errors} and a hybrid over the $D$ occurrences give a total loader error at most $\epsilon/128$.
The other error budgets may be chosen as in the direct-Cayley proof, proving \cref{eq:wsp-output} for this route as well.
All branch paths and all $J$ system matrices are fixed before applying the single clock theorem.
The module index is part of the retained direct-sum processing space; one feedback connection applies a single call to the original oracle coherently across that space.
This explains why $J$ contributes to the dimension and gates implementing the known system matrix, but not as an extra multiplicative factor in the original-query counts.

For the chosen $D$, define the target rates
\begin{align}
t_j=\frac{\log(2/\epsilon)}{\lambda_{0j}} +\frac{D}{A}\sqrt{\frac{|a_j|}{\lambda_{0j}C_j}} \sum_k\sqrt{\frac{|a_k|C_k}{\lambda_{0k}}}.
\label{eq:wsp-rates}
\end{align}
Choose any integer $T\ge\max_jt_j$ and set $r_j=\lceil T/t_j\rceil$.
Then $T/t_j\le r_j\le2T/t_j$ and
\begin{align}
\frac{D}{A}\sum_j\frac{|a_j|}{\lambda_{0j}t_j}\le1,
\quad \log(2/\epsilon)\max_j\frac1{\lambda_{0j}t_j}\le1.
\label{eq:wsp-allocation-inequalities}
\end{align}
Consequently, \cref{eq:wsp-horizon} gives $N=O(T)$ and $q_j=O(t_j)$, including the constant number of original preparation calls per specified feedback operation.
Summing $C_jt_j$ gives
\begin{align}
\sum_j C_jq_j=O\!\left[ \frac{D}{A}\left(\sum_j\sqrt{\frac{|a_j|C_j}{\lambda_{0j}}}\right)^2 +\log\frac1\epsilon\sum_j\frac{C_j}{\lambda_{0j}}\right].
\label{eq:wsp-unified-cost}
\end{align}
The two choices of $D$ prove \cref{eq:wsp-unknown-cost,eq:wsp-known-cost}.
For HamSim, replace $D$ by $D\tau$ in \cref{eq:wsp-rates,eq:wsp-allocation-inequalities,eq:wsp-unified-cost}.
The two parameter choices give the same asymptotic bounds: $(\tau,D)=(\pi/2,O((A/s_0)\log(1/\epsilon))) $ or $ (\pi A/(2s),1)$.
For \cref{eq:wsp-sharp-allocation}, Cauchy--Schwarz gives the lower bound, and equality holds for $r_j\propto\sqrt{C_j\lambda_{0j}/|a_j|}$.
This last statement concerns continuous leading-term allocation only.

Finally, for $\eta=\|\boldsymbol{b}-\tfrac12|\Psi\rangle\|$, the reverse triangle inequality gives $|\|\boldsymbol{b}\|-1/2|\le\eta$ and
\begin{align}
\left\|\frac{\boldsymbol{b}}{\|\boldsymbol{b}\|}-|\Psi\rangle\right\| \le\left\|\frac{\boldsymbol{b}}{\|\boldsymbol{b}\|}-2\boldsymbol{b}\right\| +\|2\boldsymbol{b}-|\Psi\rangle\| \le4\eta\le\epsilon.
\label{eq:wsp-postselection}
\end{align}
This also proves the stated success-probability interval.
The resource bounds follow by applying the unitary-lift implementation theorem to the full retained dimensions in \cref{eq:wsp-dimensions}, with all known system operations and forward and adjoint oracle calls counted.

\section{Linear systems, Volterra certificates, and eigenvalue estimation}
\label{app:qlsp}

\subsection{Realization and finite-query proofs}
\label{wql:app:proofs}

The proofs of \cref{wql:thm:main,wql:thm:factor} use the unitary Toeplitz lift and analytic clock theorem in \cref{thm:iir-lift,thm:iir-analytic-clock-selection}.
The rational completion in \cref{wql:lem:reciprocal} is the Reciprocal module of \cref{lem:mod-reciprocal}.

\subsubsection{The known Cayley system}

Apply \cref{lem:common-cayley} with couplings $\mathcal C_j$.
Writing $\boldsymbol h=(\boldsymbol u+\boldsymbol v)/2$ for the average load wave and $\boldsymbol q,\boldsymbol w$ for the incoming and outgoing private waves gives
\begin{align}
\boldsymbol q+\boldsymbol w=2\mathcal C\boldsymbol h,
\quad \boldsymbol u-\boldsymbol h=iK_0\boldsymbol h+ \mathcal C^\dagger(\boldsymbol w-\mathcal C\boldsymbol h).
\label{wql:eq:junction-waves}
\end{align}
The response and unitarity follow from the common junction lemma; the remaining step is the identity closure of the auxiliary load coordinate.

For the constraint network $K_0=0$, and \cref{wql:eq:sum-gram} makes $\mathsf R$ a known diagonal matrix of scalar multiples of identities.
The system matrix's block on the closed auxiliary space is $(\mathsf S_0)_{\mathcal Y\mathcal Y} =2(1+K\|\boldsymbol{\lambda}\|_1+K\|\boldsymbol{a}\|_1/\rho)^{-1}I-I$.
Its identity feedback is therefore well defined.
If $e$ denotes all remaining coordinates, the assembled system is the known unitary
\begin{align}
\mathsf S_\rho =\mathsf S_{ee}+\mathsf S_{eY} (I-\mathsf S_{YY})^{-1}\mathsf S_{Ye}.
\label{wql:eq:known-closure}
\end{align}
Norm conservation follows by cancelling the equal incoming and outgoing $\mathcal Y$ waves.
At $\Bz=\boldsymbol{1}$ the closed equations yield \cref{wql:eq:one-reflection}.
At $\Bz=\boldsymbol{1}$, the private wave is unique.
Invertibility of $H$ fixes the public vector and scalar through the reflection.
The public row then determines $\boldsymbol{y}$, and \cref{wql:eq:junction-waves} fixes the oracle waves.
Applying the known system matrices in order on their shared scalar coordinate and disjoint remaining coordinates gives the cascade.
The series connection rule gives its transfer function and preserves unitarity.

\subsubsection{Uniform-clock output and the remaining equation residual}

\begin{lemma}[Uniform clock and terminal constraint]
\label{wql:lem:terminal}
Use positive integer matrix and preparation delays $r_j,s_k\ge1$.
The input and output clocks are uniform over $N$ time slots.
Let $\boldsymbol{y}_N$ be the complete clock-selected public output, and let $\boldsymbol{e}_N$ be the terminal private output.
If $W$ is the delayed steady-state weight for a normalized fixed input $\boldsymbol{u}$, then
\begin{align}
\operatorname{Re}\langle F\boldsymbol{u},\boldsymbol{y}_N\rangle
\quad \ge1-2W/N,
\quad \|\boldsymbol{y}_N-F\boldsymbol{u}\|\le2\sqrt{W/N},
\quad \|\boldsymbol{e}_N\|\le2\sqrt{W/N}.
\label{wql:eq:uniform-estimates}
\end{align}
For each scale of the constraint cascade, its $\mathcal H$ output obeys \cref{wql:eq:sum-residual,wql:eq:sum-residual-gram}.
The normalized terminal map has an implementation with a constant number of queries of each original type, independent of the delays and number of scales.
\end{lemma}
\begin{proof}
Include the causal delay lines in a one-step unitary system.
Denote the private update contraction by $D_{\mathrm{buf}}$ and the public-output block by $B_0$.
The steady-state private vector is $\boldsymbol{g}$, with $\|\boldsymbol{g}\|^2=W$.
Drive the public input at constant amplitude for $N$ steps.
With input $\boldsymbol{u}/\sqrt N$, the private vector before step $t$ is $(I-D_{\mathrm{buf}}^t)\boldsymbol{g}/\sqrt N$.
Its terminal value and averaged public output are
\begin{align}
\boldsymbol{e}_N=(I-D_{\mathrm{buf}}^N)\boldsymbol{g}/\sqrt N,
\quad \boldsymbol{y}_N=F\boldsymbol{u}-\frac1N\sum_{t=0}^{N-1}B_0D_{\mathrm{buf}}^t\boldsymbol{g}.
\label{wql:eq:uniform-exact}
\end{align}
The cross-column identity and finite telescoping in \cref{lem:iir-uniform-terminal-residual} give
\begin{align}
\langle F\boldsymbol u,\boldsymbol y_N\rangle =1-\frac1N\langle\boldsymbol g,(I-D_{\rm buf}^N)\boldsymbol g\rangle.
\label{wql:eq:overlap-exact}
\end{align}
Contractivity and \cref{thm:iir-uniform} prove \cref{wql:eq:uniform-estimates}.
Keeping the unselected public outputs and terminal buffers gives the complete contractive output column.

Sum the local constraint equations with the output clock coefficient $N^{-1/2}$.
For a feedback connection of delay $r$, let $\mathcal T_r$ sum all its terminal outgoing buffer entries.
With zero initial private amplitude, the averaged feedback equations give
\begin{align}
\boldsymbol{q}=-iV\left(\boldsymbol{w}- \mathcal T_r\boldsymbol{e}_N/\sqrt N\right),
\quad \mathcal T_r\mathcal T_r^\dagger=rI.
\label{wql:eq:buffer-sum}
\end{align}
The private row of the known system still gives $\boldsymbol{q}+\boldsymbol{w}=2\mathcal C\boldsymbol{h}$.
Therefore
\begin{align}
\boldsymbol{w}-\mathcal C\boldsymbol{h} =iV\mathcal C\boldsymbol{h}+ \frac{I-iV}{2\sqrt N}\mathcal T_r\boldsymbol{e}_N.
\label{wql:eq:finite-private-wave}
\end{align}
Apply this identity to every feedback connection of one system matrix.
Let $P_Y$ project onto the closed load coordinate.
Form $\mathcal T$ as the direct sum of the buffer maps, using the delay of each matrix or preparation port.
Define
\begin{align}
L_\rho =\frac12P_Y\mathcal C^\dagger(I-iV)\mathcal T.
\label{wql:eq:terminal-map}
\end{align}
The closed $\mathcal Y$ row of \cref{wql:eq:junction-waves} has zero incoming--outgoing difference.
At the all-ones load it is
\begin{align}
0=iK\left(H\boldsymbol{x}_{N,\rho} -\frac{c_{N,\rho}}\rho\boldsymbol{b}\right) +\frac1{\sqrt N}L_\rho\boldsymbol{e}_{N,\rho}.
\label{wql:eq:closed-finite-row}
\end{align}
The zero system input gives $\boldsymbol{v}_{N,\rho}=2\boldsymbol{x}_{N,\rho}$, which yields \cref{wql:eq:sum-residual}.

Since $(I-iV)(I+iV)=2I$, each feedback connection contributes one half its delay multiplied by the Gram matrix of its map into $\mathcal Y$.
The matrix and preparation feedback operations contribute, respectively, $K\lambda_jr_jI/2$ and $K|a_k|s_kI/(2\rho)$.
Different feedback connections and different scale system matrices occupy orthogonal private subspaces.
Summing proves \cref{wql:eq:sum-residual-gram}.

The normalized buffer sum $\mathcal T_r/\sqrt r$ is implemented by projection onto the uniform state of the buffer address.
The factor $(I-iV)/\sqrt2$ is unitary.
One controlled oracle call computes its involution eigenvalue into a qubit; applying phase $e^{\mp i\pi/4}$ and a second controlled call erases that qubit.
The normalized remaining row is a known coordinate map with known weights.
These operations implement $L_\rho/\ell$ for any known $\ell\ge\|L_\rho\|$.
Its rejected output is retained.
A single original-oracle call acts on every copy with the occurrence label unchanged.
For buffers stored after an oracle application, first apply the inverse oracle, adding a constant number of calls.
Either convention uses $O(1)$ original calls per type and retains all preparation workspace, including amplitudes outside its zero-input subspace.
\end{proof}

\subsubsection{Reciprocal with separate matrix queries}

\begin{lemma}[Weighted matrix-only reciprocal]
\label{wql:lem:reciprocal}
Let $H=\sum_jH_j$ be Hermitian and invertible, with inverse-norm bound $\|H^{-1}\|\le K$.
Assume self-inverse component encodings $O_j=\Be{H_j/\lambda_j}$.
Choose positive integer matrix delays $r_j$ and error $0<\eta<1/4$.
A finite unitary circuit has a block-encoded operator $J$ satisfying
\begin{align}
\left\|J-\frac{H^{-1}}{4K}\right\|\le\eta.
\label{wql:eq:reciprocal-guarantee}
\end{align}
It uses no state-preparation query.
A sufficient horizon and its original matrix counts are
\begin{align}
N_1=O\!\left[ \left(K\sum_j\lambda_jr_j+\max_jr_j\right) \log\frac2\eta\right],
\quad Q_j^{\rm inv}=O(N_1/r_j),
\label{wql:eq:reciprocal-queries}
\end{align}
with constants independent of the number of terms.
\end{lemma}
\begin{proof}
Use the scalar approximation, two-coordinate unitary completion, and at most $3d$ all-pass sections of \cref{lem:mod-reciprocal,app:mod-reciprocal}.
The following argument bounds the separate matrix ports of that same completion.

Substitute $u=2KH$.
On each known active public subspace, a section has the response
\begin{align}
F_{b,a,\sigma}(\Bz) &=I-2b\,[bI+\mathcal M_\sigma(\Bz)-i\sigma aI]^{-1},
\label{wql:eq:reciprocal-factor}
\\
\mathcal M_\sigma(\Bz) &=\sum_j\left[ \lambda_j\frac{1-z_j^2}{1+z_j^2}I +i\sigma\frac{2z_j}{1+z_j^2}H_j \right],
\quad |a|,b\le(2K)^{-1},
\quad b>0,
\quad\sigma\in\{-1,1\}.
\label{wql:eq:reciprocal-load}
\end{align}
The orthogonal public subspace is transmitted unchanged.
\cref{lem:common-cayley} realizes this section with $\text{coupling }\sqrt{\lambda_j/b}(|0\rangle_j\otimes I)$, known load $K_0=-\sigma aI/b$, and feedback $-i\sigma z_jO_j$.
Apply a known minus phase on the active public subspace after the junction.
For the scalar variable $u=2KH$, the choice $\sigma=-1$ then gives the Blaschke factor $B_j(u)$ of \cref{app:mod-reciprocal}, and $\sigma=+1$ gives its inverse; the orthogonal public subspace is unchanged.
Both sign choices remain analytic in the query markers at zero; neither requires negative powers of those markers.
For a scalar signal, \cref{wql:eq:reciprocal-factor} is a shifted all-pass Cayley response; the matrix terms enter through its load.

Write $r_{\max}=\max_jr_j$ and choose a sufficiently small universal $c>0$.
For $|z-1|\le c/(K\sum_j\lambda_jr_j+r_{\max})$, elementary bounds on $z^{r_j}$ and the denominators in \cref{wql:eq:reciprocal-load} give
\begin{align}
\|\mathcal M_\sigma(z^{r_1},\ldots,z^{r_m})-i\sigma H\| \le O\!\left(\sum_j\lambda_jr_j\right)|z-1|.
\label{wql:eq:reciprocal-perturbation}
\end{align}
The constant is independent of $m$.
The gap promise gives
\begin{align}
\|[bI+i\sigma(H-aI)]^{-1}\|\le2K.
\end{align}
On the same neighborhood, a Neumann series bounds each perturbed inverse by $4K$ and each complete factor by $5$.
These bounds are uniform in the pole width.

At $z=1$, the derivative of the matrix coefficient $2z^{r_j}/(1+z^{2r_j})$ vanishes.
Thus
\begin{align}
\frac{d}{dz}\mathcal M_\sigma(z^{r_1},\ldots,z^{r_m})\bigg|_{z=1} =-\sum_j\lambda_jr_j I,
\end{align}
and each factor derivative has norm $O(K\sum_j\lambda_jr_j)$.
All factors evaluated at $\Bz=\boldsymbol{1}$ are unitary.
Differentiating their ordered product therefore bounds the group-delay norm of the completed response by $O(dK\sum_j\lambda_jr_j)$.
Its norm throughout the common neighborhood is at most $5^{3d}$.
These are operator-norm estimates; the off-boundary factors need not commute.

The analytic clock theorem gives clock normalization two and numerator error $O(\eta)$ with horizon
\begin{align}
N_1=O\!\left[ dK\sum_j\lambda_jr_j+ \left(K\sum_j\lambda_jr_j+r_{\max}\right) \left(d+\log\frac2\eta\right)\right].
\end{align}
Choose $d=O(\log(2/\eta))$ so that \cref{eq:mod-reciprocal-error} has error $O(\eta)$.
The clock-encoded operator approximates $\mathscr R_d(2KH)/2$.
Selecting its first matrix coordinate therefore approximates $H^{-1}/(4K)$.
Allocate the error $\eta$ between scalar reciprocal approximation and clock truncation to obtain \cref{wql:eq:reciprocal-guarantee}.

Apply the unitary Toeplitz lift to the complete cascade.
Sections using oracle $j$ have a common delay and occupy a direct sum on which one controlled $O_j$ call implements each scheduled feedback operation.
The integer-delay schedule gives \cref{wql:eq:reciprocal-queries}, independently of the number of sections.
The section count still enters the known-gate cost and private dimension.
\end{proof}

\subsubsection{Success probability and approximation error}

\begin{proposition}[Completion of the weighted construction]
\label{wql:prop:success}
The delays in \cref{wql:eq:delays}, a coarse horizon $N_0=\lceil4096L\rceil$, and the reciprocal of \cref{wql:lem:reciprocal} with $\eta=\epsilon/32$ give the output guarantee and query counts of \cref{wql:thm:main}.
The coarse state-preparation circuit is independent of $\epsilon$.
\end{proposition}
\begin{proof}
The rate inequalities give $W\le10L$, $\ell^2\le2L$, and $\ell=\max_\rho\|L_\rho\|$.
The terminal-map bound follows from \cref{wql:eq:sum-residual-gram}.
Let $\Pi_{\rm out}$ project onto all system output summands of the scale cascade.
The finite public output is $\boldsymbol{v}_{N_0}=\Pi_{\rm out}\boldsymbol{y}_{N_0}$.
The steady-state output obeys $\|\Pi_{\rm out}F\boldsymbol{u}\|\ge4/5$.
By \cref{wql:lem:terminal},
\begin{align}
\|\boldsymbol{v}_{N_0}-\Pi_{\rm out}F\boldsymbol{u}\| \le2\sqrt{10/4096},
\quad \|\boldsymbol{e}_{N_0}\|\le2\sqrt{10/4096}.
\end{align}
The ideal residual-correction map has norm at most $2\ell/\sqrt{N_0}$, because $\|H^{-1}\|\le K$.
Its action on the terminal vector has norm at most $4\sqrt{20}/4096$.
The known row coefficient $\beta=8\ell/\sqrt{N_0}$ obeys $\beta^2\le128/4096$.
Thus the accepted norm with an exact reciprocal is at least
\begin{align}
\frac{\frac45-2\sqrt{10/4096}-4\sqrt{20}/4096} {\sqrt{1+128/4096}}>\frac23.
\label{wql:eq:explicit-success-margin}
\end{align}
By \cref{wql:eq:sum-residual}, each scale component has the solution direction.
The ideal accepted vector factors as $|\mathrm{sol}\rangle\otimes\boldsymbol{\gamma}$, separating the solution from its scale amplitudes.
Its norm exceeds $2/3$.

The reciprocal error changes the accepted vector by at most $\eta$, since $\beta/\sqrt{1+\beta^2}$, the terminal map, and its input column are contractions.
Allocate another $\epsilon/32$ of isometry error to known-gate synthesis, including the terminal map and clock preparations.
The total accepted-vector error is at most $\epsilon/16$.
Its norm therefore remains above $1/2$ throughout the assumed precision range.
For nonzero vectors $\Bz,\Bz_0$,
\begin{align}
\left\|\frac{\Bz}{\|\Bz\|} -\frac{\Bz_0}{\|\Bz_0\|}\right\| \le\frac{2\|\Bz-\Bz_0\|}{\|\Bz_0\|}.
\end{align}
The normalized vector error is consequently less than $3\epsilon/16$.
Discarding the scale register cannot increase trace distance, so the conditional data error is at most $\epsilon$ without knowledge of the successful scalar coefficient.

The coarse horizon gives $O(t_j)$ matrix calls and $O(u_k)$ preparation calls.
For the reciprocal, the same matrix delays give $O(t_j\log(1/\epsilon))$ calls, and the terminal map adds only $O(1)$ of each type.
Summing the weighted rates proves the two costs.
The termwise Hermitian dilation preserves $K$ and $R_0$ by \cref{wql:eq:dilation-norms} and adds only a constant number of original calls per feedback unitary.
Thus the same result holds for arbitrary $A_j$.
\end{proof}

\subsubsection{Separate factor and preparation queries}
\label{wql:subsec:factor-proof}

For \cref{wql:thm:factor}, define the solution norm and inverse-amplitude ratio by $H=T^2$, $R=\|H^{-1}\boldsymbol{b}\|$, $r_0=R_0/\|\boldsymbol{a}\|_1$, and $D=\kappa/r_0\ge1$.
Use the scale sequence in \cref{wql:eq:sum-scales} with inverse-norm bound $K=\kappa$.
It brackets the solution norm and gives steady-state public transmission at least $16/25$.

On load coordinates $(\boldsymbol{x},c\mid\boldsymbol{y}_1,\boldsymbol{u},\boldsymbol{y}_2)$, use
\begin{align}
\mathcal C_{j,1}\boldsymbol{h} &=\sqrt{\sqrt\kappa\lambda_j} \begin{pmatrix}(|0\rangle_j\otimes I)\boldsymbol{x}
\\
(|0\rangle_j\otimes I)\boldsymbol{y}_1\end{pmatrix},
\quad \mathcal C_{j,2}\boldsymbol{h} =\sqrt{\sqrt{\kappa D}\lambda_j} \begin{pmatrix}(|0\rangle_j\otimes I)\boldsymbol{u}
\\
(|0\rangle_j\otimes I)\boldsymbol{y}_2\end{pmatrix},
\label{wql:eq:factor-maps}
\\
\mathcal D_k\boldsymbol{h} &=\sqrt{\kappa|a_k|/\rho} \begin{pmatrix}(|0\rangle_{w_k}\otimes I)\boldsymbol{y}_2
\\
c|0\rangle\end{pmatrix},
\quad K_0=-\sqrt D (|y_1\rangle\langle u|+|u\rangle\langle y_1|)\otimes I.
\label{wql:eq:factor-source-known}
\end{align}
Here $|y_1\rangle,|u\rangle$ are basis labels for the coordinate subspaces of the Cayley load.
Each factor edge uses the involution
\begin{align}
V_j=\begin{pmatrix}0&O_{T,j}
\\
O_{T,j}&0\end{pmatrix}.
\end{align}
Each preparation edge uses the corresponding $W_k$ from \cref{wql:eq:sum-state-maps}.
Their orthogonal direct-sum maps form $\mathcal C$.
Its Gram matrix is
\begin{align}
\mathcal C^\dagger\mathcal C =\diag\!\left( \sqrt\kappa\|\boldsymbol{\lambda}\|_1I, \frac\kappa\rho\|\boldsymbol{a}\|_1, \sqrt\kappa\|\boldsymbol{\lambda}\|_1I, \sqrt{\kappa D}\|\boldsymbol{\lambda}\|_1I, \left[\sqrt{\kappa D}\|\boldsymbol{\lambda}\|_1+ \frac\kappa\rho\|\boldsymbol{a}\|_1\right]I \right).
\label{wql:eq:factor-known-gram}
\end{align}
The known system is \cref{eq:imp-junction} with these maps and $K_0$, and its open transfer is \cref{eq:imp-response}.
At $\Bz=\boldsymbol{1}$, the load is \cref{wql:eq:factor-scalar-load} after substituting the factor operator for the scalar signal.
The last edge becomes the coupling to $\boldsymbol{b}$.
Closing all three auxiliary subspaces gives the reflection in \cref{wql:eq:one-reflection}.
Here the matrix is $H=T^2$, and the right-hand side is the vector $\boldsymbol{b}$.
The Gram matrix is strictly positive on the closed subspaces, so the known identity closure is well defined: an eigenvector at eigenvalue one with zero external output would have zero image under every $\mathcal C_j$ and hence would be zero.
The unknown operators occur only in the oracle connections.

The steady-state vectors are
\begin{align}
c=\frac1{1+R^2/\rho^2},
\quad \boldsymbol{x}=\frac c\rho H^{-1}\boldsymbol{b},
\quad \boldsymbol{u}=\sqrt{r_0}\,T\boldsymbol{x},
\quad \boldsymbol{y}_1=\frac{i}{\sqrt\kappa}T^{-1}\boldsymbol{x},
\quad \boldsymbol{y}_2=\frac{i}{\kappa}T^{-2}\boldsymbol{x}.
\label{wql:eq:factor-waves-general}
\end{align}
The inverse-norm bound gives $\|\boldsymbol{y}_i\|\le\|\boldsymbol{x}\|$.
For the remaining vector, positivity of $H=T^2$ and Cauchy--Schwarz give
\begin{align}
\|\boldsymbol{u}\|^2 =\frac{|c|^2r_0}{\rho^2}\langle\boldsymbol{b},H^{-1}\boldsymbol{b}\rangle \le\frac{|c|^2r_0}{\rho^2}\|\boldsymbol{a}\|_1 R \le\frac{|c|^2R^2}{\rho^2}=\|\boldsymbol{x}\|^2.
\label{wql:eq:factor-u-bound}
\end{align}
This uses $r_0\|\boldsymbol{a}\|_1=R_0\le R$ and is valid even when $r_0<1$.
For each factor term, the two edges together obey $p_\rho=4\|\boldsymbol{x}\|^2$ and combined weight at most $(\sqrt\kappa+\sqrt{\kappa D})\lambda_jp_\rho$.
Each preparation edge satisfies \cref{wql:eq:local-source-weight} with $K=\kappa$ and $\boldsymbol{y}=\boldsymbol{y}_2$.
After cascading,
\begin{align}
W_{T,j}\le(\sqrt\kappa+\sqrt{\kappa D})\lambda_j,
\quad W_{b,k}<4\kappa|a_k|/R_0.
\label{wql:eq:factor-weighted-action}
\end{align}
These estimates use the inverse of the sum $T$ and permit noncommuting $T_j$.

Give both copies of each factor term delay $r_j$.
Give each preparation port its delay $s_k$.
The buffer identity \cref{wql:eq:finite-private-wave} applies unchanged, including the known $K_0$ term.
Let $L_{i,\rho}$ be $\frac12P_{y_i}\mathcal C^\dagger(I-iV)\mathcal T$.
The two finite closed rows read
\begin{subequations}
\label{wql:eq:factor-finite-rows}
\begin{align}
\sqrt\kappa T\boldsymbol{x}_{N,\rho}-\sqrt D\,\boldsymbol{u}_{N,\rho} &=iL_{1,\rho}\boldsymbol{e}_{N,\rho}/\sqrt N,
\\
\sqrt{\kappa D}T\boldsymbol{u}_{N,\rho}-\kappa c_{N,\rho}\boldsymbol{b}/\rho &=iL_{2,\rho}\boldsymbol{e}_{N,\rho}/\sqrt N.
\end{align}
\end{subequations}
Eliminating $\boldsymbol{u}_{N,\rho}$ proves
\begin{align}
H\boldsymbol{v}_{N,\rho} =\frac{2c_{N,\rho}}\rho\boldsymbol{b} +\frac{2i}{\kappa\sqrt N} (L_{2,\rho}+\sqrt\kappa T L_{1,\rho})\boldsymbol{e}_{N,\rho}.
\label{wql:eq:factor-weighted-residual}
\end{align}
The Gram calculation of \cref{wql:lem:terminal} gives
\begin{subequations}
\label{wql:eq:factor-gram}
\begin{align}
L_{1,\rho}L_{1,\rho}^\dagger &=\frac{\sqrt\kappa}{2}\sum_j\lambda_jr_j I,
\\
L_{2,\rho}L_{2,\rho}^\dagger &=\frac12\left(\sqrt{\kappa D}\sum_j\lambda_jr_j +\frac\kappa\rho\sum_k|a_k|s_k\right)I,
\quad L_{1,\rho}L_{2,\rho}^\dagger=0.
\end{align}
\end{subequations}
The rows for $y_1$ and $y_2$ occupy orthogonal edge subspaces, even when they share an oracle call, so their cross Gram matrix vanishes.
\cref{wql:lem:terminal} implements the normalized rows jointly using $O(1)$ original calls of each type.

For the coarse stage choose
\begin{align}
t_j=1+(\sqrt\kappa+\sqrt{\kappa D}) \sqrt{\frac{\lambda_j}{C_j}}\sum_i\sqrt{C_i\lambda_i},
\quad u_k=1+\frac\kappa{R_0}\sqrt{\frac{|a_k|}{c_k}} \sum_l\sqrt{c_l|a_l|},
\label{wql:eq:factor-coarse-rates}
\end{align}
and use \cref{wql:eq:delays} with $L\ge\max\{1,t_j,u_k\}$.
The same square-root calculation and ceiling bounds give
\begin{align}
(\sqrt\kappa+\sqrt{\kappa D})\sum_j\lambda_jr_j\le2L,
\quad \frac\kappa{R_0}\sum_k|a_k|s_k\le2L.
\label{wql:eq:factor-rounded-sums}
\end{align}
The direct sums of terminal maps over scales have bounds $\ell_i$ satisfying $W\le10L$, $\ell_1^2\le L$, and $\ell_2^2\le2L$.
Choose the coarse horizon $N_0=\lceil2^{16}L\rceil$.

Apply \cref{wql:lem:reciprocal} to the factor operator, using inverse-norm bound $K=\sqrt\kappa$ and error $\eta=\epsilon/128$.
Write its encoded operator as $J$, with $J_0=T^{-1}/(4\sqrt\kappa)$.
Two executions with separate success ancillas and all other outputs retained give selected product $J^2$ and
\begin{align}
\|J^2-T^{-2}/(16\kappa)\| \le(\|J\|+\|J_0\|)\eta\le2\eta.
\label{wql:eq:factor-product-error}
\end{align}
Separate success ancillas enforce the block product $J^2$; reuse of the same ancilla would include paths through the other blocks of the unitary.
The orthogonal terminal rows and a known row selection implement
\begin{align}
\frac{\boldsymbol{v}_{N_0} -i\beta_1J(L_1/\ell_1)\boldsymbol{e}_{N_0} -i\beta_2J^2(L_2/\ell_2)\boldsymbol{e}_{N_0}} {\sqrt{1+\beta_1^2+\beta_2^2}},
\quad \beta_1=\frac{8\ell_1}{\sqrt{N_0}},
\quad \beta_2=\frac{32\ell_2}{\sqrt{N_0}}.
\label{wql:eq:factor-readout}
\end{align}
For exact reciprocals, the numerator is precisely the correction of \cref{wql:eq:factor-weighted-residual}.
Each scale component is proportional to $T^{-2}\boldsymbol{b}$.
Its ideal correction map has norm at most $2\sqrt{\ell_1^2+\ell_2^2}/\sqrt{N_0}$, and $\beta_1^2+\beta_2^2\le2112L/N_0$.
By \cref{wql:lem:terminal}, the ideal accepted amplitude is therefore at least
\begin{align}
\frac{\frac45-2\sqrt{10/2^{16}}-4\sqrt{30}/2^{16}} {\sqrt{1+2112/2^{16}}}>\frac34.
\label{wql:eq:factor-success-margin}
\end{align}
The joint error from $J,J^2$ is at most $\sqrt5\eta$, since the terminal rows form a contraction and the final row is normalized.
Allocate an additional $\epsilon/32$ isometry error to all known gates.
The accepted vector then differs from its ideal value by less than $\epsilon/16$.
The normalization argument of \cref{wql:prop:success} gives success probability at least $1/4$ and conditional trace-distance error at most $\epsilon$.

The coarse circuit uses $O(t_j)$ factor calls and $O(u_k)$ preparation calls.
The matrix-only reciprocal uses independently chosen delays:
\begin{align}
v_j=1+\sqrt\kappa\sqrt{\frac{\lambda_j}{C_j}} \sum_i\sqrt{C_i\lambda_i},
\quad \widehat r_j=\lceil L_1/v_j\rceil,
\quad L_1\ge\max_jv_j.
\label{wql:eq:factor-fine-rates}
\end{align}
Then \cref{wql:lem:reciprocal} gives $O(v_j\log(1/\epsilon))$ original calls, including the two-use branch.
The terminal readout adds $O(1)$ calls of each original type.
Consequently,
\begin{align}
Q_{T,j}=O\!\left(t_j+v_j\log\frac1\epsilon\right),
\quad Q_{b,k}=O(u_k).
\label{wql:eq:factor-individual}
\end{align}
Multiplying by $C_j,c_k$ and summing, with $D\ge1$, gives \cref{wql:eq:factor-matrix-cost,wql:eq:factor-state-cost}.
The separate precision schedule contributes $\sqrt\kappa\log(1/\epsilon)$ independently of the coarse $\sqrt{\kappa D}$ term.

\subsection{Implementation of the weighted linear-system solver}
\label{sec:implementation}

The known junctions in \cref{eq:imp-junction} act on repeated system indices and a finite set of port labels.
\cref{lem:small-matrix} implements their constant-dimensional factors; \cref{lem:closed-system} includes the state preparation on the port labels.
All costs use \cref{def:resource-accounting}.

\begin{lemma}[Controlled small matrices]
\label{lem:small-matrix}
A fixed-dimensional unitary acting on a bounded number of port labels, controlled by zero tests on at most $b$ other bits, has an $O(b)$-gate implementation with constant clean workspace.
A sequence of $J$ such operations costs $O(Jb)$ gates with the same work qubits reused.
\end{lemma}
\begin{proof}
A fixed-dimensional unitary has a decomposition into a constant number of two-level unitaries and phases.
After fixed relabelings of its constant-size target register, each factor is a controlled one-qubit unitary.
Conjunctions of the remaining equality and zero conditions can be computed into one clean flag using an $O(b)$-Toffoli multi-controlled NOT with a constant number of clean auxiliaries~\cite[Section 5]{Khattar2025ConditionallyCleanAncillae}.
Apply the factor controlled on the flag and uncompute the conjunction.
Negative controls require at most $O(b)$ known bit flips.
The controlled one-qubit factor and each Toffoli have constant-size decompositions in the present gate set.
All work is restored, so it is reused for the subsequent factors and scale system matrices.
The multi-controlled-NOT construction of \cite{Khattar2025ConditionallyCleanAncillae} is an exact unitary on arbitrary control states; no measurement or assumption about their amplitudes is used.
\end{proof}

\begin{lemma}[Closed constraint system matrices]
\label{lem:closed-system}
A closed direct or factor constraint system in \cref{wql:subsec:sums,wql:subsec:factor-proof} with $m$ matrix terms and $n$ prepared states has a unitary implementation using $O((m+n)b)\text{ gates}$ on a binary address of width $b$, with constant restored workspace.
\end{lemma}
\begin{proof}
Write the open system matrix from \cref{eq:imp-junction} as $S$.
Partition its coordinates into external and closed sets, denoted by $e,c$.
The feedback connection gives
\begin{align}
S_{\rm cl}=S_{ee}+S_{ec}(I-S_{cc})^{-1}S_{ce}.
\label{eq:imp-close}
\end{align}
The positive diagonal Gram matrix on the closed coordinates makes $I-S_{cc}$ invertible, as proved in \cref{lem:common-cayley}.
Unitarity of $S$ cancels the equal incoming and outgoing norms on the closed coordinate.
Thus $S_{\rm cl}$ is unitary on the external space.

The coupling maps in \cref{wql:eq:sum-matrix-maps,wql:eq:sum-state-maps,wql:eq:factor-maps} are scalar multiples of system injections or known zero-state embeddings.
Their Gram matrix and the known load $K_0$ act on a bounded number of coordinate types for each repeated system index.
Normalize each of these coupling rows by preparing its scalar port weights, using $O(m+n)$ rotations \cite{shende2006synthesis}.
In that basis, the junction and its closure reduce to bounded-dimensional matrices plus identity on the orthogonal complements.
The same scalar rotations apply to every system index.
Zero tests distinguish the signal subspaces and the scalar reference state; \cref{lem:small-matrix} costs $O(b)$ gates per controlled factor.
Preparing and unpreparing the port weights, and applying the bounded-dimensional matrices, therefore cost $O((m+n)b)$.
Inverse and controlled forms obey the same bound.
\end{proof}

\cref{lem:ordered-cascade} implements the cascade without repeating an oracle query for each scale.
\begin{lemma}[An ordered cascade with shared feedback connections]
\label{lem:ordered-cascade}
A cascade of $J$ closed constraint systems, connected through their scalar reference ports and retaining their separate system and private outputs, costs $O(J(m+n)b)$ known gates.
Occurrences of the same original oracle with the same delay use one scheduled controlled call per direction, with the occurrence label unchanged.
\end{lemma}
\begin{proof}
Embed each system on a common reference coordinate and its own system and private subspaces, then apply the systems in cascade order.
After system $j$, the reference coordinate contains the input of system $j+1$; later systems leave earlier system and private outputs unchanged.
This proves the series connection on arbitrary inputs and the gate count by \cref{lem:closed-system}.
For one original oracle $O$, its action on all matching occurrences is $I_{\rm occurrence}\otimes O$, controlled on their common residue condition.
Port routing is included in \cref{thm:iir-gates}.
The preparation feedback operation $-\left(\begin{smallmatrix}0&B_k\\B_k^\dagger&0\end{smallmatrix}\right)$ uses both original directions, with their separate constant costs.
\end{proof}

The uniform-clock and terminal-residual identities are proved in \cref{wql:lem:terminal}.
For their implementation, a delay-buffer sum is $\sqrt r$ times a normalized uniform bra, so direct clock preparation costs $O(r)$ gates.
For an involution $V^2=I$, the operation $(I-iV)/\sqrt2$ uses two controlled oracle calls: compute its eigenvalue into a qubit, apply the corresponding phase, and uncompute.
These operations retain the other output subspaces and add a constant number of original calls.

\cref{prop:qlsp-resources} combines these costs with the rational reciprocal of \cref{wql:lem:reciprocal}.
\begin{proposition}[Gate and qubit costs of the weighted solver]
\label{prop:qlsp-resources}
Denote the coarse and reciprocal horizons of \cref{wql:thm:main,wql:thm:factor} by $N_0,N_1$.
Let $J$ be the scale count and $d=O(\log(1/\epsilon))$ the reciprocal degree.
Assume the prescribed delays are at most their respective horizons.
If $w$ includes the padded system and all oracle workspaces, an implementation uses
\begin{align}
b&=w+O\!\left(\log[2J(m+n+1)(N_0+1)(N_1+d+1)^2]\right),
\label{eq:imp-bits}
\\
G_{\rm total}&\le\sum_jQ_jG_j+\sum_kQ_{b_k}G_{b_k} +O\!\left([(m+n)JN_0+mdN_1]b\log(2+b)\right),
\label{eq:imp-gates}
\\
q_{\rm total}&\le b+O(1).
\label{eq:imp-qubits}
\end{align}
Gate costs include clock preparation and terminal extraction.
The same bounds hold for the inverse circuit.
For factor access, the two reciprocal applications use separate encoding ancillas.
\end{proposition}
\begin{proof}
The coarse direct-sum space has dimension $O(2^wJ(m+n)(N_0+1))$: every delay is at most $N_0$ and each scale has a bounded number of coordinates per port.
Each reciprocal uses at most $3d$ shifted Cayley factors per matrix term.
Its clock and private spaces increase this dimension by a polynomial in $m(N_1+d+1)$; two independent invocations still contribute only the logarithmic width in \cref{eq:imp-bits}.

By \cref{lem:ordered-cascade}, coarse system operations cost $O((m+n)JN_0b)$ gates.
Each reciprocal system operation costs $O(md b)$ for the ordered Cayley factors, and occurs $N_1$ times.
The residue routing of \cref{thm:iir-gates} adds the displayed $\log(2+b)$ factor.
Direct preparation of the clocks costs $O(N_0+N_1)$ gates.
Buffer extraction costs at most $O(J(m+n)N_0b\log(2+b))$, already included in the coarse term.
Constant-size readout rotations and the two reciprocal invocations change only the absolute constants.
All given-oracle gates appear in the two query sums.
Their workspaces remain part of $w$.
\end{proof}

For one matrix and one normalized state, the direct construction has $N_0=O(\kappa)$, $J=O(\log(2/p))$, and $N_1=O(\kappa\log(1/\epsilon))$.
Its known-gate overhead is $O(\kappa[\log(2/p)+\log^2(1/\epsilon)]b\log(2+b))$.
The extra precision logarithm counts the ordered reciprocal factors; it does not multiply the matrix-query bound because their occurrences share the scheduled original oracle call.

\subsection{Support restriction and preparation stability}
\label{cert:app:support}

\subsubsection{The full oracle spaces on a supported input}
\label{cert:proof:support}

To apply \cref{cert:thm:tail}, the supported space must remain invariant under the complete circuit, including every unitary completion of the encoded matrix.
\cref{cert:lem:junction-support} constructs an invariant space that includes those completion amplitudes; \cref{cert:lem:support} applies it to the solver.

\begin{lemma}[Invariant space of a Cayley junction]
\label{cert:lem:junction-support}
Let $P$ be an orthogonal projection on the load space of a Cayley junction.
Let $K_0$ be its Hermitian load, and let $C_j$ map into the corresponding private space.
Each private output is fed back to its input through a self-inverse unitary $V_j$.
Suppose
\begin{align}
[P,K_0]=0,
\quad [P,C_j^\dagger C_j]=0,
\quad [P,C_j^\dagger V_jC_j]=0
\quad\text{for every }j.
\label{cert:eq:invariant-assumptions}
\end{align}
Then the complete junction preserves
\begin{align}
\ran P\ \oplus\ \bigoplus_j\mathcal L_{j,P},
\quad \mathcal L_{j,P} =\operatorname{span}\{C_j\ran P,\ V_jC_j\ran P\}.
\label{cert:eq:invariant-private}
\end{align}
It remains invariant after adding positive integer delays and after well-defined identity feedback on load coordinates whose coordinate projections commute with $P$.
\end{lemma}
\begin{proof}
The involution $V_j$ preserves $\mathcal L_{j,P}$.
Both terms in its spanning set are mapped by $C_j^\dagger$ into $\ran P$ by \cref{cert:eq:invariant-assumptions}.
Also, $C_j\ran P\subseteq \mathcal L_{j,P}$.
The known inverse
\begin{align}
R=(I+C^\dagger C+iK_0)^{-1},
\quad C=\bigoplus_j C_j,
\label{cert:eq:junction-R}
\end{align}
commutes with $P$.
Each block of the unitary system matrix
\begin{align}
\mathsf S_0= \left(\begin{array}{c|c} 2R-I&2RC^\dagger
\\
\hline
2CR&2CRC^\dagger-I \end{array}\right)
\label{cert:eq:invariant-system}
\end{align}
therefore maps \cref{cert:eq:invariant-private} into itself.
The subspace is reducing because the system is unitary in finite dimension.
A delay line is a direct sum of copies of the same invariant private space.
A well-defined identity closure eliminates closed coordinates by a Schur complement; every factor in that complement respects the projection.
The closed junction and the delayed circuit therefore preserve the displayed space.
\end{proof}

The restriction in \cref{cert:lem:support} applies to both the coarse constraint circuit and its matrix-only precision correction.

\begin{lemma}[QLSP on a supported singular subspace]
\label{cert:lem:support}
Let $A,U_A,B$ satisfy \cref{cert:eq:access}.
Suppose $P_R,P_L$ are orthogonal projections such that
\begin{align}
AP_R=P_LA,
\quad P_L|b\rangle=|b\rangle,
\label{cert:eq:supported-pair}
\end{align}
and $A:\ran P_R\to\ran P_L$ is bijective with inverse norm at most $K\ge2$.
Given $1\le r_0\le\|A^{-1}|b\rangle\|$, the constraint circuit of~\cref{wql:thm:main} can be run using the parameters $K,r_0$.
It uses
\begin{align}
Q_A=O\!\left(K\log\frac2\eta\right),
\quad Q_b\le C_b(1+K/r_0),
\label{cert:eq:support-count}
\end{align}
where $C_b$ is a universal constant, and has the same constant-success inverse-state guarantee with error $\eta$.
No bound on the inverse of $A$ outside these subspaces is required.
\end{lemma}
\begin{proof}
For general $A$, form the usual Hermitian dilation using the original controlled queries:
\begin{align}
H=\begin{pmatrix}0&A
\\
A^\dagger&0\end{pmatrix},
\quad O_H=\begin{pmatrix}0&U_A
\\
U_A^\dagger&0\end{pmatrix},
\quad |\beta\rangle=\begin{pmatrix}|b\rangle
\\
0\end{pmatrix}.
\label{cert:eq:hermitian-dilation}
\end{align}
Reorder the signal ancilla so that $(\langle0|\otimes I)O_H(|0\rangle\otimes I)=H$.
The projection $P_L\oplus P_R$ reduces $H$ and contains the input state.
On this subspace the inverse norm is at most $K$.
One $O_H$ call uses at most one original forward and one original inverse matrix call.
It suffices to analyze a Hermitian signal with such a reducing projection.

On each constraint system, let the load projection retain the supported subspace in both vector coordinates and retain its full scalar coordinate.
The known Gram matrices $C_j^\dagger C_j$ are scalar multiples of identities on those coordinates.
The matrix compression is an off-diagonal copy of $H$, and the preparation compression couples the scalar only to $|\beta\rangle$.
Both respect this load projection.
\cref{cert:lem:junction-support} therefore gives an invariant private space containing the complementary oracle amplitudes, including when the completion mixes the bare signal subspaces.

The reference-coordinate cascade, the uniform clock, and its terminal maps also preserve this restriction.
In particular, the terminal map has form $P_Y C^\dagger(I-iV)\mathcal T/2$, where $\mathcal T$ sums the final buffer entries.
Every factor maps the restricted private state into the supported load coordinates.
The preparation delays, action estimates, and terminal Gram bounds of \cref{lem:common-cayley,wql:lem:terminal} consequently apply with inverse scale $K$.
They give a coarse horizon $O(K)$ and $O(1+K/r_0)$ preparation queries.

For the precision stage, consider a full shifted Cayley factor in the Reciprocal module.
On its active public subspace its load is
\begin{align}
M_\sigma(z)=\frac{1-z^2}{1+z^2}I +i\sigma\frac{2z}{1+z^2}H,
\quad \sigma\in\{-1,1\}.
\label{cert:eq:supported-reciprocal-load}
\end{align}
For pole parameters in the stated range, the supported inverse obeys $|a|,b\le(2K)^{-1}$ and $b>0 $ implies $ \|(bI+i\sigma(H-aI))^{-1}\|_{\rm supp}\le2K$.
The subscript restricts the operator to the supported subspace.
There, $M_\sigma(z)-i\sigma H=O(|z-1|)$, so one common neighborhood $|z-1|\le c_1/K$ bounds each full factor by a universal constant.
The reciprocal uses $d=O(\log(1/\eta))$ factors.
Their logarithmic norm bound and derivative norm at $z=1$ are $O(d)$ and $O(Kd)$, respectively.
\cref{wql:lem:reciprocal} and the analytic-clock theorem therefore require $N_1=O(K\log(1/\eta))$ on this restriction.
All remaining real signals still have a finite unitary realization: the pole widths are nonzero, and the finite clocks use known coefficients determined by $K,\eta$.
No analytic bound on the unused signal sector is invoked.

The same terminal correction and known row normalization give an ideal accepted amplitude greater than $2/3$ by the calculation in \cref{wql:prop:success}.
Reducing the error budgets gives an accepted amplitude greater than $3/5$ and conditional error $\eta$.
Known public processing matrices act only on the reciprocal-section labels, and therefore preserve each supported copy.
The terminal vectors and all scale outputs stay in the supported subspace.
The inverse-precision circuit uses no preparation calls; all other gates and query locations are determined by the given numerical bounds.
This proves \cref{cert:eq:support-count}, with a universal $C_b$ that also covers the Hermitian dilation and terminal readout.
\end{proof}

\subsubsection{Approximate support}
\label{cert:proof:tail}
\begin{proof}[Proof of \cref{cert:thm:tail}]
Split the solution and right-hand side using the singular subspaces:
\begin{align}
\boldsymbol x_<=\Pi_{<1/K}\boldsymbol x, \boldsymbol x_\ge=(I-\Pi_{<1/K})\boldsymbol x,
\quad \boldsymbol b_<=A\boldsymbol x_<, \boldsymbol b_\ge=A\boldsymbol x_\ge.
\label{cert:eq:tail-split}
\end{align}
The right-hand-side components are orthogonal because $A$ maps disjoint right singular subspaces into disjoint left singular subspaces.
Hence
\begin{align}
t_b:=\|\boldsymbol b_<\|\le\tau(K)/K,
\quad \|\boldsymbol b_\ge\|=\sqrt{1-t_b^2},
\quad \|\boldsymbol x_\ge\|\le K\|\boldsymbol b_\ge\|\le K.
\label{cert:eq:source-tail}
\end{align}
Pythagoras and the small-tail premise give $r^2=\|\boldsymbol x_\ge\|^2+\tau(K)^2$ and $r_0\le r\le2K $ for $c\epsilon\le1/4$.
The normalized supported input vector and its solution are
\begin{align}
|b_\ge\rangle=\boldsymbol b_\ge/\|\boldsymbol b_\ge\|,
\quad \boldsymbol x^{\,\prime}=A^{-1}|b_\ge\rangle =\boldsymbol x_\ge/\|\boldsymbol b_\ge\|.
\label{cert:eq:comparison-source}
\end{align}
Their solution norm satisfies $\|\boldsymbol x^{\,\prime}\|\ge\sqrt{r^2-\tau(K)^2}\ge r_0/2$ and is at least one.
It is also at most $K$.
Thus \cref{cert:lem:support} applies with lower bound $\max\{1,r_0/2\}$.
This changes the preparation-query coefficient by at most a universal factor; enlarge $C_b$ in \cref{cert:eq:support-count} to cover that choice.

There is a unitary $B_\ge$ preparing $|b_\ge\rangle$ such that
\begin{align}
\|B_\ge-B\|\le\sqrt2\,t_b.
\label{cert:eq:preparation-rotation}
\end{align}
Indeed, $\langle b_\ge|b\rangle=\sqrt{1-t_b^2}$ is nonnegative.
A rotation in the plane spanned by the two normalized columns, extended by identity on its orthogonal complement, maps the actual input state to the supported one.
Its operator distance from identity is $\sqrt{2-2\sqrt{1-t_b^2}}\le\sqrt2t_b$.
Left-multiply $B$ by that rotation, acting on the full preparation space.
This defines a comparison oracle only.

Run the same finite circuit with $B$ and with $B_\ge$.
Its known gates, clock states, and matrix oracles are identical, since their construction uses only $K,r_0,\epsilon$.
A telescoping hybrid over the forward and inverse preparation calls gives Euclidean distance between the full output states at most
\begin{align}
q_b\|B_\ge-B\| \le C_b(1+K/r_0)\sqrt2\,\tau(K)/K
\quad \le3\sqrt2 C_b\,\tau(K)/r_0.
\label{cert:eq:hybrid}
\end{align}
Only the preparation count enters this bound.
The matrix-only precision stage, however long, is a contraction on the comparison difference.
The normalized supported solution differs from the target by at most
\begin{align}
\left\|\frac{\boldsymbol x_\ge}{\|\boldsymbol x_\ge\|} -\frac{\boldsymbol x}{r}\right\| \le 2\tau(K)/r\le2\tau(K)/r_0.
\label{cert:eq:tail-target-error}
\end{align}

Choose the supported solver's conditional error at most $\epsilon/16$, and choose, for example, any universal $c\le\min\{1/64,(1024C_b)^{-1}\}$.
\cref{cert:eq:hybrid} then gives Euclidean distance between the full output states less than $\epsilon/64$.
The supported circuit's accepted amplitude exceeds $3/5$.
The actual accepted amplitude therefore exceeds $1/2$.
Projecting onto success and normalizing changes a vector perturbation of norm $\delta$ by at most $4\delta$ when both accepted norms are at least one half.
Combining this with \cref{cert:eq:tail-target-error} and the supported solver error gives conditional trace distance less than $\epsilon$.
Partial trace over scale and work records cannot increase it.
The actual circuit uses $U_A,B$ throughout, with the counts in \cref{cert:eq:cutoff-cost}.
\end{proof}

\begin{proof}[Proof of \cref{cert:cor:factor}]
On the supported subspace, $|S|\ge1/\sqrt K$.
The factor construction of \cref{wql:cor:single,wql:subsec:factor-proof} therefore has coarse cost $O(K/\sqrt{r_0})$ and matrix-only precision cost $O(\sqrt K\log(1/\epsilon))$.
Its two factor couplings preserve that subspace.
The discarded input has norm at most $\|\mathbf1_{[0,1/K)}(H)\boldsymbol x\|/K$, so the preparation-oracle comparison in the proof of \cref{cert:thm:tail} applies with the same number of preparation calls.
Replacing $r_0$ internally by $\max\{1,r_0/2\}$ changes only constants.
\end{proof}

\subsection{Bounded-budget amplification and a separation example}
\label{cert:app:short}

\cref{cert:lem:amplification} gives a finite worst-case query budget for the amplification used in \cref{cert:prop:short}.

\begin{lemma}[Bounded-budget amplification]
\label{cert:lem:amplification}
Let a finite unitary $W$ prepare a successful vector whose norm is at least $0<a_0\le1$.
A known randomized choice of Grover lengths, followed by two bounded attempts, succeeds with probability greater than one quarter using $O(1/a_0)$ calls to $W,W^\dagger$.
Every successful branch has the same normalized system state as the successful part of $W|0\rangle$.
\end{lemma}
\begin{proof}
Let the actual successful amplitude be $\sin\theta\ge a_0$ and set $M=\lceil2/a_0\rceil$.
With probability $1/2$, use $W$ without a Grover iteration.
Otherwise choose a uniform integer $j\in\{0,\ldots,M-1\}$ and use that many Grover iterations.
Their successful probabilities average to
\begin{align}
\frac1M\sum_{j=0}^{M-1}\sin^2((2j+1)\theta) =\frac12-\frac{\sin(4M\theta)}{4M\sin(2\theta)}.
\label{cert:eq:grover-average}
\end{align}
For $\theta\le\pi/4$, we have $\sin(2\theta)\ge a_0$.
The average is therefore at least $3/8$, and the mixed experiment succeeds with probability at least $3/16$.
If $\theta>\pi/4$, its zero-iteration branch alone contributes more than $1/4$.
Two attempts therefore succeed with probability at least $1-(13/16)^2>1/4$.
Each attempt uses at most $2M-1$ calls to $W$ or its inverse.
The two-reflection dynamics changes only the coefficients of the fixed successful and unsuccessful vectors, so conditioning on success preserves the successful system-state direction.
The choices and stopping records can be retained coherently.
This is the usual two-dimensional amplitude amplification argument~\cite{brassard2002quantum}.
\end{proof}

\cref{cert:prop:separation} exhibits a fixed-degree polynomial that applies the inverse to the specified input whose required inverse cutoff parameter $K$ grows exponentially with dimension.

\begin{proposition}[A low-degree inverse approximation can improve on singular-value truncation]
\label{cert:prop:separation}
Fix $\rho>1$ and let $\alpha=1+\rho$.
Define the lower truncated shift by $T_me_j=e_{j+1} $ for $0\le j<m-1$ and $T_me_{m-1}=0$.
Define
\begin{align}
A_m=(I-\rho T_m)/\alpha,
\quad |b\rangle=e_{m-2},
\quad m\ge2.
\label{cert:eq:jordan-family}
\end{align}
Then $A_m$ is an invertible contraction.
The identity for the given input $(I-\alpha A_m)^2|b\rangle=0$ gives
\begin{align}
A_m^{-1}|b\rangle =\alpha(e_{m-2}+\rho e_{m-1}) =(2\alpha I-\alpha^2A_m)|b\rangle.
\label{cert:eq:short-jordan-inverse}
\end{align}
The degree is one; the coefficient normalization and solution norm are $2\alpha+\alpha^2$ and $\alpha\sqrt{1+\rho^2}$, respectively.
All three are independent of dimension.
Nevertheless, every cutoff discarding at most a sufficiently small fixed fraction of the normalized solution has $K=\Omega_\rho(\rho^m)$.
\end{proposition}
\begin{proof}
The contraction bound is $\|I-\rho T_m\|\le1+\rho$.
The inverse and polynomial identities follow from $T_m^2e_{m-2}=0$.
For $C_m=I-\rho T_m$, the inverse entries are
\begin{align}
[C_m^{-1}]_{ij}=\begin{cases}\rho^{i-j},&i\ge j,
\\
0,&i<j.\end{cases}
\end{align}
Put $v_i=\rho^i$ and $w_j=\rho^{-j}$.
Then
\begin{subequations}
\label{cert:eq:rank-one-inverse}
\begin{align}
\|C_m^{-1}-\boldsymbol v\boldsymbol w^\dagger\| &\le\sum_{k\ge1}\rho^{-k}=\frac1{\rho-1},
\\
\|\boldsymbol v\boldsymbol w^\dagger\| &=\frac{\rho^{m+1}(1-\rho^{-2m})}{\rho^2-1}.
\end{align}
\end{subequations}
Only upper diagonals differ in the first line.
Thus the largest singular value of $C_m^{-1}$ is $\Theta_\rho(\rho^m)$.
If $|u\rangle$ is a corresponding left singular vector, projection of its singular-vector equation onto $\boldsymbol v^\perp$ gives
\begin{align}
\|(I-\Pi_{\boldsymbol v})|u\rangle\| \le \frac{1/(\rho-1)}{\|\boldsymbol v\boldsymbol w^\dagger\|-1/(\rho-1)} =O_\rho(\rho^{-m}).
\label{cert:eq:singular-direction}
\end{align}
For sufficiently large $m$, the denominator is positive.
The vector $|u\rangle$ is the smallest right singular vector of $C_m$.
Scaling the matrix preserves this singular direction.
Its squared overlap with the normalized solution tends to
\begin{align}
\frac{\rho^{2m-4}(1+\rho^2)}{\|\boldsymbol v\|^2} \longrightarrow 1-\rho^{-4}>0.
\label{cert:eq:bad-overlap}
\end{align}
A cutoff that discards this smallest singular direction therefore has a fixed positive solution error.
Retaining it requires the inverse cutoff parameter $K$ to be at least the reciprocal of the smallest singular value, which is $\Theta_\rho(\rho^m)$.
In contrast, \cref{cert:eq:short-jordan-inverse} is a constant-query LCU action.
At $\rho=3/2$, one attempt has success probability $13/81$, so a fixed number of attempts gives constant success.
\end{proof}

A bound for a finite-query inverse approximation includes its coefficient normalization as well as its degree; both enter the cost in \cref{cert:eq:action-cost}.

\subsection{Identities for classical spectral bounds}\label{cert:app:spectral}
\subsubsection{The global inverse norm and inverse moments}
\label{cert:subsec:moments}
A known global inverse bound gives a zero solution tail once the cutoff exceeds that bound.
Thus the usual QLSP theorem is the zero-tail case.
For $q>0$, a known upper bound
\begin{align}
\mu_q\ge\big\||A|^{-q}\boldsymbol x\big\|
\label{cert:eq:inverse-moment}
\end{align}
gives the sharper alternative
\begin{align}
\tau(K)\le\frac{\mu_q}{K^q},
\quad K\ge\max\left\{2, \left(\frac{\mu_q}{c\epsilon r_0}\right)^{1/q}\right\}.
\label{cert:eq:moment-cutoff}
\end{align}
Indeed, $|A|^{-2q}\succeq K^{2q}\Pi_{<1/K}$, and taking the quadratic form on $\boldsymbol x$ proves the bound.
If $\kappa$ is also given, choose the smaller of $\max\{2,\kappa\}$ and the sufficient cutoff in \cref{cert:eq:moment-cutoff}.

For $q=1$, the relevant quantity is the forward--adjoint inverse norm:
\begin{align}
\big\||A|^{-1}\boldsymbol x\big\| =\big\|(A^\dagger)^{-1}\boldsymbol x\big\|,
\quad r^2\le\big\|(A^\dagger)^{-1}\boldsymbol x\big\| \le\|A^{-1}\|r.
\label{cert:eq:adjoint-moment}
\end{align}
The equality follows from the polar decomposition, and the lower bound from $r^2=\langle b|(A^\dagger)^{-1}\boldsymbol x\rangle$.
A bound $\mu_1$ therefore yields
\begin{align}
Q_A=O\!\left[ \max\!\left\{1,\frac{\mu_1}{\epsilon r_0}\right\} \log\frac1\epsilon\right],
\quad Q_b=O\!\left(1+\frac{\mu_1}{\epsilon r_0^2}\right).
\label{cert:eq:first-moment-cost}
\end{align}
For a nonnormal matrix, $A^{-2}|b\rangle$ is generally different from $(A^\dagger)^{-1}A^{-1}|b\rangle$ and cannot be substituted in this bound.

For positive integer $q$, the norm in \cref{cert:eq:inverse-moment} equals
\begin{align}
\big\||A|^{-q}\boldsymbol x\big\| =\begin{cases} \|(A^\dagger A)^{-q/2}\boldsymbol x\|,&q\text{ even},
\\
\|(A^\dagger)^{-1}(A^\dagger A)^{-(q-1)/2}\boldsymbol x\|, &q\text{ odd}. \end{cases}
\label{cert:eq:alternating}
\end{align}
A regularity condition can give these norms without an inverse-state preparation.
If, for some unprepared vector $\boldsymbol h$,
\begin{align}
|b\rangle=(AA^\dagger)^{(q+1)/2}\boldsymbol h,
\label{cert:eq:regularity}
\end{align}
then a singular-value decomposition gives $\||A|^{-q}\boldsymbol x\|=\|\boldsymbol h\|$.
A known bound on $\|\boldsymbol h\|$ therefore supplies \cref{cert:eq:inverse-moment}.

Higher moments can improve the dependence on accuracy.
For example, suppose a known $r_+\ge r$ obeys $r_+/r_0=O(1)$.
Assume the moment bounds satisfy, for every positive integer order,
\begin{align}
\mu_q\le r_+(Lq^a)^q,
\quad L\ge1,
\quad a\ge0.
\label{cert:eq:moment-growth}
\end{align}
Taking $q=\max\{1,\lceil\log(r_+/(c\epsilon r_0))\rceil\}$ in \cref{cert:eq:moment-cutoff} gives
\begin{align}
K=O\!\left(L\log^a\frac1\epsilon\right),
\quad Q_A=O\!\left(L\log^{a+1}\frac1\epsilon\right),
\quad Q_b=O\!\left(1+\frac L{r_0}\log^a\frac1\epsilon\right).
\label{cert:eq:moment-family-cost}
\end{align}
The constants here may depend on the fixed $a$ and the bound on $r_+/r_0$.

\subsubsection{Sectors, singular bands, and the full tail}
\label{cert:subsec:sectors}
Let $P_j$ be mutually orthogonal projections summing to identity and reducing $|A|$.
Put $\boldsymbol x_j=P_j\boldsymbol x$.
Given upper bounds
\begin{align}
w_j\ge\|\boldsymbol x_j\|^2,
\quad \mu_{j,q}\ge\big\||A|^{-q}\boldsymbol x_j\big\|,
\label{cert:eq:sector-data}
\end{align}
orthogonality and \cref{cert:eq:moment-cutoff} imply
\begin{align}
\tau(K)^2\le \sum_j\min\left\{w_j,\inf_{q\in\mathcal Q_j} \frac{\mu_{j,q}^2}{K^{2q}}\right\}.
\label{cert:eq:sector-tail}
\end{align}
If a sector has singular values at least $g_j$, replace its summand by zero when $K\ge1/g_j$.
Different moment orders can be used in different sectors.
The set $\mathcal Q_j$ contains only orders for which a bound is given.

Alternatively, suppose disjoint left-singular bands $[a_j,b_j]$ have positive lower endpoints.
Let $\pi_j$ bound the input-state weight in each band.
Their solution weights are at most $\pi_j/a_j^2$.
A sufficient tail bound is therefore
\begin{align}
\tau(K)^2\le\sum_{j:a_j<1/K}\frac{\pi_j}{a_j^2},
\label{cert:eq:source-band}
\end{align}
including any separately bounded uncovered spectrum.
A small input weight near a small singular value must be divided by the square of that singular value before it can be neglected in the solution.

For $A=\sum_i\sigma_i|u_i\rangle\langle v_i|$, the complete information is the solution-weighted measure
\begin{align}
\nu_x=\sum_i\frac{|\langle u_i|b\rangle|^2}{\sigma_i^2}\, \delta_{\sigma_i},
\quad \tau(K)^2=\nu_x([0,1/K)),
\quad \nu_x([0,1])=r^2.
\label{cert:eq:solution-measure}
\end{align}
An upper envelope for the cumulative mass determines a sufficient cutoff for \cref{cert:thm:tail}, while the finite-query inverse construction in \cref{cert:prop:short} uses polynomial relations on the input state.

\subsubsection{Positive moments of the input state}
\label{cert:subsec:positive}
Positive moments can supply a tail bound without inverse-moment data.
Define
\begin{align}
m_j=\langle b|(AA^\dagger)^j|b\rangle,
\quad m_0=1.
\label{cert:eq:positive-moments}
\end{align}
The singular-value decomposition gives, for $j\ge1$,
\begin{align}
\langle\boldsymbol x,(A^\dagger A)^j\boldsymbol x\rangle=m_{j-1}.
\label{cert:eq:moment-shift}
\end{align}
Let $h(t)=\sum_{j=0}^d h_jt^j$ be a real polynomial satisfying
\begin{align}
h(t)\ge1
\quad(0\le t\le K^{-2}),
\quad h(t)\ge0
\quad(K^{-2}\le t\le1).
\label{cert:eq:majorant}
\end{align}
Functional calculus and \cref{cert:eq:moment-shift} imply
\begin{align}
\tau(K)^2\le h_0r^2+\sum_{j=1}^d h_jm_{j-1}.
\label{cert:eq:positive-tail}
\end{align}
With interval data, take the maximum of the right-hand side over the given intervals for $r^2$ and the moments.
Negative coefficients require the appropriate lower endpoints.
The intervals must be narrow enough for this upper bound to meet \cref{cert:eq:tail-condition}.
The constant coefficient also requires an upper bound on the solution norm.

\subsubsection{Finite-query inverse approximation on a specified inputs}
\label{cert:sec:short}

A spectral cutoff is unnecessary when a known finite circuit already approximates the inverse on the particular input.
The certificate specifies both its accuracy and block-encoding normalization.

\begin{proposition}[Finite-query inverse approximation on a specified input]
\label{cert:prop:short}
Suppose a specified circuit uses $d_A$ original matrix queries and block-encodes $F/\beta$.
Its positive normalization is given.
Let $\boldsymbol y=F|b\rangle$ and suppose classical bounds give
\begin{align}
0<y_0\le\|\boldsymbol y\|,
\quad \min_{\phi\in\mathbb R} \left\|\frac{\boldsymbol y}{\|\boldsymbol y\|} -e^{i\phi}|\mathrm{sol}\rangle\right\| \le\epsilon/4.
\label{cert:eq:action-certificate}
\end{align}
Then bounded-budget amplification prepares the solution with constant success probability and conditional trace-distance error at most $\epsilon$, using
\begin{align}
Q_A=O(d_A\beta/y_0),
\quad Q_b=O(1+\beta/y_0).
\label{cert:eq:action-cost}
\end{align}
The parameters $d_A,\beta,y_0$ and the given approximation can depend on $\epsilon$; there is no additional inverse-solver precision logarithm.
\end{proposition}
\begin{proof}
One preparation followed by the matrix-only circuit has success amplitude $\|\boldsymbol y\|/\beta\ge y_0/\beta$.
The bounded-budget amplification in \cref{cert:lem:amplification} uses $O(\beta/y_0)$ calls to this circuit and its inverse, and preserves the normalized successful system vector.
The state error in \cref{cert:eq:action-certificate}, together with a separate known-gate synthesis budget, gives the stated guarantee.
\end{proof}

For real $p(t)=\sum_{j=0}^d c_jt^j$, consider the singular-value action
\begin{align}
F=A^\dagger p(AA^\dagger),
\quad g(t)=tp(t^2).
\label{cert:eq:singular-polynomial}
\end{align}
A given global bound $\beta\ge\max_{-1\le t\le1}|g(t)|$ permits a normalization-$O(\beta)$ realization by bounded polynomial singular-value transformation~\cite{Gilyen2018singular}.
It uses $2d+1$ matrix queries.
This is a finite unitary polynomial realization; its complementary outputs remain part of the unitary.
Alternatively, a linear combination of the words $A^\dagger(AA^\dagger)^j$ gives normalization $\sum_j|c_j|$ with the same maximum query length.
The former can have a smaller normalization; the latter uses only explicit products and coefficient preparations.

The positive moments in \cref{cert:eq:positive-moments} determine
\begin{subequations}
\label{cert:eq:moment-action}
\begin{align}
\langle\boldsymbol x,\boldsymbol y\rangle &=\sum_{j=0}^dc_jm_j,
\\
\|\boldsymbol y\|^2 &=\sum_{j,k=0}^dc_jc_km_{j+k+1},
\\
\min_{\phi\in\mathbb R} \left\|\frac{\boldsymbol y}{\|\boldsymbol y\|} -e^{i\phi}\frac{\boldsymbol x}{r}\right\|^2 &=2-\frac{2|\sum_{j=0}^dc_jm_j|} {r\sqrt{\sum_{j,k=0}^dc_jc_km_{j+k+1}}}.
\end{align}
\end{subequations}
These identities follow by expanding in singular vectors.
With exact moments and $r$, they test \cref{cert:eq:action-certificate}; with intervals, use a lower bound on the numerator and upper bounds on both denominator factors.
A positive lower bound on the squared norm supplies $y_0$.
This requires moments through $m_{2d+1}$ and a sufficiently accurate upper bound on the solution norm.

For an input state supported on singular values one and one half, choose $p(t)=5-4t$ and $g(t)=t\,p(t^2)=5t-4t^3$.
The odd polynomial agrees with the reciprocal at both supported singular values.
Its global magnitude is at most $5\sqrt{15}/9$.
A constant-degree, constant-normalization inverse action is therefore possible even when unused singular values are arbitrarily small.

\begin{proposition}[A polynomial identity for the given input]
\label{cert:prop:annihilator}
Let a given polynomial annihilate the input state while having a nonzero constant coefficient: \begin{align}
a(z)=\sum_{j=0}^d a_jz^j,
\quad a_0\ne0,
\quad a(A)|b\rangle=0.
\end{align} Define
\begin{align}
p(z)=\frac{1-a(z)/a_0}{z} =-\sum_{j=0}^{d-1}\frac{a_{j+1}}{a_0}z^j,
\quad \beta_p=\sum_{j=0}^{d-1}|a_{j+1}/a_0|.
\label{cert:eq:annihilator}
\end{align}
Then $p(A)|b\rangle=\boldsymbol x$.
A finite linear combination of matrix powers prepares the solution with constant success and query costs
\begin{align}
Q_A=O((d-1)\beta_p/r_0),
\quad Q_b=O(1+\beta_p/r_0).
\label{cert:eq:annihilator-cost}
\end{align}
For an approximate identity, the exact error is
\begin{align}
p(A)|b\rangle-\boldsymbol x =-a_0^{-1}A^{-1}a(A)|b\rangle.
\label{cert:eq:approximate-annihilator}
\end{align}
\end{proposition}
\begin{proof}
The numerator defining $p$ vanishes at zero, and $Ap(A)|b\rangle=(I-a(A)/a_0)|b\rangle=|b\rangle$.
The power of largest degree uses $d-1$ matrix queries.
Conditional calls at slot $k$ on coefficient indices $j\ge k$ give all powers in that same schedule, with a separate signal ancilla for each product position.
Selecting all product signals and the coefficient register gives $p(A)/\beta_p$.
\cref{cert:prop:short} applies with $y_0=r_0$.
Without exact annihilation, multiplying the defining polynomial identity by $A^{-1}$ gives \cref{cert:eq:approximate-annihilator}.
\end{proof}

A bound on $\|a(A)|b\rangle\|$ alone does not control \cref{cert:eq:approximate-annihilator} for a nonnormal or poorly conditioned matrix.
The inverse-weighted residual or an equivalent directional error bound must be part of the certificate.
\cref{cert:prop:separation} gives an example where an exact degree-one inverse action has constant normalization but every accurate inverse cutoff parameter is exponentially large.
The example separates the costs proved by the two certificate classes.

\subsection{Classical certificates for Volterra histories}
\label{vdc:app:certificates}
The results below supply the history, Green-function, and moment bounds used in \cref{sec:vde-certificates}.
The prepared input state and its forward history are
\begin{align}
\ket c=\boldsymbol a\oplus0\oplus\boldsymbol c_\perp,
\quad \boldsymbol v=G\boldsymbol a=\frac{\sqrt m}{\sqrt2 B}\boldsymbol x_{\rm hist},
\quad \boldsymbol y=P\boldsymbol v=\frac{\sqrt m}{\sqrt2 B}\boldsymbol x_{\rm e}(T).
\label{vdc:eq:source-proof}
\end{align}
\subsubsection{A continuous mean-square bound suffices}
\cref{vdc:prop:continuous} converts a continuous trajectory-energy bound to the sampled history bound used by the circuit.

Root-mean-square history norms and classical Lyapunov stability data have also been used in quantum ODE solvers \cite{Jennings2024CostLinearDE}.
The bounds here include the coordinates containing the orthogonal completion of the state preparation and the scaled final-state output of the Volterra system.

\begin{proposition}[From continuous energy to the sampled history]
\label{vdc:prop:continuous}
Suppose a known $E_2$ satisfies
\begin{align}
E_2^2\ge\frac1T\int_0^T\norm{\boldsymbol x(t)}^2\,dt.
\label{vdc:eq:continuous-energy}
\end{align}
Under the norm and mesh assumptions above, a sufficiently large universal constant times $\max\{r,E_2,h\beta\}$ is a valid $W_2$ in \cref{eq:vde-history-rms}.
\end{proposition}
\begin{proof}
On an interval of length $\Delta\le h$, the propagator in either direction has norm at most $e^{1/2}$.
Variation of constants therefore gives, for either endpoint $s$ and any point $t$ in the interval,
\begin{align}
\norm{\boldsymbol x(s)}\le e^{1/2}\norm{\boldsymbol x(t)}+e^{1/2}\Delta\beta,
\quad \norm{\boldsymbol x(s)}^2\le2e\norm{\boldsymbol x(t)}^2+2e\Delta^2\beta^2.
\label{vdc:eq:sample-energy}
\end{align}
Integrate and sum over the $m$ coarse intervals.
Their normalized sum is at most $2e(E_2^2+h^2\beta^2)$.
Using the last interval once more bounds the additional final boundary by another $2e(E_2^2+h^2\beta^2)$.
The fine stage sum is at most $2e(E_2^2+h^2\beta^2)$ by the same argument with $\Delta=h/N$.
Thus the exact sampled history divided by $\sqrt m$ has norm at most $\sqrt{6e}\,(E_2^2+h^2\beta^2)^{1/2}$.
Minkowski's inequality adds at most $\sqrt{2+1/m}\,\epsilon r/64$ for Euler error.

The first interval also gives $\norm{\boldsymbol x_0}^2/m^2\le2eE_2^2+2eh^2\beta^2$.
Consequently, $B/m=O(E_2+h\beta)$.
Increasing the universal constant enforces both floors in \cref{eq:vde-history-rms}.
\end{proof}
For example, for homogeneous evolution with $\operatorname{Re}H(t)\succeq\gamma I$ and $\gamma>0$, then
\begin{align}
\frac d{dt}\norm{\boldsymbol x(t)}^2\le-2\gamma\norm{\boldsymbol x(t)}^2,
\quad E_2\le\frac{\norm{\boldsymbol x_0}}{\sqrt{2\gamma T}}.
\label{vdc:eq:energy-dissipation}
\end{align}
Integration proves the second inequality.
The final-norm floor and preparation cost remain; decay does not permit ignoring a small endpoint.

\cref{vdc:prop:lyapunov} obtains that energy bound by differentiating a classical positive quadratic form.

\begin{proposition}[Classical Lyapunov data supply a history certificate]
\label{vdc:prop:lyapunov}
Suppose classical Lyapunov data satisfy, at every time, $\mathsf P\succ0$, $\eta>0$, and $H(t)^\dagger\mathsf P+\mathsf P H(t)\succeq2\eta I$.
For homogeneous dynamics one may choose the energy upper bound to satisfy
\begin{align}
E_2^2\le\frac{\boldsymbol x_0^\dagger\mathsf P\boldsymbol x_0}{2\eta T}.
\label{vdc:eq:lyapunov-homogeneous}
\end{align}
For an inhomogeneous equation one may choose
\begin{align}
E_2^2\le\frac{\boldsymbol x_0^\dagger\mathsf P\boldsymbol x_0}{\eta T} +\frac1{\eta^2T}\int_0^T\norm{\mathsf P\boldsymbol b(t)}^2\,dt.
\label{vdc:eq:lyapunov-forced}
\end{align}
Known upper bounds for the right-hand sides suffice.
No quantum circuit for $\mathsf P$ is required when it is used only to certify these numbers.
\end{proposition}
\begin{proof}
For $V(t)=\boldsymbol x(t)^\dagger\mathsf P\boldsymbol x(t)$, differentiation gives $V'\le-2\eta\norm{\boldsymbol x}^2+2\operatorname{Re}\boldsymbol x^\dagger\mathsf P\boldsymbol b$.
In the homogeneous case, integrate and use $V(T)\ge0$.
In the forced case use $2|\boldsymbol x^\dagger\mathsf P\boldsymbol b|\le\eta\norm{\boldsymbol x}^2+ \norm{\mathsf P\boldsymbol b}^2/\eta$ before integration.
\end{proof}
The same Lyapunov inequality gives $C\le\sqrt{\lambda_{\max}(\mathsf P)/\lambda_{\min}(\mathsf P)}$ by applying the homogeneous energy inequality to arbitrary initial vectors on each interval.
It need not give a sharp final-row bound; a two-time envelope can improve that estimate separately.

\subsubsection{Ordered Green-function bounds}
The required certificates can be reduced to scalar bounds on ordered propagation.
With $\ket u=N^{-1/2}\sum_{k=0}^{N-1}\ket k$ and $J_N=N^{-1}\sum_{\ell<k}\ket k\bra\ell$, write
\begin{align}
E_j=I+hJ_ND_j,
\quad F_j=I-h\bra uD_jE_j^{-1}\ket u,
\quad D_j=\diag_k H(t_{j,k}).
\label{vdc:eq:local}
\end{align}
Tensor identities on the system register are understood.
The boundary Green matrix is
\begin{align}
\mathcal G= \begin{pmatrix} \sqrt m I&0&0&0&\cdots
\\
\sqrt m F_0&I&0&0&\cdots
\\
\sqrt m F_1F_0&F_1&I&0&\cdots
\\
\sqrt m F_2F_1F_0&F_2F_1&F_2&I&\cdots
\\
\vdots&\vdots&\vdots&\vdots&\ddots \end{pmatrix}.
\label{vdc:eq:ordered}
\end{align}
The first column retains the initial-row scaling by $1/\sqrt m$.
For varying generators, the ordered products depend on both time indices; the first column also includes the initial-row scaling.

\cref{vdc:prop:envelope} bounds the full history and final-row inverses separately from scalar bounds on these ordered products.

\begin{proposition}[Scalar envelopes for the Green-function bounds]
\label{vdc:prop:envelope}
Let the nonnegative scalar envelope satisfy $f_{jk}\ge\norm{F_{j-1}\cdots F_k} $ for $j>k$, $f_{jj}=1$, and $f_{jk}=0 $ for $j<k$.
Set $\mathsf f=(f_{jk})$ and $\mathsf D_m=\diag(\sqrt m,1,\ldots,1)$.
One may use
\begin{subequations}
\label{vdc:eq:scalar-green}
\begin{align}
k&=2+3\sqrt2\,\norm{\mathsf f\mathsf D_m}_2,
\quad \gamma=\sqrt{2\left(mf_{m0}^2+\sum_{k=1}^{m}f_{mk}^2\right)},
\\
\norm{\mathsf f\mathsf D_m}_2 &\le\norm{\mathsf f}_2+(\sqrt m-1) \left(\sum_{j=0}^{m}f_{j0}^2\right)^{1/2},
\\
\norm{\mathsf f}_2&\le \sqrt{\left(\max_j\sum_k f_{jk}\right) \left(\max_k\sum_j f_{jk}\right)}.
\end{align}
\end{subequations}
\end{proposition}
\begin{proof}
For arbitrary stage and endpoint residuals $\boldsymbol a_j,\boldsymbol b_j$, eliminating the stage in
\begin{align}
E_j\boldsymbol Y_j-\ket u\boldsymbol X_j=\boldsymbol a_j,
\quad \boldsymbol X_{j+1}-\boldsymbol X_j+h\bra uD_j\boldsymbol Y_j=\boldsymbol b_j
\label{vdc:eq:residual-equations}
\end{align}
gives
\begin{align}
\boldsymbol X_{j+1}=F_j\boldsymbol X_j+\boldsymbol b_j-h\bra uD_jE_j^{-1}\boldsymbol a_j.
\label{vdc:eq:residual-step}
\end{align}
The initial residual $\boldsymbol e_0$ gives $\boldsymbol X_0=\sqrt m\boldsymbol e_0$.
The local inverse has norm at most two and $\norm{h\bra uD_jE_j^{-1}}\le1$.
The map from the original residual to the input coordinates of $\mathcal G$ has norm at most $\sqrt2$, with the initial-row scaling already included in that matrix.
Hence $\norm{\boldsymbol X}\le\sqrt2\norm{\mathcal G}\norm{\text{residual}}$.
Reconstructing the stages gives $\norm{\boldsymbol Y}\le2\norm{\boldsymbol X}+2\norm{\boldsymbol a}$, whence $\norm{(\boldsymbol X,\boldsymbol Y)}\le3\norm{\boldsymbol X}+2\norm{\boldsymbol a}$.
This proves the bound on $k$.
The final-row operator $PG$ needs only the last row of the boundary Green matrix.
Cauchy--Schwarz gives its stated norm bound.

For block vectors, the vector of output block norms is bounded componentwise by $\mathsf f\mathsf D_m$ times the vector of input block norms.
Hence $\norm{\mathcal G}\le\norm{\mathsf f\mathsf D_m}_2$.
Separate the extra first column to obtain the middle bound in \cref{vdc:eq:scalar-green}.
The last inequality in \cref{vdc:eq:scalar-green} is the row/column-sum bound for the spectral norm.
\end{proof}
The envelope here bounds the numerical propagators actually used.
A continuous bound $f(t,s)$ requires the corresponding numerical-error allowance before substituting $f_{jk}=f(jh,kh)$.

For example, suppose $f_{jk}\le q^{j-k}$ for $0\le q<1$.
Put $\ell=\min\{m+1,(1-q)^{-1}\}$.
Summing the geometric series gives $k=O(\ell+\sqrt{m\ell})$ and $\gamma=O(\sqrt\ell)$.
In particular, the endpoint initial-column term obeys $m q^{2m}=O(\ell)$.
Using $W_2\ge r$ gives
\begin{align}
Q_H=O\!\left([\ell+\sqrt{m\ell}\,W_2/r]\log(1/\epsilon)\right),
\quad Q_0+Q_b=O\!\left(1+\frac{B\ell}{mW_2} +\frac{B\sqrt\ell}{r\sqrt m}\right).
\label{vdc:eq:decay}
\end{align}
This improvement uses the memory length of the Green function and the driven history norm together; the temporal decay supplies information beyond the peak propagator bound.

\subsubsection{Inverse moments of the balanced history}

Every certificate in this section concerns $A_\tau$ and $\boldsymbol z_\tau$ in \cref{vdc:eq:balanced}.
A bound for the unbalanced history cannot be substituted without transforming it.
Let $\mathcal Q$ be the set of positive moment orders for which classical bounds are available.
For each $q\in\mathcal Q$, suppose a number $\mu_q$ satisfies
\begin{align}
\mu_q\ge\norm{|A_\tau|^{-q}\boldsymbol z_\tau}.
\label{vdc:eq:moments}
\end{align}
Applying the inverse-moment estimate of \cref{cert:eq:moment-cutoff} to the balanced history gives
\begin{align}
\norm{\mathbf1_{[0,1/K)}(|A_\tau|)\boldsymbol z_\tau} \le\frac{\mu_q}{K^q}.
\label{vdc:eq:markov}
\end{align}
Consequently, one may take
\begin{align}
K_\epsilon\le\min\left\{K_0,K_{\rm G}, \inf_{q\in\mathcal Q}\max\left[2, \left(\frac{\mu_q}{c\epsilon R_0}\right)^{1/q}\right]\right\},
\label{vdc:eq:best-cutoff}
\end{align}
where only available bounds are included and any chosen scale is rounded up if necessary.
For a finite list of available bounds, choose its smallest sufficient scale.
An infimum over an infinite family can be approached within any fixed multiplicative slack.

If orthogonal sectors reduce $|A_\tau|$, give bounds for each sector: $w_j\ge\norm{P_j\boldsymbol z_\tau}^2$ and $\mu_{j,q}\ge\norm{|A_\tau|^{-q}P_j\boldsymbol z_\tau}$.
Their squared tail bounds add:
\begin{align}
\operatorname{tail}(K)^2\le \sum_j\min\left\{w_j,\inf_{q\in\mathcal Q} \frac{\mu_{j,q}^2}{K^{2q}}\right\}.
\label{vdc:eq:bands}
\end{align}
A sector with certified singular gap $g_j$ contributes zero once $K\ge1/g_j$.
The orthogonal-sector calculation is \cref{cert:eq:sector-tail}; these projectors enter only the classical certificate.

Suppose $R_+\ge\norm{\boldsymbol z_\tau}$ and $R_+/R_0=O(1)$.
Assume the given bounds obey, for all positive integer orders,
\begin{align}
\mu_q\le R_+(Lq^a)^q,
\quad a\ge0.
\label{vdc:eq:moment-growth}
\end{align}
Choosing $q=\lceil\log(R_+/(c\epsilon R_0))\rceil$ proves
\begin{align}
K_\epsilon=O(L\log^a(1/\epsilon)),
\quad Q_H=O(L\log^{a+1}(1/\epsilon)),
\quad Q_0+Q_b=O\!\left(1+\frac L{R_0}\log^a\frac1\epsilon\right).
\label{vdc:eq:moment-cost}
\end{align}
A fixed supported gap gives a single precision logarithm, whereas a finite inverse moment gives an $\epsilon^{-1/q}$ cutoff.
One can always choose the global inverse-norm bound, for which no singular component is discarded if it is smaller.
These singular moment conditions describe the encoded inverse problem and are different from analyticity of $H(t)$ in complex time.

\cref{vdc:prop:adjoint} gives the first inverse moment, including the orthogonal preparation completion and final-state output weights.

\begin{proposition}[First inverse moment of the balanced system]
\label{vdc:prop:adjoint}
With the notation of \cref{vdc:eq:source-proof}, the exact first moment is
\begin{align}
\norm{|A_\tau|^{-1}\boldsymbol z_\tau} =16\left[ \norm{G^\dagger(\boldsymbol v+\tau^{-2}P^\dagger\boldsymbol y)}^2 +\tau^{-4}\norm{\boldsymbol y}^2+\norm{\boldsymbol c_\perp}^2 \right]^{1/2}.
\label{vdc:eq:adjoint}
\end{align}
\end{proposition}
\begin{proof}
Polar decomposition implies $\norm{|A_\tau|^{-1}\boldsymbol z_\tau} =\norm{(A_\tau^\dagger)^{-1}\boldsymbol z_\tau}$.
Take the adjoint of \cref{eq:vde-border-inverse} and multiply it by $\boldsymbol z_\tau$.
The three orthogonal components are $16G^\dagger(\boldsymbol v+\tau^{-2}P^\dagger\boldsymbol y)$, $16\tau^{-2}\boldsymbol y$, and $16\boldsymbol c_\perp$.
Adding their squared norms gives the formula.
\end{proof}
The mean-square bound controls the driven history, while the first inverse moment also depends on its backward response under $G^\dagger$.
The final-state output contributes a further adjoint response weighted by $\tau^{-2}$.
A smaller $W_2$ changes this weight, so a moment certificate must be recomputed rather than multiplied by an independently asserted improvement factor.
In general $\norm{|A_\tau|^{-1}\boldsymbol z_\tau}\ge\norm{\boldsymbol z_\tau}^2$ because $\bra c(A_\tau^\dagger)^{-1}\boldsymbol z_\tau=\norm{\boldsymbol z_\tau}^2$.

\begin{proof}[Proof of \cref{vdc:thm:sectors}]
Lift $\Pi$ to every time and final-state output coordinate and retain the full orthogonal completion sector.
Every known temporal coefficient and every individual signal compression preserves this lifted sector.
The supported QLSP lemma of~\cref{cert:lem:support} then constructs invariant private spaces for each given full oracle, even when its unspecified completion mixes the bare system sectors.
Thus the inverse-norm bound is needed only on this restriction.
The original normalizations $\alpha_j$ are unchanged.

In the Volterra encoding the $j$th unknown port has normalization $\lambda_j=\sqrt2 h\alpha_j/4$, while the known port has constant normalization.
Reserve a constant fraction of the delay budget for the known port.
The remaining weighted scheduling theorem has cost $O((K h\langle\boldsymbol C,\boldsymbol\alpha\rangle_{1/2}+\norm{\boldsymbol C}_1)\log(1/\epsilon))$.
The combined unnormalized right-hand side has solution norm at least $4mW_2$.
Using $K=128C_\Pi mW_2/r$ proves both bounds.
\end{proof}
A bound on the full inverse norm can replace the two propagation-based coefficients in \cref{vdc:eq:weighted}: $C_\Pi TW_2/r\ \mapsto\ O(hK_{\rm G})$ and $C_\Pi/r\ \mapsto\ O(K_{\rm G}/(mW_2))$.
This does not require a inverse cutoff parameter.

For an arbitrary inverse cutoff parameter with separate oracle rates, use the paired invariance and hypotheses on the individual input vectors in \cref{cert:eq:weighted-invariance}.
A cutoff for a bundled coefficient encoding uses \cref{vdc:thm:combined}, with the bundled oracle's gate cost.
\subsubsection{Choosing the endpoint scale}
The scale in \cref{vdc:eq:balanced} ensures a constant probability of selecting the final-state output.
Additional classical bounds can support a different scale and lower query cost.
Suppose classical bounds give $V_+\ge\norm{\boldsymbol v}$, $0<s_0\le\norm{P\boldsymbol v}$, and $c_+\ge\norm{\boldsymbol c_\perp}$.
Every
\begin{align}
0<\tau\le\min\left\{1,\frac{s_0}{2\sqrt{V_+^2+c_+^2}}\right\}
\label{vdc:eq:tau-family}
\end{align}
gives probability of selecting the final-state output at least $4/5$ and a solution-norm lower bound $4s_0/\tau$.
A fixed smaller constant lower bound can be used if desired.
For each endpoint scale, recompute the tail or moment bound of the balanced matrix and choose a certified cutoff $K_\epsilon(\tau)$.
For fixed oracle prices, the query objective is, up to universal constants,
\begin{align}
C_HK_\epsilon(\tau)\log(1/\epsilon) +C_{\rm prep}\left(1+\frac{K_\epsilon(\tau)\tau}{4s_0}\right).
\label{vdc:eq:tau-objective}
\end{align}
The classical minimization is over given certificates.
Taking $\tau$ smaller always increases the output amplitude, but need not improve this objective because the balanced singular spectrum also changes.

A reversibly computable scalar profile $\chi(t)>0$ is another independent choice.
Substituting $\boldsymbol x=\chi\boldsymbol z$ gives
\begin{align}
\boldsymbol z'=-[H+(\chi'/\chi)I]\boldsymbol z+\boldsymbol b/\chi,
\quad U_\chi(t,s)=\frac{\chi(s)}{\chi(t)}U(t,s),
\quad r_\chi=r/\chi(T).
\label{vdc:eq:profile}
\end{align}
Differentiation proves these identities.
Apply the preceding results to the transformed history and its actual preparation normalization.
The implementation includes the scalar-term arithmetic and the changed state preparations.
A matrix-valued change of coordinates additionally requires circuits for its action and inverse; a Lyapunov inequality alone serves here as classical input.

\subsubsection{A finite-query inverse approximation can bypass singular-value truncation}
\cref{vdc:prop:finite} amplifies the finite-query approximation to the full inverse history before selecting the final-state output.

\begin{proof}[Proof of \cref{vdc:prop:finite}]
One state preparation followed by the given circuit succeeds with amplitude $\norm{\boldsymbol w}/\beta_F\ge y_0/\beta_F$.
Bounded-budget amplitude amplification uses $O(\beta_F/y_0)$ calls to this circuit and its inverse and preserves the normalized successful vector.
Its physical probability of selecting the final-state output remains constant by \cref{eq:vde-endpoint-norm} and the certified error.
A constant number of attempts and a separate gate-synthesis budget prove the claim.
\end{proof}
The cost depends on the block-encoding normalization and error bound as well as the degree.
For a polynomial annihilating the input state with $a(0)\ne0$, the exact inverse action is
\begin{align}
p(z)=\frac{1-a(z)/a(0)}z,
\quad p(A_\tau)\ket c=\boldsymbol z_\tau.
\label{vdc:eq:annihilator}
\end{align}
Multiplication by $A_\tau$ proves the identity.
An approximate annihilator gives the residual $p(A_\tau)\ket c-\boldsymbol z_\tau =-a(0)^{-1}A_\tau^{-1}a(A_\tau)\ket c$.
The certificate must bound this inverse-weighted residual.

\cref{vdc:lem:powers} implements all polynomial degrees with one reusable signal register and a first-failure label.

\begin{lemma}[Matrix powers with a first-failure register]
\label{vdc:lem:powers}
Suppose controlled $U_M=\Be{M}$ uses $a$ signal qubits.
One can coherently select any power up to degree $d$ with $d$ matrix calls, one signal register, and $O(\log(d+1))$ additional label qubits, apart from reusable control arithmetic.
Therefore an LCU polynomial $p(M)=\sum_{j=0}^d c_jM^j$ has normalization $\sum_j|c_j|$ with that query count and logarithmic additional label space.
\end{lemma}
\begin{proof}
Prepare a degree label $j$ and a failure-time register $f=0$.
At slot $k$, apply the oracle only on the branch with no recorded failure and an unfinished degree.
Then record a nonzero signal by the following transposition: $U_M\text{ controlled on }\{f=0,\ j\ge k\}$ and $|0\rangle_f\leftrightarrow|k\rangle_f \text{ controlled on }\{\text{signal}\ne0,\ j\ge k\}$.
Both operations are unitary.
On a branch whose first nonzero signal occurs at slot $k$, the transposition records $f=k$.
Every subsequent slot leaves that branch unchanged, since later transpositions involve different labels.
On the branch $f=0$, the signal is zero after every slot, and the successful system action is precisely $M^j$.
Induction on $k$ proves both statements.
The complementary amplitude stays in the signal and first-failure registers.

Prepare the degree superposition with amplitudes $\sqrt{|c_j|/\sum|c_j|}$, apply the controlled powers and coefficient phases, and unprepare.
Selecting degree zero, signal zero, and $f=0$ gives the claimed polynomial block.
Binary comparisons and conditional transpositions have polynomial cost in $a+\log(d+1)$ and use reusable scratch.
A conservative extra gate bound is $O(d[a+\log(d+1)]^2)$ plus coefficient loading, with $O(a+\log(d+1))$ reusable workspace beyond the system register.
The inverse circuit reverses the recorded operations exactly.
\end{proof}
This construction prevents a linear accumulation of product signal registers.
Loading a generic length-$d$ coefficient list still costs $O(d)$ rotations, and more efficient loading requires additional structure.

\subsubsection{Three distinct improvements}
\paragraph{A transient with a small mean-square history.}
Let the lower truncated shift generate the homogeneous evolution: $S_De_j=e_{j+1}\ (j<D-1)$, $S_De_{D-1}=0$, $H=I-S_D$, $\boldsymbol x_0=e_0$, and $\boldsymbol b=0$.
Assume $D\ge8T+16$ and $T\ge1$.
The generator is accretive and $\alpha\le2$, so $C=1$.
Its solution is
\begin{align}
\boldsymbol x(t)=e^{-t}\sum_{j=0}^{D-1}\frac{t^j}{j!}e_j,
\quad \norm{\boldsymbol x(t)}=\Theta((1+t)^{-1/4}).
\label{vdc:eq:shift}
\end{align}
For the infinite sum, Parseval's identity gives
\begin{align}
e^{-2t}\sum_{j\ge0}\frac{t^{2j}}{(j!)^2} =\frac1{2\pi}\int_{-\pi}^{\pi}e^{-2t(1-\cos\theta)}\,d\theta =\Theta((1+t)^{-1/2}).
\label{vdc:eq:shift-proof}
\end{align}
The last estimate follows from $2\theta^2/\pi^2\le1-\cos\theta\le\theta^2/2$.
The omitted finite-dimensional tail is exponentially small uniformly on $[0,T]$: it is at most the square of a Poisson upper-tail probability, which obeys $\Pr\{X\ge D\}\le(e t/D)^D$ by the exponential Markov bound.
The stated choice of $D$ preserves the two-sided estimates.
Both the endpoint norm and the continuous root-mean-square norm scale as $T^{-1/4}$.
Thus \cref{vdc:prop:continuous} gives $W_2=\Theta(T^{-1/4})$.
The peak-history estimate gives $O(T^{5/4}\log(1/\epsilon))$ generator queries; the mean-square theorem gives $O(T\log(1/\epsilon))$.
Preparation queries remain $O(T^{1/4})$.
The comparison concerns the two sufficient bounds for the same family.

\paragraph{An unpopulated unstable sector.}
Take $H=H_{\rm good}\oplus(-\eta I)$, with $\operatorname{Re} H_{\rm good}\succeq0$, and all individual input vectors supported on the first summand.
The full propagator bound can grow as $e^{\eta T}$, while $C_\Pi=1$.
\cref{vdc:thm:sectors} removes this exponential factor without constructing the sector projector.
Each oracle keeps its original normalization, and the discretization estimate is evaluated on the supported sector.

\paragraph{A low-degree inverse approximation despite a long evolution time.}
Suppose the homogeneous generator is $H=\lambda I+N_0$ and the normalized initial state is $\ket\psi$.
A classical certificate and the oracle access satisfy $N_0^2\ket\psi=0$, $\bra\psi N_0\ket\psi=0$, $\norm{N_0\ket\psi}=\nu>0$, and $\Be{N_0/\nu}\text{ is given}$.
Then
\begin{align}
e^{-HT}\ket\psi=e^{-\lambda T}(I-TN_0)\ket\psi,
\quad \norm{(I-TN_0)\ket\psi}=\sqrt{1+\nu^2T^2}.
\label{vdc:eq:nilpotent}
\end{align}
The degree-one LCU has normalization $1+\nu T$, whose ratio to this norm is at most $\sqrt2$.
Thus a constant number of generator and preparation queries suffices for the normalized state, independently of $T$; rotation precision remains charged.
The truncated exponential follows directly from $N_0^2\ket\psi=0$, and orthogonality proves its norm.
This uses a stronger algebraic certificate than a history norm or singular-tail estimate.

\subsection{Details for generalized eigenvalue estimation}
\label{app:gep-details}

\subsubsection{Trial states and confidence}
\label{app:gep-trials}

The history map in \cref{eq:gep-circuit-history} has norm at most $\sqrt{N\kappa_B}$ before the boundary scale, and eigenstate output norm $\sqrt N$.
An input perturbation of size $\xi/\sqrt{\kappa_B}$ changes the normalized history by at most $2\xi$.
Its norm changes by at most a relative fraction $\xi$.
For $\xi\le1/100$, the lower bound $\rho_0$ in \cref{eq:gep-solver-parameters} remains valid.
In the main proof, use the enlarged tolerance $\tau+2\xi\le3/100$.
This still gives $\mathbb E X\ge1/5$, so the same median bound proves the input tolerance.

For a trial state, choose $B$-orthonormal generalized eigenvectors $\boldsymbol{x}_j$, and define
\begin{align}
p_j=\frac{|\boldsymbol{x}_j^\dagger B|\psi\rangle|^2} {\langle\psi|B|\psi\rangle},
\quad \boldsymbol{x}_j^\dagger B\boldsymbol{x}_k=\delta_{jk},
\quad \sum_jp_j=1.
\label{eq:gep-overlap}
\end{align}
For a degenerate eigenvalue, sum over its eigenspace.

\cref{prop:gep-trials} bounds the number of attempts needed to include a mode with known minimum metric overlap.
\begin{proposition}[Trial-state lists]
\label{prop:gep-trials}
For any given weight bound $p_j\ge p>0$, use $m=\lceil(4\kappa_B/p)\log(2/\delta)\rceil$ independent attempts of the algorithm below.
With probability at least $1-\delta$, their list contains an estimate of the corresponding eigenvalue with additive error at most $\epsilon$.
Their total costs are
\begin{align}
Q_A=O\!\left[m\left(\frac{\alpha_A}{\beta\epsilon}\right) \log\frac{\kappa_B}{p}\right],
\quad Q_B=O\!\left[m\left(\frac{R\alpha_B}{\beta\epsilon}\right) \log\frac{\kappa_B}{p}\right],
\quad Q_\psi=O(m\kappa_B).
\label{eq:gep-trial-cost}
\end{align}
The additional guarantee that every retained entry lies within $\epsilon$ of the spectrum has the costs in \cref{eq:gep-valid-list-cost}.
Neither guarantee requires an eigenvalue gap.
\end{proposition}

\begin{proof}[Proof of \cref{prop:gep-trials}]
Use the construction in \cref{eq:gep-circuit-history,eq:gep-affine}.
Every partial controlled circuit has smallest singular value at least $\kappa_B^{-1/2}$, so $\|M_N^{-1}|0,c,\psi\rangle\|\ge\sqrt{N/\kappa_B}/s =N/(\beta\kappa_B)$.
Apply \cref{eq:gep-solver-cost} with $K=2N/\beta$ and $\rho_0=N/(\beta\kappa_B)$.
This gives $Q_\psi=O(\kappa_B)$ per attempt, independently of the window error $\eta$.

Let $|\chi\rangle=B^{1/2}|\psi\rangle/ \sqrt{\langle\psi|B|\psi\rangle}$.
Its $H$-eigenmode weights are $p_j$.
The direct circuit history is obtained from the $V$ history by applying $I\otimes B^{-1/2}$ and renormalizing.
The input to that history is $|\chi\rangle$.
For any clock event $E$, the bounds on $B^{-1}$ give the elementary comparison
\begin{align}
\kappa_B^{-1}\Pr_\chi(E)\le\Pr_{\rm direct}(E) \le\kappa_B\Pr_\chi(E).
\label{eq:gep-metric-sandwich}
\end{align}
At least half the ideal history times have completed all controlled powers.
Apply \cref{eq:gep-kaiser} to the Fourier readout at these times.
Including the inverse success probability of at least $2/3$, trace-distance error $\tau$, and phase-grid randomization gives
\begin{align}
\Pr(\text{retain and }|\widehat\lambda-\lambda_j|\le\epsilon) &\ge\frac23\left(\frac{(1-\eta)p_j}{\kappa_B}-\tau\right),
\label{eq:gep-target-probability}
\\
\Pr(\text{retain and }\operatorname{dist}(\widehat\lambda, \operatorname{spec}H)>\epsilon) &\le\kappa_B\eta+\tau.
\label{eq:gep-spectrum-probability}
\end{align}
Take $\eta=1/10$ and $\tau=p/(16\kappa_B)$.
Each attempt hits the target with probability at least $p/(4\kappa_B)$.
The probability of missing it in $m$ attempts is bounded by $e^{-mp/(4\kappa_B)}\le\delta/2$.
Substituting $N=O(R/\epsilon)$ in \cref{eq:gep-solver-cost} proves \cref{eq:gep-trial-cost}.

For the additional guarantee that every retained entry is accurate, take
\begin{align}
m=\left\lceil\frac{4\kappa_B}{p}\log\frac2\delta\right\rceil,
\quad \eta=\frac{\delta}{4m\kappa_B},
\quad \tau=\frac{\delta}{4m}.
\label{eq:gep-list-budgets}
\end{align}
The parameter choices satisfy $\eta\le1/16$ and $\tau\le p/(16\kappa_B)$.
Hence the target-hit probability is again at least $p/(4\kappa_B)$.
Target omission costs at most $\delta/2$, while a union bound gives probability at most $m(\kappa_B\eta+\tau)=\delta/2$ of any accepted off-spectrum estimate.
The same QLSP substitution, now with $N=O((R/\epsilon)\log(8m\kappa_B/\delta))$, gives
\begin{subequations}
\label{eq:gep-valid-list-cost}
\begin{align}
Q_A&=O\!\left[m\left(\frac{\alpha_A}{\beta\epsilon} \log\frac{8m\kappa_B}{\delta}\right) \log\frac{8m}{\delta}\right],
\\
Q_B&=O\!\left[m\left(\frac{R\alpha_B}{\beta\epsilon} \log\frac{8m\kappa_B}{\delta}\right) \log\frac{8m}{\delta}\right],
\quad Q_\psi=O(m\kappa_B).
\end{align}
\end{subequations}
All failed and discarded attempts are included.
\end{proof}

For $r$ target eigenvalues, each with weight at least $p$, increase the confidence logarithm in the attempt count to $\log(2r/\delta)$.
The same union bound then controls all targets.
No gap is needed for inclusion; identifying a specified mode requires information that distinguishes it from the others.

The direct output selection loss can be attained with commuting matrices.
Consider the commuting example $B=\operatorname{diag}(\beta,b_{\max})$ and $A=R\operatorname{diag}(-\beta,b_{\max})$.
A Euclidean weight on the second eigenvector gives metric weight $p$ when
\begin{align}
q=\frac{\beta p}{b_{\max}(1-p)+\beta p}.
\end{align}
The ideal resolved direct output selection samples that eigenvector with probability $q=\Theta(p/\kappa_B)$ for fixed $p\in(0,1)$.
Alternatively, the ideal history filter $B^{1/2}/\sqrt{b_{\max}}$ restores the canonical metric readout.
The spectral bounds on the metric matrix put its success probability in $[\kappa_B^{-1},1]$.
Its finite implementation adds square-root synthesis and a conditioning error budget; the corresponding input-conversion cost is included in the Hermitian comparison row.

The preparation dependence can also be necessary.
Take the matrices and their Euclidean-normalized generalized eigenstates to be
\begin{align}
B=\operatorname{diag}(\kappa_B,1),
\quad A=\begin{pmatrix}0&\sqrt{\kappa_B}
\\
\sqrt{\kappa_B}&0\end{pmatrix},
\quad |\psi_\pm\rangle=\frac{(1,\pm\sqrt{\kappa_B})^{\mathsf T}}{\sqrt{1+\kappa_B}}.
\end{align}
Their generalized eigenvalues are $\pm1$.
Their overlap has magnitude $(\kappa_B-1)/(\kappa_B+1)$.
For $t$ copies, their trace distance is $\sqrt{1-[(\kappa_B-1)/(\kappa_B+1)]^{2t}}$.
Constant-success discrimination therefore requires $t=\Omega(\kappa_B)$.
The preparers rotate $|1\rangle$ through angles $\pm\arctan(\kappa_B^{-1/2})$, up to a global phase on one output.
Their unitary distance is $2/\sqrt{1+\kappa_B}$.
The query hybrid bounds final-state distance by $2q/\sqrt{1+\kappa_B}$ after $q$ controlled or inverse preparation calls.
Estimation at error below one distinguishes the outputs with constant bias, proving $q=\Omega(\sqrt{\kappa_B})$, even with $A,B$ known.

\subsubsection{Comparison under the same promises}
\label{app:gep-comparison}

All rows of \cref{tab:gep-comparison} receive the same exact generalized eigenstate and return a classical estimate.
If $\epsilon\ge\min\{R,\alpha_A/\beta\}$, returning zero suffices; the table uses the remaining regime.
The bounds below retain the normalizations independently, subject to \cref{eq:gep-promises}.

For QEP, encode $D=B^{-1}A$ at normalization $O(\alpha_A/\beta)$.
Its eigenvector matrix is $B^{-1/2}Y$, where $Y$ is a unitary eigenvector matrix of the Hermitian reduction.
Its condition number is therefore at most $\sqrt{\kappa_B}$.
The original estimator \cite[Theorem~3]{Low2024Eigenvalue} uses $O(\alpha_A\sqrt{\kappa_B}/(\beta\epsilon))$ matrix and preparation calls.
Improved QEP \cite[Theorem~6]{Low2024QLSP} uses $O(\sqrt{\kappa_B})$ preparation calls, with an additional matrix factor $\log(\kappa_B)\log\log(\kappa_B)$.
Its constant-factor norm estimate is available uniformly: rescale the Chebyshev signal to $[-1/2,1/2]$, write it as $\cos\theta$, and use
\begin{align}
\left|\sum_{l=0}^{n-1}\cos^2(l\theta)-\frac n2\right| \le\frac1{2|\sin\theta|}\le\frac1{\sqrt3}.
\end{align}
The history norm is $\Theta(\sqrt n)$; the boundary preconditioner in Eq.~(205) of that work multiplies it by $\sqrt n$.
Thus its norm is $\Theta(n)$, with no norm-search queries.

The state-preparation-optimal QLSP theorem \cite[Theorem~2]{Low2024QLSP}, for inverse success amplitude parameter $D_0$, has the matrix factor
\begin{align}
\log(2D_0)\left[\log\frac2\tau+\log\log(2D_0)\right]
\label{eq:gep-low-su-logs}
\end{align}
and $O(D_0)$ preparation calls.
For the Cayley--LCU row, $D_0=O(\sqrt{\kappa_B})$ by \cref{eq:gep-source-response,eq:gep-inverse-bound} and $\tau$ is constant.
Ordinary LCU encodes the history matrix at normalization $O(\alpha_A/R+\alpha_B)$.
The boundary scale contributes no larger term because $s\le b_{\max}\le\alpha_B$.
Multiplying by $2N/\beta$ and \cref{eq:gep-low-su-logs} gives the displayed cost to each matrix.
An ordinary solver with the optimal matrix-query order can remove these condition logarithms, but its generic preparation count then follows the full matrix-query count.

Bounded polynomial transformations \cite[Corollaries~67 and~69 and Theorem~56]{Gilyen2018singular} approximate the two operators $\beta B^{-1}/2$ and $\sqrt\beta B^{-1/2}/2$.
To error $\zeta$, their synthesized implementations use \begin{align}
O\!\left(\frac{\alpha_B}\beta\log\frac2\zeta\right)
\label{eq:gep-power-synthesis}
\end{align} $B$ queries.
An outer $q$-query algorithm can take $\zeta=O(q^{-2})$.
To check this for arbitrary encoding completions, the dilation
\begin{align}
\mathcal D(X)=\begin{pmatrix} X&\sqrt{I-XX^\dagger}
\\
\sqrt{I-X^\dagger X}&-X^\dagger \end{pmatrix}
\end{align}
satisfies the perturbation estimate $\|X-Y\|\le\zeta $ implies $ \|\mathcal D(X)-\mathcal D(Y)\|\le4\sqrt\zeta$.
Use the product and positive-square-root bounds $\|XX^\dagger-YY^\dagger\|\le2\zeta$ and $\|\sqrt P-\sqrt Q\|\le\sqrt{\|P-Q\|}$.
Polar decomposition of the off-diagonal completion blocks expresses any padded completion through this dilation and unitaries fixing the selected coordinates.
Keeping those unitaries gives a nearby completion of $Y$.
The query hybrid therefore costs at most $4q\sqrt\zeta$.
For QEP this gives the $B$-only logarithm $\log(2+\alpha_A\kappa_B/(\beta\epsilon))$, including the improved row's condition factors.
Each synthesized query uses $O(1)A$ calls.

The Hermitian reduction has encoding normalization $O(\alpha_A/\beta)$.
Constant-success phase estimation then uses $q=O(\alpha_A/(\beta\epsilon))$ queries.
Substitution in \cref{eq:gep-power-synthesis} gives its first $B$ term.
Its input must first be transformed by $B^{1/2}$.
The block $B^{1/2}/(2\sqrt{b_{\max}})$ has success at least $1/(4\kappa_B)$ and admits the same degree order in \cref{eq:gep-power-synthesis}.
Apply the bounded local Taylor theorem \cite[Corollary~66]{Gilyen2018singular} to $\tfrac12\sqrt{\alpha_Bt/b_{\max}} $ for $t\in[\beta/\alpha_B,b_{\max}/\alpha_B]$.
Its analytic margin is $\Theta(\beta/\alpha_B)$.
The absolute binomial-coefficient sum on the extended interval is bounded; a constant rescaling yields the bounded transform.
Using $\beta/2$ covers a narrow or degenerate interval.
Error $O(\kappa_B^{-1/2})$ preserves the conditional input to constant accuracy.
Amplifying this input conversion before phase estimation costs $O(\sqrt{\kappa_B})\text{ preparation calls}$ and $O((\alpha_B\sqrt{\kappa_B}/\beta)\log(\kappa_B))\text{ metric queries}$.
No output conversion is needed for a classical estimate.

Median repetition multiplies every row by $\log(2/\delta)$, keeping constant internal error.
Shao and Liu \cite{Shao2020GeneralizedEigenvalue} also give a differential-equation reduction in a broader real-diagonalizable setting, with complexity stated up to logarithmic factors.
The rows of \cref{tab:gep-comparison} use the same eigenstate promise; arbitrary trial states require the overlap analysis in \cref{prop:gep-trials}.
State-preparation scaling and input tolerance independent of $\epsilon$ were already obtained by improved QEP.
Separate scheduling preserves the different matrix coefficients in \cref{eq:gep-priced-cost}.

\subsection{Amplified generalized eigenvalue estimation}
\label{app:gepamp}
We first derive one coherent estimation attempt, then amplify its accepted component.
The QLSP construction takes an affine linear system, an inverse-norm upper bound, and a solution-norm lower bound, as required by \cref{wql:thm:main}.

\paragraph{Padded Cayley history.}
For scalar $a\in\mathbb R$ and $b>0$, the phase response and its infinite impulse response transfer function are
\begin{align}
q(a,b)=\frac{b+ia/R}{b-ia/R},
\quad G(z;a,b)=\frac1{1-zq(a,b)} =\sum_{k\ge0}z^kq(a,b)^k,
\quad |z|<1.
\label{eq:gepamp-scalar}
\end{align}
The feedback connection is the one in \cref{fig:gep-junction}.
Its coefficients supply the phases for a Kaiser-window state.
The ordered matrix substitution gives
\begin{align}
T_{\rm Cay}=(B-iA/R)^{-1}(B+iA/R)=B^{-1/2}VB^{1/2},
\quad V=(I-iH/R)^{-1}(I+iH/R).
\label{eq:gepamp-cayley}
\end{align}
Here $V$ is unitary and its eigenphases are $\theta_j=2\arctan(\lambda_j/R)$.
This identity identifies the response; the implementation below uses only an affine combination of the supplied encodings.

Choose a normalized Kaiser state $|c\rangle$ with $n$ coefficients.
The window guarantee of Berry et al.~\cite[Sections II.B.3 and IV.A]{Berry2024Rapid} allows
\begin{align}
n=O\!\left(\frac Rh\log\frac4\eta\right),
\quad \Pr(|\widehat\lambda-\lambda_j|\ge h)\le\eta
\label{eq:gepamp-window}
\end{align}
on every eigenstate.
A finite coherent Fourier readout giving this bound is described below.
By the endpoint separation in \cref{thm:gepamp-main}, the canonical state $\sum_k c_k|k\rangle V^k|\chi\rangle$ gives an estimate in $I$ with probability at least $(1-\eta)p_I$.
The probability of an estimate farther than $h$ from the entire spectrum is at most $\eta$.

Take $N=3n$ computation-time coordinates, and put $s=\beta\sqrt{\kappa_B/N}$.
On the phase register let $E_t$ project onto $k\ge t-n$ for $n+1\le t\le2n-1$, and let $E_t=0$ otherwise.
The space of equation coordinates is the direct sum of the $N$ time coordinates, each carrying the phase and system registers.
Define $M_N$ by
\begin{align}
s\boldsymbol z_0=|c\rangle|\psi\rangle,
\quad (B-iE_tA/R)\boldsymbol z_t =(B+iE_tA/R)\boldsymbol z_{t-1},
\quad 1\le t<N.
\label{eq:gepamp-history}
\end{align}
Identity tensor factors are implicit.
There are $n$ initial idle steps, $n-1$ controlled Cayley steps, and $n$ final idle steps.
For $\ell_t=\min\{n-1,\max\{0,t-n\}\}$, the solution is
\begin{align}
\boldsymbol z=s^{-1}\sum_{t=0}^{N-1}\sum_{k=0}^{n-1} c_k|t,k\rangle T_{\rm Cay}^{\min\{k,\ell_t\}}|\psi\rangle.
\label{eq:gepamp-response}
\end{align}
Let $\Pi_{\rm c}$ select $t\ge2n-1$; these $n+1$ coordinates contain completed phase estimation.
Initial padding gives
\begin{align}
\|M_N^{-1}|0,c,\psi\rangle\| \ge\frac{\sqrt{n+1}}s \ge\frac{N}{2\beta\sqrt{\kappa_B}}.
\label{eq:gepamp-solution-lower}
\end{align}
For the inverse bound, the block column of the inverse corresponding to the initial time coordinate has norm at most $\sqrt{N\kappa_B}/s=N/\beta$.
Every inverse block corresponding to a later input time is of the form
\begin{align}
(I\otimes B^{-1/2})\mathcal U_{t,u} (I-iE_u\otimes H/R)^{-1}(I\otimes B^{-1/2}),
\end{align}
where $\mathcal U_{t,u}$ is a unitary product of controlled $V$ steps.
Each block has norm at most $1/\beta$.
The row and column sums therefore give
\begin{align}
\|M_N^{-1}\|\le2N/\beta,
\quad \|F_BM_N^{-1}\|\le2N/\beta,
\quad F_B=\Pi_{\rm c}\otimes I\otimes B^{1/2}/\sqrt\beta.
\label{eq:gepamp-inverse}
\end{align}
The second estimate follows by cancelling the left inverse square root in each block; the initial block column has the same bound.

\paragraph{Rational metric reweighting.}
To compensate for the coordinate change without synthesizing $B^{1/2}$ inside each query, use the scalar responses
\begin{align}
t_\ell=\beta2^\ell,
\quad f_\ell(b)=\sqrt{t_\ell/\beta}\frac b{b+t_\ell},
\quad 0\le\ell\le L=\lceil\log_2\kappa_B\rceil.
\label{eq:gepamp-metric-response}
\end{align}
For $b\in[\beta,b_{\max}]$ these obey
\begin{align}
\frac b{4\beta}\le\sum_{\ell=0}^Lf_\ell(b)^2 \le\frac{2b}{\beta}.
\label{eq:gepamp-metric-bound}
\end{align}
For the lower bound, write $x=b/\beta$.
If $2^{j-1}<x\le2^j$ with $j\ge1$, the two terms $\ell=j-1,j$ each contribute at least $2x/9$.
At $x=1$ the term $\ell=0$ contributes $x/4$.
For the upper bound, use the telescoping inequality
\begin{align}
\frac{x^2u}{(x+u)^2} \le2x^2\!\left(\frac1{x+u}-\frac1{x+2u}\right),
\quad u>0.
\end{align}
Summing with $u=2^\ell$ proves \cref{eq:gepamp-metric-bound}.

\begin{figure}[htbp]
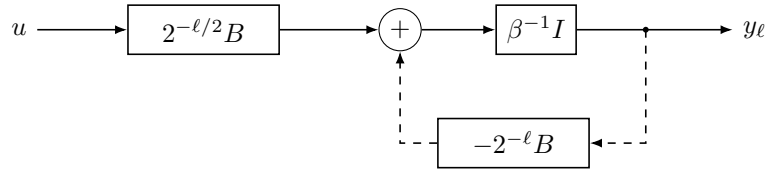

\centering
\IIRDiagram{GeneralizedEigenvalueMetric}
\caption{The rational metric reweighting, shown as a classical signal-flow relation for each $\ell$.
Closing the dashed feedback gives $y_\ell=f_\ell(B)u$, where $u=\Pi_{\rm c}\boldsymbol z$ is the completed Cayley history.
The affine equations for all branches and the history are compiled together by the QLSP IIR.
The individual matrix gains in this diagram are not implemented as normalized quantum operations.}
\label{fig:gep-metric-iir}
\end{figure}

Append auxiliary output coordinates satisfying
\begin{align}
(\beta I+2^{-\ell}B)\boldsymbol y_\ell =2^{-\ell/2}B\Pi_{\rm c}\boldsymbol z.
\label{eq:gepamp-metric-equation}
\end{align}
Their solution is $\boldsymbol y_\ell=f_\ell(B)\Pi_{\rm c}\boldsymbol z$, as shown in \cref{fig:gep-metric-iir}.
Writing $F=(f_\ell(B)\Pi_{\rm c})_\ell$ and $D=\operatorname{diag}_\ell(\beta I+2^{-\ell}B)$, the full affine matrix and its inverse are
\begin{align}
\mathcal L_N= \begin{pmatrix}M_N&0
\\
-DF&D\end{pmatrix},
\quad \mathcal L_N^{-1}= \begin{pmatrix}M_N^{-1}&0
\\
FM_N^{-1}&D^{-1}\end{pmatrix}.
\label{eq:gepamp-full}
\end{align}
Functional calculus gives $F^\dagger F\preceq2F_B^\dagger F_B$.
Thus \cref{eq:gepamp-inverse} proves the inverse-norm and solution-norm bounds
\begin{align}
\|\mathcal L_N^{-1}\|\le\frac{2\sqrt3N+1}{\beta} \le\frac{5N}{\beta}=:K,
\quad \|\mathcal L_N^{-1}|0,c,\psi\rangle\| \ge\frac{N}{2\beta\sqrt{\kappa_B}}=:\rho_0.
\label{eq:gepamp-certificates}
\end{align}
The second inequality retains the original history coordinates.

Let $b_\psi=\langle\psi|B|\psi\rangle$ and $\gamma=(n+1)/N\ge1/3$.
For a phase-register event of canonical probability $q$, the direct history has total squared norm at most $Nb_\psi/(\beta s^2)$.
Its completed event has squared norm at least $\gamma Nb_\psi q/(b_{\max}s^2)$.
Consequently the joint direct output selection probability is at least $q/(3\kappa_B)$ and at most $\kappa_B q$.
For the metric-reweighted output, \cref{eq:gepamp-metric-bound} gives
\begin{align}
\frac{\gamma Nb_\psi q}{4\beta s^2} \le\|\boldsymbol y_{\rm event}\|^2 \le\frac{2\gamma Nb_\psi q}{\beta s^2}.
\label{eq:gepamp-event}
\end{align}
Divide by the total norm of $(\boldsymbol z,\boldsymbol y)$.
Using the upper bound for both parts for the lower probability, and the lower bound for $\|\boldsymbol y\|^2$ for the upper probability, proves
\begin{align}
\frac q{20}\le\Pr_{\rm ideal}(\text{metric-reweighted output and event}) \le8q.
\label{eq:gepamp-event-probability}
\end{align}
These estimates are inequalities for phase-register projectors; they remain valid when $A$ and $B$ do not commute.

\paragraph{Compiling one coherent attempt.}
As in the exact-eigenstate construction, the coefficients of $A$ and $B$ in $M_N$ have norms at most $2/R$ and $2$, respectively.
In $\mathcal L_N$ they are bounded by $2/R$ and $4$; the added column has norm at most $(\sum_\ell2^{-\ell})^{1/2}<\sqrt2$.
The known coefficient has norm $\max\{s,\beta\}$.
The separate-family rates of the QLSP theorem therefore give, for trace-distance error $\tau$, per-attempt bounds
\begin{align}
Q_A=O\!\left[\left(1+\frac{N\alpha_A}{R\beta}\right) \log\frac2\tau\right],
\quad Q_B=O\!\left[\left(1+\frac{N\alpha_B}{\beta}\right) \log\frac2\tau\right],
\quad Q_\psi=O(\sqrt{\kappa_B}).
\label{eq:gepamp-attempt-cost}
\end{align}
The preparation count follows from $K/\rho_0=O(\sqrt{\kappa_B})$.
Each encoding of a coefficient uses only its corresponding original oracle; the readout coordinates do not contribute a factor $L$.
For example, their direct-sum $B$ operation uses one controlled $B$ encoding.

Run at most eight copies of the finite QLSP circuit and select the first successful copy.
Their combined success probability exceeds $1-(3/4)^8>2/3$.
Store every outcome and workspace coherently.
The resulting finite unitary $U$ implements one attempt and has an available inverse.
Its success-conditioned system state has trace-distance error at most $\tau$; this changes any joint event probability by at most $\tau$.
Reversibly mark the conjunction of inversion success, readout selection, and $\widehat\lambda\in I$.

Let $a$ be this marked probability and let $b$ be the probability of a marked estimate with error greater than $h$ from every eigenvalue in $I$.
The endpoint promise implies that every such bad estimate is farther than $h$ from the full spectrum.
Hence
\begin{subequations}
\label{eq:gepamp-a-b}
\begin{align}
a&\ge\frac23\left(\frac{(1-\eta)p}{20}-\tau\right),
\quad b
\quad \le8\eta+\tau
\quad \text{(metric-reweighted output)},
\\
a&\ge\frac23\left(\frac{(1-\eta)p}{3\kappa_B}-\tau\right),
\quad b\le\kappa_B\eta+\tau
\quad \text{(direct output selection)}.
\end{align}
\end{subequations}
For the metric-reweighted output choose $\eta=\tau=p\delta/4096$ and $a_0=p/40$.
For the direct output selection choose $\eta=\tau=p\delta/(4096\kappa_B^2)$ and $a_0=p/(16\kappa_B)$.
In both cases $a\ge a_0$ and $b/a\le\delta/4$.

\paragraph{Amplification and its proof.}
The following elementary implementation of amplitude amplification uses only a lower bound on success~\cite{brassard2002quantum}.
Let $\Pi$ be the marked projector, put $M=\lceil a_0^{-1/2}\rceil$, and define
\begin{align}
G=-U(I-2|0\rangle\langle0|)U^\dagger(I-2\Pi).
\end{align}
In each batch, with probability one half use $j=0$; otherwise choose $j$ uniformly from $0,\ldots,M-1$.
Apply $G^jU$ and measure the marked flag.
Stop at the first marked outcome and report its estimate.
Use $r=\lceil8\log(2/\delta)\rceil$ batches.

Writing $a=\sin^2\theta$, the marked probability after $j$ Grover steps is $\sin^2((2j+1)\theta)$.
If $a\ge1/4$, the $j=0$ branch gives batch success at least $1/8$.
For $a\in[a_0,1/4]$, the uniform branch satisfies
\begin{align}
\frac1M\sum_{j=0}^{M-1}\sin^2((2j+1)\theta) =\frac12-\frac{\sin(4M\theta)}{4M\sin(2\theta)} \ge\frac12-\frac1{4\sqrt3}>\frac14.
\label{eq:gepamp-grover-average}
\end{align}
Every batch therefore succeeds with probability at least $1/8$.
The probability of no marked outcome is at most $e^{-r/8}\le\delta/2$.
Every accepted state is proportional to the same vector $\Pi U|0\rangle$, so its bad-output probability is exactly $b/a$.
The total failure probability is at most $\delta/2+b/a<\delta$.
The algorithm uses $O(a_0^{-1/2}\log(2/\delta))$ calls to $U$ and $U^\dagger$.
Substituting \cref{eq:gepamp-window,eq:gepamp-attempt-cost} and the chosen errors proves \cref{thm:gepamp-main}.
Reflections use the actual finite unitary $U$ and its inverse.
Thus no approximation error is accumulated by replacing $U$ with an idealized circuit in the Grover steps.

\paragraph{A finite coherent Fourier readout.}
For completeness, the continuous Kaiser-window tail bound can be used without continuous randomness.
Define the phase density
\begin{align}
f_\theta(u)=\frac1{2\pi} \left|\sum_{k=0}^{n-1}c_ke^{ik(\theta-u)}\right|^2.
\end{align}
It integrates to one.
Choose its tail outside circular distance $h/(2R)$ to be at most $\eta/2$.
Embed the $n$ window coefficients into a register of power-of-two dimension $D\ge16\pi n/\eta$, with zero amplitudes on the added indices, and apply the $D$-point inverse Fourier transform.
The probability at $u_j=2\pi j/D$ is exactly $(2\pi/D)f_\theta(u_j)$.
Parseval and Cauchy--Schwarz give
\begin{align}
\int_0^{2\pi}|f'_\theta(u)|\,du\le2(n-1),
\quad \|f_\theta\|_\infty\le n/(2\pi).
\end{align}
Multiplication by the indicator of the tail adds only two jumps.
Its total variation is at most $4n$, so the Riemann-sum error is at most $8\pi n/D\le\eta/2$.
Clip the principal phase to $[-\pi/2,\pi/2]$ and apply $\widehat\lambda=R\tan(\widehat\theta/2)$, whose derivative on that interval is at most $R$.
Computing this value to error at most $h/4$ gives \cref{eq:gepamp-window}.
All window, Fourier, arithmetic, and interval-test operations are known gates.
Their synthesis tolerances can be included in $\tau$ by allocating a constant fraction of its budget.
Zero-amplitude phase indices do not increase the number of controlled Cayley steps or original-oracle queries.

\section{Downfolding and online extensions}

\subsection{The downfolding cascade circuit}
\label{app:downfolding-circuit}
\cref{app:downfolding-circuit} proves the logarithmic dependence on cascade length asserted in \cref{thm:app-gates}.

\begin{proof}
We first close the internal public coordinates explicitly.
In each sector the junction couples only to a weighted superposition of signal inputs, defined by the isometry
\begin{align}
V_i=\frac1{\sqrt{\Lambda_i}} \bigoplus_j\sqrt{\widetilde\lambda_j}\, (|0\rangle_j\otimes I)\Pi_jQ_i.
\label{eq:app-bright}
\end{align}
These isometries have mutually orthogonal ranges.
Set $u=t/(2J)$ in the junction of \cref{eq:app-junction}.
On sector $Q_i$ and the corresponding vector in the coupling range, it acts by
\begin{align}
\left(\begin{array}{c|c} c_i&-i\sqrt{1-c_i^2}
\\
\hline
\sqrt{1-c_i^2}&ic_i \end{array}\right)\otimes I_{Q_i},
\quad c_i=\frac{1-u\Lambda_i}{1+u\Lambda_i},
\label{eq:app-bright-rotation}
\end{align}
and acts as $iI$ on the orthogonal private coordinates.
For an internal sector, equate its public input and output: $\boldsymbol{g}=c_i\boldsymbol{g}-i\sqrt{1-c_i^2}\boldsymbol{v} $ for $i\ge1$.
Here $\boldsymbol{g}$ is the feedback amplitude and $\boldsymbol{v}$ is the private input.
The resulting private output is
\begin{align}
\sqrt{1-c_i^2}\boldsymbol{g}+ic_i\boldsymbol{v} =i\left(c_i-\frac{1-c_i^2}{1-c_i}\right)\boldsymbol{v}=-i\boldsymbol{v}.
\end{align}
Thus closing an internal public coordinate gives a known reflection.
The circuit applies this reflection directly; the displayed division verifies its action and requires no arithmetic involving $K^{-1}$.

Let $R$ denote the real rotation between the active public coordinates and their weighted superpositions in the ranges of $V_i$.
After alignment, the complete closed system matrix is
\begin{align}
W_{\rm sys}=R\,(I_{\rm public}\oplus P_{\rm fb}),
\quad P_{\rm fb}=i\left(I-2\sum_{i\ge1}V_iV_i^\dagger\right).
\label{eq:app-closed-work}
\end{align}
Each private phase commutes with the rotations at the other stages.
We can therefore collect them into $I_{\rm public}\oplus(I_J\otimes P_{\rm fb})$ and apply the phase operation once, with the stage register as a spectator.
The remaining product has $J$ identical real rotations, each between the public coordinate and a different stage coordinate.

The exact dyadic factorization of Chen et al.\ \cite[Lemma 10 and Proposition 11]{Chen2026GateEfficientSimulation} gives
\begin{align}
R_{J-1}\cdots R_0= \widehat W_{\rm out}\, \operatorname{Rot}(c_0^J)_{\rm public,0}\, \widehat W_{\rm in}^{\dagger}.
\label{eq:app-dyadic}
\end{align}
Each basis change uses one controlled one-qubit gate per stage-address bit.
The hatted operators fix the public coordinate and unused private modes, so the identity holds on arbitrary feedback amplitudes, as required inside the unitary lift.
Compute and retain the nested lower-bit-zero predicates until they are used, then erase them.
This takes $O(b+\log J)$ gates and $O(b)$ reusable scratch.
Including alignment, one full known-operation update costs $O(G_{\rm align}+b+\log J)$.

To count qubits, let $\mathcal L_j$ be the full original oracle space of $\Be{H_j/\lambda_j}$.
The lift acts on
\begin{align}
(\mathbb C^N\otimes\mathcal P)\ \oplus\ \bigoplus_j (\mathbb C^{r_j}\otimes\mathbb C^J\otimes\mathcal L_j).
\label{eq:app-carrier}
\end{align}
Padding this direct sum into the common data and encoding fields gives the address width \cref{eq:app-address}.
The address records the stage and delay, which are spectators during a component call.
Thus one call to component $j$ acts across its stages.
Calls to different components remain separate and are counted by $q_j$; coherently selecting among the component oracles has the cost of its actual controlled calls.

One lift has $N$ passes of known operations and $(m+1)N+\sum_jq_j^{\rm lift}$ routing operations.
Each comparison, constant addition, or interval permutation uses $O(b)$ gates and reusable scratch.
The lift count is at most $N$, and the cascade address width is bounded by $b$.
These operations therefore contribute $O(N[G_{\rm align}+(m+1)b])$ gates.
The padding makes the stage and port permutations known binary relabelings; this count uses that padded representation.

The rank-one clock uses normalized endpoint states on $N$ time slots and a constant number of padding labels.
Binary-prefix preparation, \cref{eq:qudit-clock-preparation}, computes the successive conditional amplitudes and implements each endpoint with $O(N)$ gates and $O(\log(N+1))$ qubits.
This includes the smoothed coefficients; no prepared clock state is given as an oracle.

Scratch is erased between calls and known-operation passes, so its width is the maximum requirement of these operations.
Oblivious amplitude amplification in \cref{lem:iir-cubic}, clock preparations, and reflections about the all-zero ancillary subspace change the counts only by constants.
The active-sector test needed for each reflection is included in alignment.
Combining the oracle, known-operation, routing, and clock costs proves \cref{eq:app-gates}; packing the persistent registers and reusing the scratch proves \cref{eq:app-space}.
\end{proof}

\subsection{Additional certificates for online kernels}
\label{app:td-proofs}
The following extensions can be proved by ordered-product perturbation, finite moment matching, and polynomial approximation.
They use the actual scheduled kernel $H_{k,s}$ of \cref{subsec:td-lift}; all windows consist of physical event times.
The Kalman construction uses its weighted controllability Gramian instead and needs no analytic continuation in physical time.

\paragraph{Memory stability.}
Suppose each frozen private update satisfies $\|D(t)^r\|\le C_f\rho_f^r$, with $\rho_f<1$, and successive updates differ by at most $\nu$.
For an integer $b\ge1$ with $0<q_b:=C_f\rho_f^b+\nu b(b-1)/2<1$, telescoping a block of $b$ contractions proves $\|D_k\cdots D_{k-r+1}\|\le q_b^{\lfloor r/b\rfloor}$.
An analytic continuation of the complete event matrices to a time strip, with a uniform perturbation margin preserving this decay, gives a joint bound $\|F(t,z)\|\le M$ for $|z|\le R$, where $R>1$.
Cauchy's estimate then gives the memory tail $\tau_L=MR^{-L}/(1-R^{-1})$.
For one terminal output, a uniform length-$L$ input clock and that output slot give normalization $\sqrt L$ and error at most $\tau_L$.

\paragraph{Moment-corrected clocks.}
Let $J$ be a window of $m\ge2$ consecutive output samples, rescaled to $x_k=-1+2k/(m-1)$, and let $J_0\subset J$ contain $m_0$ samples.
The uniform weights $c_k^{(0)}$ on $J_0$ average the output.
To correct this average to a target coordinate $x_*$, choose $p<m$ and set
\begin{align}
\boldsymbol v(x)=(\sqrt{2\ell+1}\,P_\ell(x))_{\ell=0}^p,
\quad H_p=\frac1m\sum_{k\in J}\boldsymbol v(x_k)\boldsymbol v(x_k)^T,
\quad \boldsymbol d=\boldsymbol v(x_*)-\sum_kc_k^{(0)}\boldsymbol v(x_k).
\label{tdreg:eq:correction-data}
\end{align}
Here $P_\ell$ is the Legendre polynomial; $H_p$ is positive definite because the $m$ distinct nodes determine a polynomial of degree $p$.
The correction $c_k^{(c)}=m^{-1}\boldsymbol v(x_k)^TH_p^{-1}\boldsymbol d$ reproduces every moment through degree $p$ and has zero total weight.
Enlarge $J_0,J$ on the left by $L-1$ samples to obtain $I_0,I$.
The overlap matrix $X_{k,s}=c_k^{(0)}\boldsymbol1_{I_0}(s)+c_k^{(c)}\boldsymbol1_I(s)$ has rank at most two and satisfies
\begin{align}
\|X\|_*\le\alpha_0+\alpha_c,
\quad \alpha_0=\sqrt{1+(L-1)/m_0},
\quad \alpha_c=\sqrt{\frac{m+L-1}{m}\boldsymbol d^TH_p^{-1}\boldsymbol d}.
\label{tdreg:eq:exact-alpha}
\end{align}
If a degree-$p$ operator polynomial approximates $F(t,1)$ on the whole window, including the target, with error $E_p$, one can prove
\begin{align}
\|\mathcal L_X[\mathcal K]-F(t_*,1)\| \le(1+\beta)\tau_L+(2+\beta)E_p,
\quad \beta=\|\boldsymbol c^{(c)}\|_1\le\alpha_c.
\label{tdreg:eq:main-error}
\end{align}
Indeed, each rectangle contains every memory age below $L$, and the corrected weights reproduce the approximating polynomial exactly.
The two rectangles prepare the clock using one additional label qubit.

\paragraph{Normalization and temporal regularity.}
For $0<\delta\le1$, choosing $m_0\ge\max\{1,(L-1)/\delta\}$ and increasing $m$ until $\alpha_c\le\delta/2$ gives normalization at most $1+\delta$.
Bounds for Legendre derivatives and the discrete Gram matrix prove the sufficient horizons
\begin{align}
N_{\rm centered}=O\!\left(L+p^2+\frac{L(p+1)^{5/2}}{\delta^{3/2}}\right),
\quad N_{\rm trailing}=O\!\left(L+p^2+\frac{L(p+1)^3}{\delta^2}\right).
\label{tdreg:eq:window-horizons}
\end{align}
Centered windows read out after the target time; trailing windows use only samples up to that time.
These are sufficient bounds for this construction, including $p=0$.

To prove the horizons, use odd, concentric windows $J_0\subset J$ with $x_*=0$ in the centered case; in the trailing case, let both windows end at $x_*=1$.
Integer rounding changes the bounds only by constants.
Orthogonality of the normalized Legendre basis gives $\frac12\int_{-1}^1\boldsymbol v(x)\boldsymbol v(x)^Tdx=I$.
For $q=\boldsymbol a^T\boldsymbol v$, the identities $\int_{-1}^1(P_\ell')^2=\ell(\ell+1)$ and Cauchy--Schwarz imply $\|q'\|_{L^2}=O((p+1)^2)\|q\|_{L^2}$.
Thus the total variation of $|q|^2$ is $O((p+1)^2)\|q\|_{L^2}^2$.
The uniform-grid quadrature error, including its endpoint weights, is consequently $O((p+1)^2/m)\|q\|_{L^2}^2$.
Taking $m\ge C(p+1)^2$ for a sufficiently large universal constant gives $\frac12I\preceq H_p\preceq\frac32I$.

The endpoint derivative bounds for Legendre polynomials give $\sup_{[-1,1]}\|\boldsymbol v'\|=O((p+1)^3)$ and $\sup_{[-1,1]}\|\boldsymbol v''\|=O((p+1)^5)$.
In the centered case the first-order terms cancel under the symmetric average, so Taylor's formula bounds $\|\boldsymbol d\|=O((p+1)^5(m_0/m)^2)$.
For the trailing average it instead gives $\|\boldsymbol d\|=O((p+1)^3m_0/m)$.
For $m\ge L$, \cref{tdreg:eq:exact-alpha} therefore has $\alpha_c=O(\|\boldsymbol d\|)$.
Choose $m_0=O(1+L/\delta)$ and then increase $m$ to satisfy $m\ge C m_0(p+1)^{5/2}/\sqrt\delta$ or $m\ge C m_0(p+1)^3/\delta$, respectively, as well as the preceding lower bounds.
These choices give $\alpha_c\le\delta/2$ and the horizons in \cref{tdreg:eq:window-horizons}, including the $L-1$ extra input slots.
Finally, $\alpha_0\le\sqrt{1+\delta}\le1+\delta/2$, proving the claimed normalization.

For an analytic response bounded by $M_t$ on a Bernstein ellipse of parameter $\chi>1$, one may use $E_p=2M_t\chi^{-p}/(\chi-1)$ \cite[Chapter 8]{Trefethen2019Approximation}.
A time strip of width $\sigma$ contains this ellipse only when the window half-width $a=(m-1)h/2$ satisfies
\begin{align}
a(\chi-\chi^{-1})/2<\sigma.
\label{tdreg:eq:ellipse-fit}
\end{align}
The analytic domain must also cover the ellipse's real extent.
Keeping these constants uniform is the condition for a polylogarithmic precision bound.
The actual ordered kernel must be checked after any change of sampling rate or grouping of events.

All input slots must be available when the circuit starts.
For a process initialized at time zero, using the infinite-history response as an intermediate comparison adds at most $\tau_{k_*+1}$ at target index $k_*$.
An unknown initial memory is included as a separate public input.
A length-$N$ window uses at most $\lceil N/r_j\rceil$ events of a periodic port; replay and inverse access are separate oracle resources.
The weights require $O(mp^2+p^3)$ classical arithmetic and generic $O(m)$ clock-preparation gates.
Endpoint state errors at most $\eta$ each add at most $2(1+\delta)\eta$ to the unnormalized output error.

\bibliographystyle{alphaUrlePrint}
\bibliography{main}

@ARTICLE{Rossi2025ModularQSP,
  title     = "Modular quantum signal processing in many variables",
  author    = "Rossi, Zane M and Ceroni, Jack L and Chuang, Isaac L",
  journal   = "Quantum",
  publisher = "Verein zur Forderung des Open Access Publizierens in den
               Quantenwissenschaften",
  volume    =  9,
  number    =  1776,
  pages     =  1776,
  month     =  "18~" # jun,
  year      =  2025
}

@ARTICLE{Laneve2025MultivariateQSP,
  title     = "On multivariate polynomials achievable with quantum signal
               processing",
  author    = "Laneve, Lorenzo and Wolf, Stefan",
  journal   = "Quantum",
  publisher = "Verein zur Forderung des Open Access Publizierens in den
               Quantenwissenschaften",
  volume    =  9,
  number    =  1641,
  pages     =  1641,
  month     =  "20~" # feb,
  year      =  2025
}

@article{Somma2025LinearMatrix,
  author = {Rolando D. Somma and Guang Hao Low and Dominic W. Berry and Ryan Babbush},
  doi = {10.1038/s41534-026-01360-6},
  title = {Quantum algorithm for linear matrix equations},
  year = {2026},
  journal = {npj Quantum Information},
  url = {https://doi.org/10.1038/s41534-026-01360-6},
  eprint = {2508.02822},
  archivePrefix = {arXiv},
  note = {Published online 21 August 2026; advance online publication},
}

@article{Harrow2009,
   author = {Aram W. Harrow and Avinatan Hassidim and Seth Lloyd},
   doi = {10.1103/PhysRevLett.103.150502},
   issn = {0031-9007},
   issue = {15},
   journal = {Physical Review Letters},
   month = {10},
   pages = {150502},
   publisher = {American Physical Society},
   title = {Quantum algorithm for linear systems of equations},
   volume = {103},
   url = {http://link.aps.org/doi/10.1103/PhysRevLett.103.150502 https://link.aps.org/doi/10.1103/PhysRevLett.103.150502},
   year = {2009}
}

@article{Berry2024Rapid,
   author = {Dominic W. Berry and Yu Tong and Tanuj Khattar and Alec White and Tae In Kim and Guang Hao Low and Sergio Boixo and Zhiyan Ding and Lin Lin and Seunghoon Lee and Garnet Kin-Lic Chan and Ryan Babbush and Nicholas C. Rubin},
   doi = {10.1103/PRXQuantum.6.020327},
   issn = {2691-3399},
   issue = {2},
   journal = {PRX Quantum},
   month = {5},
   pages = {020327},
   title = {Rapid initial-state preparation for the quantum simulation of strongly correlated molecules},
   volume = {6},
   year = {2025}
}

@article{King2025SOSSA,
  author = {Robbie King and Guang Hao Low and Ryan Babbush and Rolando D. Somma and Nicholas C. Rubin},
  doi = {10.1103/m3fj-m4rm},
  title = {Quantum simulation with sum-of-squares spectral amplification},
  year = {2026},
  journal = {Physical Review Letters},
  url = {https://doi.org/10.1103/m3fj-m4rm},
  volume = {136},
  pages = {110601},
  eprint = {2505.01528},
  archivePrefix = {arXiv},
}

@article{Eremenko2011Sign,
   author = {Alexandre Eremenko and Peter Yuditskii},
   doi = {10.1007/s11854-011-0018-7},
   issn = {0021-7670},
   issue = {1},
   journal = {Journal d'Analyse Mathématique},
   month = {9},
   pages = {285-315},
   title = {Polynomials of the best uniform approximation to sgn(x) on two intervals},
   volume = {114},
   year = {2011}
}

@article{Yagle1985SchurAlgorithm,
  author = {Yagle, Andrew E. and Levy, Bernard C.},
  title = {The Schur algorithm and its applications},
  journal = {Acta Applicandae Mathematicae},
  year = {1985},
  volume = {3},
  issn = {1572-9036},
  url = {http://dx.doi.org/10.1007/BF00047331},
  doi = {10.1007/bf00047331},
  number = {3},
  publisher = {Springer Science and Business Media LLC},
  month = May,
  pages = {255–284},
}

@article{Tong2021FastInversion,
  author = {Tong, Yu and An, Dong and Wiebe, Nathan and Lin, Lin},
  title = {Fast inversion, preconditioned quantum linear system solvers, fast Green’s-function computation, and fast evaluation of matrix functions},
  journal = {Physical Review A},
  year = {2021},
  volume = {104},
  issn = {2469-9934},
  url = {http://dx.doi.org/10.1103/PhysRevA.104.032422},
  doi = {10.1103/physreva.104.032422},
  number = {3},
  publisher = {American Physical Society (APS)},
  month = Sept,
}

@article{Low2017USA,
   author = {Guang Hao Low and Isaac L. Chuang},
   journal = {arXiv:1707.05391},
   month = {7},
   title = {Hamiltonian simulation by uniform spectral amplification},
   url = {https://arxiv.org/abs/1707.05391},
   year = {2017}
}

@article{Jin2025Schrodingerization,
  author = {Shi Jin and Nana Liu and Chuwen Ma and Yizhe Peng and Yue Yu},
  doi = {10.48550/arXiv.2505.00370},
  journal = {arXiv preprint: 2505.00370},
  month = {5},
  title = {On the Schrödingerization method for linear non-unitary dynamics with optimal dependence on matrix queries},
  year = {2025},
}

@article{an2023lchsoptimal,
  author = {Dong An and Andrew M. Childs and Lin Lin},
  doi = {10.1007/s00220-025-05509-w},
  title = {Quantum algorithm for linear non-unitary dynamics with near-optimal dependence on all parameters},
  year = {2026},
  journal = {Communications in Mathematical Physics},
  url = {https://doi.org/10.1007/s00220-025-05509-w},
  volume = {407},
  pages = {19},
  eprint = {2312.03916},
  archivePrefix = {arXiv},
}

@article{Childs2012LCU,
   author = {Andrew M Childs and Nathan Wiebe},
   city = {Paramus, NJ},
   issn = {1533-7146},
   issue = {11-12},
   journal = {Quantum Information \& Computation},
   month = {11},
   pages = {901-924},
   publisher = {Rinton Press, Incorporated},
   title = {Hamiltonian simulation using linear combinations of unitary operations},
   volume = {12},
   url = {http://dl.acm.org/citation.cfm?id=2481569.2481570},
   year = {2012}
}

@article{an2023lchs,
   author = {Dong An and Jin-Peng Liu and Lin Lin},
   doi = {10.1103/PhysRevLett.131.150603},
   issn = {0031-9007},
   issue = {15},
   journal = {Physical Review Letters},
   month = {10},
   pages = {150603},
   title = {Linear combination of Hamiltonian simulation for nonunitary dynamics with optimal state preparation Cost},
   volume = {131},
   year = {2023}
}

@article{Motlagh2024gqsp,
   author = {Danial Motlagh and Nathan Wiebe},
   doi = {10.1103/PRXQuantum.5.020368},
   issn = {2691-3399},
   issue = {2},
   journal = {PRX Quantum},
   month = {6},
   pages = {020368},
   title = {Generalized Quantum Signal Processing},
   volume = {5},
   year = {2024}
}

@article{Low2018IntPicSim,
   author = {Guang Hao Low and Nathan Wiebe},
   journal = {arXiv: 1805.00675},
   month = {5},
   title = {Hamiltonian simulation in the interaction picture},
   url = {https://arxiv.org/abs/1805.00675},
   year = {2018}
}

@inproceedings{Gilyen2018singular,
   author = {András Gilyén and Yuan Su and Guang Hao Low and Nathan Wiebe},
   city = {New York, New York, USA},
   doi = {10.1145/3313276.3316366},
   booktitle = {Proceedings of the 51st Annual ACM Symposium on Theory of Computing - STOC '19},
   pages = {193-204},
   publisher = {ACM Press},
   title = {Quantum singular value transformation and beyond: exponential improvements for quantum matrix arithmetics},
   url = {http://arxiv.org/abs/1806.01838},
   year = {2019}
}

@inproceedings{Berry2014Exponential,
   author = {Dominic W. Berry and Andrew M. Childs and Richard Cleve and Robin Kothari and Rolando D. Somma},
   city = {New York, New York, USA},
   doi = {10.1145/2591796.2591854},
   isbn = {9781450327107},
   booktitle = {Proceedings of the 46th Annual ACM Symposium on Theory of Computing - STOC '14},
   pages = {283-292},
   publisher = {ACM Press},
   title = {Exponential improvement in precision for simulating sparse Hamiltonians},
   url = {http://doi.acm.org/10.1145/2591796.2591854 http://dl.acm.org/citation.cfm?doid=2591796.2591854},
   year = {2014}
}

@article{Childs2010,
   author = {Andrew M. Childs},
   doi = {10.1007/s00220-009-0930-1},
   issn = {0010-3616},
   issue = {2},
   journal = {Communications in Mathematical Physics},
   month = {3},
   pages = {581-603},
   publisher = {Springer Nature},
   title = {On the relationship between continuous- and discrete-time quantum walk},
   volume = {294},
   url = {http://dx.doi.org/10.1007/s00220-009-0930-1 http://link.springer.com/10.1007/s00220-009-0930-1},
   year = {2010}
}

@inproceedings{Berry2015Hamiltonian,
   author = {Dominic W. Berry and Andrew M. Childs and Robin Kothari},
   city = {Washington, DC, USA},
   doi = {10.1109/FOCS.2015.54},
   isbn = {978-1-4673-8191-8},
   booktitle = {2015 IEEE 56th Annual Symposium on Foundations of Computer Science},
   month = {10},
   pages = {792-809},
   publisher = {IEEE},
   title = {Hamiltonian simulation with nearly optimal dependence on all parameters},
   url = {http://dx.doi.org/10.1109/FOCS.2015.54 http://ieeexplore.ieee.org/document/7354428/},
   year = {2015}
}

@inproceedings{Aharonov2003Adiabatic,
   author = {Dorit Aharonov and Amnon Ta-Shma},
   city = {New York, New York, USA},
   doi = {10.1145/780542.780546},
   isbn = {1581136749},
   booktitle = {Proceedings of the thirty-fifth ACM symposium on Theory of computing - STOC '03},
   pages = {20},
   publisher = {ACM Press},
   title = {Adiabatic quantum state generation and statistical zero knowledge},
   url = {http://doi.acm.org/10.1145/780542.780546},
   year = {2003}
}

@article{Lloyd1996universal,
   author = {Seth Lloyd},
   doi = {10.1126/science.273.5278.1073},
   issn = {0036-8075},
   issue = {5278},
   journal = {Science},
   month = {8},
   pages = {1073-1078},
   publisher = {American Association for the Advancement of Science},
   title = {Universal quantum simulators},
   volume = {273},
   url = {http://science.sciencemag.org/content/273/5278/1073 http://www.sciencemag.org/cgi/doi/10.1126/science.273.5278.1073},
   year = {1996}
}

@article{Feynman1982,
   author = {Richard P. Feynman},
   doi = {10.1007/BF02650179},
   issn = {0020-7748},
   issue = {6-7},
   journal = {International Journal of Theoretical Physics},
   month = {6},
   pages = {467-488},
   title = {Simulating physics with computers},
   volume = {21},
   url = {http://dx.doi.org/10.1007/BF02650179 http://link.springer.com/10.1007/BF02650179},
   year = {1982}
}

@book{Trefethen2019Approximation,
   author = {Lloyd N. Trefethen},
   city = {Philadelphia, PA},
   doi = {10.1137/1.9781611975949},
   isbn = {978-1-61197-593-2},
   month = {1},
   publisher = {Society for Industrial and Applied Mathematics},
   title = {Approximation theory and approximation practice, extended edition},
   year = {2019}
}

@article{Takahashi2010Adder,
   author = {Yasuhiro Takahashi and Seiichiro Tani and Noboru Kunihiro},
   doi = {10.26421/QIC10.9-10-12},
   issn = {15337146},
   issue = {9&10},
   journal = {Quantum Information and Computation},
   month = {9},
   pages = {872-890},
   title = {Quantum addition circuits and unbounded fan-out},
   volume = {10},
   year = {2010}
}

@inproceedings{Low2018SpectralNorm,
   author = {Guang Hao Low},
   city = {New York, New York, USA},
   doi = {10.1145/3313276.3316386},
   booktitle = {Proceedings of the 51st Annual ACM Symposium on Theory of Computing - STOC '19},
   pages = {491-502},
   publisher = {ACM Press},
   title = {Hamiltonian simulation with nearly optimal dependence on spectral norm},
   url = {http://arxiv.org/abs/1807.03967},
   year = {2019}
}

@article{Sanders2018,
   author = {Yuval R. Sanders and Guang Hao Low and Artur Scherer and Dominic W. Berry},
   doi = {10.1103/PhysRevLett.122.020502},
   issn = {0031-9007},
   issue = {2},
   journal = {Physical Review Letters},
   month = {1},
   pages = {020502},
   publisher = {American Physical Society},
   title = {Black-box quantum state preparation without arithmetic},
   volume = {122},
   url = {https://link.aps.org/doi/10.1103/PhysRevLett.122.020502},
   year = {2019}
}

@article{Berry2015Truncated,
   author = {Dominic W. Berry and Andrew M. Childs and Richard Cleve and Robin Kothari and Rolando D. Somma},
   doi = {10.1103/PhysRevLett.114.090502},
   issn = {0031-9007},
   issue = {9},
   journal = {Physical Review Letters},
   month = {3},
   pages = {090502},
   publisher = {American Physical Society},
   title = {Simulating Hamiltonian dynamics with a truncated Taylor series},
   volume = {114},
   url = {http://link.aps.org/doi/10.1103/PhysRevLett.114.090502 https://link.aps.org/doi/10.1103/PhysRevLett.114.090502},
   year = {2015}
}

@article{Low2025-tn,
  title = "Fast Quantum Simulation of Electronic Structure by Spectral
              Amplification",
  author = "Low, Guang Hao and King, Robbie and Berry, Dominic W and Han,
              Qiushi and DePrince, A Eugene and White, Alec F and Babbush, Ryan
              and Somma, Rolando D and Rubin, Nicholas C",
  number = 4,
  year = {2025},
  journal = {Physical Review X},
  doi = {10.1103/pb2g-j9cw},
  url = {https://doi.org/10.1103/pb2g-j9cw},
  volume = {15},
  pages = {041016},
}

@article{Arunachalam2019-fc,
  title = "Quantum query algorithms are completely bounded forms",
  author = "Arunachalam, Srinivasan and Bri{\"{e}}t, Jop and Palazuelos,
               Carlos",
  journal = "SIAM Journal on Computing",
  publisher = "Society for Industrial \& Applied Mathematics (SIAM)",
  volume = 48,
  number = 3,
  pages = "903--925",
  month = jan,
  year = 2019,
  issn = {1095-7111},
  url = {http://dx.doi.org/10.1137/18M117563X},
  doi = {10.1137/18m117563x},
}

@article{Low2016HamSim,
   author = {Guang Hao Low and Isaac L. Chuang},
   doi = {10.1103/PhysRevLett.118.010501},
   issn = {0031-9007},
   issue = {1},
   journal = {Physical Review Letters},
   month = {1},
   pages = {010501},
   publisher = {American Physical Society},
   title = {Optimal Hamiltonian simulation by quantum signal processing},
   volume = {118},
   url = {http://link.aps.org/doi/10.1103/PhysRevLett.118.010501},
   year = {2017}
}

@article{Low2016Qubitization,
   author = {Guang Hao Low and Isaac L. Chuang},
   doi = {doi.org/10.22331/q-2019-07-12-163},
   journal = {Quantum},
   pages = {163},
   title = {Hamiltonian simulation by Qubitization},
   volume = {3},
   year = {2019}
}

@article{Berry2014Differential,
   author = {Dominic W Berry},
   doi = {10.1088/1751-8113/47/10/105301},
   issn = {1751-8113},
   issue = {10},
   journal = {Journal of Physics A: Mathematical and Theoretical},
   month = {3},
   pages = {105301},
   title = {High-order quantum algorithm for solving linear differential equations},
   volume = {47},
   url = {http://stacks.iop.org/1751-8121/47/i=10/a=105301 http://stacks.iop.org/1751-8121/47/i=10/a=105301?key=crossref.3b213c2a3b3e8d0c4d50d617e8e8d3bb},
   year = {2014}
}

@article{Costa2021linearsystems,
   author = {Pedro C.S. Costa and Dong An and Yuval R. Sanders and Yuan Su and Ryan Babbush and Dominic W. Berry},
   doi = {10.1103/PRXQuantum.3.040303},
   issn = {2691-3399},
   issue = {4},
   journal = {PRX Quantum},
   month = {10},
   pages = {040303},
   title = {Optimal scaling quantum linear-systems solver via discrete adiabatic theorem},
   volume = {3},
   year = {2022}
}

@article{Lin2020Ground,
   author = {Lin Lin and Yu Tong},
   doi = {10.22331/q-2020-12-14-372},
   issn = {2521-327X},
   journal = {Quantum},
   month = {12},
   pages = {372},
   title = {Near-optimal ground state preparation},
   volume = {4},
   year = {2020}
}

@article{An2022Differential,
   author = {Dong An and Jin-Peng Liu and Daochen Wang and Qi Zhao},
   doi = {10.1007/s00220-025-05358-7},
   issn = {0010-3616},
   issue = {8},
   journal = {Communications in Mathematical Physics},
   month = {8},
   pages = {189},
   title = {Quantum differential equation solvers: limitations and fast-forwarding},
   volume = {406},
   year = {2025}
}

@article{Jin2024Schrodingerization,
   author = {Shi Jin and Nana Liu and Yue Yu},
   doi = {10.1103/PhysRevLett.133.230602},
   issn = {0031-9007},
   issue = {23},
   journal = {Physical Review Letters},
   month = {12},
   pages = {230602},
   title = {Quantum simulation of partial differential equations via Schrödingerization},
   volume = {133},
   year = {2024}
}

@article{Krovi2022LinearDE,
   author = {Hari Krovi},
   doi = {10.22331/q-2023-02-02-913},
   issn = {2521-327X},
   journal = {Quantum},
   month = {2},
   pages = {913},
   title = {Improved quantum algorithms for linear and nonlinear differential equations},
   volume = {7},
   year = {2023}
}

@article{Fang2023Marching,
   author = {Di Fang and Lin Lin and Yu Tong},
   doi = {10.22331/q-2023-03-20-955},
   issn = {2521-327X},
   journal = {Quantum},
   month = {3},
   pages = {955},
   title = {Time-marching based quantum solvers for time-dependent linear differential equations},
   volume = {7},
   year = {2023}
}

@article{Berry2017Differential,
   author = {Dominic W. Berry and Andrew M. Childs and Aaron Ostrander and Guoming Wang},
   doi = {10.1007/s00220-017-3002-y},
   issn = {0010-3616},
   issue = {3},
   journal = {Communications in Mathematical Physics},
   month = {12},
   pages = {1057-1081},
   title = {Quantum algorithm for linear differential equations with exponentially improved dependence on precision},
   volume = {356},
   url = {http://link.springer.com/10.1007/s00220-017-3002-y},
   year = {2017}
}

@article{Low2024QLSP,
  author = {Guang Hao Low and Yuan Su},
  doi = {10.22331/q-2026-03-23-2041},
  title = {Quantum linear system algorithm with optimal queries to initial state preparation},
  year = {2026},
  journal = {Quantum},
  url = {https://doi.org/10.22331/q-2026-03-23-2041},
  volume = {10},
  pages = {2041},
  eprint = {2410.18178},
  archivePrefix = {arXiv},
}

@article{Berry2024Dyson,
   author = {Dominic W. Berry and Pedro C. S. Costa},
   doi = {10.22331/q-2024-06-13-1369},
   issn = {2521-327X},
   journal = {Quantum},
   month = {6},
   pages = {1369},
   title = {Quantum algorithm for time-dependent differential equations using Dyson series},
   volume = {8},
   year = {2024}
}

@article{Low2024Eigenvalue,
  author = {Guang Hao Low and Yuan Su},
  doi = {10.1137/24M1689363},
  title = {Quantum eigenvalue processing},
  year = {2026},
  journal = {SIAM Journal on Computing},
  url = {https://doi.org/10.1137/24M1689363},
  volume = {55},
  pages = {135--215},
  number = {1},
}

@article{Lee2020hypercontraction,
   author = {Joonho Lee and Dominic W. Berry and Craig Gidney and William J. Huggins and Jarrod R. McClean and Nathan Wiebe and Ryan Babbush},
   doi = {10.1103/PRXQuantum.2.030305},
   issn = {2691-3399},
   issue = {3},
   journal = {PRX Quantum},
   month = {7},
   pages = {030305},
   title = {Even more efficient quantum computations of chemistry through tensor hypercontraction},
   volume = {2},
   url = {http://arxiv.org/abs/2011.03494 https://link.aps.org/doi/10.1103/PRXQuantum.2.030305},
   year = {2021}
}

@article{vonBurg2020carbon,
   author = {Vera von Burg and Guang Hao Low and Thomas Häner and Damian S. Steiger and Markus Reiher and Martin Roetteler and Matthias Troyer},
   doi = {10.1103/PhysRevResearch.3.033055},
   issn = {2643-1564},
   issue = {3},
   journal = {Physical Review Research},
   month = {7},
   pages = {033055},
   title = {Quantum computing enhanced computational catalysis},
   volume = {3},
   url = {http://arxiv.org/abs/2007.14460 https://link.aps.org/doi/10.1103/PhysRevResearch.3.033055},
   year = {2021}
}

@article{Helton1974,
  author  = {Helton, J. William},
  title   = {Discrete Time Systems, Operator Models, and Scattering Theory},
  journal = {Journal of Functional Analysis},
  volume  = {16},
  number  = {1},
  pages   = {15--38},
  year    = {1974},
  doi     = {10.1016/0022-1236(74)90069-X}
}

@ARTICLE{Low2016Methodology,
  title    = "Methodology of resonant equiangular composite quantum gates",
  author   = "Low, Guang Hao and Yoder, Theodore J and Chuang, Isaac L",
  journal  = "Physical Review. X",
  volume   =  6,
  number   =  4,
  pages    =  041067,
  month    =  "28~" # dec,
  year     =  2016
}

@inproceedings{Barnum2004QuerySDP,
  title = "Quantum query complexity and semi-definite programming",
  author = "Barnum, H and Saks, M and Szegedy, M",
  booktitle = "18th IEEE Annual Conference on Computational Complexity, 2003.
               Proceedings",
  publisher = "IEEE Comput. Soc",
  pages = "179--193",
  year = {2003},
  doi = {10.1109/CCC.2003.1214419},
  isbn = {0-7695-1879-6},
  url = {http://ieeexplore.ieee.org/document/1214419/},
}

@INPROCEEDINGS{Reichardt2011Reflections,
  title     = "Reflections for quantum query algorithms",
  author    = "Reichardt, Ben W",
  booktitle = "Proceedings of the Twenty-Second Annual ACM-SIAM Symposium on
               Discrete Algorithms",
  publisher = "Society for Industrial and Applied Mathematics",
  pages     = "560--569",
  month     =  "23~" # jan,
  year      =  2011
}

@INPROCEEDINGS{Lee2011StateConversion,
  title     = "Quantum query complexity of state conversion",
  author    = "Lee, Troy and Mittal, Rajat and Reichardt, Ben W and Spalek,
               Robert and Szegedy, Mario",
  booktitle = "2011 IEEE 52nd Annual Symposium on Foundations of Computer
               Science",
  publisher = "IEEE",
  pages     = "344--353",
  month     =  oct,
  year      =  2011
}

@article{Bennett1997QuantumOracles,
  title = "Strengths and weaknesses of quantum computing",
  author = "Bennett, Charles H and Bernstein, Ethan and Brassard, Gilles and
               Vazirani, Umesh",
  journal = "SIAM Journal on Computing",
  publisher = "Society for Industrial \& Applied Mathematics (SIAM)",
  volume = 26,
  number = 5,
  pages = "1510--1523",
  month = oct,
  year = 1997,
  doi = {10.1137/S0097539796300933},
}

@article{Beals2001polynomial,
  title = "Quantum lower bounds by polynomials",
  author = "Beals, Robert and Buhrman, Harry and Cleve, Richard and Mosca,
               Michele and de Wolf, Ronald",
  journal = "Journal of the ACM",
  publisher = "Association for Computing Machinery (ACM)",
  volume = 48,
  number = 4,
  pages = "778--797",
  month = "1~" # jul,
  year = 2001,
  issn = {1557-735X},
  url = {http://dx.doi.org/10.1145/502090.502097},
  doi = {10.1145/502090.502097},
}

@article{Martyn2021GrandUnification,
  title = {Grand Unification of Quantum Algorithms},
  year = {2021},
  author = {Martyn, John M. and Rossi, Zane M. and Tan, Andrew K. and Chuang, Isaac L.},
  arxivid = {2105.02859},
  journal = {PRX Quantum},
  doi = {10.1103/PRXQuantum.2.040203},
  url = {https://doi.org/10.1103/PRXQuantum.2.040203},
  volume = {2},
  pages = {040203},
}

@article{Yoder2014,
    title = {{Fixed-point quantum search with an optimal number of queries}},
    year = {2014},
    journal = {Physical Review Letters},
    author = {Yoder, Theodore J. and Low, Guang Hao and Chuang, Isaac L.},
    number = {21},
    month = {11},
    pages = {210501},
    volume = {113},
    publisher = {American Physical Society},
    url = {http://link.aps.org/doi/10.1103/PhysRevLett.113.210501 https://link.aps.org/doi/10.1103/PhysRevLett.113.210501},
    doi = {10.1103/PhysRevLett.113.210501},
    issn = {0031-9007}
}

@article{Rossi2022MultivariableQSP,
  author = {Rossi, Zane M. and Chuang, Isaac L.},
  title = {Multivariable quantum signal processing (M-QSP): prophecies of the two-headed oracle},
  journal = {Quantum},
  year = {2022},
  volume = {6},
  issn = {2521-327X},
  url = {http://dx.doi.org/10.22331/q-2022-09-20-811},
  doi = {10.22331/q-2022-09-20-811},
  publisher = {Verein zur Forderung des Open Access Publizierens in den Quantenwissenschaften},
  month = Sept,
  pages = {811},
}

@article{Mori2024MultivariableComment,
  author = {Mori, Hitomi and Mizuta, Kaoru and Fujii, Keisuke},
  title = {Comment on “Multivariable quantum signal processing (M-QSP): prophecies of the two-headed oracle”},
  journal = {Quantum},
  year = {2024},
  volume = {8},
  issn = {2521-327X},
  url = {http://dx.doi.org/10.22331/q-2024-10-29-1512},
  doi = {10.22331/q-2024-10-29-1512},
  publisher = {Verein zur Forderung des Open Access Publizierens in den Quantenwissenschaften},
  month = Oct,
  pages = {1512},
}

@misc{Nemeth2023MultivariateVariants,
  author = {Németh, Balázs and Kövér, Blanka and Kulcsár, Boglárka and Miklósi, Roland Botond and Gilyén, András},
  title = {On variants of multivariate quantum signal processing and their characterizations},
  year = {2023},
  doi = {10.48550/ARXIV.2312.09072},
  url = {https://arxiv.org/abs/2312.09072},
  publisher = {arXiv},
  copyright = {arXiv.org perpetual, non-exclusive license},
}

@article{Ito2026ConstructiveDecision,
  author = {Ito, Yuki and Mori, Hitomi and Sakamoto, Kazuki and Fujii, Keisuke},
  title = {Polynomial time constructive decision algorithm for multivariable quantum signal processing},
  journal = {Quantum},
  year = {2026},
  volume = {10},
  issn = {2521-327X},
  url = {http://dx.doi.org/10.22331/q-2026-05-12-2102},
  doi = {10.22331/q-2026-05-12-2102},
  publisher = {Verein zur Forderung des Open Access Publizierens in den Quantenwissenschaften},
  month = May,
  pages = {2102},
}

@article{Belovs2024TamingQuantum,
  author = {Belovs, Aleksandrs and Jeffery, Stacey and Yolcu, Duyal},
  title = {Taming Quantum Time Complexity},
  journal = {Quantum},
  year = {2024},
  volume = {8},
  issn = {2521-327X},
  url = {http://dx.doi.org/10.22331/q-2024-08-23-1444},
  doi = {10.22331/q-2024-08-23-1444},
  publisher = {Verein zur Forderung des Open Access Publizierens in den Quantenwissenschaften},
  month = Aug,
  pages = {1444},
}

@misc{Chen2026OptimaTimeDependent,
  author = {Chen, Boyang and Gao, Minbo and Wang, Xinzhao and Zhou, Shuo},
  title = {Time-Dependent Hamiltonian Simulation with Optimal Query Complexity},
  year = {2026},
  doi = {10.48550/ARXIV.2608.06094},
  url = {https://arxiv.org/abs/2608.06094},
  publisher = {arXiv},
  copyright = {arXiv.org perpetual, non-exclusive license},
}

@misc{Chen2026GateEfficientSimulation,
  author = {Chen, Boyang and Gao, Minbo and Ji, Zhengfeng and Li, Tongyang and Wang, Xinzhao and Zhou, Shuo},
  title = {Gate-Efficient Implementation of the Query-Optimal Time-Dependent Hamiltonian Simulation},
  year = {2026},
  doi = {10.48550/ARXIV.2608.30629},
  url = {https://arxiv.org/abs/2608.30629},
  publisher = {arXiv},
  copyright = {arXiv.org perpetual, non-exclusive license},
}

@misc{Jeffery2026GuidedEnergy,
  author = {Jeffery, Stacey and Witteveen, Freek},
  title = {Optimal Quantum Algorithm for Ground-State Energy Estimation with a Guiding State},
  year = {2026},
  doi = {10.48550/ARXIV.2608.24494},
  url = {https://arxiv.org/abs/2608.24494},
  publisher = {arXiv},
  copyright = {Creative Commons Attribution 4.0 International},
}

@misc{Belovs2023LasVegas,
  author = {Belovs, Aleksandrs and Yolcu, Duyal},
  title = {One-Way Ticket to Las Vegas and the Quantum Adversary},
  year = {2023},
  doi = {10.48550/ARXIV.2301.02003},
  url = {https://arxiv.org/abs/2301.02003},
  publisher = {arXiv},
  copyright = {Creative Commons Attribution 4.0 International},
}

@article{Regalia1988AllpassFilters,
  author = {Regalia, P.A. and Mitra, S.K. and Vaidyanathan, P.P.},
  title = {The digital all-pass filter: a versatile signal processing building block},
  journal = {Proceedings of the IEEE},
  year = {1988},
  volume = {76},
  issn = {1558-2256},
  url = {http://dx.doi.org/10.1109/5.3286},
  doi = {10.1109/5.3286},
  number = {1},
  publisher = {Institute of Electrical and Electronics Engineers (IEEE)},
  month = Jan,
  pages = {19–37},
}

@article{Fettweis1986WaveDigitalFilters,
  author = {Fettweis, A.},
  title = {Wave digital filters: Theory and practice},
  journal = {Proceedings of the IEEE},
  year = {1986},
  volume = {74},
  issn = {0018-9219},
  url = {http://dx.doi.org/10.1109/PROC.1986.13458},
  doi = {10.1109/proc.1986.13458},
  number = {2},
  publisher = {Institute of Electrical and Electronics Engineers (IEEE)},
  pages = {270–327},
}

@article{Ball1998UnitaryColligations,
  author = {Ball, Joseph A and Trent, Tavan T},
  title = {Unitary Colligations, Reproducing Kernel Hilbert Spaces, and Nevanlinna–Pick Interpolation in Several Variables},
  journal = {Journal of Functional Analysis},
  year = {1998},
  volume = {157},
  issn = {0022-1236},
  url = {http://dx.doi.org/10.1006/jfan.1998.3278},
  doi = {10.1006/jfan.1998.3278},
  number = {1},
  publisher = {Elsevier BV},
  month = Aug,
  pages = {1–61},
}

@inbook{Ball2016NoncommutativeSchurAgler,
  author = {Ball, Joseph A. and Marx, Gregory and Vinnikov, Victor},
  title = {Interpolation and Transfer-function Realization for the Noncommutative Schur–Agler Class},
  booktitle = {Operator Theory in Different Settings and Related Applications},
  pages = {23–116},
  publisher = {Springer International Publishing},
  year = {2018},
  isbn = {9783319625270},
  issn = {2296-4878},
  url = {http://dx.doi.org/10.1007/978-3-319-62527-0_3},
  doi = {10.1007/978-3-319-62527-0_3},
}

@article{Darlington1939NetworkSynthesis,
  author = {Darlington, S.},
  title = {Synthesis of Reactance 4‐Poles Which Produce Prescribed Insertion Loss Characteristics: Including Special Applications To Filter Design},
  journal = {Journal of Mathematics and Physics},
  year = {1939},
  volume = {18},
  issn = {0097-1421},
  url = {http://dx.doi.org/10.1002/sapm1939181257},
  doi = {10.1002/sapm1939181257},
  number = {1-4},
  publisher = {Wiley},
  month = Apr,
  pages = {257–353},
}

@article{shende2006synthesis,
  author = {V. V. Shende and S. S. Bullock and I. L. Markov},
  doi = {10.1109/TCAD.2005.855930},
  issn = {0278-0070},
  issue = {6},
  journal = {IEEE Transactions on Computer-Aided Design of Integrated Circuits and Systems},
  month = {6},
  pages = {1000-1010},
  publisher = {IEEE},
  title = {Synthesis of quantum-logic circuits},
  volume = {25},
  url = {http://ieeexplore.ieee.org/document/1629135/},
  year = {2006},
}

@article{Mizel2009DampedSearch,
  author = {Mizel, Ari},
  title = {Critically Damped Quantum Search},
  journal = {Physical Review Letters},
  year = {2009},
  volume = {102},
  issn = {1079-7114},
  url = {http://dx.doi.org/10.1103/PhysRevLett.102.150501},
  doi = {10.1103/physrevlett.102.150501},
  number = {15},
  publisher = {American Physical Society (APS)},
  month = Apr,
}

@article{Tulsi2006FixedPointSearch,
  author = {Tulsi, Tathagat and Grover, Lov K. and Patel, Apoorva},
  title = {A new algorithm for fixed point quantum search},
  journal = {Quantum Info. Comput.},
  year = {2006},
  issue_date = {September 2006},
  publisher = {Rinton Press, Incorporated},
  address = {Paramus, NJ},
  volume = {6},
  number = {6},
  issn = {1533-7146},
  month = sep,
  pages = {483–494},
  numpages = {12},
  doi = {10.5555/2011691.2011693},
}

@book{Smith2010PhysicalAudioSignal,
  author = {Smith, III, Julius O.},
  title = {Physical Audio Signal Processing},
  publisher = {W3K Publishing},
  year = {2010},
  url = {http://ccrma.stanford.edu/~jos/pasp/},
}

@book{Smith2007DigitalFilters,
  author = {Smith, III, Julius O.},
  title = {Introduction to Digital Filters with Audio Applications},
  publisher = {W3K Publishing},
  year = {2007},
  url = {https://ccrma.stanford.edu/~jos/filters/},
  isbn = {978-0-9745607-1-7},
}

@article{Childs2015LinearSystems,
  author = {Andrew M. Childs and Robin Kothari and Rolando D. Somma},
  doi = {10.1137/16M1087072},
  issn = {0097-5397},
  issue = {6},
  journal = {SIAM Journal on Computing},
  month = {1},
  pages = {1920-1950},
  title = {Quantum algorithm for systems of linear equations with exponentially improved dependence on precision},
  volume = {46},
  url = {https://epubs.siam.org/doi/10.1137/16M1087072},
  year = {2017},
}

@misc{Dalzell2024LinearSystems,
  author = {Dalzell, Alexander M.},
  title = {A shortcut to an optimal quantum linear system solver},
  year = {2024},
  doi = {10.48550/ARXIV.2406.12086},
  url = {https://arxiv.org/abs/2406.12086},
  publisher = {arXiv},
  copyright = {Creative Commons Attribution Share Alike 4.0 International},
}

@inproceedings{Ambainis2012VTAA,
  author = {Andris Ambainis},
  city = {Dagstuhl, Germany},
  doi = {10.4230/LIPIcs.STACS.2012.636},
  editor = {Christoph Dürr and Thomas Wilke},
  isbn = {978-3-939897-35-4},
  issn = {1868-8969},
  booktitle = {29th International Symposium on Theoretical Aspects of Computer Science (STACS 2012)},
  note = {Keywords: quantum computing, quantum algorithms, amplitude amplification, linear equations},
  pages = {636-647},
  publisher = {Schloss Dagstuhl – Leibniz-Zentrum für Informatik},
  title = {Variable time amplitude amplification and quantum algorithms for linear algebra problems},
  volume = {14},
  url = {https://drops-dev.dagstuhl.de/entities/document/10.4230/LIPIcs.STACS.2012.636},
  year = {2012},
}

@article{Orsucci2021PositiveSystems,
  author = {Orsucci, Davide and Dunjko, Vedran},
  title = {On solving classes of positive-definite quantum linear systems with quadratically improved runtime in the condition number},
  journal = {Quantum},
  year = {2021},
  volume = {5},
  issn = {2521-327X},
  url = {http://dx.doi.org/10.22331/q-2021-11-08-573},
  doi = {10.22331/q-2021-11-08-573},
  publisher = {Verein zur Forderung des Open Access Publizierens in den Quantenwissenschaften},
  month = Nov,
  pages = {573},
}

@misc{Low2025NonunitaryDynamics,
  author = {Low, Guang Hao and Somma, Rolando D.},
  title = {Optimal quantum simulation of linear non-unitary dynamics},
  year = {2025},
  doi = {10.48550/ARXIV.2508.19238},
  url = {https://arxiv.org/abs/2508.19238},
  publisher = {arXiv},
  copyright = {arXiv.org perpetual, non-exclusive license},
}

@article{Berry2020TimeDependentL1,
  author = {Berry, Dominic W. and Childs, Andrew M. and Su, Yuan and Wang, Xin and Wiebe, Nathan},
  title = {Time-dependent Hamiltonian simulation with {$L^1$}-norm scaling},
  journal = {Quantum},
  year = {2020},
  volume = {4},
  issn = {2521-327X},
  url = {http://dx.doi.org/10.22331/q-2020-04-20-254},
  doi = {10.22331/q-2020-04-20-254},
  publisher = {Verein zur Forderung des Open Access Publizierens in den Quantenwissenschaften},
  month = Apr,
  pages = {254},
}

@article{Trotter1959Product,
  author = {H. F. Trotter},
  doi = {10.2307/2033649},
  issn = {00029939},
  issue = {4},
  journal = {Proceedings of the American Mathematical Society},
  month = {8},
  pages = {545},
  title = {On the Product of Semi-Groups of Operators},
  volume = {10},
  url = {https://www.jstor.org/stable/2033649?origin=crossref},
  year = {1959},
}

@article{Wiebe2010,
  author = {Nathan Wiebe and Dominic Berry and Peter Høyer and Barry C Sanders},
  doi = {10.1088/1751-8113/43/6/065203},
  issn = {1751-8113},
  issue = {6},
  journal = {Journal of Physics A: Mathematical and Theoretical},
  month = {1},
  pages = {065203},
  title = {Higher order decompositions of ordered operator exponentials},
  volume = {43},
  url = {http://stacks.iop.org/1751-8121/43/i=6/a=065203 http://stacks.iop.org/1751-8121/43/i=6/a=065203?key=crossref.683e2b14741b13185a0a8e98c29ea63d},
  year = {2010},
}

@article{Low2019Multiproduct,
  author = {Guang Hao Low and Vadym Kliuchnikov and Nathan Wiebe},
  journal = {arXiv:1907.11679},
  title = {Well-conditioned multiproduct Hamiltonian simulation},
  url = {https://arxiv.org/abs/1907.11679},
  year = {2019},
}

@article{childs2019theory,
  author = {Andrew M. Childs and Yuan Su and Minh C. Tran and Nathan Wiebe and Shuchen Zhu},
  doi = {10.1103/PhysRevX.11.011020},
  issn = {2160-3308},
  issue = {1},
  journal = {Physical Review X},
  month = {2},
  pages = {011020},
  title = {Theory of Trotter Error with Commutator Scaling},
  volume = {11},
  url = {https://link.aps.org/doi/10.1103/PhysRevX.11.011020},
  year = {2021},
}

@article{Haah2019Lattice,
  author = {Jeongwan Haah and Matthew B. Hastings and Robin Kothari and Guang Hao Low},
  doi = {10.1137/18M1231511},
  issn = {0097-5397},
  issue = {6},
  journal = {SIAM Journal on Computing},
  month = {12},
  pages = {FOCS18-250-FOCS18-284},
  title = {Quantum Algorithm for Simulating Real Time Evolution of Lattice Hamiltonians},
  volume = {52},
  year = {2023},
}

@article{Kitaev1997Diamond,
  author = {A Yu Kitaev},
  doi = {10.1070/RM1997v052n06ABEH002155},
  issn = {0036-0279},
  issue = {6},
  journal = {Russian Mathematical Surveys},
  month = {12},
  pages = {1191-1249},
  title = {Quantum computations: algorithms and error correction},
  volume = {52},
  url = {http://stacks.iop.org/0036-0279/52/i=6/a=R02?key=crossref.9a3dd3d906e4a9338822c6992ae26e8a},
  year = {1997},
}

@article{sho_polynomialtime_1997,
  author = {Peter W. Shor},
  doi = {10.1137/S0097539795293172},
  issn = {0097-5397},
  issue = {5},
  journal = {SIAM Journal on Computing},
  month = {10},
  pages = {1484-1509},
  title = {Polynomial-time algorithms for prime factorization and discrete logarithms on a quantum computer},
  volume = {26},
  url = {http://epubs.siam.org/doi/10.1137/S0097539795293172},
  year = {1997},
}

@article{Berry2024Topological,
  author = {Berry, Dominic W. and Su, Yuan and Gyurik, Casper and King, Robbie and Basso, Joao and Barba, Alexander Del Toro and Rajput, Abhishek and Wiebe, Nathan and Dunjko, Vedran and Babbush, Ryan},
  title = {Analyzing Prospects for Quantum Advantage in Topological Data Analysis},
  journal = {PRX Quantum},
  year = {2024},
  volume = {5},
  issn = {2691-3399},
  url = {http://dx.doi.org/10.1103/PRXQuantum.5.010319},
  doi = {10.1103/prxquantum.5.010319},
  number = {1},
  publisher = {American Physical Society (APS)},
  month = Feb,
}

@article{Bauer2020QuantumChemistry,
  author = {Bauer, Bela and Bravyi, Sergey and Motta, Mario and Chan, Garnet Kin-Lic},
  title = {Quantum Algorithms for Quantum Chemistry and Quantum Materials Science},
  journal = {Chemical Reviews},
  year = {2020},
  volume = {120},
  issn = {1520-6890},
  url = {http://dx.doi.org/10.1021/acs.chemrev.9b00829},
  doi = {10.1021/acs.chemrev.9b00829},
  number = {22},
  publisher = {American Chemical Society (ACS)},
  month = Oct,
  pages = {12685–12717},
}

@article{Reiher2016Reaction,
  author = {Markus Reiher and Nathan Wiebe and Krysta M Svore and Dave Wecker and Matthias Troyer},
  doi = {10.1073/pnas.1619152114},
  issn = {0027-8424},
  issue = {29},
  journal = {Proceedings of the National Academy of Sciences},
  month = {7},
  pages = {7555-7560},
  publisher = {National Acad Sciences},
  title = {Elucidating reaction mechanisms on quantum computers},
  volume = {114},
  url = {https://doi.org/10.1073/pnas.1619152114 http://www.pnas.org/lookup/doi/10.1073/pnas.1619152114 http://arxiv.org/abs/1605.03590 http://dx.doi.org/10.1073/pnas.1619152114},
  year = {2017},
}

@article{Dorfler2013KronReduction,
  author = {Dorfler, Florian and Bullo, Francesco},
  title = {Kron Reduction of Graphs With Applications to Electrical Networks},
  journal = {IEEE Transactions on Circuits and Systems I: Regular Papers},
  year = {2013},
  volume = {60},
  issn = {1558-0806},
  url = {http://dx.doi.org/10.1109/TCSI.2012.2215780},
  doi = {10.1109/tcsi.2012.2215780},
  number = {1},
  publisher = {Institute of Electrical and Electronics Engineers (IEEE)},
  month = Jan,
  pages = {150–163},
}

@incollection{Dusson2021SchurPerturbation,
  author = {Dusson, Geneviève and Sigal, Israel and Stamm, Benjamin},
  title = {The Feshbach-Schur map and perturbation theory},
  year = {2021},
  doi = {10.4171/ECR/18-1/5},
  url = {https://doi.org/10.4171/ECR/18-1/5},
  publisher = {EMS Press},
  copyright = {arXiv.org perpetual, non-exclusive license},
  booktitle = {Partial Differential Equations, Spectral Theory, and Mathematical Physics: The Ari Laptev Anniversary Volume},
  editor = {Pavel Exner and Rupert L. Frank and Fritz Gesztesy and Helge Holden and Timo Weidl},
  pages = {65--88},
}

@article{Bauman2019Downfolding,
  author = {Nicholas P. Bauman and Eric J. Bylaska and Sriram Krishnamoorthy and Guang Hao Low and Nathan Wiebe and Christopher E. Granade and Martin Roetteler and Matthias Troyer and Karol Kowalski},
  doi = {10.1063/1.5094643},
  issn = {0021-9606},
  issue = {1},
  journal = {The Journal of Chemical Physics},
  month = {7},
  pages = {014107},
  title = {Downfolding of many-body Hamiltonians using active-space models: Extension of the sub-system embedding sub-algebras approach to unitary coupled cluster formalisms},
  volume = {151},
  url = {http://aip.scitation.org/doi/10.1063/1.5094643},
  year = {2019},
}

@article{Becker2001ErrorEstimation,
  author = {Becker, Roland and Rannacher, Rolf},
  title = {An optimal control approach to {\emph{a posteriori}} error estimation in finite element methods},
  journal = {Acta Numerica},
  year = {2001},
  volume = {10},
  issn = {1474-0508},
  url = {http://dx.doi.org/10.1017/S0962492901000010},
  doi = {10.1017/s0962492901000010},
  publisher = {Cambridge University Press (CUP)},
  month = May,
  pages = {1–102},
}

@article{Shuman2013GraphSignals,
  author = {Shuman, D. I. and Narang, S. K. and Frossard, P. and Ortega, A. and Vandergheynst, P.},
  title = {The emerging field of signal processing on graphs: Extending high-dimensional data analysis to networks and other irregular domains},
  journal = {IEEE Signal Processing Magazine},
  year = {2013},
  volume = {30},
  issn = {1053-5888},
  url = {http://dx.doi.org/10.1109/MSP.2012.2235192},
  doi = {10.1109/msp.2012.2235192},
  number = {3},
  publisher = {Institute of Electrical and Electronics Engineers (IEEE)},
  month = May,
  pages = {83–98},
}

@inproceedings{Zhou2003GraphLearning,
  author = {Zhou, Dengyong and Bousquet, Olivier and Lal, Thomas Navin and Weston, Jason and Sch{\"o}lkopf, Bernhard},
  title = {Learning with local and global consistency},
  booktitle = {Proceedings of the 17th International Conference on Neural Information Processing Systems},
  year = {2003},
  publisher = {MIT Press},
  address = {Cambridge, MA, USA},
  pages = {321–328},
  numpages = {8},
  location = {Whistler, British Columbia, Canada},
  series = {NIPS'03},
}

@article{Kalman1960LinearFiltering,
  author = {Kalman, R. E.},
  title = {A New Approach to Linear Filtering and Prediction Problems},
  journal = {Journal of Basic Engineering},
  year = {1960},
  volume = {82},
  issn = {0021-9223},
  url = {http://dx.doi.org/10.1115/1.3662552},
  doi = {10.1115/1.3662552},
  number = {1},
  publisher = {ASME International},
  month = Mar,
  pages = {35–45},
}

@misc{Shi2024QuantumKalman,
  author = {Shi, Hao and Zhang, Guofeng and Zhang, Ming},
  title = {A quantum algorithm for the Kalman filter using block encoding},
  year = {2024},
  doi = {10.48550/ARXIV.2404.04554},
  url = {https://arxiv.org/abs/2404.04554},
  publisher = {arXiv},
  copyright = {arXiv.org perpetual, non-exclusive license},
}

@inproceedings{Barnum2007QuerySemidefiniteProgramming,
  author = {Barnum, Howard},
  title = {{Semidefinite programming characterization and spectral adversary method for quantum complexity with noncommuting unitary queries}},
  booktitle = {Algorithms and Complexity for Continuous Problems},
  year = {2007},
  pages = {1--25},
  series = {Dagstuhl Seminar Proceedings (DagSemProc)},
  issn = {1862-4405},
  volume = {6391},
  editor = {Stephan Dahlke and Klaus Ritter and Ian H. Sloan and Joseph F. Traub},
  publisher = {Schloss Dagstuhl -- Leibniz-Zentrum f{\"u}r Informatik},
  address = {Dagstuhl, Germany},
  url = {https://drops.dagstuhl.de/entities/document/10.4230/DagSemProc.06391.3},
  urn = {urn:nbn:de:0030-drops-8769},
  doi = {10.4230/DagSemProc.06391.3},
}

@article{Hoggar1974LogarithmicConcavity,
  author = {Hoggar, S.G},
  title = {Chromatic polynomials and logarithmic concavity},
  journal = {Journal of Combinatorial Theory, Series B},
  year = {1974},
  volume = {16},
  issn = {0095-8956},
  url = {http://dx.doi.org/10.1016/0095-8956(74)90071-9},
  doi = {10.1016/0095-8956(74)90071-9},
  number = {3},
  publisher = {Elsevier BV},
  month = June,
  pages = {248–254},
}

@incollection{brassard2002quantum,
  author = {Gilles Brassard and Peter H{\o}yer and Michele Mosca and Alain Tapp},
  pages = {53--74},
  title = {Quantum amplitude amplification and estimation},
  volume = {305},
  url = {https://arxiv.org/abs/quant-ph/0005055},
  year = {2002},
  booktitle = {Quantum Computation and Information},
  series = {Contemporary Mathematics},
  publisher = {American Mathematical Society},
  doi = {10.1090/conm/305/05215},
}

@book{Simon2005OrthogonalPolynomials,
  author = {Simon, Barry},
  title = {Orthogonal Polynomials on the Unit Circle.},
  publisher = {American Mathematical Society},
  year = {2005},
  series = {American Mathematical Society Colloquium Publications},
  volume = {54},
  address = {Providence, R.I.},
  isbn = {978-0-8218-3446-6},
  pages = {1044},
}

@article{Simanek2018GeronimusWeights,
  author = {Simanek, Brian},
  title = {Universality at an endpoint for orthogonal polynomials with Geronimus-type weights},
  journal = {Proceedings of the American Mathematical Society},
  year = {2018},
  volume = {146},
  issn = {0002-9939},
  url = {http://dx.doi.org/10.1090/proc/14085},
  doi = {10.1090/proc/14085},
  number = {9},
  publisher = {American Mathematical Society (AMS)},
  month = Apr,
  pages = {3995–4007},
}

@article{Laneve2026QSPAdversary,
  eprint = {2506.20484},
  archivePrefix = {arXiv},
  author = {Laneve, Lorenzo},
  title = {An adversary bound for quantum signal processing},
  journal = {Quantum},
  year = {2026},
  volume = {10},
  issn = {2521-327X},
  url = {http://dx.doi.org/10.22331/q-2026-03-13-2025},
  doi = {10.22331/q-2026-03-13-2025},
  publisher = {Verein zur Forderung des Open Access Publizierens in den Quantenwissenschaften},
  month = Mar,
  pages = {2025},
}

@article{Campbell1922WaveFilter,
  author = {Campbell, George A.},
  title = {Physical Theory of the Electric Wave-Filter},
  journal = {Bell System Technical Journal},
  year = {1922},
  volume = {1},
  issn = {0005-8580},
  url = {http://dx.doi.org/10.1002/j.1538-7305.1922.tb00386.x},
  doi = {10.1002/j.1538-7305.1922.tb00386.x},
  number = {2},
  publisher = {Institute of Electrical and Electronics Engineers (IEEE)},
  month = Nov,
  pages = {1–32},
}

@article{Butterworth1930FilterAmplifiers,
  author = {Butterworth, Stephen},
  title = {On the Theory of Filter Amplifiers},
  journal = {Experimental Wireless \& the Wireless Engineer},
  volume = {7},
  pages = {536--541},
  year = {1930},
  month = oct,
  url = {https://www.changpuak.ch/electronics/downloads/On_the_Theory_of_Filter_Amplifiers.pdf},
}

@article{Black1934FeedbackAmplifiers,
  author = {Black, H. S.},
  title = {Stabilized Feedback Amplifiers*},
  journal = {Bell System Technical Journal},
  year = {1934},
  volume = {13},
  issn = {0005-8580},
  url = {http://dx.doi.org/10.1002/j.1538-7305.1934.tb00652.x},
  doi = {10.1002/j.1538-7305.1934.tb00652.x},
  number = {1},
  publisher = {Institute of Electrical and Electronics Engineers (IEEE)},
  month = Jan,
  pages = {1–18},
}

@article{Mason1953SignalFlow,
  author = {Mason, Samuel},
  title = {Feedback Theory-Some Properties of Signal Flow Graphs},
  journal = {Proceedings of the IRE},
  year = {1953},
  volume = {41},
  issn = {0096-8390},
  url = {http://dx.doi.org/10.1109/JRPROC.1953.274449},
  doi = {10.1109/jrproc.1953.274449},
  number = {9},
  publisher = {Institute of Electrical and Electronics Engineers (IEEE)},
  month = Sept,
  pages = {1144–1156},
}

@article{Vaidyanathan1985Lossless,
  author = {Vaidyanathan, P.},
  title = {The discrete-time bounded-real lemma in digital filtering},
  journal = {IEEE Transactions on Circuits and Systems},
  year = {1985},
  volume = {32},
  issn = {0098-4094},
  url = {http://dx.doi.org/10.1109/TCS.1985.1085815},
  doi = {10.1109/tcs.1985.1085815},
  number = {9},
  publisher = {Institute of Electrical and Electronics Engineers (IEEE)},
  month = Sept,
  pages = {918–924},
}

@article{Bartha2014QuantumTuring,
  author = {Bartha, Miklós},
  title = {Quantum Turing automata},
  journal = {Electronic Proceedings in Theoretical Computer Science},
  year = {2014},
  volume = {143},
  issn = {2075-2180},
  url = {http://dx.doi.org/10.4204/EPTCS.143.2},
  doi = {10.4204/eptcs.143.2},
  publisher = {Open Publishing Association},
  month = Mar,
  pages = {17–31},
}

@phdthesis{AndresMartinez2022Loops,
  author = {Andr{\'e}s-Mart{\'i}nez, Pablo},
  title = {Unbounded Loops in Quantum Programs: Categories and Weak While Loops},
  school = {University of Edinburgh},
  year = {2022},
  doi = {10.48550/arXiv.2212.05371},
  eprint = {2212.05371},
  archiveprefix = {arXiv},
  primaryclass = {quant-ph},
  url = {https://arxiv.org/abs/2212.05371},
}

@inproceedings{ClementPerdrix2020PBS,
  author = {Clément, Alexandre and Perdrix, Simon},
  title = {PBS-Calculus: A Graphical Language for Coherent Control of Quantum Computations},
  booktitle = {45th International Symposium on Mathematical Foundations of Computer Science (MFCS 2020)},
  year = {2020},
  doi = {10.4230/LIPICS.MFCS.2020.24},
  url = {https://drops.dagstuhl.de/entities/document/10.4230/LIPIcs.MFCS.2020.24},
  language = {en},
  journal = {LIPIcs, Volume 170, MFCS 2020},
  volume = {170},
  pages = {24:1-24:14},
  publisher = {Schloss Dagstuhl – Leibniz-Zentrum für Informatik},
  copyright = {Creative Commons Attribution 3.0 Unported license},
}

@inproceedings{Carette2021DelayedTrace,
  author = {Carette, Titouan and de Visme, Marc and Perdrix, Simon},
  title = {Graphical Language with Delayed Trace: Picturing Quantum Computing with Finite Memory},
  booktitle = {2021 36th Annual ACM/IEEE Symposium on Logic in Computer Science (LICS)},
  year = {2021},
  url = {http://dx.doi.org/10.1109/LICS52264.2021.9470553},
  doi = {10.1109/lics52264.2021.9470553},
  publisher = {IEEE},
  month = June,
  pages = {1–13},
}

@article{Gough2008QuantumFeedback,
  author = {Gough, J. E. and Gohm, R. and Yanagisawa, M.},
  title = {Linear quantum feedback networks},
  journal = {Physical Review A},
  year = {2008},
  volume = {78},
  issn = {1094-1622},
  url = {http://dx.doi.org/10.1103/PhysRevA.78.062104},
  doi = {10.1103/physreva.78.062104},
  number = {6},
  publisher = {American Physical Society (APS)},
  month = Dec,
}

@incollection{Agler1990Polydics,
  author = {Agler, Jim},
  title = {On the Representation of Certain Holomorphic Functions Defined on a Polydisc},
  booktitle = {Topics in Operator Theory: {Ernst D. Hellinger} Memorial Volume},
  publisher = {Birkh{\"a}user},
  year = {1990},
  editor = {de Branges, Louis and Gohberg, Israel and Rovnyak, James},
  series = {Operator Theory: Advances and Applications},
  volume = {48},
  pages = {47--66},
  address = {Basel},
  isbn = {978-3-7643-2532-9},
}

@article{Knese2021Bidisk,
  author = {Knese, Greg},
  title = {Kummert’s approach to realization on the bidisk},
  journal = {Indiana University Mathematics Journal},
  year = {2021},
  volume = {70},
  issn = {0022-2518},
  url = {http://dx.doi.org/10.1512/iumj.2021.70.8738},
  doi = {10.1512/iumj.2021.70.8738},
  number = {6},
  publisher = {Indiana University Mathematics Journal},
  pages = {2369–2403},
}

@misc{Jeffery2022SubroutineComposition,
  author = {Jeffery, Stacey},
  title = {Quantum Subroutine Composition},
  year = {2022},
  eprint = {2209.14146},
  archiveprefix = {arXiv},
  primaryclass = {quant-ph},
  doi = {10.48550/arXiv.2209.14146},
  url = {https://arxiv.org/abs/2209.14146},
  note = {arXiv:2209.14146},
}

@article{Redheffer1962ScatteringTransfer,
  author = {Redheffer, Raymond},
  title = {On the Relation of Transmission‐Line Theory to Scattering and Transfer},
  journal = {Journal of Mathematics and Physics},
  year = {1962},
  volume = {41},
  issn = {0097-1421},
  url = {http://dx.doi.org/10.1002/sapm19624111},
  doi = {10.1002/sapm19624111},
  number = {1-4},
  publisher = {Wiley},
  month = Apr,
  pages = {1–41},
}

@misc{Dalzell2026BeyondConditionNumber,
  author = {Dalzell, Alexander M. and Li, Jianqiang and Su, Yuan},
  title = {Faster quantum linear system solver beyond the condition number},
  year = {2026},
  doi = {10.48550/ARXIV.2607.07691},
  url = {https://arxiv.org/abs/2607.07691},
  publisher = {arXiv},
  copyright = {Creative Commons Attribution Share Alike 4.0 International},
}

@misc{Li2025BeyondConditionNumber,
  author = {Li, Jianqiang},
  title = {A New Quantum Linear System Algorithm Beyond the Condition Number and Its Application to Solving Multivariate Polynomial Systems},
  year = {2025},
  doi = {10.48550/ARXIV.2510.05588},
  url = {https://arxiv.org/abs/2510.05588},
  publisher = {arXiv},
  copyright = {arXiv.org perpetual, non-exclusive license},
}

@misc{Yang2025Dissipative,
  author = {Yang, Gengzhi and Onwunta, Akwum and An, Dong},
  title = {Quantum Differential Equation Solvers with Low State Preparation Cost: Eliminating the Time Dependence in Dissipative Equations},
  year = {2025},
  doi = {10.48550/ARXIV.2508.15170},
  url = {https://arxiv.org/abs/2508.15170},
  publisher = {arXiv},
  copyright = {arXiv.org perpetual, non-exclusive license},
}

@inproceedings{Childs2003Exponential,
  author = {Andrew M. Childs and Richard Cleve and Enrico Deotto and Edward Farhi and Sam Gutmann and Daniel A. Spielman},
  city = {New York, New York, USA},
  doi = {10.1145/780542.780552},
  isbn = {1581136749},
  booktitle = {Proceedings of the thirty-fifth ACM symposium on Theory of computing - STOC '03},
  pages = {59},
  publisher = {ACM Press},
  title = {Exponential algorithmic speedup by a quantum walk},
  url = {http://doi.acm.org/10.1145/780542.780552 http://portal.acm.org/citation.cfm?doid=780542.780552},
  year = {2003},
}

@article{Childs2009Universal,
  author = {Childs, Andrew M.},
  title = {Universal Computation by Quantum Walk},
  journal = {Physical Review Letters},
  year = {2009},
  volume = {102},
  issn = {1079-7114},
  url = {http://dx.doi.org/10.1103/PhysRevLett.102.180501},
  doi = {10.1103/physrevlett.102.180501},
  number = {18},
  publisher = {American Physical Society (APS)},
  month = May,
}

@inbook{Toffoli1980Reversible,
  author = {Toffoli, Tommaso},
  title = {Reversible computing},
  booktitle = {Automata, Languages and Programming},
  pages = {632–644},
  publisher = {Springer Berlin Heidelberg},
  year = {1980},
  isbn = {9783540393467},
  issn = {1611-3349},
  url = {http://dx.doi.org/10.1007/3-540-10003-2_104},
  doi = {10.1007/3-540-10003-2_104},
}

@article{Babbush2023ClassicalOscillators,
  author = {Babbush, Ryan and Berry, Dominic W. and Kothari, Robin and Somma, Rolando D. and Wiebe, Nathan},
  title = {Exponential Quantum Speedup in Simulating Coupled Classical Oscillators},
  journal = {Physical Review X},
  year = {2023},
  volume = {13},
  issn = {2160-3308},
  url = {http://dx.doi.org/10.1103/PhysRevX.13.041041},
  doi = {10.1103/physrevx.13.041041},
  number = {4},
  publisher = {American Physical Society (APS)},
  month = Dec,
}

@article{Somma2025ShadowSimulation,
  author = {Somma, Rolando D. and King, Robbie and Kothari, Robin and O’Brien, Thomas E. and Babbush, Ryan},
  title = {Shadow hamiltonian simulation},
  journal = {Nature Communications},
  year = {2025},
  volume = {16},
  issn = {2041-1723},
  url = {http://dx.doi.org/10.1038/s41467-025-57451-z},
  doi = {10.1038/s41467-025-57451-z},
  number = {1},
  publisher = {Springer Science and Business Media LLC},
  month = Mar,
}

@article{Mande2026QPEBounds,
  author = {Mande, Nikhil S. and Wolf, Ronald de},
  title = {Tight Bounds for Quantum Phase Estimation and Related Problems},
  journal = {Quantum},
  year = {2026},
  volume = {10},
  issn = {2521-327X},
  url = {http://dx.doi.org/10.22331/q-2026-06-15-2140},
  doi = {10.22331/q-2026-06-15-2140},
  publisher = {Verein zur Forderung des Open Access Publizierens in den Quantenwissenschaften},
  month = June,
  pages = {2140},
}

@article{Ge2019GroundStatePreparation,
  author = {Ge, Yimin and Tura, Jordi and Cirac, J. Ignacio},
  title = {Faster ground state preparation and high-precision ground energy estimation with fewer qubits},
  journal = {Journal of Mathematical Physics},
  year = {2019},
  volume = {60},
  issn = {1089-7658},
  url = {http://dx.doi.org/10.1063/1.5027484},
  doi = {10.1063/1.5027484},
  number = {2},
  publisher = {AIP Publishing},
  month = Feb,
}

@misc{Somma2026GroundLowerBounds,
  author = {Somma, Rolando D. and de Wolf, Ronald},
  title = {Optimal Lower Bound for Ground-State Energy Estimation with a Guiding State},
  year = {2026},
  doi = {10.48550/ARXIV.2608.24493},
  url = {https://arxiv.org/abs/2608.24493},
  publisher = {arXiv},
  copyright = {Creative Commons Attribution 4.0 International},
}

@article{Ando1963Contraction,
  author = {And{\^o}, Tadao},
  title = {On a Pair of Commutative Contractions},
  journal = {Acta Scientiarum Mathematicarum (Szeged)},
  year = {1963},
  volume = {24},
  number = {1--2},
  pages = {88--90},
  url = {http://acta.bibl.u-szeged.hu/13987/1/math_024_fasc_001_002_088-090.pdf},
}

@article{Abbott2024IndefiniteQuery,
  author = {Abbott, Alastair A. and Mhalla, Mehdi and Pocreau, Pierre},
  title = {Quantum query complexity of Boolean functions under indefinite causal order},
  journal = {Physical Review Research},
  year = {2024},
  volume = {6},
  issn = {2643-1564},
  url = {http://dx.doi.org/10.1103/PhysRevResearch.6.L032020},
  doi = {10.1103/physrevresearch.6.l032020},
  number = {3},
  publisher = {American Physical Society (APS)},
  month = July,
}

@article{RennerBrukner2022QubitOrders,
  author = {Renner, Martin J. and Brukner, Časlav},
  title = {Computational Advantage from a Quantum Superposition of Qubit Gate Orders},
  journal = {Physical Review Letters},
  year = {2022},
  volume = {128},
  issn = {1079-7114},
  url = {http://dx.doi.org/10.1103/PhysRevLett.128.230503},
  doi = {10.1103/physrevlett.128.230503},
  number = {23},
  publisher = {American Physical Society (APS)},
  month = June,
}

@article{Wechs2021CausalOrder,
  author = {Wechs, Julian and Dourdent, Hippolyte and Abbott, Alastair A. and Branciard, Cyril},
  title = {Quantum Circuits with Classical Versus Quantum Control of Causal Order},
  journal = {PRX Quantum},
  year = {2021},
  volume = {2},
  issn = {2691-3399},
  url = {http://dx.doi.org/10.1103/PRXQuantum.2.030335},
  doi = {10.1103/prxquantum.2.030335},
  number = {3},
  publisher = {American Physical Society (APS)},
  month = Aug,
}

@article{Araujo2014ControlledOrder,
  author = {Araújo, Mateus and Costa, Fabio and Brukner, Časlav},
  title = {Computational Advantage from Quantum-Controlled Ordering of Gates},
  journal = {Physical Review Letters},
  year = {2014},
  volume = {113},
  issn = {1079-7114},
  url = {http://dx.doi.org/10.1103/PhysRevLett.113.250402},
  doi = {10.1103/physrevlett.113.250402},
  number = {25},
  publisher = {American Physical Society (APS)},
  month = Dec,
}

@article{Chiribella2013QuantumSwitch,
  author = {Chiribella, Giulio and D’Ariano, Giacomo Mauro and Perinotti, Paolo and Valiron, Benoit},
  title = {Quantum computations without definite causal structure},
  journal = {Physical Review A},
  year = {2013},
  volume = {88},
  issn = {1094-1622},
  url = {http://dx.doi.org/10.1103/PhysRevA.88.022318},
  doi = {10.1103/physreva.88.022318},
  number = {2},
  publisher = {American Physical Society (APS)},
  month = Aug,
}

@article{Knese2010RationalInner,
  author = {Knese, G.},
  title = {Rational inner functions in the Schur-Agler class of the polydisk},
  journal = {Publicacions Matemàtiques},
  year = {2011},
  volume = {55},
  issn = {0214-1493},
  url = {http://dx.doi.org/10.5565/PUBLMAT_55211_04},
  doi = {10.5565/publmat_55211_04},
  publisher = {Universitat Autonoma de Barcelona},
  month = July,
  pages = {343–357},
}

@article{HillarNie2008Hilbert17Matrices,
  author = {Hillar, Christopher and Nie, Jiawang},
  title = {An elementary and constructive solution to Hilbert’s 17th Problem for matrices},
  journal = {Proceedings of the American Mathematical Society},
  year = {2007},
  volume = {136},
  issn = {0002-9939},
  url = {http://dx.doi.org/10.1090/S0002-9939-07-09068-5},
  doi = {10.1090/s0002-9939-07-09068-5},
  number = {1},
  publisher = {American Mathematical Society (AMS)},
  month = Oct,
  pages = {73–76},
}

@article{Artin1927Hilbert17,
  author = {Artin, Emil},
  title = {Über die Zerlegung definiter Funktionen in Quadrate},
  journal = {Abhandlungen aus dem Mathematischen Seminar der Universität Hamburg},
  year = {1927},
  volume = {5},
  issn = {1865-8784},
  url = {http://dx.doi.org/10.1007/BF02952513},
  doi = {10.1007/bf02952513},
  number = {1},
  publisher = {Springer Science and Business Media LLC},
  month = Dec,
  pages = {100–115},
}

@article{Dritschel2025TwoVariable,
  author = {Dritschel, Michael A.},
  title = {Factoring non-negative operator valued trigonometric polynomials in two variables},
  journal = {Mathematische Annalen},
  year = {2024},
  volume = {391},
  issn = {1432-1807},
  url = {http://dx.doi.org/10.1007/s00208-024-02895-9},
  doi = {10.1007/s00208-024-02895-9},
  number = {1},
  publisher = {Springer Science and Business Media LLC},
  month = June,
  pages = {515–537},
}

@article{Dritschel2004Factorization,
  author = {Dritschel, Michael A.},
  title = {On Factorization of Trigonometric Polynomials},
  journal = {Integral Equations and Operator Theory},
  year = {2004},
  volume = {49},
  issn = {0378-620X},
  url = {http://dx.doi.org/10.1007/s00020-002-1198-4},
  doi = {10.1007/s00020-002-1198-4},
  number = {1},
  publisher = {Springer Science and Business Media LLC},
  month = May,
  pages = {11–42},
}

@article{Shao2020GeneralizedEigenvalue,
  author = {Shao, Changpeng and Liu, Jin-Peng},
  title = {Solving generalized eigenvalue problems by ordinary differential equations on a quantum computer},
  year = {2022},
  issn = {1471-2946},
  url = {https://doi.org/10.1098/rspa.2021.0797},
  doi = {10.1098/rspa.2021.0797},
  number = {2262},
  journal = {Proceedings of the Royal Society A},
  volume = {478},
  pages = {20210797},
}

@article{Tkatchenko2012Dispersion,
  author = {Tkatchenko, Alexandre and DiStasio, Robert A. and Car, Roberto and Scheffler, Matthias},
  title = {Accurate and Efficient Method for Many-Body van der Waals Interactions},
  journal = {Physical Review Letters},
  year = {2012},
  volume = {108},
  issn = {1079-7114},
  url = {http://dx.doi.org/10.1103/PhysRevLett.108.236402},
  doi = {10.1103/physrevlett.108.236402},
  number = {23},
  publisher = {American Physical Society (APS)},
  month = June,
}

@article{Su2021FirstQuantization,
  author = {Su, Yuan and Berry, Dominic W. and Wiebe, Nathan and Rubin, Nicholas and Babbush, Ryan},
  title = {Fault-Tolerant Quantum Simulations of Chemistry in First Quantization},
  journal = {PRX Quantum},
  year = {2021},
  volume = {2},
  issn = {2691-3399},
  url = {http://dx.doi.org/10.1103/PRXQuantum.2.040332},
  doi = {10.1103/prxquantum.2.040332},
  number = {4},
  publisher = {American Physical Society (APS)},
  month = Nov,
}

@article{Simon2006CMVMatrices,
  author = {Simon, Barry},
  title = {CMV matrices: Five years after},
  year = {2007},
  doi = {10.1016/j.cam.2006.10.033},
  url = {https://doi.org/10.1016/j.cam.2006.10.033},
  copyright = {Assumed arXiv.org perpetual, non-exclusive license to distribute this article for submissions made before January 2004},
  journal = {Journal of Computational and Applied Mathematics},
  volume = {208},
  pages = {120--154},
  number = {1},
}

@article{Schur1917SchurAlgorithm,
  author = {Schur, J.},
  title = {Über Potenzreihen, die im Innern des Einheitskreises beschränkt sind.},
  journal = {crll},
  year = {1917},
  volume = {1917},
  issn = {1435-5345},
  url = {http://dx.doi.org/10.1515/crll.1917.147.205},
  doi = {10.1515/crll.1917.147.205},
  number = {147},
  publisher = {Walter de Gruyter GmbH},
  pages = {205–232},
}

@article{Schon2005MPS,
  author = {C. Schön and E. Solano and F. Verstraete and J. I. Cirac and M. M. Wolf},
  doi = {10.1103/PhysRevLett.95.110503},
  issn = {0031-9007},
  issue = {11},
  journal = {Physical Review Letters},
  month = {9},
  pages = {110503},
  title = {Sequential Generation of Entangled Multiqubit States},
  volume = {95},
  year = {2005},
}

@misc{Scemama2013SlaterCondon,
  author = {Scemama, Anthony and Giner, Emmanuel},
  title = {An efficient implementation of Slater-Condon rules},
  year = {2013},
  doi = {10.48550/ARXIV.1311.6244},
  url = {https://arxiv.org/abs/1311.6244},
  publisher = {arXiv},
  copyright = {arXiv.org perpetual, non-exclusive license},
}

@article{Ross2016Optimal,
  author = {Neil J Ross and Peter Selinger},
  city = {Paramus, NJ},
  issn = {1533-7146},
  issue = {11-12},
  journal = {Quantum Information \& Computation},
  month = {9},
  pages = {901-953},
  publisher = {Rinton Press, Incorporated},
  title = {Optimal Ancilla-free Clifford+T Approximation of Z-rotations},
  volume = {16},
  url = {http://dl.acm.org/citation.cfm?id=3179330.3179331},
  year = {2016},
}

@article{Recht2010NuclearNorm,
  author = {Recht, Benjamin and Fazel, Maryam and Parrilo, Pablo A.},
  title = {Guaranteed Minimum-Rank Solutions of Linear Matrix Equations via Nuclear Norm Minimization},
  journal = {SIAM Review},
  year = {2010},
  volume = {52},
  issn = {1095-7200},
  url = {http://dx.doi.org/10.1137/070697835},
  doi = {10.1137/070697835},
  number = {3},
  publisher = {Society for Industrial & Applied Mathematics (SIAM)},
  month = Jan,
  pages = {471–501},
}

@article{Prodan2005Nearsightedness,
  author = {Prodan, E. and Kohn, W.},
  title = {Nearsightedness of electronic matter},
  journal = {Proceedings of the National Academy of Sciences},
  year = {2005},
  volume = {102},
  issn = {1091-6490},
  url = {http://dx.doi.org/10.1073/pnas.0505436102},
  doi = {10.1073/pnas.0505436102},
  number = {33},
  publisher = {National Academy of Sciences},
  month = Aug,
  pages = {11635–11638},
}

@article{Olsen2010PolarizableEmbedding,
  author = {Olsen, Jógvan Magnus and Aidas, Kęstutis and Kongsted, Jacob},
  title = {Excited States in Solution through Polarizable Embedding},
  journal = {Journal of Chemical Theory and Computation},
  year = {2010},
  volume = {6},
  issn = {1549-9626},
  url = {http://dx.doi.org/10.1021/ct1003803},
  doi = {10.1021/ct1003803},
  number = {12},
  publisher = {American Chemical Society (ACS)},
  month = Oct,
  pages = {3721–3734},
}

@article{Mottonen2005StatePreparation,
  author = {Mottonen, M. and Vartiainen, J. J. and Bergholm, V. and Salomaa, M. M.},
  title = {Transformation of quantum states using uniformly controlled rotations},
  journal = {Quantum Information and Computation},
  year = {2005},
  volume = {5},
  issn = {1533-7146},
  url = {http://dx.doi.org/10.26421/QIC5.6-5},
  doi = {10.26421/qic5.6-5},
  number = {6},
  publisher = {Rinton Press},
  month = Sept,
  pages = {467–473},
}

@article{Lowdin1951PerturbationTheory,
  author = {Löwdin, Per-Olov},
  title = {A Note on the Quantum-Mechanical Perturbation Theory},
  journal = {The Journal of Chemical Physics},
  year = {1951},
  volume = {19},
  issn = {1089-7690},
  url = {http://dx.doi.org/10.1063/1.1748067},
  doi = {10.1063/1.1748067},
  number = {11},
  publisher = {AIP Publishing},
  month = Nov,
  pages = {1396–1401},
}

@misc{Li2026MultilevelCorrection,
  author = {Li, Xiantao},
  title = {Toward Efficient End-to-End Quantum Elliptic PDE Solvers: a Multilevel Correction Algorithm for Direct Observable Estimation},
  year = {2026},
  doi = {10.48550/ARXIV.2606.01270},
  url = {https://arxiv.org/abs/2606.01270},
  publisher = {arXiv},
  copyright = {arXiv.org perpetual, non-exclusive license},
}

@article{Larsson2022LargeSites,
  author = {Larsson, Henrik R. and Zhai, Huanchen and Gunst, Klaas and Chan, Garnet Kin-Lic},
  title = {Matrix Product States with Large Sites},
  journal = {Journal of Chemical Theory and Computation},
  year = {2022},
  volume = {18},
  issn = {1549-9626},
  url = {http://dx.doi.org/10.1021/acs.jctc.1c00957},
  doi = {10.1021/acs.jctc.1c00957},
  number = {2},
  publisher = {American Chemical Society (ACS)},
  month = Jan,
  pages = {749–762},
}

@article{Kirby2021CompactEncoding,
  author = {Kirby, William M. and Hadi, Sultana and Kreshchuk, Michael and Love, Peter J.},
  title = {Quantum simulation of second-quantized Hamiltonians in compact encoding},
  journal = {Physical Review A},
  year = {2021},
  volume = {104},
  issn = {2469-9934},
  url = {http://dx.doi.org/10.1103/PhysRevA.104.042607},
  doi = {10.1103/physreva.104.042607},
  number = {4},
  publisher = {American Physical Society (APS)},
  month = Oct,
}

@article{Khattar2025ConditionallyCleanAncillae,
  author = {Khattar, Tanuj and Gidney, Craig},
  title = {Rise of conditionally clean ancillae for efficient quantum circuit constructions},
  journal = {Quantum},
  year = {2025},
  volume = {9},
  issn = {2521-327X},
  url = {http://dx.doi.org/10.22331/q-2025-05-21-1752},
  doi = {10.22331/q-2025-05-21-1752},
  publisher = {Verein zur Forderung des Open Access Publizierens in den Quantenwissenschaften},
  month = May,
  pages = {1752},
}

@article{Jury2019WeakFactorization,
  author = {Jury, Michael T. and Martin, Robert T.W.},
  title = {Factorization in weak products of complete Pick spaces},
  journal = {Bulletin of the London Mathematical Society},
  year = {2018},
  volume = {51},
  issn = {1469-2120},
  url = {http://dx.doi.org/10.1112/blms.12222},
  doi = {10.1112/blms.12222},
  number = {2},
  publisher = {Wiley},
  month = Nov,
  pages = {223–229},
}

@article{Higham2005MatrixExponential,
  author = {Higham, Nicholas J.},
  title = {The Scaling and Squaring Method for the Matrix Exponential Revisited},
  journal = {SIAM Journal on Matrix Analysis and Applications},
  year = {2005},
  volume = {26},
  issn = {1095-7162},
  url = {http://dx.doi.org/10.1137/04061101X},
  doi = {10.1137/04061101x},
  number = {4},
  publisher = {Society for Industrial & Applied Mathematics (SIAM)},
  month = Jan,
  pages = {1179–1193},
}

@misc{Hastings2024PerturbationTheory,
  author = {Hastings, M. B.},
  title = {Improving Perturbation Theory with the Sum-of-Squares: Third Order},
  year = {2024},
  doi = {10.48550/ARXIV.2412.03564},
  url = {https://arxiv.org/abs/2412.03564},
  publisher = {arXiv},
  copyright = {Creative Commons Attribution 4.0 International},
}

@article{Haner2016factoring,
  author = {Thomas Häner and Martin Roetteler and Krysta M Svore},
  city = {Paramus, NJ},
  issn = {1533-7146},
  issue = {7–8},
  journal = {Quantum Info. Comput.},
  month = {6},
  pages = {673-684},
  publisher = {Rinton Press, Incorporated},
  title = {Factoring Using 2n + 2 Qubits with Toffoli Based Modular Multiplication},
  volume = {17},
  year = {2017},
}

@article{Granade2012HamiltonianLearning,
  author = {Granade, Christopher E and Ferrie, Christopher and Wiebe, Nathan and Cory, D G},
  title = {Robust online Hamiltonian learning},
  journal = {New Journal of Physics},
  year = {2012},
  volume = {14},
  issn = {1367-2630},
  url = {http://dx.doi.org/10.1088/1367-2630/14/10/103013},
  doi = {10.1088/1367-2630/14/10/103013},
  number = {10},
  publisher = {IOP Publishing},
  month = Oct,
  pages = {103013},
}

@article{Andersson1992ActiveSpacePerturbation,
  author = {Andersson, Kerstin and Malmqvist, Per-Åke and Roos, Björn O.},
  title = {Second-order perturbation theory with a complete active space self-consistent field reference function},
  journal = {The Journal of Chemical Physics},
  year = {1992},
  volume = {96},
  issn = {1089-7690},
  url = {http://dx.doi.org/10.1063/1.462209},
  doi = {10.1063/1.462209},
  number = {2},
  publisher = {AIP Publishing},
  month = Jan,
  pages = {1218–1226},
}

@article{Babbush2016Fermions,
  author = {Ryan Babbush and Dominic W. Berry and Yuval R Sanders and Ian D. Kivlichan and Artur Scherer and Annie Y. Wei and Peter J. Love and Alán Aspuru-Guzik},
  doi = {10.1088/2058-9565/aa9463},
  issn = {2058-9565},
  issue = {1},
  journal = {Quantum Science and Technology},
  month = {1},
  pages = {015006},
  title = {Exponentially more precise quantum simulation of fermions in the configuration interaction representation},
  volume = {3},
  url = {http://arxiv.org/abs/1506.01029 http://stacks.iop.org/2058-9565/3/i=1/a=015006?key=crossref.1490393ec29d3fc760dc290d30016315},
  year = {2018},
}

@article{Bartlett2007coupledcluster,
  author = {Rodney J. Bartlett and Monika Musiał},
  doi = {10.1103/RevModPhys.79.291},
  issn = {0034-6861},
  issue = {1},
  journal = {Reviews of Modern Physics},
  month = {2},
  pages = {291-352},
  publisher = {American Physical Society},
  title = {Coupled-cluster theory in quantum chemistry},
  volume = {79},
  url = {https://link.aps.org/doi/10.1103/RevModPhys.79.291 http://link.aps.org/doi/10.1103/RevModPhys.79.291},
  year = {2007},
}

@article{Belovs2026SpaceEfficientPurification,
  author = {Belovs, Aleksandrs and Jeffery, Stacey},
  title = {Space-Efficient Quantum Error Reduction without log Factors},
  journal = {Quantum},
  year = {2026},
  volume = {10},
  issn = {2521-327X},
  url = {http://dx.doi.org/10.22331/q-2026-03-23-2039},
  doi = {10.22331/q-2026-03-23-2039},
  publisher = {Verein zur Forderung des Open Access Publizierens in den Quantenwissenschaften},
  month = Mar,
  pages = {2039},
}

@article{Bergholm2005ControlledGates,
  author = {Bergholm, Ville and Vartiainen, Juha J. and Möttönen, Mikko and Salomaa, Martti M.},
  title = {Quantum circuits with uniformly controlled one-qubit gates},
  journal = {Physical Review A},
  year = {2005},
  volume = {71},
  issn = {1094-1622},
  url = {http://dx.doi.org/10.1103/PhysRevA.71.052330},
  doi = {10.1103/physreva.71.052330},
  number = {5},
  publisher = {American Physical Society (APS)},
  month = May,
}

@article{Berry2019CholeskyQubitization,
  author = {Dominic W. Berry and Craig Gidney and Mario Motta and Jarrod R. McClean and Ryan Babbush},
  journal = {arXiv preprint arXiv:1902.02134},
  month = {2},
  title = {Qubitization of Arbitrary Basis Quantum Chemistry Leveraging Sparsity and Low Rank Factorization},
  url = {http://arxiv.org/abs/1902.02134},
  year = {2019},
}

@article{Berry2026FastMultipole,
  author = {Berry, Dominic W. and Wan, Kianna and Baczewski, Andrew D. and Eklund, Elliot C. and Tikku, Arkin and Babbush, Ryan},
  title = {Quantum Simulation of Electronic Structure via Quantum Fast Multipole Method},
  journal = {PRX Quantum},
  year = {2026},
  volume = {7},
  issn = {2691-3399},
  url = {http://dx.doi.org/10.1103/b2vn-zf6c},
  doi = {10.1103/b2vn-zf6c},
  number = {3},
  publisher = {American Physical Society (APS)},
  month = Aug,
}

@article{BinandehDehaghani2025OptimalControl,
  author = {Dehaghani, Nahid Binandeh and Wisniewski, Rafal and Aguiar, A. Pedro},
  title = {Quantum-Assisted Barrier Sequential Quadratic Programming for Nonlinear Optimal Control},
  year = {2025},
  doi = {10.1109/LCSYS.2025.3648321},
  url = {https://doi.org/10.1109/LCSYS.2025.3648321},
  copyright = {Creative Commons Attribution 4.0 International},
  journal = {IEEE Control Systems Letters},
  volume = {9},
  pages = {3005--3010},
  eprint = {2510.18116},
  archivePrefix = {arXiv},
}

@article{Bravyi2011SchriefferWolff,
  author = {Bravyi, Sergey and DiVincenzo, David P. and Loss, Daniel},
  title = {Schrieffer–Wolff transformation for quantum many-body systems},
  journal = {Annals of Physics},
  year = {2011},
  volume = {326},
  issn = {0003-4916},
  url = {http://dx.doi.org/10.1016/j.aop.2011.06.004},
  doi = {10.1016/j.aop.2011.06.004},
  number = {10},
  publisher = {Elsevier BV},
  month = Oct,
  pages = {2793–2826},
}

@article{Cuccaro2004Adder,
  author = {Steven A Cuccaro and Thomas G Draper and Samuel A Kutin and David Petrie Moulton},
  journal = {arXiv preprint quant-ph/0410184},
  title = {A new quantum ripple-carry addition circuit},
  url = {https://arxiv.org/abs/quant-ph/0410184},
  year = {2004},
}

@inproceedings{Dehaghani2025QuantumLQG,
  author = {Dehaghani, Nahid Binandeh and Wisniewski, Rafal and Aguiar, A. Pedro},
  title = {Quantum Solution Framework for Finite-Horizon LQG Control via Block Encodings and QSVT},
  booktitle = {2025 IEEE International Conference on Quantum Computing and Engineering (QCE)},
  year = {2025},
  url = {https://doi.org/10.1109/QCE65121.2025.00040},
  doi = {10.1109/QCE65121.2025.00040},
  publisher = {IEEE},
  pages = {282--290},
  volume = {3},
}

@article{Ehle1973PadeApproximations,
  author = {Ehle, Byron L.},
  title = {{$A$}-Stable Methods and {Pad\'e} Approximations to the Exponential},
  journal = {SIAM Journal on Mathematical Analysis},
  year = {1973},
  volume = {4},
  issn = {1095-7154},
  url = {http://dx.doi.org/10.1137/0504057},
  doi = {10.1137/0504057},
  number = {4},
  publisher = {Society for Industrial & Applied Mathematics (SIAM)},
  month = Nov,
  pages = {671–680},
}

@article{Feshbach1958NuclearReactions,
  author = {Feshbach, Herman},
  title = {Unified theory of nuclear reactions},
  journal = {Annals of Physics},
  year = {1958},
  volume = {5},
  issn = {0003-4916},
  url = {http://dx.doi.org/10.1016/0003-4916(58)90007-1},
  doi = {10.1016/0003-4916(58)90007-1},
  number = {4},
  publisher = {Elsevier BV},
  month = Dec,
  pages = {357–390},
}

@article{Frame2018EigenvectorContinuation,
  author = {Frame, Dillon and He, Rongzheng and Ipsen, Ilse and Lee, Daniel and Lee, Dean and Rrapaj, Ermal},
  title = {Eigenvector Continuation with Subspace Learning},
  journal = {Physical Review Letters},
  year = {2018},
  volume = {121},
  issn = {1079-7114},
  url = {http://dx.doi.org/10.1103/PhysRevLett.121.032501},
  doi = {10.1103/physrevlett.121.032501},
  number = {3},
  publisher = {American Physical Society (APS)},
  month = July,
}

@book{Garcia2017BlaschkeProducts,
  author = {Garcia, Stephan Ramon and Mashreghi, Javad and Ross, William T.},
  title = {Finite Blaschke products: a survey},
  publisher = {arXiv},
  year = {2015},
  journal = {arXiv preprint 1512.05444},
  doi = {10.48550/ARXIV.1512.05444},
  url = {https://arxiv.org/abs/1512.05444},
  copyright = {arXiv.org perpetual, non-exclusive license},
}

@article{Geremia2003QuantumMagnetometry,
  author = {Geremia, JM and Stockton, John K. and Doherty, Andrew C. and Mabuchi, Hideo},
  title = {Quantum Kalman Filtering and the Heisenberg Limit in Atomic Magnetometry},
  journal = {Physical Review Letters},
  year = {2003},
  volume = {91},
  issn = {1079-7114},
  url = {http://dx.doi.org/10.1103/PhysRevLett.91.250801},
  doi = {10.1103/physrevlett.91.250801},
  number = {25},
  publisher = {American Physical Society (APS)},
  month = Dec,
}

@book{Hardy1949DivergentSeries,
  author = {Hardy, G. H.},
  title = {Divergent Series},
  year = {1949},
  publisher = {Clarendon Press},
  address = {Oxford},
  isbn = {0198533008},
}

@misc{Draper2000QuantumAddition,
  author = {Draper, Thomas G.},
  title = {Addition on a Quantum Computer},
  year = {2000},
  doi = {10.48550/ARXIV.QUANT-PH/0008033},
  url = {https://arxiv.org/abs/quant-ph/0008033},
  publisher = {arXiv},
  copyright = {Assumed arXiv.org perpetual, non-exclusive license to distribute this article for submissions made before January 2004},
}

@article{Alpay2014RationalUnitaryFunctions,
  author = {Alpay, Daniel and Jorgensen, Palle and Lewkowicz, Izchak},
  title = {Characterizations of rectangular (para)-unitary rational functions},
  year = {2016},
  issn = {1232-9274},
  url = {https://doi.org/10.7494/OpMath.2016.36.6.695},
  doi = {10.7494/OpMath.2016.36.6.695},
  number = {6},
  journal = {Opuscula Mathematica},
  volume = {36},
  pages = {695--716},
}

@article{Bauman2019ExcitedStateDownfolding,
 author={Bauman, Nicholas P. and Low, Guang Hao and Kowalski, Karol},
 title={Quantum simulations of excited states with active-space downfolded Hamiltonians},
 journal={The Journal of Chemical Physics}, volume={151}, pages={234114},
 year={2019}, doi={10.1063/1.5128103},
 eprint={1909.06404}, archivePrefix={arXiv},
 url={https://arxiv.org/abs/1909.06404}
}

@article{Eremenko2007SignApproximation,
 author={Eremenko, Alexandre and Yuditskii, Peter},
 title={Uniform approximation of {sgn} $x$ by polynomials and entire functions},
 journal={Journal d'Analyse Math{\'e}matique},
 volume={101}, pages={313--324}, year={2007},
 doi={10.1007/s11854-007-0011-3},
 note={Theorem 1. \href{https://arxiv.org/abs/math/0604324}{arXiv:math/0604324}}
}

@article{Jennings2024CostLinearDE,
 author={David Jennings and Matteo Lostaglio and Robert B. Lowrie and Sam Pallister and Andrew T. Sornborger},
 title={The cost of solving linear differential equations on a quantum computer: fast-forwarding to explicit resource counts},
 journal={Quantum}, volume={8}, pages={1553}, year={2024},
 doi={10.22331/q-2024-12-10-1553}, eprint={2309.07881}, archivePrefix={arXiv},
 url={https://quantum-journal.org/papers/q-2024-12-10-1553/}
}

@misc{Ameri2026NonMarkovian,
 author={Abtin Ameri and Arkopal Dutt and Hari Krovi},
 title={Quantum simulation of non-{Markovian} dynamical systems},
 year={2026}, eprint={2608.13533}, archivePrefix={arXiv},
 doi={10.48550/arXiv.2608.13533}, url={https://arxiv.org/abs/2608.13533}
}

@article{Berry2018FermionicEigenstates,
 author={Dominic W. Berry and M{\'a}ria Kieferov{\'a} and Artur Scherer and Yuval R. Sanders and Guang Hao Low and Nathan Wiebe and Craig Gidney and Ryan Babbush},
 title={Improved techniques for preparing eigenstates of fermionic {Hamiltonians}},
 journal={npj Quantum Information}, volume={4}, pages={22}, year={2018},
 doi={10.1038/s41534-018-0071-5}, eprint={1711.10460}, archivePrefix={arXiv}
}

@article{Low2024TradingTGates,
 author={Guang Hao Low and Vadym Kliuchnikov and Luke Schaeffer},
 title={Trading {T} gates for dirty qubits in state preparation and unitary synthesis},
 journal={Quantum}, volume={8}, pages={1375}, year={2024},
 doi={10.22331/q-2024-06-17-1375}, eprint={1812.00954}, archivePrefix={arXiv}
}

@article{Low2023TrotterSteps,
 author={Guang Hao Low and Yuan Su and Yu Tong and Minh C. Tran},
 title={Complexity of Implementing {Trotter} Steps},
 journal={PRX Quantum}, volume={4}, pages={020323}, year={2023},
 doi={10.1103/PRXQuantum.4.020323}, eprint={2211.09133}, archivePrefix={arXiv}
}

@article{Rubin2026LowerBoundCertificates,
 author={Nicholas C. Rubin and Guang Hao Low and DePrince, III, A. Eugene},
 title={Near-Frustration-Free Electronic Structure {Hamiltonian} Representations and Lower-Bound Certificates},
 journal={Journal of Chemical Theory and Computation}, volume={22}, number={17},
 pages={8751--8766}, year={2026}, doi={10.1021/acs.jctc.6c00318},
 eprint={2602.05069}, archivePrefix={arXiv}
}

@misc{Dutkiewicz2026FirstQuantizedAmplification,
 author={Alicja Dutkiewicz and Alec F. White and Guang Hao Low and DePrince, III, A. Eugene and Matthew P. Harrigan and Marika Kieferova and Ryan Babbush and Dominic W. Berry and Nicholas C. Rubin},
 title={Spectral amplification for ground-state energy estimation of electronic structure in first quantization},
 year={2026}, eprint={2607.15358}, archivePrefix={arXiv},
 doi={10.48550/arXiv.2607.15358}, url={https://arxiv.org/abs/2607.15358}
}

@article{Low2026DenserSurfaceCode,
  author = {Low, Guang Hao and Huggins, William J. and Berry, Dominic W. and Khattar, Tanuj and White, Alec F. and Rubin, Nicholas C. and Babbush, Ryan},
  title = {A Denser Planar Surface Code},
  journal = {arXiv preprint arXiv:2605.30455},
  year = {2026},
  eprint = {2605.30455},
  archivePrefix = {arXiv},
  primaryClass = {quant-ph},
  doi = {10.48550/arXiv.2605.30455}
}
\end{document}